\documentclass[prb,aps,nofootinbib,amssymb,twocolumn,superscriptaddress,10pt,floatfix]{revtex4-2}
\usepackage[T1]{fontenc}
\usepackage{graphicx}
\usepackage{xcolor}
\usepackage{dcolumn}
\usepackage{comment}
\usepackage{amsmath,amssymb,bbold,bm}
\usepackage{hyperref}
\usepackage{amsfonts}
\usepackage{float,latexsym}
\usepackage{multirow}
\usepackage{color}
\usepackage[normalem]{ulem}
\usepackage{cleveref}
\usepackage{physics}
\usepackage{tabularx}
\usepackage{array}
\usepackage[caption=false]{subfig}
\usepackage{siunitx}
\usepackage{gensymb}
\usepackage{pifont}
\usepackage{placeins}
\usepackage{mathtools}
\usepackage{longtable}
\usepackage{multirow}
\usepackage{nicematrix}
\usepackage{tikz}
\usepackage{xstring}
\usepackage{makecell}
\usepackage{xr}
\usepackage{ifthen}
\usepackage{combelow}
\usetikzlibrary{decorations.markings}
\usetikzlibrary{positioning}
\usetikzlibrary{arrows.meta}
\usepackage{tikz-feynman}
\usetikzlibrary{calc}
\tikzfeynmanset{warn luatex=false}
\let\vec\mathbf

\usepackage{chemformula}
\usepackage{centernot}

\makeatletter
\def\maketag@@@#1{\hbox{\m@th\normalfont\normalsize#1}}
\makeatother

\crefname{appendix}{Appendix}{Appendices}
\crefname{equation}{Eq.}{Eqs.}
\crefname{figure}{Fig.}{Figs.}
\crefname{table}{Table}{Tables}
\crefname{section}{Section}{Sections}
\crefname{enumi}{Point}{Points}
\creflabelformat{appendix}{[#2#1#3]}
\crefmultiformat{figure}{Figs.~#2#1\xdef\mycreffirstarg{#1}#3}%
 { and~#2#1#3}{, #2#1#3}{ and~#2#1#3}

\makeatletter     
\renewcommand\onecolumngrid{
\do@columngrid{one}{\@ne}%
\def\set@footnotewidth{\onecolumngrid}
\def\footnoterule{\kern-6pt\hrule width 1.5in\kern6pt}%
}

\makeatletter
\AtBeginDocument{\let\LS@rot\@undefined}
\makeatother 
\renewcommand{\arraystretch}{1.2}

\crefname{appendix}{Appendix}{Appendices}
\crefname{equation}{Eq.}{Eqs.}
\crefname{figure}{Fig.}{Figs.}
\crefname{table}{Table}{Tables}
\crefname{section}{Section}{Sections}
\creflabelformat{appendix}{[#2#1#3]}

\makeatletter
\AtBeginDocument{\let\LS@rot\@undefined}
\makeatother

\makeatletter
\renewcommand\onecolumngrid{\do@columngrid{one}{\@ne}\def\set@footnotewidth{\onecolumngrid}\def\footnoterule{\kern-6pt\hrule width 1.5in\kern6pt}}

\newcommand{\siSection}{appendix}

\allowdisplaybreaks

\begin{document}
\title{A New Moir\'e Platform Based on M-Point Twisting}
\author{Dumitru C\u{a}lug\u{a}ru}
   	\thanks{These authors contributed equally to this work.}
   	\affiliation{Department of Physics, Princeton University, Princeton, New Jersey 08544, USA}
   	\author{Yi Jiang}
   	\thanks{These authors contributed equally to this work.}
   	\affiliation{Donostia International Physics Center (DIPC), Paseo Manuel de Lardizábal. 20018, San Sebastián, Spain}
   	\author{Haoyu Hu}
   	\thanks{These authors contributed equally to this work.}
   	\affiliation{Donostia International Physics Center (DIPC), Paseo Manuel de Lardizábal. 20018, San Sebastián, Spain}
   	\author{Hanqi Pi}
   	\thanks{These authors contributed equally to this work.}
   	\affiliation{Donostia International Physics Center (DIPC), Paseo Manuel de Lardizábal. 20018, San Sebastián, Spain}
   	\affiliation{Beijing National Laboratory for Condensed Matter Physics, and Institute of Physics, Chinese Academy of Sciences, Beijing 100190, China}
   	\author{Jiabin Yu}
   	\affiliation{Department of Physics, University of Florida, Gainesville, FL, USA}
   	\affiliation{Department of Physics, Princeton University, Princeton, New Jersey 08544, USA}
   	\author{Maia G.~Vergniory}
    \affiliation{Donostia International Physics Center (DIPC), Paseo Manuel de Lardizábal. 20018, San Sebastián, Spain}
    \author{Jie Shan}
    \affiliation{School of Applied and Engineering Physics, Cornell University, Ithaca, NY 14850, USA}
    \affiliation{Department of Physics, Cornell University, Ithaca, NY 14850, USA}
    \affiliation{Kavli Institute at Cornell for Nanoscale Science, Ithaca, NY 14850, USA}
    \author{Claudia Felser}
    \affiliation{Max Planck Institute for Chemical Physics of Solids, N\"{o}thnitzer Str. 40, Dresden 01187, Germany}
    \author{Leslie M.~Schoop}
    \affiliation{Department of Chemistry, Princeton University, Princeton, NJ 08540}
    \author{Dmitri K.~Efetov}
    \affiliation{Faculty of Physics, Ludwig-Maximilians-University Munich, Munich 80799, Germany}
	\affiliation{Munich Center for Quantum Science and Technology (MCQST), Ludwig-Maximilians-University Munich, Munich 80799, Germany}
    \author{Kin Fai Mak}
    \affiliation{School of Applied and Engineering Physics, Cornell University, Ithaca, NY 14850, USA}
    \affiliation{Department of Physics, Cornell University, Ithaca, NY 14850, USA}
    \affiliation{Kavli Institute at Cornell for Nanoscale Science, Ithaca, NY 14850, USA}
    \author{B.~Andrei Bernevig}
	\email{bernevig@princeton.edu}
	\affiliation{Department of Physics, Princeton University, Princeton, New Jersey 08544, USA}
	\affiliation{Donostia International Physics Center, P. Manuel de Lardizabal 4, 20018 Donostia-San Sebastian, Spain}
	\affiliation{IKERBASQUE, Basque Foundation for Science, Bilbao, Spain}

\let\oldaddcontentsline\addcontentsline
\renewcommand{\addcontentsline}[3]{}

\begin{abstract}
	We introduce a new class of moir\'e systems and materials based on monolayers with triangular lattices and low-energy states at the M points of the Brillouin zone. These M-point moir\'e materials are fundamentally distinct from those derived from $\Gamma$- or K-point monolayers, featuring three time-reversal-preserving valleys related by three-fold rotational symmetry. We propose twisted bilayers of experimentally exfoliable 1T-\ch{SnSe2} and 1T-\ch{ZrS2} as realizations of this new class. Using extensive \textit{ab initio} simulations, we develop quantitative continuum models and analytically show that the corresponding M-point moir\'e Hamiltonians exhibit emergent momentum-space non-symmorphic symmetries and a kagome plane-wave lattice in momentum space. This represents the first experimentally viable realization of a projective representation of crystalline space groups in a non-magnetic system. With interactions, these materials represent six-flavor Hubbard simulators with Mott physics, as can be seen by their flat Wilson loops. Furthermore, the presence of a non-symmorphic momentum-space in-plane mirror symmetry makes some of the M-point moir\'e Hamiltonians quasi-one-dimensional in each valley, suggesting the possibility of realizing Luttinger liquid physics. We predict the twist angles at which a series of (conduction) flat bands appear, provide a faithful continuum Hamiltonian, analyze its topology and charge density and briefly discuss several aspects of the physics of this new platform. 
\end{abstract}

\maketitle

\textit{Introduction.}~Moir\'e heterostructures have recently emerged as versatile quantum simulators of archetypal condensed matter models~\cite{AND21,KEN21}. When two identical or nearly identical monolayers are twisted, the resulting moir\'e modulation of the interlayer potential gives rise to an effective \emph{moir\'e} discrete translation symmetry. In the moir\'e Brillouin zone (BZ), the moir\'e-modulated interlayer hybridization opens gaps in the folded band structure quenching the monolayer electrons' kinetic energy~\cite{CAR17}. The moir\'e system thus enters an interaction-dominated regime, providing for a tunable platform for simulating various prototypical condensed matter systems. A notable example is twisted bilayer graphene (TBG)~\cite{BIS11}, which hosts unconventional superconductors~\cite{CAO18a} and correlated insulators~\cite{CAO18} near the magic angle, and has recently been shown to simulate topological heavy-fermions~\cite{SON22,CHO23}. Transition metal dichalcogenide (TMD) heterobilayers can emulate the Hubbard model on a triangular lattice~\cite{WU18c,TAN20}, while twisted \ch{WTe2} exhibits signatures of a one-dimensional Luttinger liquid, though its theoretical description remains challenging due to the complex monolayer band structure~\cite{WAN22}. Beyond these examples, a growing body of theoretical and experimental work has explored other exotic phases in TMDs~\cite{XU20a,AND23,ZHA23c,GUO24,XIA24a,SHE24}. Furthermore, both integer and fractional Chern insulator states have been reported in moir\'e TMD~\cite{LI21d,XU23,PAR23,ZEN23,CAI23,WAN24a,JIA24,YU24a}, graphene~\cite{SHA19,SER20,CHE20b}, and graphene-boron nitride heterostructures~\cite{LU24,DON24,DON23,HER24b}.

Until now, nearly all moir\'e heterostructures have been based on twisting monolayers with triangular lattices and low-energy states near the $\Gamma$~\cite{ANG21,CLA22a} or K~\cite{BIS11,WU19b,WU18c,DEV21,ZHA24a} points, leading to systems with one or two valleys (in the two-valley case, time-reversals exchanges the valley). This work introduces a novel family of moir\'e materials by twisting monolayers with triangular lattices and low-energy states around the M point of the BZ. These M-point moir\'e systems feature \emph{three} time reversal-preserving valleys related by $C_{3z}$ rotation symmetry. Building on extensive \textit{ab-initio} calculations, we propose (among others~\cite{JIA24b}) experimentally-exfoliable twisted \ch{SnSe2}~\cite{BUS61,SU13,ZEN18,WU19c,HUA20b} and \ch{ZrS2}~\cite{LV16,MAN16,ALS23} as promising platforms for realizing M-point moir\'e heterostructures. We develop \emph{quantitative} simplified models for these systems and perform a detailed analysis of the band structure, topology, and charge density of the flat bands at the predicted small twist angles. We show analytically that M-point moir\'e Hamiltonians exhibit a new type of symmetry, termed momentum-space non-symmorphic~\cite{DE21,CHE22,ZHA23b,XIA24}. In crystallography, space groups are symmorphic or non-symmorphic depending on whether they include symmetry operations that translate the origin by a fraction of the lattice vectors. While in real-space, conventional crystalline groups can feature both symmorphic and non-symmorphic operations, in momentum space, all conventional crystalline groups exhibit only symmorphic operations. M-point moir\'e systems are the first experimentally realizable non-magnetic systems to exhibit momentum-space non-symmorphic symmetries, all without requiring an applied magnetic field in the range of thousands of Tesla~\cite{DE21,CHE22,ZHA23b}. In a single valley, these non-symmorphic symmetries can render the system effectively one-dimensional at the single-particle level, making M-point moir\'e systems prime candidates for Luttinger liquid simulators~\cite{KEN20,WAN22}. With all three valleys considered, they can realize a multi-orbital triangular lattice Hubbard model~\cite{HU24}, where valley-spin local moments couple differently along the three $C_{3z}$-related directions, in a manner reminiscent of Kitaev's honeycomb model~\cite{KIT06}.

\begin{figure}[!t]
	\centering
	\includegraphics[width=\columnwidth]{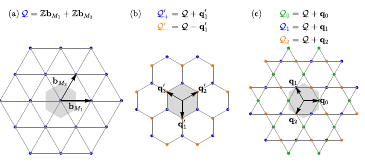}\subfloat{\label{fig:q_lattices:a}}\subfloat{\label{fig:q_lattices:b}}\subfloat{\label{fig:q_lattices:c}}\caption{Momentum-space $\vec{Q}$-lattices for twisted triangular lattice monolayers. The three panels correspond to the cases where the low-energy degrees of freedom are located at the $\Gamma$ (a), K (b), and M (c) points. In each panel, the sublattices are colored according to the legend displayed above each plot. The moir\'e BZ is shown by the gray hexagon, while the reciprocal moir\'e vectors $\vec{b}_{M_{1,2}}$, as well as the auxiliary vectors $\vec{q}_{1,2,3}$ and $\vec{q}'_{0,1,2}$ are shown by the black arrows.}
	\label{fig:q_lattices}\end{figure}

\textit{M-moir\'e models.}~For triangular monolayer lattices, the moir\'e lattice is also triangular, generated by the reciprocal lattice vectors $\vec{b}_{M_1}$ and $\vec{b}_{M_2}$ (see \cref{app:sec:SnS_SnSe_twist}). These vectors span the moir\'e reciprocal lattice $\mathcal{Q} = \mathbb{Z} \vec{b}_{M_1} + \mathbb{Z} \vec{b}_{M_2}$, as depicted in \cref{fig:q_lattices:a}. In general, the single-particle Hamiltonians of moir\'e systems take the form of a hopping model in momentum space. This arises because the moir\'e potential breaks the monolayer translation symmetry and couples momentum states that are connected by reciprocal moir\'e vectors. The single-particle moir\'e Hamiltonian typically takes the form $\mathcal{H} = \sum_{\vec{k},\vec{Q},\vec{Q}',i,j}\left[ h_{\vec{Q},\vec{Q}'} \left( \vec{k} \right) \right]_{ij} \hat{c}^\dagger_{\vec{k},\vec{Q},i} \hat{c}_{\vec{k},\vec{Q}',j}$, where $\hat{c}^\dagger_{\vec{k},\vec{Q},i}$ denote the moir\'e plane wave operators at moir\'e momentum $\vec{k}$, and $i$ denotes a combined index comprising orbital, spin, valley, layer, or other additional degrees of freedom. 

When the low-energy fermions of the monolayer are located at the $\Gamma$ point~\cite{ANG21,CLA22a}, the operators $\hat{c}^\dagger_{\vec{k},\vec{Q},i}$ carry total momentum $\vec{k} - \vec{Q}$, and the $\vec{Q}$-vectors lie on the triangular lattice shown in \cref{fig:q_lattices:a}. In the case of a monolayer with low-energy states located at the K point~\cite{BIS11,WU19b,WU18c}, the moir\'e fermions carry an additional valley index $\eta = \pm$, in which the moir\'e Hamiltonian is diagonal. The $\vec{Q}$-vectors form a honeycomb lattice, as illustrated in \cref{fig:q_lattices:b}. The moir\'e and monolayer operators are related by $\hat{c}^\dagger_{\vec{k},\vec{Q},\eta,i} = \hat{a}^\dagger_{\eta \vec{K}^l_{K} + \vec{k} - \vec{Q},l,i}$ for $\vec{Q} \in \mathcal{Q}'_{\eta l}$, where $\hat{a}^\dagger_{\vec{p},l,i}$ represents the monolayer operators from layer $l=\pm$ at momentum $\vec{p}$, and $\vec{K}^l_{K}$ is the K-point momentum of layer $l$. 

Distinctly, in M-point moir\'e materials, the $\vec{Q}$-vectors form a \emph{kagome} lattice, as shown in \cref{fig:q_lattices:c}. 
To be specific, the moir\'e operators in layer $l$ -- which, for the present case, include only an extra  spin $s = \uparrow, \downarrow$ index -- are related to the monolayer ones according to $\hat{c}^\dagger_{\vec{k},\vec{Q},s,l} = \hat{c}^\dagger_{C^{\eta}_{3z} \vec{K}^{l}_{M} + \vec{k} - \vec{Q},s,l}$, for $\vec{Q} \in \mathcal{Q}_{\eta + l}$, where $\vec{K}^{l}_{M}$ is the momentum of the monolayer M-point. The three $C_{3z}$-related valleys indexed by $\eta = 0, 1, 2$ are implicitly encoded by the kagome sublattice to which $\vec{Q}$ belongs: the valley-$\eta$ fermions are supported on the $\mathcal{Q}_{\eta \pm 1}$ sublattices (where $\eta+ l$ is taken modulo 3), as derived in \cref{app:sec:SnS_SnSe_twist}. As we will show, the kagome $\vec{Q}$-lattice leads to significantly different properties of M-moir\'e materials.  

\begin{figure}[!t]
	\centering
	\includegraphics[width=\columnwidth]{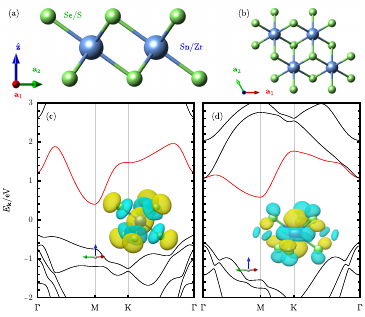}\subfloat{\label{fig:materials:a}}\subfloat{\label{fig:materials:b}}\subfloat{\label{fig:materials:c}}\subfloat{\label{fig:materials:d}}\caption{Exfoliable monolayers for M-moir\'e devices. (a) and (b) show the side and top views of the crystal structures of 1T-\ch{SnSe2} and 1T-\ch{ZrS2}. The \textit{ab initio} band structures for \ch{SnSe2} and \ch{ZrS2} are shown in (c) and (d). The lowest spinful conduction band, with minima at the M points, is highlighted in red, while the Wannier orbitals contributing to the low-energy states are shown as insets. The yellow (blue) colors correspond to the positive (negative) sign of the orbitals.}
	\label{fig:materials}\end{figure}

\textit{Materials realizations.}~We now turn to 1T-\ch{SnSe2} and 1T-\ch{ZrS2} as experimentally-exfoliable monolayers for realizing M-moir\'e heterostructures (see \cref{app:sec:DFT_single_layer}; more materials will be presented in Ref.~\cite{JIA24b}). The monolayer crystal structure of both materials is shown in \cref{fig:materials:a,fig:materials:b} and belongs to the $P\bar{3}m11'$ group, which is generated by translations, $C_{3z}$ rotations, in-plane two-fold rotations $C_{2x}$, inversion $\mathcal{I}$, and time reversal $\mathcal{T}$ symmetries. The Sn (Zr) atoms form a triangular lattice with the Se (S) atoms being located at the other $C_{3z}$-invariant Wyckoff positions above and below the Sn (Zr) plane. The \textit{ab-intio} band structures of monolayer \ch{SnSe2} and \ch{ZrS2} shown in \cref{fig:materials:c,fig:materials:d} reveal two insulators for which the conduction band minimum is located at the M-point. The first isolated Kramers-degenerate conduction band of \ch{SnSe2} is atomic, being spanned by an effective $s$-like molecular orbital centered on the Sn atom. For \ch{ZrS2}, the low-energy M-point states are contributed primarily by the $d_{z^2}$ orbitals of Zr. We also note that 1T-\ch{ZrSe2}~\cite{MAN16,ALS23}, 1T-\ch{SnS2}~\cite{XIA14,HUA22}, 1T-\ch{HfSe2}~\cite{KAN15}, and 1T-\ch{HfS2}~\cite{ZHA16,WU17,SIN19}, with a similar structure to 1T-\ch{SnSe2} and 1T-\ch{ZrS2}, as well as \ch{GaTe}~\cite{ZHA16a} with a different structure, can also serve as experimentally-exfoliable monolayers for M-point moir\'e systems~\cite{JIA24b}.

\begin{figure*}[!t]
	\centering
	\includegraphics[width=\textwidth]{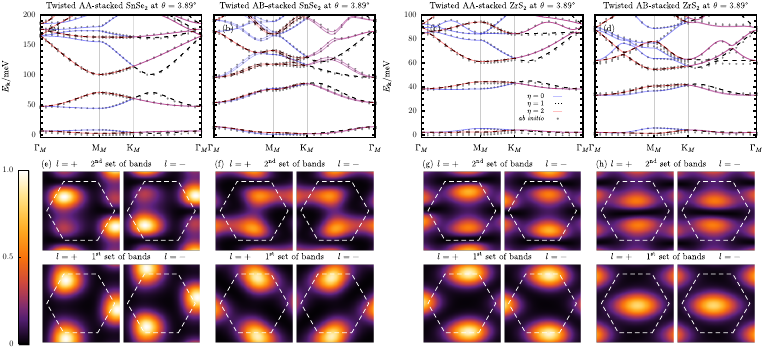}\subfloat{\label{fig:moire:a}}\subfloat{\label{fig:moire:b}}\subfloat{\label{fig:moire:c}}\subfloat{\label{fig:moire:d}}\subfloat{\label{fig:moire:e}}\subfloat{\label{fig:moire:f}}\subfloat{\label{fig:moire:g}}\subfloat{\label{fig:moire:h}}\caption{\textit{Ab initio} results for M-point moir\'e \ch{SnSe2} and \ch{ZrS2}. (a-d) present the band structures for AA-stacked (a, c) and AB-stacked (b, d) twisted \ch{SnSe2} and \ch{ZrS2} at the commensurate angle $\theta = \SI{3.89}{\degree}$. Both the \textit{ab initio} and valley-resolved continuum model band structures are shown. Below each, (e)–(h) depict the layer-resolved local density of states corresponding to the first and second sets of spinful bands in valley $\eta = 0$. The Wigner-Seitz unit cell is indicated by the dashed hexagon.}
	\label{fig:moire}\end{figure*}
\textit{Moir\'e Hamiltonians.}~Since the \ch{SnSe2} and \ch{ZrS2} monolayers lack two-fold out-of-plane rotation symmetry ($C_{2z}$), there are two distinct ways to stack and subsequently twist them by an angle $\theta$ in order to achieve a large-scale moir\'e periodicity. In the so-called AA-stacking configuration, the top ($l = +1$) and bottom ($l = -1$) layers are stacked directly on top of each other and then twisted by the layer-dependent angle $\frac{l\theta}{2}$. In contrast, for AB-stacking, the bottom layer is first rotated by \SI{180}{\degree} around the $\hat{\vec{z}}$ axis, before applying the $\frac{l\theta}{2}$ twist. As discussed in \cref{app:sec:DFT_bilayer}, the two configuration have different crystalline symmetries. While both stackings feature $C_{3z}$ and $\mathcal{T}$ symmetries, they differ in the direction of the in-plane two-fold rotation symmetry: the AA- (AB-)stacking arrangement has $C_{2x}$ ($C_{2y}$) symmetry.

We perform large-scale \textit{ab initio} calculations (which include relaxation effects) at commensurate twist angles ranging from $\SI{13.17}{\degree} \geq \theta \geq \SI{3.89}{\degree}$ (see \cref{app:sec:DFT_bilayer,app:sec:first_princ_ham_valley}), and construct two types of moir\'e Hamiltonian models for each angle and stacking configuration according to the method outlined in \cref{app:sec:fitting_method}. The first is a \emph{numerically exact} model, which accurately reproduces a large set (at least the first five sets per valley) of spinful bands in both energy and wave function. The second is an \emph{analytical} approximate continuum model capturing the dispersion and wave function of the first or first two (depending on the angle) lowest-energy spinful gapped bands (and qualitatively the higher energy spectrum) in each valley. The comprehensive results at all angles are presented in \cref{app:sec:fitted_models}. Unlike the case of $\Gamma$- or K-point twisting, \textit{ab initio} simulations are \emph{crucial} for obtaining even the correct \emph{qualitative} moir\'e Hamiltonian. The \emph{two-center first monolayer harmonic approximation} incorrectly predicts continuous translation symmetry along one direction ({\it e.g.}{}, along the $C^{\eta}_{3z} \hat{\vec{y}}$ direction in valley $\eta$) and an overall gapless spectrum, as shown in \cref{app:sec:add_sym} and in some  toy models~\cite{FUJ22}.

\Cref{fig:moire} summarizes the \textit{ab initio} results for twisted AA- and AB-stacked \ch{SnSe2} and \ch{ZrS2} at low twist angle. Both stacking configurations exhibit approximate spin $\mathrm{SU} \left( {2} \right)$ symmetry (see \cref{app:sec:fitting_method}) and feature two spinful sets of gapped bands in each of the three $C_{3z}$-related valleys, as shown in \crefrange{fig:moire:a}{fig:moire:d}. The lowest-energy set of bands has a narrow bandwidth of around $\SI{10}{\milli\electronvolt}$. Examining the local density of states (LDOS) for the lowest two bands in valley $\eta = 0$, shown in \crefrange{fig:moire:e}{fig:moire:h}, reveals that these moir\'e systems possess approximate spatial symmetries beyond the exact valley-preserving $C_{2x}$ and $C_{2y}$ symmetries expected in the AA- and AB-stacked configurations, respectively. For instance, the LDOS of the first set of spinful bands in AA-stacked \ch{SnSe2}, as well as the first two sets of bands in twisted \ch{ZrS2} feature an \emph{approximate} two-fold rotation symmetry (the second set of spinful bands in AA-stacked \ch{SnSe2} exhibits this symmetry to lesser extent). In the AB-stacked configuration, the center of the approximate $C_{2z}$ symmetry aligns with the unit cell origin, while in the AA-stacked case, the effective $\tilde{C}_{2z}$ rotation center is shifted \emph{away} from the unit cell origin and will be specified below. Moreover, the LDOS suggests the presence of an approximate in-plane mirror symmetry, $\tilde{M}_{z}$. These effective symmetries (whose origin is explained below and in \cref{app:sec:SnS_SnSe_twist_general_woGradient:approx}) prompt us to construct simplified analytical continuum models that can capture and explain these features.

\begin{figure}[!t]
	\centering
	\includegraphics[width=\columnwidth]{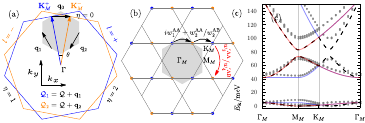}\subfloat{\label{fig:simple_model:a}}\subfloat{\label{fig:simple_model:b}}\subfloat{\label{fig:simple_model:c}}\caption{Analytical continuum M-point moir\'e models. (a) illustrates the relationship between the monolayer and moir\'e BZs, with the colored and gray hexagons representing the respective Brillouin zones. (b) details the generation of the $\vec{Q}$-lattice for the $\eta = 0$ valley and shows the hopping terms of the moir\'e potential matrix $T_{\vec{Q},\vec{Q}'}$. The band structure of the simplified moir\'e model for AA-stacked \ch{SnSe2} at $\theta = \SI{3.89}{\degree}$ is also shown. The color scheme matches that of \cref{fig:moire}.}
	\label{fig:simple_model}\end{figure}

In valley $\eta = 0$, the simplified M-point moir\'e Hamiltonian can be expressed as
{\small \begin{align}
	\left[ h_{\vec{Q},\vec{Q}'} \left( \vec{k} \right) \right]_{s l; s' l'} &=  \delta_{\vec{Q}, \vec{Q}'} \delta_{s s'} \delta_{l l'} \left[ \frac{\left( k_{x} - Q_{x} \right)^2}{2 m_x} + \frac{\left( k_{y} - Q_{y} \right)^2}{2 m_y} \right] \nonumber \\
	&+ \left[ T_{\vec{Q},\vec{Q}'} \right]_{s l; s' l'}, \qq{for} \vec{Q}^{(\prime)} \in \mathcal{Q}_{l^{(\prime)}}, \label{eqn:single_particle_hamiltonian}
\end{align}}where $m_x$ and $m_y$ are the anisotropic effective masses of \ch{SnSe2} and \ch{ZrS2}, with numerical values provided in \cref{tab:params}. As shown in \cref{fig:simple_model:a,fig:simple_model:b}, the moir\'e potential takes the form of a hopping model on two of the three sublattices of the kagome M-point $\vec{Q}$-lattice. Explicitly, the simplified Hermitian moir\'e potential tensor exhibits spin $\mathrm{SU} \left( {2} \right)$ symmetry and includes only interlayer terms, given by $\left[ T^{\text{AA}}_{\vec{Q},\vec{Q}'} \right]_{ls;(-l)s} = \left( \pm i w^{\text{AA}}_1 + w^{\text{AA}}_2 \right) \delta_{\vec{Q} \pm \vec{q}_{0}, \vec{Q}'} + w^{\prime \text{AA}}_3 \delta_{\vec{Q} \pm \left( \vec{q}_1 - \vec{q}_2 \right), \vec{Q}'}$ and $\left[ T^{\text{AB}}_{\vec{Q},\vec{Q}'} \right]_{ls;(-l)s} = w^{\text{AB}}_2 \delta_{\vec{Q} \pm \vec{q}_{0}, \vec{Q}'} + w^{\prime \text{AB}}_4 \delta_{\vec{Q} \pm \left( \vec{q}_1 - \vec{q}_2 \right), \vec{Q}'}$. The interlayer hopping parameters, obtained by fitting to the \textit{ab-initio} band structure, are listed in \cref{tab:params}. The simplified model's band structure for AA-stacked \ch{SnSe2} is shown in \cref{fig:simple_model:c}, showing excellent qualitative agreement (for such small number of parameters, with the \textit{ab-initio} results). In the simplified models, for both \ch{SnSe2} and \ch{ZrS2}, the overlap between the fitted and \textit{ab-initio} bands is larger than $95\%$ ($85\%$) with the first (second) set of spinful bands, as we show in \cref{app:sec:fitted_models}.

\begin{table}[!t]
	\begin{tabular}{|l|r r|r r r|r r|}
		\hline
		Monolayer & $m_x$ & $m_y$ & $w^{\text{AA}}_1$ & $w^{\text{AA}}_2$ & $w^{\prime \text{AA}}_3$ & $w^{\text{AB}}_2$ & $w^{\text{AB}}_4$ \\ \hline
		\ch{SnSe2} & $0.21$ & $0.73$ & $66.38$ & $88.80$ & $-18.94$ & $-77.80$ & $27.04$ \\ \hline
		\ch{ZrS2} & $0.29$ & $1.86$ & $-12.35$ & $50.50$ & $-19.83$ & $-35.88$ & $-16.88$ \\ \hline
	\end{tabular}
	\caption{Parameters of the simple moir\'e models for twisted \ch{SnSe2} and \ch{ZrS2} at $\theta = \SI{3.89}{\degree}$. The effective masses are given in units of the bare electron mass $m_e$, while the hopping amplitudes are given in units of $\si{\milli\electronvolt}$.}
	\label{tab:params}
\end{table}

\textit{Non-symmorphic momentum-space symmetries.}~The approximate symmetries inferred from the layer-resolved LDOS of the M-point moir\'e Hamiltonian are \emph{exact} symmetries in the simplified moir\'e models from \cref{eqn:single_particle_hamiltonian} (see the detailed discussion in \cref{app:sec:add_sym}). Specifically, the center of the effective two-fold rotation symmetry $\tilde{C}_{2z}$ for the AA-stacked Hamiltonian is located at $\frac{\vec{q}_{\eta}}{\abs{\vec{q}_{\eta}}^2} \arg \left( i w^{\text{AA}}_1 + w^{\text{AA}}_2 \right)$ in valley $\eta$. In contrast, the simplified AB-stacked moir\'e Hamiltonian exhibits $C_{2z}$ symmetry, with its rotation center aligned with the origin of the moir\'e unit cell. Since both models are effectively spinless (due to atomistic arguments presented in \cref{app:sec:fitted_models}) and exhibit either $\tilde{C}_{2z}\mathcal{T}$ or $C_{2z} \mathcal{T}$ symmetry in each valley, the Berry curvature of any gapped set of bands is exactly zero. Consequently, the first two set of bands of both the AA- and AB-stacked moir\'e Hamiltonians are topological trivial and, hence, Wannierizable. This is also consistent (and the result of) the bands being flat and exhibiting a large (\SI{40}{\milli\electronvolt}) gap from one another. However, the physics of these Hubbard (with interaction) bands is far from trivial in this system, as shown below.  

Unlike the $C_{2z}$ and $\tilde{C}_{2z}$ symmetries, the effective mirror $\tilde{M}_z$ symmetry has an unconventional action on the momentum-space moir\'e fermions. Specifically, $\tilde{M}_z$ acts non-symmorphically in momentum space, with $\tilde{M}_z \hat{c}^\dagger_{\vec{k},\vec{Q},s,l} \tilde{M}_z^{-1} = \hat{c}^\dagger_{\vec{k} + \vec{q}_{\eta}, \vec{Q} + \vec{q}_{\eta},s, -l}$, for $\vec{Q} \in \mathcal{Q}_{\eta + l}$. Since $\vec{q}_{0} = \frac{\vec{b}_{M_1}}{2}$, the action of $\tilde{M}_z$ can only be made conventional by folding the moir\'e BZ along $\vec{q}_{\eta}$, which would break the moir\'e translation symmetry. The non-symmorphic action of the $\tilde{M}_z$ symmetry stems from the moir\'e fermions realizing a projective representation of the symmetry group of the system. Letting $T'_{\vec{a}_{M_{1,2}}}$ denote the two moir\'e translation operators for valley $\eta = 0$ along the direct moir\'e lattice vectors $\vec{a}_{M_{1,2}}$ (with $\vec{a}_{M_i} \cdot \vec{b}_{M_j} = 2 \pi \delta_{ij}$), we find that $\commutator{T'_{\vec{a}_{M_{2}}}}{\tilde{M}_z} = \anticommutator{T'_{\vec{a}_{M_{1}}}}{\tilde{M}_z} = 0$.

It is important to note that the effective $\tilde{M}_z$ symmetry is not accidental. In the AA-stacked case, it can be shown to hold \emph{exactly} for arbitrary moir\'e harmonics within the local-stacking approximation~\cite{JUN14}. In the limit of vanishing twist angle ($\theta \to 0$), the moir\'e Hamiltonian can be constrained by the exact symmetries of the \emph{untwisted} bilayer configuration. The inversion symmetry of the untwisted AA-stacked bilayer gives rise to the $\tilde{M}_z$ symmetry of the moir\'e Hamiltonian as shown in  \cref{app:sec:SnS_SnSe_twist_general_woGradient:approx,app:sec:add_sym}. In the AB-stacked case, the true in-plane mirror symmetry of the untwisted bilayer leads to an effective inversion symmetry $\tilde{\mathcal{I}}$ of the corresponding moir\'e Hamiltonian, which also acts non-symmorphically in momentum space. In the simplified AB-stacked model, the approximate $C_{2z}$ symmetry, combined with the $\tilde{\mathcal{I}}$ symmetry, leads to an $\tilde{M}_z = C_{2z} \tilde{\mathcal{I}}$ symmetry of the system. 

\begin{figure}[!t]
	\centering
	\includegraphics[width=\columnwidth]{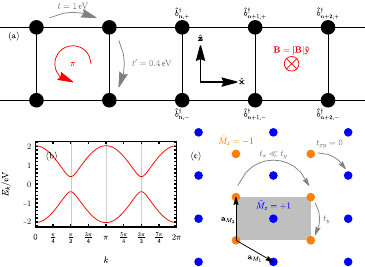}\subfloat{\label{fig:non_symmorphic:a}}\subfloat{\label{fig:non_symmorphic:b}}\subfloat{\label{fig:non_symmorphic:c}}\caption{Non-symmorphic momentum-space symmetries. (a) A ladder tight-binding model with magnetic flux that realizes the $\tilde{M}_z$ symmetry. Fermion operators and hopping amplitudes are indicated above each site (black dots). Its dispersion relation is plotted in (b). (c) schematically illustrates the quasi-one-dimensional character of atomic bands in M-point moir\'e systems for valley $\eta = 0$. Each Wannier orbital (dots) is colored according to its $\tilde{M}_{z}$ eigenvalue. The gray rectangle represents the rectangular unit cell of each $\tilde{M}_z$ symmetry sector.}
	\label{fig:non_symmorphic}\end{figure}

Projective fermion representations that realize momentum-space non-symmorphic symmetries have previously been proposed in magnetic systems~\cite{XIA24} or systems subjected to a large magnetic field (on the order of thousands of Tesla)~\cite{CHE22,ZHA23b,HER23}. M-point moir\'e materials provide the first experimentally viable realization of these symmetries in any ({\it i.e.}{} magnetic or non-magnetic) system. To better understand the origin of the momentum-space non-symmorphic action of the $\tilde{M}_{z}$ symmetry, we construct a simple one-dimensional tight-binding model that incorporates it. The resulting ladder model, shown in \cref{fig:non_symmorphic:a}, mimics the dispersion of an atomic band in the M-point moiré Hamiltonian for valley $\eta = 0$ along the $\hat{\vec{x}}$ direction (see \cref{app:sec:add_sym}). Each unit cell is threaded by a uniform perpendicular magnetic field, enclosing a $\pi$-flux. Since $\pi$- and $(-\pi)$-fluxes are equivalent, the model also respects time reversal and $\tilde{M}_{z}$ symmetry. In the Fourier-transformed basis $\hat{b}^\dagger_{k, l} = \frac{1}{\sqrt{N}}\sum_n \hat{b}^\dagger_{n,l}e^{ikn}$, the $\tilde{M}_{z}$ symmetry acts non-symmorphically as $\tilde{M}_z \hat{b}^\dagger_{k,l} \tilde{M}^{-1}_z = \hat{b}^\dagger_{k + \pi, -l}$, ensuring that the spectra of the Hamiltonian at $k$ and $k + \pi$ are identical, as shown in \cref{fig:non_symmorphic:b}.

\textit{Hubbard and Luttinger simulators.}~Within each valley, the first two sets of spinful bands in \ch{SnSe2} and \ch{ZrS2} bilayers are individually Wannierizable, with their bandwidths tunable by adjusting the twist angle. Given the excellent $\mathrm{SU} \left( {2} \right)$ symmetry, these M-point moir\'e systems become effective simulators of the Hubbard model when Coulomb interactions are included~\cite{HU24}. However, due to the additional valley degree of freedom, these systems go beyond the single-band $\mathrm{U} \left( {2} \right)$ Hubbard model, instead realizing a six-flavor $\mathrm{U} \left( {2} \right) \times \mathrm{U} \left( {2} \right) \times \mathrm{U} \left( {2} \right)$ Hubbard model.

Another key distinction from the standard Hubbard model can arise from the $\tilde{M}_z$ symmetry. In real space, $\tilde{M}_{z}$ does not change the position along the moir\'e heterostructure. As a result, the continuum moir\'e Hamiltonian can be made diagonal in the $\tilde{M}_{z}$ basis. Because $\left( T'_{\vec{a}_{M_1}} \right)^2 \left( T'_{\vec{a}_{M_2}} \right)^{-1}$ and $T'_{\vec{a}_{M_2}}$ both commute with $\tilde{M}_{z}$, each mirror sector of valley $\eta = 0$ will feature reduced translation symmetry specified by the rectangular lattice vectors $2 \vec{a}_{M_1} - \vec{a}_{M_2}$ and $\vec{a}_{M_2}$. The $T'_{\vec{a}_{M_1}}$ operator anticommutes with $\tilde{M}_z$, exchanging the two mirror sectors. The Wannier orbitals of any atomic band can therefore be split by their $\tilde{M}_z$ eigenvalues: the orbitals of each mirror-sector are displaced by $\vec{a}_{M_1}$ and form two interpenetrating rectangular lattices shown in \cref{fig:non_symmorphic:c}. Within each mirror-sector and in valley $\eta = 0$, the inter-orbitals separation is larger by a factor of $\sqrt{3}$ along the $\hat{\vec{x}}$ direction compared to the $\hat{\vec{y}}$ one. Provided that the Wannier orbital spread is approximately isotropic (as it happens for the first band of AA-stacked \ch{SnSe2}, but \emph{not} in the first band of twisted \ch{ZrS2}), this will lead to reduced hopping along $\hat{\vec{x}}$ compared to $\hat{\vec{y}}$ (see \cref{app:sec:add_sym}). As the tunneling between $\tilde{M}_z$ sectors is forbidden, the system in each valley will behave quasi-one-dimensionally, with flatter dispersion along the $C^{\eta}_{3z} \hat{\vec{x}}$ direction, effectively emulating a Luttinger model. In the three-valley system, this quasi-one-dimensional behavior causes the $\mathrm{U} \left( {2} \right) \times \mathrm{U} \left( {2} \right) \times \mathrm{U} \left( {2} \right)$ local moments to couple differently along three $C_{3z}$-related directions, similar (but not identical) to the couplings of the Kitaev model~\cite{KIT06}. 

Note, however, that quasi-one-dimensionality along the $C^{\eta}_{3z} \hat{\vec{y}}$ direction ({\it i.e.}{} flatter dispersion along the $C^{\eta}_{3z} \hat{\vec{x}}$ direction) is \emph{not} a \emph{fundamental} or generic feature of M-point moir\'e materials, as initially proposed in toy models~\cite{KAR19}. Instead, it is the presence of the effective $\tilde{M}_z$ symmetry, not previously identified, that plays a crucial role. This symmetry, \emph{along with the orbital content}, is compatible with effects opposite to those proposed in Ref.~\cite{KAR19}. 
Indeed, inspecting the monolayer masses from \cref{tab:params}, we find that $m_y > m_x$ promotes a flatter dispersion \emph{along the $C^{\eta}_{3z} \hat{\vec{y}}$ direction} rather than along the $C^{\eta}_{3z} \hat{\vec{x}}$ direction. For AA-stacked \ch{SnSe2}, while $m_y > m_x$, the difference is not substantial ($m_y \centernot{\gg} m_x$), allowing the effect of $\tilde{M}_z$ symmetry to dominate. In twisted \ch{ZrS2}, the much enhanced monolayer mass asymmetry ($m_y \gg m_x$), or alternatively the real-space orbitals elongated along the $C^{\eta}_{3z} \hat{\vec{x}}$ direction, results in a flatter dispersion along the $C^{\eta}_{3z} \hat{\vec{y}}$ direction for the first set of conduction bands, despite the system having \emph{excellent} $\tilde{M}_z$ symmetry. The effective $\tilde{M}_z$ still enforces nesting due to the approximate energy degeneracy between $\vec{k}$ and $\vec{k} + \vec{q}_{\eta}$ and further leads to the recently introduced \emph{quantum geometric nesting}~\cite{HAN24}, but not to quasi-one-dimensionality along the $C^{\eta}_{3z}\hat{\vec{y}}$ direction. The mass-enforced band flattening was also identified in \ch{BC3}~\cite{KAR23}, where $m_y \ggg m_x$, but it occurs in the opposite direction to that predicted by Ref.~\cite{KAR19}. Our nonsymmorphic $M_z$ symmetry \emph{and} constituent orbitals are thus crucial for the interacting Hamiltonian, giving rise to different directions of one- or two-dimensional physics~\cite{HU24}.

\textit{Discussion.}~We have introduced a novel platform for moir\'e materials based on monolayers with triangular lattices, where the low-energy states are located at the M points of the BZ. The presence of three $C_{3z}$-related valleys makes M-point moir\'e materials manifestly different from pre-existing $\Gamma$- and K-point twisted heterostructures. We have shown that M-point moir\'e materials can be realized in many materials \cite{PET24,JIA24b} and specifically in twisted 1T-\ch{SnSe2} and 1T-\ch{ZrS2}, both of which are experimentally exfoliable. By constructing the corresponding moir\'e Hamiltonians, we have revealed that these materials provide the first experimentally viable example of momentum-space non-symmorphic symmetry in a nonmagnetic system. The projective representations of the crystallographic space groups associated with these symmetries extend beyond current theoretical frameworks~\cite{BRA17,PO17,KRU17}, opening new avenues for discovering symmetry-protected topological phases.

When electron-electron interactions are considered, twisted \ch{SnSe2} and \ch{ZrS2} bilayers can realize strongly correlated, tunable six-flavor Hubbard models. In addition to exhibiting Mott physics and correlated insulating phases at integer fillings, these systems can spontaneously break the $\tilde{M}_z$ symmetry, potentially giving rise to various stripe phases which will be explored in future publications~\cite{HU24}. In particular, we find that multi-valley Wannier model for AA-stacked \ch{SnSe2} can be exactly solved in the strong-coupled limit at integer fillings $0 \leq \nu \leq 6$ of the lowest six flat bands. In particular, the solutions comprise a classical spin liquid ($\nu = 1$ and $\nu=5$), a valence bond solid ($\nu = 2$ and $\nu = 4$), and quantum spin liquid ($\nu = 3$)~\cite{HU24}. The perfect nesting at momentum $\vec{q}_{\eta}$ in valley $\eta$, enforced by the $\tilde{M}_z$ symmetry, further enhances the potential for novel correlated phases, as does the recently introduced quantum nesting condition~\cite{HAN24}, satisfied as a result of the same symmetry. Moreover, due to their quasi-one-dimensional nature within each valley, these materials are promising candidates for exploring Luttinger physics.

\begin{acknowledgments}

We thank Urko Petralanda, Emilia Moro\cb{s}an, Grigorii Skorupskii, Martin Claassen, Dante M. Kennes, Angel Rubio, Lede Xian, Qiaoling Xu, Luis Elcoro, and Nicolas Regnault for collaboration on two related projects~\cite{PET24,JIA24b}, as well as Siddharth A. Parameswaran for his collaboration on a separate project. H.P. and Y.J. thank Quansheng Wu, Yan Zhang, and Jiaxuan Liu for helpful discussions. We are also grateful to the authors of Ref.~\cite{MAH24} for sharing their manuscript before posting. The simulations presented in this article were performed on computational resources managed and supported by Princeton Research Computing, a consortium of groups including the Princeton Institute for Computational Science and Engineering (PICSciE) and the Office of Information Technology's High Performance Computing Center and Visualization Laboratory at Princeton University. D.C. acknowledges support from the DOE Grant No. DE-SC0016239 and the hospitality of the Donostia International Physics Center, at which this work was carried out. H.H. and Y.J. were supported by the European Research Council (ERC) under the European Union’s Horizon 2020 research and innovation program (Grant Agreement No. 101020833), as well as by the IKUR Strategy under the collaboration agreement between Ikerbasque Foundation and DIPC on behalf of the Department of Education of the Basque Government. M.G.V and H.P. were supported by the Ministry for Digital Transformation and of Civil Service of the Spanish Government through the QUANTUM ENIA project call - Quantum Spain project, and by the European Union through the Recovery, Transformation and Resilience Plan - NextGenerationEU within the framework of the Digital Spain 2026 Agenda. B.A.B. was supported by the Gordon and Betty Moore Foundation through Grant No. GBMF8685 towards the Princeton theory program, the Gordon and Betty Moore Foundation’s EPiQS Initiative (Grant No. GBMF11070), the Office of Naval Research (ONR Grant No. N00014-20-1-2303), the Global Collaborative Network Grant at Princeton University, the Simons Investigator Grant No. 404513, the BSF Israel US foundation No. 2018226, the NSF-MERSEC (Grant No. MERSEC DMR 2011750), the Simons Collaboration on New Frontiers in Superconductivity, and the Schmidt Foundation at the Princeton University. L.M.S. was supported by the Gordon and Betty Moore Foundation’s EPIQS initiative through Grants GBMF9064, the David and Lucille Packard foundation, and NSF MRSEC through the Princeton Center for Complex Materials, DMR-2011750. D.K.E. acknowledges funding from the European Research Council (ERC) under the European Union’s Horizon 2020 research and innovation program (grant agreement No. 852927), the German Research Foundation (DFG) under the priority program SPP2244 (project No. 535146365), the EU EIC Pathfinder Grant ``FLATS'' (grant agreement No. 101099139) and the Keele Foundation.

\textit{Note added}.~In Ref.~\cite{MAH24}, the authors also introduced a new platform based on M-point twisting. Where overlapping, our manuscripts agree in their conclusions. 

\textit{Data availability}:~All data generated in this study is included in the main text and Supplementary Materials. Additional data, along with any code required for reproducing the figures, are available from the authors upon reasonable request.
\end{acknowledgments}

\let\addcontentsline\oldaddcontentsline

\renewcommand{\thetable}{S\arabic{table}}
\renewcommand{\thefigure}{S\arabic{figure}}
\renewcommand{\theequation}{S\arabic{section}.\arabic{equation}}
\onecolumngrid
\pagebreak
\thispagestyle{empty}
\newpage
\begin{center}
	\textbf{\large Supplementary Materials for ``A New Moir\'e Platform Based on M-Point Twisting{}''}\\[.2cm]
\end{center}

\appendix
\renewcommand{\thesection}{\Roman{section}}
\tableofcontents
\let\oldaddcontentsline\addcontentsline
\newpage

\section{Introduction to the supplementary materials}\label{app:sec:SN_introduction}

The supplementary materials provide a self-contained, comprehensive, and pedagogical presentation of our results and the techniques employed in this work. To guide the reader through these materials, we offer an overview of each \siSection{} below.  

In \cref{app:sec:DFT_single_layer}, we begin with detailed \textit{ab initio} results for the 1T-\ch{SnSe2} and 1T-\ch{ZrS2} monolayers (referred to hereafter as \ch{SnSe2} and \ch{ZrS2}, respectively). This section discusses the crystal structure, symmetries, and low-energy physics of these materials. We then turn to their homobilayer heterostructures in \Cref{app:sec:DFT_bilayer}, presenting first-principles results for both untwisted and twisted bilayers. Key results include band structures, lattice relaxation profiles, Wilson loops, and a comparison of different van der Waals functionals.

In \cref{app:sec:SnS_SnSe_twist}, we shift to a symmetry-based analytical approach, deriving the moir\'e Hamiltonian for twisted \ch{SnSe2} and \ch{ZrS2} bilayers. We start by obtaining a Bistritzer-MacDonald model~\cite{BIS11} for the twisted heterostructures, using a two-center, first-monolayer harmonic approximation for the interlayer hopping amplitude. We then introduce the notation for the low-energy moir\'e Hamiltonian and analyze the exact crystalline symmetries of the corresponding model. \Cref{app:sec:SnS_SnSe_twist_general_woGradient} generalizes these results by deriving the most general moir\'e potential without gradient terms ({\it i.e.}{}, terms proportional to gradients of the low-energy fermionic fields) using the exact symmetries of the twisted heterostructure. An explicit symmetry-obeying parameterization of the moir\'e potential up to the first moir\'e harmonic is also provided.  Additionally, we show how the moir\'e potential can be further constrained by the symmetries of the untwisted heterostructure in the so-called ``local-stacking approximation''~\cite{JUN14}. The limit in which these constraints are imposed is referred to as the zero-twist limit.

\Cref{app:sec:add_sym} explores additional effective symmetries of the moir\'e Hamiltonian that emerge under various physically relevant limits. A key finding is the presence of momentum-space nonsymmorphic symmetries in the zero-twist limit without gradient terms, and this \siSection{} examines their implications in detail. \Cref{app:sec:add_sym} further investigates the symmetries arising in other limits of the moir\'e Hamiltonian. Finally, it analyzes the symmetries of simplified two- and three-parameter models, which, as shown in \cref{app:sec:fitted_models}, provide an excellent description of the lowest bands of the twisted heterostructures at small angles.

For completeness, \cref{app:sec:SnS_SnSe_twist_general} derives the most general form of the moir\'e potential with gradient terms, and shows how these latter terms are further constrained in the zero-twist limit. The resulting generalized moir\'e Hamiltonian (away from the zero-twist limit) will then be used in \cref{app:sec:fitted_models} to build numerically exact models for the first few moir\'e bands. Returning to \textit{ab initio} methods, \cref{app:sec:first_princ_ham_valley} details the extraction of valley-projected moir\'e Hamiltonians in the plane wave basis, starting from the corresponding first-principles Kohn-Sham Hamiltonians. This \siSection{} also provides elementary results for the valley-projected Hamiltonians, including their \textit{ab initio} band structures and local density of states at the smallest commensurate angle studied.

\Cref{app:sec:fitting_method} is a technical \siSection{} that outlines the algorithms used to obtain continuum models from the \textit{ab initio} Hamiltonians. We present both a linear least-squares extraction method, adapted from Refs.~\cite{HU23c,JIA23a,ZHA24}, as well as a nonlinear fitting one. Strategies to minimize the number of parameters in these methods are also discussed. \Cref{app:sec:simple_ham_results} then shows how analytical band structure calculations can be performed for the M-point moir\'e plane-wave Hamiltonians. The strategy employed is similar to the ``tripod''~\cite{BIS11} and related~\cite{BER21}, plane-wave truncation schemes used for twisted bilayer graphene. 

Finally, building on the concepts introduced beforehand, \cref{app:sec:fitted_models} offers a comprehensive summary of our findings. This \siSection{} begins by presenting both numerically exact and simplified moir\'e Hamiltonian models for all monolayers, stacking configurations, as well as all the commensurate angles studied. It provides the parameter values for these models and compares their band structures, local layer-resolved density of states, Berry curvature, and Wilson loops, delivering a detailed and cohesive perspective on the twisted heterostructure physics investigated in this work.

\section{First-principle results for monolayer materials}\label{app:sec:DFT_single_layer}

In this \siSection{}, we present \textit{ab initio} results for the monolayer materials considered in this work, namely \ch{SnSe2} and \ch{ZrS2}. For each material, we discuss the crystal structure, the symmetries, and their density functional theory (DFT) band structures, and additionally provide $\vec{k} \cdot \vec{p}$ Hamiltonians around the M point of the monolayer Brillouin zone (BZ).

\subsection{Monolayer \ch{SnSe2}}\label{app:sec:DFT_single_layer:snse2}

\begin{figure}[tbp]
	\centering
	\includegraphics[width=\textwidth]{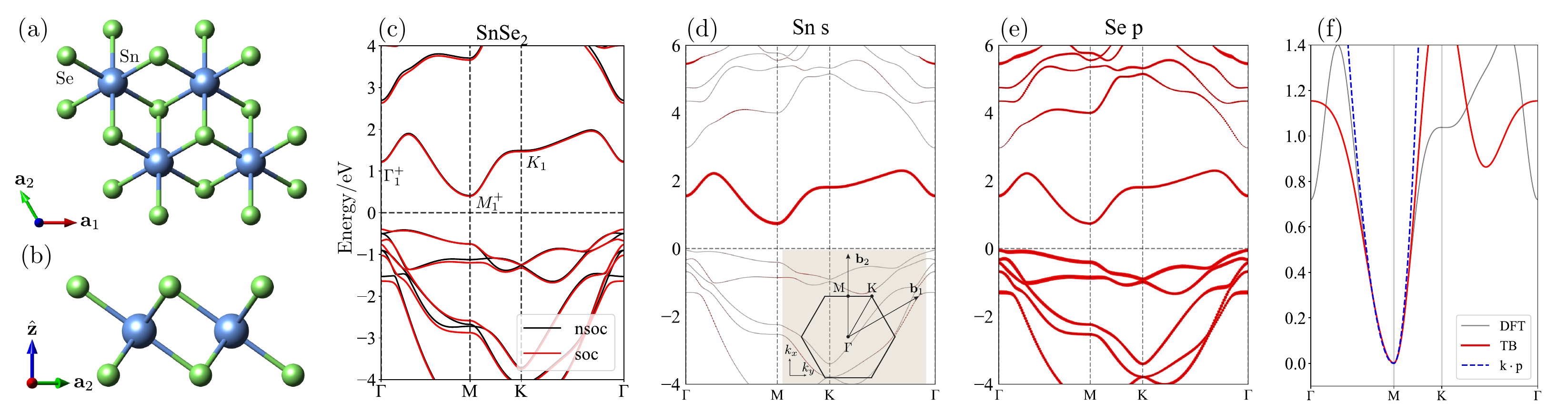}\subfloat{\label{app:fig:monolayer-bands:a}}\subfloat{\label{app:fig:monolayer-bands:b}}\subfloat{\label{app:fig:monolayer-bands:c}}\subfloat{\label{app:fig:monolayer-bands:d}}\subfloat{\label{app:fig:monolayer-bands:e}}\subfloat{\label{app:fig:monolayer-bands:f}}\caption{\textit{Ab initio} results for monolayer \ch{SnSe2}. (a) and (b) show the crystal structure of monolayer \ch{SnSe2}, from top and side views, respectively. (c) shows the \textit{ab initio} band structure of monolayer \ch{SnSe2}, with the black (red) lines denoting the bands without (with) spin-orbital coupling (SOC). The (spinless) IRREP of the M valley is marked on the plot. (d) and (e) show the orbital weights of the $s$ orbital of Sn and the $p$ orbitals of Se. The inset in (d) shows the monolayer BZ, high-symmetry momenta and reciprocal lattice vectors. The bottom conduction band is isolated with the minimum at the M point, and is contributed mainly by the $s$ orbitals of Sn and the $p$ orbitals of Se. (f) Comparison of the dispersion from the DFT result, minimal TB model defined in \cref{app:eqn:fitted_cond_band_SnSe2}, and the $\vec{k} \cdot \vec{p}$ model from \cref{app:eqn:k_dot_p_sl}. }
	\label{app:fig:monolayer-bands}
\end{figure}

The crystal structure and band dispersion for single-layer \ch{SnSe2} are shown in \cref{app:fig:monolayer-bands}. Monolayer \ch{SnSe2} crystallizes in the 1T structure within the symmetry group $P\bar{3}m11'$ [Shubnikov Space Group (SSG) 164.86], as shown in \cref{app:fig:monolayer-bands:a,app:fig:monolayer-bands:b}. The lattice vectors are given by
\begin{equation}
	\vec{a}_1=a \left(1, 0 \right), \quad 
	\vec{a}_2=a \left( -\frac{1}{2}, \frac{\sqrt{3}}{2} \right),
	\label{app:eqn:hexagonal_cell_basis}
\end{equation}
with $a$ being the lattice constant and the atoms being located at
\begin{equation}
	\label{app:eqn:atom_positions}
	\mathrm{Sn}:\ \left(0,0,0\right),\quad \mathrm{Se}:\ \left(\frac{1}{3}, \frac{2}{3}, z\right),\,\left(\frac{2}{3}, \frac{1}{3}, -z\right).  
\end{equation}
In \cref{app:eqn:atom_positions}, the first two coordinates are written in the unit cell basis $\mathbf{a}_{1,2}$, with $a=\SI{3.811}{\angstrom}$ and $z=\SI{1.528}{\angstrom}$\cite{WU19c}. The Sn atom is located at the $1a$ Wyckoff position, which features $\bar{3}m$ site symmetry (with inversion), while the Se or S atoms are positioned at $2b$ Wyckoff position, which has $3m$ symmetry (without inversion). Within a unit cell, the two $2b$ Wyckoff positions are mapped to one another by inversion. The reciprocal lattice vectors of the single-layer system are given by
\begin{equation}
	\vec{b}_1=\frac{2\pi}{a} \left( 1, \frac{1}{\sqrt{3}} \right), \quad 
	\vec{b}_2=\frac{2\pi}{a} \left( 0, \frac{2}{\sqrt{3}} \right),
\end{equation}
such that $\vec{a}_i \cdot \vec{b}_j = 2 \pi \delta_{ij}$, for $1 \leq i,j \leq 2$. 

The band structure and corresponding orbital projections of monolayer \ch{SnSe2} are shown in \crefrange{app:fig:monolayer-bands:c}{app:fig:monolayer-bands:e}. The bottom conduction band is well-isolated and mainly contributed by the $s$ orbital of Sn and the $p$ orbitals of Se, all of which have negligible spin-orbital coupling (SOC). The conduction band minima (CBM) appears at M point, and transforms as the $\bar{M}_3\bar{M}_4$ double irreducible representation (IRREP) in the spinful case. 

The isolated bottom conduction band is characterized by the $\bar{\Gamma}_4\bar{\Gamma}_5$, $\bar{M}_3\bar{M}_4$, and $\bar{K}_4\bar{K}_5$ 
(or $\Gamma_1^+$, $M_1^+$, and $K_1$ in the spinless case) IRREPs at the three high-symmetry momenta, which correspond to the elementary band representation (EBR)~\cite{BRA17} induced by $^1\bar{E}_g ^2\bar{E}_2@1a$ (or $A_{1g}@1a$ in the spinless case). This EBR is therefore induced by a (spinful) $s$ orbital located at the $1a$ position and, as such, allows us to construct a Wannier tight-binding model of the bottom spinful conduction band which features a single effective $s$ orbital at the Sn site ($1a$ position). This is also confirmed by the orbital weights shown \cref{app:fig:monolayer-bands:d,app:fig:monolayer-bands:e}, which indicate that the bottom conduction band is mainly contributed by the $s$ orbital of Sn and $p$ orbitals of Se. The latter form molecular orbitals surrounding the Sn atoms which, from a symmetry standpoint, behave as effective $s$ orbitals.

\subsubsection{Symmetries}\label{app:sec:DFT_single_layer:snse2:symmetries}

The symmetry group of monolayer \ch{SnSe2} is generated by two-dimensional translations, time-reversal symmetry $\mathcal{T}$, inversion $\mathcal{I}=\left\lbrace -1 \middle| 0  \,0 \,0 \right\rbrace$, in-plane three-fold rotation $C_{3z}=\left\lbrace 3^{+}_{001} \middle| 0 \,0 \,0 \right\rbrace$, and out of plane two-fold rotation $C_{2x}=\left\lbrace 2_{100} \middle| 0 \,0 \,0 \right\rbrace$. We now consider the action of these symmetries on the effective spinful $s$ orbital located at the $1a$ Wyckoff position which spans the bottom isolated conduction band. We let $\hat{a}^\dagger_{\vec{R},s}$ be the electron creation operator corresponding to the effective $s$ orbital in unit cell $\vec{R}$ having spin $s=\uparrow,\downarrow$ (and which is located at the unit cell origin). The corresponding momentum space operators are defined according to
\begin{equation}
	\label{app:eqn:fourier_trafo_s_SnSe2}
	\hat{a}^\dagger_{\vec{k},s} = \frac{1}{\sqrt{N}} \sum_{\vec{R}} \hat{a}^\dagger_{\vec{R},s} e^{i \vec{k} \cdot \vec{R}}.
\end{equation}

Consider $g$ to be a certain (unitary or antiunitary) symmetry operation whose action on a real space vector $\vec{r}$ is given by $g \vec{r} = R \vec{r}$, where $R$ denotes the rotation part of the transformation $g$. On the $\hat{a}^\dagger_{\vec{R},s}$ fermions, the transformation $g$ is implemented simply as
\begin{equation}
	\label{app:eqn:sym_action_real_ops}
	g \hat{a}^\dagger_{\vec{R},s_1} g^{-1} = \sum_{s_2} \left[ D^{\text{sl}} \left(g\right) \right]_{s_2 s_1} \hat{a}^\dagger_{g \vec{R},s_2},
\end{equation}
where the representation matrices $D^{\text{sl}} \left(g\right)$ for the various symmetries of the model are given explicitly by 
\begin{equation}
	\label{app:eqn:rep_matrices_syms_Snse2}
	D^{\text{sl}} \left( \mathcal{T} \right) = i s_y,  \quad
	D^{\text{sl}} \left( C_{3z} \right) = e^{-\frac{\pi i}{3} s_z}, \quad
	D^{\text{sl}} \left( C_{2x} \right) = -i  s_x, \quad
	D^{\text{sl}} \left( \mathcal{I} \right) = s_0,
\end{equation}
with $s_0$, $s_x$, $s_y$, and $s_z$ denoting the identity and Pauli matrices in the spin subspace. In momentum space, the action of $g$ reads as
\begin{equation}
	\label{app:eqn:sym_action_momentum_ops}
	g \hat{a}^\dagger_{\vec{k},s_1} g^{-1} = \sum_{s_2} \left[ D^{\text{sl}} \left(g\right) \right]_{s_2 s_1} \hat{a}^\dagger_{g \vec{k},s_2}
\end{equation}
where
\begin{equation}
	\label{app:eqn:sym_action_momentum}
	g \vec{k} = \begin{cases}
		R \vec{k} & \quad $if $ g $ is unitary$ \\
		-R \vec{k} & \quad $if $ g $ is antiunitary$
	\end{cases}.
\end{equation}

The isolated bottom conduction band exhibits an \emph{approximate} spin $\mathrm{SU} \left( {2} \right)$ symmetry due to the weak spin-orbit coupling (SOC) associated with the effective $s$ orbital, primarily contributed by the Sn $s$ and Se $p$ orbitals. To quantify the degree of $\mathrm{SU} \left( {2} \right)$ symmetry breaking, we first show that an \emph{effective} $\mathrm{SU} \left( {2} \right)$ symmetry can always be defined for a two-band system with $\mathcal{I} \mathcal{T}$ symmetry (but this \emph{effective} $\mathrm{SU} \left( {2} \right)$ symmetry does not generalize to systems with a larger number of orbitals). We then show that the global $\mathrm{SU} \left( {2} \right)$ symmetry breaking remains weak in the case of \ch{SnSe2}.

First, we consider the most general Hamiltonian featuring a single spinful band
\begin{equation}
	\label{app:eqn:general_two_band_su2}
    h \left( \vec{k} \right)=\sum_{n} h_n \left( \vec{k} \right) s_n,
\end{equation}
where $h_{n} \left( \vec{k} \right)$ are real functions of momentum $\vec{k}$. The Hamiltonian has eigenvalues $h_0 \left( \vec{k} \right)\pm \sqrt{\sum_{n=x,y,z} h_n \left( \vec{k} \right)}$. In a system with $\mathcal{I} \mathcal{T}$ symmetry, the Hamiltonian $h \left( \vec{k} \right)$ obeys $h \left( \vec{k} \right) = s_x h^{*} \left( \vec{k} \right) s_x$, which immediately implies that $h_{n} \left( 0 \right)$ (for $n = x,y,z$) rendering all bands doubly degenerate in spin, and resulting in an \emph{effective} $\mathrm{SU} \left( {2} \right)$ symmetry. It is important to note that this effective $\mathrm{SU} \left( {2} \right)$ symmetry is defined in the effective Wannier basis of the two-band model ({\it i.e.}, the basis of $h \left( \vec{k} \right)$) and may differ from the \emph{global} $\mathrm{SU} \left( {2} \right)$ symmetry.

\newcommand{\tSym}{\text{sym}}
\newcommand{\tAsym}{\text{asym}}
To evaluate the global $\mathrm{SU} \left( {2} \right)$ symmetry breaking, we consider the projector into an isolated set of bands with wavefunction $\ket{\psi_{n \vec{k},\alpha,s}}$, where $n$, $\alpha$, and $s$ are the band, orbital, and spin indices, respectively 
\begin{equation}
    \label{app:eqn:projector_SU2_snse2}
    P_{\vec{k},\alpha,s,\beta,s'} = \sum_{n} \ket{\psi_{n \vec{k},\alpha,s}} \bra{\psi_{n \vec{k},\beta,s'}}.
\end{equation}
The spin-symmetric part of the projector is defined as 
 \begin{equation}
     P_{\vec{k},\alpha,s,\beta,s'}^{\tSym} = \delta_{s s'} \frac{1}{2} \sum_{s''} P_{\vec{k},\alpha,s'',\beta,s''} 
 \end{equation}
while the spin-antiymmetric part is 
\begin{equation}
    P_{\vec{k},\alpha,s,\beta,s'}^{\tAsym} = 
    P_{\vec{k},\alpha,s,\beta,s'} - P_{\vec{k}, \alpha,s,\beta, s'}^{\tSym}
\end{equation}
The $\mathrm{SU} \left( {2} \right)$ symmetry breaking is defined as the measure of the spin-antisymmetric part
\begin{equation}
	\label{app:eqn:su2_breaking_def}
	\frac{1}{N}\sum_{\vec{k}} \frac{\norm{P_{\vec{k}}^{\tAsym}}}{\norm{P_{\vec{k}}}},
\end{equation}
where $\norm{\dots}$ denotes the Frobenius norm of a matrix. 

To numerically assess the global $\mathrm{SU} \left( {2} \right)$ symmetry breaking for the bottom conduction band in monolayer \ch{SnSe2}, we construct a Wannier tight-binding model using Sn s and Se p orbitals. From this model, we find a $9.0\%$ $\mathrm{SU} \left( {2} \right)$ symmetry breaking for the bottom conduction band, leading to the conclusion that the $\mathrm{SU} \left( {2} \right)$ symmetry is a good \emph{approximate} symmetry for the bottom conduction band. We reiterate that the bottom conduction band has a perfect \emph{effective} $\mathrm{SU} \left( {2} \right)$ symmetry, as discussed around \cref{app:eqn:general_two_band_su2}.

\subsubsection{Hamiltonian for the lowest conduction band and effective $\vec{k} \cdot \vec{p}$ model at the M valley}\label{app:sec:DFT_single_layer:snse2:model}

We now build a Wannier model for the lowest conduction band. The tight-binding (TB) Hamiltonian corresponding to the model can be written as  
\begin{equation}
	\label{app:eqn:ham_sl_snse2}
	\mathcal{H}^{\text{sl}} = \sum_{\vec{k}} \left[h^{\text{sl}} \left( \vec{k} \right)\right]_{s_1 s_2} \hat{a}^\dagger_{\vec{k},s_1} \hat{a}_{\vec{k},s_2}.
\end{equation}
In \cref{app:eqn:ham_sl_snse2}, the Hamiltonian matrix $h^{\text{sl}} \left( \vec{k} \right)$ can be obtained from Wannier90. The resulting Wannier function, which is also plotted as an inset in \cref{fig:materials:c}, is a molecular orbital formed by the $s$ orbital of Sn and the $p$ orbitals of Se and is consistent with the orbital projections in \cref{app:fig:monolayer-bands}. Its exact expression in terms of the atomic $s$ and $p$ orbitals is straightforward, but beyond the scope of this work.

We also build a simplified TB Hamiltonian with the following form, constrained by $P\bar{3}m11'$ symmetries
\begin{align}
    h^{\text{sl}} \left( \vec{k} \right) 
    =  
    & \bigg[ \epsilon_0 + 
    2t_1 \left(\cos(k_1) + \cos(k_2) + \cos(k_1+k_2) \right) + 
    2 t_2 \left( 2\cos(\frac{3}{2}k_1)\cos(\frac{1}{2}k_1+k_2) + \cos(k_1+2k_2)\right) \nonumber \\
    &  + 2 t_3 \left(\cos(2k_1) + 2\cos(k_1) \cos(k_1+2k_2)\right)
    \bigg] s_0, \label{app:eqn:fitted_cond_band_SnSe2}
\end{align}
where $s_0$ is the two-dimensional identity matrix acting on the spin subspace and $\vec{k} = k_1 \vec{b}_1 + k_2 \vec{b}_2$. The Hamiltonian matrix $h^{\text{sl}} \left( \vec{k} \right)$ includes an onsite energy $\epsilon_0$, nearest neighbor (NN) $t_1$, next NN $t_2$, and fourth NN $t_3$ hoppings. By fitting to the \textit{ab initio} dispersion, we find $\epsilon_0 = \SI{1.589}{\electronvolt}$, $t_1 = \SI{0.029}{\electronvolt}$, $t_2 =  \SI{0.115}{\electronvolt}$, and $t_3 =  \SI{-0.087}{\electronvolt}$. The large long-range hopping is due to the extended nature of the effective $s$ Wannier orbital. The fitted dispersion is shown in \cref{app:fig:monolayer-bands:f}. 

We then build an effective $\vec{k} \cdot \vec{p}$ Hamiltonian at the M valley. Letting $\delta \vec{k}$ denote the momentum deviation from the M point located at $\vec{K}_{M} = \frac12 \vec{b}_2$, the $\vec{k} \cdot \vec{p}$ expansion of the Hamiltonian matrix from \cref{app:eqn:ham_sl_snse2} around the bottom of the conduction band is given by
\begin{equation}
	\label{app:eqn:k_dot_p_sl}
	h^{\text{sl}} \left( \vec{K}_{M} + \delta \vec{k} \right) \approx \left( \frac{\delta k_x^2}{2 m_x} + \frac{\delta k_y^2}{2 m_y} + \Delta \right) s_0,
\end{equation}
where $m_x$ and $m_y$ are mass parameters, while $\Delta$ is an energy gap parameter that will be set to zero in what follows without loss of generality. By fitting to the \textit{ab initio} band structures, we obtain $m_x=0.21  \, m_e, m_y=0.73 \, m_e$ for \ch{SnSe2} at the M valley ({\it i.e.}{}, at $\vec{k} = \frac{1}{2}\mathbf{b}_2$), where $m_e$ is the electron mass. As shown in \cref{app:fig:monolayer-bands:c}, the dispersion along the $\Gamma-\mathrm{M}$ line is controlled by $m_y$ and is flatter compared to the dispersion along the $\mathrm{M}-\mathrm{K}$ line, due to the significantly larger value of $m_y$ relative to $m_x$. A comparison of the dispersion from the DFT, the minimal TB model \cref{app:eqn:fitted_cond_band_SnSe2}, and the $\vec{k} \cdot \vec{p}$ expansion is shown in \cref{app:fig:monolayer-bands:f}, all of which show excellent agreement near the M valley.

\subsection{Monolayer \ch{ZrS2} }\label{app:sec:DFT_single_layer:zrs2}

\begin{figure}[!tbp]
	\centering
	\includegraphics[width=\textwidth]{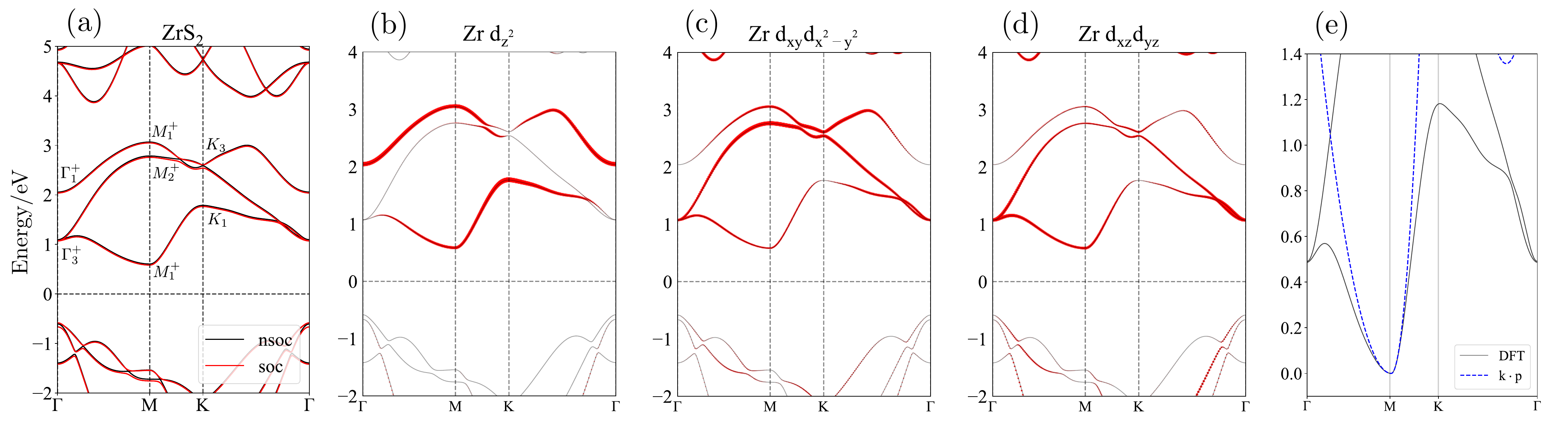}\subfloat{\label{app:fig:ZrS2-monolayer-bands:a}}\subfloat{\label{app:fig:ZrS2-monolayer-bands:b}}\subfloat{\label{app:fig:ZrS2-monolayer-bands:c}}\subfloat{\label{app:fig:ZrS2-monolayer-bands:d}}\subfloat{\label{app:fig:ZrS2-monolayer-bands:e}}\caption{\textit{Ab initio} results for monolayer \ch{ZrS2}. In (a), we compare the bands with (red) and without SOC (black). SOC has only a minor effect on the bands. The (spinless) IRREPs for the three lowest conduction bands are shown. (b)-(d) are the Zr $d$ orbital weights of the bottom three conduction bands. (e) is the comparison between the DFT dispersion and the $\vec{k} \cdot \vec{p}$ expansion near the M valley.}
	\label{app:fig:ZrS2-monolayer-bands}\end{figure}

The crystal structure of monolayer \ch{ZrS2} is the same as $1T$-\ch{SnSe2} as shown in \cref{app:fig:monolayer-bands:a}, with symmetry group $P\bar{3}m11'$. The lattice constant is $a=\SI{3.661}{\angstrom}$ and the out-of-plane displacement of S atom is $\SI{1.441}{\angstrom}$~\cite{AL-77}.

In \cref{app:fig:ZrS2-monolayer-bands}, we show the band structure of monolayer \ch{ZrS2}. Compared with \ch{SnSe2}, where there exists a single (spinful) isolated conduction band above the Fermi level, \ch{ZrS2} has three connected (spinful) conduction bands. These three bands are mainly contributed by the $d$ orbitals of Zr, with the CBM at M mostly given by the $d_{z^2}$ orbital. From the (spinless) IRREPs shown in \cref{app:fig:ZrS2-monolayer-bands:a}, the three lowest conduction bands form the EBRs $A_{1g}@1a$ and $E_g@1a$. As shown by the orbital projections of \crefrange{app:fig:ZrS2-monolayer-bands:b}{app:fig:ZrS2-monolayer-bands:d}, the CBM at M arises from the hybridization of the five $d$-orbitals of Zr. 

To build a low-energy theory of \ch{ZrS2}, we can focus exclusively on the states near the CBM at M. We observe that the IRREP of the CBM at M is $M_1^+$, which is the same as in \ch{SnSe2}. Note that $M_2^+$ -- the IRREP for the second lowest conduction band at M -- has the opposite $C_{2x}$ eigenvalue compared with $M_1^+$. We then construct an effective molecular $s$ orbital, which we also denote by $\hat{a}^\dagger_{\vec{R},s}$, in analogy to the case of \ch{SnSe2} discussed around \cref{app:eqn:fourier_trafo_s_SnSe2}. These effective $s$ orbitals are formed by a linear combination of the five $d$ orbitals of Zr, as can be seen from the orbitals weights at CBM in \crefrange{app:fig:ZrS2-monolayer-bands:b}{app:fig:ZrS2-monolayer-bands:d}. It is important to note that onsite coupling between $d_{z^2}$ and the other four $d$ orbitals is forbidden, as $d_{z^2}$ form the $A_{1g}$ IRREP while $(d_{xz},d_{yz})$ and $(d_{xy},d_{x^2-y^2})$ form the 2D $E_g$ IRREP under the point group $\bar{3}m$ of $1a$ position. However, off-site coupling is allowed, leading to hybridization at the M point. Their EBR further supports this because both $A_{1g}@1a$ and $E_g@1a$ induce $M_1^+$ IRREP at M.

The effective $s$ orbitals span the CBM at the M point. Moreover, because $\hat{a}^\dagger_{\vec{R},s}$ are effective orbitals located at the $1a$ Wyckoff position, they will have the same symmetry properties as the effective $s$ orbitals defined for \ch{SnSe2}. As such, the effective $\vec{k} \cdot \vec{p}$ Hamiltonian describing the bottom of the conduction band of \ch{ZrS2} takes the same form as the one of \ch{SnSe2} from \cref{app:eqn:k_dot_p_sl}. The corresponding masses are given by $m_x = 0.29 \, m_e$ and $m_y = 1.86 \, m_e$. 
A comparison of the DFT dispersion and $\vec{k} \cdot \vec{p}$ expansion is shown in \cref{app:fig:ZrS2-monolayer-bands:e}, which have good agreement near the M point. 
The mass along the $\hat{\vec{y}}$-direction is significantly larger than that along the $\hat{\vec{x}}$-direction, indicating flatter dispersion along the M-$\Gamma$ direction.

Similar to \ch{SnSe2}, the lowest conduction band of \ch{ZrS2} at the M valley exhibits an approximate global spin $\mathrm{SU} \left( {2} \right)$ symmetry, despite the bands originating from the $d$ orbitals of Zr. Using the global $\mathrm{SU} \left( {2} \right)$ symmetry-breaking metric defined in \cref{app:eqn:su2_breaking_def}, we find a $12.7\%$ $\mathrm{SU} \left( {2} \right)$ symmetry breaking near the M point for the lowest conduction band. At the $\Gamma$ point of the lowest conduction band, the global $\mathrm{SU} \left( {2} \right)$ symmetry breaking reaches $70.1\%$, primarily due to the weak SOC-induced splitting of the band crossings near $\Gamma$. However, being effectively a single band model, Hamiltonian describing the bottom of the conduction band exhibits perfect \emph{effective} $\mathrm{SU} \left( {2} \right)$ symmetry, as discussed around \cref{app:eqn:general_two_band_su2}.

\section{First-principle results for bilayer materials}\label{app:sec:DFT_bilayer}

Having presented the \textit{ab initio} results on the monolayer materials in \cref{app:sec:DFT_single_layer}, we now turn to the bilayer case. Specifically, for each single-layer material considered in \cref{app:sec:DFT_single_layer}, this \siSection{} discusses the symmetries and band structures of the corresponding twisted and untwisted bilayer heterostructures. 

\subsection{Bilayer \ch{SnSe2}}\label{app:sec:DFT_bilayer:SnSe2}
In this first section we show results on bilayer \ch{SnSe2}. Both the twisted and untwisted bilayer heterostructures feature two types of inequivalent stacking patterns, which we refer to as AA- and AB-stacked structures, and which will be defined below in both cases.
\begin{figure}[t]
	\centering
	\includegraphics[width=\textwidth]{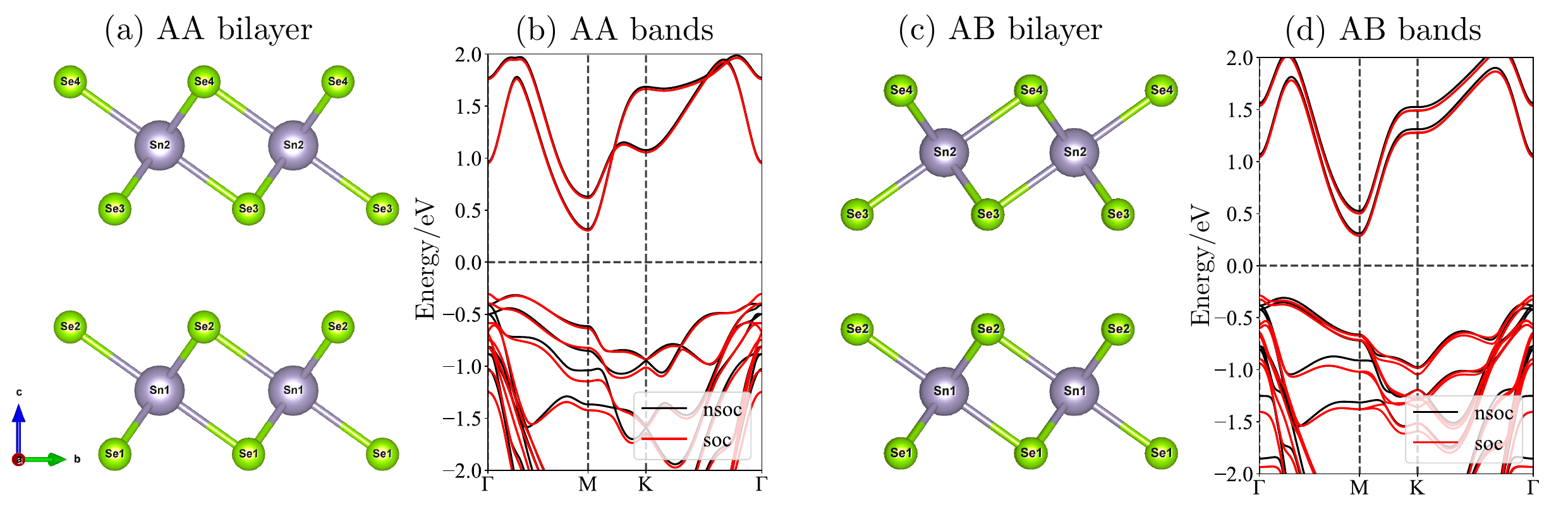}\subfloat{\label{app:fig:untwist-bilayer-bands:a}}\subfloat{\label{app:fig:untwist-bilayer-bands:b}}\subfloat{\label{app:fig:untwist-bilayer-bands:c}}\subfloat{\label{app:fig:untwist-bilayer-bands:d}}\caption{Crystal and band structures of the untwisted AA- and AB-stacked bilayer \ch{SnSe2}. (a) illustrates the crystal structure of the AA-stacked bilayer, while (b) shows the band structure with (red lines) and without (black lines) SOC. (c) and (d) are the same as (a) and (b), but for the AB-stacked bilayer.}
	\label{app:fig:untwist-bilayer-bands}\end{figure}

\subsubsection{Untwisted bilayer \ch{SnSe2}}\label{app:sec:DFT_bilayer:SnSe2:untwisted}

In the untwisted case, the crystal and band structures of the AA- and AB-stacked \ch{SnSe2} bilayers are shown in \cref{app:fig:untwist-bilayer-bands}. The symmetry group of the AA-stacked bilayer is given by $P\bar{3}m11'$ (SSG 164.86), while for the AB-stacked one, it is $P\bar{6}m21'$ (SSG 187.210). The symmetry generators in both cases comprise the $C_{3z}$ and $\mathcal{T}$ symmetries. As seen in \cref{app:fig:untwist-bilayer-bands:a,app:fig:untwist-bilayer-bands:c}, respectively, the AA-stacked bilayer additionally features inversion symmetry $\mathcal{I}$, while the AB-stacked one is symmetric under $z$-directional mirror reflections henceforth denoted by $M_z$. The resulting $\mathcal{I} \mathcal{T}$ symmetry of the AA-stacked bilayer enforces a Kramers degeneracy of its bands throughout the entire BZ, as shown in \cref{app:fig:untwist-bilayer-bands:b}. The band structure of the AB-stacked bilayer from \cref{app:fig:untwist-bilayer-bands:d} does, however, exhibit spin splitting of the valence bands. As they are mainly contributed by the $s$ orbitals of Sn and $p$ orbitals of Se, the bottom four spinful conduction bands of AB-stacked \ch{SnSe2} feature negligible SOC and spin splitting. The layer splitting in bilayer \ch{SnSe2} is $\SI{312}{\milli\electronvolt}$ for AA-stacking and $\SI{213}{\milli\electronvolt}$ for AB-stacking. 

Experimentally, the bulk \ch{SnSe2} crystals are mainly reported in the AA-stacked phase~\cite{BUS61, WU19c}. From DFT, the AA-stacked bilayer has \SI{21.6}{\milli\electronvolt} lower total energy than the AB-stacked bilayer, which means the AA-stacked bilayer is more stable thermodynamically. Nevertheless, the metastable AB-stacked configuration can still be experimentally obtained through the tear-and-stack method.

\subsubsection{Twisted bilayer \ch{SnSe2}}\label{app:sec:DFT_bilayer:SnSe2:twisted}

\begin{figure}[t]
	\centering
	\includegraphics[width=0.8\textwidth]{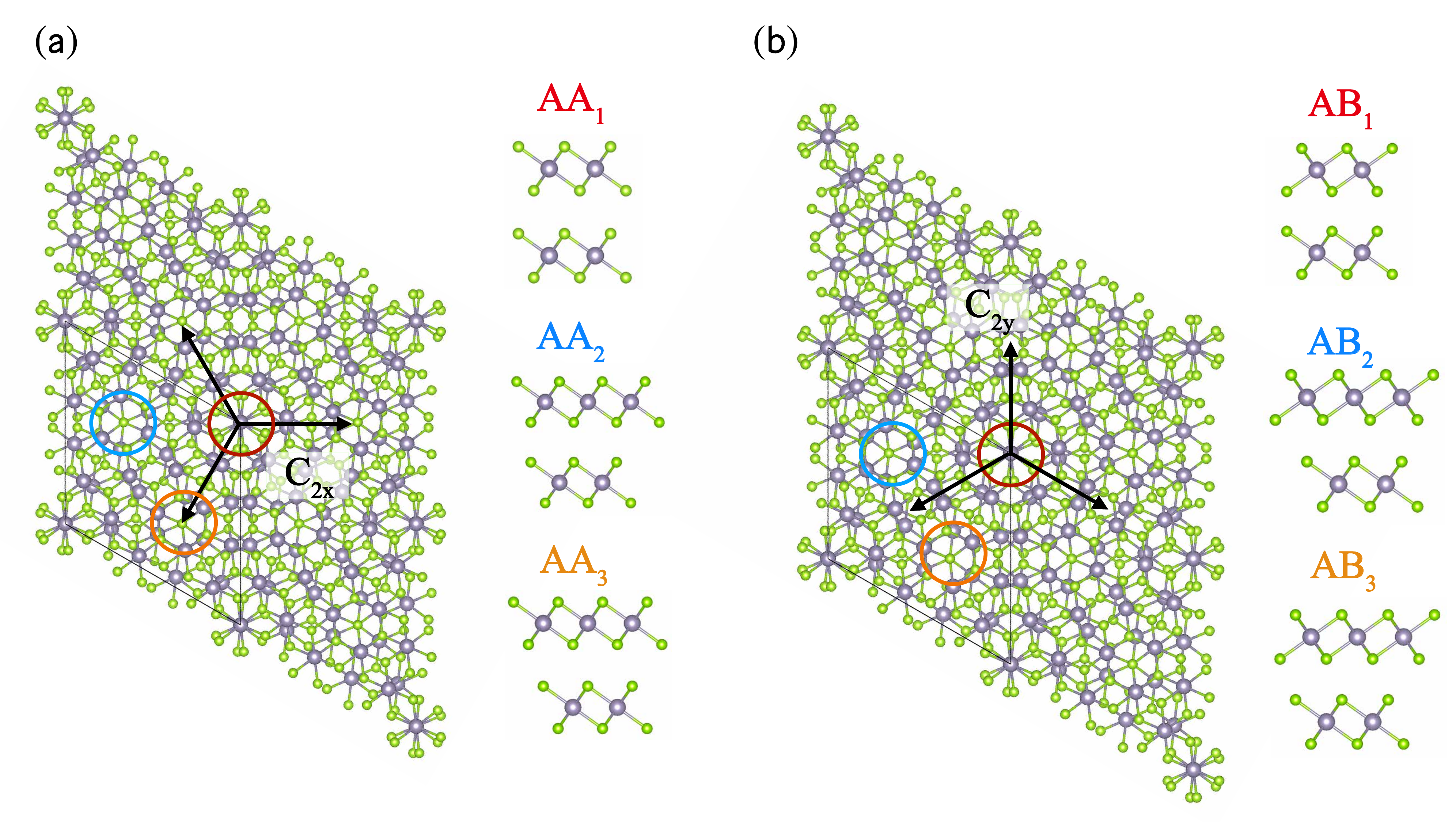}\subfloat{\label{app:fig:SnSe-AA-AB-crys:a}}\subfloat{\label{app:fig:SnSe-AA-AB-crys:b}}\caption{The crystal structure of the twisted AA- and AB-stacked bilayer \ch{SnSe2} and \ch{ZrS2} at \SI{13.17}{\degree}. The grey spheres represent Zr/Sn atoms, while the green spheres denote Sn/S atoms. The twisted AA-stacked structure is shown in (a) and is symmetric under the $P3121'$ group (SSG 149.22), generated by the $C_{3z}$, $C_{2x}$, and $\mathcal{T}$ symmetries, In contrast, the AB-stacked structure illustrated in (b) is symmetric under the $P3211'$ group (SSG 150.26), which is generated by the $C_{3z}$, $C_{2y}$, and $\mathcal{T}$ symmetries. For each structure, we use the black arrows to denote the two-fold in-plane rotation axes. Circles of different colors highlight the $C_{3z}$-symmetric local regions, {\it i.e.}{}, the triangular (red) and honeycomb (blue and orange) sites. As the twist angle decreases, the three local configurations are close to the corresponding untwisted bilayer structures shown on the side, where the subscripts $1$, $2$, and $3$ denote three types of inequivalent in-plane shifting in each stacking, as indicated in the inserted subplots. Notably, the AA$_2$ and AB$_1$ regions have neighboring Se atoms from both layers aligned at the same in-plane positions, resulting in a larger local interlayer distance (see \cref{app:fig:SnSe-AA-AB-disp} for more details).}\label{app:fig:SnSe-AA-AB-crys}\end{figure}
\begin{figure}[t]
	\centering
	\includegraphics[width=0.6\textwidth]{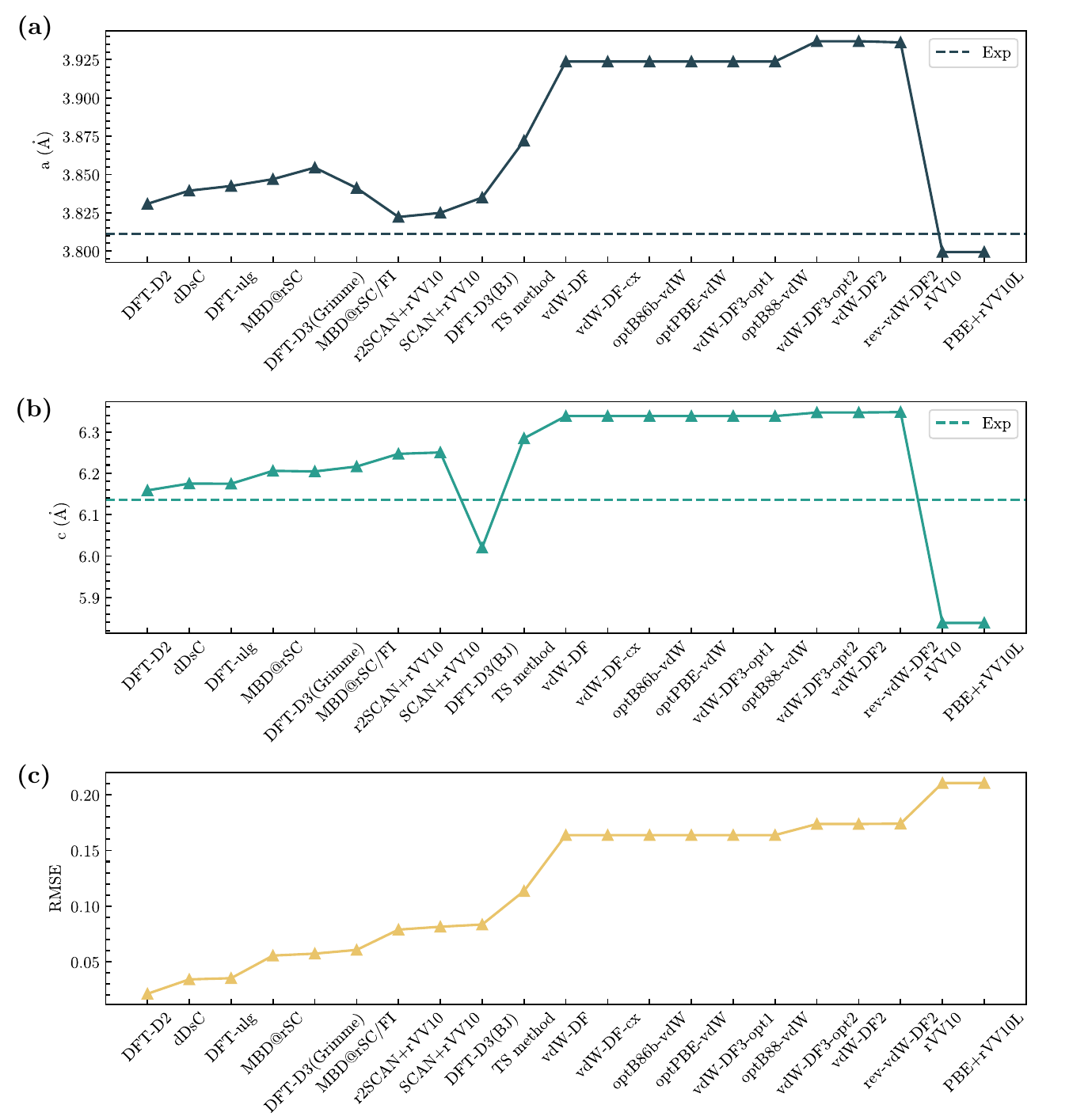}
	\subfloat{\label{app:fig:SnSe-IVDW-test:a}}\subfloat{\label{app:fig:SnSe-IVDW-test:b}}\subfloat{\label{app:fig:SnSe-IVDW-test:c}}\caption{Comparison of van der Waals (vdW) functionals for describing crystal structures of \ch{SnSe2}. (a) and (b) show the relaxed lattice constants, while (c) displays the root-mean-square error (RMSE) quantifying the discrepancy between the theoretically predicted lattice constants and the experimentally observed ones for bulk \ch{SnSe2}, using various vdW functionals implemented in VASP. The experimental values are depicted by dashed lines in (a) and (b), where ``TS'' refers to the Tkatchenko-Scheffler method. Experimental data is taken from Ref.~\cite{WU19c}. We employ the DFT-D2 method of Grimme~\cite{GRI06} to account for vdW interactions in \ch{SnSe2}, as it provides the most accurate reproduction of lattice parameters compared to experimental data.}
	\label{app:fig:SnSe-IVDW-test}
\end{figure}

Similarly to \ch{MoTe2}~\cite{YU24a,JIA24,ZHA24} (which has a different crystal structure), but unlike graphene~\cite{BIS11}, \ch{SnSe2} does not feature a $C_{2z}$ symmetry around the unit cell origin. As a result, in the twisted bilayer arrangement, just as in the untwisted case, two inequivalent stacking configurations, which we call the AA- and AB-stacked structures, can be constructed, as illustrated in \cref{app:fig:SnSe-AA-AB-crys}. We denote the two layers by $l$, where $l=+$ ($l=-$) corresponds to the top (bottom) layer. The two configurations can be obtained as follows:
\begin{enumerate}
	\item In the AA-stacking configuration shown in \cref{app:fig:SnSe-AA-AB-crys:a}, the two layers are first stacked \emph{directly} on top of one another. We then rotate layer $l$ in-plane by an angle $\frac{\theta l}{2}$ counterclockwise (with $\theta \geq 0$).
	\item In the AB-stacking configuration illustrated in \cref{app:fig:SnSe-AA-AB-crys:b}, the two layers are again first stacked \emph{directly} on top of one another. However, before twisting the two layers, the bottom layer is rotated by $\SI{180}{\degree}$ around the $\hat{\vec{z}}$ direction, with the rotation center at the origin point ({\it i.e.}, the Sn atom). Only then is each layer $l$ in-plane rotated by an angle $\frac{\theta l}{2}$ counterclockwise (with $\theta \geq 0$).
\end{enumerate}

\paragraph{\textbf{Symmetry}.}The symmetry groups of the AA- and AB-stacked twisted bilayer \ch{SnSe2} are given, respectively by the $P3121'$ (SSG 149.22) and $P3211'$ (SSG 150.26) groups. The point groups of both structures are generated by the $C_{3z}$ and in-plane two-fold symmetry $C_{2}$ operations. In the AA-stacked case, the $C_2$ axis is along the $\hat{\vec{x}}$ direction, while in the AB case, the $C_2$ axis points along the $\hat{\vec{y}}$ direction, as indicated by the black arrows in \cref{app:fig:SnSe-AA-AB-crys}. We use colored circles to denote the $C_{3z}$-symmetric local regions, {\it i.e.}{}, the triangular (red) and honeycomb (blue and orange) sites\footnote{These $C_{3z}$-symmetric local regions are the $C_{3z}$-symmetric Wyckoff position of the moir\'e unit cell.}. As the twist angle decreases, these local structures become similar to the bilayer AA/AB$_{1,2,3}$-stacking configurations shown on the right side and denoted with the same color, where the subscripts $1$, $2$, and $3$ denote three types of inequivalent in-plane shifting in each stacking. In the AA$_1$ and AB$_1$ regions, the Sn atoms in both layers are aligned directly on top of each other. In the AA$_2$ and AB$_3$ regions, the Sn atoms of the top layer are aligned with the lowermost Se atoms of the bottom layer. Lastly, in the AA$_3$ and AB$_2$ regions, the Sn atoms of the top layer are positioned above the uppermost Se atoms of the bottom layer.

\begin{figure}[t]
	\centering
	\includegraphics[width=0.8\textwidth]{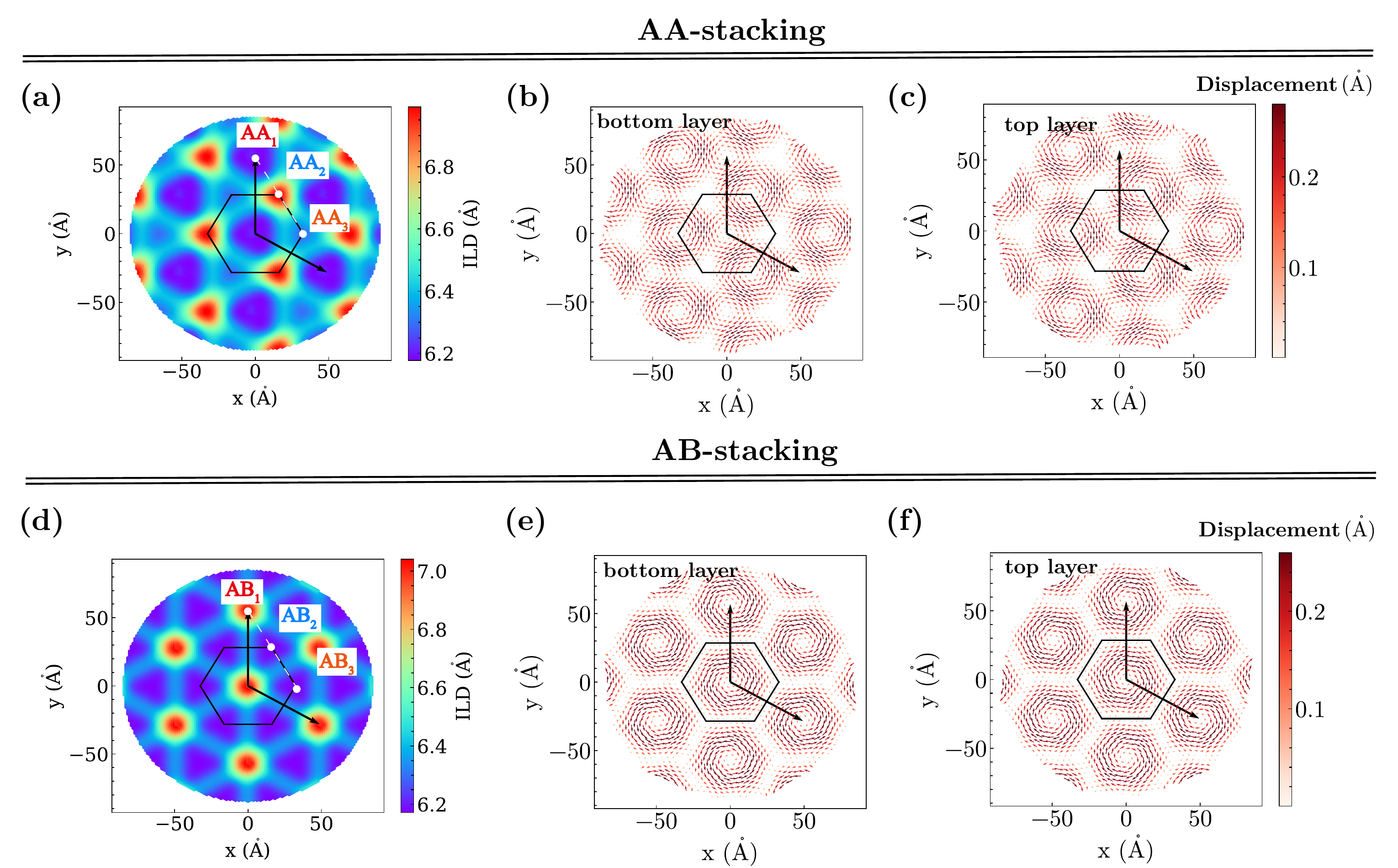}\subfloat{\label{app:fig:SnSe-AA-AB-disp:a}}\subfloat{\label{app:fig:SnSe-AA-AB-disp:b}}\subfloat{\label{app:fig:SnSe-AA-AB-disp:c}}\subfloat{\label{app:fig:SnSe-AA-AB-disp:d}}\subfloat{\label{app:fig:SnSe-AA-AB-disp:e}}\subfloat{\label{app:fig:SnSe-AA-AB-disp:f}}\caption{Lattice relaxation results of $\SI{3.89}{\degree}$ twisted AA- and AB-stacked \ch{SnSe2}. (a), (b), and (c) are interlayer distances, the intralayer displacement of the bottom and top layers for the fully-relaxed AA-stacked structure, respectively. (d), (e), and (f) are the same but for the AB-stacked structure. We note that the largest interlayer distance region is AA$_2$ (AB$_1$) in the AA- (AB-)stacked case. This agrees with the local configurations because the AA$_2$ and AB$_1$ regions have neighboring Se atoms from both layers aligned at the same in-plane positions (see \cref{app:fig:SnSe-AA-AB-crys}), resulting in a larger interlayer repulsion.}\label{app:fig:SnSe-AA-AB-disp}\end{figure}

\paragraph{\textbf{Twisted band structures}.}According to the previous researches on twisted systems, geometric relaxation significantly influences the moir\'e bands. However, large-scale \textit{ab initio} calculations require substantial computational resources and time. To reduce these costs, we initially pre-relax the twisted structure using a trained machine learning force field (MLFF), which is constructed using NequIP~\cite{BAT22} and DPmoire~\cite{LIU24}. Subsequently, we further relax the pre-relaxed structure using the Vienna \textit{ab initio} Simulation Package (VASP)~\cite{KRE96a, KRE93, KRE93a, KRE94, KRE96}. During the lattice relaxation process, we use the DFT-D2 method of Grimme~\cite{GRI06} to describe the van der Waals (vdW) corrections. This method provides lattice parameters that most closely match experimental results among 21 different vdW corrections, detailed in \cref{app:fig:SnSe-IVDW-test:a,app:fig:SnSe-IVDW-test:b}. 

\begin{figure}[t]
	\centering
	\includegraphics[width=\textwidth]{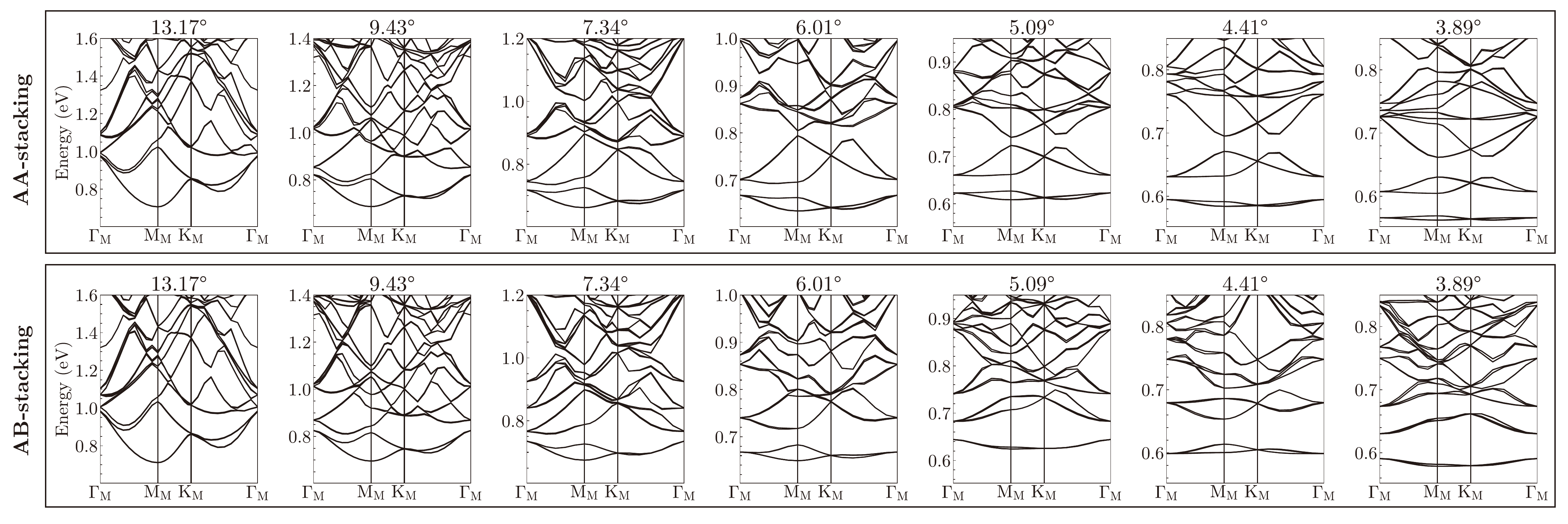}\subfloat{\label{app:fig:SnSe2-twist-bands:a}}\subfloat{\label{app:fig:SnSe2-twist-bands:b}}\subfloat{\label{app:fig:SnSe2-twist-bands:c}}\subfloat{\label{app:fig:SnSe2-twist-bands:d}}\subfloat{\label{app:fig:SnSe2-twist-bands:e}}\subfloat{\label{app:fig:SnSe2-twist-bands:f}}\subfloat{\label{app:fig:SnSe2-twist-bands:g}}\subfloat{\label{app:fig:SnSe2-twist-bands:h}}\subfloat{\label{app:fig:SnSe2-twist-bands:i}}\subfloat{\label{app:fig:SnSe2-twist-bands:j}}\subfloat{\label{app:fig:SnSe2-twist-bands:k}}\subfloat{\label{app:fig:SnSe2-twist-bands:l}}\subfloat{\label{app:fig:SnSe2-twist-bands:m}}\subfloat{\label{app:fig:SnSe2-twist-bands:n}}\caption{The \textit{ab initio} band structure of twisted AA-stacked (a-g) and AB-stacked (h-n) bilayer \ch{SnSe2} from \SI{13.17}{\degree} to \SI{3.89}{\degree}. The lowest group of conduction bands isolated from other energy bands consists of six bands stemming from the three inequivalent M valleys of monolayer \ch{SnSe2}. The bandwidths of the lowest group of six bands are listed in \cref{tab:bandwidth}.}
    \label{app:fig:SnSe2-twist-bands}
\end{figure}

In the fully-relaxed twisted AA and AB crystal structures, the interlayer distance (ILD) is defined as the distance between Sn atoms in different layers. As shown in \cref{app:fig:SnSe-AA-AB-disp:a}, the ILD reaches its maximum in the AA$_2$ region of the twisted AA configuration. This maximum is attributed to the local AA$_2$ structure, detailed in \cref{app:fig:SnSe-AA-AB-crys:a}, where the Se atoms from the top layer align directly above the Se atoms from the bottom layer, resulting in increased repulsion compared to the other local configurations. \Cref{app:fig:SnSe-AA-AB-disp:b,app:fig:SnSe-AA-AB-disp:c} illustrate the intralayer displacements of Sn atoms in the bottom and top layers, respectively, after full relaxation. The atoms rotate around the $C_{3z}$-symmetric centers, with rotation directions opposite for the top and bottom layers. We also note that the intralayer displacement is largest \emph{around} the regions with large ILD. \Crefrange{app:fig:SnSe-AA-AB-disp:d}{app:fig:SnSe-AA-AB-disp:f} display the ILD and intralayer displacements for the twisted AB configuration, which exhibit an approximately six-fold symmetric pattern, rather than just the $C_{3z}$-symmetric pattern observed in the AA configuration. This difference stems from the fact that the AB$_2$ and AB$_3$ regions are symmetrically equivalent under $C_{2y}$ and share the same Wyckoff position in the twisted AB configuration. In contrast, the two regions are inequivalent in the twisted AA configuration.

\begin{table}[t]
	\begin{tabular}{|c|c|c|c|c|c|c|c|}
	\hline
	Bandwidth ($\si{\milli\electronvolt}$) & \SI{13.17}{\degree} & \SI{9.43}{\degree} & \SI{7.34}{\degree} & \SI{6.01}{\degree} & \SI{5.09}{\degree} & \SI{4.41}{\degree} & \SI{3.89}{\degree} \\ \hline
	AA         & 318            & 136           & 64            & 33            & 19            & 11            & 7             \\ \hline
	AB         & 324            & 130           & 61            & 33            & 20            & 15            & 12             \\ \hline
	\end{tabular}
	\caption{The bandwidths of the lowest set of conduction bands of twisted AA- and AB-stacked \ch{SnSe2} for angles ranging from \SI{13.17}{\degree} to \SI{3.89}{\degree}.}
	\label{tab:bandwidth}
\end{table}

The conduction bands of twisted \ch{SnSe2} calculated using VASP are shown in \cref{app:fig:SnSe2-twist-bands}, with the twist angles ranging from $\SI{13.17}{\degree}$ to $\SI{3.89}{\degree}$. The upper row depicts the bands of twisted AA-stacked \ch{SnSe2} originating from M valleys of the untwisted bilayer structure. In these bands, we identify one or two sets of isolated moir\'e bands, each comprising six spinful bands. The moir\'e bands within each set arise from three inequivalent $C_{3z}$-related M valleys, where each valley contributes two nearly degenerate bands due to approximate $\mathrm{SU} \left( {2} \right)$ symmetry. This degeneracy among different M valleys, protected by symmetry, will be analyzed in detail in subsequent sections. As the twist angle decreases, the moir\'e bands significantly flatten, with the bandwidth of the lowest set reducing to just a few $\si{\milli\electronvolt}$, as detailed in \cref{tab:bandwidth}. The band structure of twisted AB-stacked bilayer \ch{SnSe2} shown in \crefrange{app:fig:SnSe2-twist-bands:h}{app:fig:SnSe2-twist-bands:n} is similar to the twisted AA-stacking case. When the twist angle decreases to $\SI{3.89}{\degree}$, two sets of conduction bands become isolated as shown in \crefrange{app:fig:SnSe2-twist-bands:l}{app:fig:SnSe2-twist-bands:n}. Nonetheless, the spin splitting within the same M valley is more prominent at small angles compared to the twisted AA case ({\it e.g.}{}, the splitting in the third lowest set of bands is larger in the AB case). The bandwidth of the lowest set of valence bands from $\SI{13.17}{\degree}$ to $\SI{3.89}{\degree}$ are listed in \cref{tab:bandwidth}.

\begin{figure}[t]
    \centering
    \includegraphics[width=\textwidth]{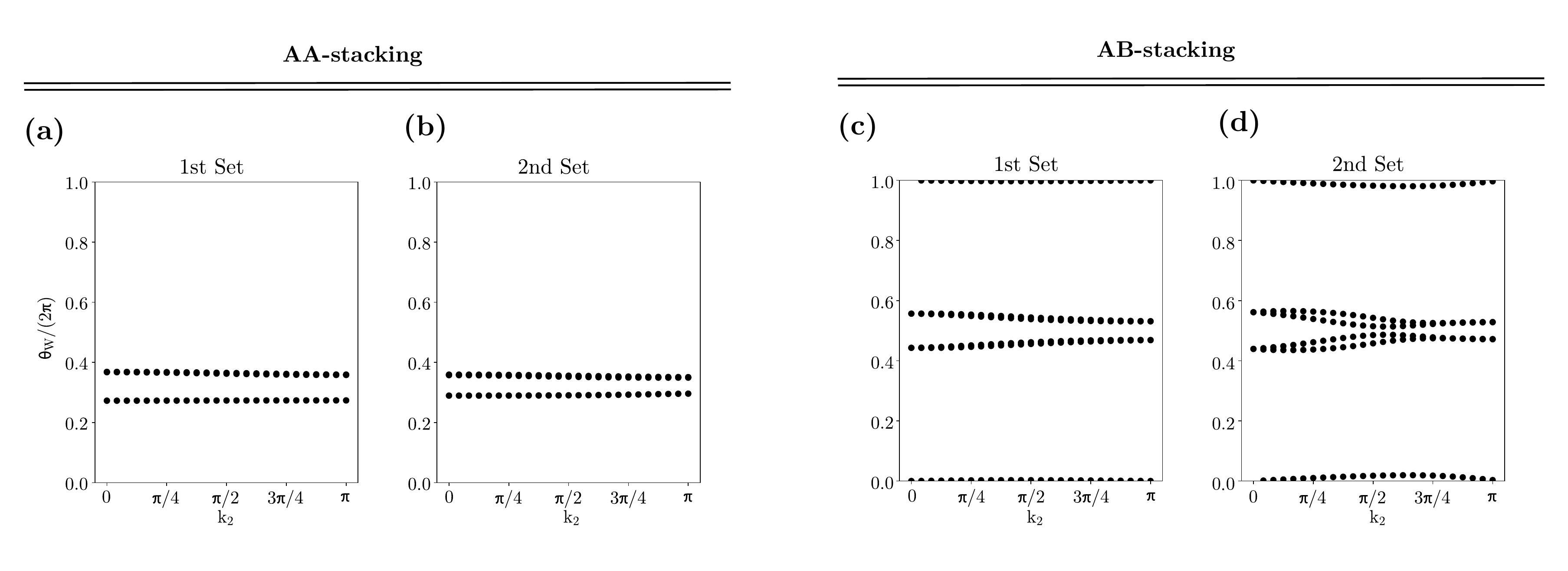}
    \caption{The Wilson loops of twisted \ch{SnSe2} at $\SI{3.89}{\degree}$, computed along the $\vec{b}_{M_1}$ direction and plotted as a function of momentum along the $\vec{b}_{M_2}$ direction. (a) and (b) show the results for the AA-stacked configuration, while (c) and (d) correspond to the AB-stacked configuration. The first and second lowest sets of conduction bands are displayed in (a) and (c), and (b) and (d), respectively.}
    \label{app:fig:twist-AA-AB-wcc}
\end{figure}

\paragraph{\textbf{Wilson loops}.}Finally, we compute the Wilson loops of the lowest conduction bands to explore their topology. As shown in \cref{app:fig:twist-AA-AB-wcc}, the two lowest isolated sets of conduction bands for both twisted AA and AB configurations exhibit no winding in their Wilson loops, indicating topologically trivial bands. However, this does not imply that the system's low-energy physics is trivial. The presence of multiple valleys and the emergence of new momentum-space non-symmorphic symmetries -- discussed in detail in \cref{app:sec:SnS_SnSe_twist_general_woGradient,app:sec:add_sym} -- introduce nontrivial physical consequences not found in other systems. Note also that the Wilson loop for the AB-stacked configuration is approximately pinned at $\theta_W = 0$ and $\theta_W = \pi$. Although no \emph{exact} symmetries enforce this behavior, it arises due to \emph{approximate} symmetries, which will be explored in \cref{app:sec:add_sym}. We further discuss the implications of these approximate symmetries in \cref{app:sec:fitted_models}.

\FloatBarrier

\subsection{Bilayer \ch{ZrS2}}\label{app:sec:DFT_bilayer:ZrS2}

This section presents the same \textit{ab initio} results for \ch{ZrS2} as discussed for \ch{SnSe2} in \cref{app:sec:DFT_bilayer:SnSe2}. For simplicity and ease of comparison, we employ similar figure layout and only highlight the differences with \ch{SnSe2}. 

\subsubsection{Untwisted bilayer \ch{ZrS2}}

\begin{figure}[t]
	\centering
	\includegraphics[width=0.4\textwidth]{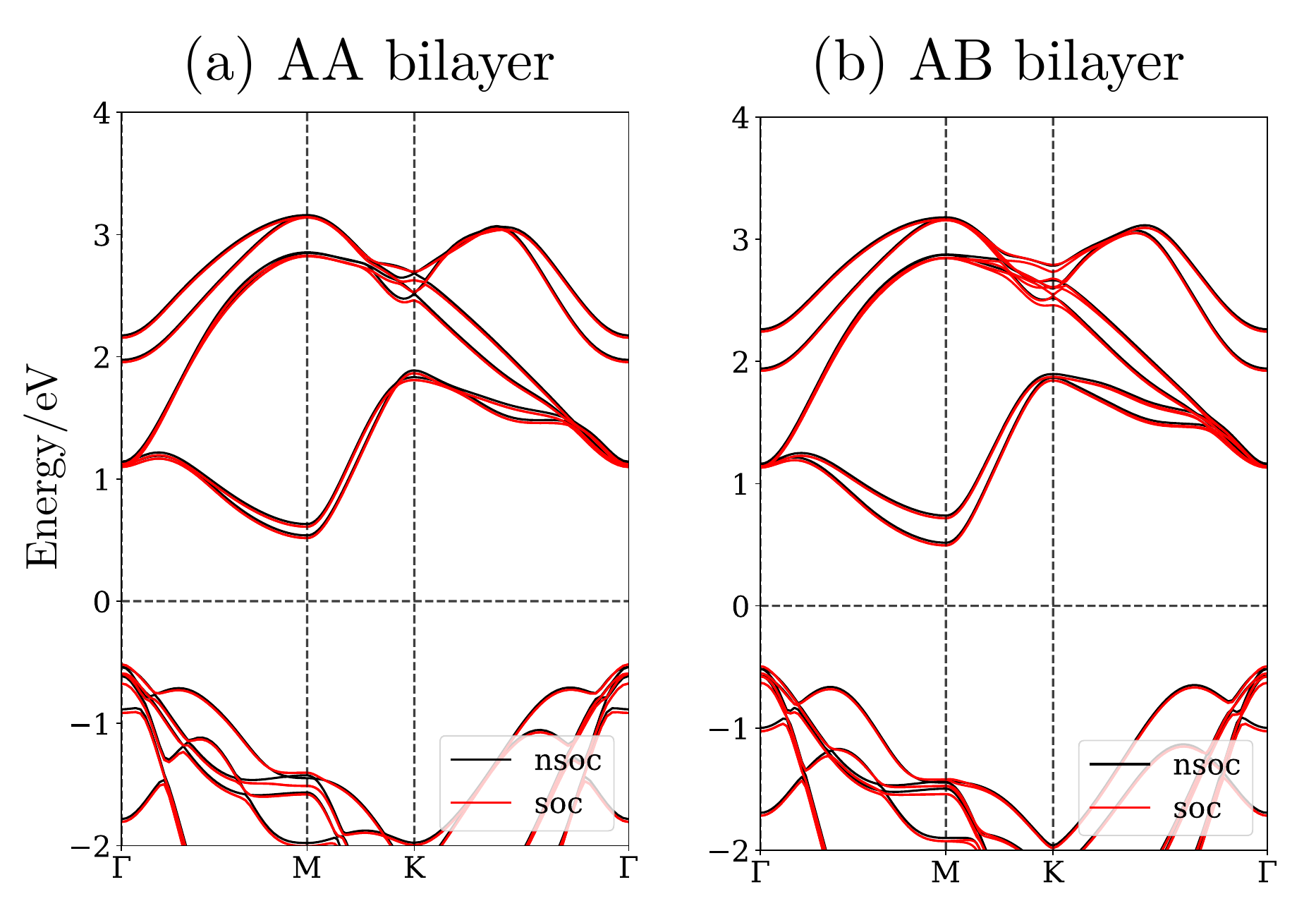}\subfloat{\label{app:fig:ZrS2-bilayer-bands:a}}\subfloat{\label{app:fig:ZrS2-bilayer-bands:b}}\caption{Band structures of the untwisted AA- and AB-stacked bilayer \ch{ZrS2}. The DFT bands obtained with (without) SOC are shown in red (black). (a) shows the band structure for the AA-stacked bilayer, while (b) shows the same for the AB-stacked configuration.}
	\label{app:fig:ZrS2-bilayer-bands}
\end{figure}

The AA- and AB-stacked untwisted bilayer \ch{ZrS2} have the same crystal structures as the untwisted bilayer \ch{SnSe2}. The corresponding band dispersions are shown in \cref{app:fig:ZrS2-bilayer-bands}. The SOC opens small hybridization gaps at the band crossings but has only minor effects on the CBM at M. The layer splitting in bilayer \ch{ZrS2} is \SI{93}{\milli\electronvolt} for the AA-stacking configuration and \SI{222}{\milli\electronvolt} for the AB-stacking one. 

\subsubsection{Twisted bilayer \ch{ZrS2}}

\begin{figure}[t]
	\centering
	\includegraphics[width=0.6\textwidth]{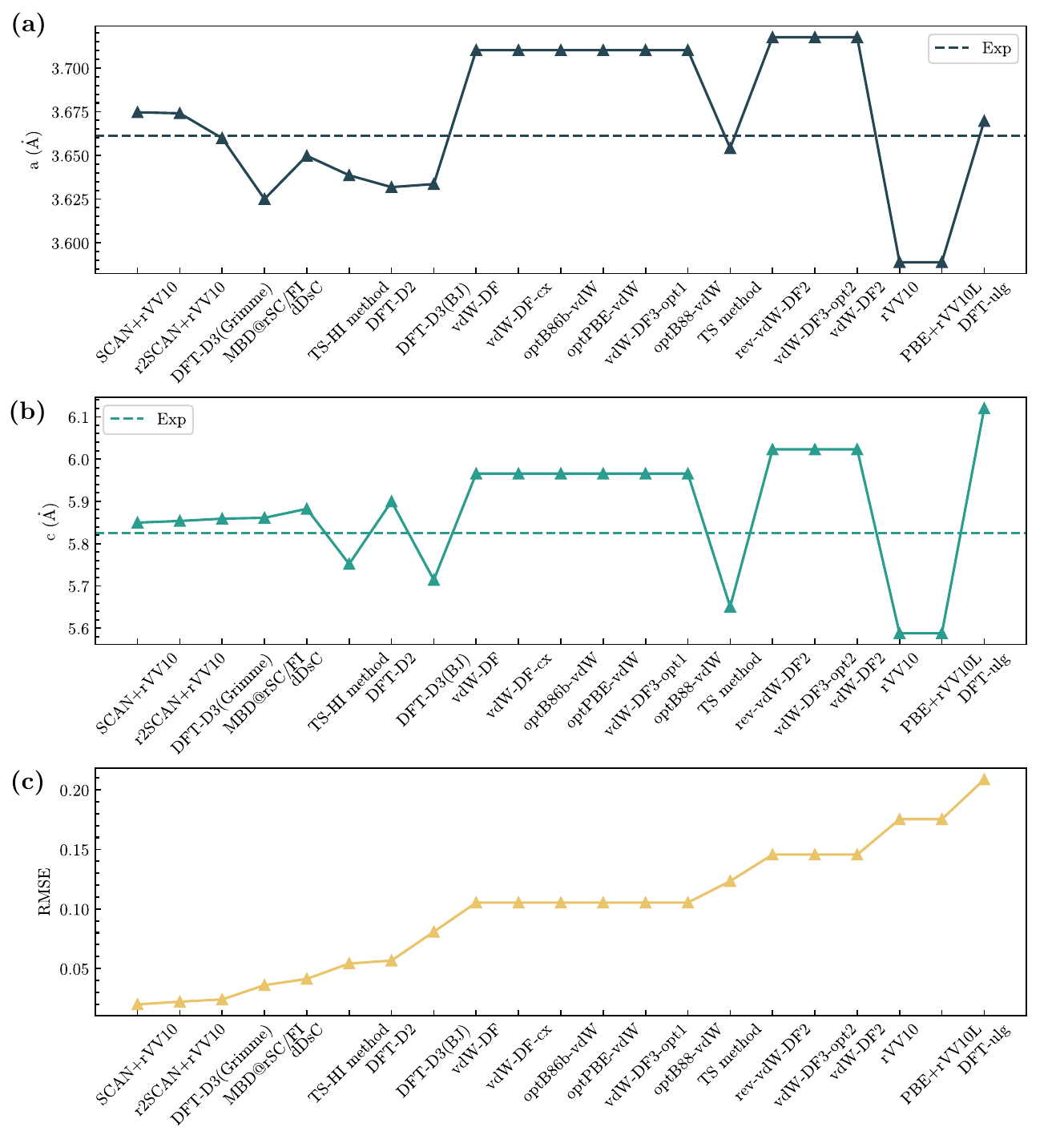}
	\subfloat{\label{app:fig:ZrS2-IVDW-test:a}}\subfloat{\label{app:fig:ZrS2-IVDW-test:b}}\subfloat{\label{app:fig:ZrS2-IVDW-test:c}}\caption{Comparison of van der Waals functionals for describing the crystal structure of \ch{ZrS2}. The layout of the figure is identical to \cref{app:fig:ZrS2-IVDW-test}, but applied to \ch{ZrS2}. Experimental data is sourced from Ref.~\cite{WU19c}.}
\label{app:fig:ZrS2-IVDW-test}
\end{figure}

\begin{figure}[t]
	\centering
	\includegraphics[width=0.8\textwidth]{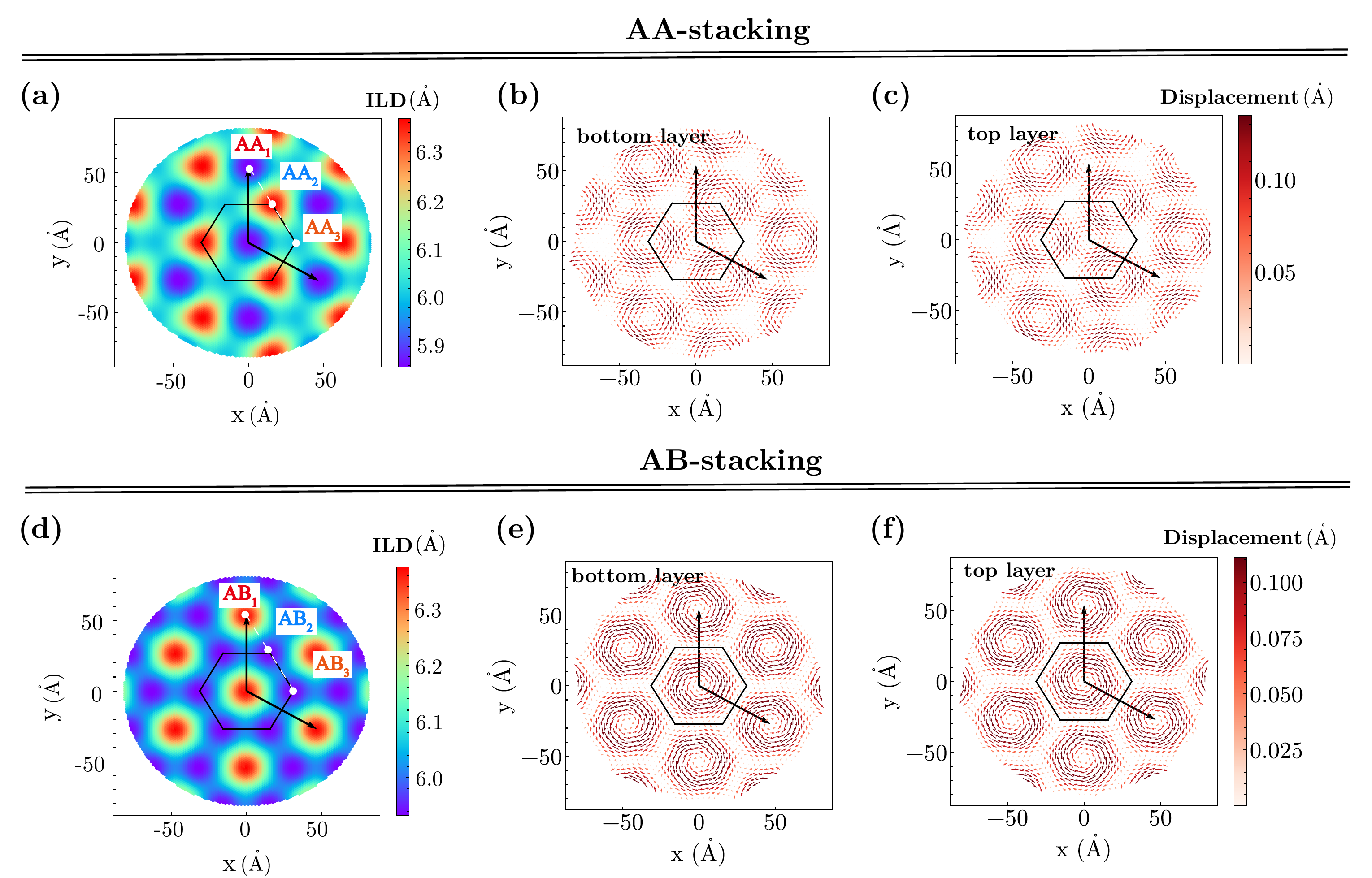}\subfloat{\label{app:fig:ZrS-AA-AB-disp:a}}\subfloat{\label{app:fig:ZrS-AA-AB-disp:b}}\subfloat{\label{app:fig:ZrS-AA-AB-disp:c}}\subfloat{\label{app:fig:ZrS-AA-AB-disp:d}}\subfloat{\label{app:fig:ZrS-AA-AB-disp:e}}\subfloat{\label{app:fig:ZrS-AA-AB-disp:f}}\caption{Lattice relaxation results of $\SI{3.89}{\degree}$ twisted AA and AB \ch{ZrS2}. (a), (b), and (c) are interlayer distances, the intralayer displacement of the bottom and top layers for the fully-relaxed AA stacking structure, respectively. (d), (e), and (f) are the same but for the AB stacking structure.}\label{app:fig:ZrS-AA-AB-disp}\end{figure}

The twisted AA- and AB-stacked bilayer \ch{ZrS2} share similar crystal structures and the same space groups as twisted bilayer \ch{SnSe2}. After evaluating 21 vdW corrections, as shown in \crefrange{app:fig:ZrS2-IVDW-test:a}{app:fig:ZrS2-IVDW-test:c}, we chose the DFT-D3 method by Grimme~\cite{GRI10} for the vdW correction in \ch{ZrS2} ({\it i.e.}, the third best one in \cref{app:fig:ZrS2-IVDW-test}). This method provides lattice constants close to experimental values and is less computationally demanding than the SCAN correction. \Cref{app:fig:ZrS-AA-AB-disp} illustrates the ILD and intralayer displacements of atoms in \ch{ZrS2} after full relaxation, showing substantial similarities with \ch{SnSe2}, although the variations in ILD and the magnitude of intralayer displacements are about half of those in \ch{SnSe2}.

\begin{figure}[t]
    \centering
    \includegraphics[width=\textwidth]{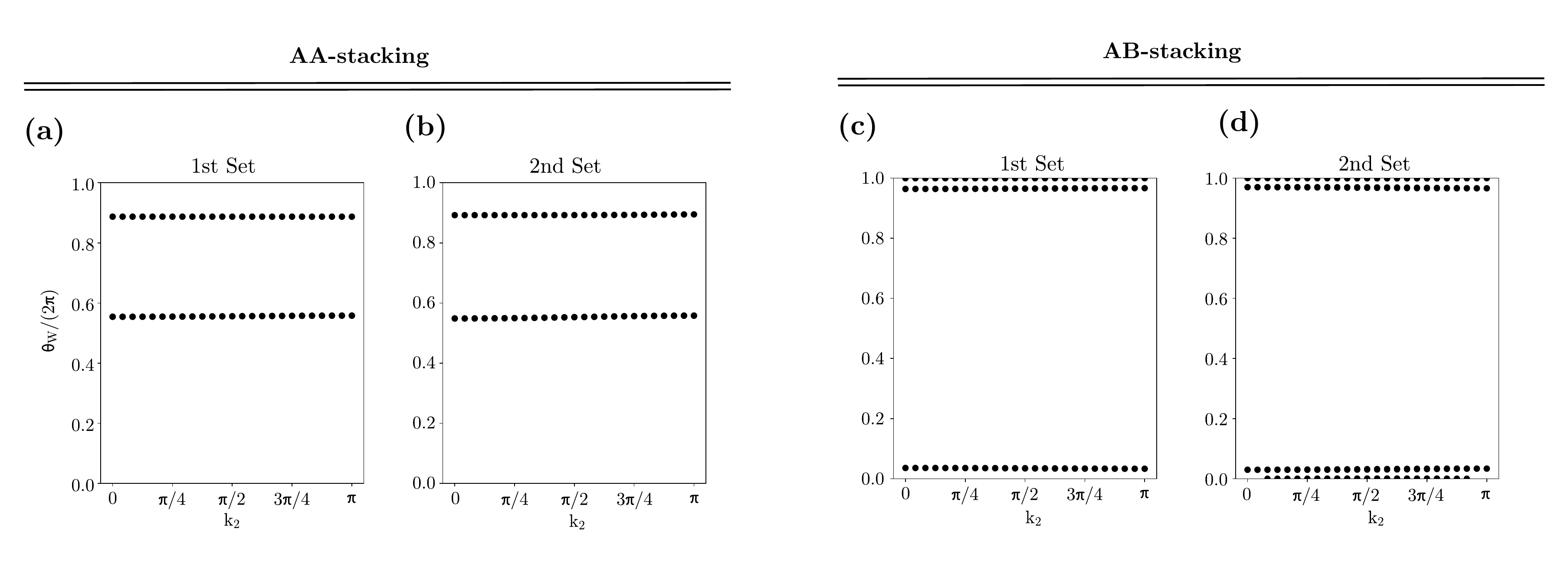}
    \caption{The Wilson loops of twisted \ch{ZrS2} at $\SI{3.89}{\degree}$. The layout of the figure is identical to \cref{app:fig:twist-AA-AB-wcc}, but applied to \ch{ZrS2}.}
    \label{app:fig:twist-ZrS2-wcc}
\end{figure}

The \textit{ab initio} band structures of \ch{ZrS2} for different angles and both stacking configurations are shown in \cref{app:fig:ZrS-AA-AB-disp}. The bands are qualitatively similar to the ones of \ch{SnSe2}, but the relevant energy scales are reduced by roughly $30\%-40\%$. The bandwidths of the lowest set of bands is given in  \cref{tab:bandwidth-ZrS2} and similarly show a reduction in the energy scale compared to \ch{SnSe2}. Finally, in \cref{app:fig:twist-ZrS2-wcc}, we also plot the Wilson loop of the first two sets of conduction bands at $\SI{3.89}{\degree}$. As is the case of \ch{SnSe2}, there is no winding in the Wilson loop spectrum. Further results at other angles for both \ch{SnSe2} and \ch{ZrS2}, as well as a discussion of the Wilson loop spectra are provided in \cref{app:sec:fitted_models}.

\begin{figure}[t]
	\centering
	\includegraphics[width=\textwidth]{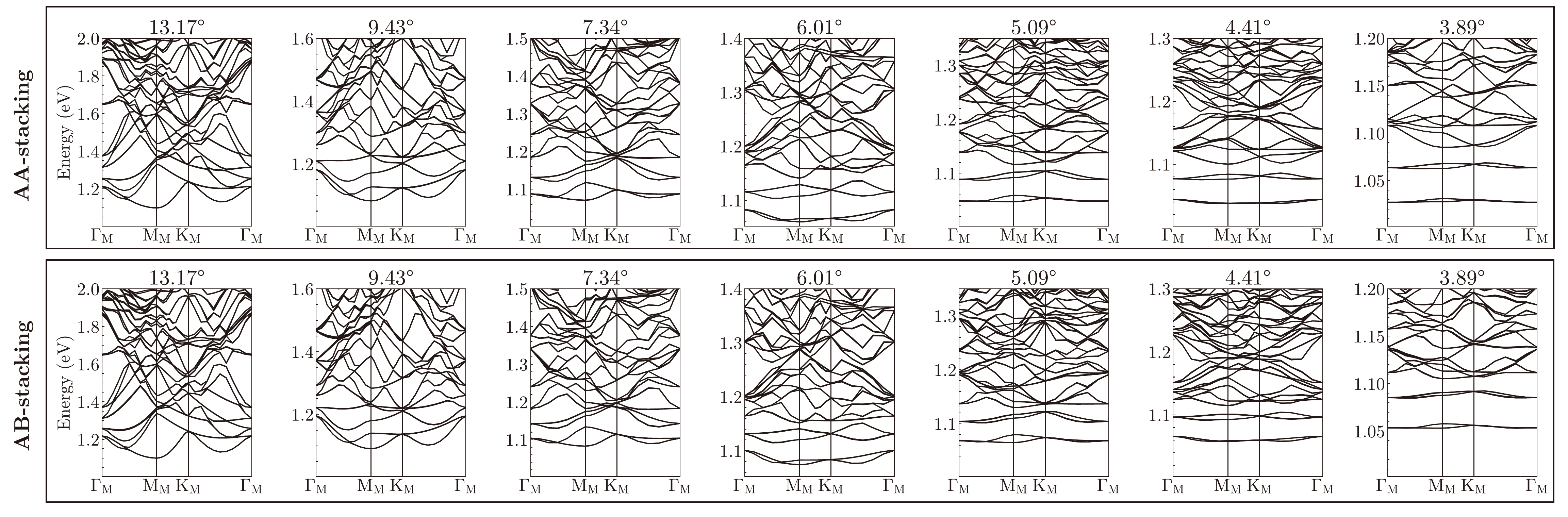}
	\caption{The \textit{ab initio} band structure of twisted AA-stacked (a-g) and AB-stacked (h-n) bilayer \ch{ZrS2} from \SI{13.17}{\degree} to \SI{3.89}{\degree}. The lowest group of conduction bands isolated from other energy bands consists of six bands stemming from the three inequivalent M valleys of monolayer \ch{ZrS2}. The bandwidths of the lowest group of six bands are listed in \cref{tab:bandwidth-ZrS2}.}
    \label{app:fig:twist-ZrS2-bands}
\end{figure}

\begin{table}[t]
	\begin{tabular}{|c|c|c|c|c|c|c|c|}
	\hline
	Bandwidth ($\si{\milli\electronvolt}$) & \SI{13.17}{\degree} & \SI{9.43}{\degree} & \SI{7.34}{\degree} & \SI{6.01}{\degree} & \SI{5.09}{\degree} & \SI{4.41}{\degree} & \SI{3.89}{\degree} \\ \hline
	AA         & 232            & 100           & 47            & 23            & 12            & 6           & 4             \\ \hline
	AB         & 237            & 105           & 51            & 26            & 14            & 8           & 5             \\ \hline
	\end{tabular}
	\caption{The bandwidths of the lowest set of conduction bands of twisted AA- and AB-stacked \ch{ZrS2} for angles ranging from \SI{13.17}{\degree} to \SI{3.89}{\degree}.}
	\label{tab:bandwidth-ZrS2}
\end{table}

\FloatBarrier

\section{Single-particle Bistritzer-MacDonald models for twisted \ch{SnSe2} and \ch{ZrS2} bilayers}\label{app:sec:SnS_SnSe_twist}

In this \siSection{}, we construct single-particle moir\'e heterostructure models for twisted \ch{SnSe2} bilayers. We start by obtaining a Bistritzer-MacDonald (BM)~\cite{BIS11} model using a two-center first monolayer harmonic approximation for the interlayer hopping amplitude. We constrain the form of the interlayer hopping amplitude using the symmetries of the twisted heterostructure and then further simplify the former using the symmetries of the untwisted bilayer. We then rewrite the resulting moir\'e Hamiltonian in the more familiar notation analogous to that used in the case of twisted bilayer graphene (TBG) by Refs.~\cite{XIE21,BER21b,LIA21,BER21a,SON21,BER21,CAL21,XIE21b} and discuss its symmetries at the general level. Importantly, this moir\'e Hamiltonian is defined on a kagome plane-wave lattice in momentum space. 

The two-center first monolayer harmonic approximation employed in this \siSection{} will be relaxed first in \cref{app:sec:SnS_SnSe_twist_general_woGradient}, where we derive the moir\'e single-particle potential without gradient terms, and then in \cref{app:sec:SnS_SnSe_twist_general}, where we additionally define and include gradient terms for the moir\'e potential. In \cref{app:sec:add_sym}, we also discuss the additional \emph{approximate} symmetries of the moir\'e Hamiltonian arising under different physically-relevant limits, and introduce simplified models, which, as shown in \cref{app:sec:fitted_models}, accurately describe twisted \ch{SnSe2} and \ch{ZrS2}.

\subsection{Deriving a BM model for twisted \ch{SnSe2} and \ch{ZrS2}}\label{app:sec:SnS_SnSe_twist:bl_model_derivation}

Starting from the single-layer model of \ch{SnSe2} from \cref{app:sec:DFT_single_layer:snse2}, we now build a BM model for the corresponding moir\'e heterostructures. We will consider both the AA- and AB-stacking configurations, which were introduced and discussed in \cref{app:sec:DFT_bilayer:SnSe2:twisted}.

\subsubsection{AA-stacking configuration}\label{app:sec:SnS_SnSe_twist:bl_model_derivation:AA}

In the context of the twisted bilayer heterostructures, we will employ the same notation as in \cref{app:sec:DFT_single_layer:snse2:model} for the fermionic operators, to which we will add a layer index. As such, $\hat{a}^\dagger_{\vec{R},s,l}$ corresponds to the Wannier orbital of the lowest conduction band\footnote{In \ch{SnSe2}, the lowest (spinful) conduction band is gapped from the rest of the spectrum and is described by a \emph{single} (spinful) Wannier orbital throughout the entire monolayer BZ. For \ch{ZrS2}, the lowest conduction band is part of a gapped group of three bands. However, we do not need to employ three Wannier orbitals because, at the M point of the monolayer BZ ({\it i.e.}{}, the low-energy electronic states relevant for the twisted heterostructure), there is a single (spinful) conduction band. This band is spanned (around the M point) by a Wannier orbital with symmetry properties analogous to an effective $s$ orbital at the $1a$ Wyckoff position, similar to \ch{SnSe2}.} and creates an electron within layer $l = \pm$ located at position $\mathcal{R}_{\theta, l} \vec{R}$, where the rotation matrix $\mathcal{R}_{\theta, l}$ is given by
\begin{equation}
	\label{app:eqn:notRrotation}
	\mathcal{R}_{\theta, l} =
	\begin{pmatrix}
		\cos \left(\frac{\theta  l}{2}\right) & -\sin \left(\frac{\theta  l}{2}\right) \\
		\sin \left(\frac{\theta  l}{2}\right) & \cos \left(\frac{\theta  l}{2}\right) \\	
	\end{pmatrix}.
\end{equation}

Correspondingly, the Fourier-transformed fermion operators for the two layers read as 
\begin{equation}
	\label{app:eqn:lattice_FFT_AA}
	\hat{a}^\dagger_{\vec{k},s,l} = \frac{1}{\sqrt{N}} \sum_{\vec{R}} \hat{a}^\dagger_{\vec{R},s,l} e^{i \vec{k} \cdot \mathcal{R}_{\theta, l} \vec{R} }.
\end{equation}
We will also find it useful to employ the Dirac bra-ket notation and define 
\begin{equation}
	\ket{\vec{k},s,l} = \hat{a}^\dagger_{\vec{k},s,l} \ket{0} \qq{and}
	\ket{\vec{R},s,l} = \hat{a}^\dagger_{\vec{R},s,l} \ket{0}.
\end{equation}

In deriving a BM model for the twisted bilayer heterostructure, we will adapt the method of Refs.~\cite{BIS11,SON19}. We start by defining the following matrix elements
\begin{equation}
	\label{app:eqn:bl_mat_elems_AA}
	\left[ h_{\text{AA}}^{\mathrm{bl}} \left( \vec{k},\vec{k}' \right) \right]_{s_1 l_1;s_2 l_2} = \mel**{\vec{k},s_1,l_1}{\mathcal{H}_{\text{AA}}}{\vec{k}',s_2,l_2},
\end{equation}
where $\mathcal{H}_{\text{AA}}$ is the (yet unknown) single-particle Hamiltonian operator of the system. The intralayer matrix elements can be approximated by the single-layer Hamiltonian from \cref{app:eqn:k_dot_p_sl} and are given by
\begin{equation}
	\label{app:eqn:intralayer_mel_AA}
	\left[ h_{\text{AA}}^{\mathrm{bl}} \left( \vec{k},\vec{k}' \right) \right]_{s_1 l;s_2 l} = \delta_{\vec{k},\vec{k}'} \left[ h^{\mathrm{sl}} \left(  \mathcal{R}^{-1}_{\theta, l} \vec{k} \right) \right]_{s_1 s_2}.
\end{equation}
This approximation ignores any intralayer moir\'e potential terms which could arise from relaxation effects~\cite{CAR18}, as it happens in the case of \ch{MoTe2}~\cite{YU24a,JIA24}.
To obtain the interlayer coupling, we first transform \cref{app:eqn:bl_mat_elems_AA} in real space to afford 
\begin{equation}
	\label{app:eqn:interlayer_mel_AA_1}
	\left[ h_{\text{AA}}^{\mathrm{bl}} \left( \vec{k},\vec{k}' \right) \right]_{s_1 l;s_2 (-l)} = \frac{1}{N} \sum_{\vec{R},\vec{R}'} e^{i \vec{k}' \cdot \mathcal{R}_{\theta, -l} \vec{R}' } e^{-i \vec{k} \cdot \mathcal{R}_{\theta, l} \vec{R}} \mel**{\vec{R},s_1,l}{\mathcal{H}_{\text{AA}}}{\vec{R}',s_2,-l}.
\end{equation}
In this \siSection{}, we will follow Ref.~\cite{BIS11} and make a two-center, tight-binding approximation for the interlayer hopping amplitude, {\it i.e.}{}
\begin{align}
	\mel**{\vec{R},s_1,l}{\mathcal{H}_{\text{AA}}}{\vec{R}',s_2,-l} &= t^{l,\text{AA}}_{s_1 s_2} \left( \mathcal{R}_{\theta, -l} \vec{R}'  - \mathcal{R}_{\theta, l}  \vec{R} \right) = \frac{1}{N \Omega_{\mathrm{sl}}} \sum_{\vec{q},\vec{g}} t^{l,\text{AA}}_{s_1 s_2} \left( \vec{q} + \vec{g} \right) e^{-i \left( \vec{q} + \vec{g} \right) \cdot \left(\mathcal{R}_{\theta, -l} \vec{R}' - \mathcal{R}_{\theta, l} \vec{R} \right) },
	\label{app:eqn:two_center_approx_AA}
\end{align}
where $\Omega_{\mathrm{sl}}$ is the surface area of the single-layer material unit cell, while $t^{l,\text{AA}}_{s_1 s_2} \left( \vec{r} \right)$ and $t^{l,\text{AA}}_{s_1 s_2} \left( \vec{q} + \vec{g} \right)$ are the tunneling amplitude between two effective $s$ Wannier orbitals located a distance $\vec{r}$ apart and its Fourier transformation, respectively. In \cref{app:eqn:two_center_approx_AA}, the sum over $\vec{q}$ runs over the momenta of the \emph{unrotated} BZ of the single-layer material, while $\vec{g}$ are the \emph{unrotated} reciprocal lattice vectors of the single-layer material. It is worth noting, however, that due to the extended nature of the effective $s$-orbitals comprising the lowest two conduction bands of \ch{SnSe2}, the two-center approximation derived here might not furnish an accurate model for the resulting moir\'e Hamiltonian. The delocalized nature of the $s$-orbitals suggest that assisted hopping will likely take place between the layers, similarly to \ch{MoTe2}, where the indirect hopping processes also give rise to additional intralayer contributions beyond \cref{app:eqn:intralayer_mel_AA}. For the sake of completeness, we first will carry out a complete derivation of the moir\'e Hamiltonian in the two-center approximation, which will then be relaxed in \cref{app:sec:SnS_SnSe_twist_general_woGradient,app:sec:SnS_SnSe_twist_general}. 

Plugging \cref{app:eqn:two_center_approx_AA} into \cref{app:eqn:interlayer_mel_AA_1}, we obtain
\begin{align}
	\left[ h_{\text{AA}}^{\mathrm{bl}} \left( \vec{k},\vec{k}' \right) \right]_{s_1 l;s_2 (-l)} &= \sum_{\substack{\vec{q},\vec{g} \\ \vec{g}', \vec{g}''}} 
	\frac{t^{l,\text{AA}}_{s_1 s_2} \left( \vec{q} + \vec{g} \right)}{\Omega_{\mathrm{sl}}} 
	\delta_{\vec{q} + \vec{g} - \vec{k}', \mathcal{R}_{\theta, -l} \vec{g}'} \delta_{\vec{k} - \vec{q} - \vec{g}, \mathcal{R}_{\theta, l} \vec{g}''} \nonumber \\
	&= \sum_{\vec{g}', \vec{g}''} \frac{t^{l,\text{AA}}_{s_1 s_2} \left( \vec{k}' +  \mathcal{R}_{\theta, -l} \vec{g}' \right)}{\Omega_{\mathrm{sl}}} 
	\delta_{\vec{k} - \vec{k}', \mathcal{R}_{\theta, l} \vec{g}'' + \mathcal{R}_{\theta, -l} \vec{g}'} \label{app:eqn:interlayer_mel_AA_2},
\end{align}
where the sum over $\vec{g}'$ and $\vec{g}''$ runs over the unrotated reciprocal lattice vectors of the monolayer material. The low-energy physics of the system near charge neutrality and for small twist angles $\theta \ll 1$ is dictated by the momentum states near the M points of the two layers. As a result, we can take $\vec{k} = C_{3z}^\eta \vec{K}^{l}_{M} + \delta \vec{k}$ and $\vec{k}' = C_{3z}^{\eta'} \vec{K}^{-l}_{M} + \delta \vec{k}'$ in \cref{app:eqn:interlayer_mel_AA_2}, where $0 \leq \eta, \eta' \leq 2$ index the three inequivalent valleys within each layer, $\abs{\delta \vec{k}}, \abs{\delta \vec{k}'} \ll \abs{\vec{K}_{M}}$ are small momentum displacements, and we have introduced the following notation for the M points of the two layers
\begin{equation}
	\vec{K}^{l}_{M} =  \mathcal{R}_{\theta, l} \vec{K}_{M}.
\end{equation}
At the same time, we assume that the Fourier transformation of the interlayer tunneling amplitude decays quickly on the single-layer reciprocal lattice scale~\cite{BIS11}, such that in the sum over $\vec{g}'$ in \cref{app:eqn:interlayer_mel_AA_2}, only the leading terms need to be considered ({\it i.e.}{} the terms for which $\abs{\vec{k}' +  \mathcal{R}_{\theta, -l} \vec{g}'} = \abs{\delta \vec{k}' +  \mathcal{R}_{\theta, -l} \left( \vec{g}' + C_{3z}^{\eta'} \vec{K}_{M} \right)} \approx \abs{ \vec{K}_{M} }$). We call the resulting approximation \emph{the two-center first monolayer harmonic approximation}. There are only two such terms, which correspond to $\vec{g}' = (-1)^n C_{3z}^{\eta'} \vec{K}_{M} - C_{3z}^{\eta'} \vec{K}_{M}$ for $n=0,1$. At the same time, since $\abs{\delta \vec{k}'} \ll \abs{\vec{K}_{M}}$, we can assume that $t^{l,\text{AA}}_{s_1 s_2} \left( \delta \vec{k}' + (-1)^n C_{3z}^{\eta'} \vec{K}^{-l}_{M} \right) \approx t^{l,\text{AA}}_{s_1 s_2} \left( (-1)^n C_{3z}^{\eta'} \vec{K}^{-l}_{M} \right) \approx t^{l,\text{AA}}_{s_1 s_2} \left( (-1)^n C_{3z}^{\eta'} \vec{K}_{M} \right) $. With these simplifications, \cref{app:eqn:interlayer_mel_AA_2} can be rewritten as
\begin{align}
	&\left[ h_{\text{AA}}^{\mathrm{bl}} \left( C_{3z}^\eta \vec{K}^{l}_{M} + \delta \vec{k} ,C_{3z}^{\eta'} \vec{K}^{-l}_{M} + \delta \vec{k}' \right) \right]_{s_1 l;s_2 (-l)} = \nonumber \\
	=& \sum_{n=0}^{1} \sum_{\vec{g}''} 
	\frac{ t^{l,\text{AA}}_{s_1 s_2} \left( (-1)^n C_{3z}^{\eta'} \vec{K}_{M} \right) }{\Omega_{\mathrm{sl}}} 
	\delta_{ C_{3z}^\eta \vec{K}^{l}_{M} + \delta \vec{k} - C_{3z}^{\eta'} \vec{K}^{-l}_{M} - \delta \vec{k}', \mathcal{R}_{\theta, l} \vec{g}'' + \mathcal{R}_{\theta, -l} \left[ (-1)^n C_{3z}^{\eta'} \vec{K}_{M} - C_{3z}^{\eta'} \vec{K}_{M} \right] } \nonumber \\
	=& \sum_{n=0}^{1}  
	\frac{ t^{l,\text{AA}}_{s_1 s_2} \left( (-1)^n C_{3z}^\eta \vec{K}_{M} \right) }{\Omega_{\mathrm{sl}}} 
	\delta_{ \delta \vec{k} - \delta \vec{k}' + C_{3z}^\eta \left( \vec{K}^{l}_{M} - \vec{K}^{-l}_{M} \right), \left( \mathcal{R}_{\theta, l} - \mathcal{R}_{\theta, -l} \right) \left[ 1 - (-1)^n \right] C_{3z}^\eta \vec{K}_{M}} \delta_{\eta \eta'} \nonumber \\
	=& \sum_{n=0}^{1}  
	\frac{ t^{l,\text{AA}}_{s_1 s_2} \left( (-1)^n C_{3z}^\eta \vec{K}_{M} \right) }{\Omega_{\mathrm{sl}}} 
	\delta_{ \delta \vec{k} - \delta \vec{k}', (-1)^n\left( \mathcal{R}_{\theta, -l} - \mathcal{R}_{\theta, l} \right) C_{3z}^\eta \vec{K}_{M}} \delta_{\eta \eta'}
	\label{app:eqn:interlayer_mel_AA_3}.
\end{align}
In going from the second to the third line of \cref{app:eqn:interlayer_mel_AA_3}, we have used the fact that $\abs{\delta \vec{k}}, \abs{\delta \vec{k}'} \ll \abs{\vec{K}_{M}}$, which implies that the $\delta$-function can only be non-zero if $\vec{g}''= - \mathcal{R}_{\theta, -l} \left[ (-1)^n C_{3z}^{\eta'} \vec{K}_{M} - C_{3z}^{\eta'} \vec{K}_{M} \right]$. Introducing the following auxiliary vectors
\begin{equation}
	\label{app:eqn:q_vecs}
	\vec{q}_\eta = C^{\eta}_{3z} \left( \vec{K}^{-}_{M} - \vec{K}^{+}_{M} \right),
	\qq{for} 0 \leq \eta \leq 2,
\end{equation}
as well as the rescaled hopping amplitude Fourier transformation
\begin{equation}
	\tilde{t}^{l,\text{AA}}_{s_1 s_2} \left( \vec{q} + \vec{g} \right) = \frac{t^{l,\text{AA}}_{s_1 s_2} \left( \vec{q} + \vec{g} \right)}{\Omega_{\mathrm{sl}}},
\end{equation}
we can rewrite the interlayer matrix elements as 
\begin{equation}
	\left[ h_{\text{AA}}^{\mathrm{bl}} \left( C_{3z}^\eta \vec{K}^{l}_{M} + \delta \vec{k} ,C_{3z}^{\eta'} \vec{K}^{-l}_{M} + \delta \vec{k}' \right) \right]_{s_1 l;s_2 (-l)} = \sum_{n=0}^{1}  
	\tilde{t}^{l,\text{AA}}_{s_1 s_2} \left( (-1)^n C_{3z}^\eta \vec{K}_{M} \right)
	\delta_{ \delta \vec{k} - \delta \vec{k}', l (-1)^n \vec{q}_\eta} \delta_{\eta \eta'}
	\label{app:eqn:interlayer_mel_AA_4},
\end{equation}
where the reader is reminded that $l$ and $s_1, s_2$ refer to the layer and spin indices, respectively. In contrast to TBG, the absence of additional phase factors in \cref{app:eqn:interlayer_mel_AA_4} arises from the fact that the effective $s$-orbitals in the monolayer material are centered at the origin of the unit cell. Furthermore, unlike TBG, within a single valley, momentum states are coupled only along a single direction (determined by $\vec{q}_{\eta}$ in valley $\eta$). As will be discussed in \cref{app:sec:add_sym:first_monolayer_harmonic}, this leads to the first monolayer harmonic model having discrete moir\'e translation symmetry in only one direction and continuous translation symmetry in the other.

\subsubsection{AB-stacking configuration}\label{app:sec:SnS_SnSe_twist:bl_model_derivation:AB}

The derivation of the single-particle Hamiltonian for the AB-stacking configuration proceeds analogously to the one shown for the AA-stacking case. This time, however, $\hat{a}^\dagger_{\vec{R},s,l}$ will create an electron within layer $l$ located at position $l \mathcal{R}_{\theta, l} \vec{R}$ (note the additional layer index specifying that the bottom layer is additionally rotated by $\SI{180}{\degree}$ around the $\hat{\vec{z}}$ direction relative to the AA-stacking configuration). The Fourier-transformed operators for the two layers $l=\pm$ are now defined by
\begin{equation}
	\label{app:eqn:lattice_FFT_AB}
	\hat{a}^\dagger_{\vec{k},s,l} = \frac{1}{\sqrt{N}} \sum_{\vec{R}} \hat{a}^\dagger_{\vec{R},s,l} e^{i l \vec{k} \cdot \mathcal{R}_{\theta, l} \vec{R} }.
\end{equation}
In this modified momentum basis, the matrix elements of the AB-stacked configurations are defined similarly to \cref{app:eqn:bl_mat_elems_AA}
\begin{equation}
	\label{app:eqn:bl_mat_elems_AB}
	\left[ h_{\text{AB}}^{\mathrm{bl}} \left( \vec{k},\vec{k}' \right) \right]_{s_1 l_1;s_2 l_2} = \mel**{\vec{k},s_1,l_1}{\mathcal{H}_{\text{AB}}}{\vec{k}',s_2,l_2},
\end{equation}
but with the intralayer matrix elements being given by
\begin{equation}
	\label{app:eqn:intralayer_mel_AB}
	\left[ h_{\text{AB}}^{\mathrm{bl}} \left( \vec{k},\vec{k}' \right) \right]_{s_1 l;s_2 l} = \delta_{\vec{k},\vec{k}'} \left[ h^{\mathrm{sl}} \left(  l\mathcal{R}^{-1}_{\theta, l} \vec{k} \right) \right]_{s_1 s_2}.
\end{equation}
Note that because the monolayer Hamiltonian $h^{\mathrm{sl}} \left(  \vec{k} \right)$ is even in $\vec{k}$, the introduction of the additional $l$ factor compared with \cref{app:eqn:intralayer_mel_AA} does not change the intralayer part of the moir\'e Hamiltonian between the AA- and AB-stacking configurations at the level of the two-center first monolayer harmonic approximation.

At the same time, the interlayer part of the Hamiltonian is given by
\begin{equation}
	\label{app:eqn:interlayer_mel_AB_1}
	\left[ h_{\text{AB}}^{\mathrm{bl}} \left( \vec{k},\vec{k}' \right) \right]_{s_1 l;s_2 (-l)} = \frac{1}{N} \sum_{\vec{R},\vec{R}'} e^{-i l \vec{k}' \cdot \mathcal{R}_{\theta, -l} \vec{R}' } e^{-i l \vec{k} \cdot \mathcal{R}_{\theta, l} \vec{R}} \mel**{\vec{R},s_1,l}{\mathcal{H}_{\text{AB}}}{\vec{R}',s_2,-l},
\end{equation}
for which the same two-center approximation as in \cref{app:eqn:two_center_approx_AA} gives
\begin{align}
	\mel**{\vec{R},s_1,l}{\mathcal{H}_{\text{AB}}}{\vec{R}',s_2,-l} &= t^{l,\text{AB}}_{s_1 s_2} \left( -l \mathcal{R}_{\theta, -l} \vec{R}'  - l \mathcal{R}_{\theta, l}  \vec{R} \right) = \frac{1}{N \Omega_{\mathrm{sl}}} \sum_{\vec{q},\vec{g}} t^{l,\text{AB}}_{s_1 s_2} \left( \vec{q} + \vec{g} \right) e^{i \left( \vec{q} + \vec{g} \right) \cdot \left(l \mathcal{R}_{\theta, -l} \vec{R}' + l \mathcal{R}_{\theta, l} \vec{R} \right) }.
	\label{app:eqn:two_center_approx_AB}
\end{align}
From \cref{app:eqn:interlayer_mel_AB_1,app:eqn:two_center_approx_AB}, we find that
\begin{align}
	\left[ h_{\text{AB}}^{\mathrm{bl}} \left( \vec{k},\vec{k}' \right) \right]_{s_1 l;s_2 (-l)} &= \sum_{\substack{\vec{q},\vec{g} \\ \vec{g}', \vec{g}''}} 
	\frac{t^{l,\text{AB}}_{s_1 s_2} \left( \vec{q} + \vec{g} \right)}{\Omega_{\mathrm{sl}}} 
	\delta_{\vec{q} + \vec{g} - \vec{k}', -l \mathcal{R}_{\theta, -l} \vec{g}'} \delta_{\vec{k} - \vec{q} - \vec{g}, l \mathcal{R}_{\theta, l} \vec{g}''} \nonumber \\
	&= \sum_{\vec{g}', \vec{g}''} \frac{t^{l,\text{AB}}_{s_1 s_2} \left( \vec{k}' - l \mathcal{R}_{\theta, -l} \vec{g}' \right)}{\Omega_{\mathrm{sl}}} 
	\delta_{\vec{k} - \vec{k}', l \mathcal{R}_{\theta, l} \vec{g}'' - l \mathcal{R}_{\theta, -l} \vec{g}'} \label{app:eqn:interlayer_mel_AB_2}.
\end{align}
As in \cref{app:sec:SnS_SnSe_twist:bl_model_derivation:AA}, we take $\vec{k} = C_{3z}^\eta \vec{K}^{l}_{M} + \delta \vec{k}$ and $\vec{k}' = C_{3z}^{\eta'} \vec{K}^{-l}_{M} + \delta \vec{k}'$ in \cref{app:eqn:interlayer_mel_AB_2}, with $0 \leq \eta, \eta' \leq 2$ and assume that the Fourier transformation of the interlayer tunneling amplitude decays quickly on the single-layer reciprocal lattice scale~\cite{BIS11}. As such, in the sum over $\vec{g}'$ from \cref{app:eqn:interlayer_mel_AB_2}, we can restrict ourselves to the leading terms for which $\abs{\vec{k}' - l  \mathcal{R}_{\theta, -l} \vec{g}'} = \abs{\delta \vec{k}' + \mathcal{R}_{\theta, -l} \left( -l \vec{g}' + C_{3z}^{\eta'} \vec{K}_{M} \right)} \approx \abs{ \vec{K}_{M} }$. There are only two such terms, which correspond to $-l \vec{g}' = (-1)^n C_{3z}^{\eta'} \vec{K}_{M} - C_{3z}^{\eta'} \vec{K}_{M}$ for $n=0,1$. At the same time, since $\abs{\delta \vec{k}'} \ll \abs{\vec{K}_{M}}$, we can assume that $t^{l,\text{AB}}_{s_1 s_2} \left( \delta \vec{k}' + (-1)^n C_{3z}^{\eta'} \vec{K}^{-l}_{M} \right) \approx t^{l,\text{AB}}_{s_1 s_2} \left( (-1)^n C_{3z}^{\eta'} \vec{K}^{-l}_{M} \right) \approx t^{l,\text{AB}}_{s_1 s_2} \left( (-1)^n C_{3z}^{\eta'} \vec{K}_{M} \right) $. With these simplifications, \cref{app:eqn:interlayer_mel_AB_2} can be rewritten as
\begin{align}
	&\left[ h_{\text{AB}}^{\mathrm{bl}} \left( C_{3z}^\eta \vec{K}^{l}_{M} + \delta \vec{k} ,C_{3z}^{\eta'} \vec{K}^{-l}_{M} + \delta \vec{k}' \right) \right]_{s_1 l;s_2 (-l)} = \nonumber \\
	=& \sum_{n=0}^{1} \sum_{\vec{g}''} 
	\frac{ t^{l,\text{AB}}_{s_1 s_2} \left( (-1)^n C_{3z}^{\eta'} \vec{K}_{M} \right) }{\Omega_{\mathrm{sl}}} 
	\delta_{ C_{3z}^\eta \vec{K}^{l}_{M} + \delta \vec{k} - C_{3z}^{\eta'} \vec{K}^{-l}_{M} - \delta \vec{k}', l \mathcal{R}_{\theta, l} \vec{g}'' + \mathcal{R}_{\theta, -l} \left[ (-1)^n C_{3z}^{\eta'} \vec{K}_{M} - C_{3z}^{\eta'} \vec{K}_{M} \right] } \nonumber \\
	=& \sum_{n=0}^{1} \sum_{\vec{g}''} 
	\frac{ t^{l,\text{AB}}_{s_1 s_2} \left( (-1)^n C_{3z}^{\eta'} \vec{K}_{M} \right) }{\Omega_{\mathrm{sl}}} 
	\delta_{ C_{3z}^\eta \vec{K}^{l}_{M} + \delta \vec{k} - C_{3z}^{\eta'} \vec{K}^{-l}_{M} - \delta \vec{k}',  \mathcal{R}_{\theta, l} \vec{g}'' + \mathcal{R}_{\theta, -l} \left[ (-1)^n C_{3z}^{\eta'} \vec{K}_{M} - C_{3z}^{\eta'} \vec{K}_{M} \right] },
	\label{app:eqn:interlayer_mel_AB_3}.
\end{align}
where, in the last line, we have changed the summation variable $\vec{g}'' \to l \vec{g}''$. \Cref{app:eqn:interlayer_mel_AB_3} is similar to the second line of \cref{app:eqn:interlayer_mel_AA_3}, so the derivation proceeds analogously, allowing us to conclude that 
\begin{equation}
	\left[ h_{\text{AB}}^{\mathrm{bl}} \left( C_{3z}^\eta \vec{K}^{l}_{M} + \delta \vec{k} ,C_{3z}^{\eta'} \vec{K}^{-l}_{M} + \delta \vec{k}' \right) \right]_{s_1 l;s_2 (-l)} = \sum_{n=0}^{1}  
	\tilde{t}^{l,\text{AB}}_{s_1 s_2}  \left( (-1)^n C_{3z}^\eta \vec{K}_{M} \right)
	\delta_{ \delta \vec{k} - \delta \vec{k}', l (-1)^n \vec{q}_\eta} \delta_{\eta \eta'}
	\label{app:eqn:interlayer_mel_AB_4},
\end{equation}
where we have defined the rescaled hopping amplitude Fourier transformation
\begin{equation}
	\tilde{t}^{l,\text{AB}}_{s_1 s_2} \left( \vec{q} + \vec{g} \right) = \frac{t^{l,\text{AB}}_{s_1 s_2} \left( \vec{q} + \vec{g} \right)}{\Omega_{\mathrm{sl}}}.
\end{equation}
Note that \cref{app:eqn:interlayer_mel_AA_4,app:eqn:interlayer_mel_AB_4} are identical in form. The difference between the interlayer Hamiltonians of the two stacking arrangements arise due to the different symmetries of the two structures, which were discussed in \cref{app:sec:DFT_bilayer:SnSe2:twisted}. The constraints imposed by these symmetries on the interlayer hopping amplitudes will be the focus of the following \cref{app:sec:SnS_SnSe_twist:symmetries_t}.

\subsection{Constraining the interlayer hopping amplitude with symmetries}\label{app:sec:SnS_SnSe_twist:symmetries_t}

Naively, the two-center interlayer hopping amplitude for both the AA- and AB-stacking configurations $\tilde{t}^{l}_{s_1s_2} \left( \pm C_{3z}^\eta \vec{K}_{M} \right)$ is characterized by $2 \left(\text{spin } s_1 \right) \times 2 \left(\text{spin } s_2 \right) \times 6 \left(\text{distinct momenta } \pm C_{3z}^\eta \vec{K}_{M} \right) \times 2 \left(\text{layers } l \right) = 48$ complex parameters (or 96 real parameters), where no Hermiticity condition was imposed. However, not all of these parameters are independent. Our goal in this section is to employ the exact symmetries of the twisted bilayer system, as well as the properties of $\tilde{t}^{l}_{s_1s_2} \left( \vec{q} + \vec{g} \right)$ to obtain the \emph{independent} components specifying the interlayer hopping amplitude. We will also show how the symmetries of the \emph{untwisted} bilayer arrangements discussed in \cref{app:sec:DFT_bilayer:SnSe2:untwisted} further provide \emph{approximate} constraints on the interlayer hopping amplitude~\cite{SCH22}.

We start by noting that due to the Hermiticity of the bilayer Hamiltonian, one must have that $\mel**{\vec{R},s_1,l}{\mathcal{H}}{\vec{R}',s_2,-l} = \mel**{\vec{R}',s_2,-l}{\mathcal{H}}{\vec{R},s_1,l}^*$, which implies that 
\begin{equation}
	\label{app:eqn:hermiticity_of_t_two_center}
	\tilde{t}^{l}_{s_1 s_2} \left( \vec{q} + \vec{g} \right) = \tilde{t}^{-l*}_{s_2 s_1} \left( \vec{q} + \vec{g} \right).
\end{equation}

Strictly speaking, for arbitrary angles and interlayer in-plane displacements (which result in an incommensurate arrangement of the two layers), our heterostructures, similar to TBG~\cite{ZOU18}, do not feature any \emph{exact} symmetries, other than time-reversal symmetry $\mathcal{T}$\footnote{For example, imposing the Hermicity condition from \cref{app:eqn:hermiticity_of_t_two_center} and the time-reversal symmetry according to \cref{app:eqn:const_T_sym} will reduce the number of real independent parameters of $\tilde{t}^{l}_{s_1s_2} \left( \pm C_{3z}^\eta \vec{K}_{M} \right)$ from 96 to 24.}. Nevertheless, for TBG at incommensurate angle, over length scales comparable with the moir\'e unit cell, but much larger than the single-layer unit cell, the $C_{6z}$ and $C_{2x}$ symmetries of the single-layer give rise to \emph{emergent} (but otherwise approximate) $C_{6z}$ and $C_{2x}$ symmetries of the twisted bilayer arrangement. Moreover, while not even an emergent symmetry of TBG at nonzero twist angles $\theta$, the $m_y$ symmetry of \emph{monolayer} graphene (corresponding to reflections across a plane perpendicular to the graphene plane) constrains the form of interlayer hopping amplitude, which is therefore only specified by two real parameters within the BM model~\cite{SCH22}.

The emergence of various symmetries in the twisted bilayer arrangement at incommensurate angles can be understood as follows. In the two-center approximation for the interlayer hopping, the $t^{l}_{s_1 s_2} \left( \vec{r} \right)$ function introduced in \cref{app:eqn:two_center_approx_AA,app:eqn:two_center_approx_AB} is independent on the twist angle $\theta$. At the same time, within the first-harmonic approximation, the interlayer matrix elements from \cref{app:eqn:interlayer_mel_AA_4,app:eqn:interlayer_mel_AB_4} only depend on the twist angle via the auxiliary vectors $\vec{q}_i$ from \cref{app:eqn:q_vecs}. As a result, one can choose a commensurate geometry for the bilayer arrangement, and use the \emph{exact} symmetries of the system to constrain the form of the $\tilde{t}^{l}_{s_1s_2} \left( \pm C_{3z}^\eta \vec{K}_{M} \right)$ tensor and determine its independent components. Finally, the parameterization of $\tilde{t}^{l}_{s_1s_2} \left( \pm C_{3z}^\eta \vec{K}_{M} \right)$ determined in the symmetric commensurate arrangement \emph{will remain} valid (within the first-harmonic approximation) for general layer displacements and small (but non-zero) twist angles. As a result, the exact symmetries of the moir\'e heterostructure at commensurate twist angles give rise to emerging symmetries of the system at generic incommensurate angles. 

An alternative approach, which we will employ below, is to consider a generic (incommensurate) twist angle $\theta$ \emph{without} any interlayer shift. This is consistent with the convention introduced around \cref{app:eqn:notRrotation}, where the twisted heterostructure is generated by rotating the two layers relative to each other around the common origins of the two monolayer lattices. This configuration retains point group symmetries about the origin, which can be used to constrain the interlayer hopping amplitude.

We will now determine the properties of the interlayer hopping amplitude $\tilde{t}^{l}_{s_1s_2} \left( \pm C_{3z}^\eta \vec{K}_{M} \right)$ arising from the symmetries of the single-layer material. We will consider both the symmetries of the system arising for a small, but \emph{non-zero} twist angle, as well as the enhanced symmetries of the bilayer system \emph{at zero twist angle}. As was discussed in \cref{app:sec:DFT_bilayer:SnSe2}, depending on the stacking arrangement, both the untwisted and twisted heterostructures will feature different symmetries, in addition to time reversal $\mathcal{T}$ symmetry:
\begin{enumerate}
    \item For AA-stacking and $\theta \neq 0$, the system features $C_{3z}$ and $C_{2x}$ symmetries which give rise to the space group $P3121'$ (SSG 149.22). In the $\theta = 0$ case, the system will additionally feature $\mathcal{I}$ symmetry, such that its symmetry group is given by $P\bar{3}m11'$ (SSG 164.86). 
    \item For AB-stacking, the system will feature $C_{3z}$ and $C_{2y}$ symmetries in the $\theta \neq 0$ case, which generate the $P3211'$ group (SSG 150.26). When $\theta = 0$, the system features $P\bar{6}m21'$ symmetry (SSG 187.210), which is generated by the $C_{3z}$, $C_{2y}$, and $M_z$ symmetries. 
\end{enumerate}

In what follows, we will first employ the symmetries of the $\theta \neq 0$ heterostructure to constrain the interlayer hopping amplitude. We will then use the additional symmetries arising in the $\theta = 0$ case to further constrain the interlayer tunneling. We expect that these latter constraints will hold approximately in the limit of small but nonvanishing twist angles. 

To this end, we let $g$ be a symmetry of the bilayer system. As a result, $\commutator{g}{ \mathcal{H}}=0$ (where $\mathcal{H}$ is the Hamiltonian of the heterostructure), which implies that
\begin{align}
	&\mel**{\vec{R},s_1,l}{\mathcal{H}}{\vec{R}',s_2,-l} = 
	\mel**{\vec{R},s_1,l}{g^{-1} g \mathcal{H}}{\vec{R}',s_2,-l} =
	\mel**{\vec{R},s_1,l}{g^{-1} \mathcal{H} g}{\vec{R}',s_2,-l} = \nonumber \\
	= & \sum_{s'_1, s'_2}
	\left[ D^{\mathrm{sl}} \left( g \right) \right]^{*}_{s'_1 s_1}
	\mel**{g \vec{R}, s'_1,\epsilon_g l}{ \mathcal{H} }{g \vec{R}', s'_2,-\epsilon_g l}^{(*)} 
	\left[ D^{\mathrm{sl}} \left( g \right) \right]_{s'_2 s_2}. \label{app:eqn:const_interlayer_mat_elem}
\end{align}
In \cref{app:eqn:const_interlayer_mat_elem}, ${}^{(*)}$ denotes a complex conjugation of the matrix elements whenever $g$ is antiunitary, while $\epsilon_g = +1$ ($\epsilon_g = -1$) if the symmetry $g$ preserves (exchanges) the two layers. Using the notation from \cref{app:eqn:two_center_approx_AA,app:eqn:two_center_approx_AB}, \cref{app:eqn:const_interlayer_mat_elem} imposes the following constraint on the two-center hopping amplitude function 
\begin{equation}
	\label{app:eqn:const_t_func_real_space}
	t^{l}_{s_1 s_2} \left( \vec{r} \right) =
	\sum_{s'_1, s'_2 }
	\left[ D^{\mathrm{sl}} \left( g \right) \right]^{*}_{s'_1 s_1}
	t^{\epsilon_g l (*)}_{s'_1 s'_2} \left( g \vec{r} \right)
	\left[ D^{\mathrm{sl}} \left( g \right) \right]_{s'_2 s_2},
\end{equation}
where $\vec{r}=\mathcal{R}_{\theta, -l} \vec{R}'  - \mathcal{R}_{\theta, l}  \vec{R}$ in the AA-stacked case or $\vec{r}=-l \mathcal{R}_{\theta, -l} \vec{R}'  - l \mathcal{R}_{\theta, l}  \vec{R}$ in the AB-stacked case. This follows directly from the definitions in \cref{app:eqn:two_center_approx_AA,app:eqn:two_center_approx_AB}. For general incommensurate twist angles, as $\vec{R}$ and $\vec{R}'$ take different discrete values in the monolayer lattice, $\vec{r}$ forms a dense subset of the entire two-dimensional space. As such, because $t^{l}_{s_1 s_2} \left( \vec{r} \right)$ is a smooth function of $\vec{r}$, \cref{app:eqn:const_t_func_real_space} will hold for \emph{any} two-dimensional vector $\vec{r}$. In momentum space, \cref{app:eqn:const_t_func_real_space} can therefore be written as
\begin{equation}
	\label{app:eqn:const_t_func_momentum_space}
	t^{l}_{s_1 s_2} \left( \vec{q} + \vec{g} \right) = \sum_{s'_1, s'_2 }
	\left[ D^{\mathrm{sl}} \left( g \right) \right]^{*}_{s'_1 s_1}
	t^{\epsilon_g l (*)}_{s'_1 s'_2} \left( g \left( \vec{q} + \vec{g} \right) \right)
	\left[ D^{\mathrm{sl}} \left( g \right) \right]_{s'_2 s_2}.
\end{equation}
\Cref{app:sec:SnS_SnSe_twist_general_woGradient:approx} provides a proof that does not rely on any assumptions about the twist angles and derives the moir\'e potential beyond the two-center first monolayer harmonic approximation.

\subsubsection{Constraints arising from the $\theta \neq 0$ symmetries}\label{app:sec:SnS_SnSe_twist:symmetries_t:exact}

Explicitly, the $C_{3z}$ and $\mathcal{T}$ symmetries, respectively, impose the following constraints on the hopping amplitude tensor for both the AA- and AB-stacking configurations
\begin{alignat}{3}
	\tilde{t}^{l}_{s_1 s_2} \left( \pm C_{3z}^\eta \vec{K}_{M} \right) &&=& 
	\sum_{s'_1, s'_2}
	\left[ D^{\mathrm{sl}} \left( C_{3z} \right) \right]^{*}_{s'_1 s_1}
	\tilde{t}^{l}_{s'_1 s'_2} \left(  \pm C_{3z}^{\eta+1}  \vec{K}_{M} \right)
	\left[ D^{\mathrm{sl}} \left( C_{3z} \right) \right]_{s'_2 s_2}, &
	\qq{with} &D^{\text{sl}} \left( C_{3z} \right) = e^{-\frac{\pi i}{3} s_z}, \label{app:eqn:const_C3_sym}\\
	\tilde{t}^{l}_{s_1 s_2} \left( \pm C_{3z}^\eta \vec{K}_{M} \right) &&=& 
	\sum_{s'_1, s'_2}
	\left[ D^{\mathrm{sl}} \left( \mathcal{T} \right) \right]^{*}_{s'_1 s_1}
	\tilde{t}^{l *}_{s'_1 s'_2} \left( \mp C_{3z}^\eta \vec{K}_{M} \right)
	\left[ D^{\mathrm{sl}} \left( \mathcal{T} \right) \right]_{s'_2 s_2}, &
	\qq{with} &D^{\text{sl}} \left( \mathcal{T} \right) = i s_y. \label{app:eqn:const_T_sym}
\end{alignat}
Depending on the stacking arrangement, we additionally have
\begin{alignat}{3}
	\tilde{t}^{l,\text{AA}}_{s_1 s_2} \left( \pm C_{3z}^\eta \vec{K}_{M} \right) &&=& 
	\sum_{s'_1, s'_2}
	\left[ D^{\mathrm{sl}} \left( C_{2x} \right) \right]^{*}_{s'_1 s_1}
	\tilde{t}^{-l,\text{AA}}_{s'_1 s'_2} \left(  \pm C_{2x} C_{3z}^\eta  \vec{K}_{M} \right)
	\left[ D^{\mathrm{sl}} \left( C_{2x} \right) \right]_{s'_2 s_2}, &
	\qq{with} &D^{\text{sl}} \left( C_{2x} \right) = -i  s_x, \label{app:eqn:const_C2x_sym_AA} \\
	\tilde{t}^{l,\text{AB}}_{s_1 s_2} \left( \pm C_{3z}^\eta \vec{K}_{M} \right) &&=& 
	\sum_{s'_1, s'_2}
	\left[ D^{\mathrm{sl}} \left( C_{2y} \right) \right]^{*}_{s'_1 s_1}
	\tilde{t}^{-l,\text{AB}}_{s'_1 s'_2} \left(  \pm C_{2y} C_{3z}^\eta  \vec{K}_{M} \right)
	\left[ D^{\mathrm{sl}} \left( C_{2y} \right) \right]_{s'_2 s_2}, &
	\qq{with} &D^{\mathrm{sl}} \left( C_{2y} \right) = -i s_y. \label{app:eqn:const_C2y_sym_AB} 
\end{alignat}
\Crefrange{app:eqn:const_C3_sym}{app:eqn:const_C2y_sym_AB} provide exact symmetry constraints on the interlayer hopping amplitude. Apart for the $C_{2y}$ symmetry, whose representation matrix is only given in \cref{app:eqn:const_C2y_sym_AB}, the representation matrices of all the other symmetries appearing in \crefrange{app:eqn:const_C3_sym}{app:eqn:const_C2y_sym_AB} are given in \cref{app:eqn:rep_matrices_syms_Snse2} and repeated here for convenience.

\subsubsection{Constraints arising from the $\theta = 0$ symmetries}\label{app:sec:SnS_SnSe_twist:symmetries_t:approx}

The zero-twist heterostructure features additional symmetries, as discussed above \cref{app:eqn:const_interlayer_mat_elem}. These symmetries further constrain the interlayer hopping amplitude for $\theta = 0$. Because $\tilde{t}^{l}_{s_1 s_2} \left( \vec{q} + \vec{g} \right)$ is taken to be approximately angle-independent, we expect that these constraints will continue to hold approximately for small, but non-zero twist angles $\theta \neq 0$. To determine the constraints imposed by the $\theta = 0$ symmetries, we can still use \cref{app:eqn:const_interlayer_mat_elem} in the limit of a small vanishing incommensurate angle, for which we find
\begin{align}
	\tilde{t}^{l,\text{AA}}_{s_1 s_2} \left( \pm C_{3z}^\eta \vec{K}_{M} \right) &= 
	\sum_{s'_1, s'_2}
	\left[ D^{\mathrm{sl}} \left( \mathcal{I} \right) \right]^{*}_{s'_1 s_1}
	\tilde{t}^{-l,\text{AA}}_{s'_1 s'_2} \left(  \mp C_{3z}^\eta  \vec{K}_{M} \right)
	\left[ D^{\mathrm{sl}} \left( \mathcal{I} \right) \right]_{s'_2 s_2}, \qq{where} D^{\text{sl}} \left( \mathcal{I} \right) = s_0, \label{app:eqn:const_I_sym_AA} \\
\tilde{t}^{l,\text{AB}}_{s_1 s_2} \left( \pm C_{3z}^\eta \vec{K}_{M} \right) &= 
	\sum_{s'_1, s'_2}
	\left[ D^{\mathrm{sl}} \left( M_z \right) \right]^{*}_{s'_1 s_1}
	\tilde{t}^{-l,\text{AB}}_{s'_1 s'_2} \left(  \pm C_{3z}^\eta  \vec{K}_{M} \right)
	\left[ D^{\mathrm{sl}} \left( M_z \right) \right]_{s'_2 s_2}, \qq{where} D^{\mathrm{sl}} \left( M_z \right) = -i s_z. \label{app:eqn:const_mz_sym_AB}
\end{align}
A more rigorous and complete derivation is presented in \cref{app:sec:SnS_SnSe_twist_general_woGradient:approx}, which also extends beyond the two-center approximation.

With the exact symmetry constraints in \cref{app:sec:SnS_SnSe_twist:symmetries_t:exact}, the interlayer hopping matrices for the AA- and AB-stacking arrangements are given by
\begin{align}
	\tilde{t}^{l,\text{AA}}\left( \pm \vec{K}_{M} \right) &= \left( \pm i l w^{\text{AA}}_1 + w^{\text{AA}}_2 \right) s_0 \pm \left(w^{\text{AA}}_4 + w^{\text{AA}}_6 e^{ \mp i l \frac{\pi}{6}}  \right) s_y + \left( i l w^{\text{AA}}_3 \pm w^{\text{AA}}_5 \right) s_z, \label{app:eqn:two_center_first_interlayer_AA} \\
	\tilde{t}^{l,\text{AB}}\left( \pm \vec{K}_{M} \right) &= w^{\text{AB}}_2 s_0 + i l w^{\text{AB}}_1 s_x \pm w^{\text{AB}}_3 s_y + i l w^{\text{AB}}_4 s_z, \label{app:eqn:two_center_first_interlayer_AB}
\end{align}
where $w^{\text{AA}}_i$ for $1 \leq i \leq 6$ and $w^{\text{AB}}_i$ for $1 \leq i \leq 4$ are real parameters characterizing the interlayer hopping amplitudes within the two-center single-harmonic approximation. The interlayer tunneling matrices at other momenta can be directly determined from \cref{app:eqn:const_C3_sym}. Finally, we note that the symmetries of the $\theta = 0$ configuration will further impose that 
\begin{equation}
    \label{app:further_constraints_simplified_model_two_center}
    w^{\text{AA}}_{i}=0, \qq{for} 3 \leq i \leq 6, \qq{and}
    w^{\text{AB}}_{i}=0, \qq{for} 3 \leq i \leq 4.
\end{equation}
In what follows, we will call the model corresponding to \cref{app:further_constraints_simplified_model_two_center} the \emph{simplified} two-center first monolayer harmonic model, while the model of \cref{app:eqn:two_center_first_interlayer_AA,app:eqn:two_center_first_interlayer_AB} will be called the \emph{full} two-center first monolayer harmonic model. It is worth noting that in the AA-stacking case, the $\mathcal{I}\mathcal{T}$ symmetry of the untwisted bilayer implies $\mathrm{SU} \left( {2} \right)$ symmetry within the simplified two-center first monolayer harmonic model, since the interlayer hopping matrix becomes spin-diagonal.

\subsection{The twisted bilayer model and its exact symmetries}\label{app:sec:SnS_SnSe_twist:bl_model}

\begin{figure}[!t]
	\centering
	\includegraphics[width=0.5\textwidth]{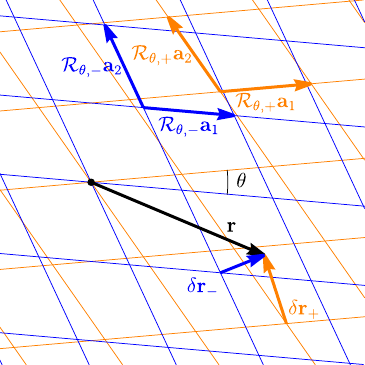}
	\caption{Deriving the moir\'e translation vectors for general twist angles $\theta$. The monolayer hexagonal lattice spanned by $\mathcal{R}_{\theta,+} \vec{a}_{1,2}$ ($\mathcal{R}_{\theta,-} \vec{a}_{1,2}$) corresponding to the top (bottom) layer is shown by the orange (blue) grid, where the rotation matrix $\mathcal{R}_{\theta,l}$ is defined in \cref{app:eqn:notRrotation}. The two lattices are rotated relative to one another by an angle $\theta$ around the common origin, which is marked by the thick black dot. A general position $\vec{r}$ is displaced by $\delta \vec{r}_{+}$ ($\delta \vec{r}_{-}$) from a nearby unit cell origin of the top (bottom) layer, as defined in \cref{app:eqn:shifts_geometric_derivation_displacement}. The relative displacement between the two layers at position $\vec{r}$ is defined and obtained around \cref{app:eqn:local_interlayer_displacement}.}
	\label{app:fig:local_displacement}
\end{figure}

From a geometric standpoint, the twisted bilayer system exhibits an effective moir\'e translation symmetry characterized by the moir\'e unit cell lattice vectors $\vec{a}_{M_1}$ and $\vec{a}_{M_2}$, which will be determined below. To derive these vectors for a general (not necessarily commensurate) twist angle $\theta$, we begin by examining \cref{app:fig:local_displacement}, which shows the two twisted monolayer lattices superimposed.

Our goal is to determine the relative displacement $\delta \vec{R} \left( \vec{r} \right)$ between the two monolayer lattices at a given position $\vec{r}$. In layer $l = \pm$, the position $\vec{r}$ is displaced from a nearby unit cell's origin of the corresponding monolayer unit cell by
\begin{align}
	\label{app:eqn:shifts_geometric_derivation_displacement}
	\delta \vec{r}_{l} &\equiv \sum_{i=1}^{2} \frac{1}{2 \pi} \left[ \left(\mathcal{R}_{\theta,l} \vec{b}_{i} \right) \cdot \vec{r} \mod 2\pi \right] \left(\mathcal{R}_{\theta,l} \vec{a}_{i} \right) \nonumber \\
	&= \sum_{i=1}^{2} \frac{1}{2 \pi} \left[ \vec{b}_{i} \cdot \mathcal{R}^{-1}_{\theta,l} \vec{r} \mod 2\pi \right] \left(\mathcal{R}_{\theta,l} \vec{a}_{i} \right),
\end{align} 
as shown in \cref{app:fig:local_displacement}. We note that $\delta \vec{r}_{l}$ is only defined up to a rotated monolayer lattice vector $\mathcal{R}_{\theta,l} \vec{a}_{i}$ (for $i=1,2$). For concreteness, we take $\delta \vec{r}_{l}$ to obey $0 \leq \delta \vec{r}_{l} \cdot \left(\mathcal{R}_{\theta,l} \vec{a}_{i} \right) < 1$, but the exact convention is unimportant in what follows.

We then imagine rotating back the two monolayer lattices ({\it i.e.}{}, layer $l$ is rotated by $-\frac{l\theta}{2}$) around the point $\vec{r}$. The two monolayer lattices will become aligned ({\it i.e.}{}, their lattice vectors will become parallel), but their origins will be displaced from one another by the \emph{relative displacement}. We define this relative displacement as the local displacement of the top layer relative to the bottom layer modulo monolayer reciprocal lattice vectors
\begin{equation}
	\delta \vec{R} \left( \vec{r} \right) \equiv \sum_{i=1}^{2} \frac{1}{2\pi}\left[ \left( \mathcal{R}^{-1}_{\theta,-} \delta \vec{r}_{-} - \mathcal{R}^{-1}_{\theta,+} \delta \vec{r}_{+} \right) \cdot \vec{b}_i \right] \vec{a}_i.
\end{equation}
The relative displacement $\delta \vec{R} \left( \vec{r} \right)$ is only defined up to a direct monolayer lattice vector.

A simpler form for the relative displacement can be derived using \cref{app:eqn:shifts_geometric_derivation_displacement}
\begin{align}
    \delta \vec{R} \left( \vec{r} \right) &= \sum_{j}^{2} \frac{1}{2 \pi} \left\lbrace \sum_{i=1}^{2} \left[ \frac{1}{2 \pi} \vec{a}_i \left( \vec{b}_i \cdot \mathcal{R}^{-1}_{\theta,-} \vec{r}  \ \mathrm{mod} \ 2 \pi \right) - \frac{1}{2 \pi} \vec{a}_i \left( \vec{b}_i \cdot \mathcal{R}^{-1}_{\theta,+} \vec{r}  \ \mathrm{mod} \ 2 \pi \right) \right] \cdot \vec{b}_{j} \right\rbrace \vec{a}_{j}\nonumber \\
    &= \sum_{i=1}^{2} \frac{1}{2 \pi} \vec{a}_i \left[ \vec{b}_i \cdot \left( \mathcal{R}^{-1}_{\theta,-} - \mathcal{R}^{-1}_{\theta,+} \right) \vec{r}  \ \mathrm{mod} \ 2 \pi \right] \nonumber \\
    &= \sum_{i=1}^{2} \frac{1}{2 \pi} \vec{a}_i \left[ \vec{b}_i \cdot \left( 2 \sin \left( \frac{\theta}{2} \right) \vec{\hat{z}} \cross \vec{r} \right)  \ \mathrm{mod} \ 2 \pi \right] \nonumber \\
    &= \sum_{i=1}^{2} \frac{1}{2 \pi} \vec{a}_i \left[ \vec{r} \cdot \left( \vec{b}_i \cross 2 \sin \left( \frac{\theta}{2} \right) \vec{\hat{z}} \right)  \ \mathrm{mod} \ 2 \pi \right]. \label{app:eqn:local_interlayer_displacement}
\end{align}
The periodicity of the moir\'e pattern arises from the periodicity of the local displacement in $\vec{r}$. For any vector $\Delta \vec{r}$ that satisfies
\begin{equation}
	\label{app:eqn:local_displacement_moire_condition}
	\Delta \vec{r} \cdot \left( \vec{b}_i \cross 2 \sin \left( \frac{\theta}{2} \right) \vec{\hat{z}} \right) \in 2 \pi \mathbb{Z} \qq{for any} i = 1,2,
\end{equation}
we must have that $\delta \vec{R} \left( \vec{r} + \Delta \vec{r} \right) = \delta \vec{R} \left( \vec{r} \right)$. The vectors $\Delta \vec{r}$ satisfying \cref{app:eqn:local_displacement_moire_condition} form the moir\'e direct lattice, which is spanned by $\vec{a}_{M_i}$ (with $i = 1,2$). To find $\vec{a}_{M_i}$, we note that \cref{app:eqn:local_displacement_moire_condition} also implies that the reciprocal moir\'e lattice is spanned by $\vec{b}_i \cross 2 \sin \left( \frac{\theta}{2} \right) \vec{\hat{z}}$, for $i = 1,2$. Using the auxiliary vectors from \cref{app:eqn:q_vecs}, we define the reciprocal moir\'e lattice vectors as
\begin{equation}
	\vec{b}_{M_1} = \vec{b}_2 \cross 2 \sin \left( \frac{\theta}{2} \right) \vec{\hat{z}} = 2 \vec{q}_0,\qquad
	\vec{b}_{M_2} = \left( \vec{b}_2 - \vec{b}_1 \right) \cross 2 \sin \left( \frac{\theta}{2} \right) \vec{\hat{z}}  = - 2 \vec{q}_2,
\end{equation}
which generate the corresponding moir\'e reciprocal lattice $\mathcal{Q} = \mathbb{Z} \vec{b}_{M_1} + \mathbb{Z} \vec{b}_{M_2}$. At the same time, the direct moir\'e lattice vectors $\vec{a}_{M_i}$ are defined to satisfy $\vec{a}_{M_i} \cdot \vec{b}_{M_j} = 2 \pi \delta_{ij}$, for $1 \leq i, j \leq 2$.

\subsubsection{Model in momentum space}\label{app:sec:SnS_SnSe_twist:bl_model:model_mom_space}

\begin{figure}[!t]
	\centering
	\includegraphics[width=1\textwidth]{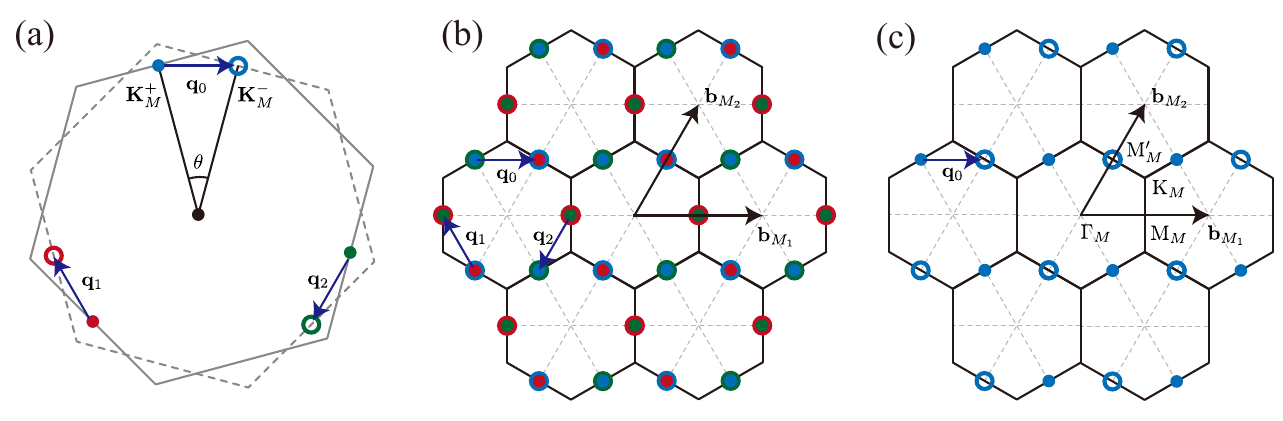}
	\subfloat{\label{app:fig:M_valley_MBZ:a}}\subfloat{\label{app:fig:M_valley_MBZ:b}}\subfloat{\label{app:fig:M_valley_MBZ:c}}\caption{The moir\'e BZ and momentum lattices generated by twisting the M valley. In (a), the gray continuous (dashed) hexagons correspond to the BZs of the top (bottom) layers. The M valleys of the two layers located at $\vec{K}_{M}^{\pm}$ are shown explicitly, together with the auxiliary $\vec{q}_{\eta}$ vectors defined in \cref{app:eqn:q_vecs}. The three inequivalent M valleys are shown in solid (hollow) blue, green, and red for the top (bottom) layers. (b) shows the three momentum lattices defined in \cref{app:eqn:three_mom_lattices_definition}, as well as the reciprocal moir\'e lattice vectors. Each lattice site corresponds to \emph{two} M valleys from different layers. For clarity, the momentum lattice corresponding to valley $\eta = 0$ is shown in (c), where we mark the high-symmetry momentum $\Gamma=(0,0), \mathrm{M}_M=\frac{1}{2}\mathbf{b}_{M_1}, \mathrm{M}_M^\prime=\frac{1}{2}\mathbf{b}_{M_2}, \mathrm{K}_M=\frac{1}{3}(\mathbf{b}_{M_1} +\mathbf{b}_{M_2})$.}
	\label{app:fig:M_valley_MBZ}
\end{figure}

We are now in a position of writing down the Hamiltonian of the twisted M-point bilayer system. To this end, we introduce \emph{three} momentum lattices, as shown in \cref{app:fig:M_valley_MBZ},
\begin{equation}
    \label{app:eqn:three_mom_lattices_definition}
    \mathcal{Q}_{n} \equiv \mathcal{Q} + \vec{q}_{n}, \qq{for} 0 \leq n \leq 2,
\end{equation}
where the $\vec{q}_{n}$ vectors were defined in \cref{app:eqn:q_vecs} and which together form a kagome lattice\footnote{It is crucial to highlight the distinction between \cref{app:eqn:three_mom_lattices_definition} and the $\mathcal{Q}'_{\pm}$ lattices that emerge when twisting the K valley (such as in TBG~\cite{SON19} or in \ch{MoTe2}~\cite{YU24a,JIA24,ZHA24}). As illustrated in \cref{fig:q_lattices}, the reciprocal moir\'e vectors are identical in both scenarios, being given by $\vec{b}_{M_i}$ for $i=1,2$. However, the three $\vec{q}^{(\prime)}_{n} = C^{n-1}_{3z} \vec{q}^{(\prime)}_{1}$ vectors are defined differently between the two cases: $\vec{q}'_{1} = \frac{1}{3} \left( \vec{b}_{M_1} - 2 \vec{b}_{M_2} \right)$, while $\vec{q}_{1} = \frac{1}{2} \left(- \vec{b}_{M_1} + \vec{b}_{M_2} \right)$. These differences in the lengths and orientations of the $\vec{q}^{(\prime)}_{n}$ vectors lead to the generation of a kagome lattice for M-point twisting, as opposed to a honeycomb $\vec{Q}$ lattice for the K-valley case.}\begin{equation}
	\label{app:eqn:total_mom_lattice_definition}
	\mathcal{Q}_{\text{tot}} \equiv \bigcup_{n=0}^{2} \mathcal{Q}_{n}.
\end{equation}
For later convenience in notation, we also extend the definitions in \cref{app:eqn:q_vecs,app:eqn:three_mom_lattices_definition} beyond $0 \leq \eta \leq 2$ and $0 \leq n \leq2$, respectively, using
\begin{equation}
    \label{app:eqn:extension_with_modulus}
    \mathcal{Q}_{n} \equiv \mathcal{Q}_{n  \ \mathrm{mod} \ 3} \qq{and} \vec{q}_n \equiv \vec{q}_{n  \ \mathrm{mod} \ 3}, \qq{for} n \in \mathbb{Z}.
\end{equation}
For every $\vec{Q} \in \mathcal{Q}_{\text{tot}}$, we also associate a sublattice factor defined as
\begin{equation}
    \zeta_{\vec{Q}} = n, \qq{for} \vec{Q} \in \mathcal{Q}_{n}, \qq{with} 0 \leq n \leq 2.
\end{equation}

Similarly to TBG~\cite{SON19}, we define the following low-energy operators
\begin{equation}
    \label{app:eqn:low_en_ops_c}
    \hat{c}^\dagger_{\vec{k}, \vec{Q}, s, l} \equiv \hat{a}^\dagger_{C^{\eta}_{3z} \vec{K}^{l}_{M} + \vec{k} - \vec{Q}, s, l}, \qq{for} \vec{Q} \in  \mathcal{Q}_{\eta + l} \qq{and} \vec{k} \in \text{MBZ},
\end{equation}
where MBZ denotes the first moir\'e BZ. The definition of the $\hat{c}^\dagger_{\vec{k},\vec{Q},s,l}$ operators can be extended outside the first MBZ through the following embedding relation
\begin{equation}
	\hat{c}^\dagger_{\vec{k} + \vec{G}, \vec{Q}, s, l} = \hat{c}^\dagger_{\vec{k}, \vec{Q} - \vec{G}, s, l}, \qq{for} \vec{G} \in \mathcal{Q}.
\end{equation}
Note that the $\hat{c}^\dagger_{\vec{k}, \vec{Q}, s, l}$ fermions do feature a layer index $l$, but not a valley $\eta$ index. This is because the valley $\eta$ associated with $\hat{c}^\dagger_{\vec{k}, \vec{Q}, s, l}$ can be directly inferred from $\vec{Q}$ and $l$: $\eta = \zeta_{\vec{Q}} - l$\footnote{This is similar to TBG~\cite{SON19}, where knowing the valley $\eta$ and the $\vec{Q}$ vector of a given low-energy fermionic operator $\hat{c}^\dagger_{\vec{k}, \vec{Q}, \alpha, \eta, s}$ fully specifies the layer $l$.}. More intuitively, the $\hat{c}^\dagger_{\vec{k}, \vec{Q}, s, l}$ from valley $\eta$ are supported on the complement of lattice $\mathcal{Q}_{\eta}$, meaning that $\vec{Q} \in \mathcal{Q}_{\text{tot}} / \mathcal{Q}_{\eta}$.

Depending on the stacking arrangement, we employ \cref{app:eqn:bl_mat_elems_AA,app:eqn:bl_mat_elems_AB} to write the Hamiltonian of the twisted bilayer heterostructure in the two-center first monolayer harmonic approximation. For AA-stacking, we have 
\begin{align}
	\mathcal{H}_{\text{AA}} =& \sum_{\delta \vec{k}, \delta \vec{k}'} \sum_{\substack{s_1, l_1, \eta_1 \\ s_2, l_2, \eta_2}}\left[ h_{\text{AA}}^{\mathrm{bl}} \left( C_{3z}^{\eta_1} \vec{K}^{l_1}_{M} + \delta \vec{k}, C_{3z}^{\eta_2} \vec{K}^{l_2}_{M} + \delta \vec{k}' \right) \right]_{s_1 l_1; s_2 l_2} \hat{a}^\dagger_{C_{3z}^{\eta_1} \vec{K}^{l_1}_{M} + \delta \vec{k}, s_1, l_1} \hat{a}_{C_{3z}^{\eta_2} \vec{K}^{l_2}_{M} + \delta \vec{k}', s_2, l_2} \nonumber \\
	=& \sum_{\delta \vec{k}} \sum_{s_1, s_2, \eta, l}\left[ h^{\mathrm{sl}} \left( C^{\eta}_{3z} \vec{K}_M + \mathcal{R}^{-1}_{\theta, l} \delta \vec{k} \right) \right]_{s_1 s_2}  \hat{a}^\dagger_{C^{\eta}_{3z} \vec{K}^{l}_{M} + \delta \vec{k}, s_1, l} \hat{a}_{C^{\eta}_{3z} \vec{K}^{l}_{M} + \delta \vec{k}, s_2, l} \nonumber \\
	&+ \sum_{\delta \vec{k}} \sum_{s_1, s_2, \eta, l}
	\sum_{m=0}^{1}  
	\tilde{t}^{l,\text{AA}}_{s_1 s_2}  \left( (-1)^m C_{3z}^\eta \vec{K}_{M} \right) \hat{a}^\dagger_{C^{\eta}_{3z} \vec{K}^{l}_{M} + \delta \vec{k} + l (-1)^m \vec{q}_\eta, s_1, l} \hat{a}_{C^{\eta}_{3z} \vec{K}^{-l}_{M} + \delta \vec{k}, s_2, -l}, \label{app:eqn:bl_ham_orig_basis_AA}
\end{align}
where the rotation matrix $\mathcal{R}_{\theta,l}$ is defined in \cref{app:eqn:notRrotation}. For AB-stacking, we similarly obtain
\begin{align}
	\mathcal{H}_{\text{AB}} =& \sum_{\delta \vec{k}} \sum_{s_1, s_2, \eta, l}\left[ h^{\mathrm{sl}} \left( l C^{\eta}_{3z} \vec{K}_M + l \mathcal{R}^{-1}_{\theta, l} \delta \vec{k} \right)  \right]_{s_1 s_2}  \hat{a}^\dagger_{C^{\eta}_{3z} \vec{K}^{l}_{M} + \delta \vec{k}, s_1, l} \hat{a}_{C^{\eta}_{3z} \vec{K}^{l}_{M} + \delta \vec{k}, s_2, l} \nonumber \\
	&+ \sum_{\delta \vec{k}} \sum_{s_1, s_2, \eta, l}
	\sum_{m=0}^{1}  
	\tilde{t}^{l,\text{AB}}_{s_1 s_2}  \left( (-1)^m C_{3z}^\eta \vec{K}_{M} \right) \hat{a}^\dagger_{C^{\eta}_{3z} \vec{K}^{l}_{M} + \delta \vec{k} + l (-1)^m \vec{q}_\eta, s_1, l} \hat{a}_{C^{\eta}_{3z} \vec{K}^{-l}_{M} + \delta \vec{k}, s_2, -l}. \label{app:eqn:bl_ham_orig_basis_AB}
\end{align}
The last term in both \cref{app:eqn:bl_ham_orig_basis_AA,app:eqn:bl_ham_orig_basis_AB} can be recast in the notation of \cref{app:eqn:low_en_ops_c} as 
\begin{align}
    & \sum_{\delta \vec{k}} \sum_{s_1, s_2, \eta, l}
	\sum_{m=0}^{1}  
	\tilde{t}^{l}_{s_1 s_2}  \left( (-1)^m C_{3z}^\eta \vec{K}_{M} \right) \hat{a}^\dagger_{C^{\eta}_{3z} \vec{K}^{l}_{M} + \delta \vec{k} + l (-1)^m \vec{q}_\eta, s_1, l} \hat{a}_{C^{\eta}_{3z} \vec{K}^{-l}_{M} + \delta \vec{k}, s_2, -l} \nonumber \\
    =& \sum_{\vec{k},\eta,l} \sum_{\vec{Q} \in \mathcal{Q}_{\eta + l}} \sum_{s_1, s_2}
	\sum_{m=0}^{1}  
	\tilde{t}^{-l}_{s_1 s_2}  \left( (-1)^m C_{3z}^\eta \vec{K}_{M} \right) \hat{c}^\dagger_{\vec{k}, \vec{Q} + l (-1)^m \vec{q}_\eta,s_1, -l} \hat{c}_{\vec{k}, \vec{Q}, s_2, l}, \label{app:eqn:converting_first_harmonic_to_Q_notation}
\end{align}
which holds for both AA- and AB-stacking and where we have made the substitution $\delta \vec{k} = \vec{k} - \vec{Q}$ as well as $l \to (-l)$. Note that if $\vec{Q} \in \mathcal{Q}_{\eta + l}$, such that $\vec{Q} = \vec{G} + \vec{q}_{\eta + l}$ (with $\vec{G} \in \mathcal{Q}$), then 
\begin{equation}
	\vec{Q} \pm \vec{q}_{\eta} =\vec{G} + \vec{q}_{\eta + l} + \vec{q}_{\eta} - \left( \vec{q}_{\eta} \mp \vec{q}_{\eta} \right) = \left[ \vec{G} - \left( \vec{q}_{\eta} \mp \vec{q}_{\eta} \right) \right] - \vec{q}_{\eta - l} \in \mathcal{Q}_{\eta - l}.
\end{equation}
This implies that both the $\hat{c}_{\vec{k}, \vec{Q}, s_2, l}$ and $\hat{c}^\dagger_{\vec{k}, \vec{Q} + l (-1)^m \vec{q}_\eta,s_1, -l}$ fermions belong to valley $\eta$. \Cref{app:eqn:converting_first_harmonic_to_Q_notation} allows us to rewrite the Hamiltonian of the entire moir\'e system as
\begin{equation}
	\mathcal{H} = \sum_{\vec{k} \in \mathrm{MBZ}} \sum_{s_1,s_2} \sum_{\vec{Q},\vec{Q}'\in \mathcal{Q}_{\text{tot}}} \left[ h_{\vec{Q},\vec{Q}'} \left( \vec{k} \right) \right]_{s_1 l_1; s_2 l_2} \hat{c}^\dagger_{\vec{k},\vec{Q},s_1,l_1} \hat{c}_{\vec{k},\vec{Q}',s_2,l_2},
\end{equation} 
where the first-quantized Hamiltonian $h_{\vec{Q},\vec{Q}'} \left( \vec{k} \right)$ is given by
\begin{equation}
    \left[ h_{\vec{Q},\vec{Q}'} \left( \vec{k} \right) \right]_{s_1 l_1; s_2 l_2} =  \delta_{\vec{Q}, \vec{Q}'} \delta_{s_1 s_2} \delta_{l_1 l_2} \eval{ \left( \frac{\delta k_{x}^2}{2 m_x} + \frac{\delta k_{y}^2}{2 m_y} \right)}_{\delta \vec{k} = C^{l_1-\zeta_{\vec{Q}}}_{3z}\mathcal{R}^{-1}_{\theta, l_1} \left( \vec{k} - \vec{Q} \right)} + \left[ T_{\vec{Q},\vec{Q}'} \right]_{s_1 l_1; s_2 l_2},\label{app:eqn:single_particle_hamiltonian}
\end{equation}
and the momentum space lattice $\mathcal{Q}_{\text{tot}}$ was defined in \cref{app:eqn:total_mom_lattice_definition}. 
The form of the first-quantized Hamiltonian from \cref{app:eqn:single_particle_hamiltonian} is generically valid even beyond the two-center first monolayer harmonic approximation. In the case of the latter, the moir\'e potential term $T_{\vec{Q},\vec{Q}'}$ can be directly determined from \cref{app:eqn:two_center_first_interlayer_AA,app:eqn:two_center_first_interlayer_AB}
\begin{equation}
    \left[ T_{\vec{Q}', \vec{Q}} \right]_{s_1 l_1; s_2 l_2} = \tilde{t}^{l_1} \left( (-1)^m C_{3z}^\eta \vec{K}_{M} \right) \delta_{\vec{Q}', \vec{Q} + l_2 (-1)^m \vec{q}_\eta} \delta_{l_1 \left( - l_2 \right)}, \qq{for} \vec{Q} \in \mathcal{Q}_{\eta + l_2},
\end{equation}
or explicitly for valley $\eta = 0$
\begin{align}
    \left[  T^{\text{AA}}_{\vec{Q}', \vec{Q}} \right]_{s_1 (-l); s_2 l} =& \delta_{\vec{Q}', \vec{Q} \pm l \vec{q}_0} \left[ \left( \mp i l w^{\text{AA}}_1 + w^{\text{AA}}_2 \right) s_0 \pm \left(w^{\text{AA}}_4 + w^{\text{AA}}_6 e^{ \pm i l \frac{\pi}{6}}  \right) s_y + \left( - i l w^{\text{AA}}_3 \pm w^{\text{AA}}_5 \right) s_z \right]_{s_1 s_2}, \label{app:eqn:moire_interlayer_AA} \\
    \left[  T^{\text{AB}}_{\vec{Q}', \vec{Q}} \right]_{s_1 (-l); s_2 l} =& \delta_{\vec{Q}', \vec{Q} \pm l \vec{q}_0} \left[w^{\text{AB}}_2 s_0 - i l w^{\text{AB}}_1 s_x \pm w^{\text{AB}}_3 s_y - i l w^{\text{AB}}_4 s_z \right]_{s_1 s_2}, \label{app:eqn:moire_interlayer_AB}
\end{align}
where $\vec{Q} \in \mathcal{Q}_{l}$. The moir\'e potential in the other valleys can be obtained using the $C_{3z}$ symmetry, as will be explained around \cref{app:eqn:symmetry_of_t}. Because the inter-valley coupling vanishes at the single-particle level, the moir\'e potential term \emph{generally} ({\it i.e.}{}, even \emph{beyond} the two-center first monolayer harmonic approximation) obeys the following property
\begin{equation}
    \label{app:eqn:valley_u1_symmetry_of_moire}
    \left[ T_{\vec{Q}', \vec{Q}} \right]_{s_1 l_1; s_2 l_2} = 0, \qq{for} \vec{Q} \in \mathcal{Q}_{\eta_1 + l_1} \qq{and} \vec{Q}' \in \mathcal{Q}_{\eta_2 + l_2}, \qq{with} \eta_1
    \neq \eta_2. 
\end{equation}

\subsubsection{Model in real space}\label{app:sec:SnS_SnSe_twist:bl_model:model_real_space}

It is also useful for the following discussion to recast the model obtained in \cref{app:sec:SnS_SnSe_twist:bl_model:model_mom_space} from momentum to real space. For this purpose, we introduce the following real-space fermion operators
\begin{align}
    \hat{\psi}^\dagger_{\eta,s,l} \left( \vec{r} \right) &= \frac{1}{\sqrt{\Omega}} \sum_{\vec{k} \in \text{MBZ}} \sum_{\vec{Q} \in \mathcal{Q}_{\eta + l}} \hat{c}^\dagger_{\vec{k},\vec{Q},s,l} e^{-i \left( \vec{k} - \vec{Q} \right) \cdot \vec{r}}, \label{app:eqn:def_real_space_fermions_to_real}\\
    \hat{c}^\dagger_{\vec{k},\vec{Q},s,l} &= \frac{1}{\sqrt{\Omega}} \int \dd[2]{r} \hat{\psi}^\dagger_{\eta,s,l} \left( \vec{r} \right) e^{i \left( \vec{k} - \vec{Q} \right) \cdot \vec{r}}, \qq{for} \vec{Q} \in \mathcal{Q}_{\eta+l},\label{app:eqn:def_real_space_fermions_to_momentum}
\end{align}
where $\Omega$ is the area of the moir\'e heterostructure. The operator $\hat{\psi}^\dagger_{\eta,s,l} \left( \vec{r} \right)$ creates a fermion of spin $s$ at position $\vec{r}$, within layer $l$ and valley $\eta$. The real space Hamiltonian then takes the form 
\begin{align}
    \mathcal{H} =& - \sum_{\eta,s,l} \int \dd[2]{r} \hat{\psi}^\dagger_{\eta,s,l}\left( \vec{r} \right) \left(C^{-\eta}_{3z} \mathcal{R}^{-1}_{\theta,l} \nabla \right)^{T} \begin{pmatrix}
        \frac{1}{2 m_x} & 0 \\
        0 & \frac{1}{2 m_y} \\
    \end{pmatrix}
    \left(C^{-\eta}_{3z} \mathcal{R}^{-1}_{\theta,l} \nabla \right) \hat{\psi}_{\eta,s,l} \left( \vec{r} \right) \nonumber \\
    &+ \sum_{\substack{\eta, s_1, s_2 \\ l_1,l_2}} \int \dd[2]{r} V^{\eta}_{s_1 l_1; s_2 l_2} \left( \vec{r} \right) \hat{\psi}^\dagger_{\eta,s_1,l_1} \left( \vec{r} \right) \hat{\psi}_{\eta,s_2,l_2} \left( \vec{r} \right),
    \label{app:eqn:real_space_ham_moire_from_first_harm}
\end{align}
with the real-space moir\'e potential being given by
\begin{equation}
	\label{app:eqn:ft_moire_potential_to_real}
    V^{\eta}_{s_1 l_1; s_2 l_2} \left( \vec{r} \right) = \sum_{\vec{G} \in \mathcal{Q}} \left[ T_{ \vec{q}_{\eta+l_1}, \vec{q}_{\eta + l_2} + \vec{G}} \right]_{s_1 l_1; s_2 l_2} e^{i \left(\vec{q}_{\eta + l_2} + \vec{G} - \vec{q}_{\eta + l_1} \right) \cdot \vec{r}},
\end{equation}
where we have employed \cref{app:eqn:extension_with_modulus}. It is easy to check that the real space moir\'e potential obeys the following periodicity property on the moir\'e lattice scale
\begin{equation}
	\label{app:eqn:periodicity_moire_simple}
	V^{\eta}_{s_1 l_1; s_2 l_2} \left( \vec{r} + \vec{R}_{M} \right) = V^{\eta}_{s_1 l_1; s_2 l_2} \left( \vec{r} \right) e^{i \left( \vec{q}_{\eta + l_2} - \vec{q}_{\eta + l_1} \right) \cdot \vec{R}_{M}}, \qq{for any} \vec{R}_{M} \in \mathbb{Z} \vec{a}_{M_1} + \mathbb{Z} \vec{a}_{M_2},
\end{equation} 
or, alternatively,
\begin{align}
	V^{\eta}_{s_1 l; s_2 l} \left( \vec{r} + \vec{R}_{M} \right) &= V^{\eta}_{s_1 l_1; s_2 l_2} \left( \vec{r} \right) e^{i \left( \vec{q}_{\eta + l_2} - \vec{q}_{\eta + l_1} \right) \cdot \vec{R}_{M}}, \label{app:eqn:periodicity_moire_simple_explicit_intra} \\
	V^{\eta}_{s_1 l; s_2 (-l)} \left( \vec{r} + \vec{R}_{M} \right) &= V^{\eta}_{s_1 l; s_2 (-l)} \left( \vec{r} \right) e^{i \left( 2 \vec{q}_{\eta - l} - \vec{q}_{\eta + l} - \vec{q}_{\eta - l} \right) \cdot \vec{R}_{M}} = V^{\eta}_{s_1 l; s_2 (-l)} \left( \vec{r} \right) e^{i \vec{q}_{\eta} \cdot \vec{R}_{M}}, \label{app:eqn:periodicity_moire_simple_explicit_inter}
\end{align} 
for any $\vec{R}_{M} \in \mathbb{Z} \vec{a}_{M_1} + \mathbb{Z} \vec{a}_{M_2}$.

\subsubsection{Exact symmetries of the model}\label{app:sec:SnS_SnSe_twist:bl_model:symmetries}

The \emph{continuum} model features a series of \emph{exact} symmetries depending on the stacking arrangement. The action of a symmetry transformation $g$ on the low-energy fermions introduced in \cref{app:eqn:low_en_ops_c} is given by
\begin{equation}
    \label{app:eqn:sym_action_moire_mom}
    g \hat{c}^\dagger_{\vec{k},\vec{Q},s_1,l_1} g^{-1} = \sum_{s_2,l_2} \left[ D(g) \right]_{s_2 l_2; s_1 l_1} \hat{c}^\dagger_{g\vec{k}, g\vec{Q}, s_2, l_2}.
\end{equation}
In both the AA- and AB-stacking arrangements, the system features $C_{3z}$ and $\mathcal{T}$ symmetries whose representation matrices are given in both cases by
\begin{equation}
    D \left( \mathcal{T} \right) = i s_y \sigma_0,  \quad
	D \left( C_{3z} \right) = e^{-\frac{\pi i}{3} s_z} \sigma_0,
\end{equation}
where here and in what follows, we define $\sigma_{a}$ (for $a=0,x,y,z$) to be the identity and the three Pauli matrices acting on the layer subspace.

Additionally, in the AA-stacked case, the system features $C_{2x}$ symmetry, while in the AB-stacked case, the system has $C_{2y}$ symmetry. The representation matrices of these symmetries read as
\begin{equation}
    D \left( C_{2x} \right) = -i s_x \sigma_x, \quad
    D \left( C_{2y} \right) = -i s_y \sigma_x.
\end{equation}

The action of these symmetries on the real space fermions defined in \cref{app:eqn:def_real_space_fermions_to_real} is given by
\begin{equation}
    \label{app:eqn:sym_action_moire_real}
    g \hat{\psi}^\dagger_{\eta_1,s_1,l_1} \left( \vec{r} \right) g^{-1} = \sum_{\eta_2, s_2,l_2} \left[ D_{\psi}(g) \right]_{\eta_2 s_2 l_2; \eta_1 s_1 l_1} \hat{\psi}^\dagger_{\eta_2, s_2, l_2} \left( g \vec{r} \right),
\end{equation}
with the corresponding real space representation matrices being given by
\begin{alignat}{4}
    D_{\psi} \left( \mathcal{T} \right) &&=& i \begin{pmatrix}
        1 & 0 & 0 \\
        0 & 1 & 0 \\
        0 & 0 & 1 \\
    \end{pmatrix} s_y \sigma_0, & \quad
	D_{\psi} \left( C_{3z} \right) &&=& \begin{pmatrix}
        0 & 0 & 1 \\
        1 & 0 & 0 \\
        0 & 1 & 0 \\
    \end{pmatrix} e^{-\frac{\pi i}{3} s_z} \sigma_0, \nonumber \\
    D_{\psi} \left( C_{2x} \right) &&=& -i \begin{pmatrix}
        1 & 0 & 0 \\
        0 & 0 & 1 \\
        0 & 1 & 0 \\
    \end{pmatrix} s_x \sigma_x, & \quad
	D_{\psi} \left( C_{2y} \right) &&=& -i \begin{pmatrix}
        1 & 0 & 0 \\
        0 & 0 & 1 \\
        0 & 1 & 0 \\
    \end{pmatrix} s_y \sigma_x, 
\end{alignat}
Finally, in both the AA- and AB-stacking cases, the model also features discrete moir\'e translation symmetry
\begin{align}
	T_{\vec{R}_M} \hat{c}^\dagger_{\vec{k},\vec{Q},s,l}  T^{-1}_{\vec{R}_M} &= \hat{c}^\dagger_{\vec{k},\vec{Q},s,l} e^{-i \left( C^{\eta}_{3z} \vec{K}^{l}_{M} + \vec{k} - \vec{Q} \right) \cdot \vec{R}_M}, \label{app:eqn:discrete_translation_moire_fermions_momentum} \\
	T_{\vec{R}_M} \hat{\psi}^\dagger_{\eta,s,l} \left( \vec{r} \right) T^{-1}_{\vec{R}_M} &= \hat{\psi}^\dagger_{\eta,s,l} \left( \vec{r} +  \vec{R}_{M} \right) e^{-i C^{\eta}_{3z} \vec{K}^{l}_{M} \cdot \vec{R}_M}, 
\end{align}
where $T_{\vec{R}_M}$ denotes the discrete moir\'e translation operator, which translates by the moir\'e lattice vector $\vec{R}_M \in \mathbb{Z} \vec{a}_{M_1} + \mathbb{Z} \vec{a}_{M_2}$. \Cref{app:eqn:discrete_translation_moire_fermions_momentum} can be proved by assuming a commensurate configuration such that $\vec{R}_M$ is also a lattice vector of the two monolayers' Bravais lattices. By definition, the action of the translation operator $T_{\vec{R}_M}$ on the monolayer Wannier orbital operators is given by 
\begin{equation}
	T_{\vec{R}_M} \hat{a}^\dagger_{\vec{R}, s, l} T^{-1}_{\vec{R}_M} = \hat{a}^\dagger_{\vec{R} + \mathcal{R}^{-1}_{\theta,l} \vec{R}_M, s, l} \qq{or} 
	T_{\vec{R}_M} \hat{a}^\dagger_{\vec{R}, s, l} T^{-1}_{\vec{R}_M} = \hat{a}^\dagger_{\vec{R} + l \mathcal{R}^{-1}_{\theta,l} \vec{R}_M, s, l},
\end{equation} 
for AA- or AB-stacking, respectively. Using the Fourier transformations from \cref{app:eqn:lattice_FFT_AA,app:eqn:lattice_FFT_AB}, we can determine the action of the translation operator on the momentum space operators for both AA- and AB-stacking
\begin{equation}
	T_{\vec{R}_M} \hat{a}^\dagger_{\vec{k}, s, l} T^{-1}_{\vec{R}_M} = \hat{a}^\dagger_{\vec{k}, s, l} e^{- i \vec{k} \cdot \vec{R}_M}.
\end{equation}
From this, \cref{app:eqn:discrete_translation_moire_fermions_momentum} directly follows using the definition in \cref{app:eqn:low_en_ops_c}. In the continuum limit, \cref{app:eqn:discrete_translation_moire_fermions_momentum} holds even when the commensuration condition is relaxed.

As mentioned already in \cref{app:sec:SnS_SnSe_twist:symmetries_t}, together with the moir\'e translation symmetry, the exact symmetries discussed in this section give rise to the $P3121'$ (SSG 149.22) and $P3211'$ (SSG 150.26) groups in the AA- and AB-stacked cases, respectively\footnote{To avoid any ambiguities, we define the rotation axes using the Cartesian coordinate system. This is because the moir\'e unit cell basis vectors and the single-layer unit cell ones are different, as explained in the beginning of \cref{app:sec:SnS_SnSe_twist:bl_model}.}. Additionally, we note that because the three M valleys are decoupled at the single-particle level, the system also features a $\mathrm{U} \left( {1} \right)\times\mathrm{U} \left( {1} \right)\times\mathrm{U} \left( {1} \right)$ valley-charge symmetry in both the AA- and AB-stacking cases. Because valley is a good quantum number, we can restrict ourselves to a single valley (which, without loss of generality, we choose to be $\eta = 0$). The single-valley system only features $C_{2x}$ ($C_{2y}$) and $\mathcal{T}$ symmetries in the AA- (AB-)stacked case, which give rise to the $P21'$ (SSG 3.2) single-valley symmetry group.

\section{Direct general derivation of the moir\'e potential without gradient terms}\label{app:sec:SnS_SnSe_twist_general_woGradient}

In \cref{app:sec:SnS_SnSe_twist:bl_model} we have derived BM models for the twisted \ch{SnSe2} and \ch{ZrS2} and heterostructures using the two-center first monolayer harmonic approximation for the interlayer tunneling. The form of the latter was derived in \cref{app:sec:SnS_SnSe_twist:symmetries_t} by employing the symmetries of the $\theta \neq 0$ or $\theta = 0$ heterostructures. In this \siSection{}, we will employ a similar strategy to \emph{directly} constrain the moir\'e potential term $T_{\vec{Q}', \vec{Q}}$ defined in \cref{app:eqn:single_particle_hamiltonian}. By doing so, not only the interlayer, but also any \emph{intralayer} moir\'e potential contributions (arising, for example, from lattice relaxation effects) can be constrained using symmetry considerations. The intralayer moir\'e potential contribution are zero in the two-center first monolayer harmonic approximation. Throughout this \siSection{}, we will also assume that the moir\'e potential has the form given by the second row of \cref{app:eqn:real_space_ham_moire_from_first_harm}. The reader is referred to the following \cref{app:sec:SnS_SnSe_twist_general} for a more general derivation that includes gradient terms in the moir\'e potential (which will also be discussed and explicitly defined therein).

Similarly to \cref{app:sec:SnS_SnSe_twist:symmetries_t:exact}, we will first use the exact symmetries of the $\theta \neq 0$ heterostructure to obtain all the symmetry allowed terms (including the intralayer terms) of the moir\'e potential up to the first harmonic (which will be defined below). We will then consider the case of small twist angles, for which the enhanced symmetries of the $\theta = 0$ heterostructure further constrain the moir\'e potential. The main results of this \siSection{} are general symmetry-obeying parameterizations of both the intralayer and interlayer moir\'e potentials.

\subsection{General form of the moir\'e potential restricted by the exact $\theta \neq 0$ symmetries of the heterostructure}\label{app:sec:SnS_SnSe_twist_general_woGradient:full}

We begin by constraining the moir\'e potential using the exact symmetries of the problem arising for non-zero twist angles. For starters, the moir\'e potential matrix must be Hermitian
\begin{equation}
    \label{app:eqn:hermiticity_of_t}
    [T_{\vec{Q}, \vec{Q}'}]_{s_1l_1;s_2l_2} = [T^{\dagger}_{\vec{Q}', \vec{Q}}]_{s_1l_1;s_2l_2},
\end{equation}
and obey the moir\'e periodicity
\begin{equation}
    \label{app:eqn:moire_periodicity_of_t}
    T_{\vec{Q}, \vec{Q}'} = T_{\vec{Q} + \vec{G}, \vec{Q}' + \vec{G}}, \qq{for} \vec{G} \in \mathcal{Q}.
\end{equation}
For every crystalline symmetry $g$ of the twisted structure, whose action on the moir\'e fermions is given by \cref{app:eqn:sym_action_moire_mom}, the moir\'e potential will satisfy
\begin{equation}
    \label{app:eqn:symmetry_of_t}
    T_{g \vec{Q}, g\vec{Q}'} = D (g) T^{(*)}_{\vec{Q}, \vec{Q}'} D^{\dagger}(g),
\end{equation}
where ${}^{(*)}$ indicates that a complex conjugation should be taken in the cases when $g$ is antiunitary. Additionally, the $\mathrm{U} \left( {1} \right) \times \mathrm{U} \left( {1} \right) \times \mathrm{U} \left( {1} \right)$ valley-charge symmetry will require that the moir\'e potential always satisfies \cref{app:eqn:valley_u1_symmetry_of_moire}. 

Using the symmetry constraints of \cref{app:eqn:hermiticity_of_t,app:eqn:moire_periodicity_of_t,app:eqn:symmetry_of_t,app:eqn:valley_u1_symmetry_of_moire} we can construct the most general form of the moir\'e potential. For simplicity, we restrict ourselves to first moir\'e harmonic terms, {\it i.e.}{} such that
\begin{equation}
    \left[ T_{\vec{Q}, \vec{Q}'} \right]_{s_1 l_1; s_2 l_2} = 0, \qq{for} \abs{\vec{Q} - \vec{Q}'} > \abs{\vec{b}_{M_1}}. 
\end{equation}
We find that compared to the two-center first monolayer harmonic approximation from \cref{app:eqn:moire_interlayer_AA,app:eqn:moire_interlayer_AB}, \cref{app:eqn:full_parameterization_AA_stacking,app:eqn:full_parameterization_AB_stacking} contains ten additional real parameters in the AA-stacked case and twelve additional ones in the AB-stacked case. Keeping the same notation as in \cref{app:eqn:full_parameterization_AA_stacking,app:eqn:full_parameterization_AB_stacking} for the parameters characterizing the first monolayer harmonic terms ({\it i.e.}{}, $w^{\text{AA}}_i$ with $1 \leq i \leq 6$ and $w^{\text{AB}}_i$ with $1 \leq i \leq 4$), we denote the additional real parameters by
\begin{equation}
	w^{\prime \text{AA}}_{i}, \qq{with} 1 \leq i \leq 10 \qq{and}
	w^{\prime \text{AB}}_{i}, \qq{with} 1 \leq i \leq 12.
\end{equation}
Depending on the stacking configuration the potential in the first moir\'e Harmonic approximation is given by
\begin{align}
    \left[T^{\text{AA}}_{\vec{q}_{- 1},\vec{q}_{+ 1}-\vec{b}_{M_2}}\right]_{s_1 (-); s_2 (+)} =&\left[\left(w^{\text{AA}}_2-i w^{\text{AA}}_1\right) s_0+\left(w^{\text{AA}}_4+\frac{1}{2} \left(\sqrt{3}+i\right) w^{\text{AA}}_6\right) s_{y}+\left(w^{\text{AA}}_5-i w^{\text{AA}}_3\right) s_{z}\right]_{s_1 s_2}\nonumber \\ 
\left[T^{\text{AA}}_{\vec{q}_{- 1},\vec{q}_{+ 1}}\right]_{s_1 (-); s_2 (+)} =&\left[s_0 w^{\prime\text{AA}}_3+\frac{1}{2} s_{x} \left(-w^{\prime\text{AA}}_8-w^{\prime\text{AA}}_9\right)+\frac{1}{2} i s_{y} \left(w^{\prime\text{AA}}_8-w^{\prime\text{AA}}_9\right)-i s_{z} w^{\prime\text{AA}}_6\right]_{s_1 s_2}\nonumber \\ 
\left[T^{\text{AA}}_{\vec{q}_{- 1},\vec{b}_{M_2}+\vec{q}_{- 1}}\right]_{s_1 (-); s_2 (-)} =&\left[s_0 \left(-i w^{\prime\text{AA}}_1+w^{\prime\text{AA}}_4-i w^{\prime\text{AA}}_7+w^{\prime\text{AA}}_{10}\right)\right]_{s_1 s_2}\nonumber \\ 
\left[T^{\text{AA}}_{\vec{q}_{+ 1},\vec{b}_{M_2}+\vec{q}_{+ 1}}\right]_{s_1 (+); s_2 (+)} =&\left[s_0 \left(-i w^{\prime\text{AA}}_1+w^{\prime\text{AA}}_4+i w^{\prime\text{AA}}_7-w^{\prime\text{AA}}_{10}\right)\right]_{s_1 s_2}\nonumber \\ 
\left[T^{\text{AA}}_{\vec{q}_{- 1},\vec{b}_{M_1}-2 \vec{b}_{M_2}+\vec{q}_{+ 1}}\right]_{s_1 (-); s_2 (+)} =&\left[s_0 w^{\prime\text{AA}}_3+\frac{1}{2} s_{x} \left(w^{\prime\text{AA}}_8+w^{\prime\text{AA}}_9\right)+\frac{1}{2} i s_{y} \left(w^{\prime\text{AA}}_8-w^{\prime\text{AA}}_9\right)-i s_{z} w^{\prime\text{AA}}_6\right]_{s_1 s_2}\nonumber \\ 
\left[T^{\text{AA}}_{\vec{q}_{- 1},\vec{b}_{M_1}-\vec{b}_{M_2}+\vec{q}_{- 1}}\right]_{s_1 (-); s_2 (-)} =&\left[s_0 \left(-i w^{\prime\text{AA}}_1+w^{\prime\text{AA}}_4+i w^{\prime\text{AA}}_7-w^{\prime\text{AA}}_{10}\right)\right]_{s_1 s_2}\nonumber \\ 
\left[T^{\text{AA}}_{\vec{q}_{- 1},\vec{b}_{M_1}-\vec{b}_{M_2}+\vec{q}_{+ 1}}\right]_{s_1 (-); s_2 (+)} =&\left[\left(w^{\text{AA}}_2+i w^{\text{AA}}_1\right) s_0+\left(-w^{\text{AA}}_4-\frac{1}{2} \left(\sqrt{3}-i\right) w^{\text{AA}}_6\right) s_{y}+\left(-w^{\text{AA}}_5-i w^{\text{AA}}_3\right) s_{z}\right]_{s_1 s_2}\nonumber \\ 
\left[T^{\text{AA}}_{\vec{q}_{+ 1},\vec{b}_{M_1}-\vec{b}_{M_2}+\vec{q}_{+ 1}}\right]_{s_1 (+); s_2 (+)} =&\left[s_0 \left(-i w^{\prime\text{AA}}_1+w^{\prime\text{AA}}_4-i w^{\prime\text{AA}}_7+w^{\prime\text{AA}}_{10}\right)\right]_{s_1 s_2}\nonumber \\ 
\left[T^{\text{AA}}_{\vec{q}_{- 1},\vec{b}_{M_1}+\vec{q}_{- 1}}\right]_{s_1 (-); s_2 (-)} =&\left[s_0 \left(w^{\prime\text{AA}}_5-i w^{\prime\text{AA}}_2\right)\right]_{s_1 s_2}\nonumber \\ 
\left[T^{\text{AA}}_{\vec{q}_{+ 1},\vec{b}_{M_1}+\vec{q}_{+ 1}}\right]_{s_1 (+); s_2 (+)} =&\left[s_0 \left(w^{\prime\text{AA}}_5-i w^{\prime\text{AA}}_2\right)\right]_{s_1 s_2} , \label{app:eqn:full_parameterization_AA_stacking}\\
    \left[T^{\text{AB}}_{\vec{q}_{- 1},\vec{q}_{+ 1}-\vec{b}_{M_2}}\right]_{s_1 (-); s_2 (+)} =&\left[w^{\text{AB}}_2 s_0-i w^{\text{AB}}_1 s_{x}+w^{\text{AB}}_3 s_{y}-i w^{\text{AB}}_4 s_{z}\right]_{s_1 s_2}\nonumber \\ 
\left[T^{\text{AB}}_{\vec{q}_{- 1},\vec{q}_{+ 1}}\right]_{s_1 (-); s_2 (+)} =&\left[s_0 \left(w^{\prime\text{AB}}_4+i w^{\prime\text{AB}}_9\right)+s_{x} \left(-w^{\prime\text{AB}}_{11}-i w^{\prime\text{AB}}_1\right)+s_{z} \left(w^{\prime\text{AB}}_5-i w^{\prime\text{AB}}_8\right)\right]_{s_1 s_2}\nonumber \\ 
\left[T^{\text{AB}}_{\vec{q}_{- 1},\vec{b}_{M_2}+\vec{q}_{- 1}}\right]_{s_1 (-); s_2 (-)} =&\left[s_0 \left(i w^{\prime\text{AB}}_2+w^{\prime\text{AB}}_6+i w^{\prime\text{AB}}_{10}+w^{\prime\text{AB}}_{12}\right)\right]_{s_1 s_2}\nonumber \\ 
\left[T^{\text{AB}}_{\vec{q}_{+ 1},\vec{b}_{M_2}+\vec{q}_{+ 1}}\right]_{s_1 (+); s_2 (+)} =&\left[s_0 \left(-i w^{\prime\text{AB}}_2+w^{\prime\text{AB}}_6+i w^{\prime\text{AB}}_{10}-w^{\prime\text{AB}}_{12}\right)\right]_{s_1 s_2}\nonumber \\ 
\left[T^{\text{AB}}_{\vec{q}_{- 1},\vec{b}_{M_1}-2 \vec{b}_{M_2}+\vec{q}_{+ 1}}\right]_{s_1 (-); s_2 (+)} =&\left[s_0 \left(w^{\prime\text{AB}}_4-i w^{\prime\text{AB}}_9\right)+s_{x} \left(w^{\prime\text{AB}}_{11}-i w^{\prime\text{AB}}_1\right)+s_{z} \left(-w^{\prime\text{AB}}_5-i w^{\prime\text{AB}}_8\right)\right]_{s_1 s_2}\nonumber \\ 
\left[T^{\text{AB}}_{\vec{q}_{- 1},\vec{b}_{M_1}-\vec{b}_{M_2}+\vec{q}_{- 1}}\right]_{s_1 (-); s_2 (-)} =&\left[s_0 \left(i w^{\prime\text{AB}}_2+w^{\prime\text{AB}}_6-i w^{\prime\text{AB}}_{10}-w^{\prime\text{AB}}_{12}\right)\right]_{s_1 s_2}\nonumber \\ 
\left[T^{\text{AB}}_{\vec{q}_{- 1},\vec{b}_{M_1}-\vec{b}_{M_2}+\vec{q}_{+ 1}}\right]_{s_1 (-); s_2 (+)} =&\left[w^{\text{AB}}_2 s_0-i w^{\text{AB}}_1 s_{x}-w^{\text{AB}}_3 s_{y}-i w^{\text{AB}}_4 s_{z}\right]_{s_1 s_2}\nonumber \\ 
\left[T^{\text{AB}}_{\vec{q}_{+ 1},\vec{b}_{M_1}-\vec{b}_{M_2}+\vec{q}_{+ 1}}\right]_{s_1 (+); s_2 (+)} =&\left[s_0 \left(-i w^{\prime\text{AB}}_2+w^{\prime\text{AB}}_6-i w^{\prime\text{AB}}_{10}+w^{\prime\text{AB}}_{12}\right)\right]_{s_1 s_2}\nonumber \\ 
\left[T^{\text{AB}}_{\vec{q}_{- 1},\vec{b}_{M_1}+\vec{q}_{- 1}}\right]_{s_1 (-); s_2 (-)} =&\left[s_0 \left(w^{\prime\text{AB}}_7+i w^{\prime\text{AB}}_3\right)\right]_{s_1 s_2}\nonumber \\ 
\left[T^{\text{AB}}_{\vec{q}_{+ 1},\vec{b}_{M_1}+\vec{q}_{+ 1}}\right]_{s_1 (+); s_2 (+)} =&\left[s_0 \left(w^{\prime\text{AB}}_7-i w^{\prime\text{AB}}_3\right)\right]_{s_1 s_2} .
    \label{app:eqn:full_parameterization_AB_stacking}
\end{align}
In \cref{app:eqn:full_parameterization_AA_stacking} we have only listed the half of the nonzero components of $T_{\vec{Q}, \vec{Q}'}$ for valley $\eta = 0$. The other nonzero components in valley $\eta = 0$ can be obtained with the Hermiticity condition from \cref{app:eqn:hermiticity_of_t}. 
The moir\'e potential in the other valleys can be obtained with the $C_{3z}$ symmetry using \cref{app:eqn:symmetry_of_t}. The additional terms are within the first moir\'e harmonic approximation, but go beyond the two-center first monolayer harmonic approximation.

\subsection{General form of the moir\'e potential restricted by the approximate $\theta = 0$ symmetries of the heterostructure}\label{app:sec:SnS_SnSe_twist_general_woGradient:approx}

Having obtained the most general form of the moir\'e potential for an arbitrary non-zero angle, we now further constrain its form in the limit of vanishing twist angle $\theta \to 0$, using the exact symmetries of the untwisted configuration. In doing so, we will make the ``local-stacking approximation''~\cite{JUN14}. We will make this approximation mathematically rigorous in \cref{app:sec:SnS_SnSe_twist_general_woGradient:approx:twisted}. Briefly, for small (but nonzero) twist angle $\theta$, we can \emph{locally} understand the heterostructure as being comprised of two monolayers which are untwisted, but are otherwise displaced by $\Delta \vec{R}$, where $\Delta \vec{R}$ is determined by the twist angle and the position within the heterostructure. We will first study the untwisted bilayer Hamiltonians $\mathcal{H} \left( \Delta \vec{R} \right)$ for different displacements $\Delta \vec{R}$, and characterize their symmetries. Then, we will illustrate how the moir\'e continuum Hamiltonian can be obtained from $\mathcal{H} \left( \Delta \vec{R} \right)$, which will finally allow us to further constrain the moir\'e potential in the limit of vanishing twist angle.

\subsubsection{Hamiltonian for the untwisted configuration}\label{app:sec:SnS_SnSe_twist_general_woGradient:approx:untwisted}

We begin by considering a family of Hamiltonians $\mathcal{H} \left( \Delta \vec{R} \right)$ describing an untwisted bilayer arrangement where the top (bottom) layer is displaced by $+\frac12 \Delta\vec{R}$ ($-\frac12 \Delta\vec{R}$). In the untwisted AA-stacked configuration, the two layers are stacked directly on top of one another for $\Delta \vec{R} = \vec{0}$, while in the AB-stacked configuration, the bottom layer is rotated around the origin by \SI{180}{\degree} relative to the top layer in the $\Delta \vec{R} = \vec{0}$ case. We let $\hat{a}^\dagger_{\vec{R},l,s}$ denote the creation operator corresponding to an electron in layer $l$ located at $\vec{r}^{\Delta \vec{R}}_{l,\vec{R}}  = \vec{R} + \frac{l}{2} \Delta \vec{R} $ ($\vec{r}^{\Delta \vec{R}}_{l,\vec{R}} = l \vec{R} + \frac{l}{2} \Delta \vec{R} $) in the AA- (AB-)stacking untwisted configuration. The Hamiltonian $\mathcal{H} \left( \Delta \vec{R} \right)$ can be written generically as
\begin{equation}
    \label{app:eqn:untwisted_snse_hamiltonian}
    \mathcal{H} \left( \Delta \vec{R} \right) = \sum_{l} \mathcal{H}^{\text{sl}}_{l} + \sum_{\substack{\vec{R}_1,l_1,s_1 \\ \vec{R}_2,l_2, s_2}} S_{s_1 l_1; s_2 l_2} \left( \Delta \vec{R}, \vec{r}^{\Delta \vec{R}}_{l_2,\vec{R}_2} - \vec{r}^{\Delta \vec{R}}_{l_1,\vec{R}_1} \right) \hat{a}^\dagger_{\vec{R}_1,s_1,l_1} \hat{a}_{\vec{R}_2,s_2,l_2}, 
\end{equation}
where the first term is the single-particle Hamiltonian of a single layer given by \cref{app:eqn:ham_sl_snse2}
\begin{equation}
	\label{app:eqn:local_staccking_definition_hsl}
    \mathcal{H}^{\text{sl}}_{l} = \sum_{\vec{k}} \left[h^{\sl} \left( \vec{k} \right) \right]_{s_1 s_2} \hat{a}^\dagger_{\vec{k},s_1,l} \hat{a}_{\vec{k},s_2,l}.
\end{equation}
The second term of \cref{app:eqn:untwisted_snse_hamiltonian} is the additional single-particle contribution arising from stacking the two layers. It contains both interlayer terms (arising from tunneling), as well as intralayer terms, arising from {\it e.g.}{} assisted hopping through the adjacent layer. Because the Hamiltonian $\mathcal{H} \left( \Delta \vec{R} \right)$ still features the discrete translation symmetry of the single layer material, for a given interlayer displacement $\Delta \vec{R}$, the amplitude of the second term $S_{s_1 l_1; s_2 l_2} \left( \Delta \vec{R}, \vec{r}^{\Delta \vec{R}}_{l_2,\vec{R}_2} - \vec{r}^{\Delta \vec{R}}_{l_1,\vec{R}_1} \right)$ only depends on the distance $\vec{r}^{\Delta \vec{R}}_{l_2,\vec{R}_2} - \vec{r}^{\Delta \vec{R}}_{l_1,\vec{R}_1}$, but not on the positions $\vec{r}^{\Delta \vec{R}}_{l_2,\vec{R}_2}$ and $\vec{r}^{\Delta \vec{R}}_{l_1,\vec{R}_1}$ separately. We will also assume that the perturbation stemming from the stacking the two layers only couples fermions that are a few lattice sites apart, which means that $S_{s_1 l_1; s_2 l_2} \left( \Delta \vec{R}, \vec{r} \right) \approx 0$ for $\vec{r} \gg \abs{\vec{a}_1}$. We will later use this condition in \cref{app:sec:SnS_SnSe_twist_general_woGradient:approx:twisted} where we derive the continuum model for the twisted moir\'e heterostructure.

\newcommand{\tBl}{\text{bl}}
In the untwisted bilayer arrangement, the tunneling between two fermions located at the same distance $\vec{r}$ is the same irrespective of whether the layers are shifted by $\Delta \vec{R}$ or by $\Delta \vec{R} + \vec{a}_{i}$ (for $1 \leq i \leq 2$), {\it i.e.}{}
\begin{equation}
    S_{s_1 l_1; s_2 l_2} \left( \Delta \vec{R} + \vec{a}_{i}, \vec{r} \right) = S_{s_1 l_1; s_2 l_2} \left( \Delta \vec{R}, \vec{r} \right), \qq{for} 1 \leq i \leq 2.
    \label{app:eq:transl_sym_local_stacking_ham}
\end{equation}
Consider now a symmetry operation $g$ of the zero-displacement ($\Delta \vec{R} = \vec{0}$) untwisted configuration, whose action on the lattice fermions is given by
\begin{equation}
    g \hat{a}^\dagger_{\vec{R},s_1,l} g^{-1} = \sum_{s_2} \left[ D^{\text{sl}}(g) \right]_{s_2 s_1} \hat{a}^\dagger_{g \vec{R}, s_2, \left( \epsilon_g l \right) },
\end{equation}
where similarly to \cref{app:eqn:const_interlayer_mat_elem}, $\epsilon_g = +1$ ($\epsilon_g = -1$) if $g$ exchanges the layers, and $D^{\text{sl}}(g)$ has been introduced in~\cref{app:eqn:sym_action_real_ops}. We also comment that certain single-layer symmetry operations, such as $C_{2x}$ or $C_{2y}$, also flip the layer index, which justify the presence of the $\epsilon_{g}$ factor. The action of the symmetry $g$ on the untwisted, but displaced bilayer Hamiltonian is given by
\begin{equation}
    \label{app:eqn:action_g_hamiltonian_untwisted}
    g \mathcal{H} \left( \Delta \vec{R} \right) g^{-1} = \mathcal{H} \left( \epsilon_{g} g \Delta \vec{R} \right).
\end{equation}
We also provide an intuitive understanding of the above equation. By acting with a symmetry operation $g$, we obtain a new Hamiltonian, $g \mathcal{H} \left( \Delta \vec{R} \right) g^{-1}$. Considering {\it e.g.}{} the AA-stacking configuration, the symmetry operation $g$ changes the position of a fermion operator $\hat{a}^\dagger_{\vec{R}, s, l}$ from $\vec{r}^{\Delta \vec{R}}_{l,\vec{R}}  = \vec{R} + \frac{l}{2} \Delta \vec{R} $ within layer $l$ to $g \vec{R} + \frac{l}{2} g \Delta \vec{R} = \vec{r}^{\epsilon_{g} g \Delta \vec{R}}_{\epsilon_{g} l ,g \vec{R}}$ within layer $\epsilon_{g} l$. Effectively, $g$ changes the interlayer displacement from $\Delta \vec{R}$ to $\epsilon_{g} g \Delta \vec{R}$, thereby showing that the two Hamiltonians $ g \mathcal{H} \left( \Delta \vec{R} \right) g^{-1}$ and $\mathcal{H} \left( \epsilon_{g} g \Delta \vec{R} \right)$ are equivalent. Since $g$ is a symmetry of the undisplaced and untwisted configuration, we have $\commutator{g}{\sum_{l} \mathcal{H}^{\sl}_l} = 0$, where $\mathcal{H}^{\sl}_l$ was defined in \cref{app:eqn:local_staccking_definition_hsl}, which means that \cref{app:eqn:untwisted_snse_hamiltonian,app:eqn:action_g_hamiltonian_untwisted} simplify to
\begin{align}
    & \sum_{\substack{\vec{R}_1, l_1, s_1, s'_1 \\ \vec{R}_2, l_2, s_2, s'_2}} S^{(*)}_{s_1 l_1; s_2 l_2} \left( \Delta \vec{R}, \vec{r}^{\Delta \vec{R}}_{l_2,\vec{R}_2} - \vec{r}^{\Delta \vec{R}}_{l_1,\vec{R}_1} \right) \left[ D^{\text{sl}}(g) \right]_{s'_1 s_1} \left[ D^{\text{sl}}(g) \right]^{*}_{s'_2 s_2} \hat{a}^\dagger_{g\vec{R}_1,s'_1,\epsilon_g l_1} \hat{a}_{g\vec{R}_2,s'_2,\epsilon_g l_2} \nonumber \\
    =& \sum_{\substack{\vec{R}_1, l_1, s_1 \\ \vec{R}_2, l_2, s_2}} S_{s_1 l_1; s_2 l_2} \left( \epsilon_g g\Delta \vec{R}, \vec{r}^{\epsilon_g g\Delta \vec{R}}_{l_2,\vec{R}_2} - \vec{r}^{\epsilon_g g\Delta \vec{R}}_{l_1,\vec{R}_1} \right) \hat{a}^\dagger_{\vec{R}_1,s_1,l_1} \hat{a}_{\vec{R}_2,s_2,l_2}. \label{app:eqn:sym_constraint_untwisted_second_quantized}
\end{align}
In \cref{app:eqn:sym_constraint_untwisted_second_quantized}, ${}^{(*)}$ indicates that a complex conjugation should be performed if $g$ is antiunitary. Equivalently, \cref{app:eqn:sym_constraint_untwisted_second_quantized} implies that
\begin{align}
    \sum_{s_1, s_2} \left[ D^{\text{sl}}(g) \right]_{s'_1 s_1} S^{(*)}_{s_1 l_1; s_2 l_2} \left( \Delta \vec{R}, \vec{r}^{\Delta \vec{R}}_{l_2,\vec{R}_2} - \vec{r}^{\Delta \vec{R}}_{l_1,\vec{R}_1} \right) \left[ D^{\text{sl}}(g) \right]^{*}_{s'_2 s_2} &= S_{s'_1 \epsilon_g l_1; s'_2 \epsilon_g l_2} \left( \epsilon_g g \Delta \vec{R}, g \left( \vec{r}^{\Delta \vec{R}}_{l_2,\vec{R}_2} - \vec{r}^{\Delta \vec{R}}_{l_1,\vec{R}_1} \right) \right) \nonumber \\
    \sum_{s_1, s_2} \left[ D^{\text{sl}}(g) \right]_{s'_1 s_1} S^{(*)}_{s_1 l_1; s_2 l_2} \left( \Delta \vec{R}, \vec{r} \right) \left[ D^{\text{sl}}(g) \right]^{*}_{s'_2 s_2} &= S_{s'_1 \epsilon_g l_1; s'_2 \epsilon_g l_2} \left( \epsilon_g g \Delta \vec{R}, g \vec{r} \right) \label{app:eqn:symmetry_moire_s}
\end{align}
Finally, we note that as a result of hermiticity, we must have that
\begin{equation}
    \label{app:eqn:hermiticity_moire_s}
    S_{s_1 l_1; s_2 l_2} \left( \Delta \vec{R}, \vec{r} \right) = S^{*}_{s_2 l_2; s_1 l_1} \left( \Delta \vec{R}, -\vec{r} \right).
\end{equation}

Because $S_{s_1 l_1; s_2 l_2} \left( \Delta \vec{R}, \vec{r} \right)$ is periodic in $\Delta \vec{R}$ (see~\cref{app:eq:transl_sym_local_stacking_ham}) for a given $\vec{r}$, it can be expanded as a Fourier series
\begin{equation}
    S_{s_1 l_1; s_2 l_2} \left( \Delta \vec{R}, \vec{r} \right)= \sum_{\vec{g}} S_{s_1 l_1; s_2 l_2} \left( \vec{g}, \vec{r} \right) e^{-i \vec{g} \cdot \Delta \vec{R}},
\end{equation}
where $\vec{g}$ runs over the reciprocal lattice vectors of the single-layer material. Additionally, for a given $\Delta \vec{R}$ $S_{s_1 l_1; s_2 l_2} \left( \Delta \vec{R}, \vec{r} \right)$ is only defined for discrete values of $\vec{r} = \vec{r}^{\Delta \vec{R}}_{l_2,\vec{R}_2} - \vec{r}^{\Delta \vec{R}}_{l_1,\vec{R}_1}$. As such, $S_{s_1 l_1; s_2 l_2} \left( \Delta \vec{R}, \vec{r} \right)$ also accepts a momentum space representation in the second variable
\begin{align}
    S_{s_1 l_1; s_2 l_2} \left( \Delta \vec{R}, \vec{r} \right) =& \frac{1}{N} \sum_{\vec{k}} S_{s_1 l_1; s_2 l_2} \left( \Delta \vec{R}, \vec{k} \right) e^{-i \vec{k} \cdot \vec{r}}, \label{app:eqn:fourier_representation_S_first_to_real}\\
    S_{s_1 l_1; s_2 l_2} \left( \Delta \vec{R}, \vec{k}\right) =&  \sum_{\vec{R}_1} S_{s_1 l_1; s_2 l_2} \left( \Delta \vec{R}, \vec{r}^{\Delta \vec{R}}_{l_2,\vec{R}_2} - \vec{r}^{\Delta \vec{R}}_{l_1,\vec{R}_1} \right) e^{i \vec{k} \cdot \left( \vec{r}^{\Delta \vec{R}}_{l_2,\vec{R}_2} - \vec{r}^{\Delta \vec{R}}_{l_1,\vec{R}_1} \right)} ,
    \label{app:eqn:fourier_representation_S_first_to_momentum}
\end{align}
with $\vec{k}$ belonging to the first BZ of the single-layer material. \Cref{app:eqn:fourier_representation_S_first_to_momentum} can be extended outside the first Brillouin zone by noting that
\begin{align}
    S_{s_1 l_1; s_2 l_2} \left( \Delta \vec{R}, \vec{k} + \vec{g} \right) =&  \sum_{\vec{R}_1} S_{s_1 l_1; s_2 l_2} \left( \Delta \vec{R}, \vec{r}^{\Delta \vec{R}}_{l_2,\vec{R}_2} - \vec{r}^{\Delta \vec{R}}_{l_1,\vec{R}_1} \right) e^{i \left( \vec{k} + \vec{g} \right) \cdot \left( \vec{r}^{\Delta \vec{R}}_{l_2,\vec{R}_2} - \vec{r}^{\Delta \vec{R}}_{l_1,\vec{R}_1} \right)} \nonumber \\
    =&  \sum_{\vec{R}_1} S_{s_1 l_1; s_2 l_2} \left( \Delta \vec{R}, \vec{r}^{\Delta \vec{R}}_{l_2,\vec{R}_2} - \vec{r}^{\Delta \vec{R}}_{l_1,\vec{R}_1} \right) e^{i \vec{k}  \cdot \left( \vec{r}^{\Delta \vec{R}}_{l_2,\vec{R}_2} - \vec{r}^{\Delta \vec{R}}_{l_1,\vec{R}_1} \right)} e^{i \vec{g} \cdot \Delta \vec{R} \frac{l_2 - l_1}{2} }\nonumber \\
    =&  S_{s_1 l_1; s_2 l_2} \left( \Delta \vec{R}, \vec{k} \right) e^{i \vec{g} \cdot \Delta \vec{R} \frac{l_2 - l_1}{2} }, \label{app:eqn:fourier_representation_S_first_extension}
\end{align}
where the phase is just the ``embedding'' factor of a typical tight-binding Hamiltonian. Finally, one can Fourier-transform over both variables and obtain
\begin{equation}
    \label{app:eqn:fourier_representation_S_both}
    S_{s_1 l_1; s_2 l_2} \left( \Delta \vec{R}, \vec{r} \right)= \frac{1}{N} \sum_{\vec{g},\vec{k}} S_{s_1 l_1; s_2 l_2} \left( \vec{g}, \vec{k} \right) e^{-i \vec{g} \cdot \Delta \vec{R}} e^{-i \vec{k} \cdot \vec{r}},
\end{equation}
which, as a result of \cref{app:eqn:fourier_representation_S_first_extension}, obeys 
\begin{equation}
	\label{app:eqn:periodicity_of_s_matrix_local_stacking}
    S_{s_1 l_1; s_2 l_2} \left( \vec{g}, \vec{k} + \vec{g}' \right)  = S_{s_1 l_1; s_2 l_2} \left( \vec{g} + \frac{l_2-l_1}{2} \vec{g}', \vec{k} \right).
\end{equation}

\subsubsection{Hamiltonian for the twisted configuration}\label{app:sec:SnS_SnSe_twist_general_woGradient:approx:twisted}

We are now in a position of obtaining the Hamiltonian for the twisted bilayer arrangement. As in \cref{app:sec:SnS_SnSe_twist:bl_model_derivation}, in the twisted configuration, the fermion $\hat{c}^\dagger_{\vec{R},s,l}$ is located at $\vec{r}_{l,\vec{R}} = R_{\theta,l} \vec{R}$ in the AA-stacked case or at $\vec{r}_{l,\vec{R}} = l R_{\theta,l} \vec{R}$ in the AB-stacked case. Consider now two fermions $\hat{c}^\dagger_{\vec{R}_1,s_1,l_1}$ and $\hat{c}^\dagger_{\vec{R}_2,s_2,l_2}$. At the single-particle level, these two fermions will only be coupled provided that $\abs{\vec{r}_{l_1,\vec{R}_1} - \vec{r}_{l_2,\vec{R}_2}}$ is not larger than a few lattice constants of the single-layer material. As explained in the beginning of \cref{app:sec:SnS_SnSe_twist:bl_model}, at a position $\vec{r}$, the two monolayer lattices are displaced by $\delta \vec{R} \left( \vec{r} \right)$, where the effective local displacement was obtained in \cref{app:eqn:local_interlayer_displacement}. Because for any two fermions $\hat{c}^\dagger_{\vec{R}_1,s_1,l_1}$ and $\hat{c}^\dagger_{\vec{R}_2,s_2,l_2}$ coupled at the single-particle level, $\abs{\vec{r}_{l_1,\vec{R}_1} - \vec{r}_{l_2,\vec{R}_2}} \sim \abs{\vec{a}_{1}} \ll \abs{\vec{a}_{M_1}}$, we will have $\delta \vec{R} \left( \vec{r}_{l_1,\vec{R}_1} \right) \approx \delta \vec{R} \left(\vec{r}_{l_2,\vec{R}_2} \right) \approx \delta \vec{R} \left(\frac{\vec{r}_{l_1,\vec{R}_1} + \vec{r}_{l_2,\vec{R}_2} }{2}\right)$. As such, we can take the single-particle term coupling these two fermions to be~\cite{JUN14}
\begin{equation}
    S_{s_1 l_1; s_2 l_2} \left( \delta \vec{R} \left(\frac{\vec{r}_{l_1,\vec{R}_1} + \vec{r}_{l_2,\vec{R}_2} }{2}\right) , \vec{r}_{l_2,\vec{R}_2} - \vec{r}_{l_1,\vec{R}_1} \right) \hat{a}^\dagger_{\vec{R}_1,s_1,l_1} \hat{a}_{\vec{R}_2,s_2,l_2},
\end{equation}
which allows us to write an expression for the Hamiltonian of the twisted bilayer heterostructure
\begin{align}
    \mathcal{H} &= \sum_{l} \mathcal{H}^{\text{sl}}_{l} + \mathcal{H}^{\text{moir\'e}}, \qq{with} \nonumber \\
    \mathcal{H}^{\text{moir\'e}} &= \sum_{\substack{\vec{R}_1,l_1,s_1 \\ \vec{R}_2,l_2, s_2}} S_{s_1 l_1; s_2 l_2} \left( \delta \vec{R} \left(\frac{\vec{r}_{l_1,\vec{R}_1} + \vec{r}_{l_2,\vec{R}_2} }{2}\right) , \vec{r}_{l_2,\vec{R}_2} - \vec{r}_{l_1,\vec{R}_1} \right) \hat{a}^\dagger_{\vec{R}_1,s_1,l_1} \hat{a}_{\vec{R}_2,s_2,l_2}.     \label{app:eqn:twisted_snse_hamiltonian_interm_1}
\end{align}
We now take a low-energy approximation for the lattice fermions, using the notation introduced in \cref{app:eqn:def_real_space_fermions_to_real}
\begin{equation}
    \hat{a}^\dagger_{\vec{R},s,l} \approx \sqrt{\Omega_0} \sum_{\eta} \int \dd[2]{r} \delta\left( \vec{r}_{l, \vec{R}} - \vec{r} \right) \hat{\psi}^\dagger_{\eta,s,l} \left( \vec{r} \right) e^{-i C^{\eta}_{3z} \vec{K}^{l}_M \cdot \vec{r}},
\end{equation}
with the aid of which the moir\'e potential part of the Hamiltonian becomes
\begin{align}
    \mathcal{H}^{\text{moir\'e}} =& \Omega_0 \sum_{\substack{\vec{R}_1,l_1,s_1,\eta_1 \\ \vec{R}_2,l_2, s_2, \eta_2}} \int \dd[2]{r_1} \dd[2]{r_2} S_{s_1 l_1; s_2 l_2} \left( \delta \vec{R} \left(\frac{\vec{r}_1 + \vec{r}_2 }{2}\right) , \vec{r}_2 - \vec{r}_{1} \right) \delta\left( \vec{r}_{l_1,\vec{R}_1} - \vec{r}_1 \right) \delta\left( \vec{r}_{l_2,\vec{R}_2} - \vec{r}_2 \right) \nonumber \\
    &\times \hat{\psi}^\dagger_{\eta_1,s_1,l_1} \left( \vec{r}_1 \right) \hat{\psi}_{\eta_2,s_2,l_2} \left( \vec{r}_2 \right) e^{-i C^{\eta_1}_{3z} \vec{K}^{l_1}_M \cdot \vec{r}_1} e^{i C^{\eta_2}_{3z} \vec{K}^{l_2}_M \cdot \vec{r}_2}. \label{app:eqn:twisted_snse_hamiltonian_interm_2}
\end{align}
We then employ the Poisson summation formula
\begin{equation}
    \sum_{\vec{R}} \delta \left( \vec{r}_{l,\vec{R}} - \vec{r} \right) = \frac{1}{\Omega_0} \sum_{\vec{g}} e^{i \mathcal{R}_{\theta,l} \vec{g} \cdot \vec{r}},
\end{equation}
and simultaneously change the integration variables to the center of mass $\vec{x}= \frac{\vec{r}_1 + \vec{r}_2}{2}$ and displacement $\vec{y} = \vec{r}_2 - \vec{r}_1$ coordinates in \cref{app:eqn:twisted_snse_hamiltonian_interm_2}
\begin{align}
    \mathcal{H}^{\text{moir\'e}} =& \frac{1}{\Omega_0} \sum_{\substack{\vec{g}_1,l_1,s_1,\eta_1 \\ \vec{g}_2,l_2, s_2, \eta_2}} \int \dd[2]{x} \dd[2]{y} S_{s_1 l_1; s_2 l_2} \left( \delta \vec{R} \left( \vec{x} \right) , \vec{y} \right) e^{-i \mathcal{R}_{\theta,l_1} \left( \vec{g}_1 + C^{\eta_1}_{3z} \vec{K}_M \right)\cdot \left( \vec{x} - \frac{\vec{y}}{2} \right)} e^{i \mathcal{R}_{\theta,l_2} \left( \vec{g}_2 + C^{\eta_2}_{3z} \vec{K}_M \right)\cdot \left( \vec{x} + \frac{\vec{y}}{2} \right)} \nonumber \\
    &\times \hat{\psi}^\dagger_{\eta_1,s_1,l_1} \left( \vec{x} - \frac{\vec{y}}{2} \right) \hat{\psi}_{\eta_2,s_2,l_2} \left( \vec{x} + \frac{\vec{y}}{2} \right) \nonumber \\
\approx & \frac{1}{\Omega_0} \sum_{\substack{\vec{g}_1,l_1,s_1,\eta_1 \\ \vec{g}_2,l_2, s_2, \eta_2}} \int \dd[2]{x} \dd[2]{y} S_{s_1 l_1; s_2 l_2} \left( \delta \vec{R} \left( \vec{x} \right) , \vec{y} \right) e^{-i \mathcal{R}_{\theta,l_1} \left( \vec{g}_1 + C^{\eta_1}_{3z} \vec{K}_M \right)\cdot \left( \vec{x} - \frac{\vec{y}}{2} \right)} e^{i \mathcal{R}_{\theta,l_2} \left( \vec{g}_2 + C^{\eta_2}_{3z} \vec{K}_M \right)\cdot \left( \vec{x} + \frac{\vec{y}}{2} \right)} \nonumber \\
    &\times \hat{\psi}^\dagger_{\eta_1,s_1,l_1} \left( \vec{x} \right) \hat{\psi}_{\eta_2,s_2,l_2} \left( \vec{x} \right) \nonumber \\
\approx & \frac{1}{\Omega_0} \sum_{\substack{\vec{g},\eta \\ l_1,s_1,l_2, s_2}} \int \dd[2]{x} \dd[2]{y} S_{s_1 l_1; s_2 l_2} \left( \delta \vec{R} \left( \vec{x} \right) , \vec{y} \right) e^{-i \mathcal{R}_{\theta,l_1} \left( \vec{g} + C^{\eta}_{3z} \vec{K}_M \right)\cdot \left( \vec{x} - \frac{\vec{y}}{2} \right)} e^{i \mathcal{R}_{\theta,l_2} \left( \vec{g} + C^{\eta}_{3z} \vec{K}_M \right)\cdot \left( \vec{x} + \frac{\vec{y}}{2} \right)} \nonumber \\
    &\times \hat{\psi}^\dagger_{\eta,s_1,l_1} \left( \vec{x} \right) \hat{\psi}_{\eta,s_2,l_2} \left( \vec{x} \right)  \nonumber \\
= & \frac{1}{\Omega_0} \sum_{\substack{\vec{g},\eta \\ l_1,s_1,l_2, s_2}} \int \dd[2]{x} \dd[2]{y} S_{s_1 l_1; s_2 l_2} \left( \delta \vec{R} \left( \vec{x} \right) , \vec{y} \right) e^{i \left( \mathcal{R}_{\theta,l_1} + \mathcal{R}_{\theta,l_2}  \right) \left( \vec{g} + C^{\eta}_{3z} \vec{K}_M \right)\cdot \frac{\vec{y}}{2} } \nonumber \\
    &\times \hat{\psi}^\dagger_{\eta,s_1,l_1} \left( \vec{x} \right) \hat{\psi}_{\eta,s_2,l_2} \left( \vec{x} \right) e^{-i \mathcal{R}_{\theta,l_1} \left( \vec{g} + C^{\eta}_{3z} \vec{K}_M \right)\cdot \vec{x} +i \mathcal{R}_{\theta,l_2} \left( \vec{g} + C^{\eta}_{3z} \vec{K}_M \right)\cdot \vec{x} }   \nonumber \\
\approx & \frac{1}{\Omega_0} \sum_{\substack{\vec{g},\eta \\ l_1,s_1,l_2, s_2}} \int \dd[2]{x} \dd[2]{y} S_{s_1 l_1; s_2 l_2} \left( \delta \vec{R} \left( \vec{x} \right) , \vec{y} \right) e^{i \left( \vec{g} + C^{\eta}_{3z} \vec{K}_M \right)\cdot \vec{y} } \nonumber \\
    &\times \hat{\psi}^\dagger_{\eta,s_1,l_1} \left( \vec{x} \right) \hat{\psi}_{\eta,s_2,l_2} \left( \vec{x} \right) e^{-i \mathcal{R}_{\theta,l_1} \left( \vec{g} + C^{\eta}_{3z} \vec{K}_M \right)\cdot \vec{x} +i \mathcal{R}_{\theta,l_2} \left( \vec{g} + C^{\eta}_{3z} \vec{K}_M \right)\cdot \vec{x} } \label{app:eqn:twisted_snse_hamiltonian_interm_3}
\end{align}
In performing the first approximation of \cref{app:eqn:twisted_snse_hamiltonian_interm_3} ({\it i.e.}{} from the first two lines to the following two), we have used the fact that $S_{s_1 l_1; s_2 l_2} \left( \delta \vec{R} \left( \vec{x} \right) , \vec{y} \right)$ is only non-zero for $\abs{\vec{y}} \sim \abs{\vec{a}_1}$, which allows us to assume $\abs{\vec{y}} \lesssim \abs{\vec{a}_1}$ under the integral. As we are interested in the long-wavelength  description of the moir\'e system, we have expanded the field operators for small $\abs{\vec{y}}$ and neglected all the terms containing derivatives of the fermionic field (since, by assumption, $\vec{y} \cdot \grad \hat{\psi}^\dagger_{\eta,s,l} \left( \vec{x} \right) \sim \frac{\abs{\vec{a}_1}}{\abs{\vec{a}_{M_1}}} \ll 1$), or equivalently speaking, we have approximated $\hat{\psi}^\dagger_{\eta,s,l}(\vec{x} \pm \vec{y}/2) \approx \hat{\psi}^\dagger_{\eta,s,l}(\vec{x})$. Additionally, from the second to the third line, we have also employed the fact the field operators are ``slow'' operators, and do not ``vary'' over the single layer lattice scale. As such, only those terms whose complex exponential factors do not significantly oscillate at the single-layer graphene scale (corresponding to $\vec{g}_1 = \vec{g}_2 = \vec{g}$ and $\eta_1 = \eta_2 = \eta$) will survive the integration over $\vec{x}$. Finally, in the last row, we have again used the fact that $S_{s_1 l_1; s_2 l_2} \left( \delta \vec{R} \left( \vec{x} \right) , \vec{y} \right)$ is only non-zero for $\abs{\vec{y}} \sim \abs{\vec{a}_{1}} \ll \abs{\vec{a}_{M_1}}$, for which we can approximate
\begin{equation}
    e^{i \mathcal{R}_{\theta,\pm} \left( \vec{g} + C^{\eta}_{3z} \vec{K}_M \right) \cdot \frac{\vec{y}}{2}} \approx e^{i \left( \vec{g} + C^{\eta}_{3z} \vec{K}_M \right) \cdot \frac{\vec{y}}{2}}.
\end{equation}
Plugging the Fourier representation from \cref{app:eqn:fourier_representation_S_both} into \cref{app:eqn:twisted_snse_hamiltonian_interm_3}, we obtain
\begin{align}
    \mathcal{H}^{\text{moir\'e}} =& \frac{1}{N \Omega_0} \sum_{\substack{\vec{g},\vec{g}',\vec{k},\eta \\ l_1,s_1,l_2, s_2}}  S_{s_1 l_1; s_2 l_2} \left( \vec{g}' , \vec{k} \right) \int \dd[2]{y} e^{-i \vec{k} \cdot \vec{y} } e^{i \left( \vec{g} + C^{\eta}_{3z} \vec{K}_M \right)\cdot \vec{y} } \nonumber \\
    &\times \int \dd[2]{x} \hat{\psi}^\dagger_{\eta,s_1,l_1} \left( \vec{x} \right) \hat{\psi}_{\eta,s_2,l_2} \left( \vec{x} \right) e^{-i \mathcal{R}_{\theta,l_1} \left( \vec{g} + C^{\eta}_{3z} \vec{K}_M \right)\cdot \vec{x} } e^{i \mathcal{R}_{\theta,l_2} \left( \vec{g} + C^{\eta}_{3z} \vec{K}_M \right)\cdot \vec{x} } e^{-i \vec{g}' \cdot \delta \vec{R} \left( \vec{x} \right)} \nonumber \\
    =& \sum_{\substack{\vec{g},\eta \\ l_1,s_1,l_2, s_2}}  S_{s_1 l_1; s_2 l_2} \left( \vec{g}, C^{\eta}_{3z} \vec{K}_M \right)  \nonumber \\
    &\times \int \dd[2]{x} \hat{\psi}^\dagger_{\eta,s_1,l_1} \left( \vec{x} \right) \hat{\psi}_{\eta,s_2,l_2} \left( \vec{x} \right) e^{i \left( \mathcal{R}_{\theta,l_2} - \mathcal{R}_{\theta,l_1} \right)  C^{\eta}_{3z} \vec{K}_M \cdot \vec{x} } e^{-i \vec{g} \cdot \delta \vec{R} \left( \vec{x} \right)} \nonumber \\
    =& \sum_{\substack{\vec{g},\eta \\ l_1,s_1,l_2, s_2}} \int \dd[2]{r}  S_{s_1 l_1; s_2 l_2} \left( \vec{g}, C^{\eta}_{3z} \vec{K}_M \right)  \hat{\psi}^\dagger_{\eta,s_1,l_1} \left( \vec{r} \right) \hat{\psi}_{\eta,s_2,l_2} \left( \vec{r} \right) e^{-i \frac{l_2 - l_1}{2} \vec{q}_\eta \cdot \vec{r} } e^{-2 i  \sin \left( \frac{\theta}{2} \right) \left( \vec{g} \cross \hat{\vec{z}} \right) \cdot \vec{r} }. 
    \label{app:eqn:twisted_snse_hamiltonian_final_moire}
\end{align}
In performing the integral over $\vec{y}$, from the first line of \cref{app:eqn:twisted_snse_hamiltonian_final_moire}, we have used the fact that the summation over $\vec{k}$ only runs over a single monolayer BZ, meaning that 
\begin{equation}
	\int \dd[2]{y} e^{-i \vec{k} \cdot \vec{y} } e^{i \left( \vec{g} + C^{\eta}_{3z} \vec{K}_M \right)\cdot \vec{y} } = N \Omega_0 \delta_{\vec{g} + C^{\eta}_{3z} \vec{K}_M,\vec{k}} = N \Omega_0 \delta_{\vec{g}, \vec{0}} \delta_{C^{\eta}_{3z} \vec{K}_M,\vec{k}}. 
\end{equation}
This completes the derivation of the moir\'e potential term. At the same time, the intralayer kinetic term can be readily expressed using the low energy operators from \cref{app:eqn:def_real_space_fermions_to_real} as 
\begin{equation}
    \label{app:eqn:twisted_snse_hamiltonian_final_kinetic}
    \sum_{l} \mathcal{H}^{\text{sl}}_{l} \approx - \sum_{\eta,s,l} \int \dd[2]{r} \hat{\psi}^\dagger_{\eta,s,l}\left( \vec{r} \right) \left(C^\eta_{3z} \mathcal{R}^{-1}_{\theta,l} \nabla \right)^{T} \begin{pmatrix}
        \frac{1}{2 m_x} & 0 \\
        0 & \frac{1}{2 m_y} \\
    \end{pmatrix}
    \left(C^\eta_{3z} \mathcal{R}^{-1}_{\theta,l} \nabla \right) \hat{\psi}_{\eta,s,l} \left( \vec{r} \right).
\end{equation}
Comparing \cref{app:eqn:twisted_snse_hamiltonian_final_kinetic,app:eqn:twisted_snse_hamiltonian_final_moire} with \cref{app:eqn:real_space_ham_moire_from_first_harm} allows us to immediately identify the moir\'e potential in real space
\begin{equation}
    V^{\eta}_{s_1 l_1; s_2 l_2} \left( \vec{r} \right) = \sum_{\vec{g}} S_{s_1 l_1; s_2 l_2} \left( \vec{g}, C^{\eta}_{3z} \vec{K}_M \right) e^{-i \frac{l_2 - l_1}{2} \vec{q}_\eta \cdot \vec{r} } e^{-2 i  \sin \left( \frac{\theta}{2} \right) \left( \vec{g} \cross \hat{\vec{z}} \right) \cdot \vec{r} },
\end{equation}
and, by Fourier inverting the Fourier transformation from \cref{app:eqn:ft_moire_potential_to_real}, we can obtain
\begin{align}
	\left[ T_{ \vec{q}_{\eta+l_1}, \vec{q}_{\eta + l_2} + \vec{G}} \right]_{s_1 l_1; s_2 l_2} &= \frac{1}{\Omega} \int \dd[2]{r}  V^{\eta}_{s_1 l_1; s_2 l_2} \left( \vec{r} \right) e^{-i \left(\vec{q}_{\eta + l_2} + \vec{G} - \vec{q}_{\eta + l_1} \right) \cdot \vec{r}} \nonumber \\
    &= \frac{1}{\Omega} \int \dd[2]{r} \sum_{\vec{g}} S_{s_1 l_1; s_2 l_2} \left( \vec{g}, C^{\eta}_{3z} \vec{K}_M \right) e^{-i \frac{l_2 - l_1}{2} \vec{q}_\eta \cdot \vec{r} } e^{-2 i  \sin \left( \frac{\theta}{2} \right) \left( \vec{g} \cross \hat{\vec{z}} \right) \cdot \vec{r} } e^{-i \left(\vec{q}_{\eta + l_2} + \vec{G} - \vec{q}_{\eta + l_1} \right) \cdot \vec{r}} \nonumber \\
    &= \frac{1}{\Omega} \int \dd[2]{r} \sum_{\vec{G}'} S_{s_1 l_1; s_2 l_2} \left( \frac{\vec{G}' \cross \hat{\vec{z}}}{2 \sin \left( \frac{\theta}{2} \right)}, C^{\eta}_{3z} \vec{K}_M \right) e^{-i \left(\vec{q}_{\eta + l_2} + \vec{G} - \vec{G}' - \vec{q}_{\eta + l_1} + \frac{l_2 - l_1}{2} \vec{q}_\eta \right) \cdot \vec{r}} \nonumber \\
    &= S_{s_1 l_1; s_2 l_2} \left( \frac{\left( \vec{q}_{\eta + l_2} + \vec{G} - \vec{q}_{\eta + l_1} + \frac{l_2 - l_1}{2} \vec{q}_\eta \right) \cross \hat{\vec{z}}}{2 \sin \left( \frac{\theta}{2} \right)}, C^{\eta}_{3z} \vec{K}_M \right) \nonumber \\
    &= S_{s_1 l_1; s_2 l_2} \left( \frac{\left[\vec{G} + \left( l_1 - l_2 \right) \vec{q}_{\eta - 1} \right] \cross \hat{\vec{z}}}{2 \sin \left( \frac{\theta}{2} \right)}, C^{\eta}_{3z} \vec{K}_M \right).\label{app:eqn:equivalence_t_matrix_s_matrix}
\end{align}
We note that for 
\begin{equation}
	\frac{\left[ \vec{G} + \left( l_1 - l_2 \right) \vec{q}_{\eta -1} \right] \cross \hat{\vec{z}}}{2 \sin \left( \frac{\theta}{2} \right)} = \vec{g},
\end{equation}
if follows from our discussion surrounding \cref{app:eqn:local_interlayer_displacement}, that 
\begin{equation}
	\vec{G} \in \mathbb{Z} \vec{b}_{M_1} + \mathbb{Z} \vec{b}_{M_2} \qq{iff} \vec{g} \in \mathbb{Z} \vec{b}_{1} + \mathbb{Z} \vec{b}_{2}.
\end{equation}
As such, we can replace the summation over $\vec{g}$ with a summation over $\vec{G}$ in the third row of \cref{app:eqn:equivalence_t_matrix_s_matrix}.

\subsubsection{Constraining the moir\'e potential}\label{app:sec:SnS_SnSe_twist_general_woGradient:approx:constraining}

Due to the one-to-one equivalence in \cref{app:eqn:equivalence_t_matrix_s_matrix}, constraining the moir\'e potential $T_{ \vec{q}_{\eta+l_1}, \vec{q}_{\eta + l_2} + \vec{G}}$ in the limit of vanishing twist angle is equivalent to constraining $S_{s_1 l_1; s_2 l_2} \left( \vec{g} , C^{\eta}_{3z} \vec{K}_M \right)$. As a result of the hermiticity property of \cref{app:eqn:hermiticity_moire_s}, the latter obeys
\begin{equation}
     S_{s_1 l_1; s_2 l_2} \left( \vec{g} , C^{\eta}_{3z} \vec{K}_M \right) = S^{*}_{s_2 l_2; s_1 l_1} \left( -\vec{g} , C^{\eta}_{3z} \vec{K}_M \right).
\end{equation}
Additionally, the constraints imposed by a crystalline symmetry $g$ in \cref{app:eqn:symmetry_moire_s} imply that 
\begin{equation}
    \label{app:eqn:symmetry_moire_s_final}
    \sum_{s_1, s_2} \left[ D^{\text{sl}}(g) \right]_{s'_1 s_1} S^{(*)}_{s_1 l_1; s_2 l_2} \left( \vec{g}, C^{\eta}_{3z} \vec{K}_M  \right) \left[ D^{\text{sl}}(g) \right]^{*}_{s'_2 s_2} = S_{s'_1 \epsilon_g l_1; s'_2 \epsilon_g l_2} \left( \epsilon_g g \vec{g}, g C^{\eta}_{3z} \vec{K}_M \right).
\end{equation}
The exact symmetries of the heterostructure in the $\theta \neq 0$ case can either be imposed through \cref{app:eqn:symmetry_of_t} or through \cref{app:eqn:symmetry_moire_s_final}: the two approaches are equivalent. However, the additional constraints arising from the symmetries of the zero-twist heterostructure can only be imposed through \cref{app:eqn:symmetry_moire_s_final}, because they do not correspond to crystalline symmetries of the twisted heterostructure. Moreover, we note that \cref{app:eqn:symmetry_moire_s_final} \emph{constrains} the moir\'e potential but does \emph{not} impose the crystalline symmetries of the untwisted heterostructure on the full moir\'e Hamiltonian. For instance, in the zero-twist limit, the AA-stacked moir\'e Hamiltonian does not acquire inversion symmetry, unlike the untwisted AA-stacked heterostructure. Instead, as we show explicitly in \cref{app:sec:add_sym:zero_twist}, it develops an effective mirror symmetry in the plane of the heterostructure.

To impose the zero-twist constraints, in practice, we first parameterize the moir\'e potential using the exact $\theta \neq 0$ symmetries, as explained in \cref{app:sec:SnS_SnSe_twist_general_woGradient:full}. The moir\'e potential can then be further simplified by imposing \cref{app:eqn:symmetry_moire_s_final} for the symmetries of the zero-twist heterostructure that are not \emph{exact} symmetries of the $\theta \neq 0$ heterostructure. To be specific in the AA-stacking case, $S_{s_1 l_1; s_2 l_2} \left( \vec{g} , C^{\eta}_{3z} \vec{K}_M \right)$ is additionally constrained by the $\mathcal{I}$ symmetry, while in the AB-stacking case, it is additionally constrained by the $M_z$ symmetry
\begin{align}
    \sum_{s_1, s_2} \left[ D^{\text{sl}}\left( \mathcal{I} \right) \right]_{s'_1 s_1} S^{\text{AA}}_{s_1 l_1; s_2 l_2} \left( \vec{g}, C^{\eta}_{3z} \vec{K}_M  \right) \left[ D^{\text{sl}}\left( \mathcal{I} \right) \right]^{*}_{s'_2 s_2} &= S^{\text{AA}}_{s'_1  (-l_1); s'_2 (-l_2)} \left( \vec{g}, - C^{\eta}_{3z} \vec{K}_M \right), \label{app:eqn:zero_twist_1st_moire_AA} \\
    \sum_{s_1, s_2} \left[ D^{\text{sl}}\left( M_z \right) \right]_{s'_1 s_1} S^{\text{AB}}_{s_1 l_1; s_2 l_2} \left( \vec{g}, C^{\eta}_{3z} \vec{K}_M  \right) \left[ D^{\text{sl}}\left( M_z \right) \right]^{*}_{s'_2 s_2} &= S^{\text{AB}}_{s'_1  (-l_1); s'_2 (-l_2)} \left( -\vec{g}, C^{\eta}_{3z} \vec{K}_M \right). \label{app:eqn:zero_twist_1st_moire_AB}
\end{align}
Imposing these constraints on the moir\'e potential derived in \cref{app:eqn:full_parameterization_AA_stacking,app:eqn:full_parameterization_AB_stacking} (as will be done explicitly in \cref{app:sec:add_sym:zero_twist}), we find that the latter is characterized by fewer tunneling parameters in this approximation
\begin{align}
    w^{\text{AA}}_{i} &= 0 \qq{for} 3 \leq i \leq 6, \nonumber \\
    w^{\prime \text{AA}}_{i} &= 0 \qq{for} 6 \leq i \leq 10, \nonumber \\
    w^{\text{AB}}_{i} &= 0 \qq{for} 3 \leq i \leq 4, \nonumber \\
    w^{\prime \text{AB}}_{i} &= 0 \qq{for} 8 \leq i \leq 12. 
\end{align}
Specifically, with the zero-twist constraints, there are only seven (nine) real tunneling parameters in the first moir\'e harmonic case for AA- (AB-)stacking. 

\section{Additional symmetries of the first moir\'e harmonic model in different limits}\label{app:sec:add_sym}

In this \siSection{}, we investigate the additional symmetries of the first moir\'e harmonic model that arise under different limits ({\it i.e.}{}, when some of its parameters vanish). For now, we will focus on identifying these limits and their corresponding symmetries, leaving the discussion of how well these limits apply to the \textit{ab initio} Hamiltonian to \cref{app:sec:fitted_models}. We begin by showing that in the zero-twist limit, the moir\'e Hamiltonian has additional symmetries in both the AA- and AB-stacked cases. Specifically, under the zero-twist constraints, the moir\'e Hamiltonian features an additional valley-preserving symmetry, whose action in momentum space is non-symmorphic: in the AA- {(AB-)stacked} case, this symmetry maps $\vec{k}$ to $\vec{k} + \vec{q}_\eta$ ($-\vec{k} + \vec{q}_\eta$) in valley $\eta$. In the AA-stacked case, this symmetry resembles an effective mirror-$z$ operation, allowing the moir\'e Hamiltonian to be block-diagonalized in real space. We also provide a simple lattice model with the same symmetry. In the AB-stacked case, the additional momentum-space non-symmorphic symmetry is equivalent to an effective inversion.

Next, we explore two further limits: the $C_{2z}$ limit, where the moir\'e Hamiltonian is $C_{2z}$ symmetric, making the AA- and AB-stacked parameterizations identical, and the $\mathrm{SU} \left( {2} \right)$ limit, where the model has $\mathrm{SU} \left( {2} \right)$ symmetry within each valley. We also investigate the additional effective symmetries of the two-center first monolayer harmonic model. Finally, we perform a symmetry analysis of the models whose moir\'e Hamiltonian is described by only three parameters and which best reproduce the \textit{ab initio} band structure at small angles of \ch{SnSe2} and \ch{ZrS2}, as shown in \cref{app:sec:fitted_models}.

\subsection{Additional symmetries in the zero-twist limit}\label{app:sec:add_sym:zero_twist}

This first section considers the consequences of \cref{app:eqn:zero_twist_1st_moire_AA,app:eqn:zero_twist_1st_moire_AB} on the moir\'e potential in more detail. Specifically, we will employ \cref{app:eqn:zero_twist_1st_moire_AA,app:eqn:zero_twist_1st_moire_AB}, as well as the mapping from \cref{app:eqn:equivalence_t_matrix_s_matrix} to explicitly constrain the moir\'e potential $T_{\vec{Q}, \vec{Q}'}$ in the zero-twist limit.

\subsubsection{Momentum-space non-symmorphic symmetries in both the AA- and AB-stacking configurations}\label{app:sec:add_sym:zero_twist:nonsym}

Starting with the AA-stacking case, we find that \cref{app:eqn:zero_twist_1st_moire_AA} becomes equivalent to
\begin{align}
	\sum_{s_1, s_2} \left[ D^{\text{sl}}\left( \mathcal{I} \right) \right]_{s'_1 s_1} S^{\text{AA}}_{s_1 l_1; s_2 l_2} \left( \vec{g}, C^{\eta}_{3z} \vec{K}_M  \right) \left[ D^{\text{sl}}\left( \mathcal{I} \right) \right]^{*}_{s'_2 s_2} &= S_{s'_1 (-l_1); s'_2 (-l_2)} \left( \vec{g}, - C^{\eta}_{3z} \vec{K}_M \right) \nonumber \\
	&= S_{s'_1 (-l_1); s'_2 (-l_2)} \left( \vec{g}, C^{\eta}_{3z} \vec{K}_M - 2 C^{\eta}_{3z} \vec{K}_M \right) \nonumber \\
	&= S^{\text{AA}}_{s'_1  (-l_1); s'_2 (-l_2)} \left( \vec{g} + \left( l_2 - l_1 \right) C^{\eta}_{3z} \vec{K}_{M}, C^{\eta}_{3z} \vec{K}_M \right), \label{app:eqn:zero_twist_AA_s_matrix_for_t_interm}
\end{align}
where we have used \cref{app:eqn:periodicity_of_s_matrix_local_stacking} in the last line. Since \cref{app:eqn:zero_twist_AA_s_matrix_for_t_interm} holds for any $\vec{g} \in \mathbb{Z} \vec{b}_1 + \mathbb{Z} \vec{b}_2$, then for given $0 \leq \eta \leq 2$, $l_1, l_2 = \pm$, and $\vec{G} \in \mathcal{Q}$, it should also hold for 
\begin{equation}
	\label{app:eqn:specific_g_vector_choice_zero_twist_symmetries}
	\vec{g} = \frac{\left( \vec{q}_{\eta + l_2} + \vec{G} - \vec{q}_{\eta + l_1} + \frac{l_2 - l_1}{2} \vec{q}_\eta \right) \cross \hat{\vec{z}}}{2 \sin \left( \frac{\theta}{2} \right)}.
\end{equation}
Upon substituting \cref{app:eqn:specific_g_vector_choice_zero_twist_symmetries} into \cref{app:eqn:zero_twist_AA_s_matrix_for_t_interm}, the latter becomes
\begin{align}
	& \sum_{s_1, s_2} \left[ D^{\text{sl}}\left( \mathcal{I} \right) \right]_{s'_1 s_1} S^{\text{AA}}_{s_1 l_1; s_2 l_2} \left( \frac{\left( \vec{q}_{\eta + l_2} + \vec{G} - \vec{q}_{\eta + l_1} + \frac{l_2 - l_1}{2} \vec{q}_\eta \right) \cross \hat{\vec{z}}}{2 \sin \left( \frac{\theta}{2} \right)}, C^{\eta}_{3z} \vec{K}_M  \right) \left[ D^{\text{sl}}\left( \mathcal{I} \right) \right]^{*}_{s'_2 s_2} \nonumber \\
	= & S^{\text{AA}}_{s'_1  (-l_1); s'_2 (-l_2)} \left( \frac{\left( \vec{q}_{\eta + l_2} + \vec{G} - \vec{q}_{\eta + l_1} + \frac{l_2 - l_1}{2} \vec{q}_\eta \right) \cross \hat{\vec{z}}}{2 \sin \left( \frac{\theta}{2} \right)} + \left( l_2 - l_1 \right) C^{\eta}_{3z} \vec{K}_{M}, C^{\eta}_{3z} \vec{K}_M \right) \nonumber \\
	= & S^{\text{AA}}_{s'_1  (-l_1); s'_2 (-l_2)} \left( \frac{\left[ \vec{q}_{\eta + l_2} + \vec{G} - \vec{q}_{\eta + l_1} + \frac{l_2 - l_1}{2} \vec{q}_\eta - \left( l_2 - l_1 \right) \vec{q}_{\eta} \right] \cross \hat{\vec{z}}}{2 \sin \left( \frac{\theta}{2} \right)}, C^{\eta}_{3z} \vec{K}_M \right) \nonumber \\
	= & S^{\text{AA}}_{s'_1  (-l_1); s'_2 (-l_2)} \left( \frac{\left( \vec{q}_{\eta + l_2} + \vec{G} - \vec{q}_{\eta + l_1} - \frac{l_2 - l_1}{2} \vec{q}_\eta \right) \cross \hat{\vec{z}}}{2 \sin \left( \frac{\theta}{2} \right)}, C^{\eta}_{3z} \vec{K}_M \right) \nonumber \\
	= & S^{\text{AA}}_{s'_1  (-l_1); s'_2 (-l_2)} \left( \frac{\left[ \vec{q}_{\eta - l_2} + \left(\vec{G} - 2 \vec{q}_{\eta - l_2} + 2 \vec{q}_{\eta - l_1} \right) - \vec{q}_{\eta - l_1} + \frac{\left( - l_2 \right) - \left(- l_1 \right)}{2} \vec{q}_\eta \right] \cross \hat{\vec{z}}}{2 \sin \left( \frac{\theta}{2} \right)}, C^{\eta}_{3z} \vec{K}_M \right), \label{app:eqn:zero_twist_AA_s_matrix_for_t}
\end{align}
where, in the last equality, we have applied $\vec{q}_{\eta + l} + \vec{q}_{\eta} = - \vec{q}_{\eta - l} = \vec{q}_{\eta-l} - 2 \vec{q}_{\eta-l}$. We now apply the one-to-one mapping between the $S_{s_1 l_1; s_2 l_2} \left( \vec{g} , C^{\eta}_{3z} \vec{K}_M \right)$ and $T_{\vec{q}_{\eta + l_1}, \vec{q}_{\eta + l_2} + \vec{G}}$ matrices from \cref{app:eqn:equivalence_t_matrix_s_matrix}, along with the representation matrix of $\mathcal{I}$ from \cref{app:eqn:rep_matrices_syms_Snse2}. Together with \cref{app:eqn:zero_twist_AA_s_matrix_for_t}, this allows us to show that in the zero-twist limit, we have
\begin{align}
	\left[ T^{\text{AA}}_{ \vec{q}_{\eta+l_1}, \vec{q}_{\eta + l_2} + \vec{G}} \right]_{s_1 l_1; s_2 l_2} = & \left[ T^{\text{AA}}_{ \vec{q}_{\eta - l_1}, \vec{q}_{\eta - l_2} + \vec{G} - 2 \vec{q}_{\eta - l_2} + 2 \vec{q}_{\eta - l_1}} \right]_{s_1 (-l_1); s_2 (-l_2)} \nonumber \\
	= & \left[ T^{\text{AA}}_{ -\vec{q}_{\eta - l_1}, -\vec{q}_{\eta - l_2} + \vec{G}} \right]_{s_1 (-l_1); s_2 (-l_2)} \nonumber \\
	= & \left[ T^{\text{AA}}_{ \vec{q}_{\eta+l_1} + \vec{q}_{\eta}, \vec{q}_{\eta + l_2} + \vec{G} + \vec{q}_{\eta}} \right]_{s_1 (-l_1); s_2 (-l_2)}.\label{app:eqn:zero_twist_AA_t_matrix}
\end{align}
Before investigating the consequences of \cref{app:eqn:zero_twist_AA_t_matrix} on the moir\'e Hamiltonian, we also consider the AB-stacking case, where evaluating \cref{app:eqn:zero_twist_1st_moire_AB} at the monolayer reciprocal vector from \cref{app:eqn:specific_g_vector_choice_zero_twist_symmetries} leads to 
\begin{align}
	&\sum_{s_1, s_2} \left[ D^{\text{sl}}\left( M_z \right) \right]_{s'_1 s_1} S^{\text{AB}}_{s_1 l_1; s_2 l_2} \left(  \frac{\left( \vec{q}_{\eta + l_2} + \vec{G} - \vec{q}_{\eta + l_1} + \frac{l_2 - l_1}{2} \vec{q}_\eta \right) \cross \hat{\vec{z}}}{2 \sin \left( \frac{\theta}{2} \right)} , C^{\eta}_{3z} \vec{K}_M  \right) \left[ D^{\text{sl}}\left( M_z \right) \right]^{*}_{s'_2 s_2} \nonumber \\ 
	= & S^{\text{AB}}_{s'_1  (-l_1); s'_2 (-l_2)} \left(  \frac{\left( - \vec{q}_{\eta + l_2} - \vec{G} + \vec{q}_{\eta + l_1} - \frac{l_2 - l_1}{2} \vec{q}_\eta \right) \cross \hat{\vec{z}}}{2 \sin \left( \frac{\theta}{2} \right)}, C^{\eta}_{3z} \vec{K}_M \right). \label{app:eqn:zero_twist_AB_s_matrix_for_t}
\end{align}
Using again \cref{app:eqn:equivalence_t_matrix_s_matrix}, \cref{app:eqn:zero_twist_AB_s_matrix_for_t} implies that in the zero-twist limit, the moir\'e potential matrix obeys
\begin{equation}
	\label{app:eqn:zero_twist_AB_t_matrix}
	\sum_{s_1, s_2} \left[ D^{\text{sl}}\left( M_z \right) \right]_{s'_1 s_1} \left[ T^{\text{AB}}_{ \vec{q}_{\eta+l_1}, \vec{q}_{\eta + l_2} + \vec{G}} \right]_{s_1 l_1; s_2 l_2} \left[ D^{\text{sl}}\left( M_z \right) \right]^{*}_{s'_2 s_2} = \left[ T^{\text{AB}}_{ -\vec{q}_{\eta+l_1} + \vec{q}_{\eta}, -\vec{q}_{\eta + l_2} - \vec{G} + \vec{q}_{\eta}} \right]_{s'_1 (-l_1); s'_2 (-l_2)},
\end{equation}
with the reader being reminded that the representation matrix of the $M_z$ symmetry was given in \cref{app:eqn:const_mz_sym_AB} and reads as $D^{\text{sl}}\left( M_z \right) = -i s_z$. 

When the zero-twist constraints are imposed, one should also ignore the relative rotation of the monolayer kinetic terms appearing in \cref{app:eqn:single_particle_hamiltonian} and approximate the first-quantized Hamiltonian $h_{\vec{Q}, \vec{Q}'} \left( \vec{k} \right)$ by
\begin{equation}
	\left[ h_{\vec{Q},\vec{Q}'} \left( \vec{k} \right) \right]_{s_1 l_1; s_2 l_2} =  \delta_{\vec{Q}, \vec{Q}'} \delta_{s_1 s_2} \delta_{l_1 l_2} \eval{ \left( \frac{\delta k_{x}^2}{2 m_x} + \frac{\delta k_{y}^2}{2 m_y} \right)}_{\delta \vec{k} = C^{l_1-\zeta_{\vec{Q}}}_{3z} \left( \vec{k} - \vec{Q} \right)} + \left[ T_{\vec{Q},\vec{Q}'} \right]_{s_1 l_1; s_2 l_2},\label{app:eqn:single_particle_hamiltonian_zero_twist}
\end{equation}
where the moir\'e potential $T_{\vec{Q},\vec{Q}'}$ matrix obeys the zero-twist constraints from \cref{app:eqn:zero_twist_AA_t_matrix,app:eqn:zero_twist_AB_t_matrix}, in that AA- and AB-stacking case, respectively. We find that \emph{in the zero-twist limit}, and depending on the stacking configuration, the moir\'e Hamiltonian obeys one of two symmetries whose action is \emph{non-symmorphic} in momentum space~\cite{CHE22,ZHA23b}
\begin{align}
	\left[ h^{\text{AA}}_{\vec{Q},\vec{Q}'} \left( \vec{k} \right) \right]_{s_1 l_1; s_2 l_2} &= \left[ h^{\text{AA}}_{\vec{Q} + \vec{q}_{\eta},\vec{Q}' + \vec{q}_{\eta}} \left( \vec{k} + \vec{q}_{\eta} \right) \right]_{s_1 (-l_1); s_2 (-l_2)}, \label{app:eqn:effective_zero_twist_mz_unitary_ham} \\
	\sum_{s_1, s_2} \left[ D^{\text{sl}}\left( M_z \right) \right]_{s'_1 s_1} \left[ h^{\text{AB}}_{\vec{Q},\vec{Q}'} \left( \vec{k} \right) \right]_{s_1 l_1; s_2 l_2} \left[ D^{\text{sl}}\left( M_z \right) \right]^{*}_{s'_2 s_2} &= \left[ h^{\text{AB}}_{ - \vec{Q} + \vec{q}_{\eta}, - \vec{Q}' + \vec{q}_{\eta}} \left( -\vec{k} + \vec{q}_{\eta} \right) \right]_{s_1 (-l_1); s (-l_2)}, \label{app:eqn:effective_zero_twist_I_unitary_ham}
\end{align}
for $\vec{Q} \in \mathcal{Q}_{\eta + l_1}, \vec{Q}' \in \mathcal{Q}_{\eta + l_2}$, with $0 \leq \eta \leq 2$. Specifically, in the AA-stacking case and in the zero-twist limit, the system is symmetric under an effective mirror-$z$ symmetry, which we denote as $\tilde{M}_z$ and whose action on the moir\'e fermions is given by
\begin{equation}
	\tilde{M}_z \hat{c}^\dagger_{\vec{k},\vec{Q},s,l} \tilde{M}_z^{-1} =  \hat{c}^\dagger_{\vec{k} + \vec{q}_{\eta}, \vec{Q} + \vec{q}_{\eta}, s,-l}, \qq{for} \vec{Q} \in \mathcal{Q}_{\eta + l}, \qq{and}
	\tilde{M}_z \hat{\psi}^\dagger_{\eta,s,l} \left( \vec{r} \right) \tilde{M}_z^{-1} = \hat{\psi}^\dagger_{\eta, s, -l} \left( \vec{r} \right). \label{app:eqn:effective_zero_twist_mz}
\end{equation}
In the zero twist limit $\commutator{\tilde{M}_z}{\mathcal{H}^{\text{AA}}} = 0$. At the same time, in the AB-stacking case, the moir\'e Hamiltonian is symmetric under an effective inversion symmetry, which we denote as $\tilde{\mathcal{I}}$, and whose action on the moir\'e fermions reads as
\begin{equation}
	\tilde{\mathcal{I}} \hat{c}^\dagger_{\vec{k},\vec{Q},s,l} \tilde{\mathcal{I}}^{-1} = (-1)^{s} \hat{c}^\dagger_{-\vec{k} + \vec{q}_{\eta}, - \vec{Q} + \vec{q}_{\eta}, s,-l}, \qq{for} \vec{Q} \in \mathcal{Q}_{\eta + l}, \qq{and}
	\tilde{\mathcal{I}} \hat{\psi}^\dagger_{\eta,s,l} \left( \vec{r} \right) \tilde{\mathcal{I}}^{-1} = (-1)^{s} \hat{\psi}^\dagger_{\eta, s, -l} \left( -\vec{r} \right), \label{app:eqn:effective_zero_twist_I}
\end{equation}
where the factor $(-1)^{s}$ is given by
\begin{equation}
	(-1)^{s} = \begin{cases}
		+1 & \qq{for} s = \uparrow \\
		-1 & \qq{for} s = \downarrow \\
	\end{cases}.
\end{equation}
In the zero-twist limit, $\commutator{\tilde{\mathcal{I}}}{\mathcal{H}^{\text{AB}}} = 0$. 

We note that both $\tilde{M}_z$ and $\tilde{\mathcal{I}}$ are termed effective symmetries because, while their action on the spatial degrees of freedom is conventional, their action on the spins does not match a conventional mirror-$z$ or inversion symmetry. Additionally, both $\tilde{M}_z$ and $\tilde{\mathcal{I}}$ preserve the valley quantum number, meaning that they enhance the single-valley symmetry group. Furthermore, when the zero-twist constraints are imposed, $\tilde{M}_z$ ensures that the spectra of the AA-stacked moir\'e Hamiltonian in valley $\eta$ at $\vec{k}$ and $\vec{k} + \vec{q}_\eta$ are identical. This can be seen from \cref{app:eqn:effective_zero_twist_mz_unitary_ham}, which implies that in the zero-twist limit, the valley-$\eta$ blocks of the Hamiltonian matrix at $\vec{k}$ and $\vec{k} + \vec{q}_{\eta}$ are unitarily related. Similarly, in the AB-stacked case, $\tilde{\mathcal{I}}$ ensures that the spectra of the moir\'e Hamiltonian in valley $\eta$ at $\vec{k}$ and $- \vec{k} + \vec{q}_\eta$ are also identical, in the zero-twist limit.

It is important to note that \cref{app:eqn:effective_zero_twist_mz_unitary_ham,app:eqn:effective_zero_twist_I_unitary_ham} hold if and only if the zero-twist constraints from \cref{app:eqn:zero_twist_1st_moire_AA,app:eqn:zero_twist_1st_moire_AB} are imposed, respectively. Consequently, the AA- and AB-stacked moiré Hamiltonians retain the momentum-space non-symmorphic symmetries $\tilde{M}_z$ and $\tilde{\mathcal{I}}$ even \emph{beyond} the first moiré harmonic level, as long as the zero-twist constraints from \cref{app:eqn:zero_twist_1st_moire_AA,app:eqn:zero_twist_1st_moire_AB} are enforced.

For AA-stacking, the $\tilde{M}_z$ symmetry preserves the spatial position and therefore allows one to define an even and odd basis for the real-space low-energy fermions. We will now discuss this in more detail in the following \cref{app:sec:add_sym:zero_twist:eo_basis}.

\subsubsection{Even-odd basis of the AA-stacked moir\'e Hamiltonian}\label{app:sec:add_sym:zero_twist:eo_basis}

As shown in \cref{app:eqn:effective_zero_twist_mz} the AA-stacked configuration the effective $\tilde{M}_z$ symmetry does not change the spatial position of the real space position of the real-space moir\'e fermion operators. Therefore, we can introduce fermion operators that diagonalize the $\tilde{M}_z$ operator at a given position
\begin{equation}
	\hat{\varphi}^\dagger_{\eta,s,p} \left( \vec{r} \right) \equiv \frac{1}{\sqrt{2}} \left(
		\hat{\psi}^\dagger_{\eta,s,+} \left( \vec{r} \right) + p \hat{\psi}^\dagger_{\eta,s,-} \left( \vec{r} \right)
	 \right),
\end{equation}
where $p= \pm 1$ denotes the parity of the $\hat{\varphi}^\dagger_{\eta,s,p} \left( \vec{r} \right)$ operator under the $\tilde{M}_z$ symmetry. The action of the $\tilde{M}_z$ symmetry and of the other (exact) symmetries of the AA-stacked moir\'e heterostructure on the $\tilde{M}_z$-symmetric states are given by
\begin{equation}
	\label{app:eqn:sym_action_moire_real_nonsym_basis}
	g \hat{\varphi}^\dagger_{\eta_1,s_1,p_1} \left( \vec{r} \right) g^{-1} = \sum_{\eta_2, s_2,p_2} \left[ D_{\varphi}(g) \right]_{\eta_2 s_2 p_2; \eta_1 s_1 p_1} \hat{\varphi}^\dagger_{\eta_2, s_2, p_2} \left( g \vec{r} \right),
\end{equation}
with the corresponding real space representation matrices reading as
\begin{alignat}{4}
	D_{\varphi} \left( \mathcal{T} \right) &&=& i \begin{pmatrix}
		1 & 0 & 0 \\
		0 & 1 & 0 \\
		0 & 0 & 1 \\
	\end{pmatrix} s_y \xi_0, & \quad
	D_{\varphi} \left( C_{3z} \right) &&=& \begin{pmatrix}
		0 & 0 & 1 \\
		1 & 0 & 0 \\
		0 & 1 & 0 \\
	\end{pmatrix} e^{-\frac{\pi i}{3} s_z} \xi_0, \nonumber \\
	D_{\varphi} \left( C_{2x} \right) &&=& -i \begin{pmatrix}
		1 & 0 & 0 \\
		0 & 0 & 1 \\
		0 & 1 & 0 \\
	\end{pmatrix} s_x \xi_z, & \quad
	D_{\varphi} \left( \tilde{M}_z \right) &&=&  \begin{pmatrix}
		1 & 0 & 0 \\
		0 & 1 & 0 \\
		0 & 0 & 1 \\
	\end{pmatrix} s_0 \xi_z, 
\end{alignat}
where $\xi_a$ (for $a = 0, x, y, z$) are the identity and the three Pauli matrices acting on the $\tilde{M}_z$-symmetric fermionic operators. Note that the layer-exchanging $C_{2x}$ symmetry also becomes diagonal in parity in the even-odd basis. Under the moir\'e translation operators, the $\tilde{M}_z$-symmetric operators transform as
\begin{align}
	T_{\vec{R}_M} \hat{\varphi}^\dagger_{\eta,s,p} \left( \vec{r} \right) T^{-1}_{\vec{R}_M} &= \hat{\psi}^\dagger_{\eta,s,+} \left( \vec{r} + \vec{R}_M \right) e^{- i C^{\eta}_{3z} \vec{K}^{+}_{M} \cdot \vec{R}_M} + p \hat{\psi}^\dagger_{\eta,s,-} \left( \vec{r} + \vec{R}_M \right) e^{- i C^{\eta}_{3z} \vec{K}^{-}_{M} \cdot \vec{R}_M} \nonumber \\
	&= \left[ \hat{\psi}^\dagger_{\eta,s,+} \left( \vec{r} + \vec{R}_M \right) + p e^{-i  C^{\eta}_{3z} \left( \vec{K}^{-}_{M} - \vec{K}^{+}_{M} \right) \cdot \vec{R}_M} \hat{\psi}^\dagger_{\eta,s,-} \left( \vec{r} + \vec{R}_M \right) \right] e^{- i C^{\eta}_{3z} \vec{K}^{+}_{M} \cdot \vec{R}_M} \nonumber \\
	&= \left( \hat{\psi}^\dagger_{\eta,s,+} \left( \vec{r} + \vec{R}_M \right) + p e^{-i \vec{q}_{\eta} \cdot \vec{R}_M} \hat{\psi}^\dagger_{\eta,s,-} \left( \vec{r} + \vec{R}_M \right) \right) e^{- i C^{\eta}_{3z} \vec{K}^{+}_{M} \cdot \vec{R}_M} \nonumber \\
	&= \hat{\varphi}^\dagger_{\eta,s,p e^{-i \vec{q}_{\eta} \cdot \vec{R}_M} } \left( \vec{r} + \vec{R}_M \right) e^{- i C^{\eta}_{3z} \vec{K}^{+}_{M} \cdot \vec{R}_M}, \qq{for} \vec{R}_M \in \mathbb{Z} \vec{a}_{M_1} + \mathbb{Z} \vec{a}_{M_2},
\end{align}
where $e^{-i \vec{q}_{\eta} \cdot \vec{R}_M} = \pm 1$. In particular, we find that
\begin{align}
	T_{C^{\eta}_{3z} \vec{a}_{M_1}} \hat{\varphi}^\dagger_{\eta,s,p} \left( \vec{r} \right) T^{-1}_{C^{\eta}_{3z} \vec{a}_{M_1}} &= \hat{\varphi}^\dagger_{\eta,s,-p} \left( \vec{r} + C^{\eta}_{3z} \vec{a}_{M_1} \right) e^{- i C^{\eta}_{3z} \vec{K}^{+}_{M} \cdot \vec{a}_{M_1}}, \label{app:eqn:mirror_fermions_translation_prop_1}\\
	T_{C^{\eta}_{3z} \vec{a}_{M_2}} \hat{\varphi}^\dagger_{\eta,s,p} \left( \vec{r} \right) T^{-1}_{C^{\eta}_{3z} \vec{a}_{M_1}} &= \hat{\varphi}^\dagger_{\eta,s,p} \left( \vec{r} + C^{\eta}_{3z} \vec{a}_{M_2} \right) e^{- i C^{\eta}_{3z} \vec{K}^{+}_{M} \cdot \vec{a}_{M_2}},
	\label{app:eqn:mirror_fermions_translation_prop_2}
\end{align}
which implies that $T_{C^{\eta}_{3z} \vec{a}_{M_1}}$ not only effects a translation of the $\tilde{M}_{z}$-symmetric fermionic operators, but also exchanges their parity (unlike a conventional mirror-$z$ symmetry). 

In the zero-twist limit, the AA-stacked moir\'e Hamiltonian becomes block-diagonal in the $\tilde{M}_{z}$-symmetric basis
\begin{align}
	\mathcal{H}^{\text{AA}} =& - \sum_{\eta,s,p} \int \dd[2]{r} \hat{\varphi}^\dagger_{\eta,s,p}\left( \vec{r} \right) \left(C^{-\eta}_{3z} \nabla \right)^{T} \begin{pmatrix}
		\frac{1}{2 m_x} & 0 \\
		0 & \frac{1}{2 m_y} \\
	\end{pmatrix}
	\left(C^{-\eta}_{3z} \nabla \right) \hat{\varphi}_{\eta,s,p} \left( \vec{r} \right) \nonumber \\
	&+ \sum_{\eta, p, s_1, s_2} \int \dd[2]{r} W^{\eta,p}_{s_1 s_2} \left( \vec{r} \right) \hat{\varphi}^\dagger_{\eta,s_1,p} \left( \vec{r} \right) \hat{\varphi}_{\eta,s_2,p} \left( \vec{r} \right),
	\label{app:eqn:even_odd_basis_AA_stacked_moire}
\end{align}
with the $\tilde{M}_{z}$-symmetric real-space moir\'e potential being given by
\begin{equation}
	W^{\eta,p}_{s_1 s_2} \left( \vec{r} \right) = V^{\eta,\text{AA}}_{s_1 +; s_2 +} \left( \vec{r} \right) + p V^{\eta,\text{AA}}_{s_1 +; s_2 -} \left( \vec{r} \right).
\end{equation}
The $\tilde{M}_{z}$-even and $\tilde{M}_{z}$-odd electrons are therefore decoupled at the single-particle level. Additionally, as a result of the $\tilde{M}_{z}$ symmetry from \cref{app:eqn:zero_twist_AA_t_matrix}, the real-space AA-stacked moir\'e potential obeys
\begin{equation}
	\label{app:eqn:zero_twist_AA_t_matrix_real_space}
	V^{\eta,\text{AA}}_{s_1 l_1; s_2 l_2} \left( \vec{r} \right) = V^{\eta,\text{AA}}_{s_1 (-l_1); s_2 (-l_2)} \left( \vec{r} \right).
\end{equation}

As a result of \cref{app:eqn:periodicity_moire_simple_explicit_intra,app:eqn:periodicity_moire_simple_explicit_inter}, the $\tilde{M}_{z}$-symmetric real-space moir\'e potential satisfies the following periodicity condition
\begin{equation}
	W^{\eta,p}_{s_1 s_2} \left( \vec{r} + 	\vec{a}^{\eta}_{W_1}  \right) = W^{\eta,p}_{s_1 s_2} \left( \vec{r} \right), \quad W^{\eta,p}_{s_1 s_2} \left( \vec{r} + \vec{a}^{\eta}_{W_2}  \right) = W^{\eta,p}_{s_1 s_2} \left( \vec{r}  \right),
\end{equation}
where we have defined the following \emph{rectangular} lattice vectors
\begin{equation}
	\vec{a}^{\eta}_{W_1} = 2 C^{\eta}_{3z} \vec{a}_{M_1} + C^{\eta}_{3z} \vec{a}_{M_2}. 
	\qq{and} 
	\vec{a}^{\eta}_{W_2} = C^{\eta}_{3z} \vec{a}_{M_2}.
\end{equation} 
Moreover, as a result of \cref{app:eqn:periodicity_moire_simple_explicit_intra,app:eqn:periodicity_moire_simple_explicit_inter}, we can show that 
\begin{equation}
	W^{\eta,+1}_{s_1 s_2} \left( \vec{r} + 	C^{\eta}_{3z} \vec{a}_{M_1}  \right) = W^{\eta,-1}_{s_1 s_2} \left( \vec{r}  \right),
\end{equation}
which enables us to rewrite the AA-stacked moir\'e Hamiltonian in the zero-twist limit as
\begin{align}
	\mathcal{H}^{\text{AA}} =& - \sum_{\eta,s,p} \int \dd[2]{r} \hat{\varphi}^\dagger_{\eta,s,p}\left( \vec{r} \right) \left(C^{-\eta}_{3z} \nabla \right)^{T} \begin{pmatrix}
		\frac{1}{2 m_x} & 0 \\
		0 & \frac{1}{2 m_y} \\
	\end{pmatrix}
	\left(C^{-\eta}_{3z} \nabla \right) \hat{\varphi}_{\eta,s,p} \left( \vec{r} \right) \nonumber \\
	&+ \sum_{\eta, s_1, s_2} \int \dd[2]{r} W^{\eta,+1}_{s_1 s_2} \left( \vec{r} \right) \left( \hat{\varphi}^\dagger_{\eta,s_1,+1} \left( \vec{r} \right) \hat{\varphi}_{\eta,s_2,+1} \left( \vec{r} \right) + \hat{\varphi}^\dagger_{\eta,s_1,-1} \left( \vec{r} + C^{\eta}_{3z} \vec{a}_{M_1} \right) \hat{\varphi}_{\eta,s_2,-1} \left( \vec{r} + C^{\eta}_{3z} \vec{a}_{M_1} \right) \right).
	\label{app:eqn:even_odd_basis_AA_stacked_moire_more_obvious}
\end{align}
Thus, within the valley $\eta$, each $\tilde{M}_{z}$-symmetry sector from the Hamiltonian in \cref{app:eqn:even_odd_basis_AA_stacked_moire} is equivalent to a problem of a quadratic fermion $\hat{\varphi}^\dagger_{\eta,s,p}\left( \vec{r} \right)$ moving in a conventional two-dimensional periodic \emph{rectangular} potential $W^{\eta,p}_{s_1 s_2} \left( \vec{r} \right)$, with periodicity defined by the lattice vectors $\vec{a}^{\eta}_{W_i}$ (for $i = 1, 2$). These vectors define a unit cell that is twice the size of the original moir\'e unit cell. Since the translation operator $T_{C^{\eta}_{3z} \vec{a}_{M_1}}$ maps between the two mirror sectors, as shown in \cref{app:eqn:mirror_fermions_translation_prop_1}, it is sufficient to focus on just one $\tilde{M}_{z}$-symmetry sector when analyzing the single-particle properties of the system.

Now, consider the scenario where the system exhibits spin $\mathrm{SU} \left( {2} \right)$ symmetry, such as the limit discussed in \cref{app:sec:add_sym:su2}. Assuming an isolated spin-degenerate band is trivial, it can support exponentially localized Wannier orbitals that respect the system's symmetries. To be specific, we focus on the $\eta = 0$ valley. In this case, the Wannier orbitals of the isolated spinful band, corresponding to different $\tilde{M}_z$ sectors, are decoupled at the single-particle level and form a rectangular lattice spanned by $\vec{a}^{0}_{W_i}$ (for $i = 1, 2$). Since the translation operator $T_{\vec{a}_{M_1}}$ maps between the two mirror sectors in the $\eta = 0$ valley, we find that the Wannier centers of the $\tilde{M}_z$-even and $\tilde{M}_z$-odd fermions are shifted by $ \vec{a}_{M_1}$. When combining the Wannier orbitals of the even and odd electrons, they form the hexagonal moir\'e lattice spanned by $\vec{a}_{M_i}$ (for $i = 1, 2$). However, if we consider only the electrons in the even (or odd) basis and assume that the Wannier orbitals have approximately isotropic spread, then the hopping along the $\hat{\vec{y}}$-direction must be much stronger than the one along the $\hat{\vec{x}}$-direction. This asymmetry arises because the distance between two neighboring Wannier orbitals along the $\hat{\vec{x}}$-direction ($\abs{\vec{a}^{0}_{W_1}}$) is larger than the distance along the $\hat{\vec{y}}$-direction ($\abs{\vec{a}^{0}_{W_2}} = \frac{\abs{\vec{a}^{0}_{W_1}}}{\sqrt{3}}$). Consequently, the system behaves effectively as one-dimensional at the single-particle level, with flat(er) dispersion along the $\hat{\vec{x}}$-direction in valley $\eta = 0$.

\subsubsection{Toy model for non-symmorphic momentum-space symmetry}\label{app:sec:add_sym:zero_twist:toy_model}
\newcommand{\tOneD}{\text{1D}}

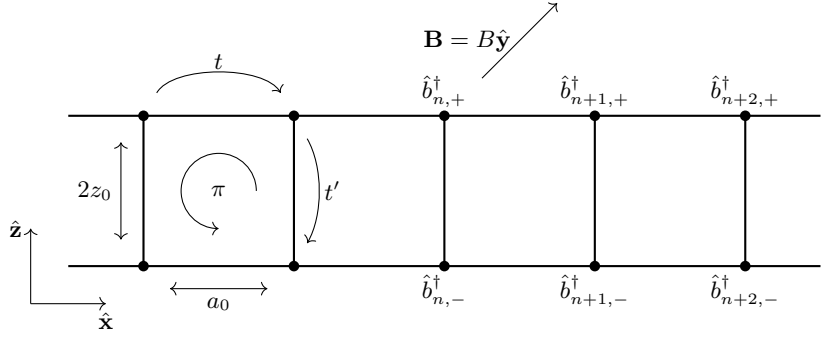
\begin{figure}
	\centering
	\begin{tikzpicture}
		
\coordinate (A0) at (-1, 1);
		\coordinate (A1) at (0, 1);
		\coordinate (A2) at (2, 1);
		\coordinate (A3) at (4, 1);
		\coordinate (A4) at (6, 1);
		\coordinate (A5) at (8, 1);
		\coordinate (A6) at (9, 1);
		
		\coordinate (B0) at (-1, -1);
		\coordinate (B1) at (0, -1);
		\coordinate (B2) at (2, -1);
		\coordinate (B3) at (4, -1);
		\coordinate (B4) at (6, -1);
		\coordinate (B5) at (8, -1);
		\coordinate (B6) at (9, -1);
		\fill (A1) circle (2pt);
		\fill (A2) circle (2pt);
		\fill (A3) circle (2pt);
		\fill (A4) circle (2pt);
		\fill (A5) circle (2pt);
		\fill (B1) circle (2pt);
		\fill (B2) circle (2pt);
		\fill (B3) circle (2pt);
		\fill (B4) circle (2pt);
		\fill (B5) circle (2pt);
		
\draw[thick] (A1) -- (B1);
		\draw[thick] (A2) -- (B2);
		\draw[thick] (A3) -- (B3);
		\draw[thick] (A4) -- (B4);
		\draw[thick] (A5) -- (B5);
		
\draw[thick] (A0) -- (A1) -- (A2) -- (A3) -- (A4) -- (A5) -- (A6);
		\draw[thick] (B0) -- (B1) -- (B2) -- (B3) -- (B4) -- (B5) -- (B6);
		
		\draw[->, shorten >= 10pt, shorten <= 10pt] (A1) to[out=60, in=120] node[above] {$t$} (A2);
		\draw[->, shorten >= 10pt, shorten <= 10pt] (A2) to[out=-60, in=60] node[right] {$t'$} (B2);
		\draw[->] (1.5,0) arc[start angle=0, end angle=270, radius=0.5];
		\node at (1,0) {$\pi$};

\node[above] at (A3) {$\hat{b}^\dagger_{n,+}$};
		\node[above] at (A4) {$\hat{b}^\dagger_{n+1,+}$};
		\node[above] at (A5) {$\hat{b}^\dagger_{n+2,+}$};
		
		\node[below] at (B3) {$\hat{b}^\dagger_{n,-}$};
		\node[below] at (B4) {$\hat{b}^\dagger_{n+1,-}$};
		\node[below] at (B5) {$\hat{b}^\dagger_{n+2,-}$};

\draw[<->, shorten >= 10pt, shorten <= 10pt] (0,-1.3) -- (2,-1.3) node[midway, below] {$a_0$};
		\draw[<->, shorten >= 10pt, shorten <= 10pt] (-0.3,1) -- (-0.3,-1) node[midway, left] {$2z_0$};
		
\draw[->] (-1.5, -1.5) -- (-0.5, -1.5) node[anchor=north] {$\hat{\vec{x}}$};
		\draw[->] (-1.5, -1.5) -- (-1.5, -0.5) node[anchor=east] {$\hat{\vec{z}}$};
		\draw[->] (4.5, 1.5) -- (5.5, 2.5) node[midway, left] {$\vec{B} = B \hat{\vec{y}}$};
	\end{tikzpicture}
	\caption{Lattice model with non-symmorphic momentum-space $\tilde{M}_{z}$ symmetry. We consider a one-dimensional ladder of fermions. The real hopping amplitudes along the $\hat{\vec{x}}$ and $\hat{\vec{z}}$ directions are given by $t$ and $t'$, respectively. Additionally, each plaquette is pierced by a $\pi$ magnetic flux stemming from a magnetic field applied along the $\hat{\vec{y}}$ direction.}
	\label{app:fig:non-symmorphic_mom_space_symmetry}  
\end{figure}

To better understand the non-symmorphic momentum-space $\tilde{M}_z$ symmetry of the AA-stacked moir\'e Hamiltonian in the zero-twist limit, we build a simple one-dimensional tight-binding toy model that also harbors such a symmetry. The connection between this toy model and the AA-stacked moir\'e Hamiltonian will be explained more carefully in \cref{app:sec:add_sym:zero_twist:connection}. We consider two chains of fermions located at
\begin{align}
	\left\lbrace \left( n a_0, z_0 \right) \big\lvert n \in \mathbb{Z} \right\rbrace \cup  \left\lbrace \left( n a_0, -z_0 \right) \big\lvert n \in \mathbb{Z} \right\rbrace 
\end{align}
and let $\hat{b}^\dagger_{n,\pm }$ create an $s$ orbital electron at position $ \left( n a_0, \pm z_0 \right) $. We then introduce a magnetic flux along the $\hat{\vec{y}}$ direction such that there is a $\pi$ magnetic flux for each small plaquette with vertices at 
\begin{align}
	\left( n a_0, z_0 \right),\quad  \left( (n+1) a_0, z_0 \right),\quad \left( (n+1) a_0, -z_0 \right),\quad \left( n a_0,-z_0 \right).
\end{align} 
We set the nearest-neighbor zero-field \emph{real} hopping amplitudes to be $t$ and $t'$ along the $\hat{\vec{x}}$ and $\hat{\vec{z}}$ directions, as depicted in \cref{app:fig:non-symmorphic_mom_space_symmetry}. In the presence of a magnetic field, a phase is introduced to all hoppings such that the product of phase factors around each plaquette equals $e^{i \pi}$ (corresponding to the $\pi$ flux induced by the magnetic field). We work in a gauge where the hopping of the bottom fermions acquires a negative sign, resulting in the following Hamiltonian for the system
\begin{align}
	\label{app:eqn:simple_nonsymorphic_one_d_model}
	\hat{H}_{\tOneD} = \sum_{n,l} t  \left( l \hat{b}^\dagger_{n,l}\hat{b}_{n+1,l} + \text{h.c.} \right) + \sum_{n} t' \left( \hat{b}^\dagger_{n,+}\hat{b}_{n,-} + \text{h.c.} \right).
\end{align}
In \cref{app:eqn:simple_nonsymorphic_one_d_model}, the additional phase factor $l = \pm$ stems from the nonzero magnetic flux piercing each plaquette. 

We now discuss the symmetries of the model. The system has translation symmetry $T_{a_0}$ and $C_{2y}$ rotation symmetry. Additionally, the system also has $\tilde{M}_z$ and $\mathcal{T}$ symmetry. This is because, under either $\tilde{M}_z$ or $\mathcal{T}$, the magnetic field (and hence the magnetic flux) will flip sign. Since a $\pi$ flux and a $-\pi$ flux are equivalent, $\tilde{M}_z$ and $\mathcal{T}$ are symmetries of the system. The actions of these symmetries on the fermions of the model are given by
\begin{alignat}{4}
	T_{a_0} \hat{b}^\dagger_{n,l}T_{a_0}^{-1} &&=&  \hat{b}^\dagger_{n+1,l}, & \quad
	\mathcal{T} \hat{b}^\dagger_{n,l} \mathcal{T}^{-1} &&=& \hat{b}^\dagger_{n,l}, \nonumber \\
	\tilde{M}_{z} \hat{b}^\dagger_{n,l} \tilde{M}_z^{-1} &&=& \hat{b}^\dagger_{n,-l}(-1)^n, & \quad
	C_{2y} \hat{b}^\dagger_{n,l}C_{2y}^{-1} &&=&  i l \hat{b}^\dagger_{-n,-l}. 
\end{alignat}
By considering the successive action of the $T_{a_0}$ and $\tilde{M}_z$ operators
\begin{align}
	&T_{a_0} \tilde{M}_z \hat{b}^\dagger_{n,l} \tilde{M}_z^{-1} T_{a_0}^{-1}=(-1)^{n} \hat{b}^\dagger_{n+1,-l}, \nonumber \\ 
	&\tilde{M}_z T_{a_0} \hat{b}^\dagger_{n,l} T _{a}^{-1}\tilde{M}_z^{-1} =(-1)^{n+1} \hat{b}^\dagger_{n+1,-l}, \nonumber\\ 
	& T_{a_0} \tilde{M}_z \hat{b}^\dagger_{n,l} \tilde{M}_z^{-1} T_{a_0}^{-1} = - \tilde{M}_z T_{a_0} \hat{b}^\dagger_{n,l} T_{a_0}^{-1}\tilde{M}_z^{-1},
\end{align}
we can see that the real-space operators $\hat{b}^\dagger_{n,l}$ form a projective representation of the space group since~\cite{CHE22,ZHA23b} 
\begin{equation}
	\anticommutator{\tilde{M}_z} {T_{a_0}} = 0,
\end{equation}
which arises from the applied magnetic $\pi$ flux.

We now discuss single-particle dispersion of the $\hat{H}_{\tOneD}$ Hamiltonian. Since $\commutator{\hat{H}_{\tOneD}}{T_{a_0}}$, we can introduced the following momentum space operators 
\begin{align}
	\hat{b}^\dagger_{k, l} = \frac{1}{\sqrt{N}}\sum_k \hat{b}^\dagger_{n,l}e^{ikn}.
\end{align}
Next, consider a certain eigenstate of the single particle Hamiltonian $\hat{H}_{1D}$ at momentum $k$
\begin{align}
	\ket{u_k} = \sum_l  u_{k,l} \hat{b}^\dagger_{k,l} \ket{0} 
\end{align}
where $u_{k,l}$ is the wave function of the state and
\begin{align}
	\hat{H}_{\tOneD} \ket{u_k} 
	= E_k \ket{u_k}. 
\end{align} 
From $\tilde{M}_z$ symmetry, we find
\begin{align}
	&\tilde{M}_z  \ket{u_k}  = \tilde{M}_z \sum_l u_{k,l} \sum_n \frac{1}{\sqrt{N}} \hat{b}^\dagger_{n,l}e^{ikn} \ket{0} \nonumber 
	\\ 
	=& \sum_l u_{k,l}\sum_n \frac{1}{\sqrt{N}}\hat{b}^\dagger_{n,-l} e^{ikn + i\pi n }\ket{0} = \sum_l u_{k,l} \hat{b}^\dagger_{k+\pi,-l} \ket{0}, 
\end{align}
meaning that $\tilde{M}_z$ maps a state with momentum $k$ to a state with momentum $k +\pi$. Since $\tilde{M}_z$ commutes with Hamiltonian,
$\tilde{M}_z \ket{u_k} $ is also a eingenstate of the Hamiltonian $\hat{H}_{\tOneD}$ with same eigenvalue $E_k$. Therefore, we conclude the spectra of the Hamiltonian at $k$ and $k+\pi$ are identical. To show this explicitly, we rewrite the Hamiltonian $\hat{H}_{\tOneD}$ in momentum space 
\begin{align}
	\hat{H}_{\tOneD} = \sum_{l,k} 2l t \cos(k) \hat{b}^\dagger_{k,l}\hat{b}_{k,l} + \sum_{l,k} t' \left( \hat{b}^\dagger_{k,+}\hat{b}_{k,-} +\text{h.c.} \right).
\end{align}
The dispersion of the Hamiltonian is given by
\begin{align}
	E_{k,\pm} = \pm \sqrt{ \left(2t\cos(k) \right)^2  + t'^2}  
\end{align}
and is seen to obey $E_{k+\pi,\pm} = E_{k,\pm}$. We note that one can also make the action of $\tilde{M}_{z}$ symmorphic in momentum space by ``folding'' the BZ so that $k$ and $k + \pi$ are identified. This is equivalent to doubling the real-space unit cell. However, adopting this convention obscures the fact that the system retains translation symmetry generated by $T_{a_0}$, rather than by $T_{2 a_0}$.

\subsubsection{Connecting the toy model and AA-stacked moir\'e Hamiltonian}\label{app:sec:add_sym:zero_twist:connection}

In the simple one-dimensional model from \cref{app:eqn:simple_nonsymorphic_one_d_model}, the non-symmorphic action of the $\tilde{M}_z$ symmetry in momentum space ({\it i.e.}{}, its anticommutation with the $T_{a_0}$ translation operator) is seen to arise from an externally applied magnetic field. To make the connection between $\tilde{M}_z$ and a magnetic field more apparent for the moir\'e AA-stacked Hamiltonian with the zero-twist constraints imposed, we can make a gauge transformation and define new operators
\begin{equation}
	\hat{\Psi}^\dagger_{\eta,s,l} \left( \vec{r} \right) = \hat{\psi}^\dagger_{\eta,s,l} \left( \vec{r} \right) e^{- i l \vec{q}_\eta \cdot \vec{r}/2}.
\end{equation}
As we show below, by expressing the AA-stacked moir\'e Hamiltonian with zero-twist constraints in this new basis, the presence of an effective magnetic field becomes explicit
\begin{align}
	\mathcal{H}^{\text{AA}} =& \sum_{\eta,s,l} \int \dd[2]{r} \hat{\Psi}^\dagger_{\eta,s,l}\left( \vec{r} \right) \left[C^{-\eta}_{3z} \left( -i \nabla - e \vec{A}^{\eta}_l \left( \vec{r} \right) \right) \right]^{T} \begin{pmatrix}
		\frac{1}{2 m_x} & 0 \\
		0 & \frac{1}{2 m_y} \\
	\end{pmatrix}
	\left[C^{-\eta}_{3z} \left( -i \nabla - e \vec{A}^{\eta}_l \left( \vec{r} \right) \right) \right] \hat{\Psi}_{\eta,s,l} \left( \vec{r} \right) \nonumber \\
	&+ \sum_{\eta, l, s_1, s_2} \int \dd[2]{r} \left(  \tilde{T}^{\eta,l}_{s_1 s_2} \left( \vec{r} \right) \hat{\Psi}^\dagger_{\eta,s_1,l} \left( \vec{r} \right) \hat{\Psi}_{\eta,s_2,(-l)} \left( \vec{r} \right) + \tilde{V}^{\eta}_{s_1 s_2} \left( \vec{r} \right) \hat{\Psi}^\dagger_{\eta,s_1,l} \left( \vec{r} \right) \hat{\Psi}_{\eta,s_2,l} \left( \vec{r} \right) \right).
	\label{app:eqn:connection_with_tight_binding_model_nonsym}
\end{align}
In \cref{app:eqn:connection_with_tight_binding_model_nonsym}, we have employed \cref{app:eqn:zero_twist_AA_t_matrix_real_space} and have also defined the following intra-, inter- and vector potentials
\begin{align}
	\tilde{T}^{\eta,l}_{s_1 s_2} \left( \vec{r} \right) &= V^{\eta,\text{AA}}_{s_1 +; s_2 -} \left( \vec{r} \right) e^{ i l \vec{q}_{\eta} \cdot \vec{r} }, \\
	\tilde{V}^{\eta}_{s_1 s_2} \left( \vec{r} \right) &= V^{\eta,\text{AA}}_{s_1 +; s_2 +}\left( \vec{r} \right), \\
	\vec{A}^{\eta}_{l} \left( \vec{r} \right) &= - \frac{l}{2 e} \vec{q}_{\eta},
\end{align}
with $e$ being the electronic charge. In this new basis, both the inter- and intra-layer potentials satisfy the moir\'e periodicity 
\begin{align}
	\tilde{T}^{\eta,l}_{s_1 s_2} \left( \vec{r} \right) &= \tilde{T}^{\eta,l}_{s_1 s_2} \left( \vec{r} + \vec{R}_M \right), \\
	\tilde{V}^{\eta}_{s_1 s_2} \left( \vec{r} \right) &= \tilde{V}^{\eta}_{s_1 s_2} \left( \vec{r} + \vec{R}_M \right),
\end{align}
for $\vec{R}_M \in \mathbb{Z} \vec{b}_{M_1} + \mathbb{Z} \vec{b}_{M_2}$. In contrast, the moir\'e potential corresponding to the $\hat{\psi}^\dagger_{\eta,s,l} \left( \vec{r} \right)$ operators introduced in \cref{app:eqn:ft_moire_potential_to_real} are only periodic up to a nontrivial phase, as shown in \cref{app:eqn:periodicity_moire_simple}. In the $\hat{\Psi}^\dagger_{\eta,s,l} \left( \vec{r} \right)$ basis, the Hamiltonian $\mathcal{H}^{\text{AA}}$ also has a coupling to an effective valley-dependent magnetic field. Within valley $\eta = 0$, the magnetic field is along the $\hat{\vec{y}}$ direction and gives rise to a flux
\begin{equation}
	\int_{\vec{r}=\vec{0}}^{\vec{r} = \vec{a}_{M_1}} e \vec{A}^{0}_{-} \left( \vec{r} \right) \cdot  \dd{\vec{r}} + 
	\int_{\vec{r}=\vec{a}_{M_1}}^{\vec{r} = \vec{0}} e \vec{A}^{0}_{+} \left( \vec{r} \right) \cdot  \dd{\vec{r}} = \pi,
\end{equation}
piercing each moir\'e unit cell from the side.

\subsection{The $C_{2z}$ symmetric limit}\label{app:sec:add_sym:c2z_limit}
Another limit that we consider is termed the $C_{2z}$ limit. In this limit, the AA- and AB-stacked moir\'e potentials have identical parameterizations and, within valley $\eta = 0$, the Hamiltonian is symmetric under $C_{2x}$, $C_{2y}$, and $C_{2z}$ symmetries. Specifically, the $C_{2z}$ limit is established whenever
\begin{align}
	w^{\text{AA}}_1 = 0, \quad
	w^{\text{AA}}_5 = 0, \quad
	w^{\text{AA}}_6 = 0, \quad
	w^{\prime \text{AA}}_1 = 0, \quad
	w^{\prime \text{AA}}_2 = 0, \quad
	w^{\prime \text{AA}}_7 = 0, \quad
	w^{\prime \text{AA}}_{8} =w^{\prime \text{AA}}_{9},\\
	w^{\text{AB}}_1 = 0, \quad
	w^{\prime \text{AB}}_1 = 0, \quad
	w^{\prime \text{AB}}_2 = 0, \quad
	w^{\prime \text{AB}}_3 = 0, \quad
	w^{\prime \text{AB}}_5 = 0, \quad
	w^{\prime \text{AB}}_9 = 0, \quad
	w^{\prime \text{AB}}_{10} =0.
\end{align} 
The representation matrices of the $C_{2x}$ and $C_{2y}$ were given in \cref{app:sec:SnS_SnSe_twist:bl_model:symmetries}. The action of the $C_{2z}$ symmetry is given by \cref{app:eqn:sym_action_moire_real,app:eqn:sym_action_moire_mom}, where the momentum- and real-space representation matrices, respectively, read as
\begin{equation}
	D \left( C_{2z} \right) = -i s_z \sigma_0 \qq{and}
	D_{\psi} \left( C_{2z} \right) = -i \begin{pmatrix}
		1 & 0 & 0 \\
		0 & 1 & 0 \\
		0 & 0 & 1 \\
	\end{pmatrix} s_z \sigma_0.
\end{equation}
In the $C_{2z}$-symmetric limit, the parameterizations of the AA- and AB-stacked moir\'e potentials are identical with the following correspondence between their parameters 
\begin{align}
	w^{\text{AA}}_{2} \leftrightarrow w^{\text{AB}}_{2}, \quad
	w^{\text{AA}}_{3} \leftrightarrow w^{\text{AB}}_{4}, \quad
	w^{\text{AA}}_{4} \leftrightarrow w^{\text{AB}}_{3}, \quad
	w^{\prime \text{AA}}_{3} \leftrightarrow w^{\prime \text{AB}}_{4}, \quad
	w^{\prime \text{AA}}_{4} \leftrightarrow w^{\prime \text{AB}}_{6}, \\
	w^{\prime \text{AA}}_{5} \leftrightarrow w^{\prime \text{AB}}_{7}, \quad
	w^{\prime \text{AA}}_{6} \leftrightarrow w^{\prime \text{AB}}_{8}, \quad
	w^{\prime \text{AA}}_{8} = w^{\prime \text{AA}}_{9} \leftrightarrow w^{\prime \text{AB}}_{11}, \quad
	w^{\prime \text{AA}}_{10} \leftrightarrow w^{\prime \text{AB}}_{12}.
\end{align}

\subsection{The $\mathrm{SU} \left( {2} \right)$ symmetric limit}\label{app:sec:add_sym:su2}

In the $\mathrm{SU} \left( {2} \right)$ symmetric limit, the moir\'e potential is spin-diagonal. This effectively enhances the $\mathrm{U} \left( {1} \right) \times \mathrm{U} \left( {1} \right) \times \mathrm{U} \left( {1} \right)$ valley-charge symmetry to a $\mathrm{U} \left( {2} \right) \times \mathrm{U} \left( {2} \right) \times \mathrm{U} \left( {2} \right)$ spin-valley-charge one. The $3 \times 4 = 12$ corresponding generators are given by
\begin{equation}
	\hat{S}_{a}^{\eta} = \sum_{\substack{\vec{k}, l, s, s'}} \sum_{\vec{Q} \in \mathcal{Q}_{\eta + l}} \left[ s_a \right]_{s_1,s_2} \hat{c}^\dagger_{\vec{k},\vec{Q},s_1,l} \hat{c}_{\vec{k},\vec{Q},s_2,l}, \qq{for} a=0,x,y,z \qq{and} \eta = 0, 1, 2.
\end{equation}
Within the first moir\'e harmonic model, this symmetry is established whenever
\begin{align}
	w^{\text{AA}}_3 = 0, \quad
	w^{\text{AA}}_4 = 0, \quad
	w^{\text{AA}}_5 = 0, \quad
	w^{\text{AA}}_6 = 0, \quad
	w^{\prime \text{AA}}_6 = 0, \quad
	w^{\prime \text{AA}}_8 = 0, \quad
	w^{\prime \text{AA}}_9 = 0, \\
	w^{\text{AB}}_1 = 0, \quad
	w^{\text{AB}}_3 = 0, \quad
	w^{\text{AB}}_4 = 0, \quad
	w^{\prime \text{AB}}_5 = 0, \quad
	w^{\prime \text{AB}}_8 = 0, \quad 
	w^{\prime \text{AB}}_{11}=0,
\end{align}
in the AA- and AB-stacked case, respectively. It is worth noting that for the AA-stacked case, the zero-twist constraints also imply the $\mathrm{SU} \left( {2} \right)$ symmetric limit. This contrasts with the AB-stacked arrangement (or even the AA-stacked case beyond the first moir\'e harmonic limit), where the zero-twist constraints \emph{do not} necessarily imply that the system has $\mathrm{SU} \left( {2} \right)$ symmetry.

\subsection{Additional symmetries in the two-center first monolayer harmonic model}\label{app:sec:add_sym:first_monolayer_harmonic}

In the two-center first monolayer harmonic model, the moir\'e potential from \cref{app:eqn:ft_moire_potential_to_real} features continuous translation symmetry along one direction. Restricting ourselves to valley $\eta=0$ without loss of generality, the real-space moir\'e potential contains only interlayer terms and is given by
\begin{align}
	V^{0, \text{AA}}_{s_1 (-l); s_2 l} \left( \vec{r} \right) &= \sum_{n = \pm 1} \left[ \left( - i n l w^{\text{AA}}_1 + w^{\text{AA}}_2 \right) s_0 + n \left(w^{\text{AA}}_4 + w^{\text{AA}}_6 e^{ i n l \frac{\pi}{6}}  \right) s_y + \left( - i l w^{\text{AA}}_3 + n w^{\text{AA}}_5 \right) s_z \right]_{s_1 s_2} e^{- i n l \vec{q}_{0} \cdot \vec{r}},  \\
	V^{0, \text{AB}}_{s_1 (-l); s_2 l} \left( \vec{r} \right) &= \sum_{n = \pm 1} \left[w^{\text{AB}}_2 s_0 - i l w^{\text{AB}}_1 s_x + n w^{\text{AB}}_3 s_y - i l w^{\text{AB}}_4 s_z \right]_{s_1 s_2} e^{- i n l \vec{q}_{0} \cdot \vec{r}}, 
\end{align}
while 
\begin{equation}
	V^{\eta}_{s_1 l; s_2 l} \left( \vec{r} \right) = 0.
\end{equation}
As a result, one finds that
\begin{equation}
	V^{0}_{s_1 l_1; s_2 l_2} \left( x, y \right) = V^{0}_{s_1 l_1; s_2 l_2} \left( x, y + \delta y \right),
\end{equation}
making the moir\'e Hamiltonian from \cref{app:eqn:real_space_ham_moire_from_first_harm} and valley $\eta = 0$ invariant with respect to translations along the $\hat{\vec{y}}$ direction. In the two-center approximation, the system is quasi-one-dimensional within each valley, having continuous translation symmetry along the $C^{\eta}_{3z} \hat{\vec{y}}$ direction in valley $\eta$.

\subsection{Additional symmetries of the simplified three-parameters models}\label{app:sec:add_sym:three_parameter}

In \cref{app:sec:fitted_models}, we show that the \textit{ab initio} band structures of AA- and AB-stacked \ch{SnSe2} and \ch{ZrS2} can be described by simplified models featuring only three parameters: $w^{\text{AA}}_{1}$, $w^{\text{AA}}_{2}$, and $w^{\prime \text{AA}}_{3}$ (in the AA-stacked case), or $w^{\text{AB}}_{2}$, $w^{\prime \text{AB}}_{3}$, and $w^{\prime \text{AB}}_{4}$ (in the AB-stacked case). In this section, we examine the symmetries of these two simplified models and briefly discuss their implications on the band structure of the corresponding moir\'e Hamiltonian.

\subsubsection{The three-parameter AA-stacked model}\label{app:sec:add_sym:three_parameter:AA}

We begin by analyzing the three-parameter moir\'e Hamiltonian for the AA-stacked configuration. On the $\vec{Q}$-lattice depicted in \cref{app:fig:M_valley_MBZ:c} corresponding to valley $\eta = 0$, $w^{\text{AA}}_{1}$ and $w^{\text{AA}}_{2}$ describe the real and imaginary parts of the NN ``hopping'' along the $\hat{\vec{x}}$ direction, while $w^{\prime \text{AA}}_{3}$ represents the NN ``hopping'' along the $\hat{\vec{y}}$ direction. This three-parameter model involves only interlayer moir\'e terms. It is already in both the zero-twist limit and the $\mathrm{SU} \left( {2} \right)$ limit, and therefore inherits all symmetries previously discussed in \cref{app:sec:add_sym:zero_twist,app:sec:add_sym:su2}.

To uncover any additional symmetries, we cast the moir\'e potential in real space to find
\begin{align}
	V^{0, \text{AA}}_{s_1 (-l); s_2 l} \left( \vec{r} \right) &= \left[ 2 \sqrt{\left( w^{\text{AA}}_1 \right)^2 + \left( w^{\text{AA}}_2 \right)^2 } \cos \left( \abs{\vec{q}_0} x + \alpha \right) + 2 w^{\prime \text{AA}}_3 \cos \left( \sqrt{3} \abs{\vec{q}_0} y \right) \right] \left[s_0\right]_{s_1 s_2}, \nonumber \\
	V^{0, \text{AA}}_{s_1 l; s_2 l} \left( \vec{r} \right) &= 0, \label{app:eqn:AA_simple_3_param_moire_potential}
\end{align}
where we have introduced a phase factor $0 \leq \alpha < 2 \pi$, such that
\begin{equation}
	e^{i \alpha} = \frac{ i w^{\text{AA}}_1 + w^{\text{AA}}_2 }{\sqrt{\left( w^{\text{AA}}_1 \right)^2 + \left( w^{\text{AA}}_2 \right)^2 }}.
\end{equation}

From \cref{app:eqn:AA_simple_3_param_moire_potential} it becomes evident that the system possesses an additional two-fold rotation symmetry, which we denote by $\tilde{C}_{2z}$, and whose center of rotation does \emph{not} coincide with the unit cell origin (located at the intersection of the $C_{2x}$ and $C_{3z}$ rotation axes). The action of this symmetry on the moir\'e fermions in real space is given by
\begin{equation}
	\label{app:eqn:sym_action_effective_C_2z}
	\tilde{C}_{2z} \hat{\psi}^\dagger_{\eta_1, s_1, l_1} \left( \vec{r} \right) \tilde{C}_{2z}^{-1} = \sum_{\eta_2, s_2, p_2} \left[ D_{\psi}\left( \tilde{C}_{2z} \right) \right]_{\eta_2 s_2 l_2; \eta_1 s_1 l_1} \hat{\psi}^\dagger_{\eta_2, s_2, l_2} \left( - \vec{r} + \frac{2 \alpha \vec{q}_\eta}{\abs{\vec{q}_\eta}^2} \right),
\end{equation}
with
\begin{equation}
	D_{\psi}\left( \tilde{C}_{2z} \right) = i \begin{pmatrix}
		1 & 0 & 0 \\
		0 & 1 & 0 \\
		0 & 0 & 1 \\
	\end{pmatrix} s_z \xi_0.
\end{equation}
Note that, within each valley $\eta$, the centers of rotation for the $\tilde{C}_{2z}$ symmetry operation are located at the generic points $\frac{\alpha \vec{q}_\eta}{\abs{\vec{q}_\eta}^2}$, $\frac{\alpha \vec{q}_\eta}{\abs{\vec{q}_\eta}^2} + \frac{\vec{a}_{M_1}}{2}$, $\frac{\alpha \vec{q}_\eta}{\abs{\vec{q}_\eta}^2} + \frac{\vec{a}_{M_2}}{2}$, and $\frac{\alpha \vec{q}_\eta}{\abs{\vec{q}_\eta}^2} + \frac{\vec{a}_{M_1} + \vec{a}_{M_2}}{2}$, none of which correspond to the unit cell origin (or any $C_{3z}$-symmetric Wyckoff position). Thus, while the symmetry group within a \emph{single} valley is enhanced to a larger crystallographic space group, the \emph{overall} symmetry group of the Hamiltonian (which incorporates all three valleys) does not \emph{necessarily} correspond to a crystallographic space group. This occurs because the $\tilde{C}_{2z}$ rotation centers vary between the three valleys. However, when one of the four $\tilde{C}_{2z}$ rotation centers coincides with one of the three $C_{3z}$ symmetric Wyckoff positions of the AA-stacked moir\'e Hamiltonian, the symmetry group of the \emph{entire} moir\'e Hamiltonian is similarly enhanced by the $\tilde{C}_{2z}$ symmetry to form a \textit{bona fide} crystallographic space group. The enhanced symmetry group of the full moir\'e Hamiltonian becomes a true crystallographic group whenever
\begin{equation}
	\label{app:eqn:correspondence_rotation_centers_general}
	\frac{\alpha \vec{q}_{0}}{\abs{\vec{q}_{0}}^2} = \frac{k}{3} \left(
	2 \vec{a}_{M_1}+ \vec{a}_{M_2} \right) + \frac{m}{2} \vec{a}_{M_1} + \frac{n}{2} \left( \vec{a}_{M_1} + \vec{a}_{M_2} \right), \qq{for} k=0,1,2 \qq{and} m,n \in \mathbb{Z}.
\end{equation}
\Cref{app:eqn:correspondence_rotation_centers_general} holds only when $m = n$, leading to the condition
\begin{equation}
	\label{app:eqn:correspondence_rotation_centers}
	\frac{\alpha \vec{q}_{0}}{\abs{\vec{q}_{0}}^2} = \frac{2 k + 3 m}{6} \left(
	2 \vec{a}_{M_1}+ \vec{a}_{M_2} \right), \qq{for} k=0,1,2 \qq{and} m \in \mathbb{Z}.
\end{equation} 
which is satisfied for $\alpha = \frac{2 \pi m}{6}$, where $m = 0, 1, \dots, 5$. Finally, we note that the $\tilde{C}{2z}$ symmetry, when combined with $\tilde{M}{z}$, generates an effective momentum-space non-symmorphic inversion symmetry, denoted as $\tilde{C}{2z} \tilde{M}{z}$. This symmetry behaves similarly to the $\tilde{\mathcal{I}}$ symmetry described in \cref{app:eqn:effective_zero_twist_I}, with the key distinction being that the inversion center of $\tilde{C}{2z} \tilde{M}{z}$ is not at the unit cell origin, but at the point $\frac{\alpha \vec{q}_\eta}{\abs{\vec{q}_\eta}^2}$.

We now turn to the topological characteristics of the three-parameter AA-stacked Hamiltonian. Due to the $\mathrm{SU} \left( {2} \right)$ symmetry present in each valley, the system can be effectively treated as spinless. Moreover, the presence of the $\tilde{C}_{2z} \mathcal{T}$ symmetry allows us to fix the gauge such that the wave function for each isolated spin-degenerate band becomes real. This gauge choice leads to a vanishing Berry curvature for each isolated spin-degenerate band. Consequently, we conclude that every isolated band in the system is topologically trivial within the three-parameter model. Because the \textit{ab initio} band structure and the one approximated by the three-parameter model are adiabatically connected, we can also conclude that each \emph{approximately} degenerate set of isolated bands of the AA-stacked moir\'e Hamiltonian are topologically trivial. As discussed at the end of \cref{app:sec:add_sym:zero_twist:eo_basis}, these bands will be quasi-one-dimensional as a result of the $\tilde{M}_z$ symmetry.

\subsubsection{The three-parameter AB-stacked model}\label{app:sec:add_sym:three_parameter:AB}

The simplified model for the AB-stacked moir\'e Hamiltonian is defined by three parameters: $w^{\text{AB}}_2$, $w^{\prime \text{AB}}_3$, and $w^{\prime \text{AB}}_4$, corresponding to the NN ``hopping'' in the $\hat{\vec{x}}$-direction, NN ``hopping'' in the $\hat{\vec{y}}$-direction, and next NN ``hopping'' in the $\hat{\vec{x}}$-direction, respectively, on the $\vec{Q}$-lattice shown in \cref{app:fig:M_valley_MBZ:c} for valley $\eta = 0$. In the AB-stacked configuration, the three-parameter model includes both interlayer and intralayer terms
\begin{align}
	V^{0, \text{AB}}_{s_1 (-l); s_2 l} \left( \vec{r} \right) &= \left[ 2 w^{\text{AB}}_{2} \cos \left( \abs{\vec{q}_0} x \right) + 2 w^{\prime \text{AB}}_{4} \cos \left( \sqrt{3} \abs{\vec{q}_0} y \right) \right] \left[s_0\right]_{s_1 s_2}, \nonumber \\
	V^{0, \text{AB}}_{s_1 l; s_2 l} \left( \vec{r} \right) &= \left[ 2 l w^{\prime \text{AB}}_{3} \sin \left( 2 \abs{\vec{q}_0} x \right) \right] \left[s_0\right]_{s_1 s_2}, \label{app:eqn:AB_simple_3_param_moire_potential}
\end{align}
This model is already in both the zero-twist and $\mathrm{SU} \left( {2} \right)$ limits, thereby inheriting all the symmetries discussed earlier in \cref{app:sec:add_sym:zero_twist,app:sec:add_sym:su2}.

To discuss the topological properties of this model, we can simplify it further by setting $w^{\prime \text{AB}}_{3}$ to zero. This is justified because: (1) in the \textit{ab initio} calculations, $w^{\prime \text{AB}}_{3}$ is smaller than $w^{\text{AB}}_2$ and $w^{\prime \text{AB}}_4$ for both \ch{SnSe2} and \ch{ZrS2}, and (2) reducing $w^{\prime \text{AB}}_{3}$ to zero still keeps the first two sets of bands individually gapped. The resulting two-parameter AB-stacked Hamiltonian reaches the $C_{2z}$ limit as outlined in \cref{app:sec:add_sym:c2z_limit}. Due to the $\mathrm{SU} \left( {2} \right)$ symmetry in each valley, the system can again be treated as effectively spinless. As with the three-parameter AA-stacked Hamiltonian in \cref{app:sec:add_sym:three_parameter:AA}, this two-parameter AB-stacked model also possesses $C_{2z} \mathcal{T}$ symmetry, which allows for a gauge choice where the wave functions of isolated spin-degenerate bands become real. This leads to vanishing Berry curvature for each isolated spin-degenerate band, confirming that all isolated bands in the system are topologically trivial. The two-parameter model is adiabatically connected to the \textit{ab initio} band structure, meaning that each approximately degenerate set of isolated bands in the AB-stacked moir\'e Hamiltonian is topologically trivial.

Furthermore, because the simplified two-parameter AB-stacked model retains both $C_{2z}$ symmetry (from the $C_{2z}$ limit) and $\tilde{\mathcal{I}}$ symmetry (from the zero-twist limit), it also exhibits $\tilde{M}_z = \tilde{\mathcal{I}} C_{2z}$ symmetry. Consequently, each gapped spin-degenerate topologically trivial band is quasi-one-dimensional, as discussed in \cref{app:sec:add_sym:zero_twist:eo_basis}, similar to the AA-stacked case.

\subsection{Summary of symmetries in different limits}\label{app:sec:add_sym:summary}

\begingroup
\renewcommand{\arraystretch}{1.5} \begin{table}[!t]
	\centering
	\resizebox{\textwidth}{!}{\begin{tabular}{|l|c|l|c|c|c|l|l|c|}
\hline
			\multirow{2}{*}{Model or limit} & \multirow{2}{*}{Stacking} & \multirow{2}{*}{Zero-parameters} & \multicolumn{3}{c|}{Symmetries} & \multicolumn{2}{c|}{Crystallographic space group} & 	\multirow{2}{*}{\makecell[tl]{Valley\\ $\mathrm{U} \left( {N} \right)$}} \\
			\cline{4-8}
			& & & $\tilde{M}_z$ & $\tilde{\mathcal{I}}$ & $C_{2z}/\tilde{C}_{2z}$ & Valley $\eta$ & All valleys & \\
\hline\hline
			\multirow{2}{*}{\makecell[tl]{First moir\'e\\harmonic}} & AA & None & \ding{55} & \ding{55} & \ding{55} & \multirow{2}{*}{$P21'$ (SSG 3.2)} & $P3121'$ (SSG 149.22) & \multirow{2}{*}{$\mathrm{U} \left( {1} \right)$} \\
			\cline{2-6}\cline{8-8}
			& AB & None & \ding{55} & \ding{55} & \ding{55} & & $P3211'$ (SSG 150.26) & \\
\hline
			\multirow{2}{*}{Zero-twist} & AA & $w^{\text{AA}}_{i}$ for $3 \leq i \leq 6$, $w^{\prime \text{AA}}_{i}$ for $6 \leq i \leq 10$ & \ding{51} & \ding{55} & \ding{55} & $Pmm21'$ (SSG 25.58) & $P\bar{6}m21'$ (SSG 187.210) & $\mathrm{U} \left( {2} \right)$ \\
			\cline{2-9}
			& AB & $w^{\text{AB}}_{i}$ for $3 \leq i \leq 4$, $w^{\prime \text{AB}}_{i}$ for $8 \leq i \leq 12$ & \ding{55} & \ding{51} & \ding{55} &  $P2/m1'$ (SSG 10.43) & $P\bar{3}m11'$ (SSG 164.86) & $\mathrm{U} \left( {1} \right)$ \\
			\hline
\multirow{2}{*}{$C_{2z}$ limit } & AA & $w^{\text{AA}}_1, w^{\text{AA}}_5, w^{\text{AA}}_6, w^{\prime \text{AA}}_1, w^{\prime \text{AA}}_2, w^{\prime \text{AA}}_7, w^{\prime \text{AA}}_{8} - w^{\prime \text{AA}}_{9}$ & \multirow{2}{*}{\ding{55}} & \multirow{2}{*}{\ding{55}} & \multirow{2}{*}{$C_{2z}$} & \multirow{2}{*}{$P2221'$ (SSG 16.2)} & \multirow{2}{*}{$P6221'$ (SSG 177.150)} & \multirow{2}{*}{$\mathrm{U} \left( {1} \right)$} \\
			\cline{2-3}
			& AB & $w^{\text{AB}}_1, w^{\prime \text{AB}}_1, w^{\prime \text{AB}}_2, w^{\prime \text{AB}}_3, w^{\prime \text{AB}}_5, w^{\prime \text{AB}}_9, w^{\prime \text{AB}}_{10}$ & & & & & & \\
			\hline
\multirow{2}{*}{$\mathrm{SU} \left( {2} \right)$ limit} & AA & $w^{\text{AA}}_3, w^{\text{AA}}_4, w^{\text{AA}}_5, w^{\text{AA}}_6, w^{\prime \text{AA}}_6, w^{\prime \text{AA}}_8, w^{\prime \text{AA}}_9$ & \multirow{2}{*}{\ding{55}} & \multirow{2}{*}{\ding{55}} & \multirow{2}{*}{\ding{55}} & \multirow{2}{*}{$P21'$ (SSG 3.2)} & $P3121'$ (SSG 149.22) & \multirow{2}{*}{$\mathrm{U} \left( {2} \right)$}  \\
			\cline{2-3} \cline{8-8}
			& AB & $w^{\text{AB}}_1, w^{\text{AB}}_3, w^{\text{AB}}_4, w^{\prime \text{AB}}_5,
			w^{\prime \text{AB}}_8, w^{\prime \text{AB}}_{11}$ & & & & & $P3211'$ (SSG 150.26) &  \\
			\hline
\multirow{2}{*}{\makecell[tl]{Two-center first\\monolayer harmonic}} & AA & $w^{\prime\text{AA}}_{i}$ for $1 \leq i \leq 10$ & \ding{55} & \ding{55} & \ding{55} & \multirow{2}{*}{\makecell[tl]{$P21'$ (SSG 3.2)\\Continuous translation\\along $C^{\eta}_{3z}\hat{\vec{y}}$}} & $P3121'$ (SSG 149.22) & \multirow{2}{*}{$\mathrm{U} \left( {1} \right)$} \\
			\cline{2-6}\cline{8-8}
			& AB & $w^{\prime\text{AA}}_{i}$ for $1 \leq i \leq 12$ & \ding{55} & \ding{55} & \ding{55} & & $P3211'$ (SSG 150.26) &  \\
			\hline
\multirow{2}{*}{Three-parameter} & AA & All \emph{except} $w^{\text{AA}}_{1}$, $w^{\text{AA}}_{2}$, and $w^{\prime \text{AA}}_{3}$ & \ding{51} & \ding{55} & $\tilde{C}_{2z}$ & $Cmmm1'$ (SSG 65.482) & $P6/mmm1'$ (SSG 191.234) & \multirow{2}{*}{$\mathrm{U} \left( {2} \right)$} \\
			\cline{2-8}
			& AB & All \emph{except} $w^{\text{AB}}_{2}$, $w^{\text{AB}}_{3}$, and $w^{\prime \text{AB}}_{4}$ & \ding{55} & \ding{51} & \ding{55} & $P2/m1'$ (SSG 10.43) & $P\bar{3}m11'$ (SSG 164.86) & \\
\hline
			Two-parameter & AB & All \emph{except} $w^{\text{AB}}_{2}$ and $w^{\prime \text{AB}}_{4}$ & \ding{51} & \ding{51} & $C_{2z}$ & $Cmmm1'$ (SSG 65.482) & $P6/mmm1'$ (SSG 191.234) & $\mathrm{U} \left( {2} \right)$ \\
			\hline
		\end{tabular}}
	\caption{Summary of the additional symmetries present in the first moir\'e harmonic model under various limits for both AA- and AB-stacking configurations. Each limit specifies the parameters set to zero, highlights any effective symmetries with non-symmorphic momentum-space action ({\it i.e.}{}, $\tilde{M}_z$ and $\tilde{\mathcal{I}}$), and indicates the presence of two-fold rotation symmetries perpendicular to the heterostructure's plane, such as $C_{2z}$ or the effective $\tilde{C}_{2z}$. The corresponding crystallographic space groups are listed assuming the non-symmorphic momentum-space symmetries act conventionally ({\it i.e.}{}, with symmorphic momentum-space actions). For the three-parameter AA-stacked model, the space group for the full model is provided when $\alpha = \frac{\pi m}{3}$, for $m = 0, 1, \dots, 5$. Lastly, the continuous symmetry group in each valley is also indicated.}
	\label{app:tab:summary_symmetries}
\end{table}
\endgroup

This section summarizes the symmetries of the first moir\'e harmonic model across different limits for AA- and AB-stacking configurations, as outlined in \cref{app:tab:summary_symmetries}. Notably, the low-energy \textit{ab initio} spectra of both \ch{SnSe2} and \ch{ZrS2} are well-approximated by simplified three-parameter and two-parameter models for the AA- and AB-stacked cases, respectively. These models exhibit enhanced symmetries, including effective symmetries acting non-symmorphically in momentum space. The table compares the various limits, listing the parameters set to zero, the presence of effective symmetries, and the relevant crystallographic space groups. These space groups are determined by assuming that the symmetries with non-symmorphic momentum-space actions act conventionally ({\it i.e.}{}, symmorphically).

\section{Direct general derivation of the moir\'e potential including gradient terms}\label{app:sec:SnS_SnSe_twist_general}

This \siSection{} further generalizes the results of the previous \cref{app:sec:SnS_SnSe_twist_general_woGradient} by considering gradient terms in the moir\'e potential. We begin by writing down the most general form of the moir\'e Hamiltonian, which includes the aforementioned gradient terms. We then show how the form of the latter can be efficiently parameterized and constrained by the exact symmetries of the $\theta \neq 0$ heterostructure, as well as by the \emph{approximate} symmetries of the $\theta = 0$ heterostructure. 

\subsection{General form of the moir\'e Hamiltonian with gradient terms}\label{app:sec:SnS_SnSe_twist_general:hamiltonian}

Following Ref.~\cite{ZHA24}, we write down the most general expression for the single-particle moir\'e Hamiltonian
\begin{align}
	\mathcal{H} =& \frac{1}{2} \sum_{\eta} \sum_{n_x, n_y} \sum_{\substack{s_1, l_1 \\ s_2, l_2}} \int \dd[2]{r} t^{\eta,n_x,n_y}_{s_1 l_1; s_2 l_2} \left( \vec{r} \right) \left[ i^{n_x + n_y} \left( \partial^{n_x}_{x} \partial^{n_y}_{y} \hat{\psi}^\dagger_{\eta,s_1,l_1} \left( \vec{r} \right) \right) \hat{\psi}_{\eta,s_2,l_2} \left( \vec{r} \right) \right. \nonumber \\
	&\left. + (-i)^{n_x + n_y} \hat{\psi}^\dagger_{\eta,s_1,l_1} \left( \vec{r} \right) \partial^{n_x}_{x} \partial^{n_y}_{y} \hat{\psi}_{\eta,s_2,l_2} \left( \vec{r} \right) 	\right], \label{app:eqn:general_moire_potential}
\end{align}
which contains terms proportional to arbitrary gradients of the low-energy field operators introduced in \cref{app:eqn:def_real_space_fermions_to_real}.
In \cref{app:eqn:general_moire_potential}, $t^{\eta,n_x,n_y}_{s_1 l_1; s_2 l_2} \left( \vec{r} \right)$ denotes the generalized moir\'e potential and obeys the following Hermiticity property
\begin{equation}
	t^{\eta,n_x,n_y}_{s_1 l_1; s_2 l_2} \left( \vec{r} \right) = 
	\left( t^{\eta,n_x,n_y}_{s_2 l_2; s_1 l_1} \left( \vec{r} \right) \right)^*.
\end{equation}
As a result, the integrand of \cref{app:eqn:general_moire_potential} (and not just the Hamiltonian as whole) is a Hermitian operator. We have also implicitly incorporated the intralayer ``kinetic'' part into the function $t^{\eta,n_x,n_y}_{s_1 l_1; s_2 l_2} \left( \vec{r} \right)$, as the distinction between the conventional ``kinetic'' and moir\'e potential fades away in the general gradient expansion case. Note that in \cref{app:eqn:general_moire_potential}, without loss of generality, only terms where all derivatives act on \emph{either} the creation \emph{or} the annihilation operators need to be considered, as we will prove below.  

To see that \cref{app:eqn:general_moire_potential} indeed represents \emph{the most general form} of the moir\'e Hamiltonian, we first notice that any term containing spatial derivatives acting on the fermionic fields has the form 
\begin{align}
	&\sum_{\eta} \sum_{\substack{s_1, l_1 \\ s_2, l_2}} \int \dd[2]{r}  t^{\prime \eta}_{s_1 l_1; s_2 l_2} \left( \vec{r} \right) i^{n_x + n_y} (-i)^{n'_x + n'_y} \left( \partial^{n_x}_{x} \partial^{n_y}_{y} \hat{\psi}^\dagger_{\eta,s_1,l_1} \left( \vec{r} \right) \right) \partial^{n'_x}_{x} \partial^{n'_y}_{y} \hat{\psi}_{\eta,s_2,l_2} \left( \vec{r} \right) \nonumber \\
	= &\sum_{\eta} \sum_{\substack{s_1, l_1 \\ s_2, l_2}} \int \dd[2]{r}  i^{n_x + n'_x + n_y + n'_y} t^{\prime \eta}_{s_1 l_1; s_2 l_2} \left( \vec{r} \right) \left( \partial^{n_x + n'_x}_{x} \partial^{n_y + n'_y}_{y} \hat{\psi}^\dagger_{\eta,s_1,l_1} \left( \vec{r} \right) \right) \hat{\psi}_{\eta,s_2,l_2} \left( \vec{r} \right) \nonumber \\
	& + \left[ \text{terms proportional to gradients of } t^{\prime \eta}_{s_1 l_1; s_2 l_2} \left( \vec{r} \right) \right]\dots \nonumber \\
	= &\frac{1}{2} \sum_{\eta} \sum_{\substack{s_1, l_1 \\ s_2, l_2}} \int \dd[2]{r} t^{\prime \eta}_{s_1 l_1; s_2 l_2} \left( \vec{r} \right) \left[ i^{n_x + n'_x + n_y + n'_y} \left( \partial^{n_x + n'_x}_{x} \partial^{n_y + n'_y}_{y} \hat{\psi}^\dagger_{\eta,s_1,l_1} \left( \vec{r} \right) \right) \hat{\psi}_{\eta,s_2,l_2} \left( \vec{r} \right) \right. \nonumber \\
	& \left. + (-i)^{n_x + n'_x + n_y + n'_y} \hat{\psi}^\dagger_{\eta,s_1,l_1} \left( \vec{r} \right) \partial^{n_x + n'_x}_{x} \partial^{n_y + n'_y}_{y} \hat{\psi}_{\eta,s_2,l_2} \left( \vec{r} \right) \right] \nonumber \\
	& + \left[ \text{terms proportional to gradients of } t^{\prime \eta}_{s_1 l_1; s_2 l_2} \left( \vec{r} \right) \right]\dots , \label{app:eqn:gradient_induction_proof}
\end{align}
where $n_x, n_y, n'_x, n'_y \in \mathbb{N}$, $t^{\prime \eta}_{s_1 l_1; s_2 l_2} \left( \vec{r} \right)$ is some moir\'e potential and the dots denote terms in which the field operators have less than $n_x + n'_x$ $x$-derivatives and less than $n_y + n'_y$ $y$-derivatives. We can therefore conclude that any term appearing in the Hamiltonian featuring spatial derivatives of the fermionic field can be written as the sum between a term where all derivatives act on the creation field and another term where the derivatives act on the annihilation field, in addition to other terms containing strictly fewer derivatives acting on the fermionic fields. Using this observation, we can straightforwardly prove (through induction), that any term having the form of the first line of \cref{app:eqn:gradient_induction_proof} can be written as a sum of pairs of the form of the last equality of \cref{app:eqn:gradient_induction_proof}: each pair is a sum between one term in which all spatial derivatives act on the creation field and a term where all spatial derivatives act on the annihilation field; each pair contains less or the same number of spatial derivatives as the original term:
\begin{align}
	&\sum_{\eta} \sum_{\substack{s_1, l_1 \\ s_2, l_2}} \int \dd[2]{r}  t^{\prime \eta}_{s_1 l_1; s_2 l_2} \left( \vec{r} \right) i^{n_x + n_y} (-i)^{n'_x + n'_y} \left( \partial^{n_x}_{x} \partial^{n_y}_{y} \hat{\psi}^\dagger_{\eta,s_1,l_1} \left( \vec{r} \right) \right) \partial^{n'_x}_{x} \partial^{n'_y}_{y} \hat{\psi}_{\eta,s_2,l_2} \left( \vec{r} \right) \nonumber \\
	= &\frac{1}{2} \sum_{\substack{ m_x \leq n_x + n'_x \\ m_y \leq n_y + n'_y}} \sum_{\eta} \sum_{\substack{s_1, l_1 \\ s_2, l_2}} \int \dd[2]{r} t^{\eta, m_x, m_y}_{s_1 l_1; s_2 l_2} \left( \vec{r} \right) \left[ i^{m_x + m_y} \left( \partial^{m_x}_{x} \partial^{m_y}_{y} \hat{\psi}^\dagger_{\eta,s_1,l_1} \left( \vec{r} \right) \right) \hat{\psi}_{\eta,s_2,l_2} \left( \vec{r} \right) \right. \nonumber \\
	& \left. + (-i)^{m_x + m_y} \hat{\psi}^\dagger_{\eta,s_1,l_1} \left( \vec{r} \right) \partial^{m_x}_{x} \partial^{m_y}_{y} \hat{\psi}_{\eta,s_2,l_2} \left( \vec{r} \right) \right], \label{app:eqn:gradient_main_proof}
\end{align}
To prove \cref{app:eqn:gradient_main_proof}, we note that \cref{app:eqn:gradient_main_proof} is trivially valid for $n_x + n'_x = n_y + n'_y = 0$. One then assumes it holds for some $n_x + n'_x = M_x$ and $n_y + n'_y = M_y$ (where $M_x, M_y \in \mathbb{N}$). Because the dots in \cref{app:eqn:gradient_induction_proof} denote terms with fewer spatial derivatives acting on the fermionic fields, it follows that if \cref{app:eqn:gradient_main_proof} holds for $n_x + n'_x = M_x$ and $n_y + n'_y = M_y$, it should also hold for $n_x + n'_x = M_x + 1$ and $n_y + n'_y = M_y$ or $n_x + n'_x = M_x$ and $n_y + n'_y = M_y + 1$, thus completing the inductive proof. In what follows, we will employ \cref{app:eqn:general_moire_potential} as the most general form of the moir\'e potential. 

As in the case where the moir\'e potential contains no gradient terms, the Hamiltonian from \cref{app:eqn:general_moire_potential} is diagonal in the valley subspace and translation-invariant at the moir\'e lattice scale. As a result of the latter property, the generalized moir\'e potential obeys 
\begin{equation}
	\label{app:eqn:periodicity_moire_generalized}
	t^{\eta,n_x,n_y}_{s_1 l_1; s_2 l_2} \left( \vec{r} + \vec{R}_{M} \right) = t^{\eta,n_x,n_y}_{s_1 l_1; s_2 l_2} \left( \vec{r} \right) e^{i \left( \vec{q}_{\eta + l_2} - \vec{q}_{\eta + l_1} \right) \cdot \vec{R}_{M}}, \qq{for any} \vec{R}_{M} \in \mathbb{Z} \vec{a}_{M_1} + \mathbb{Z} \vec{a}_{M_2}, 
\end{equation}
which is identical in form to \cref{app:eqn:periodicity_moire_simple}. This follows from \cref{app:eqn:def_real_space_fermions_to_real} which requires that
\begin{align}
	&\left( \partial^{n_x}_{x} \partial^{n_y}_{y} \hat{\psi}^\dagger_{\eta,s_1,l_1} \left( \vec{r} + \vec{R}_{M} \right) \right) \left( \partial^{m_x}_{x} \partial^{m_y}_{y} \hat{\psi}_{\eta,s_2,l_2} \left( \vec{r} + \vec{R}_{M} \right) \right) \nonumber \\
	= &\left( \partial^{n_x}_{x} \partial^{n_y}_{y} \hat{\psi}^\dagger_{\eta,s_1,l_1} \left( \vec{r} \right) \right) \left( \partial^{m_x}_{x} \partial^{m_y}_{y} \hat{\psi}_{\eta,s_2,l_2} \left( \vec{r} \right) \right) e^{-i \left( \vec{q}_{\eta + l_2} - \vec{q}_{\eta + l_1} \right) \cdot \vec{R}_{M}}, \label{app:eqn:periodicity_of_cre_des_pair}
\end{align}
for any $n_x,n_y,m_x,m_y \in \mathbb{N}$ and $\vec{R}_{M} \in \mathbb{Z} \vec{a}_{M_1} + \mathbb{Z} \vec{a}_{M_2}$. In turn, \cref{app:eqn:periodicity_moire_generalized} enables us to express the generalized moir\'e potential as the following Fourier series
\begin{align}
	t^{\eta,n_x,n_y}_{s_1 l_1; s_2 l_2} \left( \vec{r} \right) &= \sum_{\vec{G} \in \mathcal{Q}} \left[ T^{n_x,n_y}_{\vec{q}_{\eta + l_1}, \vec{q}_{\eta + l_2} + \vec{G}} \right]_{s_1 l_1; s_2 l_2} e^{i \left(\vec{q}_{\eta + l_2} + \vec{G} - \vec{q}_{\eta + l_1} \right) \cdot \vec{r}}, \label{app:eqn:ft_general_moire_potential_to_real} \\
	\left[ T^{n_x,n_y}_{\vec{q}_{\eta + l_1}, \vec{q}_{\eta + l_2} + \vec{G}} \right]_{s_1 l_1; s_2 l_2} &= \frac{1}{\Omega} \int \dd[2]{r} t^{\eta,n_x,n_y}_{s_1 l_1; s_2 l_2} \left( \vec{r} \right)  e^{-i \left(\vec{q}_{\eta + l_2} + \vec{G} - \vec{q}_{\eta + l_1} \right) \cdot \vec{r}}, \label{app:eqn:ft_general_moire_potential_to_mom}
\end{align}
where we have used a notation analogous to \cref{app:eqn:ft_moire_potential_to_real}, and cast the Hamiltonian from \cref{app:eqn:general_moire_potential} in momentum space 
\begin{equation}
	\label{app:eqn:general_moire_potential_momentum}
	\mathcal{H} = \sum_{\substack{ \vec{k}, n_x, n_y \\ \vec{Q}, \vec{Q}' \in \mathcal{Q}_{\text{tot}}}} \sum_{\substack{s_1, l_1 \\ s_2, l_2}} \frac{\left( k_x - Q_x \right)^{n_x} \left( k_y - Q_y \right)^{n_y} + \left( k_x - Q'_x \right)^{n_x} \left( k_y - Q'_y \right)^{n_y}}{2} \left[ T^{n_x,n_y}_{\vec{Q}, \vec{Q}'} \right]_{s_1 l_1; s_2 l_2} \hat{c}^\dagger_{\vec{k},\vec{Q},s_1,l_1} \hat{c}_{\vec{k},\vec{Q}',s_2,l_2}.
\end{equation}
In \cref{app:eqn:general_moire_potential_momentum} the momentum-space moir\'e potential matrix coupling \emph{any} plane wave-vectors $\vec{Q}, \vec{Q}' \in \mathcal{Q}_{\text{tot}}$ is obtained from \cref{app:eqn:ft_general_moire_potential_to_mom} through
\begin{equation}
	\label{app:eqn:general_moire_periodicity_momentum}
	\left[ T^{n_x,n_y}_{\vec{q}_{\eta + l_1} + \vec{G}_1, \vec{q}_{\eta + l_2} + \vec{G}_2} \right]_{s_1 l_1; s_2 l_2} = \left[ T^{n_x,n_y}_{\vec{q}_{\eta + l_1}, \vec{q}_{\eta + l_2} + \vec{G}_2 - \vec{G}_1} \right]_{s_1 l_1; s_2 l_2}, \qq{for any} \vec{G}_1, \vec{G}_2 \in \mathcal{Q}_0.
\end{equation}

\subsection{Restricting the generalized moir\'e potential using the exact $\theta \neq 0$ symmetries of the heterostructure}\label{app:sec:SnS_SnSe_twist_general::full}

The form of the generalized moir\'e potential matrix $T^{n_x,n_y}_{\vec{Q}, \vec{Q}'}$ can be constrained using the exact symmetries of the twisted heterostructure in the $\theta \neq 0$ case. To start with, the moir\'e Hamiltonian from \cref{app:eqn:general_moire_potential_momentum} is Hermitian, which immediately implies that
\begin{equation}
	\label{app:eqn:hermiticity_of_t_general}
	\left[ T^{n_x,n_y}_{\vec{Q}, \vec{Q}'} \right]_{s_1 l_1; s_2 l_2} = \left[ T^{n_x,n_y}_{\vec{Q}', \vec{Q}} \right]^{*}_{s_2 l_2; s_1 l_1}.
\end{equation}
At the same time, moir\'e periodicity requires that
\begin{equation}
	T^{n_x,n_y}_{\vec{Q}, \vec{Q}'} = T^{n_x,n_y}_{\vec{Q} + \vec{G}, \vec{Q}' + \vec{G}}, \qq{for} \vec{G} \in \mathcal{Q},
\end{equation}
which is equivalent to \cref{app:eqn:general_moire_periodicity_momentum}. Additionally, the $\mathrm{U} \left( {1} \right) \times \mathrm{U} \left( {1} \right) \times \mathrm{U} \left( {1} \right)$ valley-charge symmetry will require that the moir\'e potential always satisfies the analogue of \cref{app:eqn:valley_u1_symmetry_of_moire}
\begin{equation}
	\label{app:eqn:valley_u1_symmetry_of_moire_generalized}
	\left[ T^{n_x,n_y}_{\vec{Q}', \vec{Q}} \right]_{s_1 l_1; s_2 l_2} = 0, \qq{for} \vec{Q} \in \mathcal{Q}_{\eta_1 + l_1} \qq{and} \vec{Q}' \in \mathcal{Q}_{\eta_2 + l_2}, \qq{with} \eta_1 \neq \eta_2. 
\end{equation}

To obtain the constraints arising from crystalline symmetries, we note that for any symmetry $g$ and momentum $\vec{k}$
\begin{equation}
	\label{app:eqn:totally_symmetry_trafo_k}
	\left( \left[ g\vec{k} \right]_{x} \right)^{n_x} \left( \left[ g\vec{k} \right]_{y} \right)^{n_y} = \sum_{\substack{n'_x, n'_y \\ n'_x + n'_y = n_x + n_y}} U_{n'_x,n'_y;n_x,n_y} ( g ) k^{n'_x}_{x} k^{n'_y}_{y},
\end{equation}
with $\left[ g\vec{k} \right]_{x}$ denoting the $x$ component of $g \vec{k}$. In other words, under a symmetry $g$ a monomial in momentum components transforms into a polynomial function of momentum components, in which each term has the same total power in momentum as the original monomial. The corresponding matrix $U_{n_x,n_y;n'_x,n'_y} (g)$ is related to the totally symmetric representation of $g$ and can be obtained by direct construction from the cartesian representation of $g$. As such, for every crystalline symmetry $g$, the generalized moir\'e potential will obey 
\begin{equation}
	\label{app:eqn:symmetry_of_t_general}
	\sum_{\substack{n_x, n_y \\ n_x + n_y = n'_x + n'_y}} U_{n'_x,n'_y;n_x,n_y} (g) T^{n_x,n_y}_{g \vec{Q}, g\vec{Q}'} = D (g) \left(T^{n'_x, n'_y}_{\vec{Q}, \vec{Q}'} \right)^{(*)}  D^{\dagger}(g),
\end{equation}
where ${}^{(*)}$ indicates that a complex conjugation should be taken in the cases when $g$ is antiunitary. Taken together, \cref{app:eqn:hermiticity_of_t_general,app:eqn:symmetry_of_t_general} constitute a series of homogeneous linear equations for the components of $T^{n_x, n_y}_{\vec{Q}, \vec{Q}'}$. By solving the resulting system of equations, one can obtain the independent components of generalized moir\'e potential, as well as its parameterization.

\subsection{Restricting the generalized moir\'e potential using the approximate $\theta = 0$ symmetries of the heterostructure}\label{app:sec:SnS_SnSe_twist_general:approx} 

In the limit of vanishing twist angle $\theta \to 0$, the generalized moir\'e potential can be additionally constrained by the symmetries of the untwisted heterostructure. We follow the same strategy that we used to constrain the moir\'e potential without gradient terms in \cref{app:sec:SnS_SnSe_twist_general_woGradient:approx}. We start from the family of Hamiltonian $\mathcal{H} \left( \Delta \vec{R} \right)$ introduced in \cref{app:eqn:untwisted_snse_hamiltonian}, but absorb the single-particle Hamiltonian of the single-layer material into the second term by a redefinition of the $ S_{s_1 l_1; s_2 l_2} \left( \Delta \vec{R}, \vec{r}^{\Delta \vec{R}}_{l_2,\vec{R}_2} - \vec{r}^{\Delta \vec{R}}_{l_1,\vec{R}_1} \right)$ matrix, such that the Hamiltonian $\mathcal{H} \left( \Delta \vec{R} \right)$ reads as
\begin{equation}
	\label{app:eqn:untwisted_snse_hamiltonian_general}
	\mathcal{H} \left( \Delta \vec{R} \right) = \sum_{\substack{\vec{R}_1,l_1,s_1 \\ \vec{R}_2,l_2, s_2}} S_{s_1 l_1; s_2 l_2} \left( \Delta \vec{R}, \vec{r}^{\Delta \vec{R}}_{l_2,\vec{R}_2} - \vec{r}^{\Delta \vec{R}}_{l_1,\vec{R}_1} \right) \hat{a}^\dagger_{\vec{R}_1,s_1,l_1} \hat{a}_{\vec{R}_2,s_2,l_2}.
\end{equation}
Such a notation treats the single-particle Hamiltonian of the monolayer material, as well as the interlayer contribution on equal footing. 

\subsubsection{Moir\'e Hamiltonian in the local-stacking approximation with gradient terms}\label{app:sec:SnS_SnSe_twist_general:approx:twisted}

Making the same local-stacking approximation as in \cref{app:sec:SnS_SnSe_twist_general_woGradient:approx:twisted}, we find that the moir\'e Hamiltonian of the twisted configuration is given by 
\begin{align}
	\mathcal{H} =& \frac{1}{\Omega_0} \sum_{\substack{\vec{g}_1,l_1,s_1,\eta_1 \\ \vec{g}_2,l_2, s_2, \eta_2}} \int \dd[2]{x} \dd[2]{y} S_{s_1 l_1; s_2 l_2} \left( \delta \vec{R} \left( \vec{x} \right) , \vec{y} \right) e^{-i \mathcal{R}_{\theta,l_1} \left( \vec{g}_1 + C^{\eta_1}_{3z} \vec{K}_M \right)\cdot \left( \vec{x} - \frac{\vec{y}}{2} \right)} e^{i \mathcal{R}_{\theta,l_2} \left( \vec{g}_2 + C^{\eta_2}_{3z} \vec{K}_M \right)\cdot \left( \vec{x} + \frac{\vec{y}}{2} \right)} \nonumber \\
	&\times \hat{\psi}^\dagger_{\eta_1,s_1,l_1} \left( \vec{x} - \frac{\vec{y}}{2} \right) \hat{\psi}_{\eta_2,s_2,l_2} \left( \vec{x} + \frac{\vec{y}}{2} \right) \nonumber \\
\approx & \frac{1}{\Omega_0} \sum_{\substack{\vec{g},\eta \\ l_1,s_1,l_2, s_2}} \int \dd[2]{x} \dd[2]{y} S_{s_1 l_1; s_2 l_2} \left( \delta \vec{R} \left( \vec{x} \right) , \vec{y} \right) e^{-i \mathcal{R}_{\theta,l_1} \left( \vec{g} + C^{\eta}_{3z} \vec{K}_M \right)\cdot \left( \vec{x} - \frac{\vec{y}}{2} \right)} e^{i \mathcal{R}_{\theta,l_2} \left( \vec{g} + C^{\eta}_{3z} \vec{K}_M \right)\cdot \left( \vec{x} + \frac{\vec{y}}{2} \right)} \nonumber \\
	&\times \hat{\psi}^\dagger_{\eta,s_1,l_1} \left( \vec{x} - \frac{\vec{y}}{2} \right) \hat{\psi}_{\eta,s_2,l_2} \left( \vec{x} + \frac{\vec{y}}{2} \right) \nonumber \\
= & \frac{1}{2\Omega_0} \sum_{\substack{\vec{g},\eta \\ l_1,s_1,l_2, s_2}} \int \dd[2]{x} \dd[2]{y} e^{i \left( \mathcal{R}_{\theta,l_2} - \mathcal{R}_{\theta,l_1} \right) \left( \vec{g} + C^{\eta}_{3z} \vec{K}_M \right) \cdot \vec{x}} \nonumber \\
	&\times \left( S_{s_1 l_1; s_2 l_2} \left( \delta \vec{R} \left( \vec{x} - \frac{\vec{y}}{2} \right) , \vec{y} \right) \hat{\psi}^\dagger_{\eta,s_1,l_1} \left( \vec{x} - \vec{y} \right) \hat{\psi}_{\eta,s_2,l_2} \left( \vec{x} \right) e^{i \mathcal{R}_{\theta,l_1} \left( \vec{g} + C^{\eta}_{3z} \vec{K}_M \right)\cdot \vec{y}} \right. \nonumber \\  
	&+ \left. S_{s_1 l_1; s_2 l_2} \left( \delta \vec{R} \left( \vec{x} + \frac{\vec{y}}{2} \right) , \vec{y} \right) \hat{\psi}^\dagger_{\eta,s_1,l_1} \left( \vec{x} \right) \hat{\psi}_{\eta,s_2,l_2} \left( \vec{x} + \vec{y} \right) e^{i \mathcal{R}_{\theta,l_2} \left( \vec{g} + C^{\eta}_{3z} \vec{K}_M \right)\cdot \vec{y}} \right). \label{app:eqn:twisted_snse_hamiltonian_interm_general_1}
\end{align}
The first two lines of \cref{app:eqn:twisted_snse_hamiltonian_interm_general_1} is identical to the first two lines of \cref{app:eqn:twisted_snse_hamiltonian_interm_3}, with the exception that the single-layer contribution has been incorporated into the $S_{s_1 l_1; s_2 l_2} \left( \delta \vec{R} \left( \vec{x} \right), \vec{y} \right)$ matrix. From the first two lines to the following two, we used the fact that $S_{s_1 l_1; s_2 l_2} \left( \delta \vec{R} \left( \vec{x} \right), \vec{y} \right)$ is only nonvanishing for $\abs{\vec{y}} \lesssim \abs{\vec{a}_1}$, and varies slowly in $\vec{x}$ on the monolayer lattice scale, just as the low-energy fermionic fields. As a result, only the term for which the complex exponential factor does not oscillate strongly on the single-layer lattice scale will survive. Compared to \cref{app:sec:SnS_SnSe_twist_general_woGradient:approx:twisted}, we will not make any further assumption about the length-scale over which the low-energy fermion operators vary. 

We will now focus on simplifying each term of \cref{app:eqn:twisted_snse_hamiltonian_interm_general_1}. Because both terms are similar in form, we can simplify only the second one, with the first one following in the same way. To do so, we beging by noting that the fermion field operators can be expanded as 
\begin{align}
	\hat{\psi}_{\eta,s,l} \left( \vec{x} + \vec{y} \right) &= \sum_{n = 0}^{\infty} \frac{1}{n!} \left( \vec{y} \cdot \grad \right)^{n} \hat{\psi}_{\eta,s,l} \left( \vec{x} \right) \nonumber \\
	&= \sum_{n_x, n_y = 0}^{\infty} \frac{{n_x + n_y \choose n_x} y_x^{n_x} y_{y}^{n_y}}{\left(n_x + n_y \right)!} \partial_x^{n_x} \partial_y^{n_y}  \hat{\psi}_{\eta,s,l} \left( \vec{x} \right) \nonumber \\
	&= \sum_{n_x, n_y = 0}^{\infty} \frac{y_x^{n_x} y_{y}^{n_y}}{n_x! n_y!} \partial_x^{n_x} \partial_y^{n_y}  \hat{\psi}_{\eta,s,l} \left( \vec{x} \right).
\end{align}
In turn, this allows us to rewrite the second term of \cref{app:eqn:twisted_snse_hamiltonian_interm_general_1} as 
\begin{align}
	& \frac{1}{2\Omega_0} \sum_{\substack{\vec{g},\eta \\ l_1,s_1,l_2, s_2}} \int \dd[2]{x} \dd[2]{y} e^{i \left( \mathcal{R}_{\theta,l_2} - \mathcal{R}_{\theta,l_1} \right) \left( \vec{g} + C^{\eta}_{3z} \vec{K}_M \right) \cdot \vec{x}} e^{i \mathcal{R}_{\theta,l_2} \left( \vec{g} + C^{\eta}_{3z} \vec{K}_M \right)\cdot \vec{y}} \nonumber \\
	&\times S_{s_1 l_1; s_2 l_2} \left( \delta \vec{R} \left( \vec{x} + \frac{\vec{y}}{2} \right) , \vec{y} \right) \hat{\psi}^\dagger_{\eta,s_1,l_1} \left( \vec{x} \right) \hat{\psi}_{\eta,s_2,l_2} \left( \vec{x} + \vec{y} \right) = \nonumber \\
	= & \frac{1}{2\Omega_0} \sum_{\substack{\vec{g},\eta \\ l_1,s_1,l_2, s_2}} \int \dd[2]{x} \dd[2]{y} e^{i \left( \mathcal{R}_{\theta,l_2} - \mathcal{R}_{\theta,l_1} \right) \left( \vec{g} + C^{\eta}_{3z} \vec{K}_M \right) \cdot \vec{x}} e^{i \mathcal{R}_{\theta,l_2} \left( \vec{g} + C^{\eta}_{3z} \vec{K}_M \right)\cdot \vec{y}} \nonumber \\
	&\times \frac{y_x^{n_x} y_{y}^{n_y}}{n_x! n_y!} S_{s_1 l_1; s_2 l_2} \left( \delta \vec{R} \left( \vec{x} + \frac{\vec{y}}{2} \right) , \vec{y} \right) \hat{\psi}^\dagger_{\eta,s_1,l_1} \left( \vec{x} \right) \partial_x^{n_x} \partial_y^{n_y} \hat{\psi}_{\eta,s_2,l_2} \left( \vec{x} \right) \nonumber \\
	= & \frac{1}{2N\Omega_0} \sum_{\substack{\vec{g},\vec{g}',\vec{k},\eta \\ l_1,s_1,l_2, s_2}} \int \dd[2]{x} \dd[2]{y} e^{i \left( \mathcal{R}_{\theta,l_2} - \mathcal{R}_{\theta,l_1} \right) \left( \vec{g} + C^{\eta}_{3z} \vec{K}_M \right) \cdot \vec{x}} e^{i \mathcal{R}_{\theta,l_2} \left( \vec{g} + C^{\eta}_{3z} \vec{K}_M \right)\cdot \vec{y}} e^{- i \vec{g}' \cdot \delta \vec{R} \left( \vec{x} + \frac{\vec{y}}{2} \right)} e^{- i \vec{k} \cdot \vec{y}} \nonumber \\
	&\times \frac{y_x^{n_x} y_{y}^{n_y}}{n_x! n_y!} S_{s_1 l_1; s_2 l_2} \left( \vec{g}',\vec{k} \right) \hat{\psi}^\dagger_{\eta,s_1,l_1} \left( \vec{x} \right) \partial_x^{n_x} \partial_y^{n_y} \hat{\psi}_{\eta,s_2,l_2} \left( \vec{x} \right) \nonumber \\
	= & \frac{1}{2N\Omega_0} \sum_{\substack{\vec{g},\vec{g}',\vec{k},\eta \\ l_1,s_1,l_2, s_2}} \int \dd[2]{x} \dd[2]{y} e^{i \left( \mathcal{R}_{\theta,l_2} - \mathcal{R}_{\theta,l_1} \right) \left( \vec{g} + C^{\eta}_{3z} \vec{K}_M \right) \cdot \vec{x}} e^{i \mathcal{R}_{\theta,l_2} \left( \vec{g} + C^{\eta}_{3z} \vec{K}_M \right)\cdot \vec{y}} e^{- i \vec{g}' \cdot \delta \vec{R} \left( \vec{x} + \frac{\vec{y}}{2} \right)} e^{- i \vec{k} \cdot \vec{y}} \nonumber \\
	&\times \frac{(-i)^{n_x + n_y} \partial_{k_x}^{n_x} \partial_{k_y}^{n_y}}{n_x! n_y!} S_{s_1 l_1; s_2 l_2} \left( \vec{g}',\vec{k} \right) \hat{\psi}^\dagger_{\eta,s_1,l_1} \left( \vec{x} \right) \partial_x^{n_x} \partial_y^{n_y} \hat{\psi}_{\eta,s_2,l_2} \left( \vec{x} \right)  \nonumber \\
	= & \frac{1}{2N\Omega_0} \sum_{\substack{\vec{g},\vec{k},\eta \\ l_1,s_1,l_2, s_2}} \int \dd[2]{x} \dd[2]{y} e^{-i \frac{l_2 - l_1}{2} \vec{q}_{\eta} \cdot \vec{x}} e^{i \mathcal{R}_{\theta,l_2} C^{\eta}_{3z} \vec{K}_M \cdot \vec{y}} e^{- i \vec{g} \cdot \delta \vec{R} \left( \vec{x} \right)} e^{- 2 i \sin \left( \frac{\theta}{2} \right) \left( \vec{g} \cross \vec{\hat{z}} \right) \cdot \frac{\vec{y}}{2} } e^{- i \vec{k} \cdot \vec{y}} \nonumber \\
	&\times \frac{(-i)^{n_x + n_y} \partial_{k_x}^{n_x} \partial_{k_y}^{n_y}}{n_x! n_y!} S_{s_1 l_1; s_2 l_2} \left( \vec{g},\vec{k} \right) \hat{\psi}^\dagger_{\eta,s_1,l_1} \left( \vec{x} \right) \partial_x^{n_x} \partial_y^{n_y} \hat{\psi}_{\eta,s_2,l_2} \left( \vec{x} \right)  \nonumber \\
	= & \frac{1}{2} \sum_{\substack{\vec{g},\eta \\ l_1,s_1,l_2, s_2}} \int \dd[2]{x} e^{-i \frac{l_2 - l_1}{2} \vec{q}_{\eta} \cdot \vec{x}} e^{- i \vec{g} \cdot \delta \vec{R} \left( \vec{x} \right)} \hat{\psi}^\dagger_{\eta,s_1,l_1} \left( \vec{x} \right) (-i)^{n_x + n_y}  \partial_x^{n_x} \partial_y^{n_y} \hat{\psi}_{\eta,s_2,l_2} \left( \vec{x} \right) \nonumber \\
	&\times \eval{ \frac{\partial_{k_x}^{n_x} \partial_{k_y}^{n_y}}{n_x! n_y!} S_{s_1 l_1; s_2 l_2} \left( \vec{g},\vec{k} \right)}_{\vec{k} = \mathcal{R}_{\theta,l_2} C^{\eta}_{3z} \vec{K}_M - \sin \left( \frac{\theta}{2} \right) \left( \vec{g} \cross \vec{\hat{z}} \right)}, \label{app:eqn:twisted_snse_hamiltonian_interm_general_1:term_2}
\end{align}
where we have also used the Fourier representation from \cref{app:eqn:fourier_representation_S_both}. Similarly, for the first term of \cref{app:eqn:twisted_snse_hamiltonian_interm_general_1}, we find that 
\begin{align}
	& \frac{1}{2\Omega_0} \sum_{\substack{\vec{g},\eta \\ l_1,s_1,l_2, s_2}} \int \dd[2]{x} \dd[2]{y} e^{i \left( \mathcal{R}_{\theta,l_2} - \mathcal{R}_{\theta,l_1} \right) \left( \vec{g} + C^{\eta}_{3z} \vec{K}_M \right) \cdot \vec{x}} e^{i \mathcal{R}_{\theta,l_1} \left( \vec{g} + C^{\eta}_{3z} \vec{K}_M \right)\cdot \vec{y}} \nonumber \\
	&\times S_{s_1 l_1; s_2 l_2} \left( \delta \vec{R} \left( \vec{x} - \frac{\vec{y}}{2} \right) , \vec{y} \right) \hat{\psi}^\dagger_{\eta,s_1,l_1} \left( \vec{x} - \vec{y} \right) \hat{\psi}_{\eta,s_2,l_2} \left( \vec{x} \right) = \nonumber \\
	= & \frac{1}{2} \sum_{\substack{\vec{g},\eta \\ l_1,s_1,l_2, s_2}} \int \dd[2]{x} e^{-i \frac{l_2 - l_1}{2} \vec{q}_{\eta} \cdot \vec{x}} e^{- i \vec{g} \cdot \delta \vec{R} \left( \vec{x} \right)} \left( i^{n_x + n_y}  \partial_x^{n_x} \partial_y^{n_y} \hat{\psi}^\dagger_{\eta,s_1,l_1} \left( \vec{x} \right) \right) \hat{\psi}_{\eta,s_2,l_2} \left( \vec{x} \right) \nonumber \\
	&\times \eval{ \frac{\partial_{k_x}^{n_x} \partial_{k_y}^{n_y}}{n_x! n_y!} S_{s_1 l_1; s_2 l_2} \left( \vec{g},\vec{k} \right)}_{\vec{k} = \mathcal{R}_{\theta,l_1} C^{\eta}_{3z} \vec{K}_M + \sin \left( \frac{\theta}{2} \right) \left( \vec{g} \cross \vec{\hat{z}} \right)}. \label{app:eqn:twisted_snse_hamiltonian_interm_general_1:term_1}
\end{align}
Putting together \cref{app:eqn:twisted_snse_hamiltonian_interm_general_1:term_1,app:eqn:twisted_snse_hamiltonian_interm_general_1:term_2}, we find that in the local-stacking approximation, the moir\'e Hamiltonian with gradient terms can be written as
\begin{align}
	\mathcal{H} = & \frac{1}{2} \sum_{\substack{\vec{g},\eta \\ l_1,s_1,l_2, s_2}} \int \dd[2]{x} e^{-i \frac{l_2 - l_1}{2} \vec{q}_{\eta} \cdot \vec{x}} e^{- i \vec{g} \cdot \delta \vec{R} \left( \vec{x} \right)} \nonumber \\
	&\times \left[ \eval{ \frac{\partial_{k_x}^{n_x} \partial_{k_y}^{n_y}}{n_x! n_y!} S_{s_1 l_1; s_2 l_2} \left( \vec{g},\vec{k} \right)}_{\vec{k} = \mathcal{R}_{\theta,l_1} C^{\eta}_{3z} \vec{K}_M + \sin \left( \frac{\theta}{2} \right) \left( \vec{g} \cross \vec{\hat{z}} \right)} \left( i^{n_x + n_y}  \partial_x^{n_x} \partial_y^{n_y} \hat{\psi}^\dagger_{\eta,s_1,l_1} \left( \vec{x} \right) \right) \hat{\psi}_{\eta,s_2,l_2} \left( \vec{x} \right) \right. \nonumber \\
	&\left. \quad+ \eval{ \frac{\partial_{k_x}^{n_x} \partial_{k_y}^{n_y}}{n_x! n_y!} S_{s_1 l_1; s_2 l_2} \left( \vec{g},\vec{k} \right)}_{\vec{k} = \mathcal{R}_{\theta,l_2} C^{\eta}_{3z} \vec{K}_M - \sin \left( \frac{\theta}{2} \right) \left( \vec{g} \cross \vec{\hat{z}} \right)} \hat{\psi}^\dagger_{\eta,s_1,l_1} \left( \vec{x} \right) (-i)^{n_x + n_y}  \partial_x^{n_x} \partial_y^{n_y} \hat{\psi}_{\eta,s_2,l_2} \left( \vec{x} \right) \right]. \label{app:eqn:twisted_snse_hamiltonian_interm_general_2}
\end{align}
A few remarks are in order regarding \cref{app:eqn:twisted_snse_hamiltonian_interm_general_2}:
\begin{enumerate}
	\item In principle, one could obtain the moir\'e Hamiltonian, by performing multiple \textit{ab initio} simulations for \emph{untwisted} bilayer systems displaced by different positions. Effectively, this amounts to computing $S_{s_1 l_1; s_2 l_2} \left( \Delta \vec{R},\vec{r} \right)$ for multiple values of $\Delta \vec{R}$. In turn, this allows one to obtain the generalized moir\'e continuum Hamiltonian through \cref{app:eqn:twisted_snse_hamiltonian_interm_general_2}. 
	\item It is worth noting that \cref{app:eqn:twisted_snse_hamiltonian_interm_general_2} does not exactly match the form of \cref{app:eqn:general_moire_potential}. Specifically, in \cref{app:eqn:general_moire_potential}, the matrix elements multiplying the term with all the gradients acting on the creation operator and ones multiplying the term with all the gradients on the annihilation operator are the same, unlike \cref{app:eqn:twisted_snse_hamiltonian_interm_general_2}, where they are merely hermitian conjugate of one another. This difference is just a matter of notation. \Cref{app:eqn:general_moire_potential} represents one possible, but not the only, fully-general parameterization of the generalized moir\'e Hamiltonian. One could recast \cref{app:eqn:twisted_snse_hamiltonian_interm_general_2} into the form of \cref{app:eqn:general_moire_potential} by performing a series of integrations by parts, though this falls outside the scope of our current discussion.
	\item In the following \cref{app:sec:SnS_SnSe_twist_general:approx:zero_twist}, we will show that \cref{app:eqn:twisted_snse_hamiltonian_interm_general_2} can be further simplified by taking the limit $\theta \to 0$. By doing so, we will show that \cref{app:eqn:twisted_snse_hamiltonian_interm_general_2} becomes manifestly of the form of \cref{app:eqn:general_moire_potential}.
\end{enumerate}

\subsubsection{Further simplification in the zero-twist limit}\label{app:sec:SnS_SnSe_twist_general:approx:zero_twist}

In order to simplify \cref{app:eqn:twisted_snse_hamiltonian_interm_general_2} in the limit of vanishing twist angle, we now take the limit $\theta \to 0$. We cannot ignore the finite twist angle in the exponential prefactors of \cref{app:eqn:twisted_snse_hamiltonian_interm_general_2}, since $\vec{x}$ assumes values over the entire space. However, we \emph{can} impose the limit $\theta \to 0$ in the momenta at which the $S_{s_1 l_1; s_2 l_2} \left( \vec{g},\vec{k} \right)$ is being evaluated. By doing so, we directly obtain 
\begin{align}
	\mathcal{H} \approx & \frac{1}{2} \sum_{\substack{\vec{g}',\eta \\ l_1,s_1,l_2, s_2}} \sum_{n_x,n_y=0}^{\infty} \frac{\partial^{n_x}_{k_x} \partial^{n_y}_{k_y}}{n_x ! n_y !} \eval{S_{s_1 l_1; s_2 l_2} \left( \vec{g}, \vec{k} \right)}_{\vec{k} = C^{\eta}_{3z} \vec{K}_M} \nonumber \\
	&\times \int \dd[2]{r} e^{-i \frac{l_2 - l_1}{2} \vec{q}_\eta \cdot \vec{r} } e^{-2 i  \sin \left( \frac{\theta}{2} \right) \left( \vec{g} \cross \hat{\vec{z}} \right) \cdot \vec{r} } \nonumber \\
	&\times \left[ \left( i^{n_x+n_y} \partial_x^{n_x} \partial_y^{n_y} \hat{\psi}^\dagger_{\eta,s_1,l_1} \left( \vec{r} \right) \right) \hat{\psi}_{\eta,s_2,l_2} \left( \vec{r} \right) + (-i)^{n_x+n_y} \hat{\psi}^\dagger_{\eta,s_1,l_1} \left( \vec{r} \right) \partial_x^{n_x} \partial_y^{n_y} \hat{\psi}_{\eta,s_2,l_2} \left( \vec{r} \right) \right]. \label{app:eqn:twisted_snse_hamiltonian_final_moire_general}
\end{align}
Comparing \cref{app:eqn:general_moire_potential,app:eqn:twisted_snse_hamiltonian_final_moire_general}, we find that in the local-stacking approximation and with the zero-twist limit imposed, the real-space generalized moir\'e potential is given by
\begin{equation}
	t^{\eta,n_x,n_y}_{s_1 l_1; s_2 l_2} \left( \vec{r} \right) = \sum_{\vec{g}} \eval{\frac{\partial^{n_x}_{k_x} \partial^{n_y}_{k_y}}{n_x! n_y!} S_{s_1 l_1; s_2 l_2} \left( \vec{g}, \vec{k} \right)}_{\vec{k} = C^{\eta}_{3z} \vec{K}_M} e^{-i \frac{l_2 - l_1}{2} \vec{q}_\eta \cdot \vec{r} } e^{-2 i  \sin \left( \frac{\theta}{2} \right) \left( \vec{g} \cross \hat{\vec{z}} \right) \cdot \vec{r} }.
\end{equation}
To obtain the momentum-space generalized moir\'e potential, we employ \cref{app:eqn:ft_general_moire_potential_to_mom} and perform the integral in the same way as it was done in \cref{app:eqn:equivalence_t_matrix_s_matrix}
\begin{align}
	\left[ T^{n_x,n_y}_{ \vec{q}_{\eta+l_1}, \vec{q}_{\eta + l_2} + \vec{G}} \right]_{s_1 l_1; s_2 l_2} =\eval{ \frac{\partial^{n_x}_{k_x} \partial^{n_y}_{k_y}}{n_x! n_y!} S_{s_1 l_1; s_2 l_2} \left( \frac{\left[\vec{G} + \left( l_1 - l_2 \right) \vec{q}_{\eta - 1} \right] \cross \hat{\vec{z}}}{2 \sin \left( \frac{\theta}{2} \right)}, \vec{k} \right)}_{\vec{k} = C^{\eta}_{3z} \vec{K}_M}.\label{app:eqn:equivalence_t_matrix_s_matrix_general}
\end{align}
\Cref{app:eqn:equivalence_t_matrix_s_matrix_general} is a one-to-one relation between the moir\'e potential and momentum derivatives of the $S_{s_1 l_1; s_2 l_2} \left( \vec{g}, \vec{k} \right)$ matrix evaluated at the $C^{\eta}_{3z}\vec{K}_{M}$ point, akin to the one of implied by \cref{app:eqn:equivalence_t_matrix_s_matrix} for the case without gradient terms.

\subsubsection{Constraining the generalized moir\'e potential in the local-stacking approximation}\label{app:sec:SnS_SnSe_twist_general:approx:constraining}

As in \cref{app:sec:SnS_SnSe_twist_general_woGradient:approx:constraining}, we can additionally constrain the generalized moir\'e potential $T^{n_x,n_y}_{\vec{Q},\vec{Q}'}$ in the limit of vanishing twist angle by constraining the $S_{s_1 l_1; s_2 l_2} \left( \vec{g} , \vec{k} \right)$ matrix. Specifically, as a result of \cref{app:eqn:symmetry_moire_s}, under a crystalline symmetry $g$, the latter obeys 
\begin{equation}
	\label{app:eqn:symmetry_moire_s_general_k_space}
	\sum_{s_1, s_2} \left[ D^{\text{sl}}(g) \right]_{s'_1 s_1} S^{(*)}_{s_1 l_1; s_2 l_2} \left( \vec{g}, \vec{k}  \right) \left[ D^{\text{sl}}(g) \right]^{*}_{s'_2 s_2} = S_{s'_1 \epsilon_g l_1; s'_2 \epsilon_g l_2} \left( \epsilon_g g \vec{g}, g \vec{k} \right).
\end{equation}
Because the momentum derivatives of $S_{s_1 l_1; s_2 l_2} \left( \vec{g}, \vec{k}  \right)$ appear in \cref{app:eqn:equivalence_t_matrix_s_matrix_general} it is useful to derive the symmetry constrains imposed by $g$ on the partial derivatives of $S_{s_1 l_1; s_2 l_2} \left( \vec{g}, \vec{k}  \right)$ with respect to $\vec{k}$. To do so, we begin by noting that 
\begin{equation}
	\label{app:eqn:gradient_as_inverse_k}
	\frac{\partial^{n_x}_{k_x} \partial^{n_y}_{k_y}}{n_x! n_y!} k^{m_x}_x k^{m_y}_y = \frac{\partial^{n_x}_{\left[ g \vec{k} \right]_x} \partial^{n_y}_{\left[ g \vec{k} \right]_y}}{n_x! n_y!} \left(\left[ g \vec{k} \right]_x\right)^{m_x} \left(\left[ g \vec{k} \right]_y\right)^{m_y} = \delta_{n_x, m_x} \delta_{n_y, m_y}. 
\end{equation}
Because the components of $g\vec{k}$ are linear combinations of the components of $\vec{k}$, we must have that
\begin{equation}
	\label{app:eqn:trafo_totally_symmetry_partial_derivatives}
	\frac{\partial^{n_x}_{\left[ g \vec{k} \right]_x} \partial^{n_y}_{\left[ g \vec{k} \right]_y}}{n_x! n_y!} =  \sum_{\substack{n'_x, n'_y \\ n'_x + n'_y = n_x + n_y}} W_{n_x,n_y; n'_x,n'_y} ( g ) \frac{\partial^{n'_x}_{k_x} \partial^{n'_y}_{k_y}}{n'_x! n'_y!},
\end{equation}
where $W_{n_x,n_y;n'_x,n'_y} ( g )$ is a matrix which will be obtained below. Using the second equality in \cref{app:eqn:gradient_as_inverse_k}, as well as \cref{app:eqn:totally_symmetry_trafo_k}, we find that 
\begin{align}
	\delta_{n_x, m_x} \delta_{n_y, m_y} &= \sum_{\substack{n'_x, n'_y \\ n'_x + n'_y = n_x + n_y}} W_{n_x,n_y; n'_x,n'_y} ( g ) \frac{\partial^{n'_x}_{k_x} \partial^{n'_y}_{k_y}}{n'_x! n'_y!} \left( \left[ g \vec{k} \right]_x\right)^{m_x} \left(\left[ g \vec{k} \right]_y\right)^{m_y} \nonumber \\
	&= \sum_{\substack{n'_x, n'_y \\ n'_x + n'_y = n_x + n_y}} \sum_{\substack{m'_x, m'_y \\ m'_x + m'_y = m_x + m_y}} W_{n_x,n_y; n'_x,n'_y} ( g ) \frac{\partial^{n'_x}_{k_x} \partial^{n'_y}_{k_y}}{n'_x! n'_y!} U_{m'_x,m'_y;m_x,m_y} ( g ) k^{m'_x}_{x} k^{m'_y}_{y} \nonumber \\
	&= \sum_{\substack{n'_x, n'_y \\ n'_x + n'_y = n_x + n_y \\ m_x + m_y = n_x + n_y}} W_{n_x,n_y; n'_x,n'_y} ( g ) U_{n'_x,n'_y;m_x,m_y} ( g ),
\end{align}
from which we can immediately conclude that 
\begin{equation}
	\label{app:eqn:trafo_totally_symmetry_partial_derivative_UW}
	W (g) = U^{-1} (g),
\end{equation}
where the matrix $U(g)$ was defined in \cref{app:eqn:totally_symmetry_trafo_k}. As a result of \cref{app:eqn:trafo_totally_symmetry_partial_derivatives,app:eqn:trafo_totally_symmetry_partial_derivative_UW}, we can conclude that 
\begin{equation}
	\label{app:eqn:trafo_totally_symmetry_partial_derivatives_final}
	\frac{\partial^{n_x}_{k_x} \partial^{n_y}_{k_y}}{n_x! n_y!} =  \sum_{\substack{n'_x, n'_y \\ n'_x + n'_y = n_x + n_y}} U_{n_x,n_y; n'_x,n'_y} ( g ) \frac{\partial^{n'_x}_{\left[ g \vec{k} \right]_x} \partial^{n'_y}_{\left[ g \vec{k} \right]_y}}{n'_x! n'_y!},
\end{equation}
By differentiating \cref{app:eqn:symmetry_moire_s_general_k_space} with respect to $\vec{k}$ and employing \cref{app:eqn:trafo_totally_symmetry_partial_derivatives_final}, we obtain
\begin{align}
	&\sum_{s_1, s_2} \left[ D^{\text{sl}}(g) \right]_{s'_1 s_1} \frac{\partial^{n_x}_{k_x} \partial^{n_y}_{k_y}}{n_x! n_y!}S^{(*)}_{s_1 l_1; s_2 l_2} \left( \vec{g}, \vec{k}  \right) \left[ D^{\text{sl}}(g) \right]^{*}_{s'_2 s_2} \nonumber \\
	=&  \frac{\partial^{n_x}_{k_x} \partial^{n_y}_{k_y}}{n_x! n_y!} S_{s'_1 \epsilon_g l_1; s'_2 \epsilon_g l_2} \left( \epsilon_g g \vec{g}, g \vec{k} \right) \nonumber \\
	&\sum_{s_1, s_2} \left[ D^{\text{sl}}(g) \right]_{s'_1 s_1} \frac{\partial^{n_x}_{k_x} \partial^{n_y}_{k_y}}{n_x! n_y!}S^{(*)}_{s_1 l_1; s_2 l_2} \left( \vec{g}, \vec{k}  \right) \left[ D^{\text{sl}}(g) \right]^{*}_{s'_2 s_2} \nonumber \\
	=&  \sum_{\substack{n'_x, n'_y \\ n'_x + n'_y = n_x + n_y}} U_{n_x,n_y; n'_x,n'_y} ( g ) \frac{\partial^{n'_x}_{\left[ g \vec{k} \right]_x} \partial^{n'_y}_{\left[ g \vec{k} \right]_y}}{n'_x! n'_y!} S_{s'_1 \epsilon_g l_1; s'_2 \epsilon_g l_2} \left( \epsilon_g g \vec{g}, g \vec{k} \right) \nonumber \\
	&\sum_{s_1, s_2} \left[ D^{\text{sl}}(g) \right]_{s'_1 s_1} \frac{\partial^{n_x}_{k_x} \partial^{n_y}_{k_y}}{n_x! n_y!}S^{(*)}_{s_1 l_1; s_2 l_2} \left( \vec{g}, \vec{k}  \right) \left[ D^{\text{sl}}(g) \right]^{*}_{s'_2 s_2} \nonumber \\
	=&  \sum_{\substack{n'_x, n'_y \\ n'_x + n'_y = n_x + n_y}} U_{n_x,n_y; n'_x,n'_y} ( g ) \eval{\left( \frac{\partial^{n'_x}_{k'_x} \partial^{n'_y}_{k'_y}}{n'_x! n'_y!} S_{s'_1 \epsilon_g l_1; s'_2 \epsilon_g l_2} \left( \epsilon_g g \vec{g}, \vec{k}' \right) \right)}_{\vec{k}'=g \vec{k}},
\end{align}
which, upon substituting $\vec{k} = C^{\eta}_{3z} \vec{K}_M$, becomes
\begin{align}
	&\sum_{s_1, s_2} \left[ D^{\text{sl}}(g) \right]_{s'_1 s_1} \eval{ \frac{\partial^{n_x}_{k_x} \partial^{n_y}_{k_y}}{n_x! n_y!}S^{(*)}_{s_1 l_1; s_2 l_2} \left( \vec{g}, \vec{k}  \right) }_{\vec{k} = C^{\eta}_{3z} \vec{K}_M} \left[ D^{\text{sl}}(g) \right]^{*}_{s'_2 s_2} \nonumber \\
	=& \sum_{\substack{n'_x, n'_y \\ n'_x + n'_y = n_x + n_y}} U_{n_x,n_y;n'_x,n'_y} (g) \eval{\left( \frac{\partial^{n'_x}_{k_x} \partial^{n'_y}_{k_y}}{n'_x! n'_y!} S_{s'_1 \epsilon_g l_1; s'_2 \epsilon_g l_2} \left( \epsilon_g g \vec{g}, \vec{k} \right) \right)}_{\vec{k} = g C^{\eta}_{3z} \vec{K}_M}. \label{app:eqn:symmetry_moire_s_general_k_space_with_derivatives}
\end{align}

As in \cref{app:sec:SnS_SnSe_twist_general_woGradient:approx:constraining}, the exact symmetries of the heterostructure in the $\theta \neq 0$ case can either be imposed through \cref{app:eqn:symmetry_of_t_general} or through \cref{app:eqn:symmetry_moire_s_general_k_space_with_derivatives}, while the zero-twist symmetries can only be imposed indirectly through \cref{app:eqn:symmetry_moire_s_general_k_space_with_derivatives}. In practice, we first impose the exact $\theta \neq 0$ symmetries on the generalized moir\'e potential through \cref{app:eqn:symmetry_of_t_general}. One can then further simplify the generalized moir\'e potential by using \cref{app:eqn:equivalence_t_matrix_s_matrix_general,app:eqn:symmetry_moire_s_general_k_space_with_derivatives}. In the AA-stacking case, additional constraints arise from the $\mathcal{I}$ symmetry, while in the AB-stacking case, the generalized moir\'e potential can be additionally constrained by the $M_z$ symmetries.

\section{First-principle results for twisted bilayer materials with valley projection} \label{app:sec:first_princ_ham_valley}
This \siSection{} is dedicated to the valley-projected Hamiltonians of twisted \ch{SnSe2} and \ch{ZrS2}, as obtained from \textit{ab initio} methods. We begin by outlining how the valley-projected Hamiltonians are constructed from the corresponding OpenMX ones~\cite{BOK11, NEA16}. A key method employed to this end is the L\"{o}wdin orthogonalization~\cite{LOW50,CLO64, MAR97b, MAR12}, which will also be briefly reviewed. The resulting valley-projected Hamiltonians will be employed in \cref{app:sec:fitted_models} to directly obtain the parameterized continuum moir\'e Hamiltonians. Lastly, we present some preliminary \textit{ab initio} results on the valley-projected spectrum of twisted \ch{SnSe2} and \ch{ZrS2}, including the energy dispersion across a range of commensurate angles and the local density of states (LDOS) at the smallest twist angle considered, $\theta = \SI{3.89}{\degree}$. More comprehensive results at all angles, various continuum models, and further discussion are provided in \cref{app:sec:fitted_models}.

\subsection{L\"{o}wdin orthogonalization}\label{app:sec:first_princ_ham_valley:lowdin}

\begingroup
\renewcommand{\arraystretch}{1.6}
\begin{table}[!t]
	\centering
	\begin{tabular}{|l|l|l|}
\hline
		\textbf{States} & \textbf{Defined in} & \textbf{Meaning} \\
		\hline
		$\ket{\psi_{i}}$ & \makecell[l]{\cref{app:eqn:lowd_eigenstate}} & Eigenstates of $\hat{H}$ \\
		\hline
		$\ket{\phi_{i}}$ & \makecell[l]{\cref{app:sec:first_princ_ham_valley:lowdin}} & Trial states \\
		\hline
		$\ket{\bar{\phi}_{i}}$ & \makecell[l]{\cref{app:eqn:def_proj_trial_states_lowdin}} & Projected trial states \\
		\hline
		$\ket{\tilde{\phi}_{i}}$ & \makecell[l]{\cref{app:eqn:refined_lowdin_results}} & Refined trial states \\
		\hline
		$\ket{\varphi^{\beta}}$ & \makecell[l]{\cref{app:eqn:non-orthonormal_basis_ham}} & Non-orthonormal basis states \\
		\hline
\end{tabular}\caption{Summary of notation used in \cref{app:sec:first_princ_ham_valley}. Each state is listed along with the place where it is defined and a brief description of its physical meaning.}
	\label{app:tab:lowdin_notation}
\end{table}
\endgroup

We begin by reviewing L\"{o}wdin's orthogonalization method~\cite{LOW50,CLO64, MAR97b, MAR12}. Since this approach can be applied more broadly ({\it i.e.}{}, not just in the context of moir\'e materials) and to avoid keeping track of many indices, we will employ a streamlined notation in this section, which is also summarized in \cref{app:tab:lowdin_notation}. Consider a Hamiltonian operator $\hat{H}$, whose energies and corresponding eigenstates are given by $\epsilon_i$ and $\ket{\psi_i}$, respectively, such that
\begin{equation}
	\label{app:eqn:lowd_eigenstate}
	\hat{H} \ket{\psi_i} = \epsilon_i \ket{\psi_i}.
\end{equation}
In \cref{app:eqn:lowd_eigenstate}, $i$ indexes each eigenstate and 
\begin{equation}
	\bra{\psi_i} \ket{\psi_j} = \delta_{ij}. 
\end{equation} 
We now take a set of $N$ linearly independent trial states $\ket{\phi_j}$, indexed by $1 \leq j \leq N$. Our goal is to project the Hamiltonian $\hat{H}$ onto these trial states. We first assume that the Hilbert space spanned by $\ket{\phi_j}$ is \emph{exactly} identical to an $N$-dimensional eigensubspace of $\hat{H}$ and the trial states $\ket{\phi_j}$ are orthonormal. The spectrum of the projected Hamiltonian, whose matrix elements are given by $\bra{\phi_j} \hat{H} \ket{\phi_k}$ (for $1 \leq j,k \leq N$) \emph{exactly} matches the $N$-dimensional eigensubspace of $\hat{H}$. 

More often than not, the trial states do \emph{not} exactly span an $N$-dimensional eigensubspace of $\hat{H}$. For example, the $s$ orbitals of Sn in monolayer \ch{SnSe2} have a high overlap with the lowest gapped conduction band, as shown in \cref{app:fig:monolayer-bands}, but the Wannier orbitals of the band are described by \emph{effective} molecular $s$-like orbitals that also have significant weight on the Se orbitals, and not just on the $s$ orbitals of Sn. In such cases, any eigensubspace of $\hat{H}$ will not be \emph{exactly} preserved under the naive projection $\bra{\phi_j} \hat{H} \ket{\phi_k}$ (for $1 \leq j,k \leq N$).

To address this, we aim to obtain a refined set of orthonormalized states $\ket{\tilde{\phi}_j}$ that are \emph{adiabatically} connected to the original trial states $\ket{\phi_j}$, but \emph{exactly} span an $N$-dimensional eigensubspace of $\hat{H}$. This ensures that the projected spectrum of $\hat{H}$ is \emph{identical} to a particular $N$-dimensional eigensubspace of $\hat{H}$. In our \ch{SnSe2} analogy, $\ket{\phi_j}$ represent the linearly independent (but in general non-orthonormal, due to the nonzero overlaps of different trial states) $s$ orbitals of Sn, while $\ket{\tilde{\phi}_j}$ correspond to the orthonormal Wannier orbitals of the bottom conduction band, which also behave as effective $s$ orbitals located at the $1a$ Wyckoff position from a symmetry standpoint. L\"{o}wdin orthogonalization is a method used to adiabatically refine a set of trial states $\ket{\phi_j}$ (with $1 \leq j \leq N$), such that a relevant $N$-dimensional portion of the energy spectrum of $\hat{H}$ is \emph{exactly} preserved under the projection onto the resulting \emph{refined} trial states $\ket{\tilde{\phi}_j}$.

\subsubsection{Obtaining the refined trial states}\label{app:sec:first_princ_ham_valley:lowdin:refined}

To illustrate L\"{o}wdin orthogonalization, we consider a gapped $N$-dimensional eigensubspace of $\hat{H}$ spanned by the states $\ket{\psi_j}$ defined in \cref{app:eqn:lowd_eigenstate}, for $j \in \mathcal{A}$. Given the $N$ trial states $\ket{\phi_i}$, we want to obtain $N$ \emph{refined} trial states $\ket{\tilde{\phi}_i}$ (adiabatically connected to the $\ket{\phi_i}$ states), such that the eigensubspace $\mathcal{A}$ is exactly preserved under the projection of $\hat{H}$ onto the states $\ket{\tilde{\phi}_i}$. To this end, we define the projector in the eigensubspace $\mathcal{A}$ to be
\begin{equation}
	\label{app:Eqn:projector_for_lowdin}
	\hat{P}_{\mathcal{A}} = \sum_{j \in \mathcal{A}} \ket{\psi_j} \bra{\psi_j}.
\end{equation}
The projected trial states
\begin{equation}
	\label{app:eqn:def_proj_trial_states_lowdin}
	\ket{\bar{\phi}_j} = \hat{P}_{\mathcal{A}} \ket{\phi_j}, \qq{for} 1 \leq j \leq N
\end{equation}
will exactly span the eigensubspace $\mathcal{A}$ (provided that $\ket{\bar{\phi}_j}$ are linearly independent, which we assume to be the case\footnote{We also assume that the trial states $\ket{\phi_j}$ have finite overlap with the eigensubspace $\mathcal{A}$. In other words, $\hat{P}_{\mathcal{A}} \ket{\phi_j} \neq 0$.}). They are, however, not orthonormal, with their (Hermitian) overlap matrix being given by
\begin{equation}
	\label{app:eqn:lowdin_overlap_matrix}
	\mathcal{S}_{ij} \equiv \bra{\bar{\phi}_i} \ket{\bar{\phi}_j} = \bra{\phi_i} \hat{P}_{\mathcal{A}} \ket{\phi_j}, \qq{for} 1 \leq i,j \leq N.
\end{equation}
The matrix $\mathcal{S}$ is positive semi-definite, since it is the restriction of the positive semi-definite operator $\hat{P}_{\mathcal{A}}$ on the space spanned by $\ket{\phi_j}$. As we \emph{assume} $\ket{\bar{\phi}_j}$ are linearly independent and nonzero, the matrix $\mathcal{S}$ has no zero eigenvalues, meaning that it is positive \emph{definite}. This allows us to orthonormalize the projected trial state $\ket{\bar{\phi}_j}$ and obtain the \emph{refined} trial state~\cite{LOW50,CLO64, MAR97b, MAR12}
\begin{equation}
	\label{app:eqn:refined_lowdin_results}
	\ket{\tilde{\phi}_j} = \sum_{i = 1}^{N} \ket{\bar{\phi}_i} \left( \mathcal{S}^{-1/2} \right)_{i j},
\end{equation}
where $\mathcal{S}^{-1/2}$ is the \emph{unique} positive \emph{definite} matrix\footnote{In practice, $\mathcal{S}^{-1/2}$ is obtained by diagonalizing the $\mathcal{S}$ matrix as $\mathcal{S} = V D V^{\dagger}$, where $V$ is unitary and $D$ is diagonal with strictly positive entries. Defining $D^{-1/2}$ as the diagonal matrix with elements $\left( D^{-1/2} \right)_{ii} = \frac{1}{\sqrt{D_{ii}}}$, we compute $\mathcal{S}^{-1/2}$ as $\mathcal{S}^{-1/2} = V D^{-1/2} V^{\dagger}$.}, such that $\left( \mathcal{S}^{-1/2} \right)^{2} = S^{-1}$. The trial states are both orthonormal
\begin{equation}
	\bra{\tilde{\phi}_i} \ket{\tilde{\phi}_j} = \sum_{i',j' = 1}^{N} \left( \mathcal{S}^{-1/2} \right)_{i i'} \bra{\bar{\phi}_{i'}} \ket{\bar{\phi}_{j'}}  \left( \mathcal{S}^{-1/2} \right)_{j' j} = \sum_{i',j' = 1}^{N} \left( \mathcal{S}^{-1/2} \right)_{i i'} \mathcal{S}_{i' j'}  \left( \mathcal{S}^{-1/2} \right)_{j' j} = \delta_{ij}
\end{equation}
and exactly span the eigensubspace $\mathcal{A}$, by virtue of being linear combinations of the $\ket{\bar{\phi}_j}$ states. The projected Hamiltonian matrix into the $\ket{\tilde{\phi}_j}$ basis is given by $\tilde{H}_{ij} \equiv \bra{\tilde{\phi}_i} \hat{H} \ket{\tilde{\phi}_j}$

All that is left to prove is that the refined states $\ket{\tilde{\phi}_j}$ are adiabatically connected to the trial states $\ket{\phi_j}$ (meaning that $\ket{\phi_j}$ transform in the same way as $\ket{\tilde{\phi}_j}$ under the symmetries of the problem). To this end, assume that under a symmetry $g$ of $\hat{H}$ ($\commutator{\hat{H}}{g} = 0$), the trial states transform as
\begin{equation}
	\label{app:eqn:lowdin_symmetry_trafo}
	g \ket{\phi_j} = \sum_{i=1}^{N} U_{ij} \ket{\phi_i},
\end{equation}
where $U_{ij}$ is some $N$-dimensional unitary matrix. Since $\commutator{\hat{P}_{\mathcal{A}}}{g} = 0$, it follows that the projected trial states introduced in \cref{app:eqn:def_proj_trial_states_lowdin} transform in the same way as \cref{app:eqn:lowdin_symmetry_trafo} under the symmetry $g$, which implies that
\begin{equation}
	\mathcal{S}_{ij} = \sum_{i',j'=1}^{N} U^{*}_{i'i} \mathcal{S}^{(*)}_{i'j'} U_{j'j},
\end{equation}
with $^{(*)}$ denoting a complex conjugation in the cases in which $g$ is antiunitary. As a result, we find that 
\begin{align}
	g \ket{\tilde{\phi}_j} &= \sum_{i,i'=1}^{N} \ket{\bar{\phi}_{i'}} U_{i' i} \left( \mathcal{S}^{-1/2} \right)^{(*)}_{i j} \nonumber \\
	&= \sum_{i',j'=1}^{N} \ket{\bar{\phi}_{i'}}  \left( \mathcal{S}^{-1/2} \right)^{(*)}_{i' j'} U_{j' j} \nonumber \\
	&= \sum_{j'=1}^{N} \ket{\tilde{\phi}_j'} U_{j' j},
\end{align}
which means that the refined states transform in the same way as the original trial states, while spanning the $N$-dimensional eigensubspace $\mathcal{A}$ of $\hat{H}$.

Finally, it is important to reiterate the significance of L\"owdin orthogonalization: it refines a set of trial states such that a chosen eigensubspace $\mathcal{A}$ is \emph{exactly} preserved under the projection onto the refined basis. One might wonder, since L\"owdin orthogonalization employs the projector $\hat{P}_{\mathcal{A}}$ (which, as shown in \cref{app:Eqn:projector_for_lowdin}, requires explicitly obtaining the eigenstates from $\mathcal{A}$), why not simply use the states $\ket{\psi}_{j}$ (for $j \in \mathcal{A}$) to form the projected Hamiltonian directly? There are two key reasons:
\begin{itemize}
	\item Projecting $\hat{H}$ onto $\ket{\psi_{j}}$ affords the projected Hamiltonian in the \emph{band} basis, but not in a basis adiabatically connected to the trial states. For example, in the case of the moir\'e Hamiltonians, we need the projected moir\'e Hamiltonian in the plane-wave basis, not the band basis.
	\item While constructing the entire projector $\hat{P}_{\mathcal{A}}$ requires finding the states $\ket{\psi}_{j}$ (for $j \in \mathcal{A}$) via full diagonalization or shift-and-invert methods, its action on the trial states ({\it i.e.}{}, $\hat{P}_{\mathcal{A}} \ket{\phi_j}$) can be computed more efficiently using contour integration techniques~\cite{POL09}.
\end{itemize}

\subsubsection{The L\"owdin orthogonalization in a non-orthornormal basis }\label{app:sec:first_princ_ham_valley:lowdin:non-orthonormal}

The Kohn-Sham Hamiltonians used in this work, obtained through \textit{ab initio} methods, are generated with OpenMX~\cite{BOK11,NEA16}, which employs a pseudo-atomic localized basis\footnote{The pseudo-atomic basis has the same symmetry properties as atomic orbitals but is variationally optimized to reduce the computational cost for large-scale \textit{ab initio} calculations~\cite{OZA03, OZA04}}. This basis is linearly independent but is \emph{not} orthonormal. For completeness, we now show how the L\"{o}wdin orthogonalization, reviewed generally in \cref{app:sec:first_princ_ham_valley:lowdin:refined}, can be applied for such a non-orthonormal basis.

Let $\ket{\alpha}$ denote a complete non-orthonormal basis (for a lattice Hamiltonian, this would be the atomic orbital basis). The matrix elements of the Hamiltonian $\hat{H}$ in this basis, and the overlap matrix of the $\ket{\varphi^{\alpha}}$ states, are given by
\begin{align}
	H^{\alpha \beta} &\equiv \bra{\varphi^{\alpha}} \hat{H} \ket{\varphi^{\beta}}, \label{app:eqn:non-orthonormal_basis_ham}\\
	S^{\alpha \beta} &\equiv \bra{\varphi^{\alpha}} \ket{\varphi^{\beta}},
\end{align}
where $S_{\alpha \beta}$ is a positive definite matrix. The components of the eigenstates of $\hat{H}$ and those of the trial states $\ket{\phi_i}$ in the basis $\ket{\varphi^{\alpha}}$ are
\begin{align}
	\ket{\psi_i} & = \sum_{\alpha} \psi^{\alpha}_{i} \ket{\varphi^{\alpha}} , \\
	\ket{\phi_i} & = \sum_{\alpha} \phi^{\alpha}_{i} \ket{\varphi^{\alpha}} ,
\end{align}
with the vectors $\psi^{\alpha}_i$ satisfying the generalized eigenvalue equation
\begin{equation}
	\sum_{\beta} H^{\alpha \beta} \psi^{\beta}_i = \epsilon_i \sum_{\beta} S^{\alpha \beta} \psi^{\beta}_i,
\end{equation}
and being orthonormal with respect to the overlap matrix $S^{\alpha \beta}$
\begin{equation}
	\sum_{\alpha, \beta} \left(\psi^{\alpha}_i \right)^* S^{\alpha \beta} \psi^{\beta}_j = \delta_{ij}. 
\end{equation}
To perform the L\"{o}wdin orthonormalization of the $\ket{\bar{\phi}_i}$ projected trial states, we follow the same procedure as described in \cref{app:sec:first_princ_ham_valley:lowdin:refined}. The L\"{o}wdin overlap matrix, as given in \cref{app:eqn:lowdin_overlap_matrix}, is expressed as
\begin{align}
	 \mathcal{S}_{ij} = \bra{\bar{\phi}_i} \ket{\bar{\phi}_j} = \bra{\phi_i} \hat{P}_{\mathcal{A}} \ket{\phi_j} = \sum_{k \in \mathcal{A}} \bra{\phi_i} \ket{\psi_k} \bra{\psi_k} \ket{\phi_j} = \sum_{\alpha, \beta, \gamma, \delta} \sum_{k \in \mathcal{A}} \left(\phi^{\alpha}_i \right)^* S^{\alpha \beta} \psi^{\beta}_k \left( \psi^{\gamma}_k \right)^* S^{\gamma \delta} \phi^{\delta}_j.
\end{align}
Using the L\"{o}wdin overlap matrix $\mathcal{S}$, the refined trial states can then be directly obtained from \cref{app:eqn:refined_lowdin_results}. In the non-orthonormal basis $\ket{\alpha}$, the refined trial states are expressed as
\begin{equation}
	\ket{\tilde{\phi}_i} = \sum_{\alpha} \tilde{\phi}^{\alpha}_i \ket{\varphi^{\alpha}},
\end{equation}  
with 
\begin{equation}
	\tilde{\phi}^{\alpha}_j = \sum_{i = 1}^{N} \sum_{k \in \mathcal{A}} \sum_{\beta, \gamma} \psi^{\alpha}_k \left( \psi^{\beta}_k \right)^* S^{\beta \gamma} \phi^{\gamma}_i \left( \mathcal{S}^{-1/2} \right)_{i j}.
\end{equation}
Finally, the projected Hamiltonian matrix reads as
\begin{equation}
	\tilde{H}_{ij} \equiv \bra{\tilde{\phi}_i} \hat{H} \ket{\tilde{\phi}_j} = \sum_{\alpha, \beta} \left( \tilde{\phi}_i^{\alpha} \right)^* H^{\alpha \beta} \tilde{\phi}_j^{\beta}.
\end{equation}

\subsection{Obtaining the plane-wave moir\'e Hamiltonians from \textit{ab initio} simulations}\label{app:sec:first_princ_ham_valley:algorithm}

We employ the L\"{o}wdin orthogonalization method followed by projection into the refined basis, in order to construct the valley-projected Hamiltonians for the twisted bilayer systems. This procedure is carried out in two steps:
\begin{enumerate}
	\item We begin by projecting the \textit{ab initio} Kohn-Sham Hamiltonian onto a set of Wannier basis functions, which are constructed via L\"{o}wdin orthogonalization of trial atomic orbitals. For twisted \ch{SnSe2}, these trial states are the $s$ orbitals of Sn, while for twisted \ch{ZrS2}, they are the $d_{z^2}$, $d_{x^2-y^2}$, and $d_{xy}$ orbitals of Zr.
	  
	\item The resulting Wannier-projected Hamiltonian is further projected into each of the three M valleys, by employing a truncated atomic plane wave (TAPW) basis around each valley~\cite{MIA23}.
\end{enumerate}
Although these two steps can theoretically be combined into a single step that directly projects into each valley, we find that, in practice, the second step requires a careful selection of the projection eigensubspace $\mathcal{A}$ and trial wavefunctions ({\it i.e.}{} number of plane wave states). Therefore, we maintain them as two distinct steps to allow for more precise tuning of the valley projection steps. In the following, we provide a more detailed explanation of each of these two steps.

\newcommand{\vtau}{\boldsymbol{\tau}}
z
Our \text{ab initio} simulations are performed using the OpenMX~\cite{BOK11,NEA16} package, which outputs the Kohn-Sham Hamiltonian in a non-orthonormal pseudo-atomic basis. We let $\ket{\varphi_{i,l,\alpha,s} \left( \vec{R} + \vtau_{i,l,\alpha} \right)}$ denote the basis state corresponding to orbital $\alpha$ and spin $s$, located in the moir\'e unit cell $\vec{R}$. Because we work at commensurate twist angles (such that the moir\'e unit cell contains an integer number of monolayer unit cells), each basis state is additionally labeled by the integer $i$, which indexes the monolayer unit cell from layer $l$ within the moir\'e unit cell $\vec{R}$. The orbital is also displaced by $\vtau_{i,l,\alpha}$ from the moir\'e unit cell origin. In this section, we adopt the notation outlined in \cref{app:sec:first_princ_ham_valley:lowdin}, which is further summarized in \cref{app:tab:lowdin_notation} for trial, projected trial, non-orthonormal basis states, {\it etc.}{}. Additional indices such as layer, spin, and unit cell are appended as needed.

The Kohn-Sham spectrum can be determined from the OpenMX Hamiltonian and overlap matrices, which are defined by
\begin{align}
	H_{i,l,\alpha,s;j,l',\beta,s'} \left( \vec{k} \right) &= \sum_{\vec{R}} e^{i \vec{k} \cdot \left( -\vtau_{i,l,\alpha} + \vec{R} + \vtau_{j,l',\beta} \right)} \bra{\varphi_{i,l,\alpha,s} \left( \vtau_{i,l,\alpha} \right) } \hat{H}_{\text{KS}} \ket{\varphi_{i,l',\beta,s'} \left( \vec{R} + \vtau_{j,l',\beta} \right) }, \label{app:eqn:OpenMX_H} \\
	S_{i,l,\alpha,s;j,l',\beta,s'} \left( \vec{k} \right) &= \sum_{\vec{R}} e^{i \vec{k} \cdot \left( -\vtau_{i,l,\alpha} + \vec{R} + \vtau_{j,l',\beta} \right)} \bra{\varphi_{i,l,\alpha,s} \left( \vtau_{i,l,\alpha} \right) } \ket{\varphi_{i,l',\beta,s'} \left( \vec{R} + \vtau_{j,l',\beta} \right) }, \label{app:eqn:OpenMX_S}
\end{align}
where $\vec{k}$ denotes the moir\'e crystalline momentum and $\hat{H}_{\text{KS}}$ is the Kohn-Sham Hamiltonian. The \textit{ab initio} spectrum of $\hat{H}_{\text{KS}}$ can be found from $H_{i,l,\alpha,s;j,l',\beta,s'} \left( \vec{k} \right)$ and $S_{i,l,\alpha,s;j,l',\beta,s'} \left( \vec{k} \right)$ by solving a generalized eigenvalue problem.

\subsubsection{The Wannier projection step} \label{app:sec:first_princ_ham_valley:algorithm:wannier}

In the first step, which we denote as \emph{the Wannier projection step}, an appropriate low-energy eigensubspace and a set of trial Wannier orbital states must be selected. For the case of twisted bilayer \ch{SnSe2}, the spinful $s$ orbitals of Sn provide a suitable set of trial states. Note that while both Sn $s$ and Se $p$ orbitals contribute to the M valley, using only the Sn $s$ orbital is sufficient to reproduce the bands when applying the L\"{o}wdin orthogonalization method to the corresponding trial states. The low-energy projection eigensubspace is defined by the lowest set of gapped moir\'e conduction bands originating from the bottom gapped (spinful) conduction band of the monolayer. This results in a total of $N_{\text{layer}} \times N_{a} \times N_{\text{spin}} \times N_{\text{trial}} =4 N_{a}$ trial states, where $N_{a}$ is the number of monolayer unit cells within the moir\'e unit cell, $N_{\text{spin}} = 2$ is the number of spin flavors and $N_{\text{trial}} = 1$ is the number of trial orbitals within each monolayer unit cell. For \ch{ZrS2}, we choose the $d_{z^2}$, $d_{x^2-y^2}$, and $d_{xy}$ orbitals of Zr as the trial Wannier orbitals, resulting in $N_{\text{layer}} \times N_{a} \times N_{\text{spin}} \times N_{\text{trial}} =12 N_{a}$ trial states (where $N_{\text{trial}} = 3$). The low-energy projection eigensubspace in this case consists of the $12 N_{a}$ gapped moir\'e bands, originating from the three lowest gapped (spinful) conduction bands of the monolayer. 

In the L\"{o}wdin orthogonalization method, the number of states in the eigensubspace $\mathcal{A}$ must match the number of trial states. For \ch{SnSe2}, the first gapped moir\'e conduction bands arise from the monolayer's lowest-energy spinful gapped band (see \cref{app:fig:monolayer-bands}), so we include one atomic orbital trial state per monolayer unit cell. In contrast, for \ch{ZrS2}, these bands come from a set of three spinful gapped bands with significant overlap with the $d_{z^2}$, $d_{x^2-y^2}$, and $d_{xy}$ orbitals of Zr (see \cref{app:fig:ZrS2-monolayer-bands}). As such, we include three atomic orbital trial state per monolayer unit cell in \ch{ZrS2}, corresponding to the $d_{z^2}$, $d_{x^2-y^2}$, and $d_{xy}$ orbitals of Zr.

Denoting the trial states by $\ket{\phi_{i,l,\alpha,s} \left( \vec{R} + \vtau_{i,l,\alpha} \right)}$, where $1 \leq \alpha \leq N_{\text{trial}}$, we can find the refined trial states $\ket{\tilde{\phi}_{i,l,\alpha,s} \left( \vec{R} + \vtau_{i,l,\alpha} \right)}$ through the L\"{o}wdin orthogonalization procedure reviewed in \cref{app:sec:first_princ_ham_valley:lowdin}. The refined trial states are adiabatically connected to the original trial atomic orbitals. As such, they can be indexed with the same quantum numbers, with $\alpha = 1$ in the case of \ch{SnSe2} (since we include only one spinful $s$ orbital per monolayer unit cell in the trial state basis) and $\alpha = 1, 2, 3$ for \ch{ZrS2}, which, respectively, correspond to the $d_{z^2}$, $d_{x^2-y^2}$, and $d_{xy}$ orbitals of Zr. 

The Wannier projected Kohn-Sham Hamiltonian is given by
\begin{equation}
	\label{app:eqn:wannier_projected_hamiltonian}
	\tilde{H}_{i,l,\alpha,s;j,l',\beta,s'} \left( \vec{k} \right) = \bra{\tilde{\phi}_{i,l,\alpha,s} \left( \vec{k} \right)} \hat{H}_{\text{KS}} \ket{\tilde{\phi}_{j,l',\beta,s'} \left( \vec{k} \right)},
\end{equation}
where the Fourier-transformed refined trial states are given by
\begin{equation}
	\ket{\tilde{\phi}_{i,l,\alpha,s} \left( \vec{k} \right)} = \frac{1}{\sqrt{N}} \sum_{\vec{R}} e^{i \vec{k} \cdot \left( \vec{R} + \vtau_{i,l,\alpha} \right)} \ket{\tilde{\phi}_{i,l,\alpha,s} \left( \vec{R} + \vtau_{i,l,\alpha} \right)},
\end{equation}
with $N$ denoting the number of moir\'e unit cells. Because the number of trial states is significantly smaller than the number of OpenMX basis states, the Wannier projected Hamiltonian matrix $\tilde{H}_{i,l,\alpha,s;j,l',\beta,s'} \left( \vec{k} \right)$ is much easier to manipulate that the original OpenMX matrices from \cref{app:eqn:OpenMX_H,app:eqn:OpenMX_S}. Additionally, the refined trial states are orthonormal, meaning that the spectrum of $\tilde{H}_{i,l,\alpha,s;j,l',\beta,s'} \left( \vec{k} \right)$ can be found through a conventional eigenvalue problem.

\subsubsection{The valley projection step} \label{app:sec:first_princ_ham_valley:algorithm:valley}

After constructing the low-energy Wannier-projected Hamiltonian, we proceed with the valley projection. This step, while similar to the TAPW method~\cite{MIA23}, incorporates an additional L\"{o}wdin orthogonalization on the trial states, ensuring that the resulting projected Hamiltonian \emph{exactly} matches the low-energy \textit{ab initio} spectrum. The trial states used in the valley projection are written as atomic plane-wave combinations of the refined trial states from the Wannier projection step outlined in \cref{app:sec:first_princ_ham_valley:algorithm:wannier}.
\begin{align}
	\ket{\phi_{l,s} \left( \vec{k}, \vec{G}\right)} &= \frac{1}{\sqrt{N N_a}} \sum_{\vec{R}} \sum_{i} e^{i \left( \vec{k} + \vec{G} \right) \cdot \left( \vec{R} + \vtau_{i, l, \alpha, s}\right)} \ket{\tilde{\phi}_{i, l, \alpha, s} \left( \vec{R} + \vtau_{i,l,\alpha} \right)} \nonumber \\
	&= \frac{1}{\sqrt{N_a}} \sum_{i} e^{i \vec{G} \cdot \vtau_{i, l, \alpha, s} } \ket{\tilde{\phi}_{i, l, \alpha, s} \left( \vec{k} \right)}, \qq{with} \alpha = 1. \label{app:eqn:trial_valley_projection}
\end{align}
In \cref{app:eqn:trial_valley_projection}, $N_a$ is the number of monolayer unit cells in one moir\'e unit cell, $\vec{k}$ denotes the moir\'e Bloch momentum assuming values in the first moir\'e BZ, and $\vec{G}$ are the reciprocal moir\'e lattice vectors $\vec{G} \in \mathcal{Q}$. The moir\'e reciprocal vectors $\vec{G}$ characterizing the trial states are selected such that $\vec{k} + \vec{G}$ is close to one of the three M valleys. This gives a total of $N_{\text{layer}} \times N_{\text{spin}} \times N_a \times N_{\vec{G}}$ trial states used in the valley projection step, where $N_{\vec{G}}$ is the number of plane-wave states considered per valley. For each $\vec{G}$, layer $l$, and spin $s$, we select a single orbital for both \ch{SnSe2} and \ch{ZrS2}, corresponding to the effective $s$ or $d_{z^2}$ Wannier orbitals, respectively. For \ch{ZrS2}, despite the first Wannier projection step considering three $d$ orbitals of Zr, only the $d_{z^2}$ orbital (having maximal overlap with the lowest-energy monolayer states near the M point) is considered during the valley projection. After L\"{o}wdin orthogonalization, the resulting refined trial states are linear combinations of all Zr $d$ orbitals.

\newcommand{\tDFT}{\text{DFT}}

Armed with the trial states from \cref{app:eqn:trial_valley_projection}, we obtain the corresponding refined trial states $\ket{\tilde{\phi}_{l,s} \left( \vec{k}, \vec{G}\right)}$ through the L\"{o}wdin orthogonalization method reviewed in \cref{app:sec:first_princ_ham_valley:lowdin}. The low-energy projection eigensubspace used for the L\"{o}wdin orthogonalization consists of the lowest-energy states of the Wannier-projected Kohn-Sham Hamiltonian defined in \cref{app:eqn:wannier_projected_hamiltonian}. The number of states in this subspace matches the number of trial states used in the valley projection step. The moir\'e Hamiltonian expressed in the plane-wave basis from \cref{app:eqn:single_particle_hamiltonian} is then obtained by projecting the Kohn-Sham Hamiltonian into the refined trial states $\ket{\tilde{\phi}_{l,s} \left( \vec{k}, \vec{G}\right)}$ 
\begin{equation}
	\label{app:eqn:dft_to_moire}
	\left[ h^{\tDFT}_{\vec{Q}, \vec{Q}'} \left( \vec{k} \right) \right]_{s_1 l_1; s_2 l_2} = \bra{\tilde{\phi}_{l_1,s_1} \left( \bar{\vec{k}}, \vec{G} \right)} \hat{H}_{\text{KS}} \ket{\tilde{\phi}_{l_2,s_2} \left( \bar{\vec{k}}, \vec{G}' \right)}, 
\end{equation}
where 
\begin{align}
	C^{\eta}_{3z} \vec{K}_{M}^{l_1} + \vec{k} - \vec{Q} &= \bar{\vec{k}} + \vec{G}, \qq{with} \vec{Q} \in \mathcal{Q}_{\eta + l_1} \qq{and} \vec{G} \in \mathcal{Q}, \label{app:eqn:correspondence_between_DFT_and_continuum_1}\\
	C^{\eta}_{3z} \vec{K}_{M}^{l_2} + \vec{k} - \vec{Q}' &= \bar{\vec{k}} + \vec{G}', \qq{with} \vec{Q}' \in \mathcal{Q}_{\eta + l_2} \qq{and} \vec{G}' \in \mathcal{Q}, \label{app:eqn:correspondence_between_DFT_and_continuum_2}
\end{align}
for $0 \leq \eta \leq 2 $.

\subsubsection{Alternation in the mapping of M points between the monolayer and moir\'e BZs}\label{app:sec:first_princ_ham_valley:algorithm:alternation}

\begin{figure}[!t]
	\centering
	\includegraphics[width=0.4\textwidth]{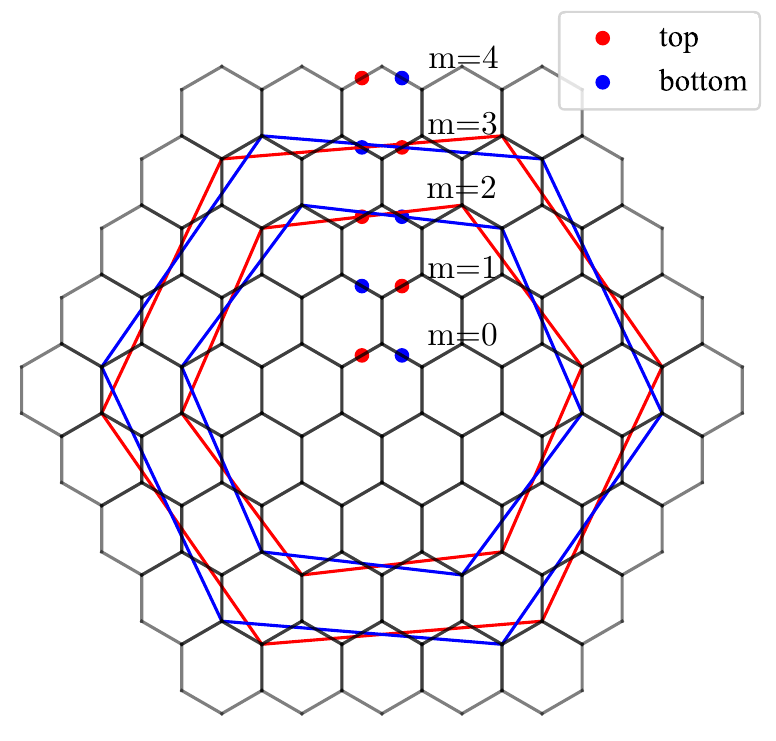}\caption{Mapping between the M valleys of the monolayer and moir\'e BZs. The red (blue) hexagons represent the top (bottom) monolayer BZs for twist indices $m=2$ (larger hexagons, with $\theta \approx \SI{13.17}{\degree}$) and $m=3$ (smaller hexagons, with $\theta \approx \SI{9.43}{\degree}$). The twist index $m$ is defined in \cref{app:eq:twist_index_m}. 
    Black hexagons denote the moir\'e BZs. The red (blue) dots correspond to the momentum sublattices $\mathcal{Q}_{+1}$ ($\mathcal{Q}_{-1}$), as defined in the continuum model and shown in  \cref{app:fig:M_valley_MBZ}.
    For twist angles with even indices $m$, the M point of layer $l$ belongs to the momentum sublattice $\mathcal{Q}_{l}$, while for odd twist angles, it is mapped to $\mathcal{Q}_{-l}$. This behavior contrasts with that of the continuum model, where the M point of layer $l$ is consistently mapped to sublattice $\mathcal{Q}_{l}$. Note that for better visualization in the plot, we maintain a fixed size for the moir\'e BZs while adjusting the size of the monolayer BZs. This approach is contrary to the actual scenario, where the monolayer BZs remain constant and the moir\'e BZs vary with different twist angles. }
	\label{app:fig:alternation_M_valley}
\end{figure}

According to \cref{app:eqn:dft_to_moire,app:eqn:wannier_projected_hamiltonian}, the low-energy spectrum of $\tilde{H}_{i,l,\alpha,s;j,l',\beta,s'} \left( \bar{\vec{k}} \right)$ should match the spectrum of $\left[ h^{\tDFT}_{\vec{Q}, \vec{Q}'} \left( \vec{k} \right) \right]_{s_1 l_1; s_2 l_2}$, provided that $\vec{k}$ and $\bar{\vec{k}}$ are related by \cref{app:eqn:correspondence_between_DFT_and_continuum_1}. We now examine this correspondence in detail for the commensurate twist angles considered in this study.

\Cref{app:fig:alternation_M_valley} illustrates the relationship between the moir\'e and twisted monolayer BZs. In our \textit{ab initio} simulations, we employ the following commensurate twist angles, as described in \cite{LOP07}:
\begin{equation}
	\label{app:eq:twist_index_m}
	\theta_m = \arccos \left( \frac{3m^2 + 3m + 1/2}{3m^2 + 3m + 1} \right), \quad m \in \mathbb{N},
\end{equation}
which are indexed by the \emph{twist index} $m$. Considering \cref{app:eqn:correspondence_between_DFT_and_continuum_1,app:eqn:correspondence_between_DFT_and_continuum_2} at $\theta = \theta_m$, we have
\begin{align}
	\begin{pmatrix}
		- \frac{l}{2} \\
		\frac{\sqrt{3}}{2} \left( 2 m + 1 \right)
	\end{pmatrix} \abs{\vec{q}_0} + C^{-\eta}_{3z}\vec{k} - \vec{q}_{l} = C^{-\eta}_{3z} \bar{\vec{k}} + \vec{G}, \qq{for} \vec{G} \in \mathcal{Q}, \nonumber \\
	-\frac{m}{2} \vec{b}_{M_1} + \left( m + \frac{1-l}{2} \right) \vec{b}_{M_2} = C^{-\eta}_{3z} \left( \bar{\vec{k}} - \vec{k} \right) + \vec{G}, \qq{for} \vec{G} \in \mathcal{Q}.
\end{align}
This demonstrates that
\begin{equation}
	\vec{k} = \begin{cases}
		\bar{\vec{k}}, \qq{for even} m \\
		\bar{\vec{k}} + \vec{q}_{\eta}, \qq{for odd} m 
	\end{cases}.
\end{equation}

In other words, for even commensurate twist angles, as defined by \cref{app:eq:twist_index_m}, the low-energy spectra of the Wannier-projected Hamiltonian from \cref{app:eqn:wannier_projected_hamiltonian} and the moir\'e Hamiltonian $h^{\tDFT}_{\vec{Q}, \vec{Q}'} \left( \vec{k} \right)$ are identical at the same momentum $\vec{k}$. For odd commensurate angles, however, the spectrum of the continuum Hamiltonian $h^{\tDFT}_{\vec{Q}, \vec{Q}'} \left( \vec{k} \right)$ in valley $\eta$ is shifted by $\vec{q}_{\eta}$ relative to the low-energy spectrum of the Wannier-projected Hamiltonian from \cref{app:eqn:wannier_projected_hamiltonian}. 

This relative shift can be understood visually by considering \cref{app:fig:M_valley_MBZ,app:fig:alternation_M_valley}. In the continuum model illustrated in \cref{app:fig:M_valley_MBZ}, the M point of layer $l$, located at $\vec{K}^{l}_{M}$, is consistently mapped to the moir\'e Bloch wave vector $\vec{k} = l \vec{q}_l$. However, for the commensurate angles defined by \cref{app:eq:twist_index_m}, this mapping to $\vec{k} = l \vec{q}_l$ holds only for even $m$. For odd $m$, the M point of layer $l$ is instead mapped to $\vec{k} = l \vec{q}_{-l}$. 

When plotting the \textit{ab initio} spectra, we always align $\bar{\vec{k}}$ along the same high-symmetry line. The spectra are then compared with those of $h_{\vec{Q}, \vec{Q}'} \left( \bar{\vec{k}} \right)$ for even commensurate angles, and with the spectra of $h_{\vec{Q}, \vec{Q}'} \left( \bar{\vec{k}} + \vec{q}_{\eta} \right)$ (where $\vec{Q}, \vec{Q}' \not\in \mathcal{Q}_{\eta}$) for odd commensurate angles.

\begin{figure}[t]
	\centering
	\includegraphics[width=\textwidth]{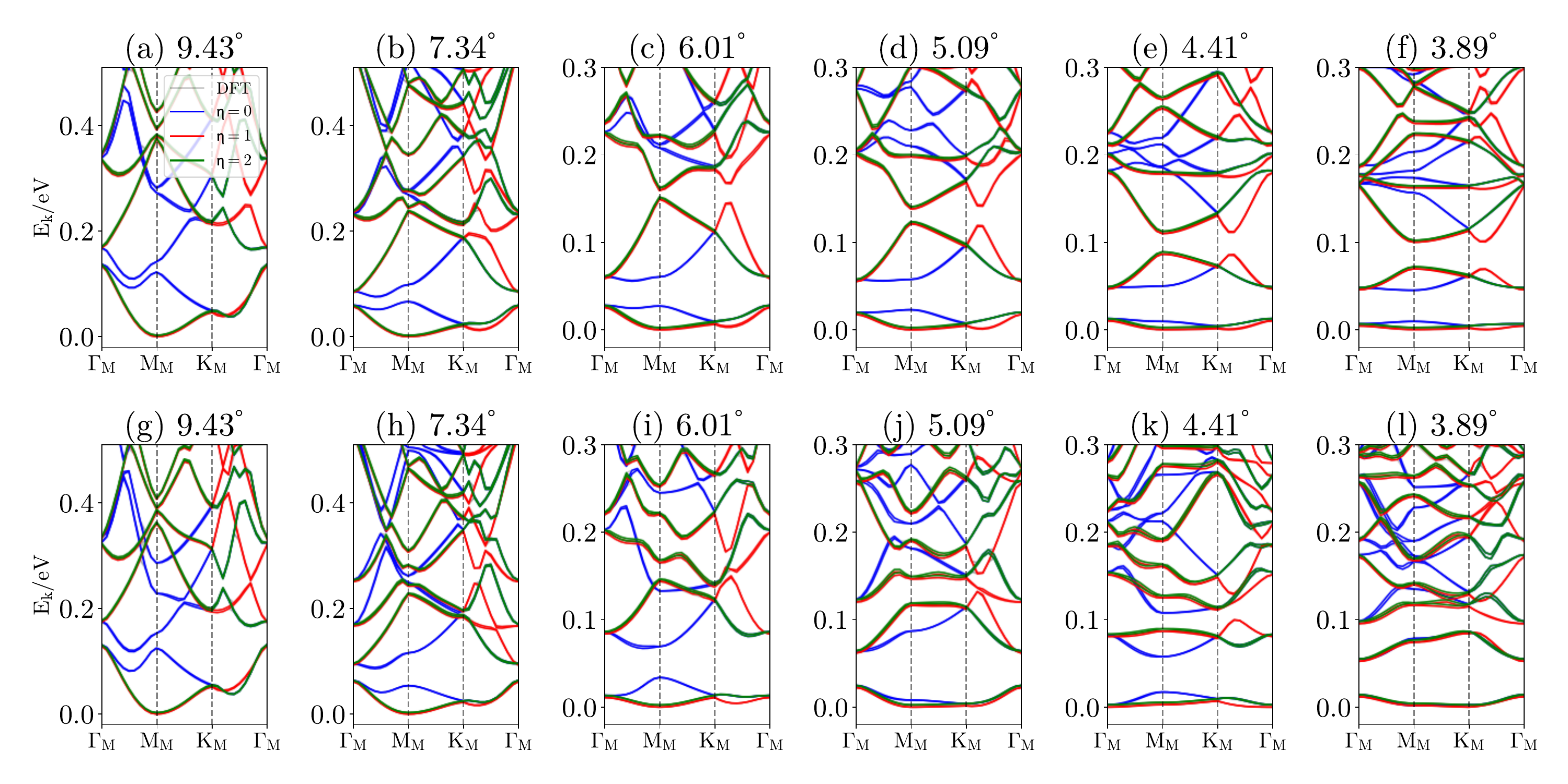}
	\caption{The valley-projected band structure of twisted bilayer \ch{SnSe2} from $\SI{9.43}{\degree}$ to $\SI{3.89}{\degree}$. (a)-(f) correspond to AA-stacking, while (g)-(l) correspond to AB-stacking. The plot shows only the bottom conduction bands, with bands from the three M valleys labeled by different colors. The high-symmetry points used in the plot are defined in \cref{app:fig:M_valley_MBZ:c}. 
}
	\label{app:fig:twist-AA-AB-valley-proj-bands-SnSe2}
\end{figure}

\subsection{Valley-projected band structure and charge density of \ch{SnSe2}}\label{app:sec:first_princ_ham_valley:SnSe2_results}

\begin{figure}[tbp]
	\centering
	\includegraphics[width=\textwidth]{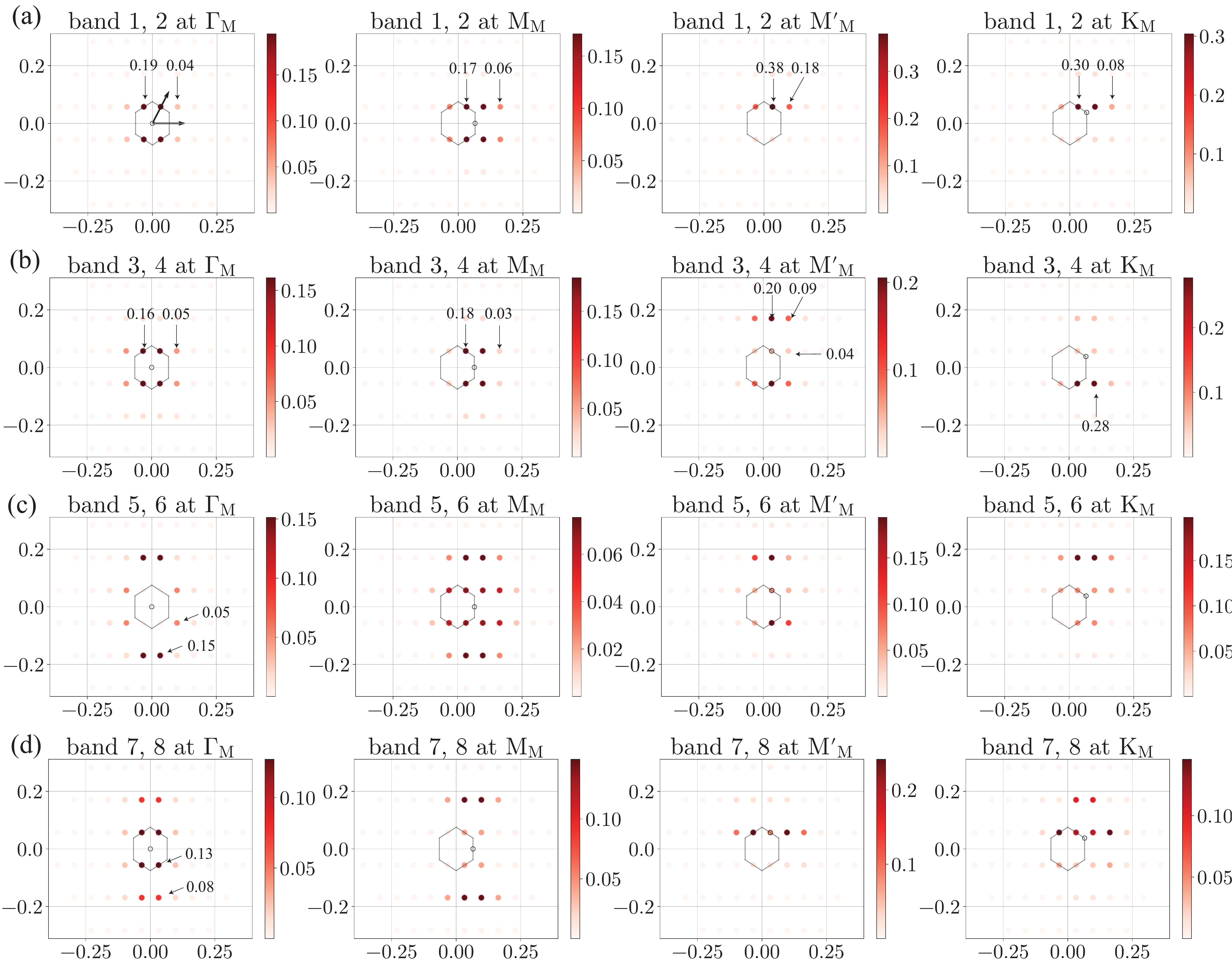}
	\caption{The \textit{ab initio} charge distribution on the $\mathbf{Q}$ lattice of the twisted AA-stacked bilayer \ch{SnSe2} at $\SI{3.89}{\degree}$, where the charge of the lowest four sets of conduction bands at $\Gamma_M$, $\mathrm{M}_M=\frac{1}{2}\mathbf{b}_{M_1}$, $\mathrm{M}_{M}^\prime=\frac{1}{2}\mathbf{b}_{M_2}$, and $\mathrm{K}_M=\frac{1}{3}(\mathbf{b}_{M_1}+\mathbf{b}_{M_2})$ of the $\eta=0$ valley are shown, where the high-symmetry points are marked by black circles in the plot. Each set contains two bands from two almost degenerate spins. The total charge has been normalized to 1 in each plot. The charge at each $\mathbf{Q}$ site is the square of the corresponding wavefunction components. Note that the charge distribution is localized at the $\mathbf{Q}$ sites close to the momentum of the wavefunction for the lowest four bands, as expected. 
	}
\label{app:fig:charge_dist_3.89_M0}
\end{figure}

\begin{figure}[tbp]
	\centering
	\includegraphics[width=\textwidth]{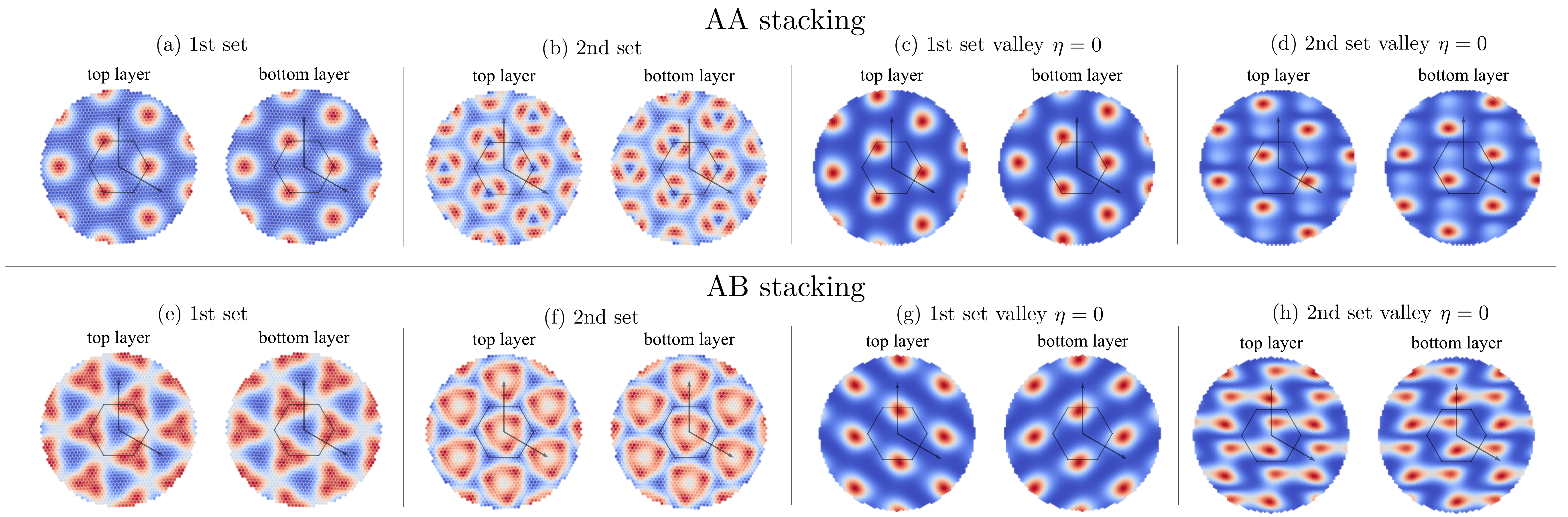}
	\caption{The \textit{ab initio} local density of states (LDOS) of the twisted both AA- and AB-stacked bilayer \ch{SnSe2} at $\SI{3.89}{\degree}$. (a)(b) are the total LDOS from three M valleys for the two lowest sets of conduction bands for AA-stacking, with the 1st set being the lowest set, while (c)(d) are the valley-projected LDOS of the valley $\eta=0$. 
		(e)-(h) are similar but for the AB-stacking. 
		In the plot, red (blue) regions have large (small) LDOS, the two black arrows denote the moir\'e unit cell bases, and the hexagon is the Wigner–Seitz cell.
	}
	\label{app:fig:LDOS_3.89_SnSe2_AA_AB}
\end{figure}

In \cref{app:fig:twist-AA-AB-valley-proj-bands-SnSe2}, we show the M valley-projected band structures for \ch{SnSe2} in the AA- and AB-stacking configurations. As the twist angle decreases from \SI{9.43}{\degree} to \SI{3.89}{\degree}, two sets of bottom conduction bands become gapped from the higher conduction bands. Each set of bands contains six bands stemming from the three M valleys, with each M valley contributing two nearly degenerate bands due to the approximate $\mathrm{SU} \left( {2} \right)$ symmetry of the model, as discussed in \cref{app:sec:DFT_single_layer:snse2:symmetries}. 

We observe that within each isolated set, the six bands split into two subsets of two and four bands, which can be understood from symmetry considerations. At small twist angles, the different M valleys are decoupled, and the symmetries of the moir\'e heterostructure relate the different valleys along different high-symmetry paths. We first analyze the AA-stacking case:
\begin{itemize}
	\item Along the $\Gamma_M-\mathrm{M}_M$ line (where $\mathrm{M}_M=\frac{1}{2}\mathbf{b}_{M_1}$), the four nearly degenerate bands stem from valleys $\eta=1, 2$. 
	The $\Gamma_M-\mathrm{M}_M$ line has $C_{2x}$ symmetry, which exchanges the valleys $\eta=1$ and $\eta = 2$, but maintains the valley $\eta=0$, leading to the degeneracy ({\it i.e.}, 4-fold and 2-fold) of bands from valleys $\eta=1,2$.
	
	\item Along the $\mathrm{M}_M-\mathrm{K}_M$ path (where $\mathrm{K}_M=\frac{1}{3}(\mathbf{b}_{M_1} +\mathbf{b}_{M_2})$), the four nearly degenerated bands also stem from the valleys $\eta=1, 2$. This is because this high-symmetry line has $C_{2x}\mathcal{T}$ symmetry, which flips the $\eta=1$ and $\eta = 2$ valleys but maintains the valley $\eta=0$.
	
	\item Along the $\mathrm{K}_M-\Gamma_M$ line, the bands of valleys $\eta=0$ and $\eta = 2$ are nearly degenerate. The $\mathrm{K}_M-\Gamma_M$ line has $C_{2, 1\bar{1}0}\mathcal{T}$ symmetry ({\it i.e.}, a twofold rotation along the $\vec{a}_{M_1}-\vec{a}_{M_2}$ direction followed by the TRS), which exchanges valleys $\eta=0$ and $\eta=2$, while keeping the $\eta=1$ invariant. 
\end{itemize}

In the AB-stacking case, the degeneracy between bands stemming from different valleys is the same but is protected by different symmetries. The $\Gamma_M-\mathrm{M}_M$, $\mathrm{M}_M-\mathrm{K}_M$, and $\mathrm{K}_M-\Gamma_M$ lines are invariant under the $C_{2y}\mathcal{T}$, $C_{2y}$, and $C_{2, 110}$ symmetries, respectively. These symmetries relate the three M valleys in the same way that the $C_{2x}$, $C_{2x}\mathcal{T}$, and $C_{2, 1\bar{1}0}\mathcal{T}$ symmetries do in the AA-stacking case. This results in the same band degeneracy in the AB-stacking case as in the AA-stacking case.

Next, we consider the charge distribution of the twisted structure. We take the $\theta = \SI{3.89}{\degree}$ in the AA-stacking case as an example. In \cref{app:fig:charge_dist_3.89_M0}, we show the charge distribution on the $\mathbf{Q}$ lattice for the lowest four sets of conduction bands of the valley $\eta=0$. Additionally, in \cref{app:fig:LDOS_3.89_SnSe2_AA_AB}~(a)(c), we show the real-space LDOS of \ch{SnSe2} at $\theta = \SI{3.89}{\degree}$. It can be seen that the lowest set of conduction bands ({\it i.e.}, \cref{app:fig:LDOS_3.89_SnSe2_AA_AB}~(a)) have their charge density located at one honeycomb site $\frac{2}{3}\vec{a}_{M_1} + \frac{1}{3}\vec{a}_{M_2}$, forming a triangular lattice. The LDOS of the second lowest set of bands ({\it i.e.}, \cref{app:fig:LDOS_3.89_SnSe2_AA_AB}~(b)) shows three $C_{3z}$-symmetric peaks surrounding the honeycomb site, resembling a distorted kagome lattice. They are not exactly at kagome sites as there is not enough symmetry. 
In \cref{app:fig:LDOS_3.89_SnSe2_AA_AB}(b)(d), we show the $\eta=0$ valley-projected LDOS. For the first set of bands, the charge is distributed close to the honeycomb site. For the second set, the charge of the valley $\eta=0$ is located near $\frac{1}{2}\vec{a}_{M_1}$, {\it i.e.}{}, one of the three kagome sites.

In \cref{app:fig:LDOS_3.89_SnSe2_AA_AB}, we present the local density of states (LDOS) for AB-stacked SnSe2 at a twist angle of $\SI{3.89}{\degree}$. It is important to note that in the AA-stacking configuration, the two layers are related by the symmetry operation $C_{2x}$, which leaves one honeycomb site invariant. As a result, the charge from both layers accumulates on the same honeycomb site. In contrast, in the AB-stacking configuration, the layers are related by $C_{2y}$, which exchanges the two honeycomb sites. Consequently, the charge from the two layers is distributed across different honeycomb sites.

\subsection{Valley-projected band structure and charge density of \ch{ZrS2}}\label{app:sec:first_princ_ham_valley:ZrS2_results}

\begin{figure}[htbp]
	\centering
	\includegraphics[width=\textwidth]{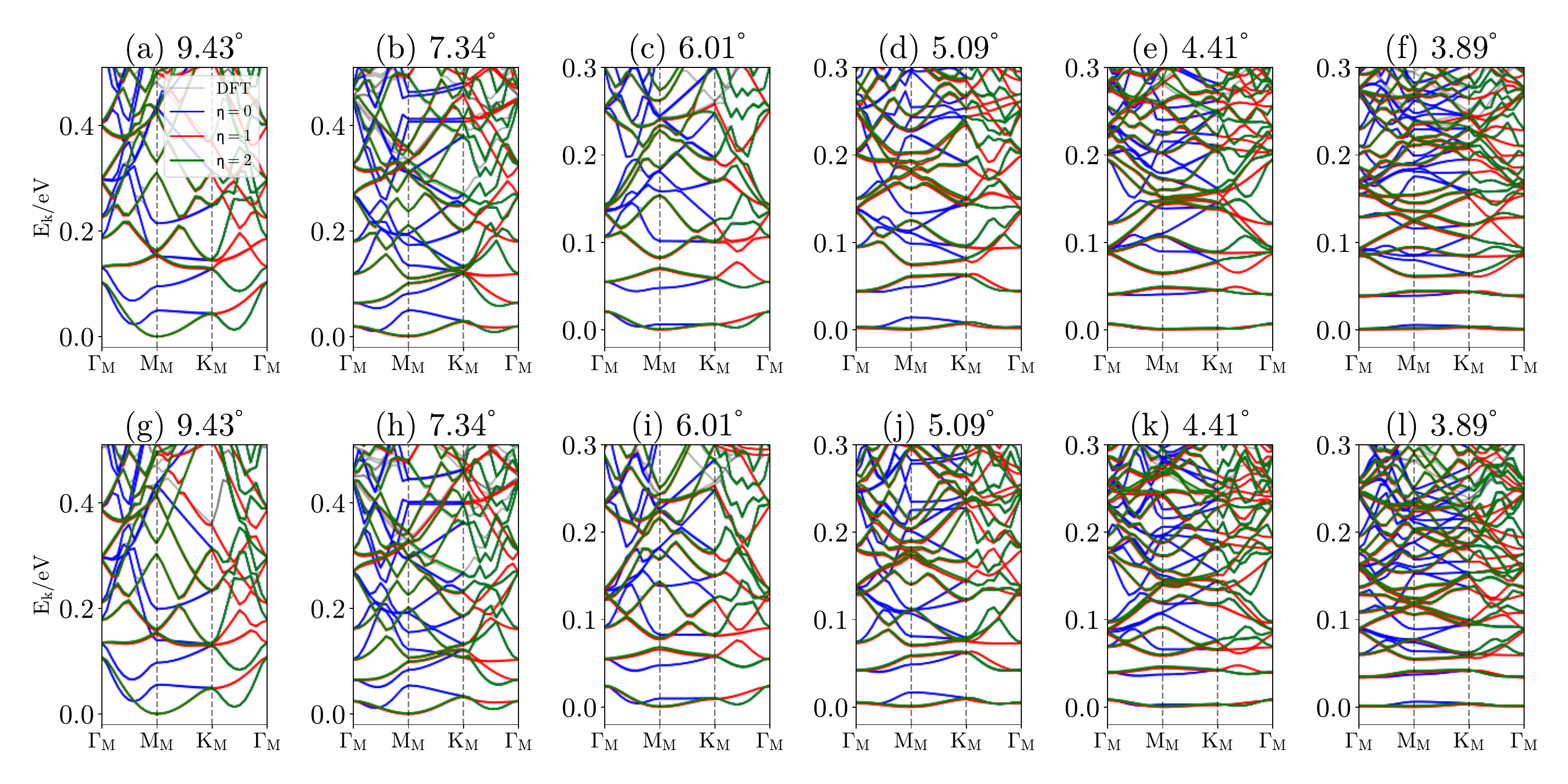}
	\caption{The (three) valley-projected band structure of twisted bilayer \ch{ZrS2} from $\SI{9.43}{\degree}$ to $\SI{3.89}{\degree}$, where (a)-(f) is for AA-stacking, while (g)-(l) is for AB-stacking. The plot shows only the bottom conduction bands, with bands from three M valleys labeled by different colors. The high-symmetry points used in the plot have the coordinates $\Gamma=(0,0), \mathrm{M}=\frac{1}{2}\mathbf{b}_{M_1}, \mathrm{K}=\frac{1}{3}(\mathbf{b}_{M_1} +\mathbf{b}_{M_2})$. 
	}
	\label{app:fig:twist-AA-AB-valley-proj-bands-ZrS2}
\end{figure}

In \cref{app:fig:twist-AA-AB-valley-proj-bands-ZrS2}, we show the (three) M valley-projected bands of \ch{ZrS2} for both AA- and AB-stacking, with the twist angle rang from 9.43\degree\ to $\SI{13.17}{\degree}$. Similar to \ch{SnSe2}, there exist two sets of isolated bottom conduction bands when the angle goes down to $\SI{3.89}{\degree}$. 

\begin{figure}[tbp]
	\centering
	\includegraphics[width=\textwidth]{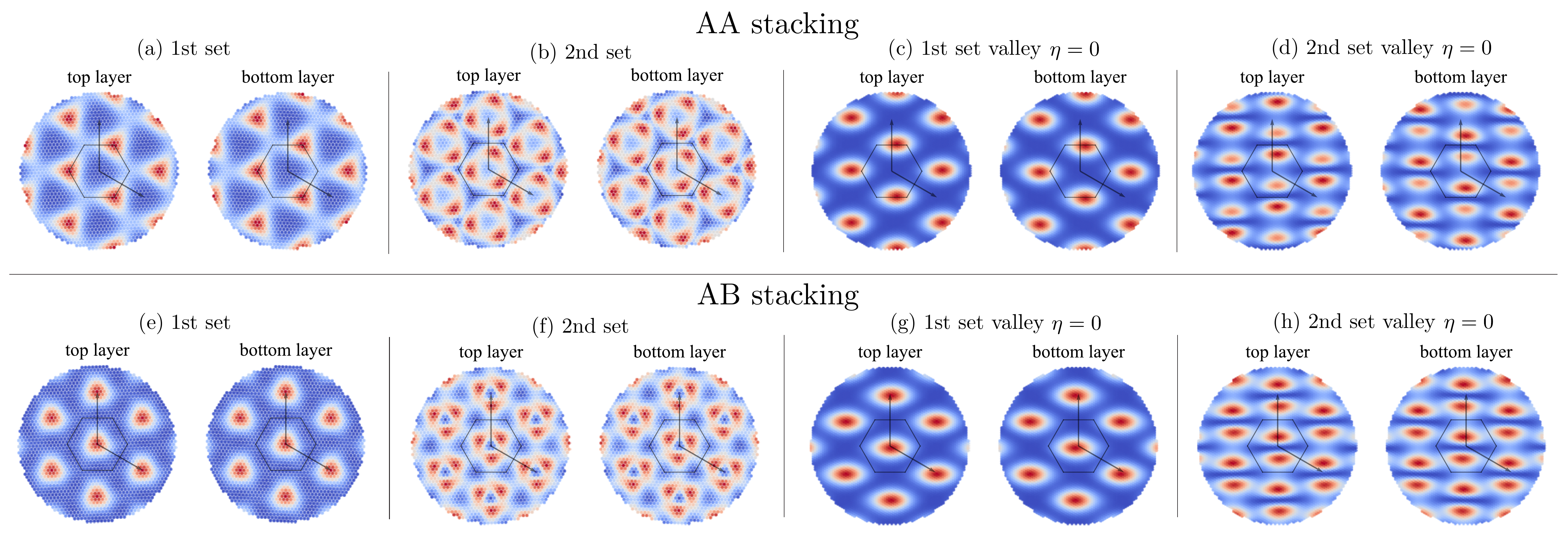}
	\caption{The \textit{ab initio} local density of states (LDOS) of the twisted both AA- and AB-stacked bilayer \ch{ZrS2} at $\SI{3.89}{\degree}$. (a)(b) are the total LDOS from three M valleys for the two lowest sets of conduction bands for AA-stacking, with the 1st set being the lowest set, while (c)(d) are the valley-projected LDOS of the valley $\eta=0$. 
		(e)-(h) are similar but for the AB-stacking. 
		In the plot, red (blue) regions have large (small) LDOS, the two black arrows denote the moir\'e unit cell bases, and the hexagon is the Wigner–Seitz cell.
	}
	\label{app:fig:LDOS_3.89_ZrS2_AA_AB}
\end{figure}

In \cref{app:fig:LDOS_3.89_ZrS2_AA_AB}(a)(c), we show the LDOS of AA-stacked \ch{ZrS2} at the $\SI{3.89}{\degree}$. The lowest set of conduction bands has charge density localized at one honeycomb site $\frac{1}{3}\vec{a}_{M_1} + \frac{2}{3}\vec{a}_{M_2}$, forming a triangular lattice. The LDOS of the second lowest set of bands shows three $C_{3z}$-symmetric peaks surrounding the same honeycomb site.  
In \cref{app:fig:LDOS_3.89_ZrS2_AA_AB}(b)(d), we show the $\eta=0$ valley-projected LDOS. The charge is distributed near one honeycomb site for the first set, while for the second set, the charge is split into two peaks near the honeycomb site. In the $\eta=0$ valley-projected LDOS, the charge distributions of two layers are related by $C_{2x}$.  

For the AB stacking shown in \cref{app:fig:LDOS_3.89_ZrS2_AA_AB}(e)-(h), the charge distribution is similar to the AA stacking case, but is centered at the triangular site $1a$. Moreover, the charge densities from two layers are related by $C_{2y}$ in the AB stacking.

\FloatBarrier
\section{Obtaining continuum model from \textit{ab initio} simulations}\label{app:sec:fitting_method}

The \textit{ab initio} moir\'e Hamiltonian from \cref{app:eqn:dft_to_moire} is computed only at a finite set of $\vec{k}$ points throughout the moir\'e BZ. However, it is often desirable to have an explicit functional form for $h_{\vec{Q}, \vec{Q}'} (\vec{k})$ that can be evaluated, either analytically or numerically, over arbitrary momentum-space $\vec{k}$-meshes. To achieve this, we can parameterize $h_{\vec{Q}, \vec{Q}'} (\vec{k})$ by enumerating all the symmetry-allowed terms (potentially including gradient terms) and then restricting them according to the \emph{exact} or \emph{approximate} symmetries of the model, as outlined in \cref{app:sec:SnS_SnSe_twist_general_woGradient,app:sec:SnS_SnSe_twist_general}. In this \siSection{}, we describe the method used to extract or \emph{fit} the numerical values for these symmetry-allowed parameters from the \textit{ab initio} moir\'e Hamiltonian $h^{\tDFT}_{\vec{Q}, \vec{Q}'} (\vec{k})$. The result is an explicit functional form of $h_{\vec{Q}, \vec{Q}'} (\vec{k})$ that can be evaluated at any arbitrary momentum in the moir\'e BZ.

Because the techniques described in this \siSection{} can be applied to \emph{any} Hamiltonian matrix, and to avoid tracking many indices, we begin by introducing a simplified generic notation. We then outline how the expansion coefficients of the analytical Hamiltonian matrix, expressed in a symmetry-allowed matrix basis, can be extracted from the \textit{ab initio} Hamiltonian. We present two methods: a linear least-squares method, similar to the one used in Refs.~\cite{HU23c,JIA23a} for conventional crystalline systems and in Ref.~\cite{ZHA24} for moir\'e models, and a nonlinear fitting method. The linear method, \emph{which does not require parameter fitting}, is effective for obtaining numerically accurate models that quantitatively reproduce most of the \textit{ab initio} spectrum, but potentially with a large number of parameters. In contrast, the nonlinear method is suited for deriving explicit Hamiltonians that quantitatively capture only the low-energy bands (with the higher energy bands reproduced qualitatively), using significantly fewer parameters.

\subsection{Notation}\label{app:sec:fitting_method:notation}
In what follows, we denote by $h^{\tDFT}_{ij} \left( \vec{k} \right)$ the \textit{ab-initio} Hamiltonian, which is evaluated at a subset of momenta in the BZ (denoted by $\mathcal{M}$). The indices $1 \leq i,j \leq \mathcal{N}$, where $\mathcal{N}$ is the dimension of $h^{\tDFT}_{ij} \left( \vec{k} \right)$, correspond to combined indices such as orbital, spin, plane wave, or other quantum numbers. Our goal is to find the \emph{analytical} explicit matrix function $h_{ij} \left( \vec{k} \right)$ that best matches $h^{\tDFT}_{ij} \left( \vec{k} \right)$ at $\vec{k} \in \mathcal{M}$ according to a cost function, which will be defined later. The matrix function $h_{ij} \left( \vec{k} \right)$ can be expanded in a predefined, symmetry-constrained basis
\begin{equation}
	\label{app:eqn:basis_expansion_fit}
	h_{ij} \left( \vec{k} \right) = \sum_{\alpha=1}^{d} \lambda_{\alpha} h_{\alpha;ij} \left( \vec{k} \right),
\end{equation}
where $d$ is the dimension of the basis and $h_{\alpha;ij} \left( \vec{k} \right)$ represent $d$ Hermitian matrix functions that obey the symmetry of the problem. The coefficients $\lambda_{\alpha}$ are the parameters of the Hamiltonian function $h_{ij} \left( \vec{k} \right)$ (which implicitly depends on them). For example, in the zero-twist limit, the AA-stacked moir\'e single-particle Hamiltonian can be expanded in the corresponding symmetrized matrix basis as
\begin{align}
	\left[ h_{\vec{Q},\vec{Q}'} \left( \vec{k} \right) \right]_{l_1 s_1; l_2 s_2} =& w^{\text{AA}}_{2} \sum_{n=0}^{2}  \left( \delta_{\vec{Q},\vec{Q}' + \vec{q}_n} \delta_{l_1 (-l_2)} \delta_{s_1 s_2} + \delta_{\vec{Q},\vec{Q}' - \vec{q}_n} \delta_{l_1 (-l_2)} \delta_{s_1 s_2} \right) \nonumber \\
	&+ \frac{1}{2 m_x} \delta_{l_1 l_2} \delta_{\vec{Q}, \vec{Q}'} \delta_{s_1 s_2} \left\lbrace \left( k_x - Q_x \right)^2 \delta_{\zeta_{\vec{Q}}-l_1,0} + \left[ \frac{ \left( k_x - Q_x \right) \sqrt{3} + \left( k_y - Q_y \right) }{2} \right]^2 \delta_{\zeta_{\vec{Q}}-l_1,2}  \right. \nonumber \\
	& \quad +\left. \left[ \frac{ \left( k_x - Q_x \right) \sqrt{3} - \left( k_y - Q_y \right) }{2} \right]^2 \delta_{\zeta_{\vec{Q}}-l_1,1} \right\rbrace + \dots, \label{app:eqn:ex_matrix_expansion}
\end{align}
with the dots denoting other terms. In \cref{app:eqn:ex_matrix_expansion}, each term comprises a parameter ({\it e.g.}{}, $w^{\text{AA}}_{2}$ and $\frac{1}{2 m_x}$) multiplying a Hermitian matrix. 

The goal of this \siSection{} is to fix the parameters of the expansion $\lambda_{\alpha}$ such that $h_{ij} \left( \vec{k} \right)$ best reproduces (according to a cost function defined below) the numerically obtained $h^{\tDFT}_{ij} \left( \vec{k} \right)$ for $\vec{k} \in \mathcal{M}$. For this purpose, it is also useful to define the eigenvalues $\epsilon_{n} \left( \vec{k} \right)$ and eigenstates $\phi_{n;j} \left( \vec{k} \right)$ of the $h^{\tDFT}_{ij} \left( \vec{k} \right)$ and $h_{ij} \left( \vec{k} \right)$ matrices according to
\begin{equation}
	\sum_{j} h^{\left( \tDFT \right)}_{ij} \left( \vec{k} \right) \phi^{\left( \tDFT \right)}_{n;j} \left( \vec{k} \right) = \epsilon^{\left( \tDFT \right)}_{n} \left( \vec{k} \right) \phi^{\left( \tDFT \right)}_{n;i} \left( \vec{k} \right),
\end{equation} 
where $n$ indexes the eigenpairs in the order of increasing eigenvalue. We will now define two types of cost functions and briefly describe the numerical methods used for their optimization.

\subsection{Linear Extraction}\label{app:sec:fitting_method:linear}

Perhaps the simplest cost function is the Frobenius norm of the matrix difference between the Hamiltonian matrix function and the \textit{ab initio} one
\begin{equation}
	\label{app:eqn:linear_cost_function_0}
	\mathcal{C}_{0} \left( \left\lbrace \lambda_{\alpha}  \right\rbrace \right) = \sum_{\vec{k} \in \mathcal{M}} \Tr\left[ \left( h \left(\vec{k} \right) - h^{\tDFT} \left(\vec{k} \right) \right)^2\right].
\end{equation}
In terms of the expansion coefficients from \cref{app:eqn:basis_expansion_fit}, \cref{app:eqn:linear_cost_function_0} takes a quadratic form
\begin{equation}
	\mathcal{C}_{0} \left( \left\lbrace \lambda_{\alpha}  \right\rbrace \right) = \sum_{\alpha,\beta=1}^{d} A^{(0)}_{\alpha \beta} \lambda_{\alpha} \lambda_{\beta} - 2 \sum_{\alpha=1}^{d} B^{(0)}_{\alpha} \lambda_{\alpha} + C^{(0)},
\end{equation}
where the real coefficients are given by
\begin{align}
	A^{(0)}_{\alpha \beta} &= \Re \left[ \sum_{\vec{k} \in \mathcal{M}} \Tr \left(h_{\alpha} \left( \vec{k} \right) h_{\beta} \left( \vec{k} \right) \right) \right], \\
	B^{(0)}_{\alpha} &= \sum_{\vec{k} \in \mathcal{M}} \Tr \left(h_{\alpha} \left( \vec{k} \right) h^{\tDFT} \left( \vec{k} \right) \right), \\
	C^{(0)} &= \sum_{\vec{k} \in \mathcal{M}} \Tr \left(h^{\tDFT} \left( \vec{k} \right) h^{\tDFT} \left( \vec{k} \right) \right).	
\end{align}
It is easy to check that the real symmetric matrix $A^{(0)}_{\alpha \beta}$ is also positive definite. Therefore, the cost function $\mathcal{C}_{0}$ admits a single minimum for 
\begin{equation}
	\lambda_{\alpha} = \left[ \left( A^{(0)} \right)^{-1} B^{(0)} \right]_{\alpha}.
\end{equation}
We refer to this method as ``linear extraction'', analogous to the linear least-squares method. In the form from \cref{app:eqn:linear_cost_function_0}, this method was employed by Ref.~\cite{ZHA24} to obtain moir\'e Hamiltonians from \textit{ab initio} simulations. 

\subsubsection{Low-energy restriction}\label{app:sec:fitting_method:linear:low_energy}
The cost function from \cref{app:eqn:linear_cost_function_0} has one shortcoming: it attempts to fit the \emph{entire} energy spectrum of $h^{\tDFT} \left(\vec{k} \right)$. This becomes clearer if we express \cref{app:eqn:linear_cost_function_0} in the eigenbasis of $h^{\tDFT} \left(\vec{k} \right)$
\begin{equation}
	\label{app:eqn:linear_cost_function_0_1}
	\mathcal{C}_{0} \left( \left\lbrace \lambda_{\alpha}  \right\rbrace \right) = \sum_{\vec{k} \in \mathcal{M}} \sum_{n} \phi^{* \tDFT}_{n;i} \left( \vec{k} \right) \left( h_{ij} \left(\vec{k} \right) - h_{ij}^{\tDFT} \left(\vec{k} \right) \right) \left( h_{jk} \left(\vec{k} \right) - h_{jk}^{\tDFT} \left(\vec{k} \right) \right) \phi^{\tDFT}_{n;k} \left( \vec{k} \right).
\end{equation}
In practice, however, we are primarily interested in the low-energy spectrum of $h^{\tDFT} \left(\vec{k} \right)$. To reduce the number of parameters in $h \left(\vec{k} \right)$ and focus on improving the match withthe low-energy spectrum, we can introduce an energy weighting function into the cost function from \cref{app:eqn:linear_cost_function_0_1}, defining a new cost function
\begin{equation}
	\label{app:eqn:linear_cost_function_1}
	\mathcal{C}_{1} \left( \left\lbrace \lambda_{\alpha}  \right\rbrace \right) = \sum_{\vec{k} \in \mathcal{M}} \sum_{n} w \left( \epsilon^{\tDFT}_n \left( \vec{k} \right) \right) \phi^{* \tDFT}_{n;i} \left( \vec{k} \right) \left( h_{ij} \left(\vec{k} \right) - h_{ij}^{\tDFT} \left(\vec{k} \right) \right) \left( h_{jk} \left(\vec{k} \right) - h_{jk}^{\tDFT} \left(\vec{k} \right) \right) \phi^{\tDFT}_{n;k} \left( \vec{k} \right),
\end{equation}
where $w \left( \epsilon \right)$ is the energy weighting function. In this work, we use a smooth step function, parameterized by an energy cutoff $\epsilon_c$ and a spread $\sigma$
\begin{equation}
	\label{app:eqn:fit_energy_weight}
	w \left( \epsilon \right) = \frac{1}{2} \left( 1 - \tanh\left(\frac{\epsilon - \epsilon_c}{\sigma} \right) \right),
\end{equation}
with the cutoff and spread chosen based on the band structure. The cost function from \cref{app:eqn:linear_cost_function_1} also takes a quadratic form in terms of the expansion coefficients from \cref{app:eqn:basis_expansion_fit}
\begin{equation}
	\mathcal{C}_{1} \left( \left\lbrace \lambda_{\alpha}  \right\rbrace \right) = \sum_{\alpha,\beta=1}^{d} A^{(1)}_{\alpha \beta} \lambda_{\alpha} \lambda_{\beta} - 2 \sum_{\alpha=1}^{d} B^{(1)}_{\alpha} \lambda_{\alpha} + C^{(1)},
\end{equation}
where the coefficients are given by
\begin{align}
	A^{(1)}_{\alpha \beta} &= \Re \left[ \sum_{\vec{k} \in \mathcal{M}} \sum_{n} w \left( \epsilon^{\tDFT}_n \left( \vec{k} \right) \right) \phi^{* \tDFT}_{n;i} \left( \vec{k} \right) h_{\alpha; ij} \left(\vec{k} \right) h_{\beta;jk} \left(\vec{k} \right) \phi^{\tDFT}_{n;k} \left( \vec{k} \right) \right], \\
	B^{(1)}_{\alpha} &= \Re \left[ \sum_{\vec{k} \in \mathcal{M}} \sum_{n} w \left( \epsilon^{\tDFT}_n \left( \vec{k} \right) \right) \phi^{* \tDFT}_{n;i} \left( \vec{k} \right) h_{\alpha; ij} \left(\vec{k} \right) h^{\tDFT}_{jk} \left(\vec{k} \right) \phi^{\tDFT}_{n;k} \left( \vec{k} \right) \right], \\
	C^{(1)} &=  \sum_{\vec{k} \in \mathcal{M}} \sum_{n} w \left( \epsilon^{\tDFT}_n \left( \vec{k} \right) \right) \phi^{* \tDFT}_{n;i} \left( \vec{k} \right) h^{\tDFT}_{ij} \left(\vec{k} \right) h^{\tDFT}_{jk} \left(\vec{k} \right) \phi^{\tDFT}_{n;k} \left( \vec{k} \right) .	
\end{align}
Since $A^{(1)}_{\alpha \beta}$ is also positive definite, the global minimum of \cref{app:eqn:linear_cost_function_1} can again be found analytically
\begin{equation}
	\lambda_{\alpha} = \left[ \left( A^{(1)} \right)^{-1} B^{(1)} \right]_{\alpha}.
\end{equation}
This method, though with a different cost function, was used by Refs.~\cite{HU23c,JIA23a} to fit \textit{ab initio} phonon and electronic spectra.

\subsubsection{Partial low-energy restriction}\label{app:sec:fitting_method:linear:low_energy_partial}

\newcommand{\tDiag}{\text{dg}}
\newcommand{\tNDiag}{\text{o-dg}}
The $\vec{Q}$-diagonal terms of the moir\'e Hamiltonian characterize the single-layer dispersion around the M point. When including $\vec{Q}$-diagonal terms of the form in \cref{app:eqn:general_moire_potential_momentum}, with a large polynomial degree corresponding to $n_x + n_y > 4$, a larger energy cutoff is required to reproduce them. Empirically, we find that a larger energy cutoff requires more moir\'e harmonics and higher-order gradients to obtain a good match with the low-energy bands. To address this, and to include $\vec{Q}$-diagonal terms with a high polynomial degree in $\vec{k}$ while keeping the number of moir\'e harmonics and their orders small, we separate the \textit{ab initio} Hamiltonian into $\vec{Q}$-diagonal and $\vec{Q}$-off-diagonal terms
\begin{equation}
	h^{\tDFT} \left( \vec{k} \right) = h^{\tDFT,\tDiag} \left( \vec{k} \right) + h^{\tDFT,\tNDiag} \left( \vec{k} \right). 
\end{equation}
Similarly, the matrix basis from \cref{app:eqn:basis_expansion_fit} can be split into $\vec{Q}$-diagonal and $\vec{Q}$-off-diagonal terms. Without loss of generality, we arrange the matrix basis so that $h_{\alpha} \left( \vec{k} \right)$ for $1\leq \alpha \leq d'$ ($d'< \alpha \leq d$) are all diagonal (off-diagonal) in $\vec{Q}$, where $1 < d' < d$ is the number of $\vec{Q}$-diagonal terms considered in the matrix expansion of $h \left( \vec{k} \right)$. The model Hamiltonian $h \left( \vec{k} \right)$ also splits into $\vec{Q}$-diagonal and $\vec{Q}$-off-diagonal terms
\begin{equation}
	h \left( \vec{k} \right) = h^{\tDiag} \left( \vec{k} \right) + h^{\tNDiag} \left( \vec{k} \right). 
\end{equation}
with
\begin{align}
	h^{\tDiag} \left( \vec{k} \right) &= \sum_{\alpha=1}^{d'} h_{\alpha} \left( \vec{k} \right), \\
	h^{\tNDiag} \left( \vec{k} \right) &= \sum_{\alpha=d'+1}^{d} h_{\alpha} \left( \vec{k} \right).  
\end{align}
The cost function \cref{app:eqn:linear_cost_function_0} can then be separated into $\vec{Q}$-diagonal and $\vec{Q}$-off-diagonal parts
\begin{equation}
	\label{app:eqn:linear_cost_function_0_separate}
	\mathcal{C}_{0} \left( \left\lbrace \lambda_{\alpha}  \right\rbrace \right) = \sum_{\vec{k} \in \mathcal{M}} \Tr\left[ \left( h^{\tDiag}  \left(\vec{k} \right) - h^{\tDFT,\tDiag} \left(\vec{k} \right) \right)^2\right] + \sum_{\vec{k} \in \mathcal{M}} \Tr\left[ \left( h^{\tNDiag}  \left(\vec{k} \right) - h^{\tDFT,\tNDiag} \left(\vec{k} \right) \right)^2\right].
\end{equation}
We can apply a similar strategy to that introduced in \cref{app:sec:fitting_method:linear:low_energy} and introduce an energy weighting function in the second term. This allows the moir\'e potential terms (which are $\vec{Q}$-off-diagonal) to be determined using only the low-energy sector, while the $\vec{Q}$-diagonal terms, characterizing the single-layer dispersion part of the moir\'e Hamiltonian, are determined using the entire \textit{ab initio} Hamiltonian. The corresponding cost function again takes a quadratic form in the fitting parameters
\begin{equation}
	\mathcal{C}_{2} \left( \left\lbrace \lambda_{\alpha}  \right\rbrace \right) = \sum_{\alpha,\beta=1}^{d} A^{(2)}_{\alpha \beta} \lambda_{\alpha} \lambda_{\beta} - 2 \sum_{\alpha=1}^{d} B^{(2)}_{\alpha} \lambda_{\alpha} + C^{(2)},
\end{equation}
where the coefficients are given by
\begin{align}
	A^{(2)}_{\alpha \beta} =& \begin{cases}
		\Re \left[ \sum_{\vec{k} \in \mathcal{M}} \Tr \left(h_{\alpha} \left( \vec{k} \right) h_{\beta} \left( \vec{k} \right) \right) \right] & 1 \leq \alpha, \beta \leq d' \\
		\Re \left[ \sum_{\vec{k} \in \mathcal{M}} \sum_{n} w \left( \epsilon^{\tDFT}_n \left( \vec{k} \right) \right) \phi^{* \tDFT}_{n;i} \left( \vec{k} \right) h_{\alpha; ij} \left(\vec{k} \right) h_{\beta;jk} \left(\vec{k} \right) \phi^{\tDFT}_{n;k} \left( \vec{k} \right) \right] & d'+1 \leq \alpha, \beta \leq d \\
		0 & \text{otherwise} \\
	\end{cases}, \\
	B^{(2)}_{\alpha} =& \begin{cases}
		\sum_{\vec{k} \in \mathcal{M}} \Tr \left(h_{\alpha} \left( \vec{k} \right) h^{\tDFT,\tDiag} \left( \vec{k} \right) \right) & 1 \leq \alpha \leq d' \\
		\Re \left[ \sum_{\vec{k} \in \mathcal{M}} \sum_{n} w \left( \epsilon^{\tDFT}_n \left( \vec{k} \right) \right) \phi^{* \tDFT}_{n;i} \left( \vec{k} \right) h_{\alpha; ij} \left(\vec{k} \right) h^{\tDFT}_{jk} \left(\vec{k} \right) \phi^{\tDFT,\tNDiag}_{n;k} \left( \vec{k} \right) \right]  & d'+1 \leq \alpha \leq d  \\
	\end{cases}, \\
	C^{(2)} =& \sum_{\vec{k} \in \mathcal{M}} \Tr \left(h^{\tDFT,\tDiag} \left( \vec{k} \right) h^{\tDFT,\tDiag} \left( \vec{k} \right) \right) \nonumber \\
	& + \sum_{\vec{k} \in \mathcal{M}} \sum_{n} w \left( \epsilon^{\tDFT}_n \left( \vec{k} \right) \right) \phi^{* \tDFT}_{n;i} \left( \vec{k} \right) h^{\tDFT,\tNDiag}_{ij} \left(\vec{k} \right) h^{\tDFT,\tNDiag}_{jk} \left(\vec{k} \right) \phi^{\tDFT}_{n;k} \left( \vec{k} \right).
\end{align}
Because the matrix $A^{(2)}$ is positive definite, the global minimum can be found analytically using 
\begin{equation}
	\lambda_{\alpha} = \left[ \left( A^{(2)} \right)^{-1} B^{(2)} \right]_{\alpha}.
\end{equation}

\subsection{Nonlinear fitting}\label{app:sec:fitting_method:nonlinear}
\newcommand{\tBand}{\text{Band}}
\newcommand{\one}{\mathbb{1}}

In the previous section, we outlined how the Hamiltonian matrix can be determined using a cost function based on the Frobenius norm of the difference between the model and \textit{ab initio} Hamiltonians. 
The linear extraction method in the previous sectoin does not require any parameter fitting because the global minimum can be found analytically.
Empirically, when the number of parameters is very small (as in the case of the first moir\'e harmonic model from \cref{app:sec:SnS_SnSe_twist_general_woGradient:full} or its enhanced-symmetry limits discussed in \cref{app:sec:add_sym}), we find that linear extraction over a large energy window yields unsatisfactory results for the low-energy bands. When the energy cutoff $\epsilon_c$ is reduced, the model Hamiltonian can exhibit ``ghost states'' -- additional bands appearing in the model Hamiltonian that are not present in the \textit{ab initio} one. This occurs because the cost function in \cref{app:eqn:linear_cost_function_1} ensures that $h \left( \vec{k} \right)$ reproduces the eigenstates of $h^{\tDFT} \left( \vec{k} \right)$ within the energy window, but does \emph{not} prevent the emergence of extra states.

To address this issue, we can design a new cost function that ensures the \emph{first} $\mathcal{N}_{\tBand}$ bands of $h \left( \vec{k} \right)$ match the \emph{first} $\mathcal{N}_{\tBand}$ bands of $h^{\tDFT} \left( \vec{k} \right)$, thereby preventing the appearance of ghost states within these bands. To achieve this, we introduce projector operators onto the first $\mathcal{N}_{\tBand}$ bands of the fitted and \textit{ab initio} Hamiltonians
\begin{equation}
	\label{app:eqn:definition_projector}
	P_{ij}^{\left( \tDFT \right)} \left( \vec{k} \right)= \sum_{m=1}^{\mathcal{N}_{\tBand}} n_m \phi^{\left( \tDFT \right)}_{m;i} \left( \vec{k} \right) \phi^{*\left( \tDFT \right)}_{m;j} \left( \vec{k} \right) = \left[ \frac{1}{2 \pi i} \oint \dd{z} \left( z \one - h^{\left( \tDFT \right)} \left( \vec{k} \right) \right)^{-1} \right]_{ij},
\end{equation}
where
\begin{equation}
	n_{m} = \begin{cases}
		1 & 1 \leq m \leq \mathcal{N}_{\tBand}	\\
		0 & \mathcal{N}_{\tBand} < m \leq \mathcal{N}
	\end{cases}.
\end{equation}

The complex contour in \cref{app:eqn:definition_projector} is taken to enclose the first $\mathcal{N}_{\tBand}$ bands of the corresponding ({\it i.e.}{}, fitted or \textit{ab initio}) Hamiltonian. It is also important to note that the projector $P \left( \vec{k} \right)$ for the fitted Hamiltonian explicitly depends on the fitting parameters $\lambda_{\alpha}$. A suitable cost function for fitting the first $\mathcal{N}_{\tBand}$ bands is
\begin{equation}
	\label{app:eqn:non_linear_cost_function}
	\mathcal{C}' \left( \left\lbrace \lambda_{\alpha}  \right\rbrace \right) = \sum_{\vec{k} \in \mathcal{M}} \Tr\left[ \left( P\left( \vec{k} \right) h\left( \vec{k} \right) - P^{\tDFT}\left( \vec{k} \right) h^{\tDFT}\left( \vec{k} \right) \right)^2 \right].
\end{equation}
Due to the nontrivial dependence of the cost function $\mathcal{C}' \left( \left\lbrace \lambda_{\alpha} \right\rbrace \right)$ on the fitting parameters, it can no longer be minimized analytically, as in the linear extraction method outlined in \cref{app:sec:fitting_method:nonlinear}. Therefore, the nonlinear fitting method requires parameter fitting. As a result, iterative numerical minimization techniques must be employed, and this approach is referred to as \emph{nonlinear fitting}. In this work, we use a Quasi-Newton method (as implemented in Mathematica) to find the local minimum of $\mathcal{C}' \left( \left\lbrace \lambda_{\alpha} \right\rbrace \right)$, starting from the initial conditions that globally minimize the $\mathcal{C} \left( \left\lbrace \lambda_{\alpha} \right\rbrace \right)$ cost function from \cref{app:eqn:linear_cost_function_0}.

\subsubsection{Analytical expression for the gradient of the cost function}\label{app:sec:fitting_method:nonlinear:gradient}

A key intermediate quantity in the Quasi-Newton method for minimization is the gradient of the cost function with respect to the fitting parameters. In principle, this gradient can be computed numerically using finite differences. However, for the cost function from \cref{app:eqn:non_linear_cost_function}, the gradient has an analytical expression that is numerically much easier to compute. In what follows, we derive an analytical formula for the gradient of the cost function. Introducing the shorthand notation
\begin{equation}
	\partial_{\alpha} \equiv \pdv{\lambda_{\alpha}},
\end{equation}
we find that 
\begin{equation}
	\label{app:eqn:nonlinear_cost_function}
	\partial_{\alpha} \mathcal{C}' \left( \left\lbrace \lambda_{\alpha}  \right\rbrace \right) = 2 \sum_{\vec{k} \in \mathcal{M}} \Tr \left[ \left( P\left( \vec{k} \right) h\left( \vec{k} \right) - P^{\tDFT}\left( \vec{k} \right) h^{\tDFT}\left( \vec{k} \right) \right) \partial_{\alpha}  \left( P\left( \vec{k} \right) h\left( \vec{k} \right) \right) \right].
\end{equation}
Since 
\begin{equation}
	\label{app:eqn:partial_projector_hamiltonian}
	\partial_{\alpha}  \left( P\left( \vec{k} \right) h\left( \vec{k} \right) \right) = \left(\partial_{\alpha}  P\left( \vec{k} \right)\right) h\left( \vec{k} \right)  + P\left( \vec{k} \right) h_{\alpha}\left( \vec{k} \right),
\end{equation}
we need to compute the partial derivative of the projector $\partial_{\alpha}  P\left( \vec{k} \right)$. This is most easily expressed using the integral expression from \cref{app:eqn:definition_projector}
\begin{equation}
	\partial_{\alpha}  P\left( \vec{k} \right) = \frac{1}{2 \pi i} \oint \dd{z} \left( z \one - h \left( \vec{k} \right) \right)^{-1} h_{\alpha} \left( \vec{k} \right) \left( z \one - h \left( \vec{k} \right) \right)^{-1},
\end{equation}
which can be cast in the band basis
\begin{align}
	\label{app:eqn:derivative_of_projector}
	\sum_{i,j} \phi^{*s}_{m;i} \left( \vec{k} \right) \left( \partial_{\alpha}  P_{ij}\left( \vec{k} \right) \right) \phi_{l;j} \left( \vec{k} \right) &= \frac{1}{2 \pi i} \sum_{i,j} \oint \dd{z}  \frac{\phi^{*s}_{m;i} \left( \vec{k} \right) h_{\alpha;ij} \left( \vec{k} \right) \phi_{l;j} \left( \vec{k} \right) }{\left(z - \epsilon_m \left( \vec{k} \right) \right) \left(z - \epsilon_l \left( \vec{k} \right) \right) } \nonumber \\
	 &=  \frac{n_m - n_l}{\epsilon_m \left( \vec{k} \right) - \epsilon_l \left( \vec{k} \right) } \sum_{i,j} \phi^{*s}_{m;i} \left( \vec{k} \right) h_{\alpha;ij} \left( \vec{k} \right) \phi_{l;j} \left( \vec{k} \right) 
\end{align}
By plugging \cref{app:eqn:derivative_of_projector} into \cref{app:eqn:partial_projector_hamiltonian}, we find
\begin{align}
	\sum_{i,j} \phi^{*s}_{m;i} \left( \vec{k} \right) \left[ \partial_{\alpha}  \left( P\left( \vec{k} \right) h\left( \vec{k} \right) \right) \right]_{ij} \phi_{l;j} \left( \vec{k} \right) &= \sum_{i,j} \left( \epsilon_{l} \left( \vec{k} \right)\phi^{*s}_{m;i} \left( \vec{k} \right) \partial_{\alpha}  P_{ij} \left( \vec{k} \right)  \phi_{l;j} \left( \vec{k} \right) + n_m \phi^{*s}_{m;i} \left( \vec{k} \right) h_{\alpha;ij} \left( \vec{k} \right)  \phi_{l;j} \left( \vec{k} \right) \right) \nonumber \\
	&= \frac{n_m \epsilon_m \left( \vec{k} \right) - n_l \epsilon_l \left( \vec{k} \right)}{\epsilon_m \left( \vec{k} \right) - \epsilon_l \left( \vec{k} \right) } \sum_{i,j} \phi^{*s}_{m;i} \left( \vec{k} \right) h_{\alpha;ij} \left( \vec{k} \right) \phi_{l;j} \left( \vec{k} \right). \label{app:eqn:partial_projector_hamiltonian_band_basis}
\end{align}
\Cref{app:eqn:partial_projector_hamiltonian_band_basis} gives the matrix elements of $\partial_{\alpha}  \left( P\left( \vec{k} \right) h\left( \vec{k} \right) \right)$ in the band basis of the fitted Hamiltonian. Plugging this into \cref{app:eqn:nonlinear_cost_function} provides an analytical expression for the gradient of the cost function, significantly improving the efficiency of the nonlinear minimization.

\subsection{Reducing the number of parameter}\label{app:sec:fitting_method:reducing_number_parameters}

In both the linear and nonlinear fitting methods, it is desirable to minimize the number of fitting parameters. To achieve this, we employ the concept of \emph{step-wise} regression. In step-wise regression, a model with $\mathcal{N}-1$ parameters is derived from one with $\mathcal{N}$ parameters by removing the ``least important'' terms. To identify these terms, we start with the optimized solution containing $\mathcal{N}$ parameters, denoted by $\left \lbrace \lambda'_{\alpha} \right \rbrace$. We then compute the cost function (without further optimization) by successively setting each of the $\mathcal{N}$ parameters to zero
\begin{equation}
	\mathcal{C}'_{\alpha} = \mathcal{C} \left( \left \lbrace \lambda_{\alpha} \right \rbrace \right) \qq{with} \lambda_{\beta} = \lambda'_{\beta} \qq{for} \beta \neq \alpha \qq{and} \lambda_{\alpha} = 0.
\end{equation}
The term $\alpha'$ corresponding to the largest value $\mathcal{C}'_{\alpha'}$ is then eliminated. The model with $\mathcal{N}-1$ parameters is optimized, resulting in an optimal solution with one fewer parameter. This process is repeated until only one parameter remains. The result is $\mathcal{N}$ models, ranging from one parameter to $\mathcal{N}$ parameters. The model with the fewest parameters and an error below a certain threshold is then selected.

Finally, we also define the relative fitting error for both the linear and nonlinear fitting methods. First, we add a chemical potential term to the \textit{ab initio} Hamiltonian
\begin{equation}
	h^{\tDFT} \left( \vec{k} \right) \to h^{\tDFT} \left( \vec{k} \right) + \mu \one,
\end{equation}
with $\mu \in \mathbb{R}$, such that
\begin{equation}
	\sum_{\vec{k} \in \mathcal{M}} \Tr  \left( h^{\tDFT} \left( \vec{k} \right) \right) = 0,
\end{equation}
for the linear extraction method, or
\begin{equation}
	\sum_{\vec{k} \in \mathcal{M}} \Tr \left( P^{\tDFT} \left( \vec{k} \right) h^{\tDFT} \left( \vec{k} \right) \right) = 0,
\end{equation}
in the nonlinear fitting method. The relative error (for a given optimized solution $\left \lbrace \lambda'_{\alpha} \right \rbrace$) is then defined as
\begin{equation}
	\label{app:eqn:relative_error_definition}
	\varepsilon = \frac{\eval{\mathcal{C} \left( \left \lbrace \lambda_{\alpha} \right \rbrace \right)}_{\lambda_{\alpha} = \lambda'_{\alpha}}}{\eval{\mathcal{C} \left( \left \lbrace \lambda_{\alpha} \right \rbrace \right)}_{\lambda_{\alpha} = 0}}.
\end{equation}

\section{Simplified Hamiltonian}\label{app:sec:simple_ham_results}
In this section, we discuss the single-particle Hamiltonian of the AA-stacking model. We will impose $\theta=0$ symmetry, which also indicates $SU(2)$ spin symmetry, in this section. Then, our Hamiltonian could be described by 
\begin{align}
    &w^{\text{AA}}_1, \quad w^{\text{AA}}_2 \nonumber\\ 
    &w^{\prime \text{AA}}_3 \nonumber\\ 
    &w^{\prime \text{AA}}_1,w^{\prime \text{AA}}_4, w^{\prime \text{AA}}_2,w^{\prime \text{AA}}_5 \nonumber\\ 
\end{align}
 $w^{\text{AA}}_1,w^{\text{AA}}_2$ characterize the inter-layer coupling between the nearest-neighbor $Q$-lattice sites. $w^{\prime \text{AA}}_3$ characterizes the inter-layer coupling between the next-nearest-neighbor $Q$-lattice sites. $w^{\prime \text{AA}}_1, w^{\prime \text{AA}}_4, w^{\prime \text{AA}}_2, w^{\prime \text{AA}}_5$
 characterizes the intra-layer coupling between the next-next-nearest-neighbor $Q$-lattice sites. 
 The single-particle Hamiltonian of the first valley can be written as 
\begin{align}
    \hat{H} =& \sum_{\vec{Q},l,s} \hat{c}^\dagger_{\vec{k},\vec{Q},s,l}
    \hat{c}_{\vec{k},\vec{Q},s,l}\left(\frac{(k_x-Q_x)^2}{2m_x} + \frac{(k_y-Q_y)^2}{2m_y}\right)  \nonumber\\ 
    &+ 
    \sum_{\vec{Q},\vec{Q}',l}\left[ \hat{c}^\dagger_{\vec{k},\vec{Q},s,l} \hat{c}_{\vec{k},\vec{Q}',s,-l}(iw^{\text{AA}}_1 +w^{\text{AA}}_2)\delta_{\vec{Q}+\vec{q}_0,\vec{Q}'}
    +\hat{c}^\dagger_{\vec{k},\vec{Q},s,l} \hat{c}_{\vec{k},\vec{Q}',s,-l}(-iw^{\text{AA}}_1 +w^{\text{AA}}_2)\delta_{\vec{Q}-\vec{q}_0,\vec{Q}'}\right]  \nonumber\\ 
    &+\sum_{\vec{Q},\vec{Q}',l,s}w^{\prime \text{AA}}_3 
  \hat{c}^\dagger_{\vec{k},\vec{Q},s,l} \hat{c}_{\vec{k},\vec{Q}',s,-l} \left[ 
    \delta_{\vec{Q}+\vec{q}_2-\vec{q}_1,\vec{Q}'}+\delta_{\vec{Q}-\vec{q}_2+\vec{q}_1,\vec{Q}'}\right] \nonumber\\ 
    & +  \sum_{\vec{Q},\vec{Q}',l,s}\hat{c}^\dagger_{\vec{k},\vec{Q},s,l} \hat{c}_{\vec{k},\vec{Q}',s,l} \left\{ (-iw^{\prime \text{AA}}_1 + w^{\prime \text{AA}}_4) 
    \left[ \delta_{\vec{Q} + \vec{q}_0-\vec{q}_1+\vec{q}_2,\vec{Q}'} + \delta_{\vec{Q}+\vec{q}_0+\vec{q}_1-\vec{q}_2,\vec{Q}'}
    \right]\right. \nonumber\\ 
    &
   \left. +(iw^{\prime \text{AA}}_1 + w^{\prime \text{AA}}_4) 
    \left[ \delta_{\vec{Q} -(\vec{q}_0-\vec{q}_1+\vec{q}_2),\vec{Q}'} + \delta_{\vec{Q} +(-\vec{q}_0-\vec{q}_1+\vec{q}_2),\vec{Q}'}
    \right] 
    \right\}  \nonumber\\ 
    &+ \sum_{\vec{k},\vec{Q},\vec{Q}',l,s}\hat{c}^\dagger_{\vec{k},\vec{Q},s,l} \hat{c}_{\vec{k},\vec{Q}',s,l}\left[ (-iw^{\prime \text{AA}}_2 +w^{\prime \text{AA}}_5)\delta_{\vec{Q} + 2\vec{q}_0, \vec{Q}'} +
    (iw^{\prime \text{AA}}_2 +w^{\prime \text{AA}}_5)\delta_{\vec{Q} - 2\vec{q}_0, \vec{Q}'} +
    \right] 
\end{align}
We focus on the single-particle Hamiltonian of the first valley $\eta=0$. 
Within the valley $\eta=0$, the single-particle Hamiltonian has $C_{2x}$ and $\mathcal{T}$ symmetry. 
With $\theta=0$ symmetry imposed, the single-particle Hamiltonian of the first valley has the following additional symmetry 
\begin{align}
    g_{\vec{q}_0} \hat{c}_{\vec{k},\vec{Q}, s, l }g_{\vec{q}_0}^{-1} = \hat{c}_{\vec{k} + \vec{q}_0 , \vec{Q} + \vec{q}_0 ,s, -l}
    \label{eq:app:q0_shift_mom_space}
\end{align}
This indicates the single-particle spectrum of the first valley at $\vec{k} $ is the same as the single-particle spectrum of the first valley at $\vec{k} + \vec{q}_0$. This translational symmetry in the momentum space is different from the translational invariance imposed by the moir\'e periodicity with wavevectors $(2,0)|\vec{q}_0| , (1,\sqrt{3})|\vec{q}_0|$

\subsection{Hamiltonian in the real-space}
We consider the Fourier transformation introduced in~\cref{app:eqn:def_real_space_fermions_to_real} which is
\begin{align}
    \hat{\psi}^\dagger_{\eta,s,l} \left( \vec{r} \right) &= \frac{1}{\sqrt{\Omega}} \sum_{\vec{k} \in \text{MBZ}} \sum_{\vec{Q} \in \mathcal{Q}_{\eta + l}} \hat{c}^\dagger_{\vec{k},\vec{Q},s,l} e^{-i \left( \vec{k} - \vec{Q} \right) \cdot \vec{r}}, 
\end{align}
The Hamiltonian of the valley $0$ in the real space can be written as
 \begin{align}
      \hat{H}=
      & \int_{\vec{r}} \sum_{l,s} \hat{\psi}^\dagger_{\eta=0,s,l}(\vec{r})\left( \frac{-\partial_{r_x}^2}{2m_x} + \frac{-\partial_{r_y}^2}{2m_y} \right) \hat{\psi}_{\eta=0,s,l}(\vec{r})\nonumber\\
      &+ 
      \int_{\vec{r}} \sum_{l,s} \hat{\psi}^\dagger_{\eta=0,s,l}(\vec{r})\hat{\psi}_{\eta=0,s,-l}(\vec{r}) T(\vec{r})
   + \int_{\vec{r}} \sum_{l,s}\hat{\psi}^\dagger_{\eta=0,s,l}(\vec{r})\hat{\psi}_{\eta=0,s,l}(\vec{r}) U(\vec{r})
    \label{eq:app:real_space_ham_val_0_simplify}
 \end{align}
 where we have introduced 
 \begin{align}
 & T(\vec{r}) = 2\sqrt{ (w^{\text{AA}}_1)^2  + (w^{\text{AA}}_2)^2}\cos(
|\vec{q}_0|r_x+\psi) + w^{\prime \text{AA}}_3 2\cos( \sqrt{3}|\vec{q}_0|r_y ) \nonumber\\ 
     &e^{i\psi } = \frac{iw^{\text{AA}}_1 +w^{\text{AA}}_2}{\sqrt{(w^{\text{AA}}_1)^2 + (w^{\text{AA}}_2)^2}/2} \nonumber\\ 
     & U(\vec{r}) = (-iw^{\prime \text{AA}}_1 + w^{\prime \text{AA}}_4) e^{-i|\vec{q}_0| r_x} \cos(\sqrt{3}|\vec{q}_0|r_y)+ (-iw^{\prime \text{AA}}_2 + w^{\prime \text{AA}}_5) e^{i2|\vec{q}_0|r_x} +\text{h.c.}
 \end{align}
and we have used 
 \begin{align}
     &\vec{q}_0 = |\vec{q}_0|(1,0) \nonumber\\ 
     &(\vec{q}_2-\vec{q}_1) = \sqrt{3}|\vec{q}_0|(0,-1)
 \end{align} 
We could also define the additional $\theta=0$ symmetry (Eq.~\ref{eq:app:q0_shift_mom_space}) in the momentum space using the real-space operator
\begin{align}
   g_{\vec{q}_0} \hat{\psi}^\dagger_{\eta=0, s,l}(\vec{r})g_{\vec{q}_0}^{-1} = \frac{1}{\sqrt{\Omega}} \sum_{\vec{k} \in \text{MBZ}}\sum_{\vec{Q} \in \mathcal{Q}_l} \hat{c}^\dagger_{\vec{k}+\vec{q}_0, \vec{Q} +\vec{q}_0, s,-l}e^{-i(\vec{k}-\vec{Q})\cdot \vec{r}} = \hat{\psi}^\dagger_{\eta=0,s,-l}(\vec{r})
\label{eq:app:q0_shift_real_space_even_odd}
\end{align} 
We can therefore take the eigenbasis of $g_{\vec{q}_0}$ which is also the even-odd basis of electron operators
 \begin{align}
     &\hat{\psi}_{\eta=0,s,e}(\vec{r}) = \frac{1}{\sqrt{2}}(
     \hat{\psi}_{\eta=0,s,+}(\vec{r})+
      \hat{\psi}_{\eta=0,s,-}(\vec{r})) \nonumber\\ 
     &\hat{\psi}_{\eta=0,s,o}(\vec{r}) = \frac{1}{\sqrt{2}}( \hat{\psi}_{\eta=0,s,+}(\vec{r})-
      \hat{\psi}_{\eta=0,s,-}(\vec{r})) 
 \end{align} 
Also, due to the $g_{\vec{q}_0}$ symmetry, two-electron operators $\hat{\psi}_{\eta=0,s,e}(\vec{r})$ and $\hat{\psi}_{\eta=0,s,o}(\vec{r})$ are decoupled at single-particle level. Then the Hamiltonian becomes 
\begin{align}
      \hat{H}=
      & \int_{\vec{r}} \sum_{s} \hat{\psi}^\dagger_{\eta=0,s,e}(\vec{r}) \left[ \frac{-\partial_{r_x}^2}{2m_x} + \frac{-\partial_{r_y}^2}{2m_y}  
      +T(\vec{r}) + U(\vec{r}) 
      \right] \hat{\psi}_{\eta=0,s,e}(\vec{r}) \nonumber\\ 
     +
     & \int_{\vec{r}} \sum_{s} \hat{\psi}^\dagger_{\eta=0,s,o}(\vec{r}) \left[ \frac{-\partial_{r_x}^2}{2m_x} + \frac{-\partial_{r_y}^2}{2m_y}  
   - T(\vec{r})+U(\vec{r})
      \right] \hat{\psi}_{\eta=0,s,o}(\vec{r}) 
 \end{align}
In this simplified limit, the single-particle Hamiltonian is simply described by a quadratic band in a periodic potential. The potentials take the following form
 \begin{align}
     V_{e/o}(\vec{r}) = \pm T(\vec{r}) + U(\vec{r})
 \end{align}
 The periodicity of the potential is described by two vectors 
 \begin{align}
     &\vec{a}_{V,1} = \frac{2\pi}{|\vec{q}_0|}(1,0)
     ,\quad 
     \vec{a}_{V,2} = \frac{2\pi}{\sqrt{3}|\vec{q}_0|}(0,1)
     \label{eq:app:av_basis_def}
     \end{align}
 with 
 \begin{align}
      V_{e/o}(\vec{r} + n\vec{a}_{V,1} + m \vec{a}_{V,2} ) = V_{e/o}(\vec{r}),\quad n,m\in \mathbb{Z}
 \end{align}
The relation between this new periodicity and the moir\'e periodicity is $\vec{a}_{V,2}=\vec{a}_{M,2}, \vec{a}_{V,1} = 2\vec{a}_{M,1}+\vec{a}_{M,2}$. In addition, we notice that 
\begin{align}
&\bm{a}_{M,1} = \frac{1}{2}(\bm{a}_{V,1}-\bm{a}_{V,2}) = \frac{\pi}{|\vec{q}_0|}(1,-\frac{1}{\sqrt{3}}),\quad 
 \bm{a}_{M,2}  = \bm{a}_{V,2} =\frac{2\pi}{\sqrt{3}|\vec{q}_0|}(0,1)
\end{align}
Under moir\'e translation, the potential follows 
\begin{align}
    &V_{e}(\vec{r}+  \bm{a}_{M,2} ) = V_{e}(\vec{r}),\quad 
    V_{o}(\vec{r}+  \bm{a}_{M,2} ) = V_{o}(\vec{r})\nonumber\\ 
    &V_{e}(\vec{r}+  \bm{a}_{M,1} ) = V_{o}(\vec{r} ),\quad 
    V_{o}(\vec{r}+  \bm{a}_{M,1} ) = V_{e}(\vec{r} )
\end{align}

\subsubsection{Symmetry properties}
We discuss the symmetry operations in the even-odd basis. First, we have zero-twist symmetry (see~\cref{eq:app:q0_shift_real_space_even_odd})
\begin{align}
   &g_{\vec{q}_0} \hat{\psi}^\dagger_{\eta=0, s,e}(\vec{r})g_{\vec{q}_0}^{-1}  = \hat{\psi}^\dagger_{\eta=0,s,e}(\vec{r}) \nonumber\\ 
   &g_{\vec{q}_0} \hat{\psi}^\dagger_{\eta=0, s,o}(\vec{r})g_{\vec{q}_0}^{-1} =  -\hat{\psi}^\dagger_{\eta=0,s,o}(\vec{r})
\end{align} 

Since the potentials are periodic, the system also has the following translational symmetry 
\begin{align}
    &T_{\bm{a}_{V,i}} \hat{\psi}^\dagger_{\eta=0,s,e} (\vec{r}) T_{\bm{a}_{V,i}}^{-1} = 
      \hat{\psi}^\dagger_{\eta=0,s,e} (\vec{r}+\bm{a}_{V,i}) \nonumber\\ 
     &T_{\bm{a}_{V,i}} \hat{\psi}^\dagger_{\eta=0,s,o} (\vec{r}) T_{\bm{a}_{V,i}}^{-1} = 
      \hat{\psi}^\dagger_{\eta=0,s,o} (\vec{r}+\bm{a}_{V,i}) 
\end{align}

We next discuss the original moir\'e translational symmetry. Under moir\'e translational transformation $T_{\bm{a}_{M,i}}$, the electron operator in the plane-wave basis transform as 
\begin{align}
    &T_{\bm{a}_{M,i}} \hat{c}^\dagger_{\vec{k}, \vec{Q}, s, l=+}  T_{\bm{a}_{M,i}}^{-1} = \hat{c}^\dagger_{\vec{k}, \vec{Q}, s, l=+}e^{-i( \vec{q}_1+ \vec{k} -\vec{Q})\cdot \bm{a}_{M,i}},\quad \vec{q}_1 \in \mathcal{Q}_1\nonumber\\ 
    &T_{\bm{a}_{M,i}} \hat{c}^\dagger_{\vec{k}, \vec{Q}, s, l=-}  T_{\bm{a}_{M,i}}^{-1} = \hat{c}^\dagger_{\vec{k}, \vec{Q}, s, l=+}e^{-i( \vec{q}_2+ \vec{k} -\vec{Q})\cdot \bm{a}_{M,i}},\quad \vec{q}_2 \in \mathcal{Q}_2
\end{align}
Then the real-space operators transform as 
\begin{align}
     &T_{\bm{a}_{M,i}} \hat{\psi}^\dagger_{\eta=0, s, l=+} (\vec{r}) T_{\bm{a}_{M,i}}^{-1} = 
      \hat{\psi}^\dagger_{\eta=0, s, l=+} (\vec{r}+\bm{a}_{M,i}) e^{-i\vec{q}_1 \cdot\bm{a}_{M,i} } \nonumber\\ 
     & T_{\bm{a}_{M,i}} \hat{\psi}^\dagger_{\eta=0, s, l=-} (\vec{r}) T_{\bm{a}_{M,i}}^{-1} = 
      \hat{\psi}^\dagger_{\eta=0, s, l=-} (\vec{r}+\bm{a}_{M,i}) e^{-i\vec{q}_2 \cdot\bm{a}_{M,i} } 
\end{align}
The real-space operators in the even-odd basis now transform as
\begin{align}
    &
    T_{\bm{a}_{M,2}} \hat{\psi}^\dagger_{\eta=0,s,e} (\vec{r})T_{\bm{a}_{M,2}}^{-1} = - \hat{\psi}^\dagger_{\eta=0,s,e} (\vec{r}+\bm{a}_{M,2}) 
    ,\quad     T_{\bm{a}_{M,2}} \hat{\psi}^\dagger_{\eta=0,s,o} (\vec{r})T_{\bm{a}_{M,2}}^{-1} =  -\hat{\psi}^\dagger_{\eta=0,s,o} (\vec{r}+\bm{a}_{M,2}) \nonumber\\ 
    &
    T_{\bm{a}_{M,1}} \hat{\psi}^\dagger_{\eta=0,s,e} (\vec{r})T_{\bm{a}_{M,1}}^{-1} = - \hat{\psi}^\dagger_{\eta=0,s,o} (\vec{r}+\bm{a}_{M,1}) 
    ,\quad     T_{\bm{a}_{M,1}} \hat{\psi}^\dagger_{\eta=0,s,o} (\vec{r})T_{\bm{a}_{M,1}}^{-1} = - \hat{\psi}^\dagger_{\eta=0,s,e} (\vec{r}+\bm{a}_{M,2}) 
    \label{eq:app:moire_trans_even_odd}
\end{align}
We can observe that $T_{\bm{a}_{M,1}} $ not only performs a translational transformation in the real space but also transforms the even (odd) orbital to the odd (even) orbital. As for $T_{\bm{a}_{M,2}}$, since $\bm{a}_{M,2} = \bm{a}_{V,2}$, then $T_{\bm{a}_{M,2}}$ is equivalent to the $T_{\bm{a}_{V,2}}$ besides an additional $-1$ sign which can be gauged away.

Finally, we discuss the effect of $C_{2x}$ symmetry. Since our simplified model has SU(2) spin symmetry, it is sufficient to consider the spinless $C_{2x}$ symmetry defined as 
\begin{align}
    \tilde{C}_{2x} \hat{c}^\dagger_{\vec{k},\vec{Q}, s,l}    \tilde{C}_{2x}^{-1} = \hat{c}^\dagger_{C_{2x}\vec{k},C_{2x}\vec{Q}, s,-l} 
\end{align}
which gives
\begin{align}
    \tilde{C}_{2x} \hat{\psi}^\dagger_{\eta=0,s,l}(\vec{r}) \tilde{C}_{2x}^{-1}
    =\hat{\psi}^\dagger_{\eta=0,s,-l}(C_{2x}\vec{r})
\end{align}
Therefore, 
\begin{align}
    &\tilde{C}_{2x} \hat{\psi}^\dagger_{\eta=0,s,e}(\vec{r}) \tilde{C}_{2x}^{-1}
    = \hat{\psi}^\dagger_{\eta=0,s,e}(C_{2x}\vec{r}) \nonumber\\ 
    &\tilde{C}_{2x} \hat{\psi}^\dagger_{\eta=0,s,o}(\vec{r}) \tilde{C}_{2x}^{-1}
    = -\hat{\psi}^\dagger_{\eta=0,s,o}(C_{2x}\vec{r})
\end{align}

In summary, the symmetry operations in the even-odd basis are
\begin{align}
    \text{Translation}:&T_{\bm{a}_{V,i}} \hat{\psi}^\dagger_{\eta=0,s,e/o} (\vec{r}) T_{\bm{a}_{V,i}}^{-1} = 
      \hat{\psi}^\dagger_{\eta=0,s,e/o} (\vec{r}+\bm{a}_{V,i}) \nonumber\\ \nonumber\\ 
      \text{Zero-twist symmetry}:& g_{\vec{q}_0} \hat{\psi}^\dagger_{\eta=0,s,e} (\vec{r}) g_{\vec{q}_0}^{-1} = 
      \hat{\psi}^\dagger_{\eta=0,s,e} (\vec{r}),\quad 
      g_{\vec{q}_0} \hat{\psi}^\dagger_{\eta=0,s,o} (\vec{r}) g_{\vec{q}_0}^{-1} = -
      \hat{\psi}^\dagger_{\eta=0,s,o} (\vec{r})
      \nonumber\\ \nonumber\\ 
      \text{Moir\'e translation}:& T_{\bm{a}_{M,1}} \hat{\psi}^\dagger_{\eta=0,s,e} (\vec{r}) T_{\bm{a}_{M,1}}^{-1} = 
      \hat{\psi}^\dagger_{\eta=0,s,o} (\vec{r} +\frac{\bm{a}_{V,1}-\bm{a}_{V,2}}{2}), \nonumber\\ 
      &
     T_{\bm{a}_{M,1}} \hat{\psi}^\dagger_{\eta=0,s,o} (\vec{r}) T_{\bm{a}_{M,1}}^{-1} = 
      \hat{\psi}^\dagger_{\eta=0,s,e} (\vec{r} +\frac{\bm{a}_{V,1}-\bm{a}_{V,2}}{2})
      \nonumber\\ \nonumber\\ 
       \text{$\tilde{C}_{2x}$}:&
       \tilde{C}_{2x} \hat{\psi}^\dagger_{\eta=0,s,e} (\vec{r}) \tilde{C}_{2x}^{-1} = 
      \hat{\psi}^\dagger_{\eta=0,s,e} (C_{2x}\vec{r}),\quad 
      \tilde{C}_{2x} \hat{\psi}^\dagger_{\eta=0,s,o} (\vec{r}) \tilde{C}_{2x}^{-1} = -
      \hat{\psi}^\dagger_{\eta=0,s,o} (C_{2x}\vec{r})
\end{align}

\subsubsection{$M_{x}$ and inversion symmetries}
We comment that, with only inter-layer first-harmonic terms $w^{\text{AA}}_1,w^{\text{AA}}_2,w^{\prime \text{AA}}_3$, the system also has mirror-$x$ and inversion symmetries. We can observe this from the real-space Hamiltonian. With only $w^{\text{AA}}_1,w^{\text{AA}}_2,w^{\prime \text{AA}}_3$, the real-space Hamiltonian takes the form of 
  \begin{align}
      \hat{H}=
      & \int_{\vec{r}} \sum_{l,s} \hat{\psi}^\dagger_{\eta=0,s,l}(\vec{r})\left( \frac{-\partial_{r_x}^2}{2m_x} + \frac{-\partial_{r_y}^2}{2m_y} \right) \hat{\psi}_{\eta=0,s,l}(\vec{r})
      + 
      \int_{\vec{r}} \sum_{l,s} \hat{\psi}^\dagger_{\eta=0,s,l}(\vec{r})\hat{\psi}_{\eta=0,s,-l}(\vec{r}) T(\vec{r})
 \end{align}
 where the inter-layer term takes the form of 
 \begin{align}
 & T(\vec{r}) = 2\sqrt{ (w^{\text{AA}}_1)^2  + (w^{\text{AA}}_2)^2}\cos(
|\vec{q}_0|r_x+\psi) + w^{\prime \text{AA}}_3 2\cos( \sqrt{3}|\vec{q}_0|r_y ) 
 \end{align}
 We can introduce new electron operators by changing the origin
 \begin{align}
     \hat{\tilde{\psi}}^\dagger_{\eta=0,s,l}(\vec{r}) = \hat{{\psi}}^\dagger_{\eta=0,s,l}\left( \vec{r} - (\frac{\psi}{|\vec{q}_0|},0)\right) 
     \label{eq:app:ele_op_shift_origin}
 \end{align}
 Then the Hamiltonian in the new basis reads
    \begin{align}
      \hat{H}=
      & \int_{\vec{r}} \sum_{l,s} \hat{\tilde{\psi}}^\dagger_{\eta=0,s,l}(\vec{r})\left( \frac{-\partial_{r_x}^2}{2m_x} + \frac{-\partial_{r_y}^2}{2m_y} \right) \hat{\tilde{\psi}}_{\eta=0,s,l}(\vec{r})
      + 
      \int_{\vec{r}} \sum_{l,s} \hat{\tilde{\psi}}^\dagger_{\eta=0,s,l}(\vec{r})\hat{\tilde{\psi}}_{\eta=0,s,-l}(\vec{r}) \tilde{T}(\vec{r}) \nonumber\\
     \tilde{T}(\vec{r})= & 2\sqrt{ (w^{\text{AA}}_1)^2  + (w^{\text{AA}}_2)^2}\cos(
|\vec{q}_0|r_x) + w^{\prime \text{AA}}_3 2\cos( \sqrt{3}|\vec{q}_0|r_y )
\label{eq:app:eff_ham_with_inv}
 \end{align}
 Since the new inter-layer term has the following properties 
 \begin{align}
     \tilde{T}(\vec{r}) =   \tilde{T}(-\vec{r}) =  \tilde{T}(C_{2y}\vec{r}) , 
 \end{align}
 the system is invariant under the following mirror-$x$ ($M_x$) and inversion $\mathcal{I}$ symmetries 
 \begin{align}
     &M_{x} \hat{\tilde{\psi}}^\dagger_{\eta=0,s,l}(\vec{r}) M_{x}^{-1} = 
     \hat{\tilde{\psi}}^\dagger_{\eta=0,s,l}(M_{y}\vec{r}) \nonumber\\ 
     &\mathcal{I} \hat{\tilde{\psi}}^\dagger_{\eta=0,s,l}(\vec{r}) \mathcal{I}^{-1}=\hat{\tilde{\psi}}^\dagger_{\eta=0,s,-l}(-\vec{r})
 \end{align}

\subsection{Hamiltonian in the even-odd basis}
\label{sec:app:ham_even_odd}
We now provide more detailed discussions on the Hamiltonian in the even-odd basis. The Hamiltonian is given below 
\begin{align}
      \hat{H}=
      & \int_{\vec{r}} \sum_{s} \hat{\psi}^\dagger_{\eta=0,s}(\vec{r}) \left[ \frac{-\partial_{r_x}^2}{2m_x} + \frac{-\partial_{r_y}^2}{2m_y}  
      +T(\vec{r}) + U(\vec{r}) 
      \right] \hat{\psi}_{\eta=0,s,e}(\vec{r}) \nonumber\\ 
     +
     & \int_{\vec{r}} \sum_{s} \hat{\psi}^\dagger_{\eta=0,s}(\vec{r}) \left[ \frac{-\partial_{r_x}^2}{2m_x} + \frac{-\partial_{r_y}^2}{2m_y}  
   - T(\vec{r})+U(\vec{r})
      \right] \hat{\psi}_{\eta=0,s,o}(\vec{r}) 
 \end{align}
We can observe the even and odd electrons are decoupled at the single-particle level, and two types of electrons are connected by translational transformation $T_{\bm{a}_{M,1}}$. Therefore, it is sufficient to focus on the even basis, when discussing the single-particle properties. 

In the even basis, the system models electrons with quadratic dispersion in the presence of a periodic potential $T(\vec{r})+U(\vec{r})$. The periodicity of the potential is characterized by $\bm{a}_{V,1}, \bm{a}_{V,2}$ (~\cref{eq:app:av_basis_def}). The corresponding reciprocal lattice vectors are 
\begin{align}
    \bm{b}_{V,1} = |\vec{q}_0|(1,0),\quad 
    ,\quad \bm{b}_{V,2} = |\vec{q}_0|(0,\sqrt{3})
\end{align}

The symmetries are 
\begin{equation}
    C_{2x},\quad \mathcal{T}
\end{equation} 
If only including $w^{\text{AA}}_1,w^{\text{AA}}_2,w^{\prime \text{AA}}_3$, the system also has 
\begin{align}
    M_x,\quad \mathcal{I}
\end{align}

We focus on the case with only $w^{\text{AA}}_1,w^{\text{AA}}_2,w^{\prime \text{AA}}_3$, the Hamiltonian can now be written as (see~\cref{{eq:app:eff_ham_with_inv}})
\begin{align}
      \hat{H}_e=
      & \int_{\vec{r}} \sum_{s} \hat{\tilde{\psi}}^\dagger_{\eta=0,s,e}(\vec{r})\left( \frac{-\partial_{r_x}^2}{2m_x} + \frac{-\partial_{r_y}^2}{2m_y} \right) \hat{\tilde{\psi}}_{\eta=0,s,e}(\vec{r})
      + 
      \int_{\vec{r}} \sum_{s} \hat{\tilde{\psi}}^\dagger_{\eta=0,s,e}(\vec{r})\hat{\tilde{\psi}}_{\eta=0,s,e}(\vec{r}) \tilde{T}(\vec{r}) \nonumber\\
     \tilde{T}(\vec{r})= & 2\sqrt{ (w^{\text{AA}}_1)^2  + (w^{\text{AA}}_2)^2}\cos(
|\vec{q}_0|r_x) + w^{\prime \text{AA}}_3 2\cos( \sqrt{3}|\vec{q}_0|r_y )
 \end{align}
 where we have introduced the following electron operator (see~\cref{eq:app:ele_op_shift_origin})
 \begin{align}
     \hat{\tilde{\psi}}^\dagger_{\eta=0,s,e}(\vec{r}) = \hat{{\psi}}^\dagger_{\eta=0,s,e}\left( \vec{r} - (\frac{\psi}{|\vec{q}_0|},0)\right) 
     \label{eq:app:shift_op_def}
 \end{align} 

 We now discuss the topological property of the model. Due to the SU(2) spin symmetry, we can also treat the system as an effective spinless system. Then from the effective $\mathcal{I}\cdot\mathcal{T}$ symmetry, we conclude the Berry curvature of each isolated band vanishes. Therefore, the isolated band of the system is topologically trivial. 
In addition, we notice that the even (or odd) basis alone forms a rectangular lattice since the corresponding periodicity is determined by $\bm{a}_{V,i=1,2}$. In addition, we also have the moir\'e translational symmetry $T_{\bm{a}_{M,1}}$ (see~\cref{eq:app:moire_trans_even_odd}) which maps even basis to the odd basis. Therefore, the Wannier centers of even and odd basis are shifted by $ \bm{a}_{M,1}$. When we combine the Wannier orbitals of the even and odd electrons, they form a hexagonal lattice.

Finally, we also comment that if we only consider the electrons of the even (or odd) basis, the hopping along the $y$-direction is much stronger than the hopping along the $x$-direction. This is due to the fact that the distance between two nearby Wannier orbitals along the $x$-direction ($|\bm{a}_{V,1}|$) is larger than the distance along the $y$-direction ($|\bm{a}_{V,2}|$). This then leads to an effective 1D system at the single-particle level with flat dispersion along $x$ direction.

\subsection{Analytical results at $\Gamma,M,K$ points}
\begin{figure}
    \centering
    \includegraphics[width=0.5\linewidth]{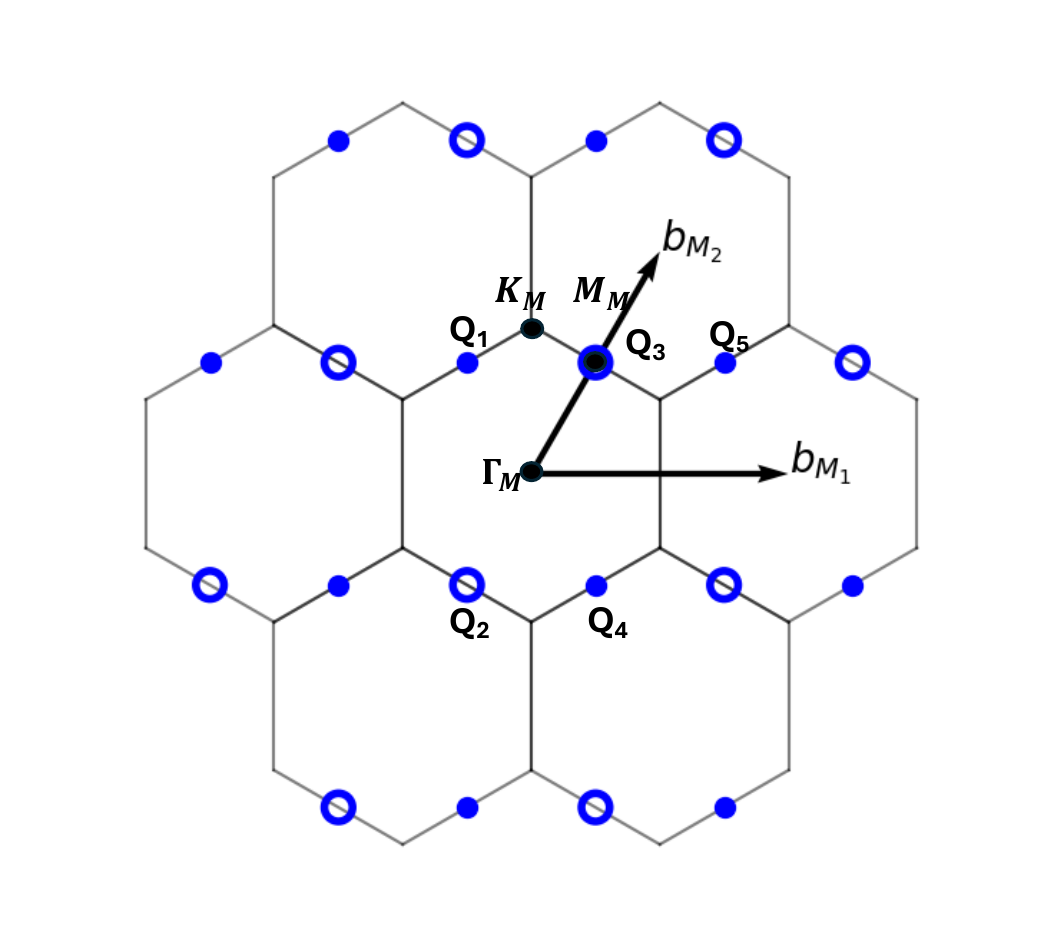}
    \caption{Labeling of high symmetry points and $\vec{Q}$-lattice.}
    \label{fig:high-sym-point}
\end{figure}
We now discuss the single-particle Hamiltonian for the simplified model with only non-zero $w^{\text{AA}}_1,w^{\text{AA}}_2,w^{\prime \text{AA}}_3$ near the following momenta (\cref{fig:high-sym-point})
 \begin{align}
     \Gamma = (0,0),\quad K = (0,\frac{2}{\sqrt{3}})|\vec{q}_0|,\quad M = (\frac{1}{2},\frac{\sqrt{3}}{2}) |\vec{q}_0|
 \end{align}
 in the Cartesian coordinates. 
 We have six $K$ points at the corner of the first moir\'e Brillouin zone: $\pm (0,\frac{\sqrt{3}}{2})|\vec{q}_0|, 
 \pm \frac{\sqrt{3}}{2}(\frac{\sqrt{3}}{2},\frac{1}{2})|\vec{q}_0|, 
 \pm \frac{\sqrt{3}}{2}(-\frac{\sqrt{3}}{2},\frac{1}{2})|\vec{q}_0|$
We note that $-K$ and $K$ are connected by $C_{2x}$ symmetry, so it is sufficient to only consider $K$. In addition,the moir\'e Brillouin zone have three $M$ points: $(1,0)|\vec{q}_0|, (\frac{1}{2},\frac{\sqrt{3}}{2})|\vec{q}_0|,(\frac{1}{2},-\frac{\sqrt{3}}{2})|\vec{q}_0|$. 
However, $(\frac{1}{2},\frac{\sqrt{3}}{2})|\vec{q}_0|$ and $(-\frac{1}{2},\frac{\sqrt{3}}{2})|\vec{q}_0|$ are connected by $C_{2x}$ symmetry, and the Hamiltonian at $(1,0)|\vec{q}_0|$ takes the same formula as the Hamiltonian at $\Gamma$ point. Therefore, it is sufficient to only consider
$M= (\frac{1}{2},\frac{\sqrt{3}}{2}) |\vec{q}_0|$. 
In addition, since we only have $C_{2x},\mathcal{T}$ symmetries for each valley, the $K$ point is no longer a high symmetry point.

 \subsubsection{$\Gamma_M$ point}
 We discuss the Hamiltonian at $\Gamma_M$ point. We only include four $Q$ points that are closest to the $\Gamma_M$ point which corresponds to the following four fermion operators 
 \begin{align}
    & \hat{c}_{\vec{p}, \vec{Q}_1,s,+},\quad \hat{c}_{\vec{p},\vec{Q}_2,s,-},\quad
\hat{c}_{\vec{p},\vec{Q}_3,s,-},\quad 
     \hat{c}_{\vec{p},\vec{Q}_4,s,+},\quad 
\end{align}
where (\cref{fig:high-sym-point})
\begin{align}
     &\vec{Q}_1=(-\frac{1}{2},\frac{\sqrt{3}}{2})|\vec{q}_0|,\quad \vec{Q}_2 =(-\frac{1}{2},-\frac{\sqrt{3}}{2})|\vec{q}_0|,\quad \vec{Q}_3=(\frac{1}{2},\frac{\sqrt{3}}{2})|\vec{q}_0|,\quad \vec{Q}_4 =  (\frac{1}{2},-\frac{\sqrt{3}}{2})|\vec{q}_0|
     \label{eq:app:def_Qpoints}
 \end{align}
 are defined in the Cartesian coordinates.

 The single-particle Hamiltonian in the basis of $ \hat{c}_{\vec{k}, \vec{Q}_1,s,+},\hat{c}_{\vec{k},\vec{Q}_2,s,-},
\hat{c}_{\vec{k},\vec{Q}_3,s,-},
     \hat{c}_{\vec{k},\vec{Q}_4,s,+} $  can be written as 
\begin{align}
    H^{\Gamma}_{\vec{p}} = 
    \begin{bmatrix}
        \epsilon^0_{\vec{p}-\vec{Q}_1} & w^{\prime \text{AA}}_3 & i w^{\text{AA}}_1 +w^{\text{AA}}_2 & 0 \\ 
        * & \epsilon^0_{\vec{p}-\vec{Q}_2} & 0 &iw^{\text{AA}}_1 + w^{\text{AA}}_2 \\
        * & * & \epsilon_{\vec{p}-\vec{Q}_3}^0 & w^{\prime \text{AA}}_3 \\ 
    * & * & * & \epsilon^0_{\vec{p}-\vec{Q}_4}
    \end{bmatrix}
    \label{eq:app:Gam_simple_Ham}
\end{align}
where $*$ denotes the complex conjugate of the corresponding matrix element. We have also introduced 
\begin{align}
    \epsilon^0_{\vec{k}} = \frac{k_x^2}{2m_x} + \frac{k_y^2}{2m_y}
\end{align}
Exactly at the $\Gamma$ point with $\vec{p} =0$, the four eigenvalues (each one is two-fold degenerate after including spin) of the Hamiltonian are, 
\begin{align}
  &E_{\Gamma}^{1} =   \epsilon_{-\vec{Q}_1}^0 - w^{\prime \text{AA}}_3 - \sqrt{ (w^{\text{AA}}_1)^2 + (w^{\text{AA}}_2)^2 } \nonumber\\ 
  &E_{\Gamma}^{2} =   \epsilon_{-\vec{Q}_1}^0 + w^{\prime \text{AA}}_3 - \sqrt{ (w^{\text{AA}}_1)^2 + (w^{\text{AA}}_2 )^2 } \nonumber\\ 
  &E_{\Gamma}^{3} =   \epsilon_{-\vec{Q}_1}^0 - w^{\prime \text{AA}}_3 + \sqrt{ (w^{\text{AA}}_1)^2 + (w^{\text{AA}}_2)^2 } \nonumber\\ 
  &E_{\Gamma}^{4} =   \epsilon_{-\vec{Q}_1}^0 + w^{\prime \text{AA}}_3 +\sqrt{ (w^{\text{AA}}_1)^2 + (w^{\text{AA}}_2 )^2 } 
  \label{eq:eigen_value_at_Gamma}
\end{align}
In practice, we find $w^{\text{AA}}_1,w^{\text{AA}}_2$ which corresponds to the coupling between the nearest-neighbor $Q$-lattice sites, are always the strongest coupling.  
The dominant $w^{\text{AA}}_1,w^{\text{AA}}_2$ terms split the four bands at the $\Gamma$ into two sets, $E_{\Gamma}^{1,2}$ and $E_{\Gamma}^{3,4}$,  where the gap between two sets of bands is $\sim 2 \sqrt{(w^{\text{AA}}_1)^2+(w^{\text{AA}}_2)^2 }$. 
The $w^{\prime \text{AA}}_3$ term will further split the spectrum into four non-degenerate bands. The energy differences between $E_{\Gamma}^1$ ($E_{\Gamma}^3$) and $E_{\Gamma}^2$ ($E_{\Gamma}^4$) are both $2|w^{\prime \text{AA}}_3|$.

\subsubsection{$K_M$ point}
 We now discuss the Hamiltonian at $K_M$ point. We only consider two $Q$ points that are closest to the $K_M$ point which corresponds to the following two fermion operators 
 \begin{align}
    & \hat{c}_{\vec{p}+K_M, \vec{Q}_1,s,+},\quad \hat{c}_{\vec{p}+K_M,\vec{Q}_3,s,-}
\end{align}
where $\vec{Q}$ have been given in ~\cref{eq:app:def_Qpoints}. 
 The single-particle Hamiltonian in the basis of $ \hat{c}_{\vec{p}+K_M, \vec{Q}_1,s,+},\hat{c}_{\vec{p}+K_M,\vec{Q}_2,s,-}$  can be written as 
\begin{align}
    H^{K_M}_{\vec{p}} = 
    \begin{bmatrix}
        \epsilon^0_{K_M+\vec{p}-\vec{Q}_1} & i w^{\text{AA}}_1 +w^{\text{AA}}_2  \\ 
       -i w^{\text{AA}}_1 +w^{\text{AA}}_2  &\epsilon^0_{K_M+\vec{p}-\vec{Q}_2}
    \end{bmatrix}
\end{align}

The Hamiltonian can be diagonalized directly, where the eigenvalues are 
\begin{align}
   & E_{K_M,\vec{p}}^1 = \frac{ m_x|\vec{q}_0|^2 + 3m_y(|\vec{q}_0|^2 +4p_x^2) + 4m_xp_y(\sqrt{3}|\vec{q}_0|+3p_y) }{24m_xm_y}
   -\frac{1}{2} \sqrt{ \frac{p_x^2 |\vec{q}_0|^2}{m_x^2} + 4(w^{\text{AA}}_1)^2 +4(w^{\text{AA}}_2)^2 }\nonumber\\ 
    &E_{K_M,\vec{p}}^2 = \frac{m_x |\vec{q}_0|^2 + 4m_xp_y(\sqrt{3}|\vec{q}_0|+3p_y)+3m_y|\vec{q}_0|^2+12m_yp_x^2}{24m_xm_y}  +\frac{1}{2} \sqrt{ \frac{p_x^2 |\vec{q}_0|^2}{m_x^2} + 4(w^{\text{AA}}_1)^2 +4(w^{\text{AA}}_2)^2 }
\end{align}
We can observe that, exactly at $K_M$ points with $p_x=p_y=0$, the eigenvalues (where each eigenvalue is two-fold degenerate after including spin) are
\begin{align}
    E_{K_M,\vec{p}=\vec{0}}^1 = \frac{ m_x|\vec{q}_0|^2 + 3m_y|\vec{q}_0|^2 }{24m_xm_y}  -  \sqrt{ (w^{\text{AA}}_1)^2 +(w^{\text{AA}}_2)^2 },\quad 
     E_{K_M,\vec{p}=\vec{0}}^2 = \frac{m_x|\vec{q}_0|^2 + 3m_y|\vec{q}_0|^2 }{24m_xm_y}  + \sqrt{ (w^{\text{AA}}_1)^2 +(w^{\text{AA}}_2)^2}
\end{align}
The energy difference between the two lowest bands is proportional to the strength of the nearest-neighbor coupling, which is $2\sqrt{ (w^{\text{AA}}_1)^2 +(w^{\text{AA}}_2)^2}$. The eigenvectors at $\vec{p}=\vec{0}$ are 
\begin{align}
    &v_{K_M,\vec{p}=0}^1 = 
    \begin{bmatrix}
        \frac{-iw^{\text{AA}}_1 -w^{\text{AA}}_2}{\sqrt{(w^{\text{AA}}_1)^2 + (w^{\text{AA}}_2)^2}} & 1 
    \end{bmatrix} \nonumber\\ 
    & v_{K_M,\vec{p}=0}^2 = 
    \begin{bmatrix}
        \frac{iw^{\text{AA}}_1+w^{\text{AA}}_2}{\sqrt{(w^{\text{AA}}_1)^2 + (w^{\text{AA}}_2)^2}} & 1 
    \end{bmatrix} 
    \label{eq:eigenvector_at_K}
\end{align}

We then perform a small $|\vec{p}|$ expansion and find 
\begin{align}
 &    E_{K_M,\vec{p}}^1 \approx  E_{K_M,\vec{p}=\vec{0}}^1 
 + \frac{p_y|\vec{q}_0|}{4\sqrt{3}m_x}
 +
  \frac{p_y^2}{4m_x}
  + \frac{p_x^2}{8m_x}
 \left[  4- \frac{|\vec{q}_0|^2}{m_x\sqrt{ (w^{\text{AA}}_1)^2 +(w^{\text{AA}}_2)^2}}
 \right]  \nonumber\\ 
 &    E_{K_M,\vec{p}}^2 \approx  E_{K_M,\vec{p}=\vec{0}}^2 
 + \frac{p_y|\vec{q}_0|}{4\sqrt{3}m_x}
 +
  \frac{p_y^2}{4m_x}
  + \frac{p_x^2}{8m_x}
 \left[  4+ \frac{|\vec{q}_0|^2}{m_x\sqrt{ (w^{\text{AA}}_1)^2 +(w^{\text{AA}}_2)^2}}
 \right]  
\end{align}

As we discussed in the~\cref{sec:app:ham_even_odd}, the dispersion of the lowest band is flat along $x$ direction due to the larger distance between Wannier orbitals. We can then estimate the dispersion along $y$ direction by comparing the energy difference of the lowest band between $\Gamma_M$ and $K_M$ point. 
We find 
\begin{align}
    E_{\Gamma_M}^{lowest} - E_{K_M}^{lowest} = &\frac{|\vec{q}_0|^2 }{8m_x}+\frac{3 |\vec{q}_0|^2}{8m_y} -|w^{\prime \text{AA}}_3|-\sqrt{(w^{\text{AA}}_1)^2 +(w^{\text{AA}}_2)^2} 
    -\bigg[ \frac{|\vec{q}_0|^2}{24m_y} - \frac{|\vec{q}_0|^2}{8m_x} - \sqrt{(w^{\text{AA}}_1)^2 +(w^{\text{AA}}_2)^2} \bigg] \nonumber\\ 
    =& \frac{|\vec{q}_0|^2}{3m_y}-|w^{\prime \text{AA}}_3|
\end{align}
Therefore, the condition for the flat-$y$ dispersion is 
\begin{align}
\label{eq:app:flat_condition}
    \frac{|\vec{q}_0|^2}{3m_y}=|w^{\prime \text{AA}}_3|
\end{align}

\subsubsection{$M_M$ point}
Finally, we discuss the Hamiltonian at $M_M$ point, where $\vec{Q}_3=M$. If we only consider the $Q$ points that are closest to the $M_M$ point, then the energy is just 
\begin{align}
    E_M = \frac{[Q_{3,x}-M_{M,x}]^2}{2m_x}+\frac{[Q_{3,y}-M_{M,y}]^2}{2m_y}=0 \,. 
\end{align} 
We can also consider the contribution from the two nearest-neighbor $Q$ points, which are (\cref{fig:high-sym-point})
\begin{align}
    \vec{Q}_1 =(-\frac{1}{2},\frac{\sqrt{3}}{2})|\vec{q}_0|,\quad \vec{Q}_5= (\frac{3}{2},\frac{\sqrt{3}}{2})|\vec{q}_0|
\end{align}
We can then take the basis of 
\begin{align}
        & \hat{c}_{\vec{p}+M, \vec{Q}_1,+},\quad \hat{c}_{\vec{p}+M,\vec{Q}_3,-},\quad
\hat{c}_{\vec{p}+M,\vec{Q}_5,+},\quad 
\end{align}
and the single-particle Hamiltonian reads \begin{align}
    H^{M_M}_{\vec{p}} = 
    \begin{bmatrix}
        \epsilon^0_{\vec{p}+M-\vec{Q}_1} &  i w^{\text{AA}}_1 +w^{\text{AA}}_2 & 0 \\ 
       - i w^{\text{AA}}_1 +w^{\text{AA}}_2 & \epsilon^0_{\vec{p}+M-\vec{Q}_3} &  i w^{\text{AA}}_1 +w^{\text{AA}}_2\\
        0 & -i w^{\text{AA}}_1 +w^{\text{AA}}_2& \epsilon_{\vec{p}+M-\vec{Q}_5}^0
    \end{bmatrix}
\end{align}

The eigenvalues exactly at $M$ point with $\vec{p}=0$ are
\begin{align}
 E_{M_M}^1 =& \frac{|\vec{q}_0|^2}{4m_x}
  - \sqrt{ \frac{|\vec{q}_0|^4}{16m_x^2} + 2\left( (w^{\text{AA}}_1)^2 +(w^{\text{AA}}_2)^2\right) }
 \nonumber\\ 
   E_{M_M}^2 =& \frac{|\vec{q}_0|^2}{2m_x} \nonumber\\
    E_{M}^3 =& \frac{|\vec{q}_0|^2}{4m_x}
  + \sqrt{ \frac{|\vec{q}_0|^4}{16m_x^2} + 2\left( (w^{\text{AA}}_1)^2 +(w^{\text{AA}}_2)^2\right) }
\end{align}

The eigenvectors at $M_M$ points are 
\begin{align}
& v_{M_M}^1 =\frac{1}{A_M^1} 
    \begin{bmatrix}
       - 2\sqrt{2}(iw^{\text{AA}}_1+w^{\text{AA}}_2) 
        & \frac{1}{2}\left( \frac{\vec{q}_0^2}{m_x} + 
        \sqrt{ \frac{|\vec{q}_0|^4}{m_x^2} + 32(w^{\text{AA}}_1)^2 + 32(w^{\text{AA}}_2)^2}\right)
        & 2\sqrt{2} (iw^{\text{AA}}_1-w^{\text{AA}}_2)
    \end{bmatrix}\nonumber\\ 
  &  v_{M_M}^2 =\frac{1}{\sqrt{2}} 
    \begin{bmatrix}
        \frac{w^{\text{AA}}_1-iw^{\text{AA}}_2}{\sqrt{(w^{\text{AA}}_1)^2+(w^{\text{AA}}_2)^2}} & 0 & \frac{w^{\text{AA}}_1+iw^{\text{AA}}_2}{\sqrt{(w^{\text{AA}}_1)^2+(w^{\text{AA}}_2)^2}}
    \end{bmatrix} \nonumber\\ 
    &v_{M_M}^3 = \frac{1}{A_M^3} 
    \begin{bmatrix}
        2\sqrt{2}(iw^{\text{AA}}_1+w^{\text{AA}}_2) 
        & \frac{1}{2}\left( \frac{\vec{q}_0^2}{m_x}- 
        \sqrt{ \frac{|\vec{q}_0|^4}{m_x^2} + 32(w^{\text{AA}}_1)^2 + 32(w^{\text{AA}}_2)^2}\right)
        & 2\sqrt{2} (iw^{\text{AA}}_1-w^{\text{AA}}_2)
    \end{bmatrix}
    \label{eq:eigenvector_at_M}
\end{align}
where $A_M^1,A_M^3$ are the normalization factor. 

\subsection{Extracting parameters}
We show that we could estimate the values of parameters by comparing the analytical results we obtained in this section and numerical DFT results. We will also compare the parameters obtained in this section with the fitted parameters given in~\cref{app:sec:fitted_models}. 
\subsubsection{Energy gap} 
We focus on the $\Gamma$ point. The eigenvalues of the analytical Hamiltonian at $\Gamma$ point are(~\cref{eq:eigen_value_at_Gamma}) 
\begin{align}
      &E_{\Gamma}^{1} = \frac{|\vec{q}_0|^2}{8m_x}+\frac{3|\vec{q}_0|^2}{8m_y} - w^{\prime \text{AA}}_3 - \sqrt{ (w^{\text{AA}}_1)^2 + (w^{\text{AA}}_2)^2 } \nonumber\\ 
  &E_{\Gamma}^{2} =  \frac{|\vec{q}_0|^2}{8m_x}+\frac{3|\vec{q}_0|^2}{8m_y} + w^{\prime \text{AA}}_3 - \sqrt{ (w^{\text{AA}}_1)^2 + (w^{\text{AA}}_2)^2 } \nonumber\\ 
  &E_{\Gamma}^{3} = \frac{|\vec{q}_0|^2}{8m_x}+\frac{3|\vec{q}_0|^2}{8m_y}- w^{\prime \text{AA}}_3 + \sqrt{ (w^{\text{AA}}_1)^2 + (w^{\text{AA}}_2)^2 } \nonumber\\ 
  &E_{\Gamma}^{4} =   \frac{|\vec{q}_0|^2}{8m_x}+\frac{3|\vec{q}_0|^2}{8m_y} + w^{\prime \text{AA}}_3 +\sqrt{ (w^{\text{AA}}_1)^2 + (w^{\text{AA}}_2 )^2 } 
\end{align}
We now consider the DFT spectrum at $\theta = 3.89^\circ $. It is important to notice that we only consider the DFT bands (the lowest three bands) whose weights are mostly located at the first $Q$ shell ($\vec{Q}_{1},\vec{Q}_2,\vec{Q}_3,\vec{Q}_4$ in~\cref{fig:high-sym-point}). The DFT spectrums are 
\begin{align}
    &E^1_{\Gamma, DFT} =0\text{meV} \nonumber\\ 
    &E^2_{\Gamma,DFT} = 41.3\text{meV}\nonumber\\ 
    &E^{3}_{\Gamma,DFT} =  160.23\text{meV} 
\end{align}
We can immediately observe the $w^{\prime \text{AA}}_3$ can be obtained by 
\begin{align}
    |w^{\prime \text{AA}}_3| = \frac{1}{2} (E_{\Gamma,DFT}^2 -E_{\Gamma,DFT}^1 ) =20.6\text{meV}
    \label{eq:app:ana_DFT_wp3}
\end{align} 
Here, the two lowest-energy states have spinless $C_{2x}$ eigenvalues $+1,-1$ respectively. In the DFT simulations, we find the lowest energy states have $C_{2x}$ eigenvalue $+1$ which indicates $w^{\prime \text{AA}}_3<0$. We also note that the current parameters approximately satisfy the flat condition given in~\cref{eq:app:flat_condition} with $|\vec{q}_0|^2/(3m_y|w^{\prime \text{AA}}_3|) \approx 0.7$

In addition, we also have 
\begin{align}
\sqrt{ \left( (w^{\text{AA}}_1)^2 +(w^{\text{AA}}_2)^2\right) } = \frac{1}{2}(E_{\Gamma,DFT}^3- E_{\Gamma,DFT}^1) \quad 
\end{align}
which gives 
\begin{align}
    \sqrt{ \left( (w^{\text{AA}}_1)^2 +(w^{\text{AA}}_2)^2\right) }  = \frac{1}{2}
    \left[   \frac{1}{2}(E_{\Gamma,DFT}^3- E_{\Gamma,DFT}^1)
    \right ] = 80.15\text{meV}
     \label{eq:app:ana_DFT_w12}
\end{align}
We thus obtain the values of $w^{\prime \text{AA}}_3, \sqrt{(w^{\text{AA}}_1)^2+(w^{\text{AA}}_2)^2}$ from DFT spectrum. In the next section, we determine the ratio $w^{\text{AA}}_1/w^{\text{AA}}_2$ by investigating the wavefunctions of the bands obtained from DFT calculations.

\subsubsection{Wavefunction} 
We now compare the wave function between the analytical model (~\cref{eq:eigenvector_at_M,eq:eigenvector_at_K}) and the DFT calculations for the lowest-energy band at $\theta= 3.89^\circ$.

At $\Gamma$ point, the DFT band eigenvector of the lowest eigenstate is 
\begin{align}
    v_{\Gamma}^1 = 
    \begin{bmatrix}
        0.45e^{-i0.86\pi} & 0.45e^{-i0.86\pi}
        & 0.45 & 0.45
    \end{bmatrix}
\end{align}
We first note that $\sum_{i=1,2,3,4}|[v_{\Gamma}^1]_i|^2 = 0.81<1 $. This is because we only consider the weight at the four nearest-neighbor $Q$ points. 

 In the DFT calculation, the lowest-energy state has $\tilde{C}_{2x}$ eigenvalue $+1$ which indicates $w^{\prime \text{AA}}_3<0$. The analytical wavefunction of the lowest band is
\begin{align}
    v_{\Gamma} =\frac{1}{2}
    \begin{bmatrix}
        \frac{-iw^{\text{AA}}_1  - w^{\text{AA}}_2 }{\sqrt{(w^{\text{AA}}_1)^2+(w^{\text{AA}}_2 )^2} } 
        & 
        \frac{-iw^{\text{AA}}_1  - w^{\text{AA}}_2 }{\sqrt{(w^{\text{AA}}_1)^2+(w^{\text{AA}}_2)^2} }   & 1 & 1 
    \end{bmatrix}
\end{align}
By fitting the phase factor, we obtain
\begin{align}
 \frac{iw^{\text{AA}}_1  + w^{\text{AA}}_2 }{\sqrt{(w^{\text{AA}}_1)^2+(w^{\text{AA}}_2 )^2} }  = e^{i0.14\pi}
 \label{eq:app:phase_w1w2_Gam}
\end{align}

We next investigate the $K$ point. The wavefunction obtained from DFT is 
\begin{align}
    v_K^1 = 
    \begin{bmatrix}
        0.59e^{-i0.72\pi} & 0.59
    \end{bmatrix}
\end{align}
where we note that $\sum_{i=1,2}|[v_K^1]_i|^2=0.70<1$. This is again because we only consider the weight at the two $Q$ points that are closest to $K$. 

Analytically, the wavefunction is 
\begin{align}
    v_{K,\vec{p}=0}^1 = 
    \begin{bmatrix}
        \frac{-iw^{\text{AA}}_1 -w^{\text{AA}}_2}{\sqrt{(w^{\text{AA}}_1)^2 + (w^{\text{AA}}_2)^2}} & 1 
    \end{bmatrix}
\end{align}
By matching the phase factor in the DFT wavefunction, we obtain
\begin{align}
     \frac{iw^{\text{AA}}_1  + w^{\text{AA}}_2 }{\sqrt{(w^{\text{AA}}_1)^2+(w^{\text{AA}}_2 )^2} }  = e^{i0.28\pi}
     \label{eq:app:phase_w1w2_K}
\end{align}

We next investigate the $M$ point. The wavefunction obtained from DFT is 
\begin{align}
    v_M^1 = 
    \begin{bmatrix}
        0.43e^{-i1.26\pi} & i0.67 & 0.43e^{i0.26\pi }
    \end{bmatrix}
\end{align}
where we note that $\sum_{i=1,2}|[v_M^1]_i|^2=0.81<1$. This is again because we only consider the three $Q$ points that are closest to the $M$. 
The analytical wavefunction is 
\begin{align}
    v_{M}^1 = \frac{1}{A_{v_{M}^1}}
    \begin{bmatrix}
       4( w^{\text{AA}}_1 -iw^{\text{AA}}_2)  & 
       i\left( \frac{|\vec{q}_0|^2}{m_x} + \sqrt{\frac{|\vec{q}_0|^4}{m_x^2} 
       + 32 \left[ (w^{\text{AA}}_1)^2 +(w^{\text{AA}}_2)^2\right]
       }
       \right) 
       & -4( w^{\text{AA}}_1 +iw^{\text{AA}}_2)
    \end{bmatrix}  
\end{align}
We compare the phase factors of the wavefunctions between DFT calculations and analytical calculations. We find 
\begin{align}
    \frac{iw^{\text{AA}}_1  + w^{\text{AA}}_2 }{\sqrt{(w^{\text{AA}}_1)^2+(w^{\text{AA}}_2 )^2} }  = e^{i0.24\pi}
    \label{eq:app:phase_w1w2_M}
\end{align}

In summary, the $ \frac{iw^{\text{AA}}_1  + w^{\text{AA}}_2 }{\sqrt{(w^{\text{AA}}_1)^2+(w^{\text{AA}}_2 )^2} } $ obtained from $\Gamma, K,M $ are $e^{i0.14\pi}, e^{i0.28\pi}, e^{i0.24\pi}$ respectively. The deviation may arise due to the simplifications (only considering a small set of $Q$ points) we made when solving the analytical model. 
We then determined the phase factor of $iw^{\text{AA}}_1+w^{\text{AA}}_2$ by taking the average value of the fitted results at $\Gamma,K,M$ (\cref{eq:app:phase_w1w2_Gam,eq:app:phase_w1w2_K,eq:app:phase_w1w2_M})
\begin{align}
    &\phi = \frac{1}{3} \left( 0.14 + 0.28 + 0.24 \right)\pi \approx 0.22\pi \nonumber\\ 
    & \frac{iw^{\text{AA}}_1  + w^{\text{AA}}_2 }{\sqrt{(w^{\text{AA}}_1)^2+(w^{\text{AA}}_2 )^2} }  = e^{i0.22\pi}
    \label{eq:app:ana_DFT_para_2}
\end{align}

\begin{figure}[t]
    \centering
    \includegraphics[width=0.8\linewidth]{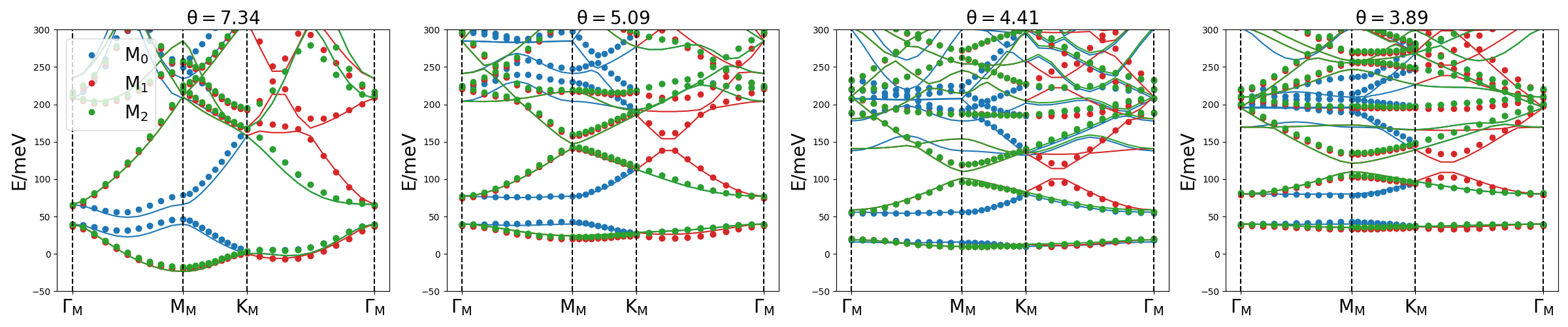}
    \caption{Disperion from the DFT (dots) and continuous model with parameters given in~\cref{tab:app:ana_fit_para}(lines). Blue, red and green denote valley-$0$, valley-$1$ and valley-$2$ respectively. }
    \label{fig:app:disp_ana}
\end{figure}

We adapt the same procedure and estimate the parameters at $\theta =3.90^\circ, 4.41^\circ, 5.09^\circ, 7.34^\circ$. 
In addition, we use the masses $m_x,m_y$ obtained from monolayer DFT calculations in the continuum model.
The results have been shown in~\cref{tab:app:ana_fit_para}. In~\cref{tab:app:ana_fit_para_ov},  we show the minimum overlapping of wavefunctions of the lowest band between the DFT model and the continuous model along the $\Gamma-M-K-\Gamma$ line. We could observe that using our estimated parameters the continuous model could reproduce the lowest band well. In~\cref{fig:app:disp_ana}, we compare the dispersion from the DFT model and the continuous model. In addition, we also observe the parameters we extract from analytical results qualitatively reproduce the fitting result shown in~\cref{app:tab:SnSe2_AA_2} for $\theta=3.89^\circ, 4.41^\circ, 5.09^\circ$. For $\theta=7.34^\circ$, the parameters considered in the fitting (~\cref{app:tab:SnSe2_AA_2}) are $w^{\text{AA}}_1,w^{\text{AA}}_2,w^{\prime \text{AA}}_1$ instead of $w^{\text{AA}}_1,w^{\text{AA}}_2,w^{\prime \text{AA}}_3$ that we used here.

\begin{table}[t]
    \centering
    \begin{tabular}{c|c|c| c | c |c}
      $\theta$ & $\sqrt{(w^{\text{AA}}_1)^2+(w^{\text{AA}}_2)^2}$ & $\text{arg}(iw^{\text{AA}}_1+w^{\text{AA}}_2)/\pi$ ($\Gamma_M,M_M,K_M$)  & $w^{\text{AA}}_1$ & $w^{\text{AA}}_2$
         & $w^{\prime \text{AA}}_3$ \\ 
         \hline  
            $7.34^\circ$ & 86.2meV &0.22 (0.16,0.28,0.24) & 54.9meV & 66.4meV &-13.1meV \\
              $5.09^\circ$ & 91.7 meV &0.25 (0.15,0.28,0.24) & 64.8meV & 64.8meV & -18.7meV\\
                  $4.41^\circ$ & 67.9meV& 0.27 (0.37,0.23,0.24) & 62.3meV & 27.0meV &-20.2meV\\ 
    $3.89^\circ$ & 80.1meV & 0.22 (0.14,0.28,0.24) & 51.1meV & 61.7meV&  -20.6meV\\ 
    \end{tabular}
    \caption{Estimation of the parameters from the analytical formula. In the third column, we show fitted phase factor $\text{arg}(iw^{\text{AA}}_1+w^{\text{AA}}_2)/\pi$ at $\Gamma_M,K_M,M_M$ in the bracket respectively.}
    \label{tab:app:ana_fit_para}
\end{table}

\begin{table}[t]
    \centering
    \begin{tabular}{c|c|c|c |c }
      $\theta$ &
       $7.34^\circ$ & 
        $5.09^\circ$ & 
         $4.41^\circ$ & 
    $3.89^\circ$  
    \\ 
    \hline 
    $|\langle \phi_{DFT}|\phi_{Cont}\rangle|$ & 0.98&0.97 &  0.93 &  0.96 
    \end{tabular}
    \caption{Minimum overlapping between the wavefunctions of continuous model and DFT model along the $\Gamma$-$M$-$K$-$\Gamma$ line. The parameters of the continuous model are given in~\cref{tab:app:ana_fit_para}. }
    \label{tab:app:ana_fit_para_ov}
\end{table}

\section{Models fitted to \textit{ab initio} simulations}\label{app:sec:fitted_models}

This \siSection{} presents detailed results on the band structure of twisted \ch{SnSe2} and \ch{ZrS2} bilayers. We begin by offering an overview of the key findings in this \siSection{}, which include the construction of various models and their comparison with the \textit{ab initio} spectra. We also present results on the LDOS, Berry curvature, and Wilson loops of the models. 

\subsection{Overview of the results}\label{app:sec:fitted_models:overview}

Starting with the \textit{ab initio} band structure computed as detailed in \cref{app:sec:DFT_bilayer}, we derive the plane-wave valley-projected Hamiltonian using the method outlined in \cref{app:sec:first_princ_ham_valley:algorithm}. From there, we construct a series of continuum models that \emph{accurately} reproduce the \textit{ab initio} spectrum of the system. The methods described in \cref{app:sec:fitting_method} are used to construct these continuum Hamiltonians. We then analyze their spectra across the full range of monolayers, stacking configurations, and commensurate angles considered in this work. A summary of all the results (detailed below) is provided in \cref{app:tab:fitting_overview}.

\subsubsection{Types of models constructed}\label{app:sec:fitted_models:overview:types_of_models}

We find it useful to construct several types of continuum models with varying levels of complexity. These models include numerically \emph{exact} ones that accurately match both the dispersion and wave functions of the first five bands, with an overlap of more than $0.999$ for the first band and over $0.95$ for the first five. To gain analytical insight and explore the \emph{emergent} symmetries of the system, we also develop simplified models, beginning with the first moiré harmonic Hamiltonian presented in \cref{app:eqn:full_parameterization_AA_stacking,app:eqn:full_parameterization_AB_stacking}, and progressively reduce the number of parameters. The five types of models we consider, in order of \emph{decreasing} complexity, are:
\begin{enumerate}
	\item \textbf{The full model}: The full model is constructed using the linear-fitting method outlined in \cref{app:sec:fitting_method:linear:low_energy_partial}. We first obtain the symmetry-obeying parameterization of the moir\'e Hamiltonian from \cref{app:eqn:general_moire_potential_momentum}, which includes gradient terms and higher-order harmonics in the moir\'e potential. For the moir\'e Hamiltonian $h_{\vec{Q}, \vec{Q}'} \left( \vec{k} \right)$, we consider moir\'e harmonics with $\abs{\vec{Q} - \vec{Q}'} \leq 4 \abs{\vec{b}_{M_1}}$, and include gradient terms up to degree $n_x + n_y = 4$ ($n_x + n_y = 10$) for the $\vec{Q} \neq \vec{Q}'$ ($\vec{Q} = \vec{Q}'$) terms. As described in \cref{app:sec:fitting_method:linear:low_energy_partial}, the $\vec{Q}$-off-diagonal terms are fitted using an energy weighting function of the form from \cref{app:eqn:fit_energy_weight}. The parameters used for the weighting function are chosen to capture the first four or five set of spinful bands within each valley and are listed in \cref{app:sec:fitted_models:param_values}. The effective masses of the monolayer are treated as fitting parameters. We reduce the number of parameters in the moiré Hamiltonian, initially exceeding a thousand, by systematically removing the least significant ones using the algorithm outlined in \cref{app:sec:fitting_method:reducing_number_parameters}. The final model retains the smallest number of parameters, ensuring that the relative error, as defined in \cref{app:eqn:relative_error_definition}, remains below $\varepsilon = 0.02 \%,\, 0.04 \%,\, 0.02 \%,\, 0.015 \%$ for AA-stacked \ch{SnSe2}, AB-stacked \ch{SnSe2}, AA-stacked \ch{ZrS2}, and AB-stacked \ch{ZrS2}, respectively. 

	\item \textbf{The full first moir\'e harmonic model}: This model is constructed using the method outlined in \cref{app:sec:fitting_method:nonlinear} and employs the first moir\'e harmonic Hamiltonian from \cref{app:eqn:full_parameterization_AA_stacking,app:eqn:full_parameterization_AB_stacking}, which includes 16 interlayer hopping parameters. The cost function from \cref{app:eqn:non_linear_cost_function} is constructed using the projectors into the first $\mathcal{N}_{\tBand} = 2$ ($\mathcal{N}_{\tBand} = 1$) sets of spinful bands within each valley for $\theta \leq \SI{6.01}{\degree}$ ($\theta > \SI{6.01}{\degree}$). The effective masses of the monolayer are also included as fitting parameters.
	
	\item \textbf{The reduced first moir\'e Harmonic model}: This model is derived from the full first moir\'e harmonic model by reducing the number of interlayer hopping parameters to just five, using the step-wise reduction method described in \cref{app:sec:fitting_method:reducing_number_parameters}.
	
	\item \textbf{The full first moir\'e harmonic model with the zero-twist constraints imposed}: This model is similar to the full first moir\'e harmonic model, but with the zero-twist constraints applied. To prevent issues related to overfitting, the effective masses of the monolayer are not treated as fitting parameters; instead, they are fixed to values extracted from the monolayer band structure in \cref{app:sec:DFT_single_layer}. Additionally, the moir\'e potential parameters with the largest absolute values, specifically $w^{\text{AA}}_1$ and $w^{\text{AA}}_2$ (or $w^{\text{AB}}_2$ for AB-stacking), are also fixed to their values obtained from the full model, rather than being treated as fitting parameters.

	\item \textbf{The reduced first moir\'e harmonic model with the zero-twist constraints imposed}: This model is derived from the full first moir\'e harmonic model, with the zero-twist constraints applied. The number of interlayer hopping parameters is reduced to just three by using the step-wise reduction method described in \cref{app:sec:fitting_method:reducing_number_parameters}.
\end{enumerate}

\begingroup
\renewcommand{\arraystretch}{1.6} \begin{table}[!t]
	\centering
	\resizebox{\textwidth}{!}{\begin{tabular}{|l|l|l|l|l|l|l|l|l|l|l|}
			\hline
			\multirow{2}{*}{Monolayer} & \multirow{2}{*}{Stacking} & \multicolumn{5}{c|}{Models} & \multirow{2}{*}{$\theta$} & \multirow{2}{*}{Spectrum} & \multirow{2}{*}{\makecell[tl]{Additional\\results}} & \multirow{2}{*}{\makecell[tl]{Wilson\\loops}} \\
			\cline{3-7} 
			& & 1 & 2 & 3 & 4 & 5 & & & & \\
			\hline
			\multirow{12}{*}{\ch{SnSe2}} & \multirow{6}{*}{AA} & \multirow{6}{*}{\cref{app:tab:linear_SnSe2_AA}} & \multirow{6}{*}{\cref{app:tab:SnSe2_AA_1}} & \multirow{6}{*}{\cref{app:tab:SnSe2_AA_2}} & \multirow{6}{*}{\cref{app:tab:SnSe2_AA_3}} & \multirow{6}{*}{\cref{app:tab:SnSe2_AA_4}} & $\SI{9.43}{\degree}$ & \makecell[tl]{\cref{app:fig:fit_BS_1}} & \makecell[tl]{\cref{app:fig:fit_2D_bands_1_1}} & \makecell[tl]{\cref{app:fig:fit_Wilson_1_WilsonX,app:fig:fit_Wilson_1_WilsonY}}\\ 
 \cline{8-8} \cline{9-11} 
 &  &  &  &  &  &  & $\SI{7.34}{\degree}$ & \makecell[tl]{\cref{app:fig:fit_BS_2}} & \makecell[tl]{\cref{app:fig:fit_2D_bands_2_1}} & \makecell[tl]{\cref{app:fig:fit_Wilson_2_WilsonX,app:fig:fit_Wilson_2_WilsonY}}\\ 
 \cline{8-8} \cline{9-11} 
 &  &  &  &  &  &  & $\SI{6.01}{\degree}$ & \makecell[tl]{\cref{app:fig:fit_BS_3}} & \makecell[tl]{\cref{app:fig:fit_2D_bands_3_1}} & \makecell[tl]{\cref{app:fig:fit_Wilson_3_WilsonX,app:fig:fit_Wilson_3_WilsonY}}\\ 
 \cline{8-8} \cline{9-11} 
 &  &  &  &  &  &  & $\SI{5.09}{\degree}$ & \makecell[tl]{\cref{app:fig:fit_BS_4}} & \makecell[tl]{\cref{app:fig:fit_2D_bands_4_1,app:fig:fit_2D_bands_4_2}} & \makecell[tl]{\cref{app:fig:fit_Wilson_4_WilsonX,app:fig:fit_Wilson_4_WilsonY}}\\ 
 \cline{8-8} \cline{9-11} 
 &  &  &  &  &  &  & $\SI{4.41}{\degree}$ & \makecell[tl]{\cref{app:fig:fit_BS_5}} & \makecell[tl]{\cref{app:fig:fit_2D_bands_5_1,app:fig:fit_2D_bands_5_2}} & \makecell[tl]{\cref{app:fig:fit_Wilson_5_WilsonX,app:fig:fit_Wilson_5_WilsonY}}\\ 
 \cline{8-8} \cline{9-11} 
 &  &  &  &  &  &  & $\SI{3.89}{\degree}$ & \makecell[tl]{\cref{app:fig:fit_BS_6}} & \makecell[tl]{\cref{app:fig:fit_2D_bands_6_1,app:fig:fit_2D_bands_6_2}} & \makecell[tl]{\cref{app:fig:fit_Wilson_6_WilsonX,app:fig:fit_Wilson_6_WilsonY}}\\ 
 \cline{2-11} 
 & \multirow{6}{*}{AB} & \multirow{6}{*}{\cref{app:tab:linear_SnSe2_AB}} & \multirow{6}{*}{\cref{app:tab:SnSe2_AB_1}} & \multirow{6}{*}{\cref{app:tab:SnSe2_AB_2}} & \multirow{6}{*}{\cref{app:tab:SnSe2_AB_3}} & \multirow{6}{*}{\cref{app:tab:SnSe2_AB_4}} & $\SI{9.43}{\degree}$ & \makecell[tl]{\cref{app:fig:fit_BS_7}} & \makecell[tl]{\cref{app:fig:fit_2D_bands_7_1}} & \makecell[tl]{\cref{app:fig:fit_Wilson_7_WilsonX,app:fig:fit_Wilson_7_WilsonY}}\\ 
 \cline{8-8} \cline{9-11} 
 &  &  &  &  &  &  & $\SI{7.34}{\degree}$ & \makecell[tl]{\cref{app:fig:fit_BS_8}} & \makecell[tl]{\cref{app:fig:fit_2D_bands_8_1}} & \makecell[tl]{\cref{app:fig:fit_Wilson_8_WilsonX,app:fig:fit_Wilson_8_WilsonY}}\\ 
 \cline{8-8} \cline{9-11} 
 &  &  &  &  &  &  & $\SI{6.01}{\degree}$ & \makecell[tl]{\cref{app:fig:fit_BS_9}} & \makecell[tl]{\cref{app:fig:fit_2D_bands_9_1}} & \makecell[tl]{\cref{app:fig:fit_Wilson_9_WilsonX,app:fig:fit_Wilson_9_WilsonY}}\\ 
 \cline{8-8} \cline{9-11} 
 &  &  &  &  &  &  & $\SI{5.09}{\degree}$ & \makecell[tl]{\cref{app:fig:fit_BS_10}} & \makecell[tl]{\cref{app:fig:fit_2D_bands_10_1}} & \makecell[tl]{\cref{app:fig:fit_Wilson_10_WilsonX,app:fig:fit_Wilson_10_WilsonY}}\\ 
 \cline{8-8} \cline{9-11} 
 &  &  &  &  &  &  & $\SI{4.41}{\degree}$ & \makecell[tl]{\cref{app:fig:fit_BS_11}} & \makecell[tl]{\cref{app:fig:fit_2D_bands_11_1,app:fig:fit_2D_bands_11_2}} & \makecell[tl]{\cref{app:fig:fit_Wilson_11_WilsonX,app:fig:fit_Wilson_11_WilsonY}}\\ 
 \cline{8-8} \cline{9-11} 
 &  &  &  &  &  &  & $\SI{3.89}{\degree}$ & \makecell[tl]{\cref{app:fig:fit_BS_12}} & \makecell[tl]{\cref{app:fig:fit_2D_bands_12_1,app:fig:fit_2D_bands_12_2}} & \makecell[tl]{\cref{app:fig:fit_Wilson_12_WilsonX,app:fig:fit_Wilson_12_WilsonY}}\\ 
 \hline 
\multirow{12}{*}{\ch{ZrS2}} & \multirow{6}{*}{AA} & \multirow{6}{*}{\cref{app:tab:linear_ZrS2_AA}} & \multirow{6}{*}{\cref{app:tab:ZrS2_AA_1}} & \multirow{6}{*}{\cref{app:tab:ZrS2_AA_2}} & \multirow{6}{*}{\cref{app:tab:ZrS2_AA_3}} & \multirow{6}{*}{\cref{app:tab:ZrS2_AA_4}} & $\SI{9.43}{\degree}$ & \makecell[tl]{\cref{app:fig:fit_BS_13}} & \makecell[tl]{\cref{app:fig:fit_2D_bands_13_1}} & \makecell[tl]{\cref{app:fig:fit_Wilson_13_WilsonX,app:fig:fit_Wilson_13_WilsonY}}\\ 
 \cline{8-8} \cline{9-11} 
 &  &  &  &  &  &  & $\SI{7.34}{\degree}$ & \makecell[tl]{\cref{app:fig:fit_BS_14}} & \makecell[tl]{\cref{app:fig:fit_2D_bands_14_1}} & \makecell[tl]{\cref{app:fig:fit_Wilson_14_WilsonX,app:fig:fit_Wilson_14_WilsonY}}\\ 
 \cline{8-8} \cline{9-11} 
 &  &  &  &  &  &  & $\SI{6.01}{\degree}$ & \makecell[tl]{\cref{app:fig:fit_BS_15}} & \makecell[tl]{\cref{app:fig:fit_2D_bands_15_1,app:fig:fit_2D_bands_15_2}} & \makecell[tl]{\cref{app:fig:fit_Wilson_15_WilsonX,app:fig:fit_Wilson_15_WilsonY}}\\ 
 \cline{8-8} \cline{9-11} 
 &  &  &  &  &  &  & $\SI{5.09}{\degree}$ & \makecell[tl]{\cref{app:fig:fit_BS_16}} & \makecell[tl]{\cref{app:fig:fit_2D_bands_16_1,app:fig:fit_2D_bands_16_2}} & \makecell[tl]{\cref{app:fig:fit_Wilson_16_WilsonX,app:fig:fit_Wilson_16_WilsonY}}\\ 
 \cline{8-8} \cline{9-11} 
 &  &  &  &  &  &  & $\SI{4.41}{\degree}$ & \makecell[tl]{\cref{app:fig:fit_BS_17}} & \makecell[tl]{\cref{app:fig:fit_2D_bands_17_1,app:fig:fit_2D_bands_17_2}} & \makecell[tl]{\cref{app:fig:fit_Wilson_17_WilsonX,app:fig:fit_Wilson_17_WilsonY}}\\ 
 \cline{8-8} \cline{9-11} 
 &  &  &  &  &  &  & $\SI{3.89}{\degree}$ & \makecell[tl]{\cref{app:fig:fit_BS_18}} & \makecell[tl]{\cref{app:fig:fit_2D_bands_18_1,app:fig:fit_2D_bands_18_2}} & \makecell[tl]{\cref{app:fig:fit_Wilson_18_WilsonX,app:fig:fit_Wilson_18_WilsonY}}\\ 
 \cline{2-11} 
 & \multirow{6}{*}{AB} & \multirow{6}{*}{\cref{app:tab:linear_ZrS2_AB}} & \multirow{6}{*}{\cref{app:tab:ZrS2_AB_1}} & \multirow{6}{*}{\cref{app:tab:ZrS2_AB_2}} & \multirow{6}{*}{\cref{app:tab:ZrS2_AB_3}} & \multirow{6}{*}{\cref{app:tab:ZrS2_AB_4}} & $\SI{9.43}{\degree}$ & \makecell[tl]{\cref{app:fig:fit_BS_19}} & \makecell[tl]{\cref{app:fig:fit_2D_bands_19_1}} & \makecell[tl]{\cref{app:fig:fit_Wilson_19_WilsonX,app:fig:fit_Wilson_19_WilsonY}}\\ 
 \cline{8-8} \cline{9-11} 
 &  &  &  &  &  &  & $\SI{7.34}{\degree}$ & \makecell[tl]{\cref{app:fig:fit_BS_20}} & \makecell[tl]{\cref{app:fig:fit_2D_bands_20_1}} & \makecell[tl]{\cref{app:fig:fit_Wilson_20_WilsonX,app:fig:fit_Wilson_20_WilsonY}}\\ 
 \cline{8-8} \cline{9-11} 
 &  &  &  &  &  &  & $\SI{6.01}{\degree}$ & \makecell[tl]{\cref{app:fig:fit_BS_21}} & \makecell[tl]{\cref{app:fig:fit_2D_bands_21_1}} & \makecell[tl]{\cref{app:fig:fit_Wilson_21_WilsonX,app:fig:fit_Wilson_21_WilsonY}}\\ 
 \cline{8-8} \cline{9-11} 
 &  &  &  &  &  &  & $\SI{5.09}{\degree}$ & \makecell[tl]{\cref{app:fig:fit_BS_22}} & \makecell[tl]{\cref{app:fig:fit_2D_bands_22_1,app:fig:fit_2D_bands_22_2}} & \makecell[tl]{\cref{app:fig:fit_Wilson_22_WilsonX,app:fig:fit_Wilson_22_WilsonY}}\\ 
 \cline{8-8} \cline{9-11} 
 &  &  &  &  &  &  & $\SI{4.41}{\degree}$ & \makecell[tl]{\cref{app:fig:fit_BS_23}} & \makecell[tl]{\cref{app:fig:fit_2D_bands_23_1,app:fig:fit_2D_bands_23_2}} & \makecell[tl]{\cref{app:fig:fit_Wilson_23_WilsonX,app:fig:fit_Wilson_23_WilsonY}}\\ 
 \cline{8-8} \cline{9-11} 
 &  &  &  &  &  &  & $\SI{3.89}{\degree}$ & \makecell[tl]{\cref{app:fig:fit_BS_24}} & \makecell[tl]{\cref{app:fig:fit_2D_bands_24_1,app:fig:fit_2D_bands_24_2}} & \makecell[tl]{\cref{app:fig:fit_Wilson_24_WilsonX,app:fig:fit_Wilson_24_WilsonY}}  \\
			\hline
		\end{tabular}}
	\caption{Summary of additional results presented in this \siSection{}. For each monolayer and stacking configuration, we list the tables that detail the corresponding continuum models. Each of the five types of models listed in \cref{app:sec:fitted_models:overview:types_of_models} -- namely: (1) the full model, (2) the full first moir\'e harmonic model, (3) the reduced first moir\'e harmonic model, (4) the full first moir\'e harmonic model with zero-twist constraints, and (5) the reduced first moir\'e harmonic model with zero-twist constraints -- are summarized in one table. For each commensurate angle, we present the valley-resolved spectra of each continuum model and compare them with the \textit{ab initio} results (referenced in the ``Spectrum'' column). Additional results are shown for each set of gapped conduction bands of the moir\'e continuum model in valley $\eta = 0$. Lastly, we compute and plot the Wilson loops for each set of gapped conduction bands along the $\vec{b}_{M_1}$ and $\vec{b}_{M_2}$ lattice vectors.}
	\label{app:tab:fitting_overview}
\end{table}
\endgroup

Before analyzing the detailed spectrum of these five models, we establish a consistent notation for the wave functions and energy eigenvalues of the moiré Hamiltonian continuum models. Specifically, we denote by $\epsilon_{n,\eta} \left(\vec{k} \right)$ and $u_{\vec{Q},s,l;\eta,n} \left( \vec{k} \right)$ the energy and wave function, respectively, of the $n$-th conduction band ($n \geq 1$) in valley $0 \leq \eta \leq 2$, such that
\begin{equation}
	\sum_{\substack{s',l' \\ \vec{Q}' \in \mathcal{Q}_{\text{tot}}}}\left[h_{\vec{Q}, \vec{Q}'} \left( \vec{k} \right) \right]_{s l;s'l'}  u_{\vec{Q}',s',l';\eta,n} \left( \vec{k} \right) = \epsilon_{n,\eta} \left(\vec{k} \right) u_{\vec{Q},s,l;\eta,n} \left( \vec{k} \right).
\end{equation}
The bands are indexed by $n$ in the order of increasing energy at every moir\'e momentum. The corresponding band-basis fermion operators are defined as
\begin{equation}
	\hat{c}^\dagger_{\vec{k},n,\eta} \equiv \sum_{\substack{s,l \\ \vec{Q} \in \mathcal{Q}_{\text{tot}}}} u_{\vec{Q},s,l;\eta,n} \left( \vec{k} \right) \hat{c}^\dagger_{\vec{k},\vec{Q},s,l}.
\end{equation}	
Due to the valley symmetry, $u_{\vec{Q},s,l;\eta,n} \left( \vec{k} \right)$ is only supported on two of the three plane-wave sublattices, so we have
\begin{equation}
	u_{\vec{Q},s,l;\eta,n} \left( \vec{k} \right) = 0, \qq{if} \vec{Q} \not\in \mathcal{Q}_{\eta + l}.
\end{equation}
Outside the first moiré BZ, the band wave functions are defined by the embedding condition
\begin{equation}
	\label{app:eqn:embedding_condition_wave_function}
	u_{\vec{Q},s,l;\eta,n} \left( \vec{k} + \vec{G} \right) = u_{\vec{Q} + \vec{G},s,l;\eta,n} \left( \vec{k} \right), \qq{for} \vec{G} \in \mathcal{Q}. 
\end{equation}
We also introduce the projector onto the $m$-th set of spinful bands in valley $\eta$ as 
\begin{equation}
	P^{m,\eta}_{\vec{Q},s,l;\vec{Q}',s',l'} \left( \vec{k} \right) \equiv \sum_{n=2m-1}^{2m} u_{\vec{Q},s,l;\eta,n} \left( \vec{k} \right) u^{*}_{\vec{Q}',s',l';\eta,n} \left( \vec{k} \right), \qq{for} m \geq 1.
\end{equation}
Additionally, we use a similar notation for the \textit{ab initio} wave function and projectors, $u^{\tDFT}_{\vec{Q},s,l;\eta,n} \left( \vec{k} \right)$ and $P^{\tDFT,m,\eta}_{\vec{Q},s,l;\vec{Q}',s',l'} \left( \vec{k} \right)$, which are directly obtained from the \textit{ab initio} moiré Hamiltonian defined in \cref{app:eqn:dft_to_moire}. 

Finally, we also define the embedding matrix
\begin{equation}
	\mathcal{V}_{\vec{Q},s,l;\vec{Q}',s',l'} \left( \vec{G} \right)  \equiv \delta_{\vec{Q}+\vec{G},\vec{Q}'} \delta_{ss'} \delta_{ll'}, \qq{for any} \vec{G} \in \mathcal{Q},
\end{equation}
such that \cref{app:eqn:embedding_condition_wave_function} becomes equivalent to
\begin{equation}
	u_{\vec{Q},s,l;\eta,n} \left( \vec{k} + \vec{G} \right) = \sum_{\substack{s',l' \\ \vec{Q}' \in \mathcal{Q}_{\text{tot}}}} \mathcal{V}_{\vec{Q},s,l;\vec{Q}',s',l'} \left( \vec{G} \right) u_{\vec{Q}',s',l';\eta,n} \left( \vec{k} \right).
\end{equation}

\subsubsection{Overlaps, zero-twist symmetries, $\mathrm{SU} \left( {2} \right)$ symmetry}\label{app:sec:fitted_models:overview:overlaps}

For the continuum models considered here, we assess the goodness of fit by computing the overlap between the fitted and \textit{ab initio} band wave function for the $m$-th set of bands
\begin{equation}
	\label{app:eqn:overlap_with_DFT_definition}
	\mathcal{O}_{m} = \min_{\substack{ 0 \leq \eta \leq 2 \\ \vec{k} \in \mathcal{M}}} \sqrt{\frac{\Tr \left( P^{m,\eta} \left( \vec{k} \right)  P^{\tDFT,m,\eta} \left( \vec{k} \right)  \right)}{2}},
\end{equation}
where $\mathcal{M}$ are the set of $\vec{k}$ point along which the \textit{ab initio} band structure is computed. These overlaps are listed in the tables from \cref{app:sec:fitted_models:param_values}. Additionally, we compare the energy spectra of the fitted continuum model and the \textit{ab initio} Hamiltonians in \cref{app:sec:fitted_models:plots}.

We also evaluate how well the emergent $\tilde{M}_z$ and $\tilde{\mathcal{I}}$ symmetries, which appear in the zero-twist limit of the AA- and AB-stacked moiré Hamiltonians without gradient terms (as detailed in \cref{app:sec:add_sym:zero_twist}), are preserved within the first set of bands of the full model moiré Hamiltonian. If these symmetries were perfectly maintained, the projectors for the first set of bands would satisfy the following conditions
\begin{align}
	P^{\text{AA},1,\eta}_{\vec{Q},s,l;\vec{Q}',s',l'} \left( \vec{k} \right) &= P^{\text{AA},1,\eta}_{\vec{Q} + \vec{q}_{\eta},s,-l;\vec{Q}' + \vec{q}_{\eta},s',-l'} \left( \vec{k} + \vec{q}_{\eta} \right) \\
	P^{\text{AB},1,\eta}_{\vec{Q},s,l;\vec{Q}',s',l'} \left( \vec{k} \right) &= (-1)^{s} (-1)^{s'} P^{\text{AB},1,\eta}_{-\vec{Q} + \vec{q}_{\eta},s,-l;-\vec{Q}' + \vec{q}_{\eta},s',-l'} \left( -\vec{k} + \vec{q}_{\eta} \right),
\end{align} 
for the AA- and AB-stacked cases, respectively. These conditions follow directly from \cref{app:eqn:effective_zero_twist_mz,app:eqn:effective_zero_twist_I}.

To quantify the degree to which these symmetries are obeyed, we first define the transformed projectors (using the full model continuum Hamiltonian) as
\begin{align}
	P^{\prime\text{AA},1,\eta}_{\vec{Q},s,l;\vec{Q}',s',l'} \left( \vec{k} \right) &\equiv P^{\text{AA},1,\eta}_{\vec{Q} + \vec{q}_{\eta},s,-l;\vec{Q}' + \vec{q}_{\eta},s',-l'} \left( \vec{k} + \vec{q}_{\eta} \right) \\
	P^{\prime\text{AB},1,\eta}_{\vec{Q},s,l;\vec{Q}',s',l'} \left( \vec{k} \right) &\equiv (-1)^{s} (-1)^{s'} P^{\text{AB},1,\eta}_{-\vec{Q} + \vec{q}_{\eta},s,-l;-\vec{Q}' + \vec{q}_{\eta},s',-l'} \left( -\vec{k} + \vec{q}_{\eta} \right),
\end{align}
and compute the following overlap metrics
\begin{align}
	\varepsilon_{\tilde{M}_z} &\equiv 1 - \sqrt{ \frac{1}{N} \sum_{\vec{k} \in \text{MBZ}} \frac{\Tr \left( P^{\prime\text{AA},1,\eta} \left( \vec{k} \right) P^{\text{AA},1,\eta} \left( \vec{k} \right) \right)}{2}}, \qq{for} \eta = 0, \\
	\varepsilon_{\tilde{\mathcal{I}}} &\equiv 1 - \sqrt{ \frac{1}{N} \sum_{\vec{k} \in \text{MBZ}} \frac{\Tr \left( P^{\prime\text{AB},1,\eta} \left( \vec{k} \right) P^{\text{AB},1,\eta} \left( \vec{k} \right) \right)}{2}}, \qq{for} \eta = 0,
\end{align}
in the AA- and AB-stacked cases, respectively. 

Finally, we compute the degree of effective $\mathrm{SU} \left( {2} \right)$ symmetry breaking for the first set of bands in a manner similar to that described in \cref{app:eqn:su2_breaking_def}. To achieve this, we define the spin-symmetric and spin-antisymmetric components of the band projectors as follows:
\begin{align}
	P^{\tSym,m,\eta}_{\vec{Q},s,l;\vec{Q}',s',l'} \left( \vec{k} \right) &\equiv \frac{\delta_{s s'}}{2} \sum_{s''} P^{m,\eta}_{\vec{Q},s'',l;\vec{Q}',s'',l'} \left( \vec{k} \right), \nonumber \\
	P^{\tAsym,m,\eta}_{\vec{Q},s,l;\vec{Q}',s',l'} \left( \vec{k} \right) &\equiv P^{m,\eta}_{\vec{Q},s,l;\vec{Q}',s',l'} \left( \vec{k} \right) - P^{\tSym,m,\eta}_{\vec{Q},s,l;\vec{Q}',s',l'} \left( \vec{k} \right).
\end{align}
The effective $\mathrm{SU} \left( {2} \right)$ symmetry breaking for the first set of bands is then quantified as
\begin{equation}
	\varepsilon_{\mathrm{SU} \left( {2} \right)} \equiv \frac{1}{N} \sum_{\vec{k} \in \text{MBZ}} \frac{\norm{P^{\tAsym,1,\eta} \left( \vec{k} \right)}}{\norm{P^{1,\eta} \left( \vec{k} \right)}}, \qq{for} \eta = 0.
\end{equation}

\subsubsection{Berry curvature, LDOS, and Wilson loops}\label{app:sec:fitted_models:overview:2dPlots}

In \cref{app:sec:fitted_models:plots:details}, we present further details on the fitted continuum models. For simplicity, each figure in \cref{app:sec:fitted_models:plots:details} follows the same layout and considers a single set of gapped moir\'e bands for each of the five continuum models listed in \cref{app:sec:fitted_models:overview:types_of_models}. For each model, we plot the dispersion of the $m$-th set of gapped bands across the entire moir\'e BZ\
\begin{equation}
	E_{\vec{k}} = \frac{\epsilon_{2m-1,\eta} \left(\vec{k} \right) + \epsilon_{2m,\eta} \left(\vec{k} \right)}{2} - E_0, \qq{for} \eta = 0.
\end{equation}
We average the energies of the two (nearly) spin-degenerate bands and add an offset $E_0$ such that $\min_{\vec{k} \in \text{MBZ}} E_{\vec{k}} = 0$. Additionally, we plot the non-abelian Berry curvature~\cite{FUK05,BRO07} for the $m$-th set of bands
\begin{equation}
	\mathcal{F}_{\vec{k}} = i \Tr \left( P^{m,\eta} \left( \vec{k} \right) \commutator{\pdv{P^{m,\eta} \left( \vec{k} \right)}{k_x}}{\pdv{P^{m,\eta} \left( \vec{k} \right)}{k_y}} \right), \qq{for} \eta = 0.
\end{equation}

We also compute and display the layer-resolved LDOS for the $m$-th set of bands within the model. Specifically, we consider the following many-body state
\begin{equation}
	\ket{ \phi^{\eta,m} } = \left( \prod_{\vec{k} \in \text{MBZ}} \hat{c}^\dagger_{\vec{k},\eta,2m-1}\hat{c}^\dagger_{\vec{k},\eta,2m} \right) \ket{0}, \qq{for} m \geq 1,
\end{equation}
which corresponds to fully filling the $m$-th set of bands in valley $\eta$. The layer-resolved LDOS is defined using the continuum real-space operators from \cref{app:eqn:def_real_space_fermions_to_real} as
\begin{align}
	\rho_{\eta,m,l} \left( \vec{r} \right) &\equiv \sum_{s} \bra{\phi^{\eta,m}} \hat{\psi}^\dagger_{\eta,s,l} \left( \vec{r} \right) \hat{\psi}_{\eta,s,l} \left( \vec{r} \right) \ket{ \phi^{\eta,m} } \nonumber \\
	&= \frac{1}{N \Omega_0} \sum_{s} \sum_{\substack{\vec{k}, \vec{k}' \in \text{MBZ} \\ \vec{Q}, \vec{Q}' \in \mathcal{Q}_{\text{tot}}}} \bra{\phi^{\eta,m}} \hat{c}^\dagger_{\vec{k},\vec{Q},s,l} e^{-i \left( \vec{k} - \vec{Q} \right) \cdot \vec{r}} \hat{c}_{\vec{k}',\vec{Q}',s,l} e^{i \left( \vec{k}' - \vec{Q}' \right) \cdot \vec{r}} \ket{ \phi^{\eta,m} } \nonumber \\
	&= \frac{1}{N \Omega_0} \sum_{s} \sum_{n = 2 m -1}^{2m} \sum_{\substack{\vec{k} \in \text{MBZ} \\ \vec{Q}, \vec{Q}' \in \mathcal{Q}_{\text{tot}}}} u^{*}_{\vec{Q},s,l;\eta,n} \left( \vec{k} \right)  u_{\vec{Q}',s,l;\eta,n} \left( \vec{k} \right)  e^{i \left( \vec{Q} - \vec{Q}' \right) \cdot \vec{r}} \nonumber \\
	&= \frac{1}{N \Omega_0} \sum_{s} \sum_{n = 2 m -1}^{2m} \sum_{\substack{\vec{k} \in \text{MBZ} \\ \vec{Q} \in \mathcal{Q}_{\eta + l} \\ \vec{G} \in \mathcal{Q}}} u^{*}_{\vec{Q}+\vec{G},s,l;\eta,n} \left( \vec{k} \right)  u_{\vec{Q},s,l;\eta,n} \left( \vec{k} \right)  e^{i \vec{G} \cdot \vec{r}}.
\end{align}
We plot the layer-resolved LDOS $\rho_{\eta,m,l} \left( \vec{r} \right)$ in valley $\eta = 0$ for each set of gapped conduction bands.

In \cref{app:sec:fitted_models:plots:wilsons_1,app:sec:fitted_models:plots:wilsons_2}, we also plot the Wilson loops for the first one or two sets of gapped moir\'e conduction bands in valley $\eta = 0$. Two Wilson loops are considered along the reciprocal moir\'e lattice vectors. Along the $\vec{b}_{M} = \vec{b}_{M_1},\vec{b}_{M_2}$ moir\'e lattice vector, the Wilson loop is defined as
\begin{equation}
	\mathcal{W}^{m}_{\vec{b}_M} \left( \vec{k} \right) = \mathcal{V} \left( -\vec{b}_{M} \right) \lim_{N_{\mathcal{W}} \to \infty} \prod_{j}^{N_{\mathcal{W}} \leftarrow 0} P^{m,\eta} \left( \vec{k} + \frac{j}{N_{\mathcal{W}}} \vec{b}_{M} \right), \qq{for} \eta = 0 
\end{equation}
where the ordered product of projectors (in the order indicated by the arrow in $\prod_{j}^{N_{\mathcal{W}}}$) is given by
\begin{equation}
	\prod_{j}^{N_{\mathcal{W}} \leftarrow 0} P^{m,\eta} \left( \vec{k} + \frac{j}{N_{\mathcal{W}}} \vec{b}_{M} \right) \equiv P^{m,\eta} \left( \vec{k} + \vec{b}_{M} \right) \dots P^{m,\eta} \left( \vec{k} + \frac{2}{N_{\mathcal{W}}} \vec{b}_{M} \right) P^{m,\eta} \left( \vec{k} + \frac{1}{N_{\mathcal{W}}} \vec{b}_{M} \right) P^{m,\eta} \left( \vec{k} \right).
\end{equation}
The nonzero eigenvalues of the Wilson matrix are phases denoted by $e^{i \theta_{W}}$. In \cref{app:sec:fitted_models:plots:wilsons_1} (\cref{app:sec:fitted_models:plots:wilsons_2}), we plot the phase of the Wilson loop $\mathcal{W}^{m}_{\vec{b}_{M_1}} \left( k_2 \vec{b}_{M_2} \right)$ [$\mathcal{W}^{m}_{\vec{b}_{M_2}} \left( k_1 \vec{b}_{M_1} \right)$] as a function of $0 \leq k_2 \leq 1$ ($0 \leq k_1 \leq 1$).

\subsection{Parameter values}\label{app:sec:fitted_models:param_values}
\subsubsection{Full models}\label{app:sec:fitted_models:param_values:full}
\begin{table}[H]
\centering
\begin{tabular}{|r|r|r|r|r|r|r|r|r|r|r|r|}
\hline
$\theta$ & \makecell[tl]{Nonzero\\ parameters} & \makecell[tl]{Total\\ parameters} & $\epsilon_c/\si{\milli\electronvolt}$ & $\sigma/\si{\milli\electronvolt}$ & $\mathcal{O}_1$ & $\mathcal{O}_2$ & $\mathcal{O}_3$ & $\mathcal{O}_4$ & $\mathcal{O}_5$ & $\varepsilon_{\mathrm{SU}(2)} (\%)$ & $\varepsilon_{\tilde{M}_z} (\%)$\\
\hline
\SI{9.43}{\degree} & $697$ & $2016$ & $500$ & $100$ & $0.9993$ & $0.9821$ & $0.9816$ & $0.9917$ & $0.9778$ & $1.25$ & $1.33$\\
\hline
\SI{7.34}{\degree} & $298$ & $2016$ & $450$ & $100$ & $0.9992$ & $0.9922$ & $0.9831$ & $0.9832$ & $0.9919$ & $1.02$ & $1.81$\\
\hline
\SI{6.01}{\degree} & $285$ & $2016$ & $375$ & $75$ & $0.9990$ & $0.9963$ & $0.9898$ & $0.9898$ & $0.9856$ & $0.52$ & $3.20$\\
\hline
\SI{5.09}{\degree} & $143$ & $3570$ & $300$ & $75$ & $0.9998$ & $0.9994$ & $0.9876$ & $0.9875$ & $0.9909$ & $0.12$ & $4.82$\\
\hline
\SI{4.41}{\degree} & $193$ & $3570$ & $300$ & $75$ & $0.9997$ & $0.9996$ & $0.9970$ & $0.9957$ & $0.9834$ & $0.67$ & $7.02$\\
\hline
\SI{3.89}{\degree} & $182$ & $3570$ & $300$ & $75$ & $0.9998$ & $0.9996$ & $0.9962$ & $0.9972$ & $0.9940$ & $0.60$ & $8.94$\\ 
\hline \end{tabular}
\caption{Details of the full continuum model for twisted AA-stacked bilayer \ch{SnSe2}. The total and nonzero parameters used in the fitting procedure are listed, along with the energy weighting function parameters. Also shown are the minimal overlaps for the first five groups of bands and the percentages of $\mathrm{SU}(2)$ and $\tilde{M}_z$ symmetry breaking in the first group of bands.}
\label{app:tab:linear_SnSe2_AA}
\end{table}
\begin{table}[H]
\centering
\begin{tabular}{|r|r|r|r|r|r|r|r|r|r|r|r|}
\hline
$\theta$ & \makecell[tl]{Nonzero\\ parameters} & \makecell[tl]{Total\\ parameters} & $\epsilon_c/\si{\milli\electronvolt}$ & $\sigma/\si{\milli\electronvolt}$ & $\mathcal{O}_1$ & $\mathcal{O}_2$ & $\mathcal{O}_3$ & $\mathcal{O}_4$ & $\mathcal{O}_5$ & $\varepsilon_{\mathrm{SU}(2)} (\%)$ & $\varepsilon_{\tilde{\mathcal{I}}_z} (\%)$\\
\hline
\SI{9.43}{\degree} & $358$ & $2010$ & $500$ & $100$ & $0.9993$ & $0.9671$ & $0.8988$ & $0.9017$ & $0.9509$ & $3.02$ & $0.59$\\
\hline
\SI{7.34}{\degree} & $293$ & $2010$ & $450$ & $100$ & $0.9995$ & $0.9908$ & $0.9897$ & $0.9897$ & $0.9760$ & $3.36$ & $1.58$\\
\hline
\SI{6.01}{\degree} & $632$ & $2592$ & $375$ & $75$ & $0.9996$ & $0.9605$ & $0.9603$ & $0.9966$ & $0.9964$ & $3.67$ & $3.48$\\
\hline
\SI{5.09}{\degree} & $171$ & $3558$ & $300$ & $75$ & $0.9996$ & $0.9926$ & $0.9922$ & $0.9901$ & $0.9899$ & $3.82$ & $5.16$\\
\hline
\SI{4.41}{\degree} & $250$ & $3558$ & $300$ & $75$ & $0.9997$ & $0.9966$ & $0.9832$ & $0.9843$ & $0.9715$ & $3.89$ & $7.29$\\
\hline
\SI{3.89}{\degree} & $309$ & $3558$ & $300$ & $75$ & $0.9998$ & $0.9982$ & $0.9901$ & $0.9901$ & $0.9827$ & $4.47$ & $8.85$\\ 
\hline \end{tabular}
\caption{Details of the full continuum model for twisted AB-stacked bilayer \ch{SnSe2}. The total and nonzero parameters used in the fitting procedure are listed, along with the energy weighting function parameters. Also shown are the minimal overlaps for the first five groups of bands and the percentages of $\mathrm{SU}(2)$ and $\tilde{\mathcal{I}}_z$ symmetry breaking in the first group of bands.}
\label{app:tab:linear_SnSe2_AB}
\end{table}
\begin{table}[H]
\centering
\begin{tabular}{|r|r|r|r|r|r|r|r|r|r|r|r|}
\hline
$\theta$ & \makecell[tl]{Nonzero\\ parameters} & \makecell[tl]{Total\\ parameters} & $\epsilon_c/\si{\milli\electronvolt}$ & $\sigma/\si{\milli\electronvolt}$ & $\mathcal{O}_1$ & $\mathcal{O}_2$ & $\mathcal{O}_3$ & $\mathcal{O}_4$ & $\mathcal{O}_5$ & $\varepsilon_{\mathrm{SU}(2)} (\%)$ & $\varepsilon_{\tilde{M}_z} (\%)$\\
\hline
\SI{9.43}{\degree} & $141$ & $1044$ & $375$ & $75$ & $0.9994$ & $0.9592$ & $0.9592$ & $0.9871$ & $0.9833$ & $0.05$ & $3.08$\\
\hline
\SI{7.34}{\degree} & $117$ & $2016$ & $338$ & $75$ & $0.9996$ & $0.9904$ & $0.9584$ & $0.9583$ & $0.9797$ & $0.00$ & $2.68$\\
\hline
\SI{6.01}{\degree} & $197$ & $2016$ & $281$ & $56$ & $0.9996$ & $0.9858$ & $0.9806$ & $0.9889$ & $0.9863$ & $0.12$ & $2.67$\\
\hline
\SI{5.09}{\degree} & $56$ & $3570$ & $225$ & $56$ & $0.9995$ & $0.9983$ & $0.9221$ & $0.9609$ & $0.9221$ & $0.00$ & $2.02$\\
\hline
\SI{4.41}{\degree} & $79$ & $3570$ & $225$ & $56$ & $0.9993$ & $0.9994$ & $0.9857$ & $0.9822$ & $0.9842$ & $0.00$ & $1.57$\\
\hline
\SI{3.89}{\degree} & $69$ & $3570$ & $225$ & $56$ & $0.9998$ & $0.9997$ & $0.9991$ & $0.9988$ & $0.9968$ & $0.00$ & $1.23$\\ 
\hline \end{tabular}
\caption{Details of the full continuum model for twisted AA-stacked bilayer \ch{ZrS2}. The total and nonzero parameters used in the fitting procedure are listed, along with the energy weighting function parameters. Also shown are the minimal overlaps for the first five groups of bands and the percentages of $\mathrm{SU}(2)$ and $\tilde{M}_z$ symmetry breaking in the first group of bands.}
\label{app:tab:linear_ZrS2_AA}
\end{table}
\begin{table}[H]
\centering
\begin{tabular}{|r|r|r|r|r|r|r|r|r|r|r|r|}
\hline
$\theta$ & \makecell[tl]{Nonzero\\ parameters} & \makecell[tl]{Total\\ parameters} & $\epsilon_c/\si{\milli\electronvolt}$ & $\sigma/\si{\milli\electronvolt}$ & $\mathcal{O}_1$ & $\mathcal{O}_2$ & $\mathcal{O}_3$ & $\mathcal{O}_4$ & $\mathcal{O}_5$ & $\varepsilon_{\mathrm{SU}(2)} (\%)$ & $\varepsilon_{\tilde{\mathcal{I}}_z} (\%)$\\
\hline
\SI{9.43}{\degree} & $169$ & $1038$ & $375$ & $75$ & $0.9993$ & $0.9979$ & $0.9786$ & $0.9785$ & $0.9971$ & $2.21$ & $3.49$\\
\hline
\SI{7.34}{\degree} & $157$ & $2010$ & $338$ & $75$ & $0.9995$ & $0.9884$ & $0.9769$ & $0.9769$ & $0.9824$ & $2.11$ & $3.71$\\
\hline
\SI{6.01}{\degree} & $171$ & $2010$ & $281$ & $56$ & $0.9991$ & $0.9948$ & $0.9935$ & $0.9933$ & $0.9953$ & $2.17$ & $3.35$\\
\hline
\SI{5.09}{\degree} & $61$ & $3558$ & $225$ & $56$ & $0.9996$ & $0.9971$ & $0.9902$ & $0.9902$ & $0.9921$ & $2.27$ & $3.11$\\
\hline
\SI{4.41}{\degree} & $73$ & $3558$ & $225$ & $56$ & $0.9994$ & $0.9974$ & $0.9938$ & $0.9940$ & $0.9954$ & $2.24$ & $3.08$\\
\hline
\SI{3.89}{\degree} & $63$ & $3558$ & $225$ & $56$ & $0.9996$ & $0.9992$ & $0.9941$ & $0.9939$ & $0.9946$ & $2.22$ & $2.78$\\ 
\hline \end{tabular}
\caption{Details of the full continuum model for twisted AB-stacked bilayer \ch{ZrS2}. The total and nonzero parameters used in the fitting procedure are listed, along with the energy weighting function parameters. Also shown are the minimal overlaps for the first five groups of bands and the percentages of $\mathrm{SU}(2)$ and $\tilde{\mathcal{I}}_z$ symmetry breaking in the first group of bands.}
\label{app:tab:linear_ZrS2_AB}
\end{table} \subsubsection{First moir\'e harmonic models}\label{app:sec:fitted_models:param_values:first_harmonic}
\begin{table}[H]
\centering
\resizebox{\textwidth}{!}{\begin{tabular}{|r|r|r|r|r|r|r|r|r|r|r|r|r|r|r|r|r|r|r|r|r|}
\hline
$\theta$ & $m_x$ & $m_y$ & $w^{\text{AA}}_1$ & $w^{\text{AA}}_2$ & $w^{\text{AA}}_3$ & $w^{\text{AA}}_4$ & $w^{\text{AA}}_5$ & $w^{\text{AA}}_6$ & $w^{\prime\text{AA}}_1$ & $w^{\prime\text{AA}}_2$ & $w^{\prime\text{AA}}_3$ & $w^{\prime\text{AA}}_4$ & $w^{\prime\text{AA}}_5$ & $w^{\prime\text{AA}}_6$ & $w^{\prime\text{AA}}_7$ & $w^{\prime\text{AA}}_8$ & $w^{\prime\text{AA}}_9$ & $w^{\prime\text{AA}}_{10}$ & $\mathcal{O}_{1}$ & $\mathcal{O}_{2}$\\
\hline
\SI{9.43}{\degree} & $0.22$ & $0.67$ & $69.21$ & $69.30$ & $-0.93$ & $1.31$ & $-1.79$ & $-2.44$ & $-5.75$ & $14.50$ & $-10.89$ & $-4.19$ & $28.17$ & $0.64$ & $20.72$ & $-0.73$ & $0.09$ & $14.65$ & $0.9933$ & $0.1505$\\
\hline
\SI{7.34}{\degree} & $0.22$ & $0.66$ & $56.40$ & $70.11$ & $-0.69$ & $-0.27$ & $-1.39$ & $-0.51$ & $-7.18$ & $21.62$ & $-7.49$ & $1.88$ & $33.60$ & $0.23$ & $23.88$ & $-0.41$ & $-0.12$ & $11.79$ & $0.9964$ & $0.9185$\\
\hline
\SI{6.01}{\degree} & $0.25$ & $0.75$ & $55.57$ & $71.10$ & $-0.55$ & $-0.42$ & $-0.88$ & $-0.23$ & $-5.21$ & $11.11$ & $-6.73$ & $3.49$ & $26.81$ & $-0.40$ & $13.98$ & $0.21$ & $0.59$ & $10.17$ & $0.9956$ & $0.9825$\\
\hline
\SI{5.09}{\degree} & $0.21$ & $0.64$ & $63.13$ & $80.70$ & $-0.43$ & $-0.87$ & $-0.70$ & $0.14$ & $-3.86$ & $13.41$ & $-11.58$ & $2.53$ & $27.13$ & $-0.27$ & $19.80$ & $0.12$ & $0.35$ & $18.32$ & $0.9976$ & $0.9891$\\
\hline
\SI{4.41}{\degree} & $0.23$ & $0.67$ & $57.21$ & $89.38$ & $-0.52$ & $-1.84$ & $-0.66$ & $1.15$ & $-2.78$ & $7.64$ & $-15.79$ & $1.90$ & $33.60$ & $-0.83$ & $25.51$ & $0.58$ & $-0.37$ & $20.36$ & $0.9982$ & $0.9913$\\
\hline
\SI{3.89}{\degree} & $0.24$ & $0.63$ & $65.29$ & $98.14$ & $-0.58$ & $-2.29$ & $-0.56$ & $1.54$ & $-0.28$ & $-2.08$ & $-18.70$ & $4.69$ & $33.32$ & $-1.01$ & $29.64$ & $0.42$ & $-0.47$ & $28.22$ & $0.9986$ & $0.9944$\\ 
\hline \end{tabular}}
\caption{Parameter values of the full first moir\'e harmonic model for twisted AA-stacked bilayer \ch{SnSe2}. The interlayer hopping parameters are given in units of \si{\milli\electronvolt}, while the effective masses are given in units of the bare electron mass $m_e$. The minimal overlaps with the first two groups of bands are listed at the end.}
\label{app:tab:SnSe2_AA_1}
\end{table}
\begin{table}[H]
\centering
\begin{tabular}{|r|r|r|r|r|r|r|r|r|r|r|}
\hline
$\theta$ & $m_x$ & $m_y$ & $w^{\text{AA}}_1$ & $w^{\text{AA}}_2$ & $w^{\prime\text{AA}}_1$ & $w^{\prime\text{AA}}_3$ & $w^{\prime\text{AA}}_7$ & $w^{\prime\text{AA}}_{10}$ & $\mathcal{O}_{1}$ & $\mathcal{O}_{2}$\\
\hline
\SI{9.43}{\degree} & $0.20$ & $0.69$ & $84.89$ & $72.39$ & $-14.08$ & N/A & $19.86$ & $16.33$ & $0.9891$ & $0.1527$\\
\hline
\SI{7.34}{\degree} & $0.20$ & $0.66$ & $88.78$ & $78.32$ & N/A & $-16.21$ & $24.61$ & $18.47$ & $0.9945$ & $0.8733$\\
\hline
\SI{6.01}{\degree} & $0.27$ & $0.75$ & $71.97$ & $63.54$ & N/A & $-15.43$ & $11.87$ & $13.92$ & $0.9928$ & $0.9645$\\
\hline
\SI{5.09}{\degree} & $0.25$ & $0.64$ & $79.24$ & $69.33$ & N/A & $-18.53$ & $16.82$ & $23.18$ & $0.9958$ & $0.9860$\\
\hline
\SI{4.41}{\degree} & $0.28$ & $0.67$ & $74.11$ & $66.00$ & N/A & $-21.13$ & $18.32$ & $26.02$ & $0.9965$ & $0.9912$\\
\hline
\SI{3.89}{\degree} & $0.27$ & $0.63$ & $77.12$ & $69.40$ & N/A & $-23.98$ & $19.69$ & $34.35$ & $0.9973$ & $0.9928$\\ 
\hline \end{tabular}
\caption{Parameter values of the reduced first moir\'e harmonic model for twisted AA-stacked bilayer \ch{SnSe2}. The interlayer hopping parameters are given in units of \si{\milli\electronvolt}, while the effective masses are given in units of the bare electron mass $m_e$. The minimal overlaps with the first two groups of bands are listed at the end.}
\label{app:tab:SnSe2_AA_2}
\end{table}
\begin{table}[H]
\centering
\begin{tabular}{|r|r|r|r|r|r|r|r|r|r|r|r|}
\hline
$\theta$ & $m_x$ & $m_y$ & $w^{\text{AA}}_1$ & $w^{\text{AA}}_2$ & $w^{\prime\text{AA}}_1$ & $w^{\prime\text{AA}}_2$ & $w^{\prime\text{AA}}_3$ & $w^{\prime\text{AA}}_4$ & $w^{\prime\text{AA}}_5$ & $\mathcal{O}_{1}$ & $\mathcal{O}_{2}$\\
\hline
\SI{9.43}{\degree} & $0.21$ & $0.73$ & $66.22$ & $65.91$ & $-6.95$ & $44.29$ & $-10.74$ & $-11.33$ & $34.62$ & $0.9688$ & $0.0267$\\
\hline
\SI{7.34}{\degree} & $0.21$ & $0.73$ & $63.73$ & $73.71$ & $-0.55$ & $31.16$ & $-7.92$ & $3.69$ & $30.77$ & $0.9834$ & $0.3233$\\
\hline
\SI{6.01}{\degree} & $0.21$ & $0.73$ & $65.09$ & $77.81$ & $-1.94$ & $17.10$ & $-9.61$ & $4.56$ & $28.55$ & $0.9835$ & $0.8716$\\
\hline
\SI{5.09}{\degree} & $0.21$ & $0.73$ & $65.58$ & $81.52$ & $-5.15$ & $14.95$ & $-9.62$ & $2.92$ & $24.88$ & $0.9805$ & $0.9079$\\
\hline
\SI{4.41}{\degree} & $0.21$ & $0.73$ & $66.85$ & $85.67$ & $-2.94$ & $14.28$ & $-11.06$ & $4.74$ & $26.71$ & $0.9767$ & $0.9226$\\
\hline
\SI{3.89}{\degree} & $0.21$ & $0.73$ & $66.38$ & $88.80$ & $-4.50$ & $11.04$ & $-7.99$ & $5.71$ & $27.25$ & $0.9722$ & $0.9280$\\ 
\hline \end{tabular}
\caption{Parameter values of the first moir\'e harmonic model with the zero-twist constraints imposed for twisted AA-stacked bilayer \ch{SnSe2}. The interlayer hopping parameters are given in units of \si{\milli\electronvolt}, while the effective masses are given in units of the bare electron mass $m_e$. The minimal overlaps with the first two groups of bands are listed at the end.}
\label{app:tab:SnSe2_AA_3}
\end{table}
\begin{table}[H]
\centering
\begin{tabular}{|r|r|r|r|r|r|r|r|r|}
\hline
$\theta$ & $m_x$ & $m_y$ & $w^{\text{AA}}_1$ & $w^{\text{AA}}_2$ & $w^{\prime\text{AA}}_1$ & $w^{\prime\text{AA}}_3$ & $\mathcal{O}_{1}$ & $\mathcal{O}_{2}$\\
\hline
\SI{9.43}{\degree} & $0.21$ & $0.73$ & $66.22$ & $65.91$ & $-12.22$ & N/A & $0.9762$ & $0.1121$\\
\hline
\SI{7.34}{\degree} & $0.21$ & $0.73$ & $63.73$ & $73.71$ & $-14.35$ & N/A & $0.9818$ & $0.2888$\\
\hline
\SI{6.01}{\degree} & $0.21$ & $0.73$ & $65.09$ & $77.81$ & N/A & $-15.34$ & $0.9719$ & $0.8747$\\
\hline
\SI{5.09}{\degree} & $0.21$ & $0.73$ & $65.58$ & $81.52$ & N/A & $-18.03$ & $0.9667$ & $0.9033$\\
\hline
\SI{4.41}{\degree} & $0.21$ & $0.73$ & $66.85$ & $85.67$ & N/A & $-18.94$ & $0.9603$ & $0.9101$\\
\hline
\SI{3.89}{\degree} & $0.21$ & $0.73$ & $66.38$ & $88.80$ & N/A & $-18.94$ & $0.9526$ & $0.9058$\\ 
\hline \end{tabular}
\caption{Parameter values of the reduced first moir\'e harmonic model with the zero-twist constraints imposed for twisted AA-stacked bilayer \ch{SnSe2}. The interlayer hopping parameters are given in units of \si{\milli\electronvolt}, while the effective masses are given in units of the bare electron mass $m_e$. The minimal overlaps with the first two groups of bands are listed at the end.}
\label{app:tab:SnSe2_AA_4}
\end{table}
\begin{table}[H]
\centering
\resizebox{\textwidth}{!}{\begin{tabular}{|r|r|r|r|r|r|r|r|r|r|r|r|r|r|r|r|r|r|r|r|r|}
\hline
$\theta$ & $m_x$ & $m_y$ & $w^{\text{AB}}_1$ & $w^{\text{AB}}_2$ & $w^{\text{AB}}_3$ & $w^{\text{AB}}_4$ & $w^{\prime\text{AB}}_1$ & $w^{\prime\text{AB}}_2$ & $w^{\prime\text{AB}}_3$ & $w^{\prime\text{AB}}_4$ & $w^{\prime\text{AB}}_5$ & $w^{\prime\text{AB}}_6$ & $w^{\prime\text{AB}}_7$ & $w^{\prime\text{AB}}_8$ & $w^{\prime\text{AB}}_9$ & $w^{\prime\text{AB}}_{10}$ & $w^{\prime\text{AB}}_{11}$ & $w^{\prime\text{AB}}_{12}$ & $\mathcal{O}_{1}$ & $\mathcal{O}_{2}$\\
\hline
\SI{9.43}{\degree} & $0.21$ & $0.70$ & $-3.53$ & $-92.77$ & $-1.14$ & $-1.01$ & $-0.32$ & $7.30$ & $-10.85$ & $22.00$ & $0.55$ & $-3.73$ & $-16.22$ & $-0.78$ & $-1.30$ & $1.88$ & $0.28$ & $-8.51$ & $0.9951$ & $0.1228$\\
\hline
\SI{7.34}{\degree} & $0.21$ & $0.67$ & $-5.32$ & $-81.96$ & $-0.63$ & $-0.52$ & $0.88$ & $7.84$ & $-14.37$ & $30.99$ & $-0.52$ & $-4.01$ & $-29.54$ & $-0.51$ & $-4.36$ & $4.23$ & $0.41$ & $-19.91$ & $0.9981$ & $0.9476$\\
\hline
\SI{6.01}{\degree} & $0.26$ & $0.77$ & $-4.43$ & $-78.43$ & $-0.50$ & $-0.21$ & $-0.67$ & $9.24$ & $-14.94$ & $35.88$ & $0.37$ & $-7.71$ & $-10.59$ & $-0.36$ & $-4.01$ & $3.94$ & $-0.34$ & $-20.41$ & $0.9984$ & $0.9737$\\
\hline
\SI{5.09}{\degree} & $0.25$ & $0.69$ & $-3.74$ & $-76.24$ & $-0.38$ & $-0.13$ & $0.04$ & $10.47$ & $-16.87$ & $39.31$ & $-0.06$ & $-7.32$ & $-14.66$ & $-0.82$ & $-6.04$ & $5.22$ & $-0.53$ & $-25.37$ & $0.9990$ & $0.9857$\\
\hline
\SI{4.41}{\degree} & $0.24$ & $0.77$ & $-3.92$ & $-78.61$ & $-0.09$ & $-0.14$ & $-0.21$ & $12.13$ & $-18.58$ & $46.65$ & $-1.01$ & $-10.52$ & $-17.25$ & $-0.90$ & $-8.31$ & $6.05$ & $-0.95$ & $-33.02$ & $0.9986$ & $0.9918$\\
\hline
\SI{3.89}{\degree} & $0.23$ & $0.72$ & $-3.65$ & $-84.94$ & $-0.23$ & $-0.01$ & $1.37$ & $14.01$ & $-19.46$ & $57.03$ & $-1.78$ & $-15.89$ & $-20.29$ & $-0.87$ & $-12.01$ & $7.77$ & $-1.27$ & $-40.84$ & $0.9986$ & $0.9933$\\ 
\hline \end{tabular}}
\caption{Parameter values of the full first moir\'e harmonic model for twisted AB-stacked bilayer \ch{SnSe2}. The interlayer hopping parameters are given in units of \si{\milli\electronvolt}, while the effective masses are given in units of the bare electron mass $m_e$. The minimal overlaps with the first two groups of bands are listed at the end.}
\label{app:tab:SnSe2_AB_1}
\end{table}
\begin{table}[H]
\centering
\begin{tabular}{|r|r|r|r|r|r|r|r|r|r|r|r|}
\hline
$\theta$ & $m_x$ & $m_y$ & $w^{\text{AB}}_2$ & $w^{\prime\text{AB}}_2$ & $w^{\prime\text{AB}}_4$ & $w^{\prime\text{AB}}_7$ & $w^{\prime\text{AB}}_9$ & $w^{\prime\text{AB}}_{10}$ & $w^{\prime\text{AB}}_{12}$ & $\mathcal{O}_{1}$ & $\mathcal{O}_{2}$\\
\hline
\SI{9.43}{\degree} & $0.21$ & $0.71$ & $-99.12$ & $7.01$ & $17.85$ & N/A & N/A & $0.95$ & $-8.09$ & $0.9934$ & $0.0347$\\
\hline
\SI{7.34}{\degree} & $0.21$ & $0.67$ & $-79.41$ & $10.23$ & $26.93$ & $-30.92$ & N/A & N/A & $-21.74$ & $0.9973$ & $0.2354$\\
\hline
\SI{6.01}{\degree} & $0.26$ & $0.79$ & $-78.22$ & N/A & $25.70$ & N/A & $-9.53$ & $8.85$ & $-18.37$ & $0.9873$ & $0.9576$\\
\hline
\SI{5.09}{\degree} & $0.25$ & $0.69$ & $-86.65$ & N/A & $29.20$ & N/A & $-11.63$ & $10.71$ & $-24.76$ & $0.9908$ & $0.9687$\\
\hline
\SI{4.41}{\degree} & $0.25$ & $0.77$ & $-90.78$ & N/A & $31.02$ & N/A & $-14.82$ & $12.95$ & $-31.33$ & $0.9918$ & $0.9820$\\
\hline
\SI{3.89}{\degree} & $0.24$ & $0.72$ & $-62.10$ & N/A & $31.80$ & $-20.94$ & N/A & $4.11$ & $-35.03$ & $0.9926$ & $0.9844$\\ 
\hline \end{tabular}
\caption{Parameter values of the reduced first moir\'e harmonic model for twisted AB-stacked bilayer \ch{SnSe2}. The interlayer hopping parameters are given in units of \si{\milli\electronvolt}, while the effective masses are given in units of the bare electron mass $m_e$. The minimal overlaps with the first two groups of bands are listed at the end.}
\label{app:tab:SnSe2_AB_2}
\end{table}
\begin{table}[H]
\centering
\begin{tabular}{|r|r|r|r|r|r|r|r|r|r|r|r|r|r|}
\hline
$\theta$ & $m_x$ & $m_y$ & $w^{\text{AB}}_1$ & $w^{\text{AB}}_2$ & $w^{\prime\text{AB}}_1$ & $w^{\prime\text{AB}}_2$ & $w^{\prime\text{AB}}_3$ & $w^{\prime\text{AB}}_4$ & $w^{\prime\text{AB}}_5$ & $w^{\prime\text{AB}}_6$ & $w^{\prime\text{AB}}_7$ & $\mathcal{O}_{1}$ & $\mathcal{O}_{2}$\\
\hline
\SI{9.43}{\degree} & $0.21$ & $0.73$ & $-3.34$ & $-81.08$ & $-0.23$ & $5.64$ & $-11.77$ & $18.66$ & $0.59$ & $-2.58$ & $-46.32$ & $0.9873$ & $0.1235$\\
\hline
\SI{7.34}{\degree} & $0.21$ & $0.73$ & $-5.07$ & $-76.32$ & $0.91$ & $11.93$ & $-16.68$ & $31.30$ & $-0.36$ & $-11.43$ & $-53.45$ & $0.9816$ & $0.4100$\\
\hline
\SI{6.01}{\degree} & $0.21$ & $0.73$ & $-5.10$ & $-82.56$ & $0.91$ & $8.22$ & $-21.15$ & $44.07$ & $0.59$ & $-13.71$ & $-5.30$ & $0.9791$ & $0.8462$\\
\hline
\SI{5.09}{\degree} & $0.21$ & $0.73$ & $-4.74$ & $-77.11$ & $-0.51$ & $8.14$ & $-21.77$ & $26.14$ & $0.17$ & $0.76$ & $-12.81$ & $0.9780$ & $0.7112$\\
\hline
\SI{4.41}{\degree} & $0.21$ & $0.73$ & $-5.27$ & $-82.84$ & $0.09$ & $8.39$ & $-23.13$ & $33.39$ & $-0.92$ & $-3.57$ & $-3.48$ & $0.9725$ & $0.8738$\\
\hline
\SI{3.89}{\degree} & $0.21$ & $0.73$ & $-4.18$ & $-77.80$ & $1.45$ & $9.45$ & $-24.71$ & $35.10$ & $-2.15$ & $-5.85$ & $-10.28$ & $0.9689$ & $0.8643$\\ 
\hline \end{tabular}
\caption{Parameter values of the first moir\'e harmonic model with the zero-twist constraints imposed for twisted AB-stacked bilayer \ch{SnSe2}. The interlayer hopping parameters are given in units of \si{\milli\electronvolt}, while the effective masses are given in units of the bare electron mass $m_e$. The minimal overlaps with the first two groups of bands are listed at the end.}
\label{app:tab:SnSe2_AB_3}
\end{table}
\begin{table}[H]
\centering
\begin{tabular}{|r|r|r|r|r|r|r|r|r|}
\hline
$\theta$ & $m_x$ & $m_y$ & $w^{\text{AB}}_2$ & $w^{\prime\text{AB}}_2$ & $w^{\prime\text{AB}}_3$ & $w^{\prime\text{AB}}_4$ & $\mathcal{O}_{1}$ & $\mathcal{O}_{2}$\\
\hline
\SI{9.43}{\degree} & $0.21$ & $0.73$ & $-81.08$ & $5.16$ & N/A & $16.17$ & $0.9847$ & $0.3528$\\
\hline
\SI{7.34}{\degree} & $0.21$ & $0.73$ & $-76.32$ & $9.30$ & N/A & $17.62$ & $0.9868$ & $0.7436$\\
\hline
\SI{6.01}{\degree} & $0.21$ & $0.73$ & $-82.56$ & $6.86$ & N/A & $26.48$ & $0.9793$ & $0.7251$\\
\hline
\SI{5.09}{\degree} & $0.21$ & $0.73$ & $-77.11$ & $6.75$ & N/A & $27.89$ & $0.9725$ & $0.8429$\\
\hline
\SI{4.41}{\degree} & $0.21$ & $0.73$ & $-82.84$ & N/A & $-23.30$ & $29.10$ & $0.9616$ & $0.8734$\\
\hline
\SI{3.89}{\degree} & $0.21$ & $0.73$ & $-77.80$ & N/A & $-23.05$ & $27.32$ & $0.9571$ & $0.8644$\\ 
\hline \end{tabular}
\caption{Parameter values of the reduced first moir\'e harmonic model with the zero-twist constraints imposed for twisted AB-stacked bilayer \ch{SnSe2}. The interlayer hopping parameters are given in units of \si{\milli\electronvolt}, while the effective masses are given in units of the bare electron mass $m_e$. The minimal overlaps with the first two groups of bands are listed at the end.}
\label{app:tab:SnSe2_AB_4}
\end{table}
\begin{table}[H]
\centering
\resizebox{\textwidth}{!}{\begin{tabular}{|r|r|r|r|r|r|r|r|r|r|r|r|r|r|r|r|r|r|r|r|r|}
\hline
$\theta$ & $m_x$ & $m_y$ & $w^{\text{AA}}_1$ & $w^{\text{AA}}_2$ & $w^{\text{AA}}_3$ & $w^{\text{AA}}_4$ & $w^{\text{AA}}_5$ & $w^{\text{AA}}_6$ & $w^{\prime\text{AA}}_1$ & $w^{\prime\text{AA}}_2$ & $w^{\prime\text{AA}}_3$ & $w^{\prime\text{AA}}_4$ & $w^{\prime\text{AA}}_5$ & $w^{\prime\text{AA}}_6$ & $w^{\prime\text{AA}}_7$ & $w^{\prime\text{AA}}_8$ & $w^{\prime\text{AA}}_9$ & $w^{\prime\text{AA}}_{10}$ & $\mathcal{O}_{1}$ & $\mathcal{O}_{2}$\\
\hline
\SI{9.43}{\degree} & $0.27$ & $1.71$ & $-9.99$ & $59.31$ & $0.20$ & $-0.99$ & $0.59$ & $0.16$ & $4.16$ & $-0.01$ & $-28.51$ & $-7.62$ & $-8.33$ & $-0.02$ & $-14.09$ & $0.06$ & $-0.12$ & $10.02$ & $0.9918$ & $0.7264$\\
\hline
\SI{7.34}{\degree} & $0.26$ & $1.54$ & $-8.58$ & $58.70$ & $0.31$ & $-1.28$ & $0.58$ & $0.57$ & $2.11$ & $-7.45$ & $-30.82$ & $-5.22$ & $-1.75$ & $-0.10$ & $-12.86$ & $-0.06$ & $0.03$ & $6.48$ & $0.9964$ & $0.4586$\\
\hline
\SI{6.01}{\degree} & $0.27$ & $2.18$ & $-1.11$ & $57.43$ & $0.17$ & $-0.83$ & $0.42$ & $0.28$ & $2.53$ & $-18.40$ & $-25.16$ & $-5.48$ & $3.53$ & $-0.08$ & $-1.82$ & $-0.05$ & $0.01$ & $2.57$ & $0.9910$ & $0.9610$\\
\hline
\SI{5.09}{\degree} & $0.29$ & $1.93$ & $-4.55$ & $53.26$ & $0.18$ & $-0.89$ & $0.43$ & $0.36$ & $0.54$ & $-13.73$ & $-27.01$ & $-5.61$ & $3.96$ & $-0.10$ & $-0.60$ & $-0.08$ & $0.05$ & $2.06$ & $0.9932$ & $0.9746$\\
\hline
\SI{4.41}{\degree} & $0.30$ & $1.99$ & $-3.37$ & $47.05$ & $0.09$ & $-0.55$ & $0.33$ & $0.10$ & $1.60$ & $-12.53$ & $-23.00$ & $-3.00$ & $4.59$ & $-0.06$ & $0.24$ & $-0.04$ & $0.04$ & $1.29$ & $0.9942$ & $0.9792$\\
\hline
\SI{3.89}{\degree} & $0.31$ & $1.88$ & $-5.08$ & $44.80$ & $0.08$ & $-0.52$ & $0.30$ & $0.12$ & $0.99$ & $-10.78$ & $-21.47$ & $-1.56$ & $4.98$ & $-0.05$ & $0.64$ & $-0.03$ & $0.04$ & $0.64$ & $0.9953$ & $0.9821$\\ 
\hline \end{tabular}}
\caption{Parameter values of the full first moir\'e harmonic model for twisted AA-stacked bilayer \ch{ZrS2}. The interlayer hopping parameters are given in units of \si{\milli\electronvolt}, while the effective masses are given in units of the bare electron mass $m_e$. The minimal overlaps with the first two groups of bands are listed at the end.}
\label{app:tab:ZrS2_AA_1}
\end{table}
\begin{table}[H]
\centering
\begin{tabular}{|r|r|r|r|r|r|r|r|r|r|r|r|r|}
\hline
$\theta$ & $m_x$ & $m_y$ & $w^{\text{AA}}_1$ & $w^{\text{AA}}_2$ & $w^{\prime\text{AA}}_1$ & $w^{\prime\text{AA}}_2$ & $w^{\prime\text{AA}}_3$ & $w^{\prime\text{AA}}_4$ & $w^{\prime\text{AA}}_7$ & $w^{\prime\text{AA}}_{10}$ & $\mathcal{O}_{1}$ & $\mathcal{O}_{2}$\\
\hline
\SI{9.43}{\degree} & $0.27$ & $1.68$ & $-13.23$ & $54.74$ & N/A & N/A & $-23.56$ & N/A & $-13.88$ & $12.41$ & $0.9920$ & $0.5733$\\
\hline
\SI{7.34}{\degree} & $0.25$ & $1.52$ & $-12.82$ & $56.38$ & N/A & N/A & $-26.49$ & N/A & $-12.76$ & $8.09$ & $0.9967$ & $0.4636$\\
\hline
\SI{6.01}{\degree} & $0.29$ & $2.21$ & N/A & $46.76$ & $1.97$ & $-19.86$ & $-18.86$ & N/A & N/A & $2.43$ & $0.9880$ & $0.9634$\\
\hline
\SI{5.09}{\degree} & $0.31$ & $1.93$ & $-4.51$ & $42.32$ & N/A & $-13.45$ & $-20.04$ & N/A & N/A & $1.96$ & $0.9922$ & $0.9750$\\
\hline
\SI{4.41}{\degree} & $0.37$ & $2.03$ & $-2.54$ & $32.78$ & $1.43$ & $-11.12$ & $-19.24$ & N/A & N/A & N/A & $0.9940$ & $0.9787$\\
\hline
\SI{3.89}{\degree} & $0.33$ & $1.89$ & $-5.57$ & $37.46$ & N/A & $-10.63$ & $-23.86$ & $-2.90$ & N/A & N/A & $0.9956$ & $0.9816$\\ 
\hline \end{tabular}
\caption{Parameter values of the reduced first moir\'e harmonic model for twisted AA-stacked bilayer \ch{ZrS2}. The interlayer hopping parameters are given in units of \si{\milli\electronvolt}, while the effective masses are given in units of the bare electron mass $m_e$. The minimal overlaps with the first two groups of bands are listed at the end.}
\label{app:tab:ZrS2_AA_2}
\end{table}
\begin{table}[H]
\centering
\begin{tabular}{|r|r|r|r|r|r|r|r|r|r|r|r|}
\hline
$\theta$ & $m_x$ & $m_y$ & $w^{\text{AA}}_1$ & $w^{\text{AA}}_2$ & $w^{\prime\text{AA}}_1$ & $w^{\prime\text{AA}}_2$ & $w^{\prime\text{AA}}_3$ & $w^{\prime\text{AA}}_4$ & $w^{\prime\text{AA}}_5$ & $\mathcal{O}_{1}$ & $\mathcal{O}_{2}$\\
\hline
\SI{9.43}{\degree} & $0.29$ & $1.86$ & $-3.53$ & $52.58$ & $8.38$ & $0.41$ & $-21.25$ & $-4.59$ & $20.03$ & $0.9603$ & $0.3280$\\
\hline
\SI{7.34}{\degree} & $0.29$ & $1.86$ & $-6.57$ & $52.53$ & $-0.44$ & $-13.58$ & $-23.08$ & $-2.49$ & $-5.92$ & $0.9797$ & $0.7335$\\
\hline
\SI{6.01}{\degree} & $0.29$ & $1.86$ & $-10.15$ & $50.51$ & $-0.31$ & $-3.44$ & $-22.68$ & $-3.22$ & $-0.73$ & $0.9866$ & $0.5615$\\
\hline
\SI{5.09}{\degree} & $0.29$ & $1.86$ & $-8.91$ & $51.29$ & $-0.60$ & $-8.21$ & $-24.06$ & $-2.56$ & $3.63$ & $0.9893$ & $0.8360$\\
\hline
\SI{4.41}{\degree} & $0.29$ & $1.86$ & $-12.00$ & $50.10$ & $-0.36$ & $-4.32$ & $-24.36$ & $-3.63$ & $2.20$ & $0.9930$ & $0.9341$\\
\hline
\SI{3.89}{\degree} & $0.29$ & $1.86$ & $-12.35$ & $50.50$ & $-0.59$ & $-5.11$ & $-23.99$ & $-2.78$ & $4.28$ & $0.9948$ & $0.9679$\\ 
\hline \end{tabular}
\caption{Parameter values of the first moir\'e harmonic model with the zero-twist constraints imposed for twisted AA-stacked bilayer \ch{ZrS2}. The interlayer hopping parameters are given in units of \si{\milli\electronvolt}, while the effective masses are given in units of the bare electron mass $m_e$. The minimal overlaps with the first two groups of bands are listed at the end.}
\label{app:tab:ZrS2_AA_3}
\end{table}
\begin{table}[H]
\centering
\begin{tabular}{|r|r|r|r|r|r|r|r|}
\hline
$\theta$ & $m_x$ & $m_y$ & $w^{\text{AA}}_1$ & $w^{\text{AA}}_2$ & $w^{\prime\text{AA}}_3$ & $\mathcal{O}_{1}$ & $\mathcal{O}_{2}$\\
\hline
\SI{9.43}{\degree} & $0.29$ & $1.86$ & $-3.53$ & $52.58$ & $-20.13$ & $0.9612$ & $0.0374$\\
\hline
\SI{7.34}{\degree} & $0.29$ & $1.86$ & $-6.57$ & $52.53$ & $-20.73$ & $0.9754$ & $0.7307$\\
\hline
\SI{6.01}{\degree} & $0.29$ & $1.86$ & $-10.15$ & $50.51$ & $-18.86$ & $0.9864$ & $0.5679$\\
\hline
\SI{5.09}{\degree} & $0.29$ & $1.86$ & $-8.91$ & $51.29$ & $-20.61$ & $0.9856$ & $0.8371$\\
\hline
\SI{4.41}{\degree} & $0.29$ & $1.86$ & $-12.00$ & $50.10$ & $-19.38$ & $0.9907$ & $0.9340$\\
\hline
\SI{3.89}{\degree} & $0.29$ & $1.86$ & $-12.35$ & $50.50$ & $-19.83$ & $0.9913$ & $0.9659$\\ 
\hline \end{tabular}
\caption{Parameter values of the reduced first moir\'e harmonic model with the zero-twist constraints imposed for twisted AA-stacked bilayer \ch{ZrS2}. The interlayer hopping parameters are given in units of \si{\milli\electronvolt}, while the effective masses are given in units of the bare electron mass $m_e$. The minimal overlaps with the first two groups of bands are listed at the end.}
\label{app:tab:ZrS2_AA_4}
\end{table}
\begin{table}[H]
\centering
\resizebox{\textwidth}{!}{\begin{tabular}{|r|r|r|r|r|r|r|r|r|r|r|r|r|r|r|r|r|r|r|r|r|}
\hline
$\theta$ & $m_x$ & $m_y$ & $w^{\text{AB}}_1$ & $w^{\text{AB}}_2$ & $w^{\text{AB}}_3$ & $w^{\text{AB}}_4$ & $w^{\prime\text{AB}}_1$ & $w^{\prime\text{AB}}_2$ & $w^{\prime\text{AB}}_3$ & $w^{\prime\text{AB}}_4$ & $w^{\prime\text{AB}}_5$ & $w^{\prime\text{AB}}_6$ & $w^{\prime\text{AB}}_7$ & $w^{\prime\text{AB}}_8$ & $w^{\prime\text{AB}}_9$ & $w^{\prime\text{AB}}_{10}$ & $w^{\prime\text{AB}}_{11}$ & $w^{\prime\text{AB}}_{12}$ & $\mathcal{O}_{1}$ & $\mathcal{O}_{2}$\\
\hline
\SI{9.43}{\degree} & $0.27$ & $1.79$ & $-2.06$ & $-53.12$ & $0.72$ & $-0.37$ & $-0.76$ & $-4.99$ & $-19.64$ & $-23.74$ & $0.14$ & $6.57$ & $0.73$ & $-0.20$ & $-8.79$ & $1.05$ & $0.25$ & $-3.23$ & $0.9984$ & $0.9832$\\
\hline
\SI{7.34}{\degree} & $0.26$ & $1.63$ & $-2.13$ & $-50.84$ & $0.62$ & $-0.28$ & $-0.87$ & $-4.13$ & $-14.24$ & $-26.51$ & $0.20$ & $5.52$ & $2.19$ & $-0.15$ & $-7.79$ & $0.51$ & $0.22$ & $-1.29$ & $0.9989$ & $0.8526$\\
\hline
\SI{6.01}{\degree} & $0.26$ & $2.27$ & $-2.23$ & $-47.30$ & $0.52$ & $-0.24$ & $-0.74$ & $-2.90$ & $-12.45$ & $-23.06$ & $0.55$ & $5.16$ & $1.07$ & $-0.19$ & $-6.42$ & $1.26$ & $0.18$ & $-0.87$ & $0.9994$ & $0.9937$\\
\hline
\SI{5.09}{\degree} & $0.24$ & $1.93$ & $-2.09$ & $-47.60$ & $0.50$ & $-0.18$ & $-0.88$ & $-2.90$ & $-11.75$ & $-25.31$ & $0.59$ & $6.22$ & $0.23$ & $-0.16$ & $-6.65$ & $1.73$ & $0.23$ & $-0.22$ & $0.9993$ & $0.9983$\\
\hline
\SI{4.41}{\degree} & $0.25$ & $2.05$ & $-2.06$ & $-44.76$ & $0.40$ & $-0.15$ & $-0.78$ & $-2.55$ & $-10.58$ & $-24.51$ & $0.48$ & $5.72$ & $0.64$ & $-0.11$ & $-5.89$ & $1.38$ & $0.20$ & $0.20$ & $0.9996$ & $0.9986$\\
\hline
\SI{3.89}{\degree} & $0.26$ & $1.91$ & $-1.74$ & $-38.37$ & $0.38$ & $-0.10$ & $-0.73$ & $-2.27$ & $-9.09$ & $-22.99$ & $0.46$ & $4.20$ & $0.05$ & $-0.08$ & $-5.29$ & $1.10$ & $0.19$ & $0.95$ & $0.9998$ & $0.9989$\\ 
\hline \end{tabular}}
\caption{Parameter values of the full first moir\'e harmonic model for twisted AB-stacked bilayer \ch{ZrS2}. The interlayer hopping parameters are given in units of \si{\milli\electronvolt}, while the effective masses are given in units of the bare electron mass $m_e$. The minimal overlaps with the first two groups of bands are listed at the end.}
\label{app:tab:ZrS2_AB_1}
\end{table}
\begin{table}[H]
\centering
\begin{tabular}{|r|r|r|r|r|r|r|r|r|r|r|}
\hline
$\theta$ & $m_x$ & $m_y$ & $w^{\text{AB}}_2$ & $w^{\prime\text{AB}}_2$ & $w^{\prime\text{AB}}_3$ & $w^{\prime\text{AB}}_4$ & $w^{\prime\text{AB}}_6$ & $w^{\prime\text{AB}}_9$ & $\mathcal{O}_{1}$ & $\mathcal{O}_{2}$\\
\hline
\SI{9.43}{\degree} & $0.27$ & $1.84$ & $-52.19$ & $-4.67$ & N/A & $-23.64$ & $7.33$ & $-7.77$ & $0.9968$ & $0.9756$\\
\hline
\SI{7.34}{\degree} & $0.26$ & $1.63$ & $-49.34$ & $-4.07$ & N/A & $-26.72$ & $5.42$ & $-7.32$ & $0.9978$ & $0.7689$\\
\hline
\SI{6.01}{\degree} & $0.27$ & $2.28$ & $-45.42$ & $-2.68$ & N/A & $-22.31$ & $4.29$ & $-4.95$ & $0.9986$ & $0.9921$\\
\hline
\SI{5.09}{\degree} & $0.25$ & $1.94$ & $-44.99$ & $-2.63$ & N/A & $-23.80$ & $4.83$ & $-4.49$ & $0.9986$ & $0.9965$\\
\hline
\SI{4.41}{\degree} & $0.24$ & $2.06$ & $-44.25$ & $-2.15$ & N/A & $-23.38$ & $4.76$ & $-4.13$ & $0.9989$ & $0.9970$\\
\hline
\SI{3.89}{\degree} & $0.24$ & $1.89$ & $-38.95$ & $-1.58$ & $-9.96$ & $-17.26$ & N/A & $-3.94$ & $0.9993$ & $0.9980$\\ 
\hline \end{tabular}
\caption{Parameter values of the reduced first moir\'e harmonic model for twisted AB-stacked bilayer \ch{ZrS2}. The interlayer hopping parameters are given in units of \si{\milli\electronvolt}, while the effective masses are given in units of the bare electron mass $m_e$. The minimal overlaps with the first two groups of bands are listed at the end.}
\label{app:tab:ZrS2_AB_2}
\end{table}
\begin{table}[H]
\centering
\begin{tabular}{|r|r|r|r|r|r|r|r|r|r|r|r|r|r|}
\hline
$\theta$ & $m_x$ & $m_y$ & $w^{\text{AB}}_1$ & $w^{\text{AB}}_2$ & $w^{\prime\text{AB}}_1$ & $w^{\prime\text{AB}}_2$ & $w^{\prime\text{AB}}_3$ & $w^{\prime\text{AB}}_4$ & $w^{\prime\text{AB}}_5$ & $w^{\prime\text{AB}}_6$ & $w^{\prime\text{AB}}_7$ & $\mathcal{O}_{1}$ & $\mathcal{O}_{2}$\\
\hline
\SI{9.43}{\degree} & $0.29$ & $1.86$ & $-2.24$ & $-43.99$ & $1.02$ & $-0.99$ & $-1.68$ & $-14.32$ & $1.25$ & $0.15$ & $3.96$ & $0.9647$ & $0.1384$\\
\hline
\SI{7.34}{\degree} & $0.29$ & $1.86$ & $-1.48$ & $-41.81$ & $-0.61$ & $-3.66$ & $-24.46$ & $-24.00$ & $0.30$ & $6.21$ & $20.96$ & $0.9782$ & $0.0654$\\
\hline
\SI{6.01}{\degree} & $0.29$ & $1.86$ & $-1.64$ & $-39.74$ & $-0.85$ & $-3.00$ & $-12.24$ & $-20.71$ & N/A & $3.85$ & $-2.23$ & $0.9829$ & $0.5894$\\
\hline
\SI{5.09}{\degree} & $0.29$ & $1.86$ & $-1.60$ & $-39.03$ & $-0.83$ & $-3.44$ & $-11.50$ & $-21.69$ & $0.63$ & $3.28$ & $0.33$ & $0.9857$ & $0.7632$\\
\hline
\SI{4.41}{\degree} & $0.29$ & $1.86$ & $-1.66$ & $-37.32$ & $-0.74$ & $-2.66$ & $-10.08$ & $-22.30$ & $0.46$ & $4.60$ & $0.46$ & $0.9876$ & $0.8792$\\
\hline
\SI{3.89}{\degree} & $0.29$ & $1.86$ & $-1.57$ & $-35.88$ & $-0.76$ & $-3.33$ & $-9.46$ & $-21.20$ & $0.48$ & $3.41$ & $0.80$ & $0.9876$ & $0.9323$\\ 
\hline \end{tabular}
\caption{Parameter values of the first moir\'e harmonic model with the zero-twist constraints imposed for twisted AB-stacked bilayer \ch{ZrS2}. The interlayer hopping parameters are given in units of \si{\milli\electronvolt}, while the effective masses are given in units of the bare electron mass $m_e$. The minimal overlaps with the first two groups of bands are listed at the end.}
\label{app:tab:ZrS2_AB_3}
\end{table}
\begin{table}[H]
\centering
\begin{tabular}{|r|r|r|r|r|r|r|r|r|}
\hline
$\theta$ & $m_x$ & $m_y$ & $w^{\text{AB}}_2$ & $w^{\prime\text{AB}}_3$ & $w^{\prime\text{AB}}_4$ & $w^{\prime\text{AB}}_6$ & $\mathcal{O}_{1}$ & $\mathcal{O}_{2}$\\
\hline
\SI{9.43}{\degree} & $0.29$ & $1.86$ & $-43.99$ & N/A & $-15.04$ & $0.08$ & $0.9613$ & $0.0177$\\
\hline
\SI{7.34}{\degree} & $0.29$ & $1.86$ & $-41.81$ & N/A & $-26.84$ & $7.98$ & $0.9743$ & $0.0457$\\
\hline
\SI{6.01}{\degree} & $0.29$ & $1.86$ & $-39.74$ & $-12.11$ & $-16.73$ & N/A & $0.9787$ & $0.4941$\\
\hline
\SI{5.09}{\degree} & $0.29$ & $1.86$ & $-39.03$ & $-11.46$ & $-18.08$ & N/A & $0.9838$ & $0.7649$\\
\hline
\SI{4.41}{\degree} & $0.29$ & $1.86$ & $-37.32$ & $-10.29$ & $-16.73$ & N/A & $0.9859$ & $0.8809$\\
\hline
\SI{3.89}{\degree} & $0.29$ & $1.86$ & $-35.88$ & $-10.05$ & $-17.03$ & N/A & $0.9871$ & $0.9317$\\ 
\hline \end{tabular}
\caption{Parameter values of the reduced first moir\'e harmonic model with the zero-twist constraints imposed for twisted AB-stacked bilayer \ch{ZrS2}. The interlayer hopping parameters are given in units of \si{\milli\electronvolt}, while the effective masses are given in units of the bare electron mass $m_e$. The minimal overlaps with the first two groups of bands are listed at the end.}
\label{app:tab:ZrS2_AB_4}
\end{table} \subsection{Numerical results}\label{app:sec:fitted_models:plots}
\subsubsection{Band structures along high-symmetry lines}\label{app:sec:fitted_models:plots:bs_hsl}
\begin{figure}[H]
\centering
\includegraphics[width=\textwidth]{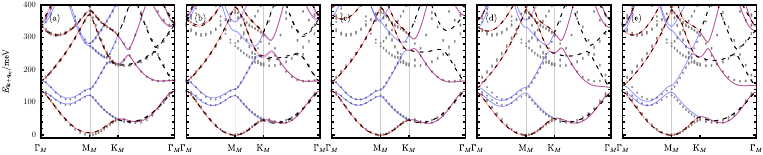}
\subfloat{\label{app:fig:fit_BS_1:a}}\subfloat{\label{app:fig:fit_BS_1:b}}\subfloat{\label{app:fig:fit_BS_1:c}}\subfloat{\label{app:fig:fit_BS_1:d}}\subfloat{\label{app:fig:fit_BS_1:e}}\caption{Model band structures for twisted AA-stacked bilayer \ch{SnSe2} at $\theta = \SI{9.43}{\degree}$. We consider the full continuum model (a), the full first moir\'e harmonic model (b), the reduced first moir\'e harmonic model (c), the first moir\'e harmonic model with the zero-twist constraints imposed (d), and the reduced first moir\'e harmonic model with the zero-twist constraints imposed (e). The bands of valleys $\eta = 0, 1, 2$ are shown by the blue, dashed black, and red lines, while the \textit{ab initio} band structure is shown by the gray dots.}
\label{app:fig:fit_BS_1}
\end{figure}
\begin{figure}[H]
\centering
\includegraphics[width=\textwidth]{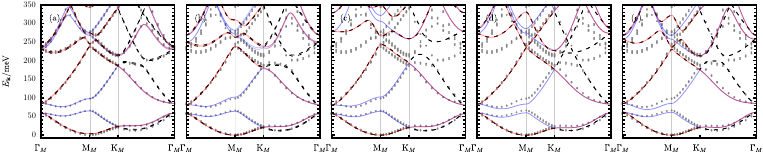}
\subfloat{\label{app:fig:fit_BS_2:a}}\subfloat{\label{app:fig:fit_BS_2:b}}\subfloat{\label{app:fig:fit_BS_2:c}}\subfloat{\label{app:fig:fit_BS_2:d}}\subfloat{\label{app:fig:fit_BS_2:e}}\caption{Model band structures for twisted AA-stacked bilayer \ch{SnSe2} at $\theta = \SI{7.34}{\degree}$. We consider the full continuum model (a), the full first moir\'e harmonic model (b), the reduced first moir\'e harmonic model (c), the first moir\'e harmonic model with the zero-twist constraints imposed (d), and the reduced first moir\'e harmonic model with the zero-twist constraints imposed (e). The bands of valleys $\eta = 0, 1, 2$ are shown by the blue, dashed black, and red lines, while the \textit{ab initio} band structure is shown by the gray dots.}
\label{app:fig:fit_BS_2}
\end{figure}
\begin{figure}[H]
\centering
\includegraphics[width=\textwidth]{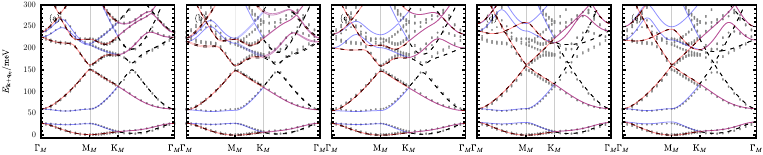}
\subfloat{\label{app:fig:fit_BS_3:a}}\subfloat{\label{app:fig:fit_BS_3:b}}\subfloat{\label{app:fig:fit_BS_3:c}}\subfloat{\label{app:fig:fit_BS_3:d}}\subfloat{\label{app:fig:fit_BS_3:e}}\caption{Model band structures for twisted AA-stacked bilayer \ch{SnSe2} at $\theta = \SI{6.01}{\degree}$. We consider the full continuum model (a), the full first moir\'e harmonic model (b), the reduced first moir\'e harmonic model (c), the first moir\'e harmonic model with the zero-twist constraints imposed (d), and the reduced first moir\'e harmonic model with the zero-twist constraints imposed (e). The bands of valleys $\eta = 0, 1, 2$ are shown by the blue, dashed black, and red lines, while the \textit{ab initio} band structure is shown by the gray dots.}
\label{app:fig:fit_BS_3}
\end{figure}
\begin{figure}[H]
\centering
\includegraphics[width=\textwidth]{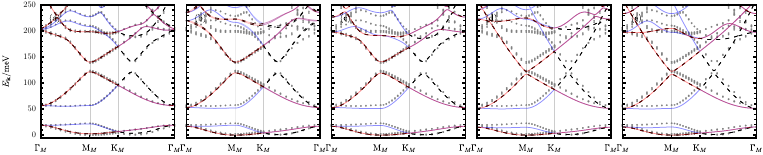}
\subfloat{\label{app:fig:fit_BS_4:a}}\subfloat{\label{app:fig:fit_BS_4:b}}\subfloat{\label{app:fig:fit_BS_4:c}}\subfloat{\label{app:fig:fit_BS_4:d}}\subfloat{\label{app:fig:fit_BS_4:e}}\caption{Model band structures for twisted AA-stacked bilayer \ch{SnSe2} at $\theta = \SI{5.09}{\degree}$. We consider the full continuum model (a), the full first moir\'e harmonic model (b), the reduced first moir\'e harmonic model (c), the first moir\'e harmonic model with the zero-twist constraints imposed (d), and the reduced first moir\'e harmonic model with the zero-twist constraints imposed (e). The bands of valleys $\eta = 0, 1, 2$ are shown by the blue, dashed black, and red lines, while the \textit{ab initio} band structure is shown by the gray dots.}
\label{app:fig:fit_BS_4}
\end{figure}
\begin{figure}[H]
\centering
\includegraphics[width=\textwidth]{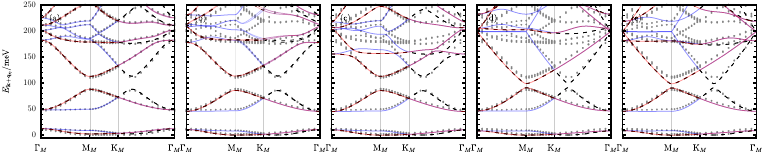}
\subfloat{\label{app:fig:fit_BS_5:a}}\subfloat{\label{app:fig:fit_BS_5:b}}\subfloat{\label{app:fig:fit_BS_5:c}}\subfloat{\label{app:fig:fit_BS_5:d}}\subfloat{\label{app:fig:fit_BS_5:e}}\caption{Model band structures for twisted AA-stacked bilayer \ch{SnSe2} at $\theta = \SI{4.41}{\degree}$. We consider the full continuum model (a), the full first moir\'e harmonic model (b), the reduced first moir\'e harmonic model (c), the first moir\'e harmonic model with the zero-twist constraints imposed (d), and the reduced first moir\'e harmonic model with the zero-twist constraints imposed (e). The bands of valleys $\eta = 0, 1, 2$ are shown by the blue, dashed black, and red lines, while the \textit{ab initio} band structure is shown by the gray dots.}
\label{app:fig:fit_BS_5}
\end{figure}
\begin{figure}[H]
\centering
\includegraphics[width=\textwidth]{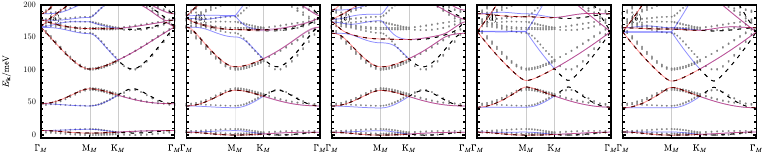}
\subfloat{\label{app:fig:fit_BS_6:a}}\subfloat{\label{app:fig:fit_BS_6:b}}\subfloat{\label{app:fig:fit_BS_6:c}}\subfloat{\label{app:fig:fit_BS_6:d}}\subfloat{\label{app:fig:fit_BS_6:e}}\caption{Model band structures for twisted AA-stacked bilayer \ch{SnSe2} at $\theta = \SI{3.89}{\degree}$. We consider the full continuum model (a), the full first moir\'e harmonic model (b), the reduced first moir\'e harmonic model (c), the first moir\'e harmonic model with the zero-twist constraints imposed (d), and the reduced first moir\'e harmonic model with the zero-twist constraints imposed (e). The bands of valleys $\eta = 0, 1, 2$ are shown by the blue, dashed black, and red lines, while the \textit{ab initio} band structure is shown by the gray dots.}
\label{app:fig:fit_BS_6}
\end{figure}
\begin{figure}[H]
\centering
\includegraphics[width=\textwidth]{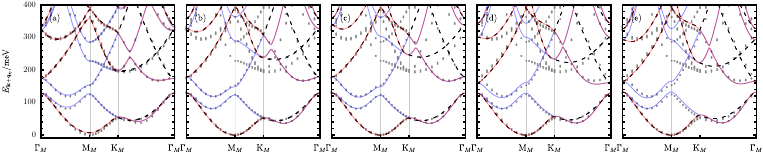}
\subfloat{\label{app:fig:fit_BS_7:a}}\subfloat{\label{app:fig:fit_BS_7:b}}\subfloat{\label{app:fig:fit_BS_7:c}}\subfloat{\label{app:fig:fit_BS_7:d}}\subfloat{\label{app:fig:fit_BS_7:e}}\caption{Model band structures for twisted AB-stacked bilayer \ch{SnSe2} at $\theta = \SI{9.43}{\degree}$. We consider the full continuum model (a), the full first moir\'e harmonic model (b), the reduced first moir\'e harmonic model (c), the first moir\'e harmonic model with the zero-twist constraints imposed (d), and the reduced first moir\'e harmonic model with the zero-twist constraints imposed (e). The bands of valleys $\eta = 0, 1, 2$ are shown by the blue, dashed black, and red lines, while the \textit{ab initio} band structure is shown by the gray dots.}
\label{app:fig:fit_BS_7}
\end{figure}
\begin{figure}[H]
\centering
\includegraphics[width=\textwidth]{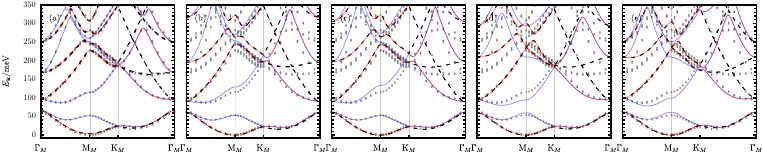}
\subfloat{\label{app:fig:fit_BS_8:a}}\subfloat{\label{app:fig:fit_BS_8:b}}\subfloat{\label{app:fig:fit_BS_8:c}}\subfloat{\label{app:fig:fit_BS_8:d}}\subfloat{\label{app:fig:fit_BS_8:e}}\caption{Model band structures for twisted AB-stacked bilayer \ch{SnSe2} at $\theta = \SI{7.34}{\degree}$. We consider the full continuum model (a), the full first moir\'e harmonic model (b), the reduced first moir\'e harmonic model (c), the first moir\'e harmonic model with the zero-twist constraints imposed (d), and the reduced first moir\'e harmonic model with the zero-twist constraints imposed (e). The bands of valleys $\eta = 0, 1, 2$ are shown by the blue, dashed black, and red lines, while the \textit{ab initio} band structure is shown by the gray dots.}
\label{app:fig:fit_BS_8}
\end{figure}
\begin{figure}[H]
\centering
\includegraphics[width=\textwidth]{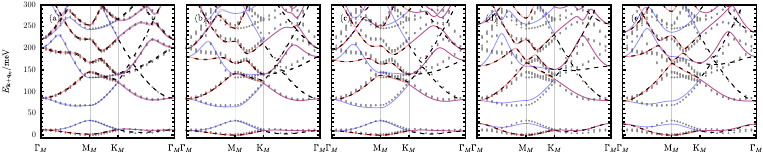}
\subfloat{\label{app:fig:fit_BS_9:a}}\subfloat{\label{app:fig:fit_BS_9:b}}\subfloat{\label{app:fig:fit_BS_9:c}}\subfloat{\label{app:fig:fit_BS_9:d}}\subfloat{\label{app:fig:fit_BS_9:e}}\caption{Model band structures for twisted AB-stacked bilayer \ch{SnSe2} at $\theta = \SI{6.01}{\degree}$. We consider the full continuum model (a), the full first moir\'e harmonic model (b), the reduced first moir\'e harmonic model (c), the first moir\'e harmonic model with the zero-twist constraints imposed (d), and the reduced first moir\'e harmonic model with the zero-twist constraints imposed (e). The bands of valleys $\eta = 0, 1, 2$ are shown by the blue, dashed black, and red lines, while the \textit{ab initio} band structure is shown by the gray dots.}
\label{app:fig:fit_BS_9}
\end{figure}
\begin{figure}[H]
\centering
\includegraphics[width=\textwidth]{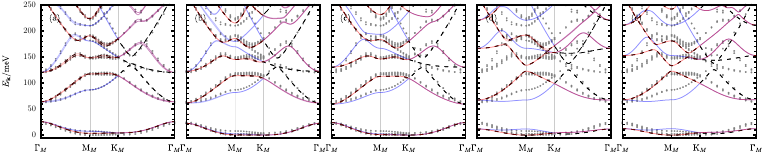}
\subfloat{\label{app:fig:fit_BS_10:a}}\subfloat{\label{app:fig:fit_BS_10:b}}\subfloat{\label{app:fig:fit_BS_10:c}}\subfloat{\label{app:fig:fit_BS_10:d}}\subfloat{\label{app:fig:fit_BS_10:e}}\caption{Model band structures for twisted AB-stacked bilayer \ch{SnSe2} at $\theta = \SI{5.09}{\degree}$. We consider the full continuum model (a), the full first moir\'e harmonic model (b), the reduced first moir\'e harmonic model (c), the first moir\'e harmonic model with the zero-twist constraints imposed (d), and the reduced first moir\'e harmonic model with the zero-twist constraints imposed (e). The bands of valleys $\eta = 0, 1, 2$ are shown by the blue, dashed black, and red lines, while the \textit{ab initio} band structure is shown by the gray dots.}
\label{app:fig:fit_BS_10}
\end{figure}
\begin{figure}[H]
\centering
\includegraphics[width=\textwidth]{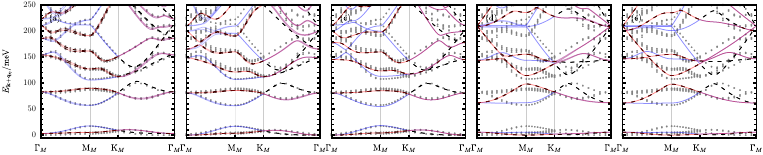}
\subfloat{\label{app:fig:fit_BS_11:a}}\subfloat{\label{app:fig:fit_BS_11:b}}\subfloat{\label{app:fig:fit_BS_11:c}}\subfloat{\label{app:fig:fit_BS_11:d}}\subfloat{\label{app:fig:fit_BS_11:e}}\caption{Model band structures for twisted AB-stacked bilayer \ch{SnSe2} at $\theta = \SI{4.41}{\degree}$. We consider the full continuum model (a), the full first moir\'e harmonic model (b), the reduced first moir\'e harmonic model (c), the first moir\'e harmonic model with the zero-twist constraints imposed (d), and the reduced first moir\'e harmonic model with the zero-twist constraints imposed (e). The bands of valleys $\eta = 0, 1, 2$ are shown by the blue, dashed black, and red lines, while the \textit{ab initio} band structure is shown by the gray dots.}
\label{app:fig:fit_BS_11}
\end{figure}
\begin{figure}[H]
\centering
\includegraphics[width=\textwidth]{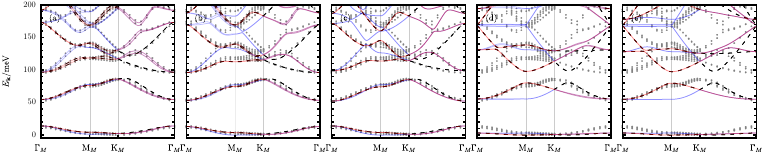}
\subfloat{\label{app:fig:fit_BS_12:a}}\subfloat{\label{app:fig:fit_BS_12:b}}\subfloat{\label{app:fig:fit_BS_12:c}}\subfloat{\label{app:fig:fit_BS_12:d}}\subfloat{\label{app:fig:fit_BS_12:e}}\caption{Model band structures for twisted AB-stacked bilayer \ch{SnSe2} at $\theta = \SI{3.89}{\degree}$. We consider the full continuum model (a), the full first moir\'e harmonic model (b), the reduced first moir\'e harmonic model (c), the first moir\'e harmonic model with the zero-twist constraints imposed (d), and the reduced first moir\'e harmonic model with the zero-twist constraints imposed (e). The bands of valleys $\eta = 0, 1, 2$ are shown by the blue, dashed black, and red lines, while the \textit{ab initio} band structure is shown by the gray dots.}
\label{app:fig:fit_BS_12}
\end{figure}
\begin{figure}[H]
\centering
\includegraphics[width=\textwidth]{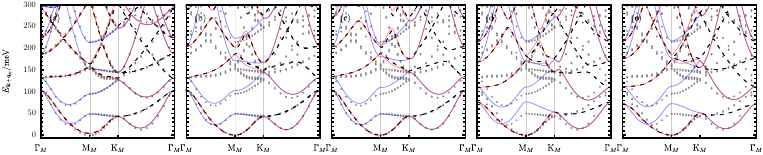}
\subfloat{\label{app:fig:fit_BS_13:a}}\subfloat{\label{app:fig:fit_BS_13:b}}\subfloat{\label{app:fig:fit_BS_13:c}}\subfloat{\label{app:fig:fit_BS_13:d}}\subfloat{\label{app:fig:fit_BS_13:e}}\caption{Model band structures for twisted AA-stacked bilayer \ch{ZrS2} at $\theta = \SI{9.43}{\degree}$. We consider the full continuum model (a), the full first moir\'e harmonic model (b), the reduced first moir\'e harmonic model (c), the first moir\'e harmonic model with the zero-twist constraints imposed (d), and the reduced first moir\'e harmonic model with the zero-twist constraints imposed (e). The bands of valleys $\eta = 0, 1, 2$ are shown by the blue, dashed black, and red lines, while the \textit{ab initio} band structure is shown by the gray dots.}
\label{app:fig:fit_BS_13}
\end{figure}
\begin{figure}[H]
\centering
\includegraphics[width=\textwidth]{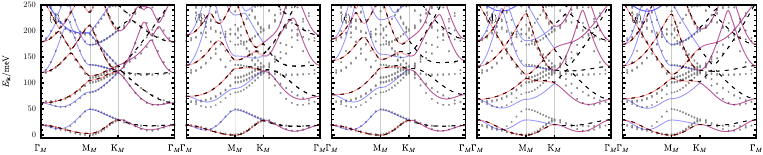}
\subfloat{\label{app:fig:fit_BS_14:a}}\subfloat{\label{app:fig:fit_BS_14:b}}\subfloat{\label{app:fig:fit_BS_14:c}}\subfloat{\label{app:fig:fit_BS_14:d}}\subfloat{\label{app:fig:fit_BS_14:e}}\caption{Model band structures for twisted AA-stacked bilayer \ch{ZrS2} at $\theta = \SI{7.34}{\degree}$. We consider the full continuum model (a), the full first moir\'e harmonic model (b), the reduced first moir\'e harmonic model (c), the first moir\'e harmonic model with the zero-twist constraints imposed (d), and the reduced first moir\'e harmonic model with the zero-twist constraints imposed (e). The bands of valleys $\eta = 0, 1, 2$ are shown by the blue, dashed black, and red lines, while the \textit{ab initio} band structure is shown by the gray dots.}
\label{app:fig:fit_BS_14}
\end{figure}
\begin{figure}[H]
\centering
\includegraphics[width=\textwidth]{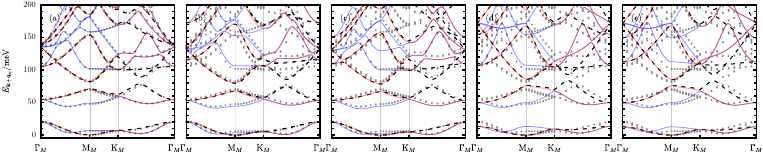}
\subfloat{\label{app:fig:fit_BS_15:a}}\subfloat{\label{app:fig:fit_BS_15:b}}\subfloat{\label{app:fig:fit_BS_15:c}}\subfloat{\label{app:fig:fit_BS_15:d}}\subfloat{\label{app:fig:fit_BS_15:e}}\caption{Model band structures for twisted AA-stacked bilayer \ch{ZrS2} at $\theta = \SI{6.01}{\degree}$. We consider the full continuum model (a), the full first moir\'e harmonic model (b), the reduced first moir\'e harmonic model (c), the first moir\'e harmonic model with the zero-twist constraints imposed (d), and the reduced first moir\'e harmonic model with the zero-twist constraints imposed (e). The bands of valleys $\eta = 0, 1, 2$ are shown by the blue, dashed black, and red lines, while the \textit{ab initio} band structure is shown by the gray dots.}
\label{app:fig:fit_BS_15}
\end{figure}
\begin{figure}[H]
\centering
\includegraphics[width=\textwidth]{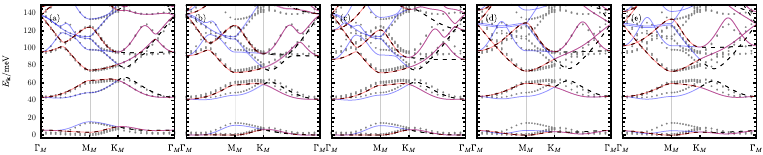}
\subfloat{\label{app:fig:fit_BS_16:a}}\subfloat{\label{app:fig:fit_BS_16:b}}\subfloat{\label{app:fig:fit_BS_16:c}}\subfloat{\label{app:fig:fit_BS_16:d}}\subfloat{\label{app:fig:fit_BS_16:e}}\caption{Model band structures for twisted AA-stacked bilayer \ch{ZrS2} at $\theta = \SI{5.09}{\degree}$. We consider the full continuum model (a), the full first moir\'e harmonic model (b), the reduced first moir\'e harmonic model (c), the first moir\'e harmonic model with the zero-twist constraints imposed (d), and the reduced first moir\'e harmonic model with the zero-twist constraints imposed (e). The bands of valleys $\eta = 0, 1, 2$ are shown by the blue, dashed black, and red lines, while the \textit{ab initio} band structure is shown by the gray dots.}
\label{app:fig:fit_BS_16}
\end{figure}
\begin{figure}[H]
\centering
\includegraphics[width=\textwidth]{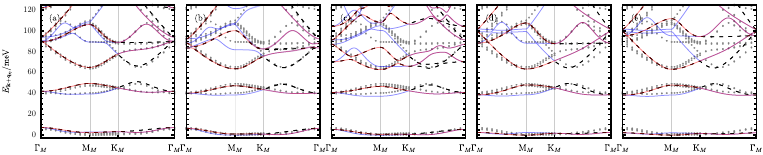}
\subfloat{\label{app:fig:fit_BS_17:a}}\subfloat{\label{app:fig:fit_BS_17:b}}\subfloat{\label{app:fig:fit_BS_17:c}}\subfloat{\label{app:fig:fit_BS_17:d}}\subfloat{\label{app:fig:fit_BS_17:e}}\caption{Model band structures for twisted AA-stacked bilayer \ch{ZrS2} at $\theta = \SI{4.41}{\degree}$. We consider the full continuum model (a), the full first moir\'e harmonic model (b), the reduced first moir\'e harmonic model (c), the first moir\'e harmonic model with the zero-twist constraints imposed (d), and the reduced first moir\'e harmonic model with the zero-twist constraints imposed (e). The bands of valleys $\eta = 0, 1, 2$ are shown by the blue, dashed black, and red lines, while the \textit{ab initio} band structure is shown by the gray dots.}
\label{app:fig:fit_BS_17}
\end{figure}
\begin{figure}[H]
\centering
\includegraphics[width=\textwidth]{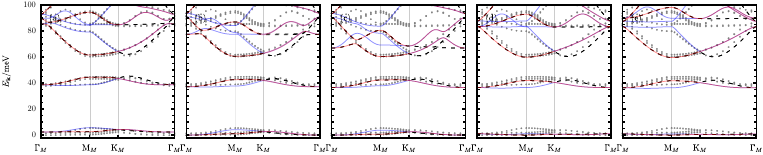}
\subfloat{\label{app:fig:fit_BS_18:a}}\subfloat{\label{app:fig:fit_BS_18:b}}\subfloat{\label{app:fig:fit_BS_18:c}}\subfloat{\label{app:fig:fit_BS_18:d}}\subfloat{\label{app:fig:fit_BS_18:e}}\caption{Model band structures for twisted AA-stacked bilayer \ch{ZrS2} at $\theta = \SI{3.89}{\degree}$. We consider the full continuum model (a), the full first moir\'e harmonic model (b), the reduced first moir\'e harmonic model (c), the first moir\'e harmonic model with the zero-twist constraints imposed (d), and the reduced first moir\'e harmonic model with the zero-twist constraints imposed (e). The bands of valleys $\eta = 0, 1, 2$ are shown by the blue, dashed black, and red lines, while the \textit{ab initio} band structure is shown by the gray dots.}
\label{app:fig:fit_BS_18}
\end{figure}
\begin{figure}[H]
\centering
\includegraphics[width=\textwidth]{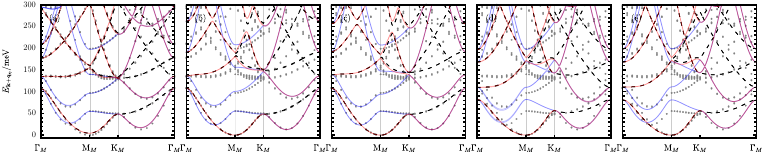}
\subfloat{\label{app:fig:fit_BS_19:a}}\subfloat{\label{app:fig:fit_BS_19:b}}\subfloat{\label{app:fig:fit_BS_19:c}}\subfloat{\label{app:fig:fit_BS_19:d}}\subfloat{\label{app:fig:fit_BS_19:e}}\caption{Model band structures for twisted AB-stacked bilayer \ch{ZrS2} at $\theta = \SI{9.43}{\degree}$. We consider the full continuum model (a), the full first moir\'e harmonic model (b), the reduced first moir\'e harmonic model (c), the first moir\'e harmonic model with the zero-twist constraints imposed (d), and the reduced first moir\'e harmonic model with the zero-twist constraints imposed (e). The bands of valleys $\eta = 0, 1, 2$ are shown by the blue, dashed black, and red lines, while the \textit{ab initio} band structure is shown by the gray dots.}
\label{app:fig:fit_BS_19}
\end{figure}
\begin{figure}[H]
\centering
\includegraphics[width=\textwidth]{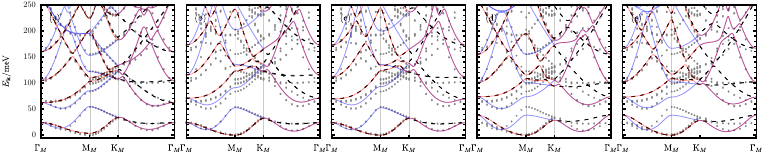}
\subfloat{\label{app:fig:fit_BS_20:a}}\subfloat{\label{app:fig:fit_BS_20:b}}\subfloat{\label{app:fig:fit_BS_20:c}}\subfloat{\label{app:fig:fit_BS_20:d}}\subfloat{\label{app:fig:fit_BS_20:e}}\caption{Model band structures for twisted AB-stacked bilayer \ch{ZrS2} at $\theta = \SI{7.34}{\degree}$. We consider the full continuum model (a), the full first moir\'e harmonic model (b), the reduced first moir\'e harmonic model (c), the first moir\'e harmonic model with the zero-twist constraints imposed (d), and the reduced first moir\'e harmonic model with the zero-twist constraints imposed (e). The bands of valleys $\eta = 0, 1, 2$ are shown by the blue, dashed black, and red lines, while the \textit{ab initio} band structure is shown by the gray dots.}
\label{app:fig:fit_BS_20}
\end{figure}
\begin{figure}[H]
\centering
\includegraphics[width=\textwidth]{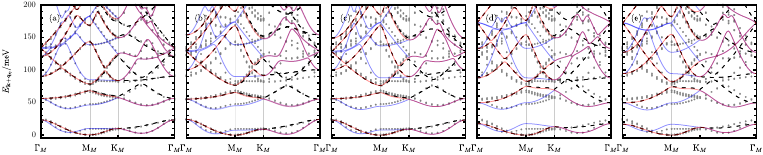}
\subfloat{\label{app:fig:fit_BS_21:a}}\subfloat{\label{app:fig:fit_BS_21:b}}\subfloat{\label{app:fig:fit_BS_21:c}}\subfloat{\label{app:fig:fit_BS_21:d}}\subfloat{\label{app:fig:fit_BS_21:e}}\caption{Model band structures for twisted AB-stacked bilayer \ch{ZrS2} at $\theta = \SI{6.01}{\degree}$. We consider the full continuum model (a), the full first moir\'e harmonic model (b), the reduced first moir\'e harmonic model (c), the first moir\'e harmonic model with the zero-twist constraints imposed (d), and the reduced first moir\'e harmonic model with the zero-twist constraints imposed (e). The bands of valleys $\eta = 0, 1, 2$ are shown by the blue, dashed black, and red lines, while the \textit{ab initio} band structure is shown by the gray dots.}
\label{app:fig:fit_BS_21}
\end{figure}
\begin{figure}[H]
\centering
\includegraphics[width=\textwidth]{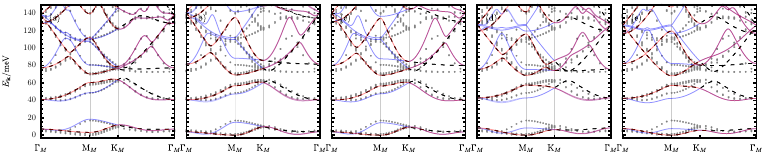}
\subfloat{\label{app:fig:fit_BS_22:a}}\subfloat{\label{app:fig:fit_BS_22:b}}\subfloat{\label{app:fig:fit_BS_22:c}}\subfloat{\label{app:fig:fit_BS_22:d}}\subfloat{\label{app:fig:fit_BS_22:e}}\caption{Model band structures for twisted AB-stacked bilayer \ch{ZrS2} at $\theta = \SI{5.09}{\degree}$. We consider the full continuum model (a), the full first moir\'e harmonic model (b), the reduced first moir\'e harmonic model (c), the first moir\'e harmonic model with the zero-twist constraints imposed (d), and the reduced first moir\'e harmonic model with the zero-twist constraints imposed (e). The bands of valleys $\eta = 0, 1, 2$ are shown by the blue, dashed black, and red lines, while the \textit{ab initio} band structure is shown by the gray dots.}
\label{app:fig:fit_BS_22}
\end{figure}
\begin{figure}[H]
\centering
\includegraphics[width=\textwidth]{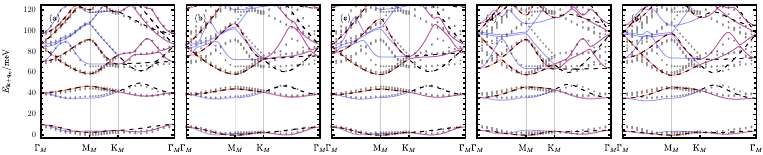}
\subfloat{\label{app:fig:fit_BS_23:a}}\subfloat{\label{app:fig:fit_BS_23:b}}\subfloat{\label{app:fig:fit_BS_23:c}}\subfloat{\label{app:fig:fit_BS_23:d}}\subfloat{\label{app:fig:fit_BS_23:e}}\caption{Model band structures for twisted AB-stacked bilayer \ch{ZrS2} at $\theta = \SI{4.41}{\degree}$. We consider the full continuum model (a), the full first moir\'e harmonic model (b), the reduced first moir\'e harmonic model (c), the first moir\'e harmonic model with the zero-twist constraints imposed (d), and the reduced first moir\'e harmonic model with the zero-twist constraints imposed (e). The bands of valleys $\eta = 0, 1, 2$ are shown by the blue, dashed black, and red lines, while the \textit{ab initio} band structure is shown by the gray dots.}
\label{app:fig:fit_BS_23}
\end{figure}
\begin{figure}[H]
\centering
\includegraphics[width=\textwidth]{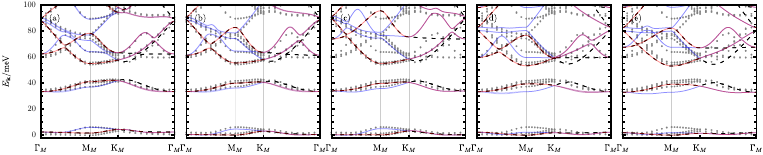}
\subfloat{\label{app:fig:fit_BS_24:a}}\subfloat{\label{app:fig:fit_BS_24:b}}\subfloat{\label{app:fig:fit_BS_24:c}}\subfloat{\label{app:fig:fit_BS_24:d}}\subfloat{\label{app:fig:fit_BS_24:e}}\caption{Model band structures for twisted AB-stacked bilayer \ch{ZrS2} at $\theta = \SI{3.89}{\degree}$. We consider the full continuum model (a), the full first moir\'e harmonic model (b), the reduced first moir\'e harmonic model (c), the first moir\'e harmonic model with the zero-twist constraints imposed (d), and the reduced first moir\'e harmonic model with the zero-twist constraints imposed (e). The bands of valleys $\eta = 0, 1, 2$ are shown by the blue, dashed black, and red lines, while the \textit{ab initio} band structure is shown by the gray dots.}
\label{app:fig:fit_BS_24}
\end{figure} \subsubsection{Spectra of the first gapped conduction bands}\label{app:sec:fitted_models:plots:details}
\begin{figure}[H]
\centering
\includegraphics[width=\textwidth]{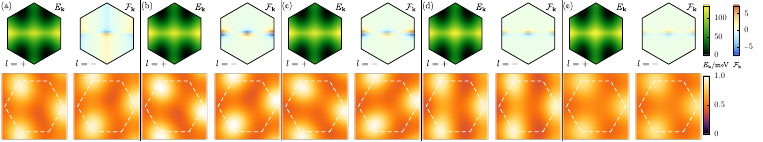}
\subfloat{\label{app:fig:fit_2D_bands_1_1:a}}\subfloat{\label{app:fig:fit_2D_bands_1_1:b}}\subfloat{\label{app:fig:fit_2D_bands_1_1:c}}\subfloat{\label{app:fig:fit_2D_bands_1_1:d}}\subfloat{\label{app:fig:fit_2D_bands_1_1:e}}\caption{The first group of conduction bands of twisted AA-stacked bilayer \ch{SnSe2} at $\theta = \SI{9.43}{\degree}$. We consider the full continuum model (a), the full first moir\'e harmonic model (b), the reduced first moir\'e harmonic model (c), the first moir\'e harmonic model with the zero-twist constraints imposed (d), and the reduced first moir\'e harmonic model with the zero-twist constraints imposed (e). Within each panel, we plot the dispersion of the (approximately or exactly) degenerate group of bands ($E_{\vec{k}}$) and their non-abelian Berry curvature ($\mathcal{F}_{\vec{k}}$) throughout the first moir\'e BZ (black hexagon). Additionally, we plot the real space LDOS in layer $l= \pm$ within the moir\'e unit cell (dashed hexagon).}
\label{app:fig:fit_2D_bands_1_1}
\end{figure}
\begin{figure}[H]
\centering
\includegraphics[width=\textwidth]{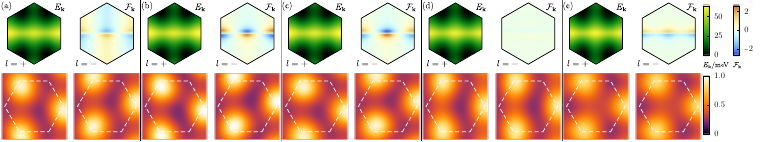}
\subfloat{\label{app:fig:fit_2D_bands_2_1:a}}\subfloat{\label{app:fig:fit_2D_bands_2_1:b}}\subfloat{\label{app:fig:fit_2D_bands_2_1:c}}\subfloat{\label{app:fig:fit_2D_bands_2_1:d}}\subfloat{\label{app:fig:fit_2D_bands_2_1:e}}\caption{The first group of conduction bands of twisted AA-stacked bilayer \ch{SnSe2} at $\theta = \SI{7.34}{\degree}$. We consider the full continuum model (a), the full first moir\'e harmonic model (b), the reduced first moir\'e harmonic model (c), the first moir\'e harmonic model with the zero-twist constraints imposed (d), and the reduced first moir\'e harmonic model with the zero-twist constraints imposed (e). Within each panel, we plot the dispersion of the (approximately or exactly) degenerate group of bands ($E_{\vec{k}}$) and their non-abelian Berry curvature ($\mathcal{F}_{\vec{k}}$) throughout the first moir\'e BZ (black hexagon). Additionally, we plot the real space LDOS in layer $l= \pm$ within the moir\'e unit cell (dashed hexagon).}
\label{app:fig:fit_2D_bands_2_1}
\end{figure}
\begin{figure}[H]
\centering
\includegraphics[width=\textwidth]{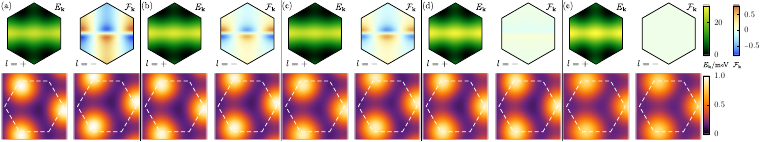}
\subfloat{\label{app:fig:fit_2D_bands_3_1:a}}\subfloat{\label{app:fig:fit_2D_bands_3_1:b}}\subfloat{\label{app:fig:fit_2D_bands_3_1:c}}\subfloat{\label{app:fig:fit_2D_bands_3_1:d}}\subfloat{\label{app:fig:fit_2D_bands_3_1:e}}\caption{The first group of conduction bands of twisted AA-stacked bilayer \ch{SnSe2} at $\theta = \SI{6.01}{\degree}$. We consider the full continuum model (a), the full first moir\'e harmonic model (b), the reduced first moir\'e harmonic model (c), the first moir\'e harmonic model with the zero-twist constraints imposed (d), and the reduced first moir\'e harmonic model with the zero-twist constraints imposed (e). Within each panel, we plot the dispersion of the (approximately or exactly) degenerate group of bands ($E_{\vec{k}}$) and their non-abelian Berry curvature ($\mathcal{F}_{\vec{k}}$) throughout the first moir\'e BZ (black hexagon). Additionally, we plot the real space LDOS in layer $l= \pm$ within the moir\'e unit cell (dashed hexagon).}
\label{app:fig:fit_2D_bands_3_1}
\end{figure}
\begin{figure}[H]
\centering
\includegraphics[width=\textwidth]{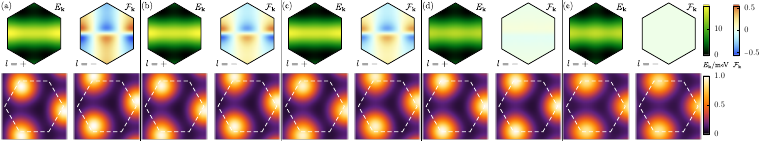}
\subfloat{\label{app:fig:fit_2D_bands_4_1:a}}\subfloat{\label{app:fig:fit_2D_bands_4_1:b}}\subfloat{\label{app:fig:fit_2D_bands_4_1:c}}\subfloat{\label{app:fig:fit_2D_bands_4_1:d}}\subfloat{\label{app:fig:fit_2D_bands_4_1:e}}\caption{The first group of conduction bands of twisted AA-stacked bilayer \ch{SnSe2} at $\theta = \SI{5.09}{\degree}$. We consider the full continuum model (a), the full first moir\'e harmonic model (b), the reduced first moir\'e harmonic model (c), the first moir\'e harmonic model with the zero-twist constraints imposed (d), and the reduced first moir\'e harmonic model with the zero-twist constraints imposed (e). Within each panel, we plot the dispersion of the (approximately or exactly) degenerate group of bands ($E_{\vec{k}}$) and their non-abelian Berry curvature ($\mathcal{F}_{\vec{k}}$) throughout the first moir\'e BZ (black hexagon). Additionally, we plot the real space LDOS in layer $l= \pm$ within the moir\'e unit cell (dashed hexagon).}
\label{app:fig:fit_2D_bands_4_1}
\end{figure}
\begin{figure}[H]
\centering
\includegraphics[width=\textwidth]{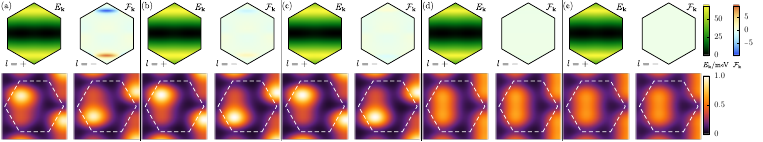}
\subfloat{\label{app:fig:fit_2D_bands_4_2:a}}\subfloat{\label{app:fig:fit_2D_bands_4_2:b}}\subfloat{\label{app:fig:fit_2D_bands_4_2:c}}\subfloat{\label{app:fig:fit_2D_bands_4_2:d}}\subfloat{\label{app:fig:fit_2D_bands_4_2:e}}\caption{The second group of conduction bands of twisted AA-stacked bilayer \ch{SnSe2} at $\theta = \SI{5.09}{\degree}$. We consider the full continuum model (a), the full first moir\'e harmonic model (b), the reduced first moir\'e harmonic model (c), the first moir\'e harmonic model with the zero-twist constraints imposed (d), and the reduced first moir\'e harmonic model with the zero-twist constraints imposed (e). Within each panel, we plot the dispersion of the (approximately or exactly) degenerate group of bands ($E_{\vec{k}}$) and their non-abelian Berry curvature ($\mathcal{F}_{\vec{k}}$) throughout the first moir\'e BZ (black hexagon). Additionally, we plot the real space LDOS in layer $l= \pm$ within the moir\'e unit cell (dashed hexagon).}
\label{app:fig:fit_2D_bands_4_2}
\end{figure}
\begin{figure}[H]
\centering
\includegraphics[width=\textwidth]{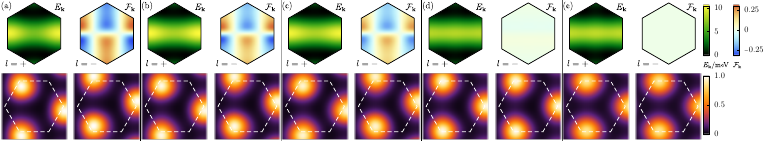}
\subfloat{\label{app:fig:fit_2D_bands_5_1:a}}\subfloat{\label{app:fig:fit_2D_bands_5_1:b}}\subfloat{\label{app:fig:fit_2D_bands_5_1:c}}\subfloat{\label{app:fig:fit_2D_bands_5_1:d}}\subfloat{\label{app:fig:fit_2D_bands_5_1:e}}\caption{The first group of conduction bands of twisted AA-stacked bilayer \ch{SnSe2} at $\theta = \SI{4.41}{\degree}$. We consider the full continuum model (a), the full first moir\'e harmonic model (b), the reduced first moir\'e harmonic model (c), the first moir\'e harmonic model with the zero-twist constraints imposed (d), and the reduced first moir\'e harmonic model with the zero-twist constraints imposed (e). Within each panel, we plot the dispersion of the (approximately or exactly) degenerate group of bands ($E_{\vec{k}}$) and their non-abelian Berry curvature ($\mathcal{F}_{\vec{k}}$) throughout the first moir\'e BZ (black hexagon). Additionally, we plot the real space LDOS in layer $l= \pm$ within the moir\'e unit cell (dashed hexagon).}
\label{app:fig:fit_2D_bands_5_1}
\end{figure}
\begin{figure}[H]
\centering
\includegraphics[width=\textwidth]{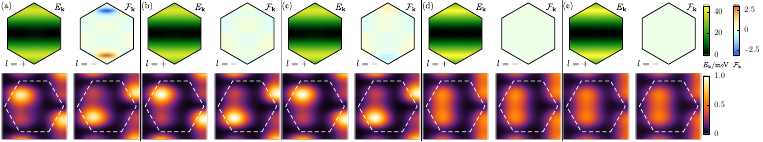}
\subfloat{\label{app:fig:fit_2D_bands_5_2:a}}\subfloat{\label{app:fig:fit_2D_bands_5_2:b}}\subfloat{\label{app:fig:fit_2D_bands_5_2:c}}\subfloat{\label{app:fig:fit_2D_bands_5_2:d}}\subfloat{\label{app:fig:fit_2D_bands_5_2:e}}\caption{The second group of conduction bands of twisted AA-stacked bilayer \ch{SnSe2} at $\theta = \SI{4.41}{\degree}$. We consider the full continuum model (a), the full first moir\'e harmonic model (b), the reduced first moir\'e harmonic model (c), the first moir\'e harmonic model with the zero-twist constraints imposed (d), and the reduced first moir\'e harmonic model with the zero-twist constraints imposed (e). Within each panel, we plot the dispersion of the (approximately or exactly) degenerate group of bands ($E_{\vec{k}}$) and their non-abelian Berry curvature ($\mathcal{F}_{\vec{k}}$) throughout the first moir\'e BZ (black hexagon). Additionally, we plot the real space LDOS in layer $l= \pm$ within the moir\'e unit cell (dashed hexagon).}
\label{app:fig:fit_2D_bands_5_2}
\end{figure}
\begin{figure}[H]
\centering
\includegraphics[width=\textwidth]{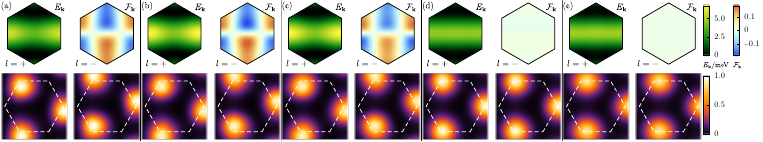}
\subfloat{\label{app:fig:fit_2D_bands_6_1:a}}\subfloat{\label{app:fig:fit_2D_bands_6_1:b}}\subfloat{\label{app:fig:fit_2D_bands_6_1:c}}\subfloat{\label{app:fig:fit_2D_bands_6_1:d}}\subfloat{\label{app:fig:fit_2D_bands_6_1:e}}\caption{The first group of conduction bands of twisted AA-stacked bilayer \ch{SnSe2} at $\theta = \SI{3.89}{\degree}$. We consider the full continuum model (a), the full first moir\'e harmonic model (b), the reduced first moir\'e harmonic model (c), the first moir\'e harmonic model with the zero-twist constraints imposed (d), and the reduced first moir\'e harmonic model with the zero-twist constraints imposed (e). Within each panel, we plot the dispersion of the (approximately or exactly) degenerate group of bands ($E_{\vec{k}}$) and their non-abelian Berry curvature ($\mathcal{F}_{\vec{k}}$) throughout the first moir\'e BZ (black hexagon). Additionally, we plot the real space LDOS in layer $l= \pm$ within the moir\'e unit cell (dashed hexagon).}
\label{app:fig:fit_2D_bands_6_1}
\end{figure}
\begin{figure}[H]
\centering
\includegraphics[width=\textwidth]{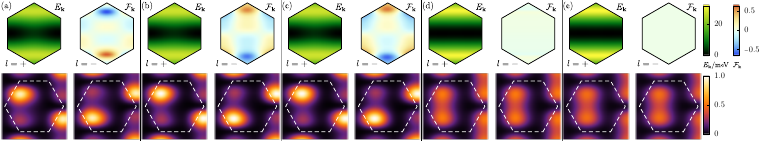}
\subfloat{\label{app:fig:fit_2D_bands_6_2:a}}\subfloat{\label{app:fig:fit_2D_bands_6_2:b}}\subfloat{\label{app:fig:fit_2D_bands_6_2:c}}\subfloat{\label{app:fig:fit_2D_bands_6_2:d}}\subfloat{\label{app:fig:fit_2D_bands_6_2:e}}\caption{The second group of conduction bands of twisted AA-stacked bilayer \ch{SnSe2} at $\theta = \SI{3.89}{\degree}$. We consider the full continuum model (a), the full first moir\'e harmonic model (b), the reduced first moir\'e harmonic model (c), the first moir\'e harmonic model with the zero-twist constraints imposed (d), and the reduced first moir\'e harmonic model with the zero-twist constraints imposed (e). Within each panel, we plot the dispersion of the (approximately or exactly) degenerate group of bands ($E_{\vec{k}}$) and their non-abelian Berry curvature ($\mathcal{F}_{\vec{k}}$) throughout the first moir\'e BZ (black hexagon). Additionally, we plot the real space LDOS in layer $l= \pm$ within the moir\'e unit cell (dashed hexagon).}
\label{app:fig:fit_2D_bands_6_2}
\end{figure}
\begin{figure}[H]
\centering
\includegraphics[width=\textwidth]{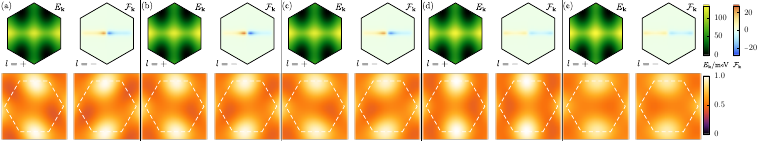}
\subfloat{\label{app:fig:fit_2D_bands_7_1:a}}\subfloat{\label{app:fig:fit_2D_bands_7_1:b}}\subfloat{\label{app:fig:fit_2D_bands_7_1:c}}\subfloat{\label{app:fig:fit_2D_bands_7_1:d}}\subfloat{\label{app:fig:fit_2D_bands_7_1:e}}\caption{The first group of conduction bands of twisted AB-stacked bilayer \ch{SnSe2} at $\theta = \SI{9.43}{\degree}$. We consider the full continuum model (a), the full first moir\'e harmonic model (b), the reduced first moir\'e harmonic model (c), the first moir\'e harmonic model with the zero-twist constraints imposed (d), and the reduced first moir\'e harmonic model with the zero-twist constraints imposed (e). Within each panel, we plot the dispersion of the (approximately or exactly) degenerate group of bands ($E_{\vec{k}}$) and their non-abelian Berry curvature ($\mathcal{F}_{\vec{k}}$) throughout the first moir\'e BZ (black hexagon). Additionally, we plot the real space LDOS in layer $l= \pm$ within the moir\'e unit cell (dashed hexagon).}
\label{app:fig:fit_2D_bands_7_1}
\end{figure}
\begin{figure}[H]
\centering
\includegraphics[width=\textwidth]{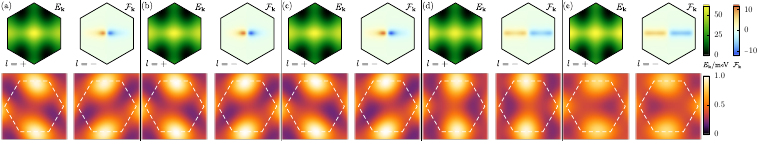}
\subfloat{\label{app:fig:fit_2D_bands_8_1:a}}\subfloat{\label{app:fig:fit_2D_bands_8_1:b}}\subfloat{\label{app:fig:fit_2D_bands_8_1:c}}\subfloat{\label{app:fig:fit_2D_bands_8_1:d}}\subfloat{\label{app:fig:fit_2D_bands_8_1:e}}\caption{The first group of conduction bands of twisted AB-stacked bilayer \ch{SnSe2} at $\theta = \SI{7.34}{\degree}$. We consider the full continuum model (a), the full first moir\'e harmonic model (b), the reduced first moir\'e harmonic model (c), the first moir\'e harmonic model with the zero-twist constraints imposed (d), and the reduced first moir\'e harmonic model with the zero-twist constraints imposed (e). Within each panel, we plot the dispersion of the (approximately or exactly) degenerate group of bands ($E_{\vec{k}}$) and their non-abelian Berry curvature ($\mathcal{F}_{\vec{k}}$) throughout the first moir\'e BZ (black hexagon). Additionally, we plot the real space LDOS in layer $l= \pm$ within the moir\'e unit cell (dashed hexagon).}
\label{app:fig:fit_2D_bands_8_1}
\end{figure}
\begin{figure}[H]
\centering
\includegraphics[width=\textwidth]{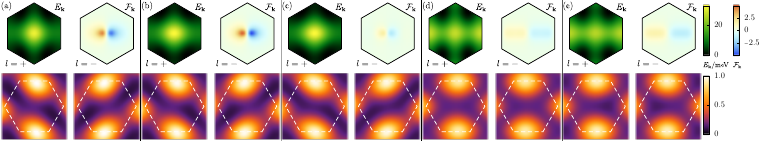}
\subfloat{\label{app:fig:fit_2D_bands_9_1:a}}\subfloat{\label{app:fig:fit_2D_bands_9_1:b}}\subfloat{\label{app:fig:fit_2D_bands_9_1:c}}\subfloat{\label{app:fig:fit_2D_bands_9_1:d}}\subfloat{\label{app:fig:fit_2D_bands_9_1:e}}\caption{The first group of conduction bands of twisted AB-stacked bilayer \ch{SnSe2} at $\theta = \SI{6.01}{\degree}$. We consider the full continuum model (a), the full first moir\'e harmonic model (b), the reduced first moir\'e harmonic model (c), the first moir\'e harmonic model with the zero-twist constraints imposed (d), and the reduced first moir\'e harmonic model with the zero-twist constraints imposed (e). Within each panel, we plot the dispersion of the (approximately or exactly) degenerate group of bands ($E_{\vec{k}}$) and their non-abelian Berry curvature ($\mathcal{F}_{\vec{k}}$) throughout the first moir\'e BZ (black hexagon). Additionally, we plot the real space LDOS in layer $l= \pm$ within the moir\'e unit cell (dashed hexagon).}
\label{app:fig:fit_2D_bands_9_1}
\end{figure}
\begin{figure}[H]
\centering
\includegraphics[width=\textwidth]{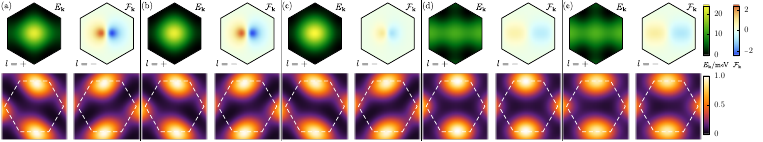}
\subfloat{\label{app:fig:fit_2D_bands_10_1:a}}\subfloat{\label{app:fig:fit_2D_bands_10_1:b}}\subfloat{\label{app:fig:fit_2D_bands_10_1:c}}\subfloat{\label{app:fig:fit_2D_bands_10_1:d}}\subfloat{\label{app:fig:fit_2D_bands_10_1:e}}\caption{The first group of conduction bands of twisted AB-stacked bilayer \ch{SnSe2} at $\theta = \SI{5.09}{\degree}$. We consider the full continuum model (a), the full first moir\'e harmonic model (b), the reduced first moir\'e harmonic model (c), the first moir\'e harmonic model with the zero-twist constraints imposed (d), and the reduced first moir\'e harmonic model with the zero-twist constraints imposed (e). Within each panel, we plot the dispersion of the (approximately or exactly) degenerate group of bands ($E_{\vec{k}}$) and their non-abelian Berry curvature ($\mathcal{F}_{\vec{k}}$) throughout the first moir\'e BZ (black hexagon). Additionally, we plot the real space LDOS in layer $l= \pm$ within the moir\'e unit cell (dashed hexagon).}
\label{app:fig:fit_2D_bands_10_1}
\end{figure}
\begin{figure}[H]
\centering
\includegraphics[width=\textwidth]{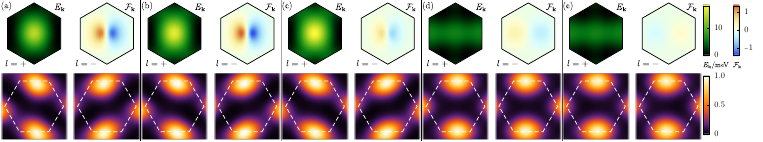}
\subfloat{\label{app:fig:fit_2D_bands_11_1:a}}\subfloat{\label{app:fig:fit_2D_bands_11_1:b}}\subfloat{\label{app:fig:fit_2D_bands_11_1:c}}\subfloat{\label{app:fig:fit_2D_bands_11_1:d}}\subfloat{\label{app:fig:fit_2D_bands_11_1:e}}\caption{The first group of conduction bands of twisted AB-stacked bilayer \ch{SnSe2} at $\theta = \SI{4.41}{\degree}$. We consider the full continuum model (a), the full first moir\'e harmonic model (b), the reduced first moir\'e harmonic model (c), the first moir\'e harmonic model with the zero-twist constraints imposed (d), and the reduced first moir\'e harmonic model with the zero-twist constraints imposed (e). Within each panel, we plot the dispersion of the (approximately or exactly) degenerate group of bands ($E_{\vec{k}}$) and their non-abelian Berry curvature ($\mathcal{F}_{\vec{k}}$) throughout the first moir\'e BZ (black hexagon). Additionally, we plot the real space LDOS in layer $l= \pm$ within the moir\'e unit cell (dashed hexagon).}
\label{app:fig:fit_2D_bands_11_1}
\end{figure}
\begin{figure}[H]
\centering
\includegraphics[width=\textwidth]{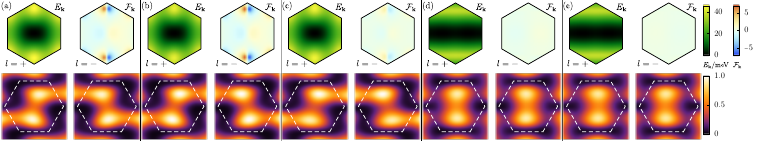}
\subfloat{\label{app:fig:fit_2D_bands_11_2:a}}\subfloat{\label{app:fig:fit_2D_bands_11_2:b}}\subfloat{\label{app:fig:fit_2D_bands_11_2:c}}\subfloat{\label{app:fig:fit_2D_bands_11_2:d}}\subfloat{\label{app:fig:fit_2D_bands_11_2:e}}\caption{The second group of conduction bands of twisted AB-stacked bilayer \ch{SnSe2} at $\theta = \SI{4.41}{\degree}$. We consider the full continuum model (a), the full first moir\'e harmonic model (b), the reduced first moir\'e harmonic model (c), the first moir\'e harmonic model with the zero-twist constraints imposed (d), and the reduced first moir\'e harmonic model with the zero-twist constraints imposed (e). Within each panel, we plot the dispersion of the (approximately or exactly) degenerate group of bands ($E_{\vec{k}}$) and their non-abelian Berry curvature ($\mathcal{F}_{\vec{k}}$) throughout the first moir\'e BZ (black hexagon). Additionally, we plot the real space LDOS in layer $l= \pm$ within the moir\'e unit cell (dashed hexagon).}
\label{app:fig:fit_2D_bands_11_2}
\end{figure}
\begin{figure}[H]
\centering
\includegraphics[width=\textwidth]{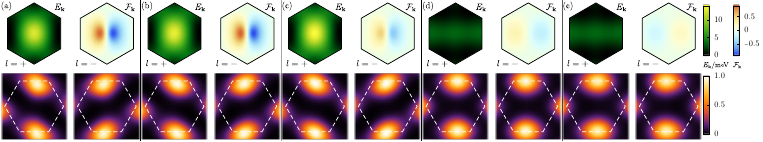}
\subfloat{\label{app:fig:fit_2D_bands_12_1:a}}\subfloat{\label{app:fig:fit_2D_bands_12_1:b}}\subfloat{\label{app:fig:fit_2D_bands_12_1:c}}\subfloat{\label{app:fig:fit_2D_bands_12_1:d}}\subfloat{\label{app:fig:fit_2D_bands_12_1:e}}\caption{The first group of conduction bands of twisted AB-stacked bilayer \ch{SnSe2} at $\theta = \SI{3.89}{\degree}$. We consider the full continuum model (a), the full first moir\'e harmonic model (b), the reduced first moir\'e harmonic model (c), the first moir\'e harmonic model with the zero-twist constraints imposed (d), and the reduced first moir\'e harmonic model with the zero-twist constraints imposed (e). Within each panel, we plot the dispersion of the (approximately or exactly) degenerate group of bands ($E_{\vec{k}}$) and their non-abelian Berry curvature ($\mathcal{F}_{\vec{k}}$) throughout the first moir\'e BZ (black hexagon). Additionally, we plot the real space LDOS in layer $l= \pm$ within the moir\'e unit cell (dashed hexagon).}
\label{app:fig:fit_2D_bands_12_1}
\end{figure}
\begin{figure}[H]
\centering
\includegraphics[width=\textwidth]{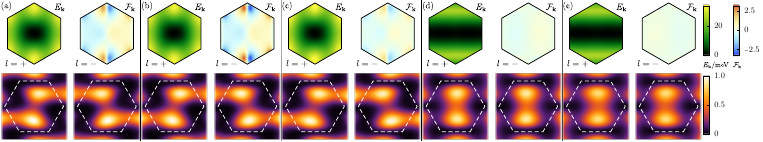}
\subfloat{\label{app:fig:fit_2D_bands_12_2:a}}\subfloat{\label{app:fig:fit_2D_bands_12_2:b}}\subfloat{\label{app:fig:fit_2D_bands_12_2:c}}\subfloat{\label{app:fig:fit_2D_bands_12_2:d}}\subfloat{\label{app:fig:fit_2D_bands_12_2:e}}\caption{The second group of conduction bands of twisted AB-stacked bilayer \ch{SnSe2} at $\theta = \SI{3.89}{\degree}$. We consider the full continuum model (a), the full first moir\'e harmonic model (b), the reduced first moir\'e harmonic model (c), the first moir\'e harmonic model with the zero-twist constraints imposed (d), and the reduced first moir\'e harmonic model with the zero-twist constraints imposed (e). Within each panel, we plot the dispersion of the (approximately or exactly) degenerate group of bands ($E_{\vec{k}}$) and their non-abelian Berry curvature ($\mathcal{F}_{\vec{k}}$) throughout the first moir\'e BZ (black hexagon). Additionally, we plot the real space LDOS in layer $l= \pm$ within the moir\'e unit cell (dashed hexagon).}
\label{app:fig:fit_2D_bands_12_2}
\end{figure}
\begin{figure}[H]
\centering
\includegraphics[width=\textwidth]{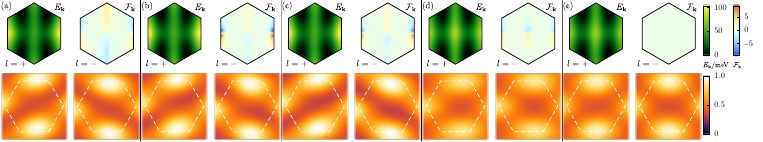}
\subfloat{\label{app:fig:fit_2D_bands_13_1:a}}\subfloat{\label{app:fig:fit_2D_bands_13_1:b}}\subfloat{\label{app:fig:fit_2D_bands_13_1:c}}\subfloat{\label{app:fig:fit_2D_bands_13_1:d}}\subfloat{\label{app:fig:fit_2D_bands_13_1:e}}\caption{The first group of conduction bands of twisted AA-stacked bilayer \ch{ZrS2} at $\theta = \SI{9.43}{\degree}$. We consider the full continuum model (a), the full first moir\'e harmonic model (b), the reduced first moir\'e harmonic model (c), the first moir\'e harmonic model with the zero-twist constraints imposed (d), and the reduced first moir\'e harmonic model with the zero-twist constraints imposed (e). Within each panel, we plot the dispersion of the (approximately or exactly) degenerate group of bands ($E_{\vec{k}}$) and their non-abelian Berry curvature ($\mathcal{F}_{\vec{k}}$) throughout the first moir\'e BZ (black hexagon). Additionally, we plot the real space LDOS in layer $l= \pm$ within the moir\'e unit cell (dashed hexagon).}
\label{app:fig:fit_2D_bands_13_1}
\end{figure}
\begin{figure}[H]
\centering
\includegraphics[width=\textwidth]{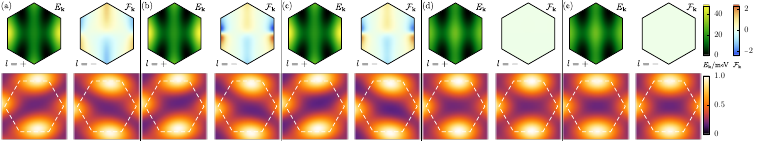}
\subfloat{\label{app:fig:fit_2D_bands_14_1:a}}\subfloat{\label{app:fig:fit_2D_bands_14_1:b}}\subfloat{\label{app:fig:fit_2D_bands_14_1:c}}\subfloat{\label{app:fig:fit_2D_bands_14_1:d}}\subfloat{\label{app:fig:fit_2D_bands_14_1:e}}\caption{The first group of conduction bands of twisted AA-stacked bilayer \ch{ZrS2} at $\theta = \SI{7.34}{\degree}$. We consider the full continuum model (a), the full first moir\'e harmonic model (b), the reduced first moir\'e harmonic model (c), the first moir\'e harmonic model with the zero-twist constraints imposed (d), and the reduced first moir\'e harmonic model with the zero-twist constraints imposed (e). Within each panel, we plot the dispersion of the (approximately or exactly) degenerate group of bands ($E_{\vec{k}}$) and their non-abelian Berry curvature ($\mathcal{F}_{\vec{k}}$) throughout the first moir\'e BZ (black hexagon). Additionally, we plot the real space LDOS in layer $l= \pm$ within the moir\'e unit cell (dashed hexagon).}
\label{app:fig:fit_2D_bands_14_1}
\end{figure}
\begin{figure}[H]
\centering
\includegraphics[width=\textwidth]{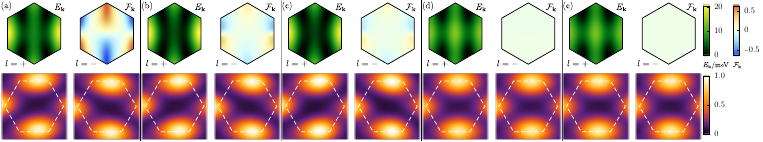}
\subfloat{\label{app:fig:fit_2D_bands_15_1:a}}\subfloat{\label{app:fig:fit_2D_bands_15_1:b}}\subfloat{\label{app:fig:fit_2D_bands_15_1:c}}\subfloat{\label{app:fig:fit_2D_bands_15_1:d}}\subfloat{\label{app:fig:fit_2D_bands_15_1:e}}\caption{The first group of conduction bands of twisted AA-stacked bilayer \ch{ZrS2} at $\theta = \SI{6.01}{\degree}$. We consider the full continuum model (a), the full first moir\'e harmonic model (b), the reduced first moir\'e harmonic model (c), the first moir\'e harmonic model with the zero-twist constraints imposed (d), and the reduced first moir\'e harmonic model with the zero-twist constraints imposed (e). Within each panel, we plot the dispersion of the (approximately or exactly) degenerate group of bands ($E_{\vec{k}}$) and their non-abelian Berry curvature ($\mathcal{F}_{\vec{k}}$) throughout the first moir\'e BZ (black hexagon). Additionally, we plot the real space LDOS in layer $l= \pm$ within the moir\'e unit cell (dashed hexagon).}
\label{app:fig:fit_2D_bands_15_1}
\end{figure}
\begin{figure}[H]
\centering
\includegraphics[width=\textwidth]{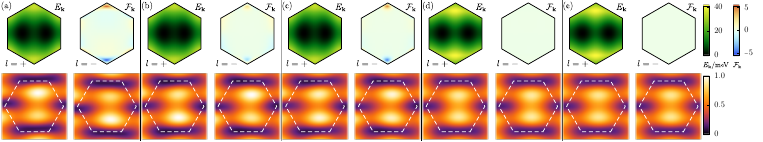}
\subfloat{\label{app:fig:fit_2D_bands_15_2:a}}\subfloat{\label{app:fig:fit_2D_bands_15_2:b}}\subfloat{\label{app:fig:fit_2D_bands_15_2:c}}\subfloat{\label{app:fig:fit_2D_bands_15_2:d}}\subfloat{\label{app:fig:fit_2D_bands_15_2:e}}\caption{The second group of conduction bands of twisted AA-stacked bilayer \ch{ZrS2} at $\theta = \SI{6.01}{\degree}$. We consider the full continuum model (a), the full first moir\'e harmonic model (b), the reduced first moir\'e harmonic model (c), the first moir\'e harmonic model with the zero-twist constraints imposed (d), and the reduced first moir\'e harmonic model with the zero-twist constraints imposed (e). Within each panel, we plot the dispersion of the (approximately or exactly) degenerate group of bands ($E_{\vec{k}}$) and their non-abelian Berry curvature ($\mathcal{F}_{\vec{k}}$) throughout the first moir\'e BZ (black hexagon). Additionally, we plot the real space LDOS in layer $l= \pm$ within the moir\'e unit cell (dashed hexagon).}
\label{app:fig:fit_2D_bands_15_2}
\end{figure}
\begin{figure}[H]
\centering
\includegraphics[width=\textwidth]{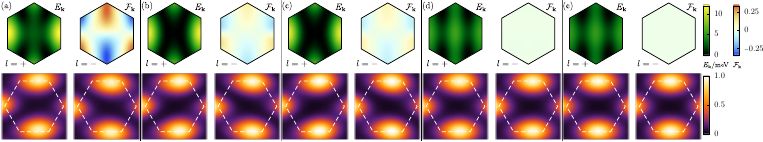}
\subfloat{\label{app:fig:fit_2D_bands_16_1:a}}\subfloat{\label{app:fig:fit_2D_bands_16_1:b}}\subfloat{\label{app:fig:fit_2D_bands_16_1:c}}\subfloat{\label{app:fig:fit_2D_bands_16_1:d}}\subfloat{\label{app:fig:fit_2D_bands_16_1:e}}\caption{The first group of conduction bands of twisted AA-stacked bilayer \ch{ZrS2} at $\theta = \SI{5.09}{\degree}$. We consider the full continuum model (a), the full first moir\'e harmonic model (b), the reduced first moir\'e harmonic model (c), the first moir\'e harmonic model with the zero-twist constraints imposed (d), and the reduced first moir\'e harmonic model with the zero-twist constraints imposed (e). Within each panel, we plot the dispersion of the (approximately or exactly) degenerate group of bands ($E_{\vec{k}}$) and their non-abelian Berry curvature ($\mathcal{F}_{\vec{k}}$) throughout the first moir\'e BZ (black hexagon). Additionally, we plot the real space LDOS in layer $l= \pm$ within the moir\'e unit cell (dashed hexagon).}
\label{app:fig:fit_2D_bands_16_1}
\end{figure}
\begin{figure}[H]
\centering
\includegraphics[width=\textwidth]{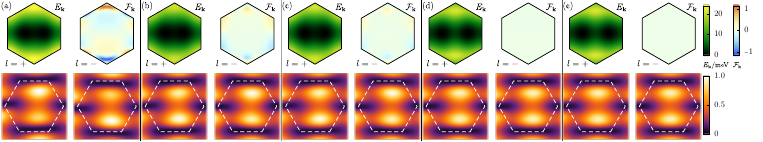}
\subfloat{\label{app:fig:fit_2D_bands_16_2:a}}\subfloat{\label{app:fig:fit_2D_bands_16_2:b}}\subfloat{\label{app:fig:fit_2D_bands_16_2:c}}\subfloat{\label{app:fig:fit_2D_bands_16_2:d}}\subfloat{\label{app:fig:fit_2D_bands_16_2:e}}\caption{The second group of conduction bands of twisted AA-stacked bilayer \ch{ZrS2} at $\theta = \SI{5.09}{\degree}$. We consider the full continuum model (a), the full first moir\'e harmonic model (b), the reduced first moir\'e harmonic model (c), the first moir\'e harmonic model with the zero-twist constraints imposed (d), and the reduced first moir\'e harmonic model with the zero-twist constraints imposed (e). Within each panel, we plot the dispersion of the (approximately or exactly) degenerate group of bands ($E_{\vec{k}}$) and their non-abelian Berry curvature ($\mathcal{F}_{\vec{k}}$) throughout the first moir\'e BZ (black hexagon). Additionally, we plot the real space LDOS in layer $l= \pm$ within the moir\'e unit cell (dashed hexagon).}
\label{app:fig:fit_2D_bands_16_2}
\end{figure}
\begin{figure}[H]
\centering
\includegraphics[width=\textwidth]{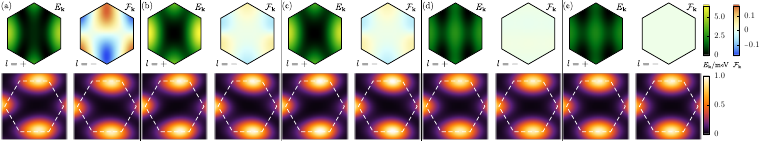}
\subfloat{\label{app:fig:fit_2D_bands_17_1:a}}\subfloat{\label{app:fig:fit_2D_bands_17_1:b}}\subfloat{\label{app:fig:fit_2D_bands_17_1:c}}\subfloat{\label{app:fig:fit_2D_bands_17_1:d}}\subfloat{\label{app:fig:fit_2D_bands_17_1:e}}\caption{The first group of conduction bands of twisted AA-stacked bilayer \ch{ZrS2} at $\theta = \SI{4.41}{\degree}$. We consider the full continuum model (a), the full first moir\'e harmonic model (b), the reduced first moir\'e harmonic model (c), the first moir\'e harmonic model with the zero-twist constraints imposed (d), and the reduced first moir\'e harmonic model with the zero-twist constraints imposed (e). Within each panel, we plot the dispersion of the (approximately or exactly) degenerate group of bands ($E_{\vec{k}}$) and their non-abelian Berry curvature ($\mathcal{F}_{\vec{k}}$) throughout the first moir\'e BZ (black hexagon). Additionally, we plot the real space LDOS in layer $l= \pm$ within the moir\'e unit cell (dashed hexagon).}
\label{app:fig:fit_2D_bands_17_1}
\end{figure}
\begin{figure}[H]
\centering
\includegraphics[width=\textwidth]{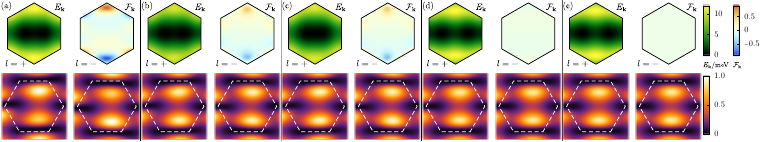}
\subfloat{\label{app:fig:fit_2D_bands_17_2:a}}\subfloat{\label{app:fig:fit_2D_bands_17_2:b}}\subfloat{\label{app:fig:fit_2D_bands_17_2:c}}\subfloat{\label{app:fig:fit_2D_bands_17_2:d}}\subfloat{\label{app:fig:fit_2D_bands_17_2:e}}\caption{The second group of conduction bands of twisted AA-stacked bilayer \ch{ZrS2} at $\theta = \SI{4.41}{\degree}$. We consider the full continuum model (a), the full first moir\'e harmonic model (b), the reduced first moir\'e harmonic model (c), the first moir\'e harmonic model with the zero-twist constraints imposed (d), and the reduced first moir\'e harmonic model with the zero-twist constraints imposed (e). Within each panel, we plot the dispersion of the (approximately or exactly) degenerate group of bands ($E_{\vec{k}}$) and their non-abelian Berry curvature ($\mathcal{F}_{\vec{k}}$) throughout the first moir\'e BZ (black hexagon). Additionally, we plot the real space LDOS in layer $l= \pm$ within the moir\'e unit cell (dashed hexagon).}
\label{app:fig:fit_2D_bands_17_2}
\end{figure}
\begin{figure}[H]
\centering
\includegraphics[width=\textwidth]{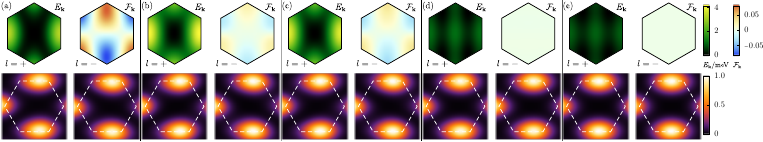}
\subfloat{\label{app:fig:fit_2D_bands_18_1:a}}\subfloat{\label{app:fig:fit_2D_bands_18_1:b}}\subfloat{\label{app:fig:fit_2D_bands_18_1:c}}\subfloat{\label{app:fig:fit_2D_bands_18_1:d}}\subfloat{\label{app:fig:fit_2D_bands_18_1:e}}\caption{The first group of conduction bands of twisted AA-stacked bilayer \ch{ZrS2} at $\theta = \SI{3.89}{\degree}$. We consider the full continuum model (a), the full first moir\'e harmonic model (b), the reduced first moir\'e harmonic model (c), the first moir\'e harmonic model with the zero-twist constraints imposed (d), and the reduced first moir\'e harmonic model with the zero-twist constraints imposed (e). Within each panel, we plot the dispersion of the (approximately or exactly) degenerate group of bands ($E_{\vec{k}}$) and their non-abelian Berry curvature ($\mathcal{F}_{\vec{k}}$) throughout the first moir\'e BZ (black hexagon). Additionally, we plot the real space LDOS in layer $l= \pm$ within the moir\'e unit cell (dashed hexagon).}
\label{app:fig:fit_2D_bands_18_1}
\end{figure}
\begin{figure}[H]
\centering
\includegraphics[width=\textwidth]{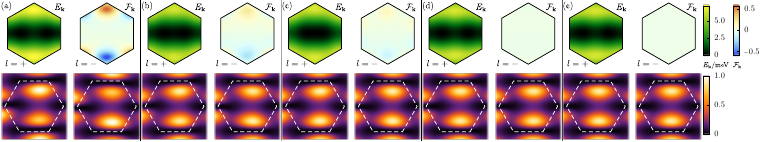}
\subfloat{\label{app:fig:fit_2D_bands_18_2:a}}\subfloat{\label{app:fig:fit_2D_bands_18_2:b}}\subfloat{\label{app:fig:fit_2D_bands_18_2:c}}\subfloat{\label{app:fig:fit_2D_bands_18_2:d}}\subfloat{\label{app:fig:fit_2D_bands_18_2:e}}\caption{The second group of conduction bands of twisted AA-stacked bilayer \ch{ZrS2} at $\theta = \SI{3.89}{\degree}$. We consider the full continuum model (a), the full first moir\'e harmonic model (b), the reduced first moir\'e harmonic model (c), the first moir\'e harmonic model with the zero-twist constraints imposed (d), and the reduced first moir\'e harmonic model with the zero-twist constraints imposed (e). Within each panel, we plot the dispersion of the (approximately or exactly) degenerate group of bands ($E_{\vec{k}}$) and their non-abelian Berry curvature ($\mathcal{F}_{\vec{k}}$) throughout the first moir\'e BZ (black hexagon). Additionally, we plot the real space LDOS in layer $l= \pm$ within the moir\'e unit cell (dashed hexagon).}
\label{app:fig:fit_2D_bands_18_2}
\end{figure}
\begin{figure}[H]
\centering
\includegraphics[width=\textwidth]{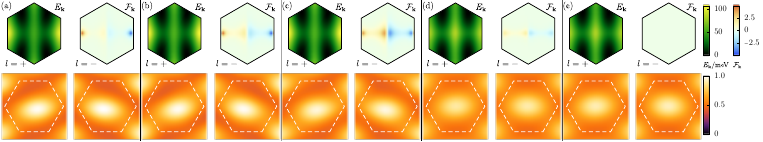}
\subfloat{\label{app:fig:fit_2D_bands_19_1:a}}\subfloat{\label{app:fig:fit_2D_bands_19_1:b}}\subfloat{\label{app:fig:fit_2D_bands_19_1:c}}\subfloat{\label{app:fig:fit_2D_bands_19_1:d}}\subfloat{\label{app:fig:fit_2D_bands_19_1:e}}\caption{The first group of conduction bands of twisted AB-stacked bilayer \ch{ZrS2} at $\theta = \SI{9.43}{\degree}$. We consider the full continuum model (a), the full first moir\'e harmonic model (b), the reduced first moir\'e harmonic model (c), the first moir\'e harmonic model with the zero-twist constraints imposed (d), and the reduced first moir\'e harmonic model with the zero-twist constraints imposed (e). Within each panel, we plot the dispersion of the (approximately or exactly) degenerate group of bands ($E_{\vec{k}}$) and their non-abelian Berry curvature ($\mathcal{F}_{\vec{k}}$) throughout the first moir\'e BZ (black hexagon). Additionally, we plot the real space LDOS in layer $l= \pm$ within the moir\'e unit cell (dashed hexagon).}
\label{app:fig:fit_2D_bands_19_1}
\end{figure}
\begin{figure}[H]
\centering
\includegraphics[width=\textwidth]{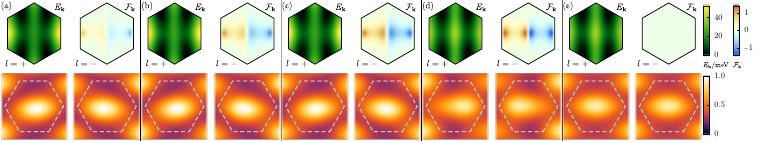}
\subfloat{\label{app:fig:fit_2D_bands_20_1:a}}\subfloat{\label{app:fig:fit_2D_bands_20_1:b}}\subfloat{\label{app:fig:fit_2D_bands_20_1:c}}\subfloat{\label{app:fig:fit_2D_bands_20_1:d}}\subfloat{\label{app:fig:fit_2D_bands_20_1:e}}\caption{The first group of conduction bands of twisted AB-stacked bilayer \ch{ZrS2} at $\theta = \SI{7.34}{\degree}$. We consider the full continuum model (a), the full first moir\'e harmonic model (b), the reduced first moir\'e harmonic model (c), the first moir\'e harmonic model with the zero-twist constraints imposed (d), and the reduced first moir\'e harmonic model with the zero-twist constraints imposed (e). Within each panel, we plot the dispersion of the (approximately or exactly) degenerate group of bands ($E_{\vec{k}}$) and their non-abelian Berry curvature ($\mathcal{F}_{\vec{k}}$) throughout the first moir\'e BZ (black hexagon). Additionally, we plot the real space LDOS in layer $l= \pm$ within the moir\'e unit cell (dashed hexagon).}
\label{app:fig:fit_2D_bands_20_1}
\end{figure}
\begin{figure}[H]
\centering
\includegraphics[width=\textwidth]{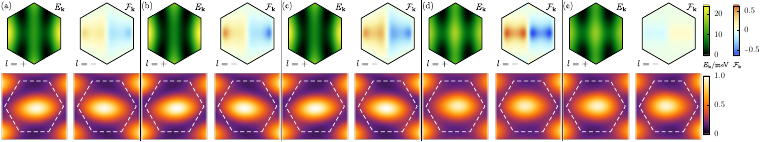}
\subfloat{\label{app:fig:fit_2D_bands_21_1:a}}\subfloat{\label{app:fig:fit_2D_bands_21_1:b}}\subfloat{\label{app:fig:fit_2D_bands_21_1:c}}\subfloat{\label{app:fig:fit_2D_bands_21_1:d}}\subfloat{\label{app:fig:fit_2D_bands_21_1:e}}\caption{The first group of conduction bands of twisted AB-stacked bilayer \ch{ZrS2} at $\theta = \SI{6.01}{\degree}$. We consider the full continuum model (a), the full first moir\'e harmonic model (b), the reduced first moir\'e harmonic model (c), the first moir\'e harmonic model with the zero-twist constraints imposed (d), and the reduced first moir\'e harmonic model with the zero-twist constraints imposed (e). Within each panel, we plot the dispersion of the (approximately or exactly) degenerate group of bands ($E_{\vec{k}}$) and their non-abelian Berry curvature ($\mathcal{F}_{\vec{k}}$) throughout the first moir\'e BZ (black hexagon). Additionally, we plot the real space LDOS in layer $l= \pm$ within the moir\'e unit cell (dashed hexagon).}
\label{app:fig:fit_2D_bands_21_1}
\end{figure}
\begin{figure}[H]
\centering
\includegraphics[width=\textwidth]{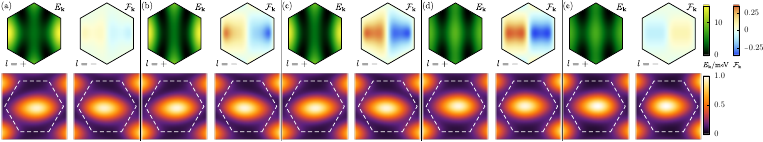}
\subfloat{\label{app:fig:fit_2D_bands_22_1:a}}\subfloat{\label{app:fig:fit_2D_bands_22_1:b}}\subfloat{\label{app:fig:fit_2D_bands_22_1:c}}\subfloat{\label{app:fig:fit_2D_bands_22_1:d}}\subfloat{\label{app:fig:fit_2D_bands_22_1:e}}\caption{The first group of conduction bands of twisted AB-stacked bilayer \ch{ZrS2} at $\theta = \SI{5.09}{\degree}$. We consider the full continuum model (a), the full first moir\'e harmonic model (b), the reduced first moir\'e harmonic model (c), the first moir\'e harmonic model with the zero-twist constraints imposed (d), and the reduced first moir\'e harmonic model with the zero-twist constraints imposed (e). Within each panel, we plot the dispersion of the (approximately or exactly) degenerate group of bands ($E_{\vec{k}}$) and their non-abelian Berry curvature ($\mathcal{F}_{\vec{k}}$) throughout the first moir\'e BZ (black hexagon). Additionally, we plot the real space LDOS in layer $l= \pm$ within the moir\'e unit cell (dashed hexagon).}
\label{app:fig:fit_2D_bands_22_1}
\end{figure}
\begin{figure}[H]
\centering
\includegraphics[width=\textwidth]{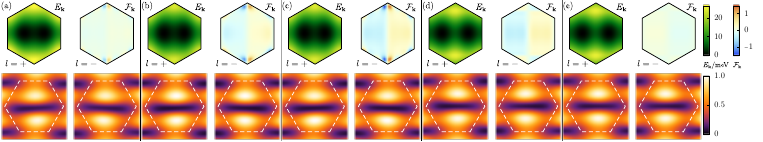}
\subfloat{\label{app:fig:fit_2D_bands_22_2:a}}\subfloat{\label{app:fig:fit_2D_bands_22_2:b}}\subfloat{\label{app:fig:fit_2D_bands_22_2:c}}\subfloat{\label{app:fig:fit_2D_bands_22_2:d}}\subfloat{\label{app:fig:fit_2D_bands_22_2:e}}\caption{The second group of conduction bands of twisted AB-stacked bilayer \ch{ZrS2} at $\theta = \SI{5.09}{\degree}$. We consider the full continuum model (a), the full first moir\'e harmonic model (b), the reduced first moir\'e harmonic model (c), the first moir\'e harmonic model with the zero-twist constraints imposed (d), and the reduced first moir\'e harmonic model with the zero-twist constraints imposed (e). Within each panel, we plot the dispersion of the (approximately or exactly) degenerate group of bands ($E_{\vec{k}}$) and their non-abelian Berry curvature ($\mathcal{F}_{\vec{k}}$) throughout the first moir\'e BZ (black hexagon). Additionally, we plot the real space LDOS in layer $l= \pm$ within the moir\'e unit cell (dashed hexagon).}
\label{app:fig:fit_2D_bands_22_2}
\end{figure}
\begin{figure}[H]
\centering
\includegraphics[width=\textwidth]{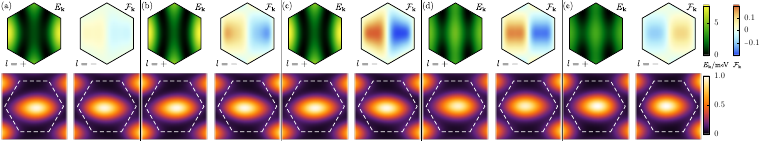}
\subfloat{\label{app:fig:fit_2D_bands_23_1:a}}\subfloat{\label{app:fig:fit_2D_bands_23_1:b}}\subfloat{\label{app:fig:fit_2D_bands_23_1:c}}\subfloat{\label{app:fig:fit_2D_bands_23_1:d}}\subfloat{\label{app:fig:fit_2D_bands_23_1:e}}\caption{The first group of conduction bands of twisted AB-stacked bilayer \ch{ZrS2} at $\theta = \SI{4.41}{\degree}$. We consider the full continuum model (a), the full first moir\'e harmonic model (b), the reduced first moir\'e harmonic model (c), the first moir\'e harmonic model with the zero-twist constraints imposed (d), and the reduced first moir\'e harmonic model with the zero-twist constraints imposed (e). Within each panel, we plot the dispersion of the (approximately or exactly) degenerate group of bands ($E_{\vec{k}}$) and their non-abelian Berry curvature ($\mathcal{F}_{\vec{k}}$) throughout the first moir\'e BZ (black hexagon). Additionally, we plot the real space LDOS in layer $l= \pm$ within the moir\'e unit cell (dashed hexagon).}
\label{app:fig:fit_2D_bands_23_1}
\end{figure}
\begin{figure}[H]
\centering
\includegraphics[width=\textwidth]{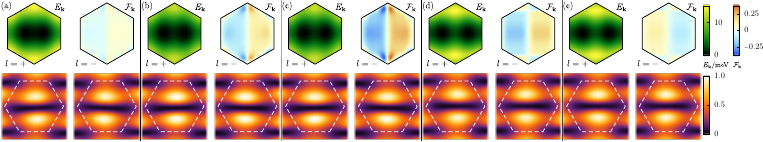}
\subfloat{\label{app:fig:fit_2D_bands_23_2:a}}\subfloat{\label{app:fig:fit_2D_bands_23_2:b}}\subfloat{\label{app:fig:fit_2D_bands_23_2:c}}\subfloat{\label{app:fig:fit_2D_bands_23_2:d}}\subfloat{\label{app:fig:fit_2D_bands_23_2:e}}\caption{The second group of conduction bands of twisted AB-stacked bilayer \ch{ZrS2} at $\theta = \SI{4.41}{\degree}$. We consider the full continuum model (a), the full first moir\'e harmonic model (b), the reduced first moir\'e harmonic model (c), the first moir\'e harmonic model with the zero-twist constraints imposed (d), and the reduced first moir\'e harmonic model with the zero-twist constraints imposed (e). Within each panel, we plot the dispersion of the (approximately or exactly) degenerate group of bands ($E_{\vec{k}}$) and their non-abelian Berry curvature ($\mathcal{F}_{\vec{k}}$) throughout the first moir\'e BZ (black hexagon). Additionally, we plot the real space LDOS in layer $l= \pm$ within the moir\'e unit cell (dashed hexagon).}
\label{app:fig:fit_2D_bands_23_2}
\end{figure}
\begin{figure}[H]
\centering
\includegraphics[width=\textwidth]{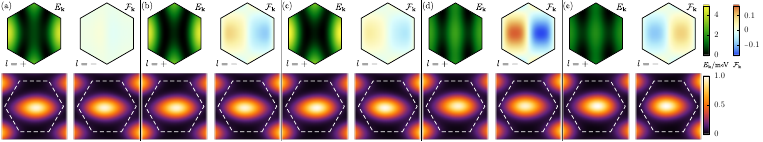}
\subfloat{\label{app:fig:fit_2D_bands_24_1:a}}\subfloat{\label{app:fig:fit_2D_bands_24_1:b}}\subfloat{\label{app:fig:fit_2D_bands_24_1:c}}\subfloat{\label{app:fig:fit_2D_bands_24_1:d}}\subfloat{\label{app:fig:fit_2D_bands_24_1:e}}\caption{The first group of conduction bands of twisted AB-stacked bilayer \ch{ZrS2} at $\theta = \SI{3.89}{\degree}$. We consider the full continuum model (a), the full first moir\'e harmonic model (b), the reduced first moir\'e harmonic model (c), the first moir\'e harmonic model with the zero-twist constraints imposed (d), and the reduced first moir\'e harmonic model with the zero-twist constraints imposed (e). Within each panel, we plot the dispersion of the (approximately or exactly) degenerate group of bands ($E_{\vec{k}}$) and their non-abelian Berry curvature ($\mathcal{F}_{\vec{k}}$) throughout the first moir\'e BZ (black hexagon). Additionally, we plot the real space LDOS in layer $l= \pm$ within the moir\'e unit cell (dashed hexagon).}
\label{app:fig:fit_2D_bands_24_1}
\end{figure}
\begin{figure}[H]
\centering
\includegraphics[width=\textwidth]{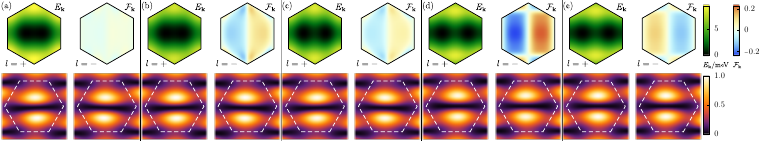}
\subfloat{\label{app:fig:fit_2D_bands_24_2:a}}\subfloat{\label{app:fig:fit_2D_bands_24_2:b}}\subfloat{\label{app:fig:fit_2D_bands_24_2:c}}\subfloat{\label{app:fig:fit_2D_bands_24_2:d}}\subfloat{\label{app:fig:fit_2D_bands_24_2:e}}\caption{The second group of conduction bands of twisted AB-stacked bilayer \ch{ZrS2} at $\theta = \SI{3.89}{\degree}$. We consider the full continuum model (a), the full first moir\'e harmonic model (b), the reduced first moir\'e harmonic model (c), the first moir\'e harmonic model with the zero-twist constraints imposed (d), and the reduced first moir\'e harmonic model with the zero-twist constraints imposed (e). Within each panel, we plot the dispersion of the (approximately or exactly) degenerate group of bands ($E_{\vec{k}}$) and their non-abelian Berry curvature ($\mathcal{F}_{\vec{k}}$) throughout the first moir\'e BZ (black hexagon). Additionally, we plot the real space LDOS in layer $l= \pm$ within the moir\'e unit cell (dashed hexagon).}
\label{app:fig:fit_2D_bands_24_2}
\end{figure} \subsubsection{Wilson loops of the first gapped conduction bands along $\vec{b}_{M_1}$}\label{app:sec:fitted_models:plots:wilsons_1}
\begin{figure}[H]
\centering
\includegraphics[width=\textwidth]{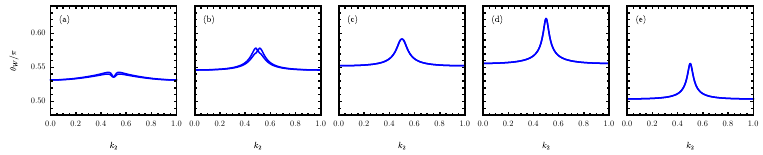}
\subfloat{\label{app:fig:fit_Wilson_1_WilsonX:a}}\subfloat{\label{app:fig:fit_Wilson_1_WilsonX:b}}\subfloat{\label{app:fig:fit_Wilson_1_WilsonX:c}}\subfloat{\label{app:fig:fit_Wilson_1_WilsonX:d}}\subfloat{\label{app:fig:fit_Wilson_1_WilsonX:e}}\caption{Wilson loops for the first set of conduction bands of twisted AA-stacked bilayer \ch{SnSe2} at $\theta = \SI{9.43}{\degree}$. The Wilson loop is computed along $\vec{b}_{M_1}$ We consider the full continuum model (a), the full first moir\'e harmonic model (b), the reduced first moir\'e harmonic model (c), the first moir\'e harmonic model with the zero-twist constraints imposed (d), and the reduced first moir\'e harmonic model with the zero-twist constraints imposed (e).}
\label{app:fig:fit_Wilson_1_WilsonX}
\end{figure}
\begin{figure}[H]
\centering
\includegraphics[width=\textwidth]{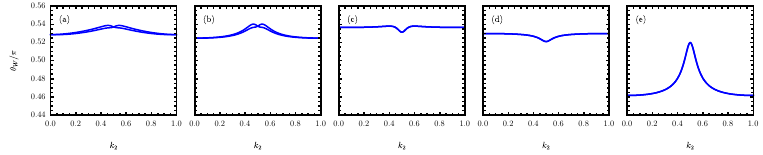}
\subfloat{\label{app:fig:fit_Wilson_2_WilsonX:a}}\subfloat{\label{app:fig:fit_Wilson_2_WilsonX:b}}\subfloat{\label{app:fig:fit_Wilson_2_WilsonX:c}}\subfloat{\label{app:fig:fit_Wilson_2_WilsonX:d}}\subfloat{\label{app:fig:fit_Wilson_2_WilsonX:e}}\caption{Wilson loops for the first set of conduction bands of twisted AA-stacked bilayer \ch{SnSe2} at $\theta = \SI{7.34}{\degree}$. The Wilson loop is computed along $\vec{b}_{M_1}$ We consider the full continuum model (a), the full first moir\'e harmonic model (b), the reduced first moir\'e harmonic model (c), the first moir\'e harmonic model with the zero-twist constraints imposed (d), and the reduced first moir\'e harmonic model with the zero-twist constraints imposed (e).}
\label{app:fig:fit_Wilson_2_WilsonX}
\end{figure}
\begin{figure}[H]
\centering
\includegraphics[width=\textwidth]{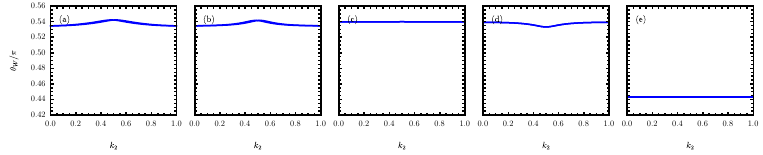}
\subfloat{\label{app:fig:fit_Wilson_3_WilsonX:a}}\subfloat{\label{app:fig:fit_Wilson_3_WilsonX:b}}\subfloat{\label{app:fig:fit_Wilson_3_WilsonX:c}}\subfloat{\label{app:fig:fit_Wilson_3_WilsonX:d}}\subfloat{\label{app:fig:fit_Wilson_3_WilsonX:e}}\caption{Wilson loops for the first set of conduction bands of twisted AA-stacked bilayer \ch{SnSe2} at $\theta = \SI{6.01}{\degree}$. The Wilson loop is computed along $\vec{b}_{M_1}$ We consider the full continuum model (a), the full first moir\'e harmonic model (b), the reduced first moir\'e harmonic model (c), the first moir\'e harmonic model with the zero-twist constraints imposed (d), and the reduced first moir\'e harmonic model with the zero-twist constraints imposed (e).}
\label{app:fig:fit_Wilson_3_WilsonX}
\end{figure}
\begin{figure}[H]
\centering
\includegraphics[width=\textwidth]{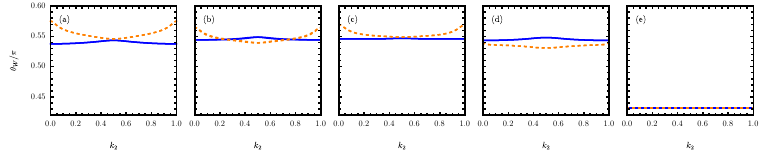}
\subfloat{\label{app:fig:fit_Wilson_4_WilsonX:a}}\subfloat{\label{app:fig:fit_Wilson_4_WilsonX:b}}\subfloat{\label{app:fig:fit_Wilson_4_WilsonX:c}}\subfloat{\label{app:fig:fit_Wilson_4_WilsonX:d}}\subfloat{\label{app:fig:fit_Wilson_4_WilsonX:e}}\caption{Wilson loops for the first two sets of conduction bands of twisted AA-stacked bilayer \ch{SnSe2} at $\theta = \SI{5.09}{\degree}$. The Wilson loop is computed along $\vec{b}_{M_1}$ We consider the full continuum model (a), the full first moir\'e harmonic model (b), the reduced first moir\'e harmonic model (c), the first moir\'e harmonic model with the zero-twist constraints imposed (d), and the reduced first moir\'e harmonic model with the zero-twist constraints imposed (e). The blue (dashed orange) lines correspond to the first (second) set of conduction bands.}
\label{app:fig:fit_Wilson_4_WilsonX}
\end{figure}
\begin{figure}[H]
\centering
\includegraphics[width=\textwidth]{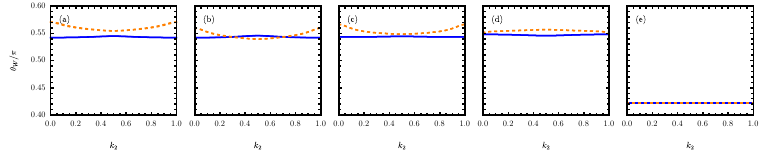}
\subfloat{\label{app:fig:fit_Wilson_5_WilsonX:a}}\subfloat{\label{app:fig:fit_Wilson_5_WilsonX:b}}\subfloat{\label{app:fig:fit_Wilson_5_WilsonX:c}}\subfloat{\label{app:fig:fit_Wilson_5_WilsonX:d}}\subfloat{\label{app:fig:fit_Wilson_5_WilsonX:e}}\caption{Wilson loops for the first two sets of conduction bands of twisted AA-stacked bilayer \ch{SnSe2} at $\theta = \SI{4.41}{\degree}$. The Wilson loop is computed along $\vec{b}_{M_1}$ We consider the full continuum model (a), the full first moir\'e harmonic model (b), the reduced first moir\'e harmonic model (c), the first moir\'e harmonic model with the zero-twist constraints imposed (d), and the reduced first moir\'e harmonic model with the zero-twist constraints imposed (e). The blue (dashed orange) lines correspond to the first (second) set of conduction bands.}
\label{app:fig:fit_Wilson_5_WilsonX}
\end{figure}
\begin{figure}[H]
\centering
\includegraphics[width=\textwidth]{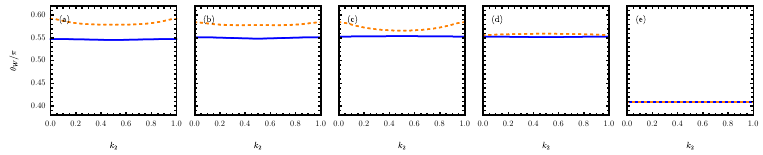}
\subfloat{\label{app:fig:fit_Wilson_6_WilsonX:a}}\subfloat{\label{app:fig:fit_Wilson_6_WilsonX:b}}\subfloat{\label{app:fig:fit_Wilson_6_WilsonX:c}}\subfloat{\label{app:fig:fit_Wilson_6_WilsonX:d}}\subfloat{\label{app:fig:fit_Wilson_6_WilsonX:e}}\caption{Wilson loops for the first two sets of conduction bands of twisted AA-stacked bilayer \ch{SnSe2} at $\theta = \SI{3.89}{\degree}$. The Wilson loop is computed along $\vec{b}_{M_1}$ We consider the full continuum model (a), the full first moir\'e harmonic model (b), the reduced first moir\'e harmonic model (c), the first moir\'e harmonic model with the zero-twist constraints imposed (d), and the reduced first moir\'e harmonic model with the zero-twist constraints imposed (e). The blue (dashed orange) lines correspond to the first (second) set of conduction bands.}
\label{app:fig:fit_Wilson_6_WilsonX}
\end{figure}
\begin{figure}[H]
\centering
\includegraphics[width=\textwidth]{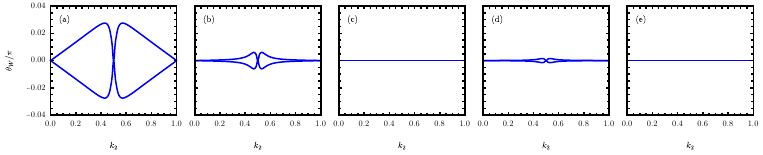}
\subfloat{\label{app:fig:fit_Wilson_7_WilsonX:a}}\subfloat{\label{app:fig:fit_Wilson_7_WilsonX:b}}\subfloat{\label{app:fig:fit_Wilson_7_WilsonX:c}}\subfloat{\label{app:fig:fit_Wilson_7_WilsonX:d}}\subfloat{\label{app:fig:fit_Wilson_7_WilsonX:e}}\caption{Wilson loops for the first set of conduction bands of twisted AB-stacked bilayer \ch{SnSe2} at $\theta = \SI{9.43}{\degree}$. The Wilson loop is computed along $\vec{b}_{M_1}$ We consider the full continuum model (a), the full first moir\'e harmonic model (b), the reduced first moir\'e harmonic model (c), the first moir\'e harmonic model with the zero-twist constraints imposed (d), and the reduced first moir\'e harmonic model with the zero-twist constraints imposed (e).}
\label{app:fig:fit_Wilson_7_WilsonX}
\end{figure}
\begin{figure}[H]
\centering
\includegraphics[width=\textwidth]{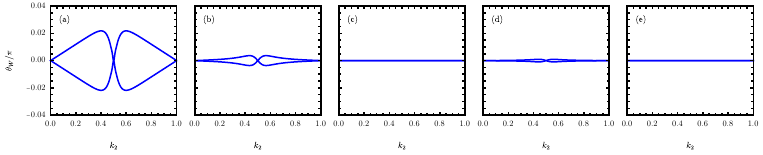}
\subfloat{\label{app:fig:fit_Wilson_8_WilsonX:a}}\subfloat{\label{app:fig:fit_Wilson_8_WilsonX:b}}\subfloat{\label{app:fig:fit_Wilson_8_WilsonX:c}}\subfloat{\label{app:fig:fit_Wilson_8_WilsonX:d}}\subfloat{\label{app:fig:fit_Wilson_8_WilsonX:e}}\caption{Wilson loops for the first set of conduction bands of twisted AB-stacked bilayer \ch{SnSe2} at $\theta = \SI{7.34}{\degree}$. The Wilson loop is computed along $\vec{b}_{M_1}$ We consider the full continuum model (a), the full first moir\'e harmonic model (b), the reduced first moir\'e harmonic model (c), the first moir\'e harmonic model with the zero-twist constraints imposed (d), and the reduced first moir\'e harmonic model with the zero-twist constraints imposed (e).}
\label{app:fig:fit_Wilson_8_WilsonX}
\end{figure}
\begin{figure}[H]
\centering
\includegraphics[width=\textwidth]{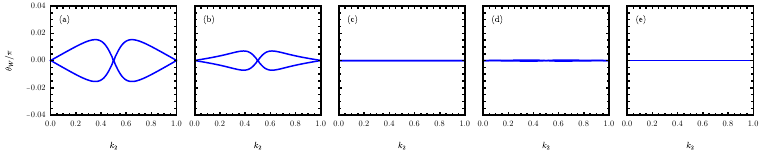}
\subfloat{\label{app:fig:fit_Wilson_9_WilsonX:a}}\subfloat{\label{app:fig:fit_Wilson_9_WilsonX:b}}\subfloat{\label{app:fig:fit_Wilson_9_WilsonX:c}}\subfloat{\label{app:fig:fit_Wilson_9_WilsonX:d}}\subfloat{\label{app:fig:fit_Wilson_9_WilsonX:e}}\caption{Wilson loops for the first set of conduction bands of twisted AB-stacked bilayer \ch{SnSe2} at $\theta = \SI{6.01}{\degree}$. The Wilson loop is computed along $\vec{b}_{M_1}$ We consider the full continuum model (a), the full first moir\'e harmonic model (b), the reduced first moir\'e harmonic model (c), the first moir\'e harmonic model with the zero-twist constraints imposed (d), and the reduced first moir\'e harmonic model with the zero-twist constraints imposed (e).}
\label{app:fig:fit_Wilson_9_WilsonX}
\end{figure}
\begin{figure}[H]
\centering
\includegraphics[width=\textwidth]{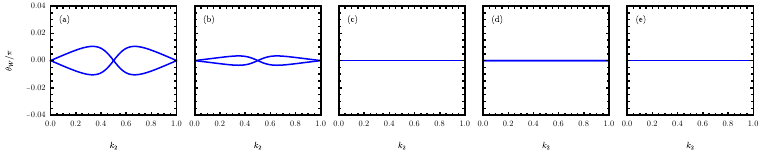}
\subfloat{\label{app:fig:fit_Wilson_10_WilsonX:a}}\subfloat{\label{app:fig:fit_Wilson_10_WilsonX:b}}\subfloat{\label{app:fig:fit_Wilson_10_WilsonX:c}}\subfloat{\label{app:fig:fit_Wilson_10_WilsonX:d}}\subfloat{\label{app:fig:fit_Wilson_10_WilsonX:e}}\caption{Wilson loops for the first set of conduction bands of twisted AB-stacked bilayer \ch{SnSe2} at $\theta = \SI{5.09}{\degree}$. The Wilson loop is computed along $\vec{b}_{M_1}$ We consider the full continuum model (a), the full first moir\'e harmonic model (b), the reduced first moir\'e harmonic model (c), the first moir\'e harmonic model with the zero-twist constraints imposed (d), and the reduced first moir\'e harmonic model with the zero-twist constraints imposed (e).}
\label{app:fig:fit_Wilson_10_WilsonX}
\end{figure}
\begin{figure}[H]
\centering
\includegraphics[width=\textwidth]{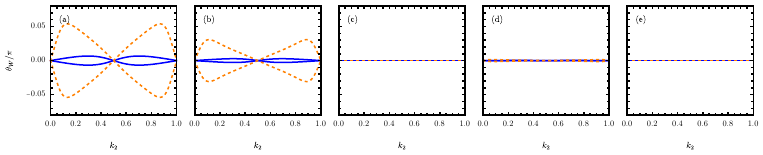}
\subfloat{\label{app:fig:fit_Wilson_11_WilsonX:a}}\subfloat{\label{app:fig:fit_Wilson_11_WilsonX:b}}\subfloat{\label{app:fig:fit_Wilson_11_WilsonX:c}}\subfloat{\label{app:fig:fit_Wilson_11_WilsonX:d}}\subfloat{\label{app:fig:fit_Wilson_11_WilsonX:e}}\caption{Wilson loops for the first two sets of conduction bands of twisted AB-stacked bilayer \ch{SnSe2} at $\theta = \SI{4.41}{\degree}$. The Wilson loop is computed along $\vec{b}_{M_1}$ We consider the full continuum model (a), the full first moir\'e harmonic model (b), the reduced first moir\'e harmonic model (c), the first moir\'e harmonic model with the zero-twist constraints imposed (d), and the reduced first moir\'e harmonic model with the zero-twist constraints imposed (e). The blue (dashed orange) lines correspond to the first (second) set of conduction bands.}
\label{app:fig:fit_Wilson_11_WilsonX}
\end{figure}
\begin{figure}[H]
\centering
\includegraphics[width=\textwidth]{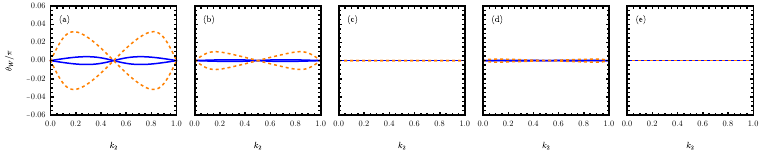}
\subfloat{\label{app:fig:fit_Wilson_12_WilsonX:a}}\subfloat{\label{app:fig:fit_Wilson_12_WilsonX:b}}\subfloat{\label{app:fig:fit_Wilson_12_WilsonX:c}}\subfloat{\label{app:fig:fit_Wilson_12_WilsonX:d}}\subfloat{\label{app:fig:fit_Wilson_12_WilsonX:e}}\caption{Wilson loops for the first two sets of conduction bands of twisted AB-stacked bilayer \ch{SnSe2} at $\theta = \SI{3.89}{\degree}$. The Wilson loop is computed along $\vec{b}_{M_1}$ We consider the full continuum model (a), the full first moir\'e harmonic model (b), the reduced first moir\'e harmonic model (c), the first moir\'e harmonic model with the zero-twist constraints imposed (d), and the reduced first moir\'e harmonic model with the zero-twist constraints imposed (e). The blue (dashed orange) lines correspond to the first (second) set of conduction bands.}
\label{app:fig:fit_Wilson_12_WilsonX}
\end{figure}
\begin{figure}[H]
\centering
\includegraphics[width=\textwidth]{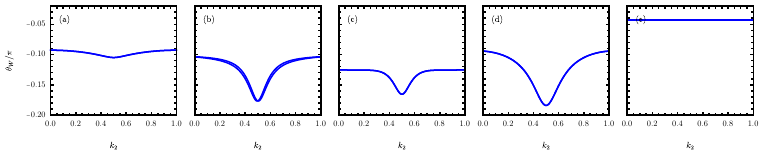}
\subfloat{\label{app:fig:fit_Wilson_13_WilsonX:a}}\subfloat{\label{app:fig:fit_Wilson_13_WilsonX:b}}\subfloat{\label{app:fig:fit_Wilson_13_WilsonX:c}}\subfloat{\label{app:fig:fit_Wilson_13_WilsonX:d}}\subfloat{\label{app:fig:fit_Wilson_13_WilsonX:e}}\caption{Wilson loops for the first set of conduction bands of twisted AA-stacked bilayer \ch{ZrS2} at $\theta = \SI{9.43}{\degree}$. The Wilson loop is computed along $\vec{b}_{M_1}$ We consider the full continuum model (a), the full first moir\'e harmonic model (b), the reduced first moir\'e harmonic model (c), the first moir\'e harmonic model with the zero-twist constraints imposed (d), and the reduced first moir\'e harmonic model with the zero-twist constraints imposed (e).}
\label{app:fig:fit_Wilson_13_WilsonX}
\end{figure}
\begin{figure}[H]
\centering
\includegraphics[width=\textwidth]{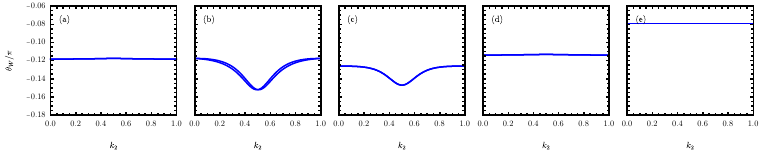}
\subfloat{\label{app:fig:fit_Wilson_14_WilsonX:a}}\subfloat{\label{app:fig:fit_Wilson_14_WilsonX:b}}\subfloat{\label{app:fig:fit_Wilson_14_WilsonX:c}}\subfloat{\label{app:fig:fit_Wilson_14_WilsonX:d}}\subfloat{\label{app:fig:fit_Wilson_14_WilsonX:e}}\caption{Wilson loops for the first set of conduction bands of twisted AA-stacked bilayer \ch{ZrS2} at $\theta = \SI{7.34}{\degree}$. The Wilson loop is computed along $\vec{b}_{M_1}$ We consider the full continuum model (a), the full first moir\'e harmonic model (b), the reduced first moir\'e harmonic model (c), the first moir\'e harmonic model with the zero-twist constraints imposed (d), and the reduced first moir\'e harmonic model with the zero-twist constraints imposed (e).}
\label{app:fig:fit_Wilson_14_WilsonX}
\end{figure}
\begin{figure}[H]
\centering
\includegraphics[width=\textwidth]{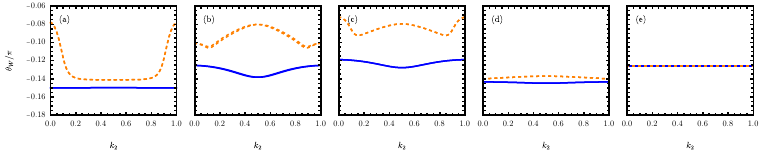}
\subfloat{\label{app:fig:fit_Wilson_15_WilsonX:a}}\subfloat{\label{app:fig:fit_Wilson_15_WilsonX:b}}\subfloat{\label{app:fig:fit_Wilson_15_WilsonX:c}}\subfloat{\label{app:fig:fit_Wilson_15_WilsonX:d}}\subfloat{\label{app:fig:fit_Wilson_15_WilsonX:e}}\caption{Wilson loops for the first two sets of conduction bands of twisted AA-stacked bilayer \ch{ZrS2} at $\theta = \SI{6.01}{\degree}$. The Wilson loop is computed along $\vec{b}_{M_1}$ We consider the full continuum model (a), the full first moir\'e harmonic model (b), the reduced first moir\'e harmonic model (c), the first moir\'e harmonic model with the zero-twist constraints imposed (d), and the reduced first moir\'e harmonic model with the zero-twist constraints imposed (e). The blue (dashed orange) lines correspond to the first (second) set of conduction bands.}
\label{app:fig:fit_Wilson_15_WilsonX}
\end{figure}
\begin{figure}[H]
\centering
\includegraphics[width=\textwidth]{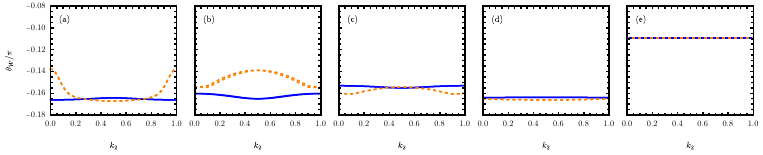}
\subfloat{\label{app:fig:fit_Wilson_16_WilsonX:a}}\subfloat{\label{app:fig:fit_Wilson_16_WilsonX:b}}\subfloat{\label{app:fig:fit_Wilson_16_WilsonX:c}}\subfloat{\label{app:fig:fit_Wilson_16_WilsonX:d}}\subfloat{\label{app:fig:fit_Wilson_16_WilsonX:e}}\caption{Wilson loops for the first two sets of conduction bands of twisted AA-stacked bilayer \ch{ZrS2} at $\theta = \SI{5.09}{\degree}$. The Wilson loop is computed along $\vec{b}_{M_1}$ We consider the full continuum model (a), the full first moir\'e harmonic model (b), the reduced first moir\'e harmonic model (c), the first moir\'e harmonic model with the zero-twist constraints imposed (d), and the reduced first moir\'e harmonic model with the zero-twist constraints imposed (e). The blue (dashed orange) lines correspond to the first (second) set of conduction bands.}
\label{app:fig:fit_Wilson_16_WilsonX}
\end{figure}
\begin{figure}[H]
\centering
\includegraphics[width=\textwidth]{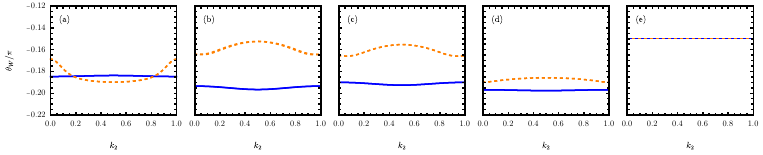}
\subfloat{\label{app:fig:fit_Wilson_17_WilsonX:a}}\subfloat{\label{app:fig:fit_Wilson_17_WilsonX:b}}\subfloat{\label{app:fig:fit_Wilson_17_WilsonX:c}}\subfloat{\label{app:fig:fit_Wilson_17_WilsonX:d}}\subfloat{\label{app:fig:fit_Wilson_17_WilsonX:e}}\caption{Wilson loops for the first two sets of conduction bands of twisted AA-stacked bilayer \ch{ZrS2} at $\theta = \SI{4.41}{\degree}$. The Wilson loop is computed along $\vec{b}_{M_1}$ We consider the full continuum model (a), the full first moir\'e harmonic model (b), the reduced first moir\'e harmonic model (c), the first moir\'e harmonic model with the zero-twist constraints imposed (d), and the reduced first moir\'e harmonic model with the zero-twist constraints imposed (e). The blue (dashed orange) lines correspond to the first (second) set of conduction bands.}
\label{app:fig:fit_Wilson_17_WilsonX}
\end{figure}
\begin{figure}[H]
\centering
\includegraphics[width=\textwidth]{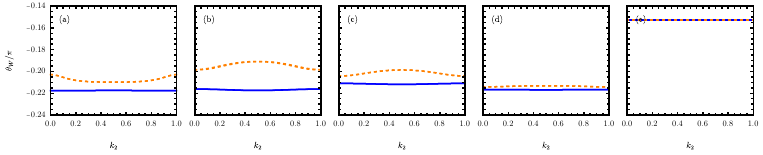}
\subfloat{\label{app:fig:fit_Wilson_18_WilsonX:a}}\subfloat{\label{app:fig:fit_Wilson_18_WilsonX:b}}\subfloat{\label{app:fig:fit_Wilson_18_WilsonX:c}}\subfloat{\label{app:fig:fit_Wilson_18_WilsonX:d}}\subfloat{\label{app:fig:fit_Wilson_18_WilsonX:e}}\caption{Wilson loops for the first two sets of conduction bands of twisted AA-stacked bilayer \ch{ZrS2} at $\theta = \SI{3.89}{\degree}$. The Wilson loop is computed along $\vec{b}_{M_1}$ We consider the full continuum model (a), the full first moir\'e harmonic model (b), the reduced first moir\'e harmonic model (c), the first moir\'e harmonic model with the zero-twist constraints imposed (d), and the reduced first moir\'e harmonic model with the zero-twist constraints imposed (e). The blue (dashed orange) lines correspond to the first (second) set of conduction bands.}
\label{app:fig:fit_Wilson_18_WilsonX}
\end{figure}
\begin{figure}[H]
\centering
\includegraphics[width=\textwidth]{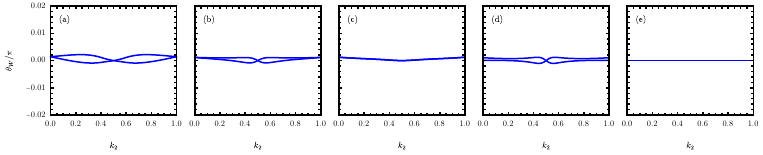}
\subfloat{\label{app:fig:fit_Wilson_19_WilsonX:a}}\subfloat{\label{app:fig:fit_Wilson_19_WilsonX:b}}\subfloat{\label{app:fig:fit_Wilson_19_WilsonX:c}}\subfloat{\label{app:fig:fit_Wilson_19_WilsonX:d}}\subfloat{\label{app:fig:fit_Wilson_19_WilsonX:e}}\caption{Wilson loops for the first set of conduction bands of twisted AB-stacked bilayer \ch{ZrS2} at $\theta = \SI{9.43}{\degree}$. The Wilson loop is computed along $\vec{b}_{M_1}$ We consider the full continuum model (a), the full first moir\'e harmonic model (b), the reduced first moir\'e harmonic model (c), the first moir\'e harmonic model with the zero-twist constraints imposed (d), and the reduced first moir\'e harmonic model with the zero-twist constraints imposed (e).}
\label{app:fig:fit_Wilson_19_WilsonX}
\end{figure}
\begin{figure}[H]
\centering
\includegraphics[width=\textwidth]{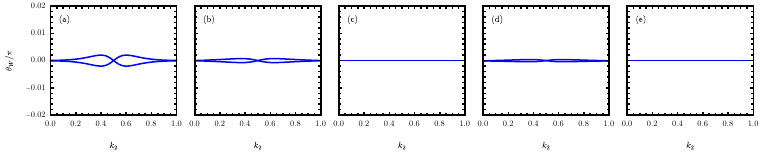}
\subfloat{\label{app:fig:fit_Wilson_20_WilsonX:a}}\subfloat{\label{app:fig:fit_Wilson_20_WilsonX:b}}\subfloat{\label{app:fig:fit_Wilson_20_WilsonX:c}}\subfloat{\label{app:fig:fit_Wilson_20_WilsonX:d}}\subfloat{\label{app:fig:fit_Wilson_20_WilsonX:e}}\caption{Wilson loops for the first set of conduction bands of twisted AB-stacked bilayer \ch{ZrS2} at $\theta = \SI{7.34}{\degree}$. The Wilson loop is computed along $\vec{b}_{M_1}$ We consider the full continuum model (a), the full first moir\'e harmonic model (b), the reduced first moir\'e harmonic model (c), the first moir\'e harmonic model with the zero-twist constraints imposed (d), and the reduced first moir\'e harmonic model with the zero-twist constraints imposed (e).}
\label{app:fig:fit_Wilson_20_WilsonX}
\end{figure}
\begin{figure}[H]
\centering
\includegraphics[width=\textwidth]{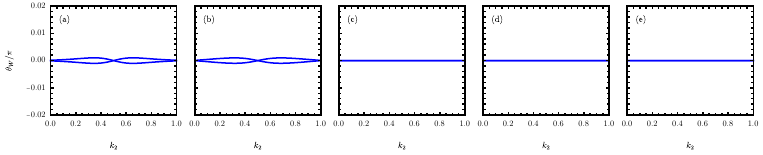}
\subfloat{\label{app:fig:fit_Wilson_21_WilsonX:a}}\subfloat{\label{app:fig:fit_Wilson_21_WilsonX:b}}\subfloat{\label{app:fig:fit_Wilson_21_WilsonX:c}}\subfloat{\label{app:fig:fit_Wilson_21_WilsonX:d}}\subfloat{\label{app:fig:fit_Wilson_21_WilsonX:e}}\caption{Wilson loops for the first set of conduction bands of twisted AB-stacked bilayer \ch{ZrS2} at $\theta = \SI{6.01}{\degree}$. The Wilson loop is computed along $\vec{b}_{M_1}$ We consider the full continuum model (a), the full first moir\'e harmonic model (b), the reduced first moir\'e harmonic model (c), the first moir\'e harmonic model with the zero-twist constraints imposed (d), and the reduced first moir\'e harmonic model with the zero-twist constraints imposed (e).}
\label{app:fig:fit_Wilson_21_WilsonX}
\end{figure}
\begin{figure}[H]
\centering
\includegraphics[width=\textwidth]{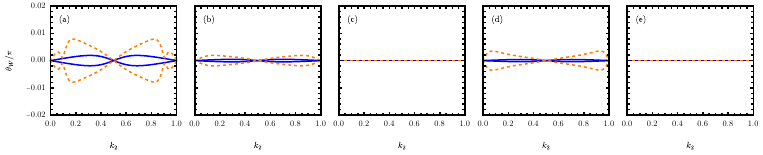}
\subfloat{\label{app:fig:fit_Wilson_22_WilsonX:a}}\subfloat{\label{app:fig:fit_Wilson_22_WilsonX:b}}\subfloat{\label{app:fig:fit_Wilson_22_WilsonX:c}}\subfloat{\label{app:fig:fit_Wilson_22_WilsonX:d}}\subfloat{\label{app:fig:fit_Wilson_22_WilsonX:e}}\caption{Wilson loops for the first two sets of conduction bands of twisted AB-stacked bilayer \ch{ZrS2} at $\theta = \SI{5.09}{\degree}$. The Wilson loop is computed along $\vec{b}_{M_1}$ We consider the full continuum model (a), the full first moir\'e harmonic model (b), the reduced first moir\'e harmonic model (c), the first moir\'e harmonic model with the zero-twist constraints imposed (d), and the reduced first moir\'e harmonic model with the zero-twist constraints imposed (e). The blue (dashed orange) lines correspond to the first (second) set of conduction bands.}
\label{app:fig:fit_Wilson_22_WilsonX}
\end{figure}
\begin{figure}[H]
\centering
\includegraphics[width=\textwidth]{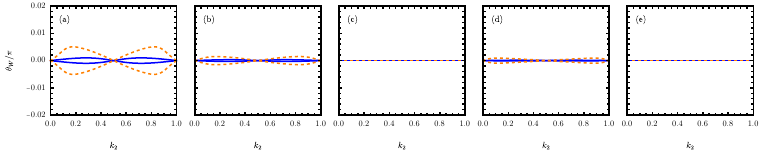}
\subfloat{\label{app:fig:fit_Wilson_23_WilsonX:a}}\subfloat{\label{app:fig:fit_Wilson_23_WilsonX:b}}\subfloat{\label{app:fig:fit_Wilson_23_WilsonX:c}}\subfloat{\label{app:fig:fit_Wilson_23_WilsonX:d}}\subfloat{\label{app:fig:fit_Wilson_23_WilsonX:e}}\caption{Wilson loops for the first two sets of conduction bands of twisted AB-stacked bilayer \ch{ZrS2} at $\theta = \SI{4.41}{\degree}$. The Wilson loop is computed along $\vec{b}_{M_1}$ We consider the full continuum model (a), the full first moir\'e harmonic model (b), the reduced first moir\'e harmonic model (c), the first moir\'e harmonic model with the zero-twist constraints imposed (d), and the reduced first moir\'e harmonic model with the zero-twist constraints imposed (e). The blue (dashed orange) lines correspond to the first (second) set of conduction bands.}
\label{app:fig:fit_Wilson_23_WilsonX}
\end{figure}
\begin{figure}[H]
\centering
\includegraphics[width=\textwidth]{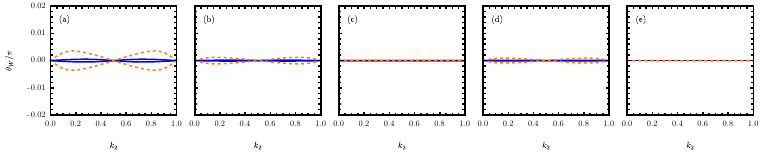}
\subfloat{\label{app:fig:fit_Wilson_24_WilsonX:a}}\subfloat{\label{app:fig:fit_Wilson_24_WilsonX:b}}\subfloat{\label{app:fig:fit_Wilson_24_WilsonX:c}}\subfloat{\label{app:fig:fit_Wilson_24_WilsonX:d}}\subfloat{\label{app:fig:fit_Wilson_24_WilsonX:e}}\caption{Wilson loops for the first two sets of conduction bands of twisted AB-stacked bilayer \ch{ZrS2} at $\theta = \SI{3.89}{\degree}$. The Wilson loop is computed along $\vec{b}_{M_1}$ We consider the full continuum model (a), the full first moir\'e harmonic model (b), the reduced first moir\'e harmonic model (c), the first moir\'e harmonic model with the zero-twist constraints imposed (d), and the reduced first moir\'e harmonic model with the zero-twist constraints imposed (e). The blue (dashed orange) lines correspond to the first (second) set of conduction bands.}
\label{app:fig:fit_Wilson_24_WilsonX}
\end{figure} \subsubsection{Wilson loops of the first gapped conduction bands along $\vec{b}_{M_2}$}\label{app:sec:fitted_models:plots:wilsons_2}
\begin{figure}[H]
\centering
\includegraphics[width=\textwidth]{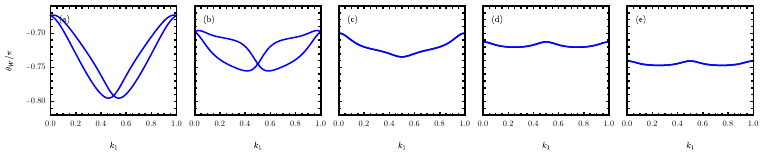}
\subfloat{\label{app:fig:fit_Wilson_1_WilsonY:a}}\subfloat{\label{app:fig:fit_Wilson_1_WilsonY:b}}\subfloat{\label{app:fig:fit_Wilson_1_WilsonY:c}}\subfloat{\label{app:fig:fit_Wilson_1_WilsonY:d}}\subfloat{\label{app:fig:fit_Wilson_1_WilsonY:e}}\caption{Wilson loops for the first set of conduction bands of twisted AA-stacked bilayer \ch{SnSe2} at $\theta = \SI{9.43}{\degree}$. The Wilson loop is computed along $\vec{b}_{M_2}$ We consider the full continuum model (a), the full first moir\'e harmonic model (b), the reduced first moir\'e harmonic model (c), the first moir\'e harmonic model with the zero-twist constraints imposed (d), and the reduced first moir\'e harmonic model with the zero-twist constraints imposed (e).}
\label{app:fig:fit_Wilson_1_WilsonY}
\end{figure}
\begin{figure}[H]
\centering
\includegraphics[width=\textwidth]{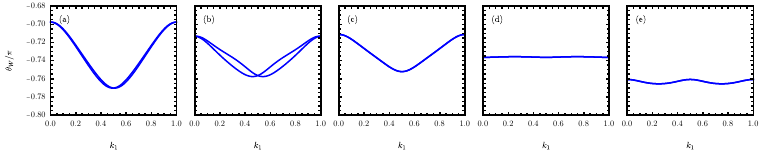}
\subfloat{\label{app:fig:fit_Wilson_2_WilsonY:a}}\subfloat{\label{app:fig:fit_Wilson_2_WilsonY:b}}\subfloat{\label{app:fig:fit_Wilson_2_WilsonY:c}}\subfloat{\label{app:fig:fit_Wilson_2_WilsonY:d}}\subfloat{\label{app:fig:fit_Wilson_2_WilsonY:e}}\caption{Wilson loops for the first set of conduction bands of twisted AA-stacked bilayer \ch{SnSe2} at $\theta = \SI{7.34}{\degree}$. The Wilson loop is computed along $\vec{b}_{M_2}$ We consider the full continuum model (a), the full first moir\'e harmonic model (b), the reduced first moir\'e harmonic model (c), the first moir\'e harmonic model with the zero-twist constraints imposed (d), and the reduced first moir\'e harmonic model with the zero-twist constraints imposed (e).}
\label{app:fig:fit_Wilson_2_WilsonY}
\end{figure}
\begin{figure}[H]
\centering
\includegraphics[width=\textwidth]{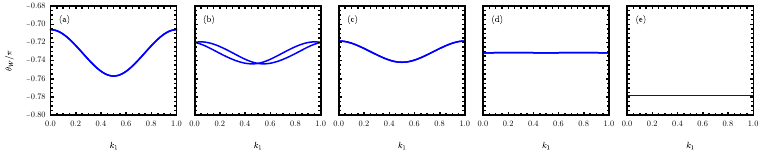}
\subfloat{\label{app:fig:fit_Wilson_3_WilsonY:a}}\subfloat{\label{app:fig:fit_Wilson_3_WilsonY:b}}\subfloat{\label{app:fig:fit_Wilson_3_WilsonY:c}}\subfloat{\label{app:fig:fit_Wilson_3_WilsonY:d}}\subfloat{\label{app:fig:fit_Wilson_3_WilsonY:e}}\caption{Wilson loops for the first set of conduction bands of twisted AA-stacked bilayer \ch{SnSe2} at $\theta = \SI{6.01}{\degree}$. The Wilson loop is computed along $\vec{b}_{M_2}$ We consider the full continuum model (a), the full first moir\'e harmonic model (b), the reduced first moir\'e harmonic model (c), the first moir\'e harmonic model with the zero-twist constraints imposed (d), and the reduced first moir\'e harmonic model with the zero-twist constraints imposed (e).}
\label{app:fig:fit_Wilson_3_WilsonY}
\end{figure}
\begin{figure}[H]
\centering
\includegraphics[width=\textwidth]{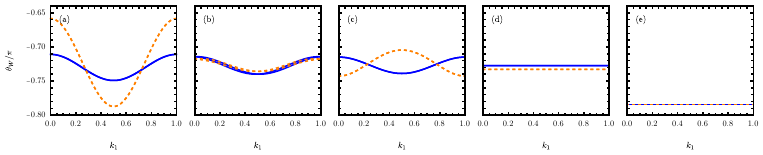}
\subfloat{\label{app:fig:fit_Wilson_4_WilsonY:a}}\subfloat{\label{app:fig:fit_Wilson_4_WilsonY:b}}\subfloat{\label{app:fig:fit_Wilson_4_WilsonY:c}}\subfloat{\label{app:fig:fit_Wilson_4_WilsonY:d}}\subfloat{\label{app:fig:fit_Wilson_4_WilsonY:e}}\caption{Wilson loops for the first two sets of conduction bands of twisted AA-stacked bilayer \ch{SnSe2} at $\theta = \SI{5.09}{\degree}$. The Wilson loop is computed along $\vec{b}_{M_2}$ We consider the full continuum model (a), the full first moir\'e harmonic model (b), the reduced first moir\'e harmonic model (c), the first moir\'e harmonic model with the zero-twist constraints imposed (d), and the reduced first moir\'e harmonic model with the zero-twist constraints imposed (e). The blue (dashed orange) lines correspond to the first (second) set of conduction bands.}
\label{app:fig:fit_Wilson_4_WilsonY}
\end{figure}
\begin{figure}[H]
\centering
\includegraphics[width=\textwidth]{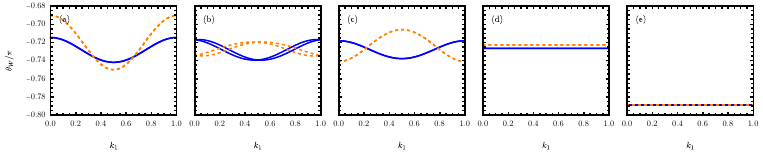}
\subfloat{\label{app:fig:fit_Wilson_5_WilsonY:a}}\subfloat{\label{app:fig:fit_Wilson_5_WilsonY:b}}\subfloat{\label{app:fig:fit_Wilson_5_WilsonY:c}}\subfloat{\label{app:fig:fit_Wilson_5_WilsonY:d}}\subfloat{\label{app:fig:fit_Wilson_5_WilsonY:e}}\caption{Wilson loops for the first two sets of conduction bands of twisted AA-stacked bilayer \ch{SnSe2} at $\theta = \SI{4.41}{\degree}$. The Wilson loop is computed along $\vec{b}_{M_2}$ We consider the full continuum model (a), the full first moir\'e harmonic model (b), the reduced first moir\'e harmonic model (c), the first moir\'e harmonic model with the zero-twist constraints imposed (d), and the reduced first moir\'e harmonic model with the zero-twist constraints imposed (e). The blue (dashed orange) lines correspond to the first (second) set of conduction bands.}
\label{app:fig:fit_Wilson_5_WilsonY}
\end{figure}
\begin{figure}[H]
\centering
\includegraphics[width=\textwidth]{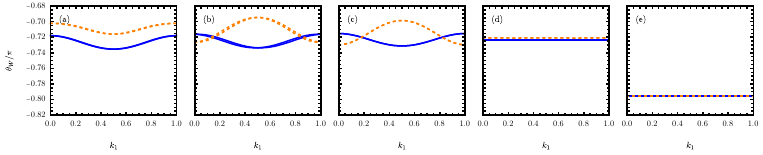}
\subfloat{\label{app:fig:fit_Wilson_6_WilsonY:a}}\subfloat{\label{app:fig:fit_Wilson_6_WilsonY:b}}\subfloat{\label{app:fig:fit_Wilson_6_WilsonY:c}}\subfloat{\label{app:fig:fit_Wilson_6_WilsonY:d}}\subfloat{\label{app:fig:fit_Wilson_6_WilsonY:e}}\caption{Wilson loops for the first two sets of conduction bands of twisted AA-stacked bilayer \ch{SnSe2} at $\theta = \SI{3.89}{\degree}$. The Wilson loop is computed along $\vec{b}_{M_2}$ We consider the full continuum model (a), the full first moir\'e harmonic model (b), the reduced first moir\'e harmonic model (c), the first moir\'e harmonic model with the zero-twist constraints imposed (d), and the reduced first moir\'e harmonic model with the zero-twist constraints imposed (e). The blue (dashed orange) lines correspond to the first (second) set of conduction bands.}
\label{app:fig:fit_Wilson_6_WilsonY}
\end{figure}
\begin{figure}[H]
\centering
\includegraphics[width=\textwidth]{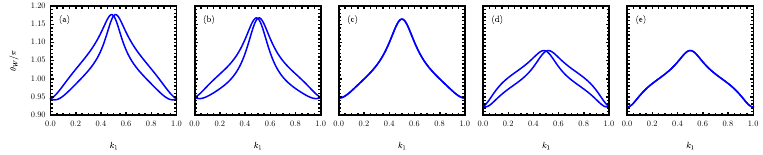}
\subfloat{\label{app:fig:fit_Wilson_7_WilsonY:a}}\subfloat{\label{app:fig:fit_Wilson_7_WilsonY:b}}\subfloat{\label{app:fig:fit_Wilson_7_WilsonY:c}}\subfloat{\label{app:fig:fit_Wilson_7_WilsonY:d}}\subfloat{\label{app:fig:fit_Wilson_7_WilsonY:e}}\caption{Wilson loops for the first set of conduction bands of twisted AB-stacked bilayer \ch{SnSe2} at $\theta = \SI{9.43}{\degree}$. The Wilson loop is computed along $\vec{b}_{M_2}$ We consider the full continuum model (a), the full first moir\'e harmonic model (b), the reduced first moir\'e harmonic model (c), the first moir\'e harmonic model with the zero-twist constraints imposed (d), and the reduced first moir\'e harmonic model with the zero-twist constraints imposed (e).}
\label{app:fig:fit_Wilson_7_WilsonY}
\end{figure}
\begin{figure}[H]
\centering
\includegraphics[width=\textwidth]{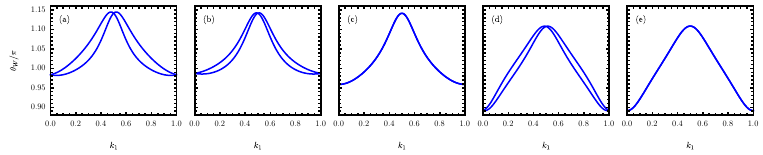}
\subfloat{\label{app:fig:fit_Wilson_8_WilsonY:a}}\subfloat{\label{app:fig:fit_Wilson_8_WilsonY:b}}\subfloat{\label{app:fig:fit_Wilson_8_WilsonY:c}}\subfloat{\label{app:fig:fit_Wilson_8_WilsonY:d}}\subfloat{\label{app:fig:fit_Wilson_8_WilsonY:e}}\caption{Wilson loops for the first set of conduction bands of twisted AB-stacked bilayer \ch{SnSe2} at $\theta = \SI{7.34}{\degree}$. The Wilson loop is computed along $\vec{b}_{M_2}$ We consider the full continuum model (a), the full first moir\'e harmonic model (b), the reduced first moir\'e harmonic model (c), the first moir\'e harmonic model with the zero-twist constraints imposed (d), and the reduced first moir\'e harmonic model with the zero-twist constraints imposed (e).}
\label{app:fig:fit_Wilson_8_WilsonY}
\end{figure}
\begin{figure}[H]
\centering
\includegraphics[width=\textwidth]{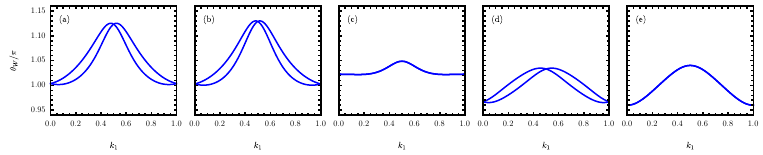}
\subfloat{\label{app:fig:fit_Wilson_9_WilsonY:a}}\subfloat{\label{app:fig:fit_Wilson_9_WilsonY:b}}\subfloat{\label{app:fig:fit_Wilson_9_WilsonY:c}}\subfloat{\label{app:fig:fit_Wilson_9_WilsonY:d}}\subfloat{\label{app:fig:fit_Wilson_9_WilsonY:e}}\caption{Wilson loops for the first set of conduction bands of twisted AB-stacked bilayer \ch{SnSe2} at $\theta = \SI{6.01}{\degree}$. The Wilson loop is computed along $\vec{b}_{M_2}$ We consider the full continuum model (a), the full first moir\'e harmonic model (b), the reduced first moir\'e harmonic model (c), the first moir\'e harmonic model with the zero-twist constraints imposed (d), and the reduced first moir\'e harmonic model with the zero-twist constraints imposed (e).}
\label{app:fig:fit_Wilson_9_WilsonY}
\end{figure}
\begin{figure}[H]
\centering
\includegraphics[width=\textwidth]{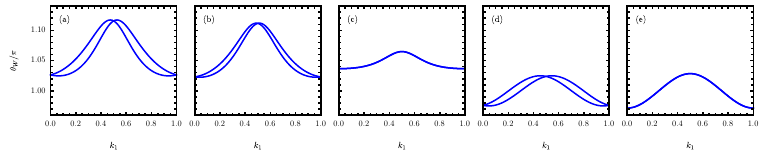}
\subfloat{\label{app:fig:fit_Wilson_10_WilsonY:a}}\subfloat{\label{app:fig:fit_Wilson_10_WilsonY:b}}\subfloat{\label{app:fig:fit_Wilson_10_WilsonY:c}}\subfloat{\label{app:fig:fit_Wilson_10_WilsonY:d}}\subfloat{\label{app:fig:fit_Wilson_10_WilsonY:e}}\caption{Wilson loops for the first set of conduction bands of twisted AB-stacked bilayer \ch{SnSe2} at $\theta = \SI{5.09}{\degree}$. The Wilson loop is computed along $\vec{b}_{M_2}$ We consider the full continuum model (a), the full first moir\'e harmonic model (b), the reduced first moir\'e harmonic model (c), the first moir\'e harmonic model with the zero-twist constraints imposed (d), and the reduced first moir\'e harmonic model with the zero-twist constraints imposed (e).}
\label{app:fig:fit_Wilson_10_WilsonY}
\end{figure}
\begin{figure}[H]
\centering
\includegraphics[width=\textwidth]{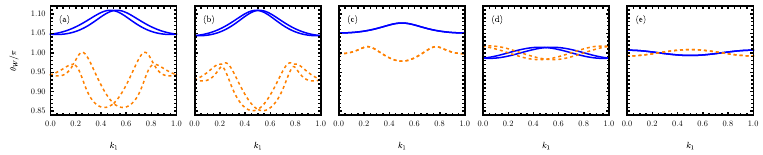}
\subfloat{\label{app:fig:fit_Wilson_11_WilsonY:a}}\subfloat{\label{app:fig:fit_Wilson_11_WilsonY:b}}\subfloat{\label{app:fig:fit_Wilson_11_WilsonY:c}}\subfloat{\label{app:fig:fit_Wilson_11_WilsonY:d}}\subfloat{\label{app:fig:fit_Wilson_11_WilsonY:e}}\caption{Wilson loops for the first two sets of conduction bands of twisted AB-stacked bilayer \ch{SnSe2} at $\theta = \SI{4.41}{\degree}$. The Wilson loop is computed along $\vec{b}_{M_2}$ We consider the full continuum model (a), the full first moir\'e harmonic model (b), the reduced first moir\'e harmonic model (c), the first moir\'e harmonic model with the zero-twist constraints imposed (d), and the reduced first moir\'e harmonic model with the zero-twist constraints imposed (e). The blue (dashed orange) lines correspond to the first (second) set of conduction bands.}
\label{app:fig:fit_Wilson_11_WilsonY}
\end{figure}
\begin{figure}[H]
\centering
\includegraphics[width=\textwidth]{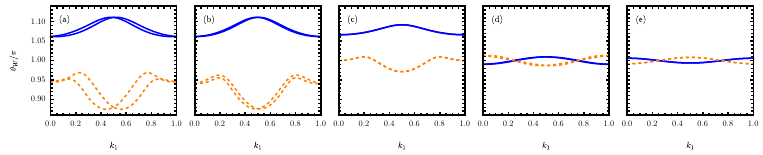}
\subfloat{\label{app:fig:fit_Wilson_12_WilsonY:a}}\subfloat{\label{app:fig:fit_Wilson_12_WilsonY:b}}\subfloat{\label{app:fig:fit_Wilson_12_WilsonY:c}}\subfloat{\label{app:fig:fit_Wilson_12_WilsonY:d}}\subfloat{\label{app:fig:fit_Wilson_12_WilsonY:e}}\caption{Wilson loops for the first two sets of conduction bands of twisted AB-stacked bilayer \ch{SnSe2} at $\theta = \SI{3.89}{\degree}$. The Wilson loop is computed along $\vec{b}_{M_2}$ We consider the full continuum model (a), the full first moir\'e harmonic model (b), the reduced first moir\'e harmonic model (c), the first moir\'e harmonic model with the zero-twist constraints imposed (d), and the reduced first moir\'e harmonic model with the zero-twist constraints imposed (e). The blue (dashed orange) lines correspond to the first (second) set of conduction bands.}
\label{app:fig:fit_Wilson_12_WilsonY}
\end{figure}
\begin{figure}[H]
\centering
\includegraphics[width=\textwidth]{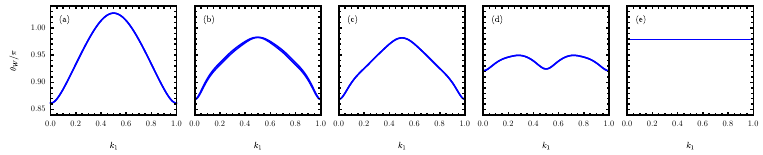}
\subfloat{\label{app:fig:fit_Wilson_13_WilsonY:a}}\subfloat{\label{app:fig:fit_Wilson_13_WilsonY:b}}\subfloat{\label{app:fig:fit_Wilson_13_WilsonY:c}}\subfloat{\label{app:fig:fit_Wilson_13_WilsonY:d}}\subfloat{\label{app:fig:fit_Wilson_13_WilsonY:e}}\caption{Wilson loops for the first set of conduction bands of twisted AA-stacked bilayer \ch{ZrS2} at $\theta = \SI{9.43}{\degree}$. The Wilson loop is computed along $\vec{b}_{M_2}$ We consider the full continuum model (a), the full first moir\'e harmonic model (b), the reduced first moir\'e harmonic model (c), the first moir\'e harmonic model with the zero-twist constraints imposed (d), and the reduced first moir\'e harmonic model with the zero-twist constraints imposed (e).}
\label{app:fig:fit_Wilson_13_WilsonY}
\end{figure}
\begin{figure}[H]
\centering
\includegraphics[width=\textwidth]{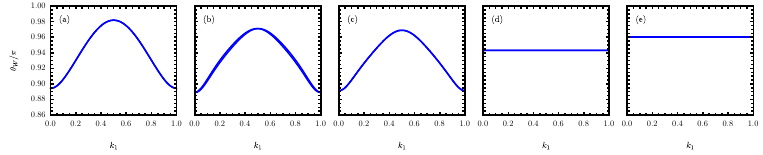}
\subfloat{\label{app:fig:fit_Wilson_14_WilsonY:a}}\subfloat{\label{app:fig:fit_Wilson_14_WilsonY:b}}\subfloat{\label{app:fig:fit_Wilson_14_WilsonY:c}}\subfloat{\label{app:fig:fit_Wilson_14_WilsonY:d}}\subfloat{\label{app:fig:fit_Wilson_14_WilsonY:e}}\caption{Wilson loops for the first set of conduction bands of twisted AA-stacked bilayer \ch{ZrS2} at $\theta = \SI{7.34}{\degree}$. The Wilson loop is computed along $\vec{b}_{M_2}$ We consider the full continuum model (a), the full first moir\'e harmonic model (b), the reduced first moir\'e harmonic model (c), the first moir\'e harmonic model with the zero-twist constraints imposed (d), and the reduced first moir\'e harmonic model with the zero-twist constraints imposed (e).}
\label{app:fig:fit_Wilson_14_WilsonY}
\end{figure}
\begin{figure}[H]
\centering
\includegraphics[width=\textwidth]{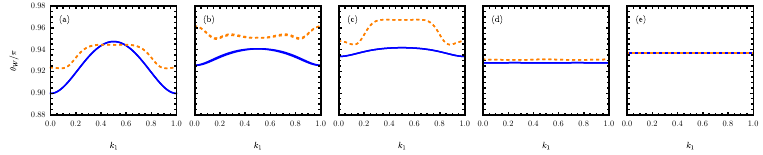}
\subfloat{\label{app:fig:fit_Wilson_15_WilsonY:a}}\subfloat{\label{app:fig:fit_Wilson_15_WilsonY:b}}\subfloat{\label{app:fig:fit_Wilson_15_WilsonY:c}}\subfloat{\label{app:fig:fit_Wilson_15_WilsonY:d}}\subfloat{\label{app:fig:fit_Wilson_15_WilsonY:e}}\caption{Wilson loops for the first two sets of conduction bands of twisted AA-stacked bilayer \ch{ZrS2} at $\theta = \SI{6.01}{\degree}$. The Wilson loop is computed along $\vec{b}_{M_2}$ We consider the full continuum model (a), the full first moir\'e harmonic model (b), the reduced first moir\'e harmonic model (c), the first moir\'e harmonic model with the zero-twist constraints imposed (d), and the reduced first moir\'e harmonic model with the zero-twist constraints imposed (e). The blue (dashed orange) lines correspond to the first (second) set of conduction bands.}
\label{app:fig:fit_Wilson_15_WilsonY}
\end{figure}
\begin{figure}[H]
\centering
\includegraphics[width=\textwidth]{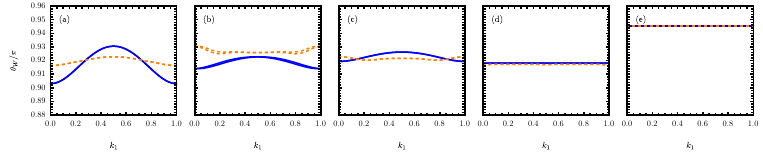}
\subfloat{\label{app:fig:fit_Wilson_16_WilsonY:a}}\subfloat{\label{app:fig:fit_Wilson_16_WilsonY:b}}\subfloat{\label{app:fig:fit_Wilson_16_WilsonY:c}}\subfloat{\label{app:fig:fit_Wilson_16_WilsonY:d}}\subfloat{\label{app:fig:fit_Wilson_16_WilsonY:e}}\caption{Wilson loops for the first two sets of conduction bands of twisted AA-stacked bilayer \ch{ZrS2} at $\theta = \SI{5.09}{\degree}$. The Wilson loop is computed along $\vec{b}_{M_2}$ We consider the full continuum model (a), the full first moir\'e harmonic model (b), the reduced first moir\'e harmonic model (c), the first moir\'e harmonic model with the zero-twist constraints imposed (d), and the reduced first moir\'e harmonic model with the zero-twist constraints imposed (e). The blue (dashed orange) lines correspond to the first (second) set of conduction bands.}
\label{app:fig:fit_Wilson_16_WilsonY}
\end{figure}
\begin{figure}[H]
\centering
\includegraphics[width=\textwidth]{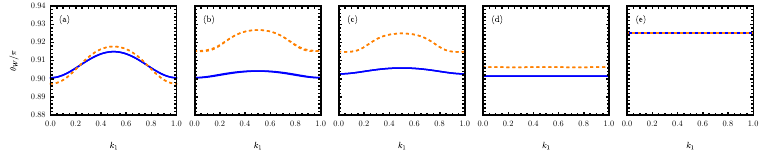}
\subfloat{\label{app:fig:fit_Wilson_17_WilsonY:a}}\subfloat{\label{app:fig:fit_Wilson_17_WilsonY:b}}\subfloat{\label{app:fig:fit_Wilson_17_WilsonY:c}}\subfloat{\label{app:fig:fit_Wilson_17_WilsonY:d}}\subfloat{\label{app:fig:fit_Wilson_17_WilsonY:e}}\caption{Wilson loops for the first two sets of conduction bands of twisted AA-stacked bilayer \ch{ZrS2} at $\theta = \SI{4.41}{\degree}$. The Wilson loop is computed along $\vec{b}_{M_2}$ We consider the full continuum model (a), the full first moir\'e harmonic model (b), the reduced first moir\'e harmonic model (c), the first moir\'e harmonic model with the zero-twist constraints imposed (d), and the reduced first moir\'e harmonic model with the zero-twist constraints imposed (e). The blue (dashed orange) lines correspond to the first (second) set of conduction bands.}
\label{app:fig:fit_Wilson_17_WilsonY}
\end{figure}
\begin{figure}[H]
\centering
\includegraphics[width=\textwidth]{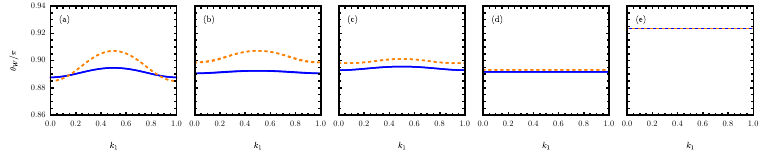}
\subfloat{\label{app:fig:fit_Wilson_18_WilsonY:a}}\subfloat{\label{app:fig:fit_Wilson_18_WilsonY:b}}\subfloat{\label{app:fig:fit_Wilson_18_WilsonY:c}}\subfloat{\label{app:fig:fit_Wilson_18_WilsonY:d}}\subfloat{\label{app:fig:fit_Wilson_18_WilsonY:e}}\caption{Wilson loops for the first two sets of conduction bands of twisted AA-stacked bilayer \ch{ZrS2} at $\theta = \SI{3.89}{\degree}$. The Wilson loop is computed along $\vec{b}_{M_2}$ We consider the full continuum model (a), the full first moir\'e harmonic model (b), the reduced first moir\'e harmonic model (c), the first moir\'e harmonic model with the zero-twist constraints imposed (d), and the reduced first moir\'e harmonic model with the zero-twist constraints imposed (e). The blue (dashed orange) lines correspond to the first (second) set of conduction bands.}
\label{app:fig:fit_Wilson_18_WilsonY}
\end{figure}
\begin{figure}[H]
\centering
\includegraphics[width=\textwidth]{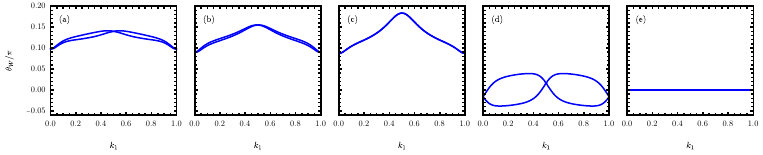}
\subfloat{\label{app:fig:fit_Wilson_19_WilsonY:a}}\subfloat{\label{app:fig:fit_Wilson_19_WilsonY:b}}\subfloat{\label{app:fig:fit_Wilson_19_WilsonY:c}}\subfloat{\label{app:fig:fit_Wilson_19_WilsonY:d}}\subfloat{\label{app:fig:fit_Wilson_19_WilsonY:e}}\caption{Wilson loops for the first set of conduction bands of twisted AB-stacked bilayer \ch{ZrS2} at $\theta = \SI{9.43}{\degree}$. The Wilson loop is computed along $\vec{b}_{M_2}$ We consider the full continuum model (a), the full first moir\'e harmonic model (b), the reduced first moir\'e harmonic model (c), the first moir\'e harmonic model with the zero-twist constraints imposed (d), and the reduced first moir\'e harmonic model with the zero-twist constraints imposed (e).}
\label{app:fig:fit_Wilson_19_WilsonY}
\end{figure}
\begin{figure}[H]
\centering
\includegraphics[width=\textwidth]{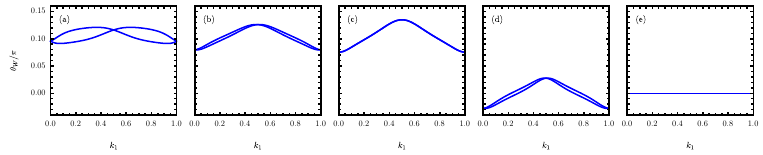}
\subfloat{\label{app:fig:fit_Wilson_20_WilsonY:a}}\subfloat{\label{app:fig:fit_Wilson_20_WilsonY:b}}\subfloat{\label{app:fig:fit_Wilson_20_WilsonY:c}}\subfloat{\label{app:fig:fit_Wilson_20_WilsonY:d}}\subfloat{\label{app:fig:fit_Wilson_20_WilsonY:e}}\caption{Wilson loops for the first set of conduction bands of twisted AB-stacked bilayer \ch{ZrS2} at $\theta = \SI{7.34}{\degree}$. The Wilson loop is computed along $\vec{b}_{M_2}$ We consider the full continuum model (a), the full first moir\'e harmonic model (b), the reduced first moir\'e harmonic model (c), the first moir\'e harmonic model with the zero-twist constraints imposed (d), and the reduced first moir\'e harmonic model with the zero-twist constraints imposed (e).}
\label{app:fig:fit_Wilson_20_WilsonY}
\end{figure}
\begin{figure}[H]
\centering
\includegraphics[width=\textwidth]{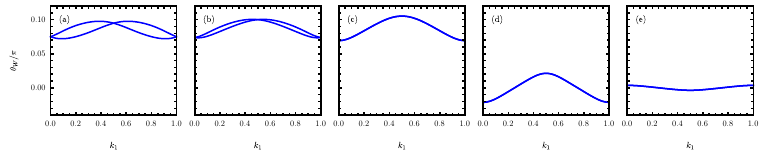}
\subfloat{\label{app:fig:fit_Wilson_21_WilsonY:a}}\subfloat{\label{app:fig:fit_Wilson_21_WilsonY:b}}\subfloat{\label{app:fig:fit_Wilson_21_WilsonY:c}}\subfloat{\label{app:fig:fit_Wilson_21_WilsonY:d}}\subfloat{\label{app:fig:fit_Wilson_21_WilsonY:e}}\caption{Wilson loops for the first set of conduction bands of twisted AB-stacked bilayer \ch{ZrS2} at $\theta = \SI{6.01}{\degree}$. The Wilson loop is computed along $\vec{b}_{M_2}$ We consider the full continuum model (a), the full first moir\'e harmonic model (b), the reduced first moir\'e harmonic model (c), the first moir\'e harmonic model with the zero-twist constraints imposed (d), and the reduced first moir\'e harmonic model with the zero-twist constraints imposed (e).}
\label{app:fig:fit_Wilson_21_WilsonY}
\end{figure}
\begin{figure}[H]
\centering
\includegraphics[width=\textwidth]{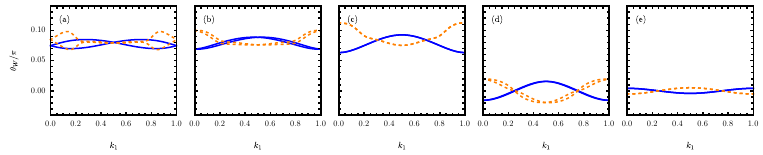}
\subfloat{\label{app:fig:fit_Wilson_22_WilsonY:a}}\subfloat{\label{app:fig:fit_Wilson_22_WilsonY:b}}\subfloat{\label{app:fig:fit_Wilson_22_WilsonY:c}}\subfloat{\label{app:fig:fit_Wilson_22_WilsonY:d}}\subfloat{\label{app:fig:fit_Wilson_22_WilsonY:e}}\caption{Wilson loops for the first two sets of conduction bands of twisted AB-stacked bilayer \ch{ZrS2} at $\theta = \SI{5.09}{\degree}$. The Wilson loop is computed along $\vec{b}_{M_2}$ We consider the full continuum model (a), the full first moir\'e harmonic model (b), the reduced first moir\'e harmonic model (c), the first moir\'e harmonic model with the zero-twist constraints imposed (d), and the reduced first moir\'e harmonic model with the zero-twist constraints imposed (e). The blue (dashed orange) lines correspond to the first (second) set of conduction bands.}
\label{app:fig:fit_Wilson_22_WilsonY}
\end{figure}
\begin{figure}[H]
\centering
\includegraphics[width=\textwidth]{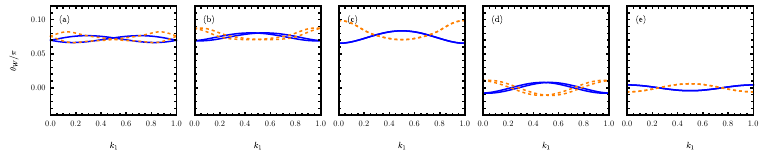}
\subfloat{\label{app:fig:fit_Wilson_23_WilsonY:a}}\subfloat{\label{app:fig:fit_Wilson_23_WilsonY:b}}\subfloat{\label{app:fig:fit_Wilson_23_WilsonY:c}}\subfloat{\label{app:fig:fit_Wilson_23_WilsonY:d}}\subfloat{\label{app:fig:fit_Wilson_23_WilsonY:e}}\caption{Wilson loops for the first two sets of conduction bands of twisted AB-stacked bilayer \ch{ZrS2} at $\theta = \SI{4.41}{\degree}$. The Wilson loop is computed along $\vec{b}_{M_2}$ We consider the full continuum model (a), the full first moir\'e harmonic model (b), the reduced first moir\'e harmonic model (c), the first moir\'e harmonic model with the zero-twist constraints imposed (d), and the reduced first moir\'e harmonic model with the zero-twist constraints imposed (e). The blue (dashed orange) lines correspond to the first (second) set of conduction bands.}
\label{app:fig:fit_Wilson_23_WilsonY}
\end{figure}
\begin{figure}[H]
\centering
\includegraphics[width=\textwidth]{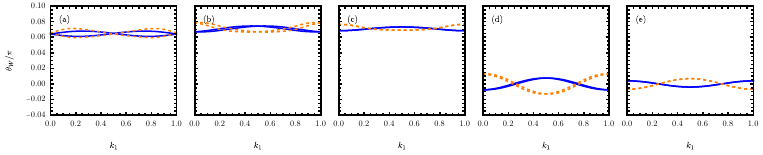}
\subfloat{\label{app:fig:fit_Wilson_24_WilsonY:a}}\subfloat{\label{app:fig:fit_Wilson_24_WilsonY:b}}\subfloat{\label{app:fig:fit_Wilson_24_WilsonY:c}}\subfloat{\label{app:fig:fit_Wilson_24_WilsonY:d}}\subfloat{\label{app:fig:fit_Wilson_24_WilsonY:e}}\caption{Wilson loops for the first two sets of conduction bands of twisted AB-stacked bilayer \ch{ZrS2} at $\theta = \SI{3.89}{\degree}$. The Wilson loop is computed along $\vec{b}_{M_2}$ We consider the full continuum model (a), the full first moir\'e harmonic model (b), the reduced first moir\'e harmonic model (c), the first moir\'e harmonic model with the zero-twist constraints imposed (d), and the reduced first moir\'e harmonic model with the zero-twist constraints imposed (e). The blue (dashed orange) lines correspond to the first (second) set of conduction bands.}
\label{app:fig:fit_Wilson_24_WilsonY}
\end{figure} 

\end{document}